\documentclass{book}

\usepackage{epsfig,verbatim,cite,amsmath,amssymb,subfigure,color}
\usepackage{enumitem}

\renewcommand{\vec}[1]{\mbox{${\bf #1}$}}
\renewcommand{\v}[1]{{{\bf #1}}}
\newcommand{\dyad}[1]{\mbox{$\overline{ \v{#1} }$}}

\def\dyad #1{\v{\overline{#1}}}

\def\dyad #1{\v{\overline{#1}}}
\def\dyadg #1{{\boldsymbol{\overline{#1}}}}

\def\be{\begin{equation} }
\def\ee{\end{equation} }

\def\dyad #1{\v{\overline{#1}}}

\newcount\timehh\newcount\timemm
\timehh=\time \divide\timehh by 60 \timemm=\time \count255=\timehh
\multiply\count255 by -60 \advance\timemm by \count255
\def\draftheader{\slshape Revised on \today\ at
\ifnum\timehh<10 0\fi\number\timehh\,:\,\ifnum\timemm<10 0\fi\number\timemm}%

\newcommand{\vg} [1]{\mbox{\boldmath $#1$}}

\def\dyad #1{\v{\overline{#1}}}
\def\dyadg #1{\boldsymbol{\overline{#1}}}

\def\LHS{left-hand side}
\def\RHS{right-hand side}

\newcount\timehh\newcount\timemm
\timehh=\time \divide\timehh by 60 \timemm=\time \count255=\timehh
\multiply\count255 by -60 \advance\timemm by \count255
\def\draftheader{\slshape Revised on \today\ at
\ifnum\timehh<10 0\fi\number\timehh\,:\,\ifnum\timemm<10 0\fi\number\timemm}%

\def\dyad #1{\v{\overline{#1}}}

\def\dyadg #1{\boldsymbol{\overline{#1}}}

\newcount\timehh\newcount\timemm
\timehh=\time \divide\timehh by 60 \timemm=\time \count255=\timehh
\multiply\count255 by -60 \advance\timemm by \count255
\def\draftheader{\slshape Revised on \today\ at
\ifnum\timehh<10 0\fi\number\timehh\,:\,\ifnum\timemm<10 0\fi\number\timemm}%

\renewcommand{\v}[1]{{{\bf #1}}}
\def\dyad #1{\v{\overline{#1}}}

\newcount\timehh\newcount\timemm
\timehh=\time \divide\timehh by 60 \timemm=\time \count255=\timehh
\multiply\count255 by -60 \advance\timemm by \count255
\def\draftheader{\slshape Revised on \today\ at
\ifnum\timehh<10 0\fi\number\timehh\,:\,\ifnum\timemm<10 0\fi\number\timemm}%

\def\div{\nabla\cdot}

\newcount\timehh\newcount\timemm
\timehh=\time \divide\timehh by 60 \timemm=\time \count255=\timehh
\multiply\count255 by -60 \advance\timemm by \count255
\def\draftheader{\slshape Revised on \today\ at
\ifnum\timehh<10 0\fi\number\timehh\,:\,\ifnum\timemm<10 0\fi\number\timemm}%

\def\eeqa{\end{eqnarray}}
\renewcommand{\v}[1]{{{\bf #1}}}

\def\dyad #1{\v{\overline{#1}}}
\def\dyadg #1{{\boldsymbol{\overline{#1}}}}
\def\vg #1{{\boldsymbol{#1}}}

\def\be{\begin{equation} }
\def\ee{\end{equation} }

\def\dyad #1{\v{\overline{#1}}}
\def\RHS{right-hand side}

\usepackage{imakeidx}
\makeindex[columns=2, title=Index, intoc, columnsep=4em, options= -s index_style.ist] 

\begin{document}

\def\booktitle{\textsc{Lectures on\\ Theory of Microwave and Optical Waveguides}
}
\def\smallbooktitle{\textsc{Theory of Microwave and Optical Waveguides}
}

\pagestyle{empty}

\author{\textsc{Weng Cho Chew}
}
\title{\bf\booktitle}
\date{Fall 2015\footnote{Preface and dedication updated \today}}

\maketitle

\cleardoublepage

\hbox{}

\vskip 2 in
\begin{center}

{\bf\Huge Dedication}

\vskip 2 in

 {\Large\it }

{\Large\it To the Memory of Shun Lien Chuang }

 {\Large\it }

\end{center}

\cleardoublepage

\frontmatter

\pagenumbering{roman}

\pagestyle{myheadings}

\markboth{\rm\smallbooktitle}{\rm\textsc{Contents}}

\tableofcontents

\cleardoublepage


\def\chaptitle{\textsc{Preface}}

\markboth{\rm\smallbooktitle}{\rm\chaptitle}

\chapter*{Preface}

\addcontentsline{toc}{chapter}{Preface}

I first taught this course, “Theory of Guided Waves”, shortly after I joined U of Illinois in 1985.  After two semesters of undergraduate electromagnetics, and one semester of a graduate course in electromagnetics, I started to develop a new course at Illinois in Spring 1987 in “Waves and Fields in Inhomogeneous Media”.  My interest in developing this course stemmed from my four years at Schlumberger-Doll Research (SDR) from 1981-1985.  That period included some of the most enriching years of my life, as I got to work with physicists, chemists, mathematicians, in addition to working with fabulous engineers.  The free-wheeling style of research at SDR was wonderful for imbibing new knowledge.  The leadership at SDR knew the importance of electromagnetics and wave physics.  Top researchers such as James Wait, Don Dudley, Emil Wolf, Adrianus de Hoop, Hans Blok, Peter M van den Berg, Jin Au Kong, plus many others, were frequent visitors to SDR as consultants.  I had greatly benefited from interacting with these renown scientists, as well as with my fellow scientists at SDR.

I started to teach “Theory of Guided Waves" course during the Fall of 1987.  This course was originally taught by Paul Klock at Illinois.  Since Paul was retiring, someone had to continue the tradition of teaching this course, and the onus was upon me to do so.  We can regard this course as complementary to the “Waves and Fields in Inhomogeneous Media” course as many advanced topics that cannot be covered there can be visited in this course.

Electromagnetics has had a great tradition at Illinois.  The leadership at Illinois realized that electromagnetics was an indispensable part of electrical engineering.  They recruited Georges A.  Deschamps from ATT Lab in 1958 to lead the electromagnetics research at Illinois, not long after John Bardeen had joined Illinois in 1951, and later, Nick Holonyak in 1954.

Under the leadership of Deschamps, the electromagnetics research at Illinois reached new heights and was  placed on the map.   Notable researchers were Yuen Tze Lo, Raj Mittra, Paul Mayes, and Shung-Wu Lee.  At its peak, together with younger recruits such as Shun-Lien Chuang (unfortunately, he passed away during the Spring of 2014), me, Jose Schutt-Aine, Jian-Ming Jin, Eric Michielssen, and Jennifer Bernhard, the total number of students/researchers in the electromagnetics group was over 70.     There was a weekly electromagnetics seminar that students, professors, and visitors gave talks to the group.  It was the most exciting time of electromagnetics research at Illinois.

It was under such a fertile ground that new courses could be developed.  New courses were taken by electromagnetics students as well as students from the remote/ionospheric sensing group and the optics/photonic group.  The remote/ionospheric sensing group consisted of Kung-Chieh Yeh, Chao-Han Liu, Erhan Kudeki, Steve Franke, George Swenson, Gary Swenson, Chet Gardner, and Jonathan Makela.   So on and off, I taught “Theory of Guided Waves” for about 30 years at Illinois.

Even though these lecture notes were compiled over a 30-year period, I did not feel a compulsion to publish them as a book.  First, there was a formidable tome of Robert E. Collin “Field Theory of Guided Waves” that I felt that these lecture notes would overlap with.  So I tried to look for inspiration in the direction of optical waveguides, and microwave waveguides, and solitons.  The chapters on these topics can be thought of as beyond what Collin had talked about his tome.  They might be my claim to modernity.   In recent years, I have been interested in quantum electromagnetics.  If I have a chance to teach this course again, I will include them in the course materials.

Guided waves have captivated the interest of many over the years, and even up to today.  John Scott Russell observed a soliton wave in 1834.  Hans Bethe was awarded the Nobel Prize for his numerous contributions, including the waveguide Bethe coupling work.  Bragg scattering, a Nobel Prize winning work, was indirectly discussed in the scattering by periodic structure.  Charles Kao was awarded the Nobel Prize for his seminal work on the optical fiber.   Due to the need in communications, microwave integrated circuits and high-frequency circuits have grown in importance in recent years.   This need has spawned the electronic design and automation (EDA) industry, which has great needs for solving highly complex problems encountered in computer chip and computer circuit designs.  Solutions to these problems are needed  to overcome and understand interference and compatibility (electromagnetic interference and compatibility) issues.  With the advent of 5G wireless communications, biomedical electromagnetics, quantum computing, quantum communications as well as quantum sensing, there are no ends to the needs for electromagnetic expertise from nanometer length scales to galactic length scales.  The emerging quantum technologies calls for our need to educate students from classical to quantum phenomena and increase their quantum awareness.  These will be the callings for the future generations.

In developing these lecture notes, I had benefited greatly from my interaction with my colleagues at Illinois as well as several written communications with Robert E Collin.  In addition, discussions with students and researchers at Illinois were greatly appreciated.  Many of the figures in these notes were provided by students over the years, especially in the term projects they had worked on.  More often than not, I have lost track of their original contributors.  Nevertheless, I like to thank them for their contributions.  I did remember that Fernando Teixeira, Kaladhar Radhakrishnan, and Andy Greenwood provided some of the excellent figures for these notes.

\vskip 1cm
{\noindent\it Weng Cho Chew}

{\noindent\it Summer, 2021}

{\noindent\it Purdue University}

{\noindent\it This preface was written prior to the publication of the archival copy of these lecture notes.  The contents of the lecture notes were last updated December 2015.}

\vskip 1cm






\vskip 0.5cm

\cleardoublepage

\pagenumbering{arabic}

\numberwithin{equation}{section}

\mainmatter



\def\v #1{{\bf #1}}
\def\vg #1{{\boldsymbol #1}}
\def\dyad#1{\overline {\bf #1}}
\def\dyadg#1{\overline {\vg #1}}
\def\beq{\begin{equation}}\def\eeq{\end{equation}}
\def\tinf{\text{\it inf\,}}\def\^{\hat}
\def\cal#1{\mathcal{#1}}
\def\ed{\end{document}}
\def\dalpha{\dyadg\alpha}
\def\dbeta{\dyadg\beta}
\def\Rg{\Re g}
\def\l{\label}


\def\chaptitle{Preliminary Background}
\chapter{\chaptitle}

\markboth{\smallbooktitle}{\chaptitle}





\section{{{ Introduction}}}

\setcounter{equation}{0} \setcounter{figure}{0}

\index{Wave}
Waveguiding phenomena occur naturally or are man made. For instance,
waveguides are a fundamental component of radio wave, microwave and
optical circuits\cite{COLLIND,COLLINE,MITTRA&LEE,LEVIN,MARCUVITZE}.
They are indispensable in modern technology in the radio frequency
to the optical frequency range. They are used in the
telecommunications as well as in wireless communications, for
example, in the design of a cell phone.  The purpose of a waveguide
is to guide the energy of a wave through a channel or a path with
little attenuation. Waveguides are also used to prevent interference
between two electromagnetic signals.

The precursor to electromagnetic waveguides were acoustic waveguides
as acoustic wave theory, being scalar, was well established before
electromagnetic theory\cite{RAY2}. Since acoustic waves are
longitudinal waves. They can be guided as a longitudinal mode in a
hollow tube for all frequencies. As a result, tubes of acoustic
waveguides of different lengths have been used as musical
instruments since ancient times.  The first analysis of
electromagnetic guided wave was probably done by Lord
Rayleigh\cite{RAY1}.

As we shall see later, a simple way to guide electromagnetic wave
for all frequencies, is to use two metallic conductors, usually an
inner one and an outer one as in a coaxial cable. As optical fiber
guides a mode for all frequencies too, but as shall be shown, when
the frequency is very low, the mode's energy is weakly trapped
inside the fiber, making it impractical as a waveguide for extremely
low frequencies. Since most sources are finite in extent, e.g.,
antennas, they generate spherical waves in the far-field which
decays algebraically. However, waveguides, by confining the energy
of the wave to a tube or a line, can cause a wave to traverse great
distances with little attenuation. An example is an optical fiber,
which can guide a signal with less than 0.3 dB/Km of
attenuation\cite{HECHT}.

Other emerging waveguiding technologies are plasmonic waveguides at
optical frequencies in nano-optics, or guiding waves using a chain
of nano particles\cite{NOVOTNY_HECHT}.   As nanoelectronic devices
are getting smaller, their dimensions are approaching the
wavelengths of electron wave functions.  The propagation of electron
waves in a channel can be viewed as a waveguiding
problem\cite{DATTAS}.

\begin{figure}[htb]
 \begin{center}
  \hfill\includegraphics[width=4.5truein]{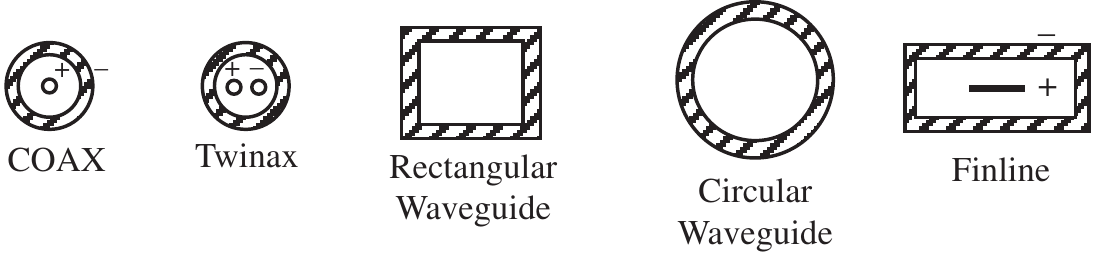}\hfill
 \end{center}
 \caption{Examples of closed waveguides.}
 \label{fg 1.1}
\end{figure}

\begin{figure}[htb]
 \begin{center}
  \hfill\includegraphics[width=4.5truein]{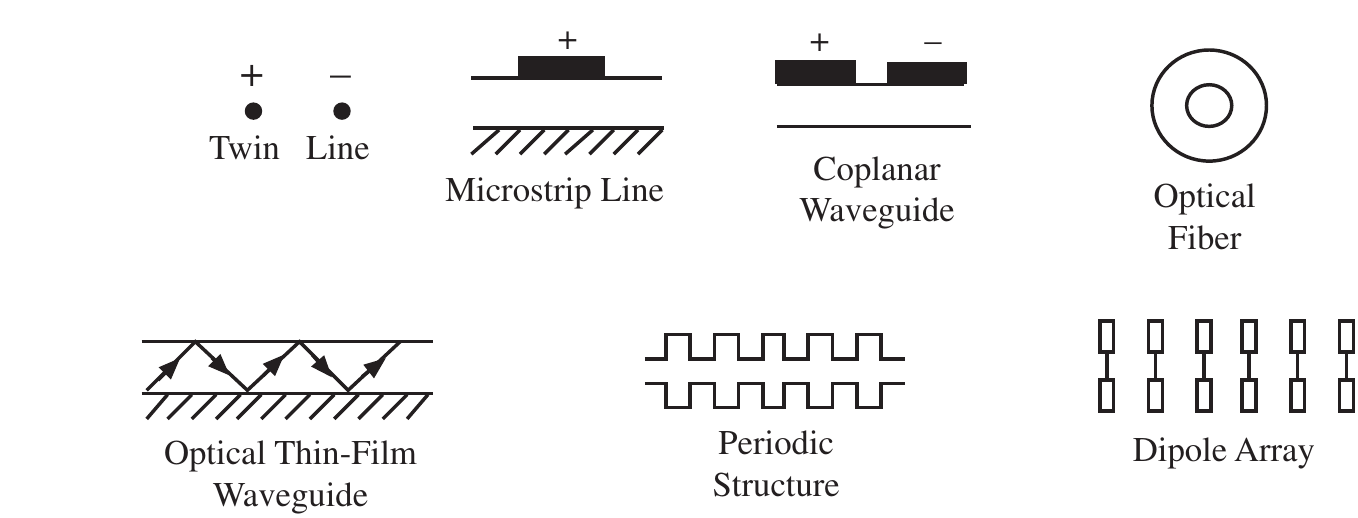}\hfill
 \end{center}
 \caption{Examples of open waveguides.}
 \label{fg 1.2}
\end{figure}

There are two main types of waveguides:  the closed waveguide and
the open waveguide.  In a closed waveguide, the electromagnetic
energy is completely trapped within metallic walls. The only way
to gain access to the energy is to tap holes in the waveguide
wall. Hence, it transmits signals with very good shielding and
very little interference from other signals. Figure \ref{fg 1.1}
shows some examples of closed waveguides. Notice that a closed
waveguide can be of one or more conductors. On the other hand, an
open waveguide\index{Waveguide} allows its field to permeate all of space, even
though most of the energy is still trapped and localized around
the guidance structure. As shown in Figure \ref{fg 1.2}, an open
waveguide is either a multi-conductor waveguide or a dielectric
waveguide. It is usually easier to fabricate an open waveguide.
However, as a result of their openness, such waveguides usually
radiate at discontinuities and bends.

Because open waveguides radiate at discontinuities and bends, some
of them are even used as antennas\cite{OLINER}.  There are also
modes that are weakly guided by an open waveguide, i.e., it radiates
as it is being guided. Examples of such modes are the leaky modes.
An antenna built using such a mode is known as a leaky wave antenna.

The analysis of waveguides requires a basic understanding of
electromagnetic theory. We will review our basic electromagnetic
theory in the following section.

\section{{{ History of Electricity and Magnetism}}}

Humans are exposed to electromagnetic phenomena on a daily basis.
Light wave is an electromagnetic phenomenon, so is lightning.
Lodestone is probably the first human experience with something
magnetic. Ancient Chinese knew about the magnetic properties of
lodestones, and made compasses out of them. Static electricity was a
phenomenon popularly demonstrated in European courts to entertain
the nobilities. But it was not until 1771-1773 that serious
experiments were done on static electricity by Henry Cavendish
(1731-1810). To this day, the Cavendish Laboratory stands in the
University of Cambridge in England to the honor of
Cavendish\cite{CAVENDISH}.

\index{Faraday's law}
Faraday's law was formulated by Michael Faraday (1791-1867) to
describe the fact that a changing magnetic flux, linked to a
metallic loop, will induce a voltage in the loop\cite{FARADAY}. This
fact can be used to design generators that produce electricity for
our homes. A multi-turn coil can be immersed in the \index{Field!magnetic}magnetic \index{Field}field
of a permanent magnet, and rotated rapidly. A voltage is then
induced in the coil, which can be tapped to deliver electricity for
a large number of applications.  Conversely, a DC current in a
static magnetic field experiences a force due to Lorentz force law.
This idea can be used to design a motor.  In fact, a DC motor was
invented by William Sturgeon in 1832\cite{CHAUETAL}.

\index{Ampere's Law}
Ampere's law was later formulated by Andr\'e Mari\'e Amp\`ere
(1775-1836) who stipulated that a wire carrying a current produces a
magnetic field\cite{AMPERE}. Moreover, the magnetic field is
produced according to the right-hand rule. (Note: The stipulation
that the magnetic field goes from the north pole of a bar magnet to
its south pole is entirely by convention. Hence, the right-hand rule
in Ampere's law is also entirely by convention. Also, the concept of
right-handedness and left-handedness is hard to describe to an
extra-terrestrial creature living in another universe who has never
seen a human before. Try that for yourself\cite{FEYNMAN}!)

\index{Gauss' law}
Gauss' law by Carl Friedrich Gauss (1777-1855) describes that if a
charge generating an \index{Field!electric}electric field is enclosed by a surface $S$,
the sum of the total flux flowing through the surface is equal to
the total charge contained within the surface\cite{GAUSS}. If the
surface does not enclose any charge, the sum of the total charge
through the surface is equal to zero. Coulomb's law can be derived
from Gauss' law.

The above period represented the era during which the understanding
of electromagnetism was incomplete.  Nevertheless, technology using
electricity and magnetism was prevalent.  As soon as Alessandro
Volta invented the battery, Ampere developed telegraphy in the early
1800s.  In fact, submarine cables were laid during a large part of
the nineteenth century by the British empire around the world to
enable telegraphic communication.  So it was quite well known that
wave phenomena existed on telegraphic lines before the completion of
Maxwell's theory as we shall discuss next.

Electromagnetic theory was completely formulated by the work of
James Clerk Maxwell (1831-1879)\cite{MAXWELL}. In 1864, he put forth
the theory that there should be a term, called the displacement
current term, to be added to Ampere's law. The work completed
electromagnetic theory and it was proven mathematically that
electromagnetic wave was a possible electromagnetic phenomenon.
Consequently, it was realized that light waves were electromagnetic
waves. Because of this important discovery, electromagnetic theory
is also known as Maxwell's theory, and the set of equations is also
known as Maxwell's equations. However, when Maxwell first wrote down
the complete form of electromagnetic theory, it was in some 20
equations.  It was Oliver Heaviside who recast those equations in
their present succinct form. Rightfully, these equations should be
called the Maxwell-Heaviside equations\cite{HEAVISIDE}.

In 1888, Heinrich Rudolf Hertz (1857-1894) performed an experiment
to verify the existence of electromagnetic wave. Two spheres in
close proximity to each other were used as capacitors to store
electric charges. The charges generate an electric field. A rapid
discharge of the electric charge causes the electric field to
collapse, producing an electromagnetic wave. The wave has both
electric and magnetic field in it.  Therefore, a wire loop, via
Faraday's law, can be linked to the time varying magnetic flux,
producing a voltage. This voltage creates a spark in a gap left in
the loop, even when the loop is at a distance from the spheres. To
his honor, the unit for frequency, which was cycles/second, is now
named Hertz. The term megahertz (MHz), or gigahertz (GHz) now
adorns the spec sheets of most computers.

In 1901, Guglielmo Marchese Marconi (1874-1937) successfully
transmitted an electromagnetic signal across the Atlantic Ocean from
Cornwall, England to Saint John, Newfoundland in North
America\cite{MARCONI}. Many nay sayers predicted that he would be
doomed to failure as the earth surface is curved.  Fortunately, the
ionosphere in the outer atmosphere acted like a mirror, and the
electromagnetic waves bounced back to earth. It was in fact a
serendipitous experiment. It was after his experiments that wireless
telegraphy was established.   However, Marconi never received a
patent for his invention.  The patent for telecommunication was
claimed by Nikola Tesla, who had the idea before Marconi.

Since then, electromagnetic theory has spurred the development of
myriads of technologies, many of which are electrical engineering
related.  Some of the more prominent ones are the development of the
radar, various antennas for telecommunication, remote sensing
systems, lasers and optics and more recently, wireless
communications, computer chip design, and electromagnetic
compatibility and electromagnetic interference. The advent of
quantum technologies as seen in quantum optics, quantum computers,
quantum communications, Casimir force in MEMS/NEMS, quantum
transport in electronic devices, photonics, will also dwell on
classical electromagnetics in combination with modern physics
concepts.

As of this date, electromagnetic theory continues to help in the
conception, analysis, and design of many new technologies. Hence,
Maxwell's equations are solved over and again for many of these
analysis tasks. As a consequence, much research has gone into
developing methods to impact many analyses in science and
engineering.

\section{{{ Maxwell's Equations}}}

Soon after the advent of Maxwell's\index{Maxwell's equations} theory, much analysis was
performed with Maxwell's equations.  By 1897, Lord Rayleigh had
already studied the propagation of electromagnetic \index{Wave}waves through
tubes. There was then much knowledge on propagation and guidance of
acoustic waves. Hence, analogue between acoustic waves and
electromagnetic waves were drawn as much as possible, although
acoustic waves are scalar while electromagnetic waves are vector in
nature.

In vector notation, and MKS units, Maxwell's equations are given as

\begin{equation}
\nabla\times\v E(\v r, t)=-\frac {\partial }{\partial t}\v B(\v r,
t), \label{eq121}
\end{equation}
\begin{equation}
\nabla\times\v H(\v r, t)=\frac {\partial }{\partial t}\v D(\v r,
t)+\v J(\v r,t), \label{eq122}
\end{equation}\begin{equation}
\nabla\cdot\v B(\v r, t)=0, \label{eq123}
\end{equation}\begin{equation}
\nabla\cdot\v D(\v r, t)=\rho (\v r, t). \label{eq124}
\end{equation}
where $\v E$ is the electric field\index{Field!magnetic} in volts/m, $\v H$ is the
magnetic field\index{Field!electric} in amperes/m, $\v D$ is the electric flux\index{Electric flux density} in
coulombs/m$^2$, $\v B$ is the magnetic flux\index{Magnetic flux density} in webers/m$^2$, $\v
J(\v r, t)$ is the current density\index{Current density} in amperes/m$^2$, and $\rho (\v
r, t)$ is the charge density\index{Charge density} in coulombs/m$^3$. For time varying
electromagnetic fields, only two of the four Maxwell's equations are
independent. Equations (\ref{eq123}) and (\ref{eq124}) can be
derived from Equations (\ref{eq121}) and (\ref{eq122}) by using the
continuity equation\index{Continuity equation}:
\begin{equation}
\nabla\cdot\v J(\v r, t)+\frac {\partial\rho (\v r, t)}{\partial
t}=0. \label{eq125}
\end{equation}

If we assume that $\v A(\v r,t)={\Re e \left\{ {\v A(\v r)
e^{-i\omega t} } \right\} }$, for all $\v E(\v r,t)$, $\v H(\v
r,t)$, $\v J(\v r,t)$, and $\rho (\v r,t) = \Re e \left\{\rho (\v
r)e^{-i\omega t}\right\}$; namely, the fields are time harmonic, the
above equations become,
\begin{equation}
\nabla\times\v E(\v r)=i\omega\v B(\v r), \label{eq126}
\end{equation}\begin{equation}
\nabla\times\v H(\v r)=-i\omega\v D(\v r)+\v J(\v r),
\label{eq127}
\end{equation}\begin{equation}
\nabla\cdot\v B(\v r)=0, \label{eq128}
\end{equation}\begin{equation}
\nabla\cdot\v D(\v r)=\rho (\v r). \label{eq129}
\end{equation}

\index{Medium}
The electric and magnetic fluxes are related to the electric and
magnetic fields via the constitutive relations\index{Constitutive relations}, the most general
of which are
\begin{equation}
\v D=\dyad{\vg\epsilon}\cdot\v E+\dyad{\vg\xi}\cdot\v H,
\label{eq1210}
\end{equation}\begin{equation}
\v B=\dyad{\vg\mu}\cdot\v H+\dyad{\vg\zeta}\cdot\v E,
\label{eq1211}
\end{equation}
where $\dyad{\vg\epsilon}$, $\dyad{\vg\xi}$, $\dyad{\vg\mu}$ and
$\dyad{\vg\zeta}$ are tensors. It is also the constitutive
relations that characterize the medium we are describing. A medium
with the above constitutive relations is known as a \emph{bianisotropic}
medium\index{Medium!bianisotropic}. A more commonly encountered medium is an \emph{anisotropic}
medium\index{Medium!anisotropic} with the constitutive relations
\begin{equation}
\v D=\dyad{\vg\epsilon}\cdot\v E, \label{eq1212}
\end{equation}\begin{equation}
\v B=\dyad{\vg\mu}\cdot\v H. \label{eq1213}
\end{equation}
When $\dyad{\vg\epsilon}$, $\dyad{\vg\xi}$, $\dyad{\vg\mu}$ and
$\dyad{\vg\zeta}$ are functions of space, the medium is also known
as an inhomogeneous medium\index{Medium!inhomogeneous}. When they are functions of frequency,
the medium is frequency dispersive\index{Medium!frequency dispersive}. When they are functions of
wavelength, it is spatially dispersive. For an isotropic medium,
the constitutive relations simply become
\begin{equation}
\v D=\epsilon\v E,\qquad\v B=\mu\v H. \label{eq1214}
\end{equation}
In free-space\index{Free space}, $\epsilon =\epsilon _0=8.854\times 10^{-12}$
farad/m, $\mu=\mu_0=4\pi\times 10^{-7}$ henry/m. The constant $c =
\frac 1{\sqrt {\mu _0\epsilon _0}}$ is related to the velocity of
light\index{Speed of light}, which has been very accurately measured.  The unit of meter
is defined such that $c$ is exactly equal to 299,792,458 m/s.  The
value of $\mu _0$ is assigned to be $4\pi\times 10^{-7}$ henry/m
while the value of $\epsilon _0$ is calculated from $c$.

\section {{{ Wave Equation}}}
\index{Wave equation}

For an anisotropic, inhomogeneous medium, Maxwell's equations for
time-harmonic fields\index{Field!time-harmonic} could be written as
\begin{equation}
\nabla\times\v E(\v r)=i\omega\dyad{\vg\mu}\cdot\v H(\v r),
\label{eq1215}
\end{equation}\begin{equation}
\nabla\times\v H(\v r)=-i\omega\dyad{\vg\epsilon}\cdot\v E(\v
r)+\v J(\v r), \label{eq1216}
\end{equation}\begin{equation}
\nabla\cdot\dyad{\vg\mu}\cdot\v H(\v r)=0, \label{eq1217}
\end{equation}\begin{equation}
\nabla\cdot\dyad{\vg\epsilon }\cdot\v E(\v r)=\rho (\v r).
\label{eq1218}
\end{equation}
If we take the curl of $\dyad{\vg\mu}^{-1}\cdot (\ref{eq1215})$,
we obtain, via the use of (\ref{eq1216}), that
\begin{equation}
\nabla\times\dyad{\vg\mu}^{-1}\cdot\nabla\times\v E(\v r)-\omega
^2\dyad{\vg\epsilon }\cdot\v E(\v r)=i\omega\v J(\v r).
\label{eq1219}
\end{equation}
Similarly, we can show that
\begin{equation}
\nabla\times\dyad{\vg\epsilon }^{-1}\cdot\nabla\times\v H(\v
r)-\omega ^2\dyad{\vg\mu }\cdot\v H(\v
r)=\nabla\times\dyad{\vg\epsilon}^{-1}\cdot\v J(\v r).
\label{eq1220}
\end{equation}
Equations (\ref{eq1219}) and (\ref{eq1220}) are two vector wave
equations governing the solutions of electromagnetic fields in an
inhomogeneous, anisotropic medium\index{Medium!inhomogeneous anisotropic}. Here, $\dyad{\vg\mu}$ and
$\dyad{\vg\epsilon} $ are functions of positions; hence, they do
not commute with the $\nabla$ operator. Also, for time-varying
fields\index{Field!time-varying}, $\v E$ and $\v H$ are derivable from each other; only one
of the two equations (\ref{eq1219}) and (\ref{eq1220}) is
necessary to fully describe the electromagnetic fields.

For an isotropic medium\index{Medium!isotropic}, (\ref{eq1219}) and (\ref{eq1220}) reduce
to
\begin{equation}
\nabla\times\mu ^{-1}\nabla\times\v E(\v r)-\omega ^2\epsilon\v
E(\v r)=i\omega\v J(\v r), \label{eq1221}
\end{equation}\begin{equation}
\nabla\times\epsilon ^{-1}\nabla\times\v H(\v r)-\omega ^2\mu\v
H(\v r)=\nabla\times\epsilon ^{-1}\v J(\v r). \label{eq1222}
\end{equation}
For electrodynamics\index{Electrodynamics}, either one of the above equations is
self-contained. We can derive the phenomena of dynamic
electromagnetic fields by just studying one of them.  However, when
$\omega\rightarrow 0$, these equations are not solvable, and we have
to invoke all four of Maxwell's equations when solving static
problems.

\section {{{ Boundary Conditions}}}

We cannot find a unique solution to a partial differential
equation unless we specify the boundary conditions\index{Boundary conditions} as well.
Equations (\ref{eq1219}) to (\ref{eq1222}) are vector wave
equations\index{Wave equation!vector} whose solutions we will seek over and over again. One common
method of solving the above equations is to find the solutions in
each of the homogeneous regions that constitute the inhomogeneity,
provided that the inhomogeneity is piecewise constant.  The unique
solution is then obtained by matching the boundary conditions at
the interface.

Since either Equation (\ref{eq1219} ) or (\ref{eq1220} ) is
sufficient in describing electromagnetic fields, the boundary
conditions must be buried in them.  Therefore, we can derive the
boundary conditions\index{Boundary conditions} from them.  To do this, we integrate
(\ref{eq1219}) about a small area between the interface of two
media. Invoking Stokes' theorem, we have
\begin{equation}
\oint\limits _C d\v l\cdot (\dyad {\vg\mu
}^{-1}\cdot\nabla\times\v E)-\omega ^2\int\limits _A d\v
S\cdot\dyad{\vg\epsilon}\cdot\v E=i\omega\int\limits_A d\v
S\cdot\v J. \label{eq1223}
\end{equation}

\begin{figure}[htb]
\begin{center}
\hfil\includegraphics[width=3.0truein]{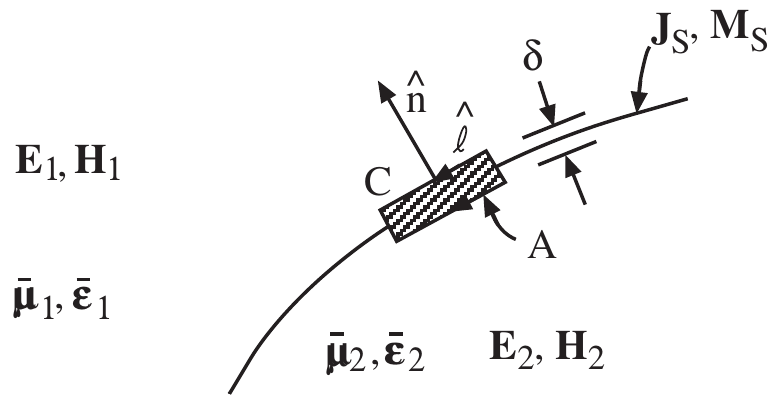}\hfil
\end{center}
\caption{Boundary conditions at an interface.}\label{fg13}
\end{figure}

%
Letting $\delta\rightarrow 0$, the surface integral on the
left-hand side of the above equation vanishes. Assuming that we
have a current sheet $\v J_s$, we can show that
\begin{equation}
\^ n\times (\dyad{\vg\mu}_1^{-1}\cdot\nabla\times\v E_1)-\^
n\times (\dyad{\vg\mu}_2^{-1}\cdot\nabla\times\v E_2)=i\omega\v
J_s. \label{eq1224}
\end{equation}
Since $\nabla\times\v E=i\omega\dyad {\vg\mu }\cdot\v H$, we have
\begin{equation}
\^ n\times\v H_1-\^ n\times\v H_2=\v J_s. \label{eq1225}
\end{equation}
Performing the same analysis for Equation (\ref{eq1220}), we
arrive at
\begin{equation}
\^ n\times\v E_1-\^ n\times\v E_2=0. \label{eq1226}
\end{equation}
Equations (\ref{eq1225}) and (\ref{eq1226}) are the important
boundary conditions we will use over and again.

The boundary condition (\ref{eq1225}) can also be gleaned from
(\ref{eq127}). If $\v J(\v r)$ is a current sheet at on interface,
represented by a delta function singularity, then this singularity
must be from the normal derivative of the tangential component of
the magnetic field. From this fact we can derive (\ref{eq1225}).
By the some token, (\ref{eq1226}) can be derived from
(\ref{eq128}).

\section{{{ Reciprocity Theorem}}}
\index{Reciprocity theorem}
\index{Symmetry}
\index{Theorem}

If we have two sources $\v J_1$ and $\v J_2$ radiating in an
anisotropic, inhomogeneous medium, and $\v J_1$ produces the field
$\v E_1$, $\v J_2$ produces the field $\v E_2$, the reciprocity
theorem\index{Theorem!reciprocity} requires that for a reciprocal medium\index{Medium!reciprocal},
\begin{equation}
\langle\v E_1, \v J_2\rangle=\langle\v E_2, \v J_1\rangle,
\label{eq1227}
\end{equation}
where $\langle\v A, \v B\rangle$ stands for $\int d\v r\v A\cdot\v
B$.  This theorem is derivable from Equation (\ref{eq1219}) with
constraints on $\dyad{\vg\epsilon }$ and $\dyad{\vg\mu }$.  When
the source $\v J_1$ is radiating, the field $\v E_1$ satisfies the
equation
\begin{equation}
\nabla\times\dyad {\vg\mu }^{-1}\cdot\nabla\times\v
E_1-\omega^2\dyad {\vg\epsilon }\cdot\v E_1=i\omega\v J_1.
\label{eq1228}
\end{equation}
When $\v J_2$ is radiating, the field $\v E_2$ satisfies the
equation
\begin{equation}
\nabla\times\dyad {\vg\mu }^{-1}\cdot\nabla\times\v E_2-\omega
^2\dyad {\vg\epsilon }\cdot\v E_2=i\omega\v J_2, \label{eq1229}
\end{equation}
where $\dyad{\vg\mu }$ and $\dyad {\vg\epsilon }$ in
(\ref{eq1228}) and (\ref{eq1229}) represent the same medium.
Dot-multiplying (\ref{eq1228}) by $\v E_2$ and integrating, and
(\ref{eq1229}) by $\v E_1$ and integrating, we have
\begin{equation}
\langle\v E_2, \nabla\times\dyad {\vg\mu }^{-1}\cdot\nabla\times\v
E_1\rangle-\omega ^2\langle\v E_2, \dyad {\vg\epsilon }\cdot\v
E_1\rangle=i\omega\langle\v E_2, \v J_1\rangle, \label{eq1230}
\end{equation}
\begin{equation}
\langle\v E_1, \nabla\times\dyad{\vg\mu }^{-1}\cdot\nabla\times\v
E_2\rangle-\omega ^2\langle\v E_1, \dyad {\vg\epsilon }\cdot\v
E_2\rangle=i\omega\langle\v E_1, \v J_2\rangle. \label{eq1231}
\end{equation}
Since
\begin{equation}
\langle\v E_2, \nabla\times\dyad {\vg\mu }^{-1}\cdot\nabla\times\v
E_1\rangle=\int\limits _V d\v r\v E_2\cdot\nabla\times\dyad
{\vg\mu }^{-1}\cdot\nabla\times \v E_1, \label{eq1232}
\end{equation}
we can use the identity
\begin{align}
\nabla\cdot(\v A\times\v B)=\v B\cdot\nabla\times\v A-\v
A\cdot\nabla\times\v B
\end{align}
and Gauss' theorem to get
\begin{align}
\langle\v E_2, \nabla\times\dyad {\vg\mu }^{-1}\cdot\nabla\times\v
E_1\rangle &=\int\limits _V d\v r(\nabla\times\v E_2)\cdot\dyad
{\vg\mu }^{-1}\cdot(\nabla\times\v E_1)\\ &+\int\limits _S dS\^
n\cdot(\dyad {\vg\mu }^{-1}\cdot\nabla\times\v E_1)\times\v E_2\nonumber\\
\int\limits _V d\v r(\nabla\times\v E_2)\cdot\dyad {\vg\mu
}^{-1}\cdot(\nabla\times\v E_1)\\ &+i\omega\int\limits _S dS\^
n\cdot(\v H_1)\times\v E_2,
 \label{eq1233}
\end{align}
where $V$ and $S$ are a volume and a surface tending to infinity.
When $S\rightarrow\infty$, $\dyad {\vg\mu }$ becomes isotropic and
homogeneous. Furthermore, the solutions to the vector wave
equation become plane waves\index{Wave!plane}.  Hence, $\nabla\rightarrow i\v k$,
and we have
\begin{equation}
(\dyad {\vg\mu }^{-1}\cdot\nabla\times\v E_1)\times\left.\v
E_2\right|_{\v r\epsilon s}=i\mu _0^{-1}(\v k\times\v E_1)\times\v
E_2=-i\mu _0^{-1}\v k(\v E_1\cdot\v E_2). \label{eq1234}
\end{equation}
where we have assumed that $\v k\cdot\v E_2=0$.
In this manner, the surface integral in (\ref{eq1233}) is symmetric
about $\v E_1$ and $\v E_2$. If $\dyad {\vg\mu }^{-1}$ is symmetric,
then the first integral on the right-hand side of (\ref{eq1233}) is
also symmetric about $\v E_1$ and $\v E_2$. Hence, if $\dyad {\vg\mu
}^{-1}$ is symmetric, the first term of (\ref{eq1230}) and
(\ref{eq1231}) are equal. If $\dyad {\vg\epsilon }$ is also
symmetric, then the second term of (\ref{eq1230}) and (\ref{eq1231})
are also equal. Therefore, we deduce that (\ref{eq1227}) is
satisfied or that reciprocity holds when
\begin{equation}
\dyad {\vg\mu }=\dyad {\vg\mu }^t,\qquad\dyad {\vg\epsilon }=\dyad
{\vg\epsilon }^t. \label{eq1235}
\end{equation}
In other words, $\dyad {\vg\mu}$ and $\dyad {\vg\epsilon}$ are
symmetric (if $\dyad {\vg\mu }$ is symmetric, $\dyad {\vg\mu
}^{-1}$ is symmetric). The condition expressed in Equation
(\ref{eq1235}) is necessary for an anisotropic medium\index{Medium!anisotropic} to be
reciprocal medium\index{Medium!reciprocal}. It also follows that all isotropic media\index{Medium!isotropic} are
reciprocal.

The integral defined in Equation (\ref{eq1227}) is also known as a
\emph{reaction}. It could be thought of as a generalized measurement. In
words, the reciprocity theorem states that for a reciprocal
medium, {\it the $\v E$-field due to $\v J_1$ measured by $\v J_2$
is the same as the $\v E$-field due to $\v J_2$ measured by $\v
J_1$.}

\begin{figure}[htb]
\begin{center}
\hfil\includegraphics[width=4.0truein]{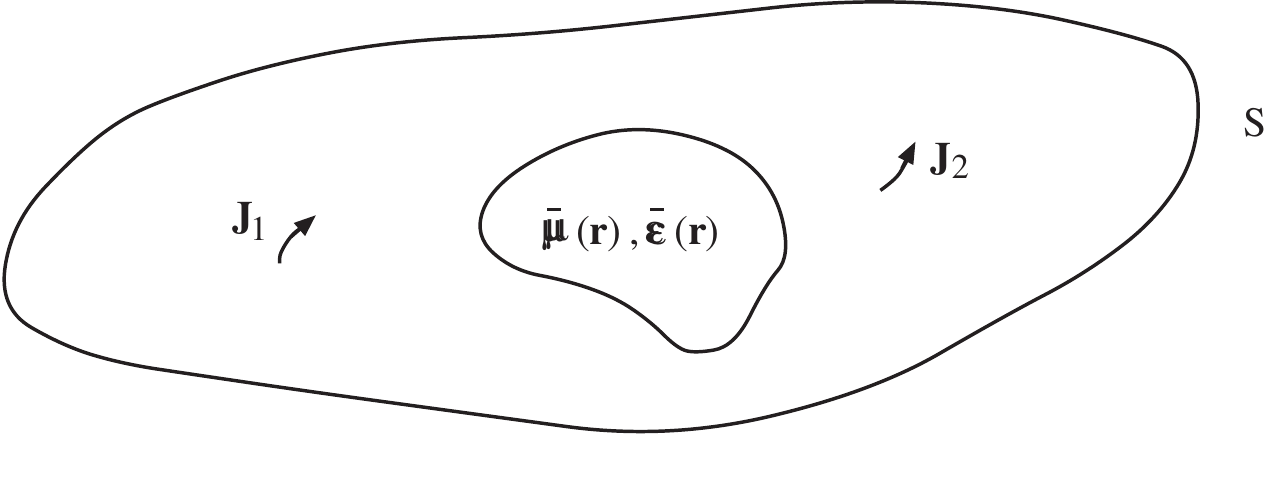}\hfil
\end{center}
\caption{Proof of Reciprocity.}\label{fg14}
\end{figure}


Examples of non-reciprocal media are plasma and ferrite media
biased by a magnetic field.  A medium can be lossy and still be
reciprocal.

In electromagnetics, it is customary to add a fictitious magnetic
current\index{Magnetic current density} $\v M$ to Faraday's law\index{Faraday's law} such that
\begin{equation}
\nabla\times\v E=-i\omega\v D-\v M \label{eq1236}
\end{equation}
A reciprocity theorem that can be derived involving magnetic
current is
\begin{equation}
\langle\v E_1,\v J_2\rangle-\langle\v H_1,\v M_2\rangle=\langle\v
E_2,\v J_1\rangle-\langle\v H_2,\v M_1\rangle \label{eq1237a}
\end{equation}
in replacement of (\ref{eq1227}).

Reciprocity theorem is deeply related to the symmetry\index{Maxwell's equations!symmetry} of
differential operators related to Maxwell's equations. For example,
we can express (\ref{eq1228}) and (\ref{eq1229}) as
\begin{equation}
\cal D\v E_1=\v J_1 \label{eq1237b}
\end{equation}
\begin{equation}
\cal D\v E_2=\v J_2 \label{eq1237c}
\end{equation}
Where $\cal D$ is the pertinent differential operators. Then,
$\langle\v E_2,\v J_1\rangle=\langle\v E_1,\v J_2\rangle$ implies
\begin{equation}
\left\langle\v E_2,\cal D\v E_1\right\rangle=\left\langle\v E_1,\cal D\v
E_2\right\rangle \label{eq1237d}
\end{equation}
The above is the analogue of
\begin{equation}
\v a^t\cdot\dyad A\cdot\v b=\v b^t\cdot\dyad A\cdot\v a
\label{eq1237e}
\end{equation}
which implies the $\dyad A=\dyad A^t$ or $\dyad A$ is symmetric.
Hence, (\ref{eq1237d}) implies that $\cal D$ is symmetric, and
this is possible only if (\ref{eq1235}) is satisfied.

The symmetry of $\cal D$ is the deeper underlying reason for the
reciprocity theorem.  For media that are reciprocal, the symmetry
of the electromagnetic equations will give rise to a number of
operators that are also \index{Symmetry}symmetrical such as the impedance and
admittance matrices, as we shall learn later.

\subsection{Lorentz Reciprocity Theorem}

If the volume integrals in \eqref{eq1230} and \eqref{eq1231} are
taken over a finite volume, then on subtracting the two equations,
making use of \eqref{eq1233}, and assuming the symmetry of the
permeability and permittivity tensors, we arrive at the general case
of the reciprocity theorem\index{Theorem!Lorentz reciprocity},
\begin{align}
\langle\v E_2, \v J_1\rangle-\langle\v E_1, \v J_2\rangle= \oint_S
dS \hat n\cdot \left(\v E_1\times\v H_2-\v E_2\times\v H_1\right)
\end{align}
When the volume does not enclose the sources, we arrive at
\begin{align}
\oint_S dS \hat n\cdot \left(\v E_1\times\v H_2\right)=\oint_S dS
\hat n\cdot\left(\v E_2\times\v H_1\right)
\end{align}
The above is generally known as the Lorentz reciprocity theorem.  It
is useful in waveguides\index{Waveguide} when sources are not involved.

\section{{{ Energy Conservation}}}

Energy conservation\index{Energy conservation} in electromagectics is defined by the Poynting
theorem\index{Theorem!Poynting}. Poynting theorem holds for the time domain as well as the
frequency domain. The theorem in the time domain is actually quite
different from that in the frequency domain. We shall present first
the time domain version.

\subsection{ Time Domain Poynting Theorem}

The time domain Poynting theorem\index{Theorem!time domain Poynting}, sometimes known as the real
Poynting theorem\index{Theorem!real Poynting}, governs the conservation of instantaneous energy
for electromagnetic field\index{Energy conservation}. To derive it, we start with
\begin{equation}
\begin{split}
\nabla\cdot\left[\v E(\v r, t)\times\v H(\v r, t)\right] & =\v H\cdot\nabla\times\v E-\v E\cdot\nabla\times\v H\\
& =-\v H\cdot\frac{\partial\v B}{\partial t}-\v
E\cdot\frac{\partial\v D}{\partial t}-\v E\cdot\v J
\end{split}
\label{eq161}
\end{equation}
Defining the Poynting vector
\begin{equation}
\v S\left(\v r, t\right)=\v E\left(\v r, t\right)\times\v
H\left(\v r, t\right) \label{eq162}
\end{equation}
we have
\begin{equation}
\nabla\cdot\v S(\v r, t)=-\left(\v H\cdot\frac{\partial\v
B}{\partial t}+\v E\cdot\frac{\partial\v D}{\partial t}\right)-\v
E\cdot\v J \label{eq163}
\end{equation}
For free space where $\v B=\mu_0\v H$, $\v D=\epsilon_0\v E$, we can
show that
\begin{equation}
\v H\cdot\frac{\partial\v B}{\partial t}=\v
H\cdot\mu_0\frac{\partial\v H}{\partial
t}=\frac{1}{2}\mu_0\frac{\partial}{\partial t}\v H\cdot\v H
\label{eq164}
\end{equation}
and similarly, for the electric flux term, we have
\begin{equation}
\nabla\cdot\v S(\v r,t)=-\frac{\partial}{\partial
t}\frac{1}{2}(\mu_0\v H\cdot\v H+\epsilon_0\v E\cdot\v E)-\v
E\cdot\v J \label{eq165}
\end{equation}
Then term
\begin{equation}
{W_{T}}=\frac{1}{2}(\mu_0\v H\cdot\v H+\epsilon_0\v E\cdot\v E)
\label{eq166}
\end{equation}
corresponds to the total energy\index{Total energy stored} stored in the magnetic field and
the electric field. When $\frac{\partial}{\partial t}W_{T}$ is
positive, it corresponds or contributing the negative term to
$\nabla\cdot\v S$ implying a influx of power at a point. The last
term $\v E\cdot\v J$ corresponds to power absorbed or generated by
the current $\v J$. When $\v E\cdot\v J$ is positive, the current
$\v J$ is absorptive. This is true of a conductive medium\index{Medium!conductive} where
$\v J=\sigma\v E$.

Please note that the above derivation that leads to expression
\eqref{eq166} is not valid for material media\index{Medium!material}.  All material media
have to be frequency dispersive\index{Medium!frequency dispersive}, and hence, in the time domain, the
constitutive relations are denoted by time convolutions.  As a
curious fact, the above can be generalized to inhomogeneous,
anisotropic, reciprocal media.

\subsection{ Frequency Domain Poynting Theorem}

The frequency domain Poynting theorem\index{Theorem!frequency domain Poynting} governs energy conservation
for complex power\index{Complex power}. Hence, it is also known as the complex Poynting
theorem\index{Theorem!complex Poynting} \cite{WFIMD}. We start with
\begin{equation}
\begin{split}
\nabla\cdot(\v E\times\v H^*) &=i\omega\v H^*\cdot\v B-i\omega\v E\cdot\v D^*-\v E\cdot\v J^*\\
&=i\omega\v H^*\cdot\dyad\mu\cdot\v H-i\omega\v
E\cdot\dyad\epsilon^*\cdot\v E^*-\v E\cdot\v J^*\\
\end{split}
\label{eq167}
\end{equation}
Defining the complex Poynting vector\index{Complex Poynting vector}
\begin{equation}
\tilde{\v S}(\v r)=\v E(\v r)\times\v H^*(\v r) \label{eq168}
\end{equation}
the above becomes
\begin{equation}
\nabla\cdot\tilde{\v S}=i\omega(\v H^*\cdot\dyad\mu\cdot\v H-\v
E\cdot\dyad\epsilon^*\cdot\v E^*)-\v E\cdot\v J^* \label{eq169}
\end{equation}
For a source-free region, this becomes
\begin{equation}
\nabla\cdot\tilde{\v S}=i\omega(\v H^*\cdot\dyad\mu\cdot\v H-\v
E\cdot\dyad\epsilon\cdot\v E^*) \label{eq1610}
\end{equation}
If
\begin{equation}
\v H^*\cdot\dyad\mu\cdot\v H=\v E\cdot\dyad\epsilon\cdot\v E^*
\label{eq1611}
\end{equation}
in a region, then the right-hand side is zero, and there is no net
power flux into or out of the region. This occurs at resonance in
a cavity\index{Resonance in cavity}.

\subsection{ Complex Power}

The complex Poynting theorem is quite different from the real
Poynting theorem. It can be shown that half the real part of the
complex Poynting vector is the time average\index{Time average of Poynting vector} of the instantaneous
Poynting vector, viz., [see problem 1.2]
\begin{equation}
\langle\v S(\v r, t)\rangle=\frac{1}{2}\Re e\left[\tilde{\v S}(\v
r)\right] \label{eq1612}
\end{equation}
The part that corresponds to the stored energy in the complex
Poynting theorem is the difference of the magnetic energy\index{Magnetic energy} and
electric energy\index{Electric energy} stored, whereas that in the real Poynting theorem is
the sum of the two. This is because the imaginary part of the
complex power is reactive power [see problem 1.3]. Reactive power in
a time harmonic system corresponds to power that flows into a
system, and later flows out of a system. Hence, its time average is
zero.

Notice that in a resonance system\index{Resonance system} or circuit such as the LC tank
circuit, the stored magnetic energy and electric energy are equal
to each other, and they exchange with each other.   Since the
reactive power is the difference in the store magnetic and
electric energy, it is zero in this case.  Therefore, when an LC
tank circuit is at resonance, there is no need for an external
supply of reactive power\index{Reactive power}.

\begin{figure}[htb]
\begin{center}
\hfil\includegraphics[width=3.0truein]{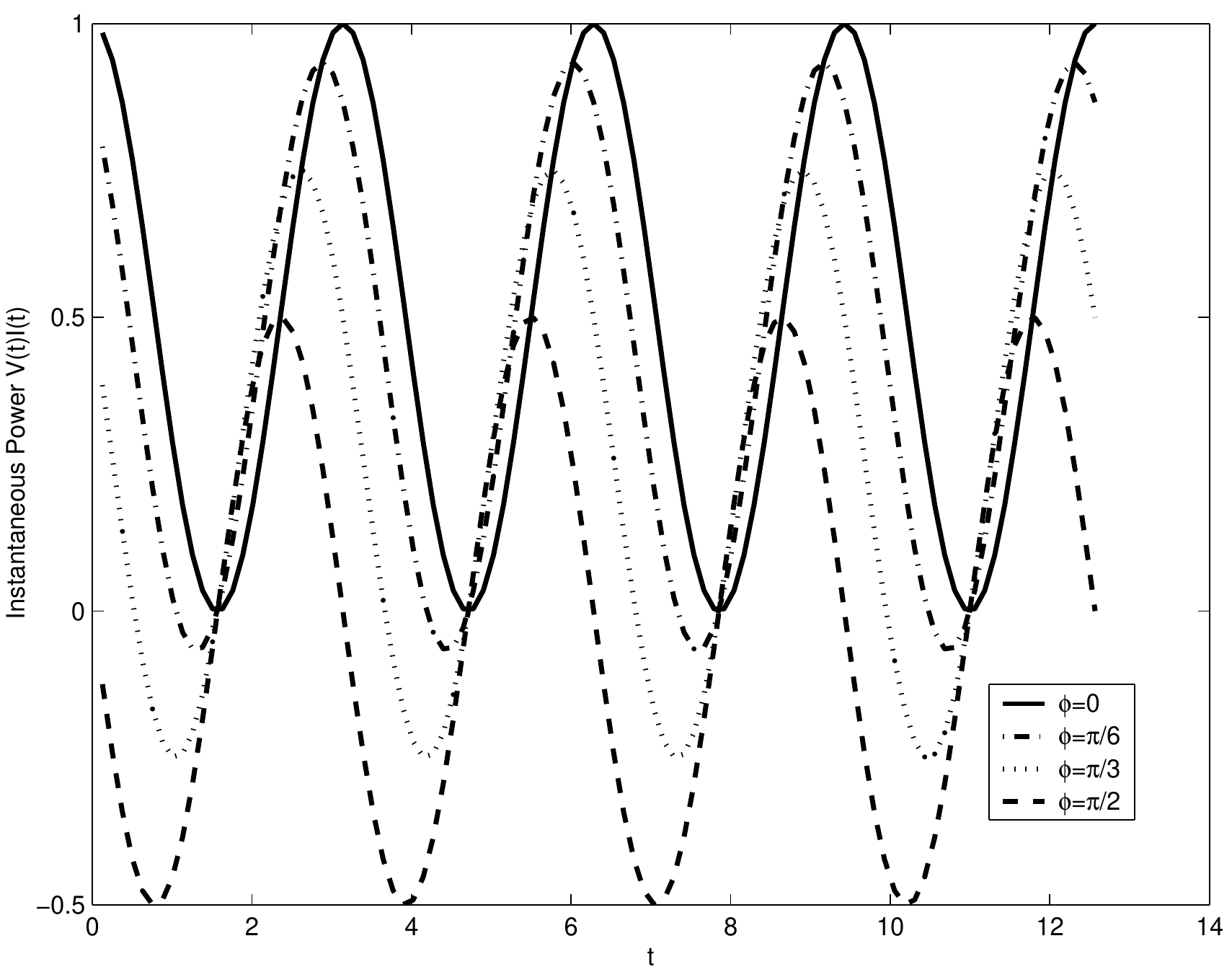}\hfil
\end{center}
\caption{Plots of $P(t)=V(t)I(t)$ versus time $t$ where
$V(t)=\cos(t)$, and $I(t)=\cos(t+\phi)$ for various values of
$\phi$.}\label{fg14a}
\end{figure}


Even though no net power is delivered in the reactive power, a power
utility company will still charge its customers for the use of this
power for two reasons: First, not all reactive power is retrievable
as it has to be sent over power lines that have conductive losses.
Second, the power company has to maintain a generator that can
absorb the oscillation in the total power caused by the presence of
reactive power.   Figure {\ref{fg14a}} shows that the instantaneous
power can be negative as well as positive when there is a phase
shift between the voltage and the current in a circuit.

\subsection{Lossless Conditions}

For an isotropic medium\index{Medium!isotropic}, the conditions for it to be lossless are
that $\Im m(\mu)=0$, and $\Im m (\epsilon ) = 0$ where ``$\Im m$''
implies ``imaginary part.'' However, the condition for an
anisotropic medium\index{Medium!anisotropic} is quite different. We can derive the general
lossless condition from energy conservation.

For a lossless medium\index{Medium!lossless}, for energy conservation, and from the complex
Poynting theorem, we require that
\begin{equation}
\Re e \int\limits _V dV\nabla\cdot(\v E\times\v H^*)=\Re e
\oint\limits _ S d\v S\cdot (\v E\times\v H^*)=0, \label{eq1237}
\end{equation}
since $\Re e[\v E \times\v H^*]$ corresponds to time average power
flow. The above implies that
\begin{equation}
\Re e \left\{ i\omega\int\limits _V dV(\v H^*\cdot\dyad {\vg\mu
}\cdot\v H-\v E\cdot\dyad {\vg\epsilon }^*\cdot\v E^*)\right\}=0.
\label{eq1238}
\end{equation}
A sufficient condition for arbitrary $V$ is to require that $\v
H^*\cdot\dyad {\vg\mu }\cdot\v H$ and $\v E\cdot\dyad
{\vg\epsilon}^*\cdot\v E^*$ to be purely real or their conjugates
to be themselves, i.e.,
\begin{equation}
(\v H^*\cdot\dyad {\vg\mu }\cdot\v H)^*=\v H\cdot\dyad {\vg\mu
}^*\cdot\v H^*=\v H^*\cdot\dyad {\vg\mu }^\dag\cdot\v H=\v
H^*\cdot\dyad {\vg\mu }\cdot\v H. \label{eq1239}
\end{equation}
Therefore, $\dyad {\vg\mu }^\dag=\dyad {\vg\mu }$. Similarly the
condition on $\v E\cdot\dyad {\vg\epsilon }^*\cdot\v E^*$ to be
purely real is $\dyad {\vg\epsilon }^\dag=\dyad {\vg\epsilon }$.
Consequently, the lossless conditions for an anisotropic medium\index{Lossless condition} is
\begin{equation}
\dyad{\vg\epsilon }=\dyad {\vg\epsilon }^\dag,\qquad \dyad{\mu
}=\dyad {\vg\mu }^\dag. \label{eq1240}
\end{equation}
In other words, the permittivity tensor\index{Permittivity tensor} and the permeability
tensor\index{Permeability tensor} have to be Hermitian.

\section{Energy Density in Dispersive Medium}

\newcommand\Ev{\mathbf{E}}
\newcommand\Hv{\mathbf{H}}
\newcommand\Jv{\mathbf{J}}

\newcommand\zerov{\boldsymbol{0}}
\newcommand\Lambdav{\boldsymbol{\Lambda}}

\newcommand\mud{\overline{\mathbf{\mu}}}
\newcommand\epsd{\overline{\mathbf{\epsilon}}}


In the following derivation, we assume that $\Ev$, $\v H$, and $\v
J$ have $e^{-i\omega t}$ time dependence, where $\omega$ is a
complex frequency\cite{HAUS}.   Then
\begin{align}
\nabla \cdot [\Ev(t) \times \Hv^{*}(t)]&=\Hv^{*}(t)\cdot\nabla\times\Ev(t)-\Ev(t)\cdot\nabla\times\Hv^{*}(t)\nonumber\\
                                 &=\Hv^{*}(t)\cdot[i\omega\mud\cdot\Hv(t)]-\Ev(t)\cdot [i\omega^{*}\epsd^{*}\cdot\Ev^{*}(t)+\Jv^{*}(t)]\nonumber\\
                                 &=i\omega\Hv^{*}(t)\cdot\mud\cdot\Hv(t)-i\omega^{*}\Ev(t)\cdot\epsd^{*}\cdot\Ev^{*}(t)-\Ev(t)\cdot\Jv^{*}(t)
                                 \label{Eq1}
\end{align}
Next, we let $\omega=\omega'+i\omega''$ where $\omega'$ and
$\omega''$ are real numbers.  Then
\begin{align}
       \nabla \cdot [\Ev(t) \times
       \Hv^{*}(t)]&=i(\omega'+i\omega'')\Hv^{*}(t)\cdot\mud(\omega'+i\omega'')\cdot\Hv(t)\nonumber\\
                                 &\quad -i(\omega'-i\omega'')\Ev(t)\cdot\epsd^{*}(\omega'+i\omega'')\cdot\Ev^{*}(t)-\Ev(t)\cdot\Jv^{*}(t)
\label{Eq1a}
\end{align}
Ordinarily, if $\omega$ is pure real, the time dependence would have
canceled in the above, but because $\omega$ is complex, each of the
above terms has time dependence of $\exp(2\omega''t)$.
 Assuming that $ \omega''\ll \omega'$, we can Taylor expand the
\RHS\ to get
\begin{align}
\nabla \cdot (\Ev \times \Hv^{*})& \doteq i(\omega'+i\omega'')\Hv^{*}\cdot\left[\mud(\omega')+i\omega''\frac{\partial}{\partial\omega'}\mud(\omega')\right]\cdot\Hv\nonumber\\
                                 &\quad
                                 -i(\omega'-i\omega'')\Ev\cdot\left[\epsd^{*}(\omega')-i\omega''\frac{\partial}{\partial\omega'}\epsd^{*}(\omega')\right]\cdot\Ev^{*}-\Ev\cdot\Jv^{*}
\end{align}
Collecting leading order and first order terms, we have
\begin{align}
       \nabla \cdot (\Ev \times \Hv^{*})& \doteq  i\omega'[\Hv^{*}\cdot\mud(\omega')\cdot\Hv-\Ev\cdot\epsd^{*}(\omega')\cdot\Ev^{*}]\nonumber\\
                                 &\quad -\omega''\left\{\Hv^{*}\cdot\left[\mud(\omega')+\omega'\frac{\partial}{\partial\omega'}\mud(\omega')\right]\cdot\Hv\right.\nonumber\\
                                 &\quad +\left.\Ev\cdot\left[\epsd^{*}(\omega')+\omega'\frac{\partial}{\partial\omega'}\epsd^*(\omega')\right]\cdot\Ev\right\}-\Ev\cdot\Jv^{*}
\nonumber\\
&=  i\omega'[\Hv^{*}\cdot\mud(\omega')\cdot\Hv-\Ev\cdot\epsd^{*}(\omega')\cdot\Ev^{*}]\nonumber\\
                                 &\quad
                                 -\omega''\left\{\Hv^{*}\cdot\frac{\partial}{\partial\omega'}[\omega'\mud(\omega')]\cdot\Hv\right.
                                 +\left.\Ev\cdot\frac{\partial}{\partial\omega'}[\omega'\epsd^*(\omega')]\cdot\Ev\right\}-\Ev\cdot\Jv^{*}
\label{Eq2}
\end{align}
%
For lossless media\index{Medium!lossless}, the first term is purely imaginary, while the
second term is purely real. The first term corresponds to reactive
power.  The second term comes about because for a complex
exponential of the form
$e^{-i(\omega'+i\omega'')t}=e^{-i\omega'+\omega''t}$, the field
strength is growing with $e^{2\omega''t}$ time dependence. If we
take the real part of \eqref{Eq2}, and focussing on the part of
space where $\v J=0$, we have
\begin{align}
\nabla \cdot \frac{1}{2}\Re e(\Ev \times \Hv^{*})&=
-\frac{1}{2}\omega''\left\{\Hv^{*}\cdot
\frac{\partial}{\partial\omega'}\omega'\mud(\omega')\cdot\Hv
                                 +\Ev\cdot\frac{\partial}{\partial\omega'}\omega'\epsd^{*}(\omega')\cdot\Ev^{*}\right\}\nonumber\\
                                 &=-\frac{\partial}{\partial t}W_T
\label{Eq4}
\end{align}
The above has the physical meaning that the divergence of the
time-average real power flow on the \LHS\ is due to the time
variation of the energy density on the \RHS.  The energy density has
a time dependence of $e^{2\omega''t}$.  Consequently, we identify
the energy density\index{Energy density} for dispersive media\index{Medium!dispersive} as
\begin{align}
W_T=\frac{1}{4}\left\{\Hv^{*}\cdot
\frac{\partial}{\partial\omega'}\omega'\mud(\omega')\cdot\Hv
                                 +\Ev\cdot\frac{\partial}{\partial\omega'}\omega'\epsd^{*}(\omega')\cdot\Ev^{*}\right\}
\label{Eq4a}
\end{align}
When the medium is free space, we have
\begin{align}
W_T=\frac{1}{4}\left\{\Hv^{*}\cdot \mu_0\Hv
                                 +\Ev\cdot\epsilon_0\Ev^{*}\right\}
\label{Eq4b}
\end{align}
which agrees with what we have derived from time-domain Poynting
theorem.

\section{{{ Symmetries in Electromagnetics}}}

Symmetries\index{Maxwell's equations!symmetry} play an important role in the solutions of Maxwell's
equations.  They can be used for simplifying solutions to Maxwell's
equations, or they can be used to derive new solutions.
Alternatively, they can be used to predict how solutions behave once
symmetry is broken.

In solid state physics, symmetry is used to understand the
propagation of electronic waves in crystalline structures which have
a high degree of symmetry.  Due to the symmetries, group theory\index{Group theory} can
be used to analyze the physical characteristics of waves in
crystalline structures\cite{FALICOV,TUNG}.  Unlike electronic waves,
electromagnetic waves, for most applications, propagate in
nonsymmetric structures. Hence, the exploitation of symmetry in
electromagnetics has not reached the level in solid state physics.
However, there are still a few symmetries we can exploit, especially
in waveguides and resonators which usually have some degrees of
symmetry associated with them.

Some of the obvious symmetries are translational symmetry\index{Symmetry!translational} and
rotational symmetry\index{Symmetry!rotational}.  Translational symmetry exploits the fact that
Maxwell's equations are invariant after a translation in space.
Rotational symmetry implies that a solution of Maxwell's equations
remains a solution after rotation. Other symmetries are
time-reversal symmetry and reflection symmetry that we shall discuss
next.\footnote{Some of these symmetries have been used successfully
in computational electromagnetics to expedite numerical solutions of
Maxwell's equations\cite{FEACEM}. See also discussions in \cite[p.
268]{JACKSON}.}

\subsection{{{ Time Reversal Symmetry}}}

In a lossless environment, solutions to Maxwell's equations are time
reversible\index{Symmetry!time-reversal} in the same medium.  That is if a solution is found, and
if we change $t$ to $-t$, the solution remains a valid solution to
Maxwell's equations within the same medium. This is like playing a
movie backward.  However, the right-hand rule becomes the left-hand
rule in the movie playback.

It is clear that the solution to the wave equation\index{Wave equation} is time
reversible. However, when we have a lossy wave equation, the
solution decays forward in time, but grows backward in time.
Therefore, the solution is not time reversible, namely, the
reverse-time solution is a solution to an active medium (amplifying
medium like a laser cavity) but the forward-time solution
corresponds to a lossy medium\index{Medium!lossy}.  For the same reason, solutions to
Maxwell's equations are not time reversible in a lossy medium.

To obtain a time-reversed field, we let $t\rightarrow -t$.  For
example, a time-reversed $\v B$ field is $\v B(\v r, -t)$. Then
the time derivative of this time-reversed $\v B$ field is
\begin{equation}
\frac{\partial}{\partial t}\v B(\v r,
-t)=-\frac{\partial}{\partial t'}\v B(\v r, t')
\end{equation}
which is the negative of the original time derivative. Consequently,
when time-reversed fields are substituted back into Maxwell's
equations, they can be written as
\begin{equation}
\nabla\times\v E(\v r, t')=\frac {\partial }{\partial t'}\v B(\v
r, t'), \label{}
\end{equation}
\begin{equation}
\nabla\times\v H(\v r, t')=-\frac {\partial }{\partial t'}\v D(\v
r, t')+\v J(\v r,t'), \label{}
\end{equation}\begin{equation}
\nabla\cdot\v B(\v r, t')=0, \label{}
\end{equation}\begin{equation}
\nabla\cdot\v D(\v r, t')=\rho (\v r, t'). \label{}
\end{equation}
We will retrieve the original Maxwell's equations if the signs of
$\v H$, $\v B$, and $\v J$ are reversed. The need to reverse these
quantities is also necessary for energy conservation in Poynting
theorem. The sign of $\v E\times \v H$ and has to change for
time-reversed solution to reflect that the energy flow has to change
direction.  Note that the sources $\v J$ and $\rho$ are also time
reversed.

Alternatively, we can change the sign of $\v E$, $\v D$, and
$\rho$.  But the convention is to change the signs of $\v H$, $\v
B$, and $\v J$.  A positive charge, when moving through space,
remains a positive charge when time-reversed.  However, it
produces a current of opposite polarity.

Since time always occurs as $\exp(-i\omega t)$ in the frequency
domain, replacing $t$ with $-t$ is the same as replacing $i$ with
$-i$.  Hence, a time reversed field is obtained by conjugating the
frequency-domain field. If a time-harmonic field is represented by
its phasor, then the conjugate of the phasor represents a
time-reversed solution as it can be easily shown that if
\begin{align}
\v E(\v r,t)=\Re e \left[\v E(\v r,\omega)e^{-i\omega t}\right]
\end{align}
then
\begin{align}
\Re e \left[\v E^*(\v r,\omega)e^{-i\omega t}\right]=\Re e \left[\v
E(\v r,\omega)e^{i\omega t}\right]=\v E(\v r,-t)
\end{align}
The design of phase-conjugate mirror was in vogue in optics to
create a time-reversed optical
field\cite{YARIV_CM,AGARWAL_ETAL,CHEW_HABASHY}.

\subsection{{{ Reflection Symmetry}}}

It was believed once that all laws of physics can be replicated in
the mirror world, namely, laws of physics remain the same under
reflection.  This is known as the \emph{conservation of parity}\index{Parity conservation}. However,
it is now known that some laws of physics do not satisfy parity
conservation\cite{LEE&YANG}.  However, the law of electromagnetics
satisfies parity conservation.  We just need to replace a right-hand
rule with a left-hand rule for the reflected solution, namely, the
solution in the mirror world.

A symmetry closely related to reflection symmetry\index{Symmetry!reflection} is inversion
symmetry\cite{FALICOV}.  In inversion, we let $\v r\rightarrow -\v
r$, or in detail, $x\rightarrow -x$, $y\rightarrow -y$, and
$z\rightarrow -z$.  A reflected function or object can always be
obtained from an inverted function or object by a rotation.  For
instance, if we have a mirror in the $xy$ plane, and we put an
object in front of the mirror, the reflected object will have
$z\rightarrow -z$ with its $xy$ coordinates unchanged.  However, this
can also be obtained by first inverting the object, followed by a
$180$ degree rotation about the $z$ axis.  Since a rotation of a
solution is still a solution to Maxwell's equations, we will just
discuss what inversion does to a solution.

When we have a vector field such as $\v E(\v r, t)$, we assume that
the direction of the field also change after inversion by replacing
$\hat x\rightarrow -\hat x$, $\hat y\rightarrow -\hat y$, and $\hat
z\rightarrow -\hat z$. Hence, a vector field under inversion becomes
$-\v E(-\v r, t)$. If we take the curl of this inverted field\index{Field!inverted}, we
have
\begin{equation}
\nabla\times[-\v E(-\v r, t)]=\nabla'\times\v E(\v r', t)
\end{equation}
after we let $\v r'=-\v r$, and under this change of variables,
$\nabla=-\nabla'$.  We can substitute these inverted fields into
Maxwell's equations\index{Maxwell's equations} to see if these equations retain their original
forms.

Consequently, Maxwell's equations under substitution of inverted
fields, and with a change of variables $\v r'=-\v r$, become
\begin{equation}
\nabla'\times\v E(\v r', t)=\frac {\partial }{\partial t}\v B(\v
r', t), \label{}
\end{equation}
\begin{equation}\label{a1}
\nabla'\times\v H(\v r', t)=-\frac {\partial }{\partial t}\v D(\v
r', t)-\v J(\v r',t),
\end{equation}\begin{equation}
\nabla'\cdot\v B(\v r', t)=0, \label{}
\end{equation}\begin{equation}
\nabla'\cdot\v D(\v r', t)=\rho (\v r', t). \label{}
\end{equation}
However, the above is not the original Maxwell's equations.  The
original Maxwell's equations can be retrieved if we can change the
signs of $\v B$ and $\v H$.  This is understandable, since the
right-hand rule becomes a left-hand rule in the mirror or reflected
world (since the reflected world is related to the inverted world by
just a rotation). Hence, a change of the signs of $\v B$ and $\v H$
will convert the left-hand rule back to the right-hand rule.

\subsection{{{ Polar Vectors and Pseudovectors}}}

A word is in order about polar vectors\index{Polar vector} versus pseudovectors\index{Pseudovector} (also
known as axial vectors)\cite{JACKSON}.  A polar vector (or vector)
changes sign under inversion, but a pseudovector does not.  For
instance, if $\v A$ and $\v B$ are polar vectors, they will change
sign under inversion.  However, a vector $\v C=\v A\times\v B$ will
not change sign under inversion if the cross product is defined with
with the right-hand rule in the original right-handed coordinate
system.  It will change sign if the left-hand rule is used.  Hence,
$\v C$ is a pseudovector because it does not follow the sign-change
rule of the polar vectors under inversion. In electromagnetics, we
can regard $\v H$ and $\v B$ as pseudovectors that do not change
sign under inversion. In this case, Maxwell's equations are
invariant under inversion or reflection.

By the same token, pseudoscalars\index{Pseudoscalar} exist.  The scalar $\v a\cdot\v
b\times\v c$, where $\v a$, $\v b$, and $\v c$ are polar vectors,
is a pseudoscalar which changes sign under inversion.

If we define $\v B$ and $\v H$ to be pseudovectors instead, and they
do not change sign under inversion, then we do not have to change
the sign of $\v B$ and $\v H$.

\section{{{ Green's Function}}}

The Green's function\index{Green's function} to a wave equation is the solution when the
source is a point source\cite{GREEN,WFIMD}.  When we know the
solution to the wave equation due to a point source, the solution
due to a general source can be obtained by the principle of linear
superposition. This is because the wave equation is linear, and a
general source could be thought of as a superposition of point
sources.

For example, if we need to find the solution to the following
equation,
\begin{equation}
(\nabla ^2+k^2)\psi (\v r)=S(\v r), \label{eq1241}
\end{equation}
we can first find the Green's function which is the solution to
the following equation,
\begin{equation}
(\nabla ^2+k^2)g(\v r-\v r')=-\delta (\v r-\v r'). \label{eq1242}
\end{equation}
If we know $g(\v r-\v r'),\quad\psi (\v r)$ can be found formally.
Multiplying (\ref{eq1241}) by $g(\v r-\v r')$ and (\ref{eq1242})
by $\psi (\v r)$, and integrating over volume, and subtracting, we
obtain
\begin{equation}
\int\limits _V d\v r[g(\v r-\v r')\nabla ^2\psi (\v r)-\psi (\v
r)\nabla ^2g(\v r-\v r')] = \int\limits _Vg(\v r-\v r')S(\v r)d\v
r+\psi (\v r'). \label{eq1243}
\end{equation}
Noting that $g\nabla ^2\psi -\psi\nabla
^2g=\nabla\cdot(g\nabla\psi -\psi\nabla g)$, we can rewrite the
left-hand side, using Gauss' divergence theorem as
\begin{equation}
\oint\limits _S d\v S\cdot(g\nabla\psi -\psi\nabla g)=\int\limits
_V g(\v r-\v r')S(\v r)d\v r+\psi (\v r'). \label{eq1244}
\end{equation}
When $S\rightarrow \infty$, all fields look like plane waves\index{Wave!plane}, and
we can replace $\nabla\rightarrow i\v k$. The left-hand side of
(\ref{eq1244}) then vanishes, and we have
\begin{equation}
\psi (\v r)=-\int\limits _V d\v r'g(\v r-\v r')S (\v r').
\label{eq1245}
\end{equation}
Hence the solution of (\ref{eq1241}) can be written as an integral
superposition of the solution of (\ref{eq1242}).  In fact, we can
invoke the principle of linear superposition to arrive at the above
as well.

To find the solution of Equation (\ref{eq1242}), we solve it in
spherical coordinates with the origin at $\v r'$. Then, it becomes
\begin{equation}
(\nabla ^2+k^2)g(\v r)=-\delta (\v r)=-\delta (x)\delta (y)\delta
(z). \label{eq1246}
\end{equation}
For $r\ne 0$, the homogeneous, spherically symmetric\index{Wave!spherical} solution to
(\ref{eq1246}) is
\begin{equation}
g(\v r)=C\frac {e^{+ikr}}{r}+D\frac {e^{-ikr}}r. \label{eq1247}
\end{equation}

\begin{figure}[htb]
\begin{center}
\hfil\includegraphics[width=3.0truein]{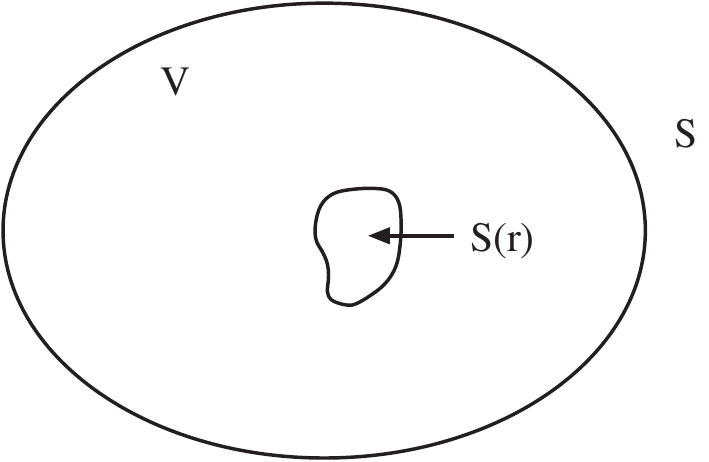}\hfil
\end{center}
\caption{The radiation of a source $S(\v r)$ in a volume $V$.}\label{fg15}
\end{figure}


Physical grounds require that we have only outgoing solutions;
hence,
\begin{equation}
g(\v r)=C\frac {e^{ikr}}r. \label{eq1248}
\end{equation}
We can match the constant $C$ to the singularity at the origin by
substituting (\ref{eq1248}) into (\ref{eq1246}), and integrating
Equation (\ref{eq1246}) over a small volume about the origin.
\begin{equation}
\int\limits _{\Delta V} dV\nabla\cdot\nabla\frac
{Ce^{ikr}}r+\int\limits _{\Delta V} dVk^2\frac {Ce^{ikr}}r=-1.
\label{eq1249}
\end{equation}
The second integral vanishes when $\Delta V\rightarrow 0$, because
$dV = 4\pi r^2dr$.  We can convert the first integral in
(\ref{eq1249}) into a surface integral using Gauss' theorem, and
obtain
\begin{equation}
\lim _{r\rightarrow 0}4\pi r^2\frac {\partial }{\partial r}C\frac
{e^{ikr}}r=-1, \label{eq1250}
\end{equation}
or that $C =1/4\pi$.  Therefore, in general \index{Green's function!point source}
\begin{equation}
g(\v r-\v r)=\frac {e^{ik|\v r-\v r'|}}{4\pi|\v r-\v r'|}
\label{eq1251}
\end{equation}
The solution to (\ref{eq1241}), from Equation (\ref{eq1245}) is
then
\begin{equation}
\psi (\v r)=-\int\limits _V d\v r'\frac {e^{ik|\v r-\v r'|}}{4\pi
|\v r-\v r'|}S(\v r'). \label{eq1252}
\end{equation}
Equation (\ref{eq1252}) is a convolutional integral, a consequence
of the principle of linear superposition.  The above Green's
function is the one that satisfies the \emph{radiation condition}\index{Radiation condition}.  Hence,
the linearly superposed solution also satisfies the radiation
condition.

For the vector wave equation in a homogeneous, isotropic medium\index{Medium!homogeneous isotropic},
the equation is
\begin{equation}
\nabla\times\nabla\times\v E(\v r)-k^2\v E(\v r)=i\omega\mu\v J(\v
r).        \ \label{eq1253}
\end{equation}

By using the fact that $\nabla\times\nabla\times\v E=-\nabla ^2\v
E+\nabla\nabla\cdot\v E$, and that $\nabla\cdot\v E=\frac
1{\epsilon}\rho=\frac 1{i\omega\epsilon }\nabla\cdot\v J$, we can
rewrite (\ref{eq1253}) as
\begin{equation}
\nabla ^2\v E(\v r)+k^2\v E(\v r)=-i\omega\mu \left(\dyad I+\frac
{\nabla\nabla }{k^2}\right)\cdot\v J(\v r). \label{eq1254}
\end{equation}
There are three scalar wave equations\index{Wave equation!scalar} embedded in the above
equation. We can solve each of them in the manner of Equation
(\ref{eq1245}), and we have
\begin{equation}
\v E(\v r)=i\omega\mu\int\limits _V d\v r'g(\v r'-\v r)
\left(\dyad I+\frac {\nabla'\nabla'}{k^2}\right)\cdot\v J(\v r').
\label{eq1255}
\end{equation}
It can be shown that
\begin{equation}
\int\limits _V d\v r'g(\v r-\v r')\nabla' f(\v
r')=\nabla\int\limits _V d\v r'g(\v r-\v r')f(\v r'),
\label{eq1256}
\end{equation}\begin{equation}                                                 \int\limits _V d\v r'g(\v r-\v r')\nabla'\cdot\v F(\v r')=\nabla\cdot\int\limits _V d\v r'g(\v r-\v r')\v F(\v r'),
\label{eq1257}
\end{equation}
by using the vector identities $\nabla gf=f\nabla g+g\nabla
f,\quad\nabla\cdot g\v F=g\nabla\cdot\v F+(\nabla g)\cdot\v F$,
and that $\nabla' g(\v r-\v r')=-\nabla g(\v r-\v r')$.  Hence,
Equation (\ref{eq1255}) can be rewritten as
\begin{equation}
\v E(\v r)=i\omega\mu \left( \dyad I+\frac {\nabla\nabla
}{k^2}\right)\cdot\int\limits _V d\v r'g(\v r-\v r')\v J(\v r').
\label{eq1258}
\end{equation}

Sometimes, Equation (\ref{eq1258}) is written as
\begin{equation}
\v E(\v r)=i\omega\mu\int\limits _V d\v r'\dyad G(\v r, \v
r')\cdot\v J(\v r'), \label{eq1259}
\end{equation}
where
\begin{equation}
\dyad G(\v r, \v r')=\left(\dyad I+\frac {\nabla\nabla
}{k^2}\right) g(\v r-\v r'), \label{eq1260}
\end{equation}
is a dyad known as the \emph{dyadic Green's function}\index{Green's function!dyadic}.  It has to be used
with caution, since Equation (\ref{eq1259}), with the $\nabla\nabla$
operator inside the integration, has to be clarified since it does
not converge uniformly when $\v r$ is also in the source region
occupied by $\v J(\v r)$.  Hence, it is only a convenient notation
when the observation point is outside the source region.


\section{Uniqueness Theorem}

The \emph{uniqueness theorem}\index{Theorem!uniqueness} provides conditions under which the solution to the wave equation\index{Wave equation} is unique.
This is especially important because the solutions to a problem
should not be indeterminate.  These conditions under which a
solution to a wave equation is unique are the boundary conditions
and the radiation condition.  Uniqueness also allows one to
construct solutions by inspections; if a candidate solution
satisfies the conditions of uniqueness, it is the unique solution.
Because of its simplicity, the scalar wave equation shall be
examined first for easier insight into this problem.

\subsection {Scalar Wave Equation}

Given a scalar wave equation with a source term on the right-hand side, we
shall derive the conditions under which a solution is unique.
First, assume that there are two different solutions to the \emph{scalar
wave equation}\index{Wave equation!scalar}, namely,
\begin{align}
[\nabla ^2+k^2(\v r)]\, \phi _1(\v r)=s(\v r), \label {ENUM1a}
\end{align}
\begin{align}
[\nabla ^2+k^2(\v r)]\, \phi _2(\v r)=s(\v r), \label {ENUM1b}
\end{align}
where $k^2(\v r)$ includes inhomogeneities of finite extent.  Then,
on subtracting the two equations, we have
\begin{align}
[\nabla ^2+k^2(\v r)]\,\delta \phi (\v r)=0, \label {ENUM2}
\end{align}
where $\delta \phi (\v r)=\phi _1(\v r)-\phi _2(\v r)$.  Note that
the solution is unique if and only if $\delta \phi =0$ for all $\v
r$.

Then, after multiplying \eqref{ENUM2} by $\delta\phi ^*$,
integrating over volume, and using the vector identity $\nabla
\cdot\psi\v A=\v A\cdot\nabla\psi +\psi\nabla\cdot\v A$, we have
\begin{align}
\int \limits _S\^ n\cdot (\delta \phi ^*\nabla\delta\phi )\,
dS-\int\limits _V|\nabla\delta\phi |^2dV+\int\limits
_Vk^2|\delta\phi |^2dV=0, \label {ENUM3}
\end{align}
where $\^ n$ is a unit normal to the surface $S$. Then, the
imaginary part of the above equation is
\begin{align}
{\Im m }\int \limits _S\^ n\cdot (\delta\phi ^*\nabla\delta\phi )\,
dS+ \int\limits _V{\Im m } (k^2)|\delta\phi |^2dV=0. \label {ENUM4}
\end{align}
Hence,  if $ {\Im m } [k^2(\v r)]\neq 0$ in $V$, and
\begin{enumerate}[label=(\roman*)]
\item $\delta\phi =0$ or $\^ n\cdot\nabla\delta\phi =0$
on $S$,
\item $\delta \phi =0$ on part of $S$ and $\^ n\cdot\nabla\delta\phi =0$
on the rest of $S$, or
\item $\delta\phi+\alpha\^ n\cdot\nabla\delta\phi=0$ on $S$, where
$\alpha$ is real, \footnote{The author is grateful to J. Mamou for
pointing out this case.}
\end{enumerate}
\noindent then the first integral above vanishes, and we have
\begin{align}
\int \limits _V{\Im m}[k^2(\v r)]|\delta\phi |^2dV=0.\label {ENUM5}
\end{align}
\noindent Since $|\delta\phi |^2$ is positive definite for
$\delta\phi\neq 0$, and $\Im m (k^2)\neq 0$ in $V$,\footnote{More
specifically, $\Im m [k^2(\v r)]>0$, $\forall$ $\v r\in V$, or $\Im
m [k^2 (\v r)]<0$, $\forall$ $\v r\in V$.} the above is only
possible if $\delta\phi =0$ everywhere inside $V$. Also, in the
third case above, $\alpha$ can vary on the surface $S$.  It can also
be chosen so that the first two cases are the special cases of the
third case.

Therefore, in order to guarantee uniqueness, so that $\phi _1=\phi
_2$ in $V$, the above conditions are equivalent to either
\begin{enumerate}[label=(\roman*)]
\item $\phi _1=\phi _2$ on
$S$ or $\^ n\cdot\nabla\phi _1=\^ n\cdot\nabla\phi _2$ on $S$,
\item $\phi _1=\phi _2$
on one part of $S$, and $\^ n\cdot\nabla\phi _1=\^ n\cdot\nabla\phi
_2$ on the rest of $S$, or
\item $\phi_1+\alpha \hat n\cdot\nabla \phi_1=
\phi_2+\alpha \hat n\cdot\nabla \phi_2$, where $\alpha$ is real.
\end{enumerate}
\noindent The specification of $\phi$ on $S$ is also known as the
{\textbf{\emph{Dirichlet}}} boundary condition\index{Boundary conditions!Dirichlet}, while the
specification of $\^ n\cdot\nabla\phi$, namely, the normal
derivative, is also known as the {\textbf{\emph{Neumann}}} boundary
condition\index{Boundary conditions!Neumann}.  The third is the reactive impedance boundary condition.
In words, the uniqueness theorem says that if two solutions satisfy
the same Dirichlet or Neumann boundary condition or a mixture
thereof on $S$, or the reactive impedance boundary condition\index{Boundary conditions!impedance}, the
two solutions must be identical.

Notice that the difference solution, $\delta \phi$ satisfies the
boundary conditions above \eqref{ENUM5} are all lossless
(non-dissipative or non-gain) boundary conditions.  When $\Im m
[k^2(\v r)]\ne 0$, and when such boundary conditions are satisfied
by the difference solution, \eqref{ENUM5} implies that only trivial
solution $\delta\phi=0$ exists.  In other words, no time-harmonic
difference solution can exist in such media with loss or gain.

     When ${\Im m }(k^2)=0$, i.e., when $k^2$ is real, the condition
$\delta\phi =0$ or $\^ n\cdot\nabla\delta\phi =0$ on $S$ in
\eqref{ENUM3} does not necessarily lead to $\delta\phi =0$ in $V$,
or uniqueness. The reason is that solutions for $\delta\phi =\phi
_1-\phi _2$ where
\begin{align}
\int \limits _V|\nabla\delta\phi |^2\,dV=\int \limits
_Vk^2|\delta\phi |^2\,dV \label {ENUM6}
\end{align}
can exist.  These are the resonance solutions\index{Resonance solution} in the volume $V$.
These resonance solutions are the homogeneous solutions\footnote
{``Homogeneous solutions'' is a mathematical parlance for solutions
to \eqref{ENUM1a} without the source term.} to the wave equation
\eqref{ENUM1a} at the real resonance frequencies of the volume $V$.
Because the medium is lossless, they are time harmonic solutions
which satisfies the boundary conditions, and hence, can be added to
the particular solution of \eqref{ENUM1a}.  In fact, the particular
solution usually becomes infinite at these resonance frequencies if
$S(\v r)\neq 0$.

Equation \eqref{ENUM6} implies the balance of two energies.  In the
case of acoustic waves, for example, it represents the balance of
the kinetic energy and the potential energy in a volume $V$.
When ${\Im m }(k^2)\neq 0$, however, the resonance solutions of the
volume $V$ are exponentially decaying with time for a lossy medium\index{Medium!lossy}
$\left[{\Im m }(k^2)>0\right]$, and they are exponentially growing
with time for an active medium $\left[{\Im m }(k^2)<0\right]$.  But
if only time harmonic solutions $\phi _1$ and $\phi _2$ are
permitted in (1), these resonance solutions are automatically
eliminated from the class of permissible solutions.  Hence, for a
lossy medium $[ {\Im m } (k^2)>0]$ or an active medium\index{Medium!active} $[{\Im m }
(k^2)<0]$, the uniqueness of the solution is guaranteed if we
consider only time harmonic solutions where $\omega$ is real,
namely, two solutions will be identical if they have the same
boundary conditions for $\phi $ and $\^ n\cdot\nabla\phi$ on
$S$.\footnote{The nonuniqueness associated with the resonance
solution for a lossless medium can be eliminated if we consider time
domain solutions.  In the time domain, we can set up an initial
value problem in time, e.g., by requiring all fields be zero for
$t<0$; thus, the nonuniqueness problem can be removed via the
causality requirement\index{Causality}.  The resonance solution, being time harmonic,
is noncausal.}

When $S\rightarrow \infty$ or $V\rightarrow \infty$, the
number of resonance frequencies of $V$ becomes denser.  In fact,
when $S\rightarrow \infty $, the resonance frequencies of $V$ become
a continuum implying that any real frequency could be the resonant
frequency of $V$. Hence, if the medium is lossless, the uniqueness
of the solution is not guaranteed at any frequency, even with
appropriate boundary conditions on $S$ at infinity, as a result of
the presence of the continuum of resonance frequencies.  One remedy
then is to introduce a small loss.  With this small loss [$ {\Im m }
(k)>0$], the solution is either exponentially small when
$r\rightarrow\infty$ (if a solution corresponds to an outgoing wave,
$e^{ikr}$), or exponentially large when $r\rightarrow \infty$ (if a
solution corresponds to an incoming wave, $e^{-ikr}$).  Now, if the
solution is exponentially small, namely, keeping only the outgoing
wave solutions, it is clear that the surface integral term in
\eqref{ENUM4} vanishes when $S\rightarrow \infty$, and the
uniqueness of the solution is guaranteed.  This manner of imposing
the outgoing wave condition at infinity is also known as the
{\textbf{\emph{Sommerfeld radiation condition}}}\index{Sommerfeld radiation condition} \cite[p.
188]{SOMMERFELD}.  This radiation condition can be used in the limit
of a vanishing loss for an unbounded medium\index{Medium!unbounded} to guarantee uniqueness.

The uniqueness of the solution to the Helmholtz wave equation\index{Wave equation!uniqueness} is
similar to the uniqueness of the solution to the the matrix equation
\begin{align}
\dyad A\cdot\v x =\v b\label{mat1}
\end{align}
If a solution to the equation
\begin{align}
\dyad A\cdot\v x_N = 0\label{mat2}
\end{align}
exists, then the solution to the first equation is not unique. $x_N$
is the null-space solution to the matrix.

 The right-hand side of \eqref{mat1} is the driving term.  If the driving
term to Helmholtz wave equation is zero, and yet, a solution exists,
it is usually called the resonance solution.\footnote{This is called
the homogeneous solution in mathematical parlance. The solution that
corresponds to the driving term on the right-hand side is called the
inhomogeneous solution.} The resonance solution is equivalent to the
null-space solution in matrix theory.

\subsection {Vector Wave Equation}

Similar to the uniqueness conditions for the scalar wave equation,
analogous conditions for the vector wave equation\index{Wave equation!vector} can also be
derived. First, assume that there are two different solutions to a
vector wave Equation, i.e.,
\begin{align}
\nabla\times\dyadg \mu ^{\, -1}\cdot\nabla\times\v E_1(\v r)-\omega
^2\, \dyadg \epsilon\cdot\v E_1(\v r)=\v S(\v r), \label {ENUM7a}
\end{align}
\begin{align}
\nabla\times\dyadg \mu ^{\, -1}\cdot\nabla\times\v E_2(\v r)-\omega
^2\, \dyadg \epsilon\cdot\v E_2(\v r)=\v S(\v r), \label {ENUM7b}
\end{align}
where $\v S(\v r)=i\omega\,\v J(\v r)-\nabla\times\dyadg \mu ^{\,
-1}\cdot\v M(\v r)$ corresponds to a source of finite extent.
Similarly, $\dyadg \mu$ and $\dyadg \epsilon$ correspond to an
inhomogeneity of finite extent.  Subtracting \eqref{ENUM7a} from
\eqref{ENUM7b} then yields
\begin{align}
\nabla\times\dyadg \mu ^{\, -1}\cdot\nabla\times\delta \v E-\omega
^2\, \dyadg \epsilon\cdot\delta\v E=0, \label {ENUM8}
\end{align}
where $\delta\v E=\v E_1-\v E_2$.  The solution is unique if and
only if $\delta \v E=0$.  Next, on multiplying the above by
$\delta\v E^*$, integrating over volume $V$, and using the vector
identity $\v A\cdot\nabla\times\v B=-\nabla\cdot (\v A\times\v B)+\v
B\cdot\nabla\times\v A$, we have
\begin{align}
 -\int \limits _S\^ n\cdot (\delta \v E^*\times\dyadg \mu ^{\,
-1}\cdot\nabla\times\delta\v E)\, dS+\int \limits
_V\nabla\times\delta\v
E^*\cdot\dyadg \mu ^{\, -1}\cdot\nabla\times\delta\v E\, dV\nonumber\\
\shoveright{-\omega ^2\int \limits
_V\delta\v E^*\cdot\dyadg \epsilon\cdot\delta\v E\, dV=0. }\nonumber\\
\vspace{-15pt}\label{ENUM9}
\end{align}
Since $\nabla\times\delta\v E=i\omega\,\dyadg \mu \cdot\delta \v H$,
the above can be rewritten as
\begin{align}
i\omega\int \limits _S\^ n\cdot (\delta\v E^*\times\delta\v H)\,
dS+\omega ^2\int \limits _V(\delta\v H^*\cdot\dyadg \mu ^\dag
\cdot\delta\v H-\delta\v E^*\cdot\dyadg \epsilon\cdot\delta\v E)\,
dV=0.\qquad \label {ENUM10}
\end{align}
Then, taking the imaginary part of \eqref{ENUM10} yields
\begin{align}
{\Im m } \left\{i\omega \int \limits _S\^ n\cdot (\delta\v E^*\times
\delta\v
H)\, dS\right\}\nonumber\\
\shoveright{-\frac {i\omega ^2}{2}\int \limits _V[\delta\v H^*\cdot
(\dyadg \mu ^\dag -\dyadg \mu )\cdot\delta\v H+\delta\v E^*\cdot
(\dyadg \epsilon ^\dag -\dyadg
\epsilon )\cdot\delta \v E]\, dV=0. \quad}\\
\vspace{-15pt} \, \label {ENUM11}
\end{align}

But if the medium is not lossless (either lossy or active), then
$\dyadg \mu ^\dag \neq \dyadg \mu$ and $\dyadg \epsilon ^\dag \neq
\dyadg \epsilon$, and the second integral in \eqref{ENUM11} may not
be zero. Moreover, if
\begin{enumerate}[label=(\roman*)]
\item $\^ n\times\delta\v E=0$ or $\^ n\times\delta\v H=0$ on $S$,
\item $\^ n\times\delta \v E=0$ on one part of $S$ and $\^
n\times\delta\v H=0$ on the rest of $S$, or
\item $\delta \v H-i\zeta \^n\times \delta\v E=0$ on $S$, where
$\zeta$ is a real number,
\end{enumerate}
\noindent then the first integral in \eqref{ENUM11} vanishes. The
above corresponds to lossless boundary conditions\index{Boundary conditions!lossless} for the difference
field. The third case corresponds to a lossless reactive impedance
boundary condition\index{Boundary conditions!lossless reactive impedance}. \footnote{A more complicated boundary condition
for the third case may be designed.}  Again, $\zeta$ can vary on
$S$, and the first two cases can be made special cases of the third
case.

The above implies that,
\begin{align}
\frac {\omega ^2}2\int \limits _V[\delta\v H^*\cdot i(\dyadg \mu
^\dag -\dyadg \mu )\cdot \delta\v H+\delta\v E^*\cdot i(\dyadg
\epsilon ^\dag -\dyadg \epsilon )\cdot\delta\v E]\, dV=0. \label
{ENUM12}
\end{align}
In the above, $i(\dyadg \mu ^\dag -\dyadg \mu )$ and
$i(\dyadg\epsilon ^\dag -\dyadg \epsilon )$ are Hermitian matrices.
Moreover, the integrand will be positive definite if both $\dyadg
\mu$ and $\dyadg \epsilon$ are lossy, and the integrand will be
negative definite if both $\dyadg \mu $ and $\dyadg \epsilon $ are
active.  Hence, the only way for \eqref{ENUM12} to be satisfied is
for $\delta\v E=0$ and $\delta\v H=0$, or that $\v E_1=\v E_2$ and
$\v H_1=\v H_2$, implying uniqueness.

Consequently, in order for uniqueness to be guaranteed, either
\begin{enumerate}[label=(\roman*)]
\item $\^ n\times\v E_1=\^ n\times\v E_2$ on $S$ or $\^ n\times\v
H_1=\^ n\times\v H_2$ on $S$,
\item $\^ n\times\v E_1=\^ n\times\v E_2$ on a part
of $S$ while $\^ n\times\v H_1=\^ n\times\v H_2$ on the rest of $S$,
or
\item $ \v H_1-i\zeta \^n\times \v E_1=\v H_2-i\zeta \^n\times \v
E_2$ on $S$.
\end{enumerate}
\noindent In other words, if two solutions satisfy the same boundary
conditions for tangential $\v E$ or tangential $\v H$, or a mixture
thereof on $S$, or the same reactive boundary condition, the two
solutions must be identical.

     Again, the requirement for a nonlossless condition is to eliminate the real
resonance solutions which could otherwise be time harmonic,
homogeneous solutions to \eqref{ENUM7a} satisfying the boundary
conditions. For example, if the appropriate boundary conditions for
$\delta \v E$ and $\delta\v H$ are imposed so that the first term of
\eqref{ENUM10} is zero, then
\begin{align}
\int\limits _V(\delta\v H^*\cdot\dyadg \mu^\dag \cdot\delta\v
H-\delta\v E^*\cdot\dyadg \epsilon\cdot\delta\v E)\, dV=0. \label
{ENUM13}
\end{align}
The above does not imply that $\delta\v E$ or $\delta\v H$ equals
zero, because at resonances, a perfect balance between the energy
stored in the electric field and the energy stored in the magnetic
field is maintained.  As a result, the left-hand side of the above
could vanish without having $\delta\v E$ and $\delta\v H$ be zero,
which is necessary for uniqueness.  But away from the resonances of
the volume $V$, the energy stored in the electric field is not equal
to that stored in the magnetic field.  Hence, in order for
\eqref{ENUM13} to be satisfied, $\delta\v E$ and $\delta \v H$ have
to be zero since each term in \eqref{ENUM13} is positive definite
for lossless media due to the Hermitian nature of $\dyadg\mu$ and
$\dyadg \epsilon $.

     When $V\rightarrow\infty$, as in the scalar wave equation case,
some loss has to be imposed to guarantee uniqueness.  This is the
same as requiring the wave to be outgoing at infinity, namely, the
radiation condition. Again, the radiation condition can be imposed
for an unbounded medium with vanishing loss to guarantee uniqueness.


\section{Transformation Matrices for Microwave Circuits}

\subsection{Impedance and Admittance Matrices}

A general microwave circuit consists of many ports.  A convenient
way to characterize an $N$-port network is to describe the network
in terms of impedance matrices\index{Impedance matrix} or admittance matrices\index{Admittance matrix}\cite{COLLINE}.
For example, if the $N$-port network can be characterized by a pair
of voltage and current at each port, then a column vector of
voltages can be defined and also a column vector of currents.  We
can write down a relationship between the voltages and the currents
as
\begin{equation}
\v V= \bmatrix V_1\\ V_2\\ \vdots\\ V_{N} \endbmatrix =
\bmatrix  Z_{11} & Z_{12} & \cdots & Z_{1N}  \\
Z_{21} & Z_{22} & \cdots & Z_{2N} \\
\vdots & \vdots & \ddots & \vdots \\
Z_{N1} & Z_{N2} & \cdots & Z_{NN}
 \endbmatrix
\bmatrix I_1\\ I_2\\ \vdots\\ I_{N} \endbmatrix = \dyad Z\cdot\v I
\, \label{eq131}
\end{equation}
By the same token, we can express
\begin{equation}
\v I=\dyad Y\cdot\v V\, \label{eq132}
\end{equation}
where $\dyad Y$ is the admittance matrix. For reciprocal circuits\index{Reciprocal circuit},
it can be shown that $\dyad Z$ and $\dyad Y$ are symmetric
matrices.  For lossless circuits, it can be shown that these
matrices have pure imaginary elements.

For a two port network which is reciprocal, there are three
independent matrix elements.  Therefore, a two-port network can
often be modeled by a T or a $\Pi$ equivalent circuit.

\begin{figure}[htb]
\begin{center}
\hfil\includegraphics[width=4.5truein]{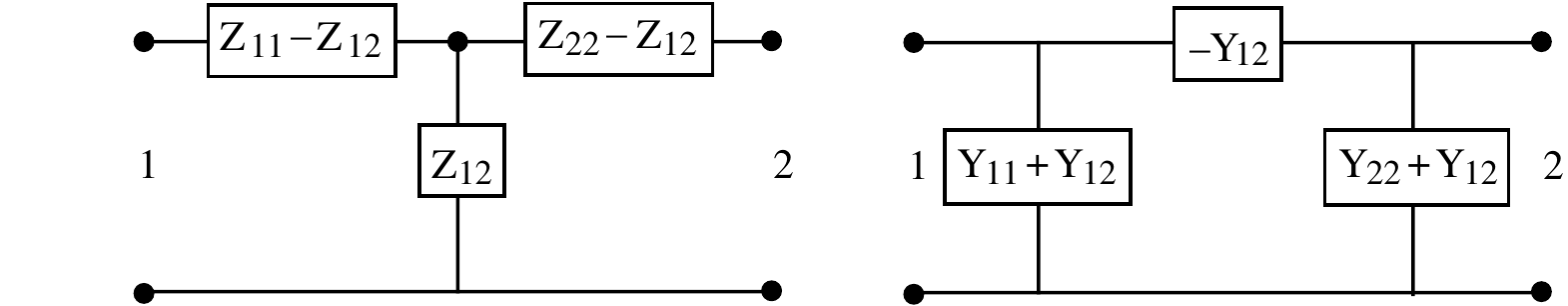}\hfil
\end{center}
\caption{The T and $\Pi$ equivalent circuits for a two-port
circuit.}\label{16}
\end{figure}


\subsection { Scattering Matrices}\index{Scattering matrix}

For high frequencies, it is more pertinent to think about waves.
Then at each port, we can define an incident and a reflected wave.
For instance, we can define an incident voltage wave $V^+$ and a
reflected voltage wave $V^-$.  A relationship can then be written
between the reflected waves at all the ports to the incident waves
at all the ports.
\begin{equation}
\v V= \bmatrix V_1^-\\ V_2^-\\ \vdots\\ V_{N}^-\\ \endbmatrix =
\bmatrix S_{11} & S_{12} & \cdots & S_{1N}\\
S_{21} & S_{22} & \cdots & S_{2N}\\
\vdots & \vdots & \ddots & \vdots\\
S_{N1} & S_{N2} & \cdots & S_{NN}\\ \endbmatrix \bmatrix V_1^+\\
V_2^+\\ \vdots\\ V_{N}^+\\ \endbmatrix = \dyad S\cdot\v V^+ \,
\label{eq133}
\end{equation}
It can be proved that $\dyad S$ has to be symmetric for reciprocal
circuits, and that it has to be unitary if the circuit is
lossless.

\subsection { Chain Matrices}

When one needs to cascade a series of two port networks, it is
more convenient to work with chain matrices or transmission
matrices.  A voltage and current transmission matrix\index{Transmission matrix} relates the
voltage and current at one port to the voltage and current at the
second port.

\begin{figure}[htb]
\begin{center}
\hfil\includegraphics[width=3.0truein]{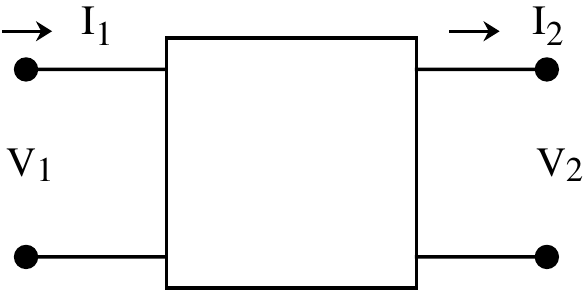}\hfil
\end{center}
\caption{A diagram for defining the voltage and current
transmission matrix.}\label{17}
\end{figure}


Written explicitly, we have
\begin{equation}
\bmatrix V_1\\I_1\\ \endbmatrix= \bmatrix A_1 & B_1\\C_1 & D_1\\
\endbmatrix \bmatrix V_2\\I_2\\ \endbmatrix \, \label{eq134}
\end{equation}
Notice that the current at Port 2 is flowing out of the port
rather than into the port.  In this manner, if we have a second
transmission matrix of a second network that relates $V_2, I_2$ to
$V_3, I_3$, viz.,
\begin{equation}
\bmatrix V_2\\I_2\\ \endbmatrix= \bmatrix A_2 & B_2\\C_2 & D_2\\
\endbmatrix \bmatrix V_3\\I_3\\ \endbmatrix \, \label{eq135}
\end{equation}
Hence, when these two networks are cascaded together, the
resultant transmission matrix is the product of the two matrices
\begin{equation}
\bmatrix V_2\\I_2\\ \endbmatrix= \bmatrix A_1 & B_1\\C_1 & D_1\\
\endbmatrix \bmatrix A_2 & B_2\\C_2 & D_2\\ \endbmatrix \bmatrix
V_3\\I_3\\ \endbmatrix \, \label{eq136}
\end{equation}

For a two port network, it can be shown that
\begin{subequations}
\begin{equation}
A=Z_{11}/Z_{12}, \quad B=(Z_{11}Z_{22}-Z_{12}^2)/Z_{12},
\label{eq137a}
\end{equation}
\begin{equation}
C=1/Z_{12}, \quad D=Z_{22}/Z_{12}, \label{eq137b}
\end{equation}
\end{subequations}
for a reciprocal network.  It is also readily verified that
\begin{equation}
AD-BC=1\, \label{eq138}
\end{equation}
for this case.  Hence, the determinant of a chain matrix\index{Chain matrix} is always
1 for a reciprocal network\index{Reciprocal network}.

\vfil\eject

\centerline{\bf Exercises for Chapter 1 }

\vskip 12pt \noindent {\bf Problem 1-1:} \noindent The fundamental
units in electromagnetics can be considered to be meter, kilogram,
second, and coulomb.
\begin{itemize}
\item [(a)] Show that 1 volt, which is 1 watt/amp, has the
dimension of (kilogram meter$^2$)/(coulomb sec$^2$). \item [(b)]
From Maxwell's equations, show that $\mu_0$ has the dimension of
(second volt)/(meter amp), and hence, its dimension is (kilogram
meter)/coulomb$^2$ in the more fundamental units.
\item [(c)] If we assign the value of $\mu_0$ to be $4\pi$ instead
of $4\pi\times 10^{-7}$, what would be the unit of coulomb in this
new assignment compared to the old unit?  What would be the
present value of 1 volt and 1 amp in this new assignment?
\end{itemize}

\noindent {\bf Problem 1-2:} \noindent Show that for two
time-harmonic functions,
\begin{align}\langle A(\v r, t) B(\v r, t)
\rangle =\frac 1{2} {\Re e} [A(\v r)B^\ast (\v r)],
\end{align}
where $A(\v r)$ and $B(\v r)$ are the phasors of $A(\v r, t)$ and
$B(\v r, t)$.The angular brackets above imply time averaging.

\vskip 8pt \noindent {\bf Problem 1-3:} \noindent Assume that a
voltage is time harmonic, i.e., $V(t)=V_0\cos \omega t$, and that
a current $I(t)=I_I\cos \omega t+ I_Q\sin \omega t$, i.e., it
consist of an in-phase and a quadrature component.
\begin{itemize}
\item [(a)] Find the instantaneous power due to this voltage and current, viz.,$V(t)I(t)$.
\item [(b)] Find the phaser representations of the voltage and current, and hence the complex power due to this voltage and current.
\item [(c)] Establish a relationship between the real part and reactive part of the complex power to the instantaneous power.
\item [(d)] Show that the reactive power is due to the quadrature component of the current, which is related to a time-varying part of the instantaneous power with zero-time average.
\end{itemize}

\noindent {\bf Problem 1-4:} \noindent
 For a scalar-wave equation, $\nabla\cdot
\epsilon ^{-1}(\v r)\nabla\phi (\v r)+k^2\phi (\v r)=S(\v r)$:
\begin{itemize}
\item [(a)] Show that a reciprocal
relationship $\langle \phi _1(\v r), S_2(\v r)\rangle =\langle
\phi _2(\v r), S_1(\v r) \rangle $ exists.
\item [(b)] What is the boundary condition satisfied by $\phi$ at an interface
where $\epsilon (\v r)$ has a step discontinuity?
\end{itemize}

\noindent {\bf Problem 1-5:} \noindent
\begin{itemize}
\item[(a)] Prove that for reciprocal circuits, the impedance matrix and the admittance matrix are symmetric.
\item[(b)] Prove that for lossless circuits, the impedance matrix and the admittance matrix have imaginary elements.
\end{itemize}

\noindent {\bf Problem 1-6:} \noindent
\begin{itemize}
\item[(a)] Prove that for reciprocal circuits, the scattering matrix is symmetric.
\item[(b)] Prove that for lossless circuits, the scattering matrix  is unitary.
\item[(c)] Prove that for reciprocal circuits, the determinant of the chain matrix is always equal to one.
\end{itemize}

\bibliographystyle{plain}
\bibliography{emt,matrix,fmm,math,fastsum,parallel,cs,my}

\def\chaptitle{Hollow Waveguides}

\chapter{\chaptitle}

\markboth{\smallbooktitle}{\chaptitle}






\def\v #1{{\bf #1}}
\def\vg #1{{\boldsymbol #1}}
\def\dyad#1{\overline {\bf #1}}
\def\dyadg#1{\overline {\vg #1}}
\def\beq{\begin{equation}}\def\eeq{\end{equation}}
\def\tinf{\text{\it inf\,}}\def\^{\hat}
\def\cal#1{\mathcal{#1}}
\def\ed{\end{document}}
\def\dalpha{\dyadg\alpha}
\def\dbeta{\dyadg\beta}
\def\Rg{\Re g}
\def\l{\label}
\def\sinc{\text{sinc}}


\numberwithin{equation}{section}







Much work has been written on hollow waveguides\index{Hollow waveguides}
\cite{LAMONT,MARCUVITZS,BUDDEN,COLLINF,COLLING,KONG,POZAR}. Theory
on guided waves dates back even to earlier dates as alluded to in
Chapter 1.  Hollow metallic waveguides, when filled with air, allow
for high power microwave transmission.  However, a hollow
electromagnetic waveguide\index{Waveguide!hollow}, unlike an acoustic waveguide, has a lower
\index{Cutoff frequency}cutoff frequency of operation.  The way to counter this is to use a
two-conductor waveguide such as a coaxial waveguide.  The support of
the inner conductor in such waveguides requires the filling of the
waveguide with dielectric materials.  It will be shown that when a
waveguide is homogeneously filled with materials, the theory is
essentially the same as that for one which is hollow.

\section{{{ General Uniform Cylindrical Waveguides }}}

\setcounter{equation}{0} \setcounter{figure}{0}

General uniform cylindrical waveguides\index{Waveguide!uniform cylindrical} include transmission lines,
and hollow waveguides, as well as multi-conductor waveguides. If
the regions between the conductors are filled with a homogeneous
material, the waveguide can support \index{Mode} purely transverse electric
(TE)\index{Mode!TE}, transverse magnetic (TM)\index{Mode!TM} or transverse electromagnetic (TEM)\index{Mode!TEM}
modes. If the wave is TEM, the waveguide is operating in the
transmission line mode. It can be shown that a hollow waveguide
cannot support a TEM wave. Therefore, a transmission line needs to
have at least two conductors.

\begin{figure}[htb]
\begin{center}
\hfil\includegraphics[width=3.0truein]{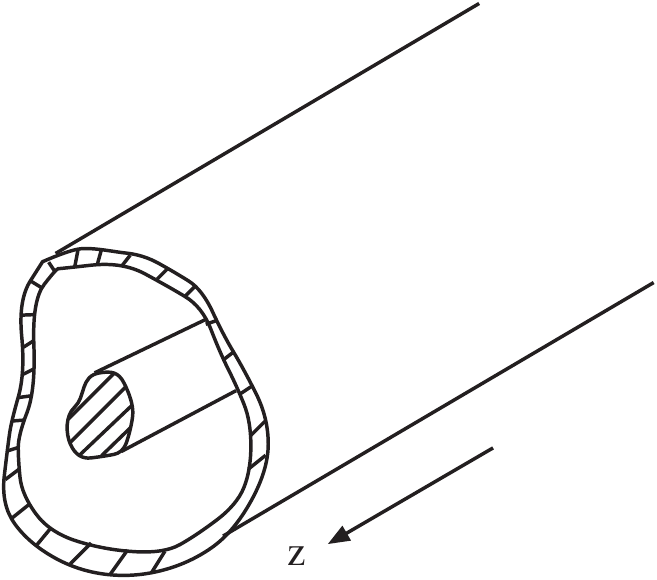}\hfil
\end{center}
\caption{Cross-section of a general uniform cylindrical
waveguide.}\label{fg211}
\end{figure}


Because the fields of a general uniform cylindrical waveguide can
be decomposed into TE and TM types, we can characterize the TE
wave with the $\hat z$-component of the magnetic field or $H_z$,
since $H_z\ne 0$ for this type of wave. Similarly, we can
characterize the TM wave with the $\hat z$-component of the
magnetic field or $E_z$ since $E_z\ne 0$ for this type of wave.

We shall derive the equations governing the $E_z$ and $H_z$
components of the fields in a general, uniform cylindrical
waveguide filled with a homogeneous material. We shall call such
waveguides hollow waveguides. We let \cite{KONG}
\begin{equation}
\v E=\v E_s+\^ zE_z,\qquad\v H=\v H_s+\^zH_z, \label{eq21}
\end{equation}
where the subscript $s$ represents the transverse to $z$
components. Substituting the above equations into Maxwell's
equations, we have
\begin{equation}
\left (\nabla _s+\^ z\frac {\partial }{\partial z}\right )\times
(\v E_s+\^ zE_z)=i\omega\mu (\v H_s+\^ zH_z), \label{eq22}
\end{equation}
\begin{equation}
\left (\nabla _s+\^ z\frac {\partial }{\partial z}\right )\times
(\v H_s+\^ zH_z)=-i\omega\epsilon (\v E_s+\^ zE_z). \label{eq23}
\end{equation}
Equating the $s$ components of (\ref{eq22}) and (\ref{eq23}), we
have
\begin{equation}
\nabla _s\times\^ zE_z+\frac {\partial }{\partial z}\^ z\times\v
E_s=i\omega\mu \v H_s, \label{eq24}
\end{equation}\begin{equation}
\nabla _s\times\^ zH_z+\frac {\partial }{\partial z}\^ z\times\v
H_s=-i\omega\epsilon \v E_s. \label{eq25}
\end{equation}
Substituting for $\v E_s$ from (\ref{eq25}) into (\ref{eq24}), we
have
\begin{equation}
\omega ^2\mu\epsilon\v H_s=-i\omega\epsilon\nabla _s\times\^
zE_z+\frac {\partial }{\partial z}\^ z\times \left( \nabla
_s\times\^ zH_z+\frac {\partial }{\partial z}\^ z\times\v H_s
\right). \label{eq26}
\end{equation}
Using the vector identities  $\^ z\times\nabla _s\times\^ z=\nabla
_s, \quad\^z\times\^ z\times\v H_s=-\v H_s$, and assuming that the
field has $e^{\pm ik_zz}$ dependence, so that $\frac
{\partial^2}{\partial z^2}\rightarrow -k_z^2$, we can rewrite
(\ref{eq26}) as
\begin{equation}
\v H_s=\frac 1{k^2-k_z^2}\left [\frac {\partial }{\partial
z}\nabla _sH_z+i\omega\epsilon\^ z\times\nabla _sE_z\right ],
\label{eq27}
\end{equation}
where $k^2 = \omega ^2\mu\epsilon $. By the same token, we have
\begin{equation}
\v E_s=\frac 1{k^2-k_z^2}\left [\frac {\partial }{\partial
z}\nabla _sE_z-i\omega\mu\^ z\times\nabla _sH_z\right ].
\label{eq28}
\end{equation}
Equations (\ref{eq27}) and (\ref{eq28}) allow us to derive all the
other components of the field in a waveguide once we know the $z$
components of the field.

If we equate the $z$ components of (\ref{eq22}) and (\ref{eq23}),
we have
\begin{equation}
\nabla _s\times\v E_s=i\omega\mu\hat z H_z, \label{eq29}
\end{equation}
\begin{equation}
\nabla _s\times\v H_s=-i\omega\epsilon\hat z E_z. \label{eq210}
\end{equation}
Substituting (\ref{eq27}) and (\ref{eq28}) into (\ref{eq29}) and
(\ref{eq210}), we have
\begin{equation}
(\nabla _s^2+k_s^2)H_z=0, \qquad\text {for}\quad \text {TE}\quad
\text {waves}, \label{eq211}
\end{equation}
\begin{equation}
(\nabla _s^2+k_s^2)E_z=0, \qquad\text {for}\quad \text {TM}\quad
\text {waves}, \label{eq212}
\end{equation}
where $k_s^2 = k^2 - k_z^2$.  Therefore, $H_z$ and $E_z$ satisfy a
two-dimensional scalar wave equation\index{Wave equation!scalar} also known as the \emph{
reduced wave equation}\index{Wave equation!reduced}. Once $E_z$ and $H_z$ are solved for from
(\ref{eq211}) and (\ref{eq212}), we can derive all the other field
components using (\ref{eq27}) and (\ref{eq28}).

In the above, \eqref{eq211} and \eqref{eq212} clearly show that
$H_z$ and $E_z$ satisfy the wave equation independently. The $E_z$
component satisfies the homogeneous Dirichlet boundary condition\index{Boundary conditions!waveguide},
$E_z=0$ on the waveguide wall.  The general boundary condition for
the TE field is that $\hat n\times\v E_s=0$ on the waveguide wall.
From \eqref{eq28}, we can show that this is equivalent to $\hat
n\cdot\nabla H_z=0$ on the waveguide wall, which is the homogeneous
Neumann boundary condition. The boundary condition does not couple
the TE and TM waves implying that they can exist independently of
each other.  This substantiates our assumption in the very
beginning.

It is to be noted that an alternative way of solving the above
problem is to define two scalar potentials $\psi_h$ and $\psi_e$,
and let
\begin{align}
\v E^{TE}=\nabla\times\hat z \psi_h\\
\v H^{TM}=\nabla\times\hat z \psi_e
\end{align}
It can be easily shown that $H_z$ from the TE field is proportional
to $\psi_h$ and $E_z$ from the TM field is proportional to $\psi_e$.
Hence, this method of solution is completely equivalent to our
previous method of solution.

We can envision that a wave is guided in a waveguide because the
wave is bouncing around the wall of the waveguide.  This is the
bouncing wave picture of the wave. In this case, the $\v k$ vector
is not pointing in the $z$ direction completely, and $k_z<k$. The
mode is either TE or TM. However, it is also possible to have a wave
directly transmitted through a waveguide, such as in the
transmission line. In this case, the $\v k$ vector is pointing
entirely in the $z$ direction, and $k_z=k$.  We shall study this
mode for transmission lines.

\section {{{ Wave Impedance}}}

With the above results, we can define wave impedance\index{Wave impedance} concepts in a
hollow waveguide.  Just as the intrinsic impedance in a homogeneous
medium relates the electric field to a magnetic field, we may relate
the transverse components\index{Field!transverse components} of $\v E$ and $\v H$ in a waveguide by a
wave impedance.

For a TE mode\index{Mode!TE}, the transverse components are
\begin{equation}
\v H_s=\frac {ik_z}{k_s^2}\nabla _sH_z, \qquad\v E_s=\frac
{-i\omega\mu }{k_s^2}\^ z\times\nabla _sH_z. \label{eq25-1}
\end{equation}
Therefore, the wave impedance is
\begin{equation}
Z^{TE}=\frac {\^ z\times\v E_s}{\v H_s}=\frac {\omega\mu }{k_z}.
\label{eq25-2}
\end{equation}
For a TM mode\index{Mode!TM}, the transverse components are
\begin{equation}
\v H_s=\frac {i\omega\epsilon }{k_s^2}\^ z\times\nabla _sE_z,
\qquad\v E_s=\frac {ik_z}{k_s^2}\nabla _sE_z. \label{eq25-3}
\end{equation}
Therefore, the wave impedance is
\begin{equation}
Z^{TM}=\frac {\^ z\times\v E_s}{\v H_s}=\frac {k_z}{\omega\epsilon
}. \label{25-4}
\end{equation}

The above is valid for a general cylindrical waveguide which is
homogeneously filled. It is useful in deriving equivalent
transmission line models for a waveguide.

\section{{{ Transmission Line Theory}}}
\index{Transmission line}

\begin{figure}[htb]
\begin{center}
\includegraphics[width=3.0truein]{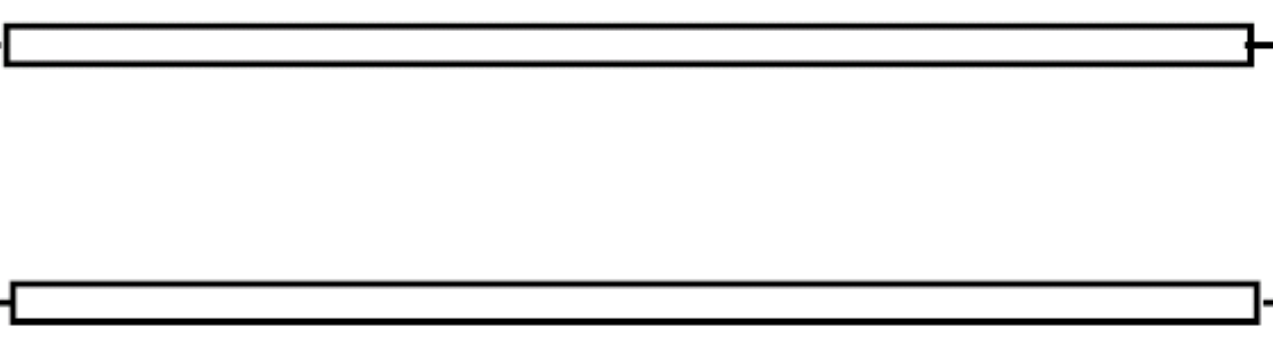}
\includegraphics[width=3.0truein]{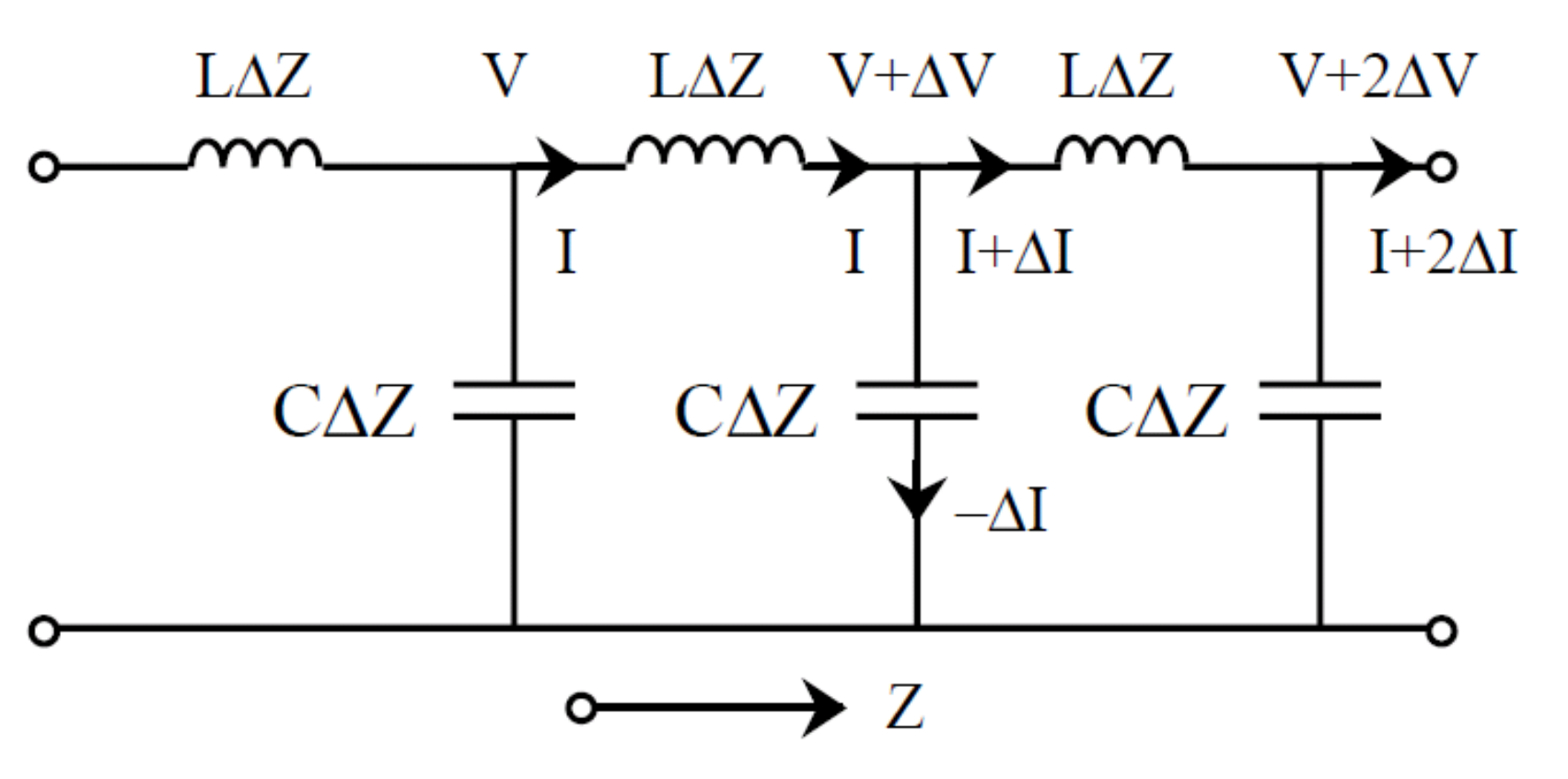}
\end{center}
\caption{Two pieces of parallel wires form a transmission line, even
when they are sinuous in nature.  A wire has inductance, and
capacitance exists between two pieces of metal.  Hence, circuit
model used to derive the telegraphers
equations.}\label{TelegraphersModel}
\end{figure}

The propagation of waves on the transmission line\index{Transmission line!theory} was first
formulated in terms of telegrapher's equations\index{Telegrapher's equations}.  Telegraphy was in
use in the early 1800s even before the completion of Maxwell's
equations in 1864.  The telegraphers equations can be derived using
circuit theory (see Fig. \ref{TelegraphersModel}), and they are
valid even for meandering lines. The transmission line can be
thought of as consisting of a sequence of coupled $L$ and $C$ tank
circuits. Each tank circuit forms a simple harmonic oscillator\index{Harmonic oscillator}.  But
on a transmission line, these harmonic oscillators are coupled
together. It is through the coupling of these harmonic oscillators
that a wave can propagate on a transmission line.

\subsection{ TEM Mode of a Transmission Line}

In a transmission line, a TEM (transverse electromagnetic) mode\index{Mode!TEM}
can propagate. For TEM waves, both $E_z$ and $H_z$ are zero.  By
looking at (\ref{eq27}) and (\ref{eq28}), $\v H_s$ and $\v E_s$
will be non-zero only if $k_z = k$. Therefore, all TEM waves, or
TEM modes in a waveguide have $e^{ikz}$ dependence.  Furthermore,
from (\ref{eq29}) and (\ref{eq210}), we conclude that for TEM
waves
\begin{equation}
\nabla _s\times\v E_s=0, \qquad\nabla _s\times\v H_s=0.
\label{eq22-1}
\end{equation}
In other words, $\v E_s$ is electrostatic in the $xy$ plane while
$\v H_s$ is magnetostatic in the $xy$ plane.  Hence, we can let
\begin{equation}
\v E_s=-\nabla _s\phi _s(x, y)e^{ikz}, \qquad\v H_s=-\nabla_s\psi
_s(x,y)e^{ikz}. \label{eq22-2}
\end{equation}
Since $\nabla\cdot\v E_s=0$ and $\nabla \cdot\v H_s=0$ inside the
waveguide, $\phi _s$ and $\psi _s$ satisfy Laplace equations
\begin{equation}
\nabla _s^2\phi _s(x, y)=0, \qquad\nabla _s^2\psi _s(x, y)=0.
\label{eq22-3}
\end{equation}
If we have perfect electric conductors, the boundary conditions
are $\hat n\times\v E_s=0$, and $\^ n\cdot\v H_s=0$ on the
metallic surfaces. These boundary conditions translate to
\begin{equation}
\phi _s=\text {constant}, \label{eq22-4}
\end{equation}
\begin{equation}
\^ n\cdot\nabla _s\psi _s=\frac {\partial }{\partial n}\psi _s=0.
\label{eq22-5}
\end{equation}
Equation (\ref{eq22-4}) is known as the {\it Dirichlet} boundary
condition while (\ref{eq22-5}) is the {\it Neumann} boundary
condition. The constants in (\ref{eq22-4}) are the potentials on
the conductors which may be different for different conductors.
$\psi_s$ is a multi-value function because magnetic field always
goes in a loop and ends on itself.  In order to avoid dealing with
multi-value functions, it is a lot easier to solve for $\phi_s$.
Alternatively, one can solve for the magnetic field using a vector
potential.

\begin{figure}[htb]
\begin{center}
\hfil\includegraphics[width=3.0truein]{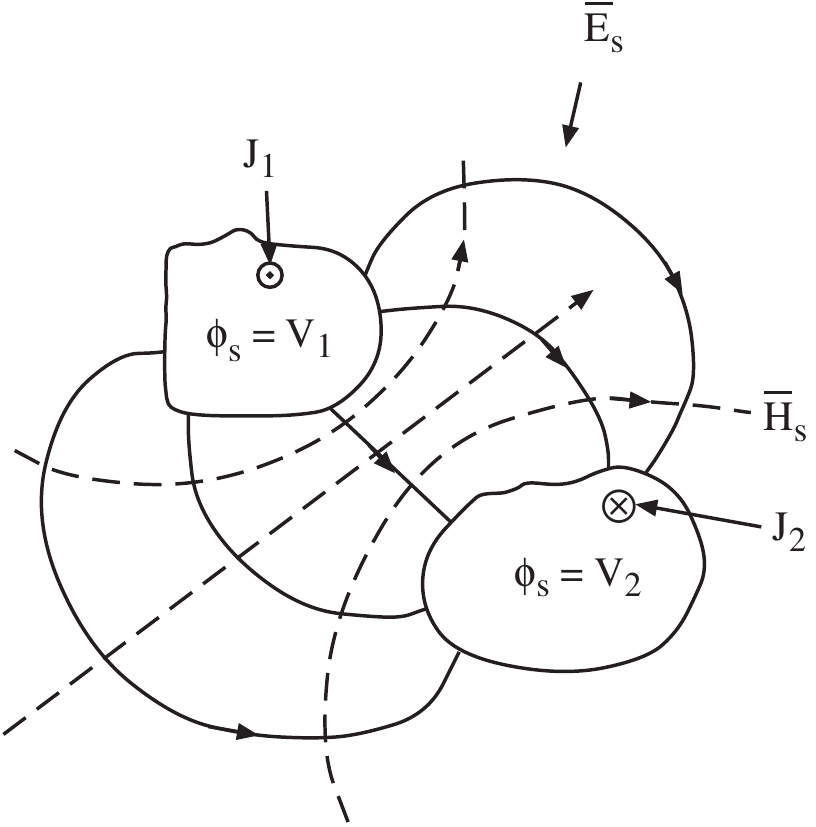}\hfil
\end{center}
\caption{TEM mode in a transmission line.}\label{fgTEM}
\end{figure}


At this point, it seems that $\phi _s$ and $\psi _s$ are
decoupled, and hence, the electric field and the magnetic field
are independent of each other. This could not be true, as the
coupling is expressed in Equations (\ref{eq24}) and (\ref{eq25})
(if we set $E_z = H_z = 0$ for discussing TEM modes). Hence, the
coupling of the fields is only through the $\^ z$-variation of the
fields, i.e.,
\begin{equation}
\frac  {\partial }{\partial z}\^ z\times\v E_s=i\omega\mu\v H_s,
\label{eq22-6}
\end{equation}
\begin{equation}
\frac {\partial }{\partial z}\^ z\times\v H_s=-i\omega\epsilon\v
E_s. \label{eq22-7}
\end{equation}
Furthermore, from the above, we deduce that $\v H_s$ and $\v E_s$
are mutually orthogonal in the TEM mode. Since the fields have
$e^{ikz}$ dependence, we conclude that
\begin{equation}
\^ z\times\v E_s=\sqrt {\frac {\mu }{\epsilon }}\v H_s\equiv\eta\v H_s,
\label{eq22-8}
\end{equation}
if we assume that the wave is only propagating in one direction.
Then $\eta$ is also known as the {\it intrinsic impedance}\index{Intrinsic impedance} of the
medium, and all TEM waves satisfy (\ref{eq22-8}).

In \eqref{eq22-3}, the electrostatic and the magnetostatic problems
are seemingly decoupled from each other, but the fields are coupled
via their $z$ variation, as indicated by the above equations.  In
\eqref{eq22-8}, it says that once we know the electrostatic field,
the magnetostatic field can be easily derived.  Hence, we need only
to solve the electrostatic problem to fully characterize the TEM
solution.

\subsubsection{Derivation of the Telegrapher Equations}

We can integrate Equation (\ref{eq22-6}) about a line contour
around one of the conductors to obtain \index{Telegrapher's equations!derivation}
\begin{equation}
\frac {\partial }{\partial z}\oint _C\^ z\times\v E_s\cdot d\v
l=i\omega\mu\oint _C\v H_s\cdot d\v l. \label{eq22-9}
\end{equation}
By Ampere's law, we have $\oint _C\v H_s\cdot d\v l=I$, the total
current on one of the conductors. For the left-hand side, we have
\begin{equation}
\begin{split}
\oint _C\^ z\times\v E_s\cdot d\v l &=\oint _C d\v l\times\hat{z}\cdot\dyad E_s\\
&=\frac{Q}{\epsilon},
\end{split}
\label{eq22-10}
\end{equation}
Since $d\v l\times\hat {z}$ is an outward normal to $C$ and Gauss'
theorem can be invoked.
Therefore, Equation (\ref{eq22-9}) becomes
\begin{equation}
\frac d{dz}Q=i\omega\mu\epsilon I. \label{eq22-13}
\end{equation}
Since the transverse field is purely static, we can define $Q =
CV$ where $C$ is the capacitance per unit length and $V = V_1 -
V_2$.  Hence, (\ref{eq22-13}) becomes
\begin{equation}
\frac d{dz}V=i\omega\frac {\mu\epsilon }C I. \label{eq22-14}
\end{equation}

Since ${\mu\epsilon }/C$ has the dimension of henry per meter, we
can define $L= {\mu\epsilon }/C$, an inductance per unit length,
and (\ref{eq22-14}) becomes

\begin{figure}[htb]
\begin{center}
\hfil\includegraphics[width=3.0truein]{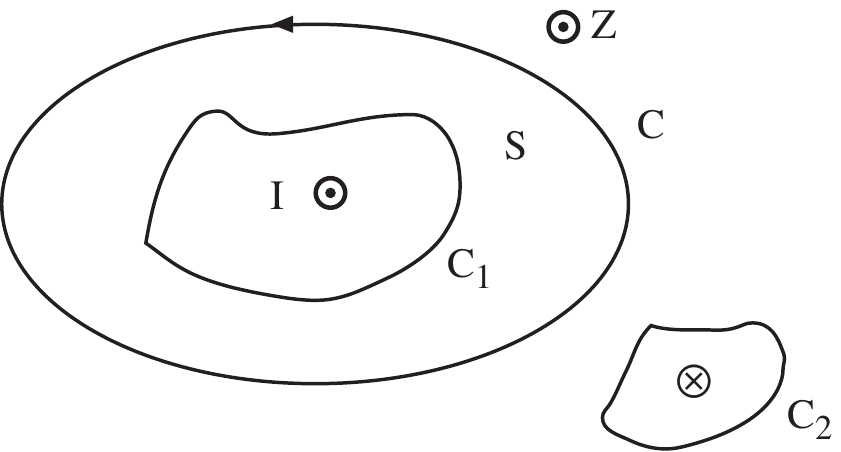}\hfil
\end{center}
\caption{Derivation of the telegrapher's equations for a
transmission line.}\label{fgDerivation}
\end{figure}


\begin{equation}
\frac {dV}{dz}=i\omega LI. \label{eq22-15}
\end{equation}
With similar manipulations to  $\^ z\times (\ref{eq22-7})$, we
obtain
\begin{equation}
\frac {dI}{dz}=i\omega CV. \label{eq22-16}
\end{equation}
Equations (\ref{eq22-15}) and (\ref{eq22-16}) are the telegrapher's
equations (also known as telegrapher's equations) for a transmission
line, which can also be derived from a circuits viewpoint. The
velocity of the wave on the line is given by
\begin{equation}
v=\frac 1{\sqrt {\mu\epsilon }}=\frac 1{\sqrt {LC}}.
\label{eq22-17}
\end{equation}
Notice that in the above, we need only to solve the electrostatic
problem to obtain $C$, and $L$ can be derived from $C$.   There is
no need to solve the magnetostatic problem.

\subsubsection{Characteristic Impedance versus Intrinsic Impedance}

Since the fields, and hence the voltage and current, have
$e^{ikz}$ dependence,where $k = \omega\sqrt {\mu\epsilon
}=\omega\sqrt {LC}$, we deduce either from (\ref{eq22-15}) or
(\ref{eq22-16}) that
\begin{equation}
\frac V{I}=\sqrt {\frac {L}C}=Z_0, \label{eq22-18}
\end{equation}
if the wave is only propagating in the positive $z$ direction.
$Z_0$ is also known as the {\it characteristic impedance}\index{Characteristic impedance} of a
transmission line. Since $C$ and $L$ are dependent on the geometry
of the transmission line, $Z_0$ is a geometry dependent impedance.
This is unlike $\eta$, the intrinsic impedance \index{Intrinsic impedance}. It can be easily
shown from (\ref{eq22-15}) and (\ref{eq22-16}) that
\begin{equation}
\frac {d^2V}{dz^2}+\omega ^2LCV=0, \label{eq22-19}
\end{equation}
\begin{equation}
\frac {d^2I}{dz^2}+\omega ^2LCI=0, \label{eq22-20}
\end{equation}
which are one-dimensional scalar wave equations\index{Wave equation!scalar} (or Helmholtz wave
equations)\index{Helmholtz equation}.

\subsubsection{Energy Density and Power Flow}

The time average energy\index{Energy density} stored per unit length in a transmission
line for a single propagating wave is given by
\begin{equation}
\langle W_e\rangle =\frac 1{4}\epsilon\int _S\v E_s\cdot\v E_s^*
dS=\frac 1{4} \epsilon\int_S(\nabla _s\phi _s)^2 dS,
\label{eq22-21}
\end{equation}
\begin{equation}
\langle W_m\rangle =\frac 1{4}\mu\int _S\v H_s\cdot\v H_s^*
dS=\frac 1{4}\mu\int _S(\nabla _s\psi _s)^2 dS, \label{eq22-22}
\end{equation}
where $\langle W_e\rangle $ and $\langle W_m\rangle $ are the time
average energy stored in the electric field and the magnetic field
respectively. Using the fact that $\nabla\cdot (\phi\nabla\phi
)=(\nabla\phi )^2+\phi\nabla ^2\phi$, we can write
\begin{equation}
\langle W_e\rangle =\frac 1{4}\epsilon\oint _{C_1+C_2}\phi\frac
{\partial } {\partial n}\phi dl=\frac 1{4}(V_1-V_2)Q=\frac
1{4}CV^2, \label{eq22-23}
\end{equation}
\begin{figure}[htb]
\begin{center}
\hfil\includegraphics[width=3.0truein]{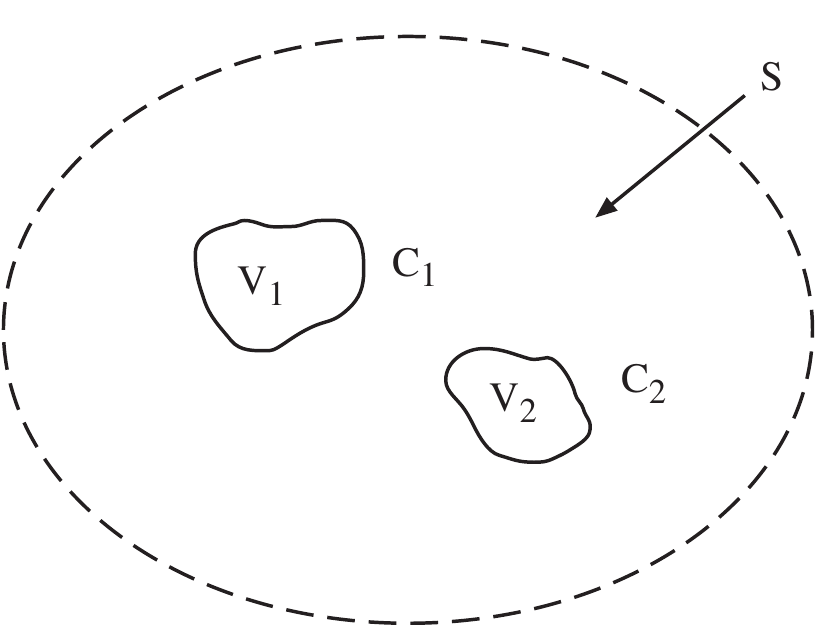}\hfil
\end{center}
\caption{Power flow in a transmission line.}\label{fgPower}
\end{figure}
%
where $V = V_1 - V_2, \quad Q = CV$. The above could also be
derived from circuit theory.  As we have shown before, $|E_s|
=\sqrt {\frac {\mu }{\epsilon }}H_s|$, hence, the time average
energy stored in the magnetic field is
\begin{equation}
\langle W_m\rangle =\langle W_e\rangle =\frac 1{4}CV^2.
\label{eq22-24}
\end{equation}

Since $V^2=Z_0^2I^2=\frac L{C}I^2$, and $\mu\epsilon =LC$, we can
also write
\begin{equation}
\langle W_m\rangle =\frac 1{4}LI^2. \label{eq22-25}
\end{equation}
Equation (\ref{eq22-25}) can also be established by circuit
theory.  It also establishes our definition of $L$ as an
inductance per unit length\index{Inductance}.

The time average power flow\index{Transmission line!time average power flow} down a transmission line is given by
\begin{equation}
\langle P\rangle =\frac 1{2}\Re e \int _S dS\^ z\cdot(\v
E_s\times\v H_s^*). \label{eq22-26}
\end{equation}
Since $\v H_s = \sqrt {\frac {\epsilon }{\mu }}\^ z\times\v E_s$
from (\ref{eq22-20}), we have
\begin{equation}
\langle P\rangle =\frac 1{2}\sqrt {\frac {\epsilon }{\mu }}\int _S
dS|\v E_s|^2=2v\langle W_e\rangle =v\langle W_e+W_m\rangle .
\label{eq22-27}
\end{equation}
Hence, the time average stored energy $\langle W_e + W_m\rangle $
moving at velocity $v$, contributes to power flow.  Equations
(\ref{eq22-15}) and (\ref{eq22-16}) can also be derived from a
circuit model.

\subsection{ Lossy Transmission Lines} \index{Transmission line!lossy}

Since the telegrapher's equations have strictly circuit theory
interpretation, using circuit theory concept, the extension to a
lossy transmission line is straight forward: we replace the series
impedance per unit length $-i\omega L$ by $-i\omega L+ R$, and the
shunt admittance per unit length $-i\omega C$ by $-i\omega C + G$
where $R$ is the series resistance per unit length in the conductor,
while $G$ is the shunt conductance per unit length in the insulator.
The telegrapher's equations then become

\begin{figure}[htb]
\hspace{0.5in}\includegraphics[width=6.0truein]{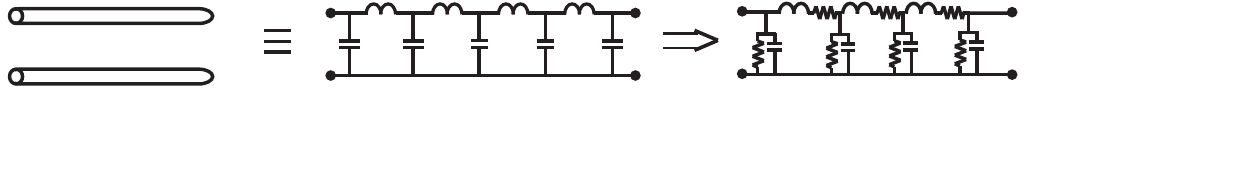}\\
\caption{Circuits equivalence of a transmission line.}\label{Circuits}
\end{figure}


\begin{equation}
\frac {dV}{dz}=(i\omega L-R)I=-ZI, \label{eq22-28}
\end{equation}
\begin{equation}
\frac {dI}{dz}=(i\omega C-G)V=-YV. \label{eq22-29}
\end{equation}
The characteristic impedance is now
\begin{equation}
Z_0=\sqrt {\frac {Z}Y}, \label{eq22-30}
\end{equation}
while the propagation constant becomes
\begin{equation}
k^2=-(i\omega L-R)(i\omega C-G), \qquad k=\omega\sqrt
{LC\left(1+\frac {iR}{\omega L}\right)\left(1+\frac {iG}{\omega
C}\right)}. \label{eq22-31}
\end{equation}
Hence $k$ is complex and the wave $e^{ikz}$ is attenuating.

Strictly speaking, when we have a lossy transmission line due to
conductor loss, a pure \index{Wave!TEM}TEM wave cannot exist.  This is because the
axial current flow meets a resistance, and hence, an axial
component of the electric field is necessary now to drive a
current in the conductor. Therefore, the field is only \index{Quasi-TEM field}quasi-TEM.
However, the conductor loss can be thought of as a small
perturbation of the perfect conductor case, and the
electromagnetic field in the lossy line will not be vastly
different from that of a lossless case.

The \index{Shunt conductance}shunt conductance $G$ in a lossy line can be found as follows.
If the capacitance per unit length between two conductors is given
by the formula
\begin{equation}
C=\epsilon K, \label{eq22-32}
\end{equation}
where $K$ is a geometry dependent factor, the shunt admittance
would be given by $Y = -i\omega C=-i\omega\epsilon K$. If now the
dielectric medium is lossy so that $\epsilon =\epsilon' +\frac
{i\sigma }{\omega }$, then the shunt admittance is given by
\begin{equation}
Y=-i\omega\epsilon' K+\sigma K. \label{eq22-33}
\end{equation}
Hence, we identify $G=\sigma K$.  Note that the derivations in
(\ref{eq22-6}) to (\ref{eq22-18}) hold true even if $\epsilon$ is
complex.  For this reason, $Y$ in (\ref{eq22-33}) is exact.

The \index{Series resistance}series resistance $R$ can be found by calculating the
resistance of the conductor in a perturbative manner when it is
lossy.  The \index{Skin effect} skin-effect will confine the current to flow only on
the surface of the conductor. Since the  \index{Skin depth} skin depth in a conductor
is $\delta =\sqrt {\frac 2{\omega\mu\sigma }}$, the current is
confined to flow in a thinner region at higher frequencies, hence,
increasing this series resistance.

Another way of calculating transmission line loss is via a
perturbation argument and the use of energy conservation.  If a
transmission line is lossy such that $k = k' + ik''$, and


\begin{equation}
V, I\sim e^{ik'z-k''z}, \label{eq22-34}
\end{equation}
Then, the power flow in a line, which is proportional to $|V|^2$
or $|I|^2$ is
\begin{equation}
P\sim e^{-2k''z}. \label{eq22-35}
\end{equation}
By energy conservation,
\begin{equation}
\frac {dP}{dz}=-P_d=-2k''P, \label{eq22-36}
\end{equation}
where $P_d$ is the power dissipated per unit length on the line.
Therefore, the \index{Attenuation constant}attenuation constant $k''$ can be derived to be
\begin{equation}
k''=\frac {P_d}{2P}, \label{eq22-37}
\end{equation}
if we know $P_d$. We can assume $P$ to be close to that of a
lossless line in using (\ref{eq22-37}) in a perturbative concept.

\subsubsection{Absence of TEM Mode in a Hollow Waveguide}
\index{Absence of TEM mode}

\begin{figure}[htb]
\begin{center}
\hfil\includegraphics[width=2.8truein]{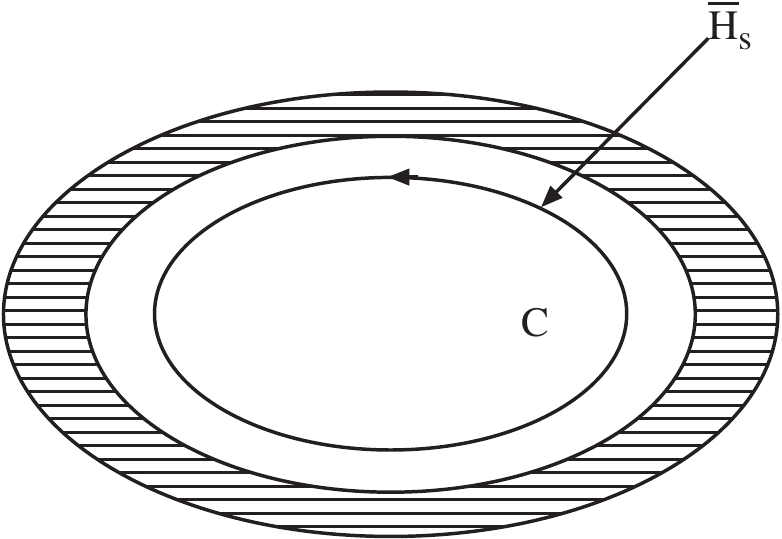}\hfil
\end{center}
\caption{Absence of TEM mode in a hollow, enclosed
waveguide.}\label{Absence}
\end{figure}

Before ending this section, we would like to prove by contradiction
that a hollow waveguide (i.e. without a center conductor) cannot
support a TEM mode as follows. If we assume that it does, then the
magnetic field has to end on itself due to the absence of magnetic
charges. It is clear that $\oint _C\v H_s\cdot d\v l\ne 0$ about any
closed contour following the magnetic field lines. This is clearly
in violation of Equation (\ref{eq22-1}) for a TEM mode which implies
that $\oint\v H_s\cdot d\v l=0$ if $C$ does not enclose any
conducting current. These two results are contradictory implying the
absence of a TEM mode in a hollow waveguide.

\section{{{ TE and TM Modes ($H$ and $E$ Modes)}}}

\subsection{Mode Orthogonality}
\index{Mode orthogonality}

\subsubsection{Mode Orthogonality for Axial Fields}
\index{Field!axial}
As shown previously, for TE and TM waves in a cylindrical waveguide,
we characterize the waves by $H_z$ and $E_z$ respectively.  The
equations governing these two scalar field components are given by
(\ref{eq211}) and (\ref{eq212}). In order to obtain a unique
solution to (\ref{eq211}) and (\ref{eq212}), we have to specify the
boundary conditions for $H_z$ and $E_z$. For a metallic waveguide,
we need to impose the boundary condition that $\^ n\times\v E=0$ on
the metallic surface. This requires $E_z = 0$ on the metallic
surface. From Equation (\ref{eq28}), we see that if $E_z= 0$, and
furthermore, if  $\^ n\cdot\nabla _sH_z =\frac {\partial}{\partial
n}H_z=0$, then $\^ n\times\v E=0$ on the metal surface. Hence, the
equations governing the \index{Mode!TE} \index{Mode!TM} TE and TM modes are

\begin{equation}
(\nabla _s^2+k_s^2)H_z=0, \qquad\frac {\partial }{\partial
n}H_z=0\quad \text {on}\quad S, \quad \text {TE}, \label{eq23-1}
\end{equation}
\begin{equation}
(\nabla _s^2+k_s^2)E_z=0, \qquad E_z=0\quad \text {on}\quad S, \quad
\text {TM}, \label{eq23-2}
\end{equation}
where $S$ is the metallic surface. The homogeneous boundary
conditions in Equation (\ref{eq23-1}) is the \index{Boundary conditions!homogeneous Neumann}homogeneous Neumann
boundary condition, while that for Equation (\ref{eq23-2}) is the
\index{Boundary conditions!homogeneous Dirichlet}homogeneous Dirichlet boundary condition. They are sufficient to
uniquely determine the solutions to the partial differential
equations.

For a closed waveguide, solutions exist for (\ref{eq23-1}) and
(\ref{eq23-2}) at only discrete values of $k_s^2$.  Furthermore,
$k_s^2$ is purely real because (\ref{eq23-1}) and (\ref{eq23-2}) are
self-adjoint problems (see Problem 2.6).  This property is
independent of the homogeneous material filling the waveguide, and
hence is true even for a lossy isotropic material.  It can be shown
easily that different solutions of (\ref{eq23-1}) or (\ref{eq23-2})
corresponding to different $k_{is}^2$ are orthogonal, i.e.,
\begin{equation}
\int\limits _SH_{iz}H_{jz} dS=C_{ih}\delta _{ij}, \qquad \text {TE},
\label{eq23-3}
\end{equation}
\begin{equation}
\int\limits _SE_{iz}E_{jz} dS=C_{ie}\delta _{ij}, \qquad \text {TM},
\label{eq23-4}
\end{equation}
where the integration is over $S$, the cross-section of the
waveguide. We can prove the above assertion quite easily, e.g., by
writing down
\begin{equation}
(\nabla _s^2+k_{is}^2)\psi _{iz}=0, \label{eq23-5}
\end{equation}
\begin{equation}
(\nabla _s^2+k_{js}^2)\psi _{jz}=0, \label{eq23-6}
\end{equation}
where $\psi _z$ in this case can be either $H_z$ or $E_z$.
Multiplying the first equation by $\psi _{jz}$ and the second
equation by $\psi _{iz}$, subtracting the two equations, and
integrating over $S$, we have
$$
(k_{is}^2-k_{js}^2)\int\limits _S\psi _{iz}\psi _{jz} dS=\int\limits
_S (\psi _{iz}\nabla _s^2\psi _{jz}-\psi _{jz}\nabla_s^2\psi _{iz})
dS,
$$
\begin{equation}
\text {Gauss' theorem}\longrightarrow =\oint _C\^ n\cdot (\psi
_{iz}\nabla _s\psi _{jz}-\psi _{jz}\nabla _s\psi _{iz}) dl.
\label{eq23-7}
\end{equation}
The above usage of Gauss' theorem is also known as \index{Theorem!Green's}Green's theorem.
With either a homogeneous Neumann, Dirichlet or mixed boundary
condition (Neumann on one part of $C$, and Dirichlet on the other
parts), the right-hand side of (\ref{eq23-7}) vanishes, and we have
\begin{equation}
(k_{is}^2-k_{js}^2)\int\limits _S\psi _{iz}\psi _{jz} dS=0.
\label{eq23-8}
\end{equation}
For $i\ne j$, we have
\begin{equation}
\int\limits _S\psi _{iz}\psi _{jz} dS=0. \label{eq23-9}
\end{equation}
Since $k_{js}^2$ can be shown to be pure real [see Problem 2-6],
$\psi_{jz}^*$ is also a solution of (\ref{eq23-6}).
Hence, we can
further say that $\int\limits _S\psi _{iz}\psi _{jz}^* dS=0, \quad
i\ne j$, since $\psi _{jz}^*$ is also a solution to (\ref{eq23-6}).
The property described by (\ref{eq23-3}) and (\ref{eq23-4}), is
known as the  mode orthogonality of the axial components of the
field.  The real-value of $k_js^2$ is related to that the operator
$\nabla_s^2$ is a Hermitian operator under appropriate boundary
condition of the field.

A point is in order regarding the orthogonality relations listed in
\eqref{eq23-3} and \eqref{eq23-4}.  When $i=j$ in \eqref{eq23-3},
\begin{align}
\int_S H_{iz}H_{iz}=C
\end{align}
It is possible that $C=0$ if $H_{iz}$ is a complex function.  Hence,
it is prudent to rewrite the orthogonality relations as
\begin{equation}
\int\limits _SH_{iz}H^*_{jz} dS=C_{ih}\delta _{ij}, \qquad \text
{TE}, \label{eq23-3a}
\end{equation}
\begin{equation}
\int\limits _SE_{iz}E^*_{jz} dS=C_{ie}\delta _{ij}, \qquad \text
{TM}, \label{eq23-4a}
\end{equation}
In this manner, $C_{ih}$ and $C_{ie}$ are ensured to be positive
real values, and they can be used to normalize the modes yielding
the orthonormal relations that
\begin{equation}
\int\limits _SH_{iz}H^*_{jz} dS=\delta _{ij}, \qquad \text {TE},
\label{eq23-3aa}
\end{equation}
\begin{equation}
\int\limits _SE_{iz}E^*_{jz} dS=\delta _{ij}, \qquad \text {TM},
\label{eq23-4aa}
\end{equation}

\subsubsection{Mode Orthogonality for Transverse Fields}
\index{Mode orthogonality!transverse field}

The transverse components of $\v E$ or $\v H$ fields are also
orthogonal.  This property can be proven a number of ways.  One way
is to relate their orthogonality to the orthogonality of the scalar
wave functions mentioned before.  However, to demonstrate that the
orthogonality is also related to the symmetry of the differential
equation, which will prove this property as follows: the $\v
E$-field of either the TE or the TM mode of a waveguide satisfies
\begin{equation}
\nabla\times\nabla\times\v E_i-k^2\v E_i=0. \label{eq23-10}
\end{equation}
Assuming that the field has $e^{ik_zz}$ dependence, and extracting
the transverse component of the above equation, we have, for any two
distinct modes,
\begin{equation}
\nabla _s\times\nabla _s\times\v E_{is}-\nabla _s\nabla_s\cdot\v
E_{is}-k_{is}^2\v E_{is}=0, \label{eq23-11}
\end{equation}
\begin{equation}
\nabla _s\times\nabla _s\times\v E_{js}-\nabla _s\nabla_s\cdot\v
E_{js}-k_{js}^2\v E_{js}=0, \label{eq23-12}
\end{equation}
where $k_{is}^2 = k^2 - k_{iz}^2, \quad k_{js}^2 = k^2 - k_{jz}^2$.
Dot multiplying (\ref{eq23-11}) by $\v E_{js}$, (\ref{eq23-12}) by
$\v E_{is}$, and subtracting, we have, after integrating,
\begin{equation}
\begin{split}
(k_{js}^2-k_{is}^2) & \int\limits _S\v E_{is}\cdot\v E_{js}dS \\
&=\int\limits _S dS(\v E_{js}\cdot\nabla _s\times\nabla _s\times\v
E_{is}-\v E_{is}\cdot\nabla
_s\times\nabla _s\times\v E_{js})\\
& -\int\limits _S dS(\v E_{js}\cdot\nabla _s\nabla _s\cdot\v
E_{is}-\v E_{is} \cdot\nabla_s\nabla _s\cdot\v E_{js}).
\end{split}
\label{eq23-13}
\end{equation}
By noting that $\nabla\cdot(\v A \times \v B)=\v B
\cdot\nabla\times\v A -\v A\cdot \nabla\times\v B,$ and hence,
\begin{equation}
\begin{split}
-\nabla _s\cdot (\v E_{js}\times\nabla _s & \times\v E_{is}-\v E_{is}\times\nabla_s\times\v E_{js})\\
&=\v E_{js}\cdot \nabla _s\times\nabla _s\times\v E_{is}-\v
E_{is}\cdot\nabla _s\times\nabla _s\times\v E_{js},
\end{split}
\label{eq23-14}
\end{equation}
and that $\nabla \cdot(\v A \phi)=\phi \nabla \cdot \v A +\v A
\cdot\nabla \phi,$ and hence,
\begin{equation}
\nabla _s\cdot(\v E_{js}\nabla _s\cdot\v E_{is}-\v E_{is}\nabla
_s\cdot\v E_{js})=\v E_{js}\cdot\nabla _s\nabla _s\cdot\v E_{is}-\v
E_{is}\cdot\nabla _s\nabla_s\cdot\v E_{js}, \label{eq23-15}
\end{equation}
we can convert the right-hand side of (\ref{eq23-13}) into line
integrals using Gauss' theorem in two dimensions, giving
\begin{equation}
\begin{split}
(k_{js}^2-k_{is}^2) &\int\limits _S\v E_{is}\cdot\v E_{js}dS\\
&=-\oint _C dl ~ \^ n\cdot [\v E_{js}\times(\nabla _s\times\v
E_{is})-\v E_{is}
\times(\nabla_s\times\v E_{js})]\\
& -\oint _C dl ~ \^ n\cdot [\v E_{js}\nabla _s\cdot\v E_{is}-\v
E_{is}\nabla _s\cdot\v E_{js}].
\end{split}
\label{eq23-16}
\end{equation}
The right-hand side of (\ref{eq23-16}) vanishes by virtue of the
boundary condition. Since $\hat n \cdot [\v E_{js} \times (\nabla_s
\times \v E_{is})] = \hat n\times \v E_{js} \cdot \nabla_s \times \v
E_{is}$,\footnote {This follows from $\v a\cdot\v b \times\v c=\v c
\cdot\v a \times\v b=\v b\cdot\v c \times\v a.$} $\hat n \times \v
E_s = 0$ implies the zero of the first integral on the right hand
size of (\ref{eq23-16}). Furthermore, $\v {\nabla }_s\cdot\v
E_{is}=-ik_zE_z=0$ on the waveguide wall implies the zero of the
second term in the right-hand side of (\ref{eq23-16}). Therefore,
\begin{equation}
\int\limits _S\v E_{is}\cdot\v E_{js} dS=0, \qquad i\ne j,
\label{eq23-17}
\end{equation}
for any two distinct modes with different $k_s^2$, irrespective of
whether they are TE or TM modes. A similar proof follows for $\v
H_s$, i.e.,
\begin{equation}
\int\limits _S\v H_{is}\cdot\v H_{js} dS=0, \qquad i\ne j.
\label{eq23-18}
\end{equation}
We can further show that
\begin{equation}
\int\limits _S\v F_{is}\cdot\v F_{js}^* ~ dS=0, \quad i\ne j,
\label{eq23-19}
\end{equation}
where $\v F$ is either $\v E$ or $\v H$.  This is because $\v E^*$
or $\v H^*$ is also a solution of (\ref{eq23-11}).  Moreover, it is
prudent to write the orthogonality relations of these fields as
\begin{equation}
\int\limits _S\v F_{is}\cdot\v F_{js}^* ~ dS=C_j\delta_{ij}, \quad
i\ne j, \label{eq23-19a}
\end{equation}
so that $C_j$ is guaranteed to be positive real and these functions
can be orthonormalized.

The orthogonality of modes in (\ref{eq23-17}) follows from the fact
that the differential operator in (\ref{eq23-11}) is symmetric with
the defined boundary conditions, i.e.,
\begin{equation}
\langle \v E_{js}, (\nabla _s\times\nabla _s\times~ -~ \nabla _s
\nabla _s\cdot )\v E_{is}\rangle =\langle \v E_{is}, (\nabla
_s\times\nabla _s\times ~-~ \nabla _s\nabla _s\cdot )\v
E_{js}\rangle . \label{eq23-20}
\end{equation}
It is analogous to the fact that eigenvectors of a symmetric matrix
with distinct eigenvalues are orthogonal.  Moreover, since these
operators are real, they are also Hermitian operators with real
eigenvalues $k_{js}^2$.

When the medium is inhomogeneous, (\ref{eq23-11}) is not valid for
describing the field, and the differential operators are not
symmetric anymore. For an inhomogeneously filled waveguide,
(\ref{eq23-17}) and (\ref{eq23-18}) are not true in general.

In the preceding proof, we can also decompose the transverse fields
into their TE and TM components, and express them in terms of $E_z$
and $H_z$.  The orthogonality of the transverse fields can hence be
related to the orthogonality of the axial fields.

\subsubsection{Mode Orthogonality for Reaction}
\index{Mode orthogonality!reaction}

A more general orthogonality condition which we shall derive later,
and is true even for inhomogeneously filled waveguides is
\begin{equation}
\int\limits _S(\v E_{is}\times\v H_{js})\cdot\^ z dS=0, \qquad i\ne
j. \label{eq23-21}
\end{equation}
The above is the \index{Reaction}reaction as is used in the Lorentz reciprocity
theorem.  To prove \eqref{eq23-21} for hollow waveguides, we first
write down the equations satisfied by $\v E_{is}$ and $\v H_{js}$,
i.e.,
\begin{equation}
\nabla _s\times\nabla _s\times\v E_{is}-\nabla _s\nabla _s\cdot\v
E_{is}-k_{is}^2\v E_{is}=0, \label{eq23-22}
\end{equation}
\begin{equation}
\nabla _s\times\nabla _s\times \v H_{js}-\nabla _s\nabla _s\cdot\v
H_{js}-k_{js}^2\v H_{js}=0. \label{eq23-23}
\end{equation}
We cross multiply (\ref{eq23-22}) by $\v H_{js}$ and (\ref{eq23-23})
by $\v E_{is}$. Upon subtraction and integration, we have
\begin{equation}
\begin{split}
( & k_{js}^2-k_{is}^2)\int\limits _S(\v E_{is}\times\v H_{js})\cdot\^ z dS \\
& =\int\limits _S[(\nabla _s\times\nabla _s\times\v E_{is})\times\v
H_{js}-\v
E_{is}\times (\nabla _s\times\nabla _s\times\v H_{js})]\cdot\^ z dS\\
& -\int\limits _S[(\nabla _s\nabla _s\cdot\v E_{is})\times \v
H_{js}-\v E_{is} \times (\nabla _s\nabla _s\cdot\v H_{js})]\cdot\^ z
dS.
\end{split}
\label{eq23-24}
\end{equation}
Using the fact that
\begin{equation}
\begin{split}
(\nabla _s\times\nabla _s\times\v E_s)\times\v H_s \cdot \^ z & =(\v
H_s\times\^ z)\cdot
(\nabla _s\times\nabla_s\times\v E_s)\\
& =\nabla _s\cdot [(\nabla _s\times\v
E_s)\times(\v H_s\times\^ z)]\\
& \qquad\quad +(\nabla _s\times\v E_s)\cdot\nabla _s\times (\v
H_s\times\^ z)
\end{split}
\label{eq23-25}
\end{equation}
and that
\begin{equation}
\begin{split}
\^ z\cdot\v E_s\times\nabla _s\nabla _s\cdot\v H_s & =(\^ z\times \v
E_s)\cdot\nabla _s\nabla _s\cdot\v H_s\\
& =\nabla _s\cdot (\nabla _s\cdot\v H_s\^ z\times\v E_s)-\nabla
_s\cdot\v H_s\nabla _s\cdot\^ z\times \v E_s,
\end{split}
\label{eq23-26}
\end{equation}
plus the fact that $\nabla _s\times (\v H_s\times \^ z)=-\^ z\nabla
_s\cdot\v H_s$, and that $\nabla _s\cdot\^ z\times\v E_s=-\^
z\cdot\nabla _s\times\v E_s$, it can be seen that the last terms in
(\ref{eq23-25}) and (\ref{eq23-26}) are identical except for a sign
difference. Therefore, after using Gauss' theorem,
\begin{equation}
\begin{split}
\int \limits _SdS[(\nabla _s \times\nabla _s\times\v E_s)\times \v H_s+\v E_s  &\times\nabla_s\nabla _s\cdot\v H_s]\cdot\^ z = \oint \limits _Cdl\hat n \cdot [(\nabla_s\times\v E_s)\\
& \times (\v H_s\times\^ z)+(\nabla _s\cdot\v H_s)\^ z\times\v E_s].
\end{split}
\label{eq23-27}
\end{equation}
Since $\nabla _s\times \v E_s=i\omega\mu\v H_z$ and $\nabla
_s\cdot\v H_s=-ik_zH_z$, the right-hand side vanishes if we have
either an electric wall or a magnetic wall or a mixture thereof.
Similarly, the other terms on the right hand side of (\ref{eq23-24})
vanish by the same argument. Therefore, in general,
\begin{equation}
\int\limits _S(\v E_{is}\times\v H_{js})\cdot\^ zdS=0, \qquad i\neq
j, \label{23-28}
\end{equation}
for any two distinct modes with different propagation constants
$k_z$ or $k_s$.

Also, the transverse fields can be related to the axial fields, and
their orthogonality can also be related to the orthogonality of the
axial fields.  We shall show later a more general proof of the above
using Lorentz reciprocity theorem.  This proof is even valid for
inhomogeneously-filled waveguides.

\subsubsection{Power Orthogonality}
\index{Power orthogonality}

Since, $\v H_{js}^*$ is also a solution to (\ref{eq23-23}), we have
\begin{equation}
\int\limits _S(\v E_{is}\times\v H_{js}^*)\cdot\^ z dS=0, \qquad
i\ne j, \qquad\text {power orthogonality}. \label{eq23-29}
\end{equation}
Since $\v E\times\v H^*$ represents the complex Poynting vector,
Equation (\ref{eq23-29}) implies that the power flow in a waveguide
is independently carried by each mode.  Cross interactions between
the $\v E$ and $\v H$ fields of two different modes do not result in
power flow as testified by Equation (\ref{eq23-29}).

The above orthogonality principles assume that the modes have
distinct eigenvalues $k_{is}^2$ or distinct axial propagation
constants $k_{iz}^2$. When $k_{is}^2$ for two different modes are
the same, the modes are called degenerate. If there are $N$
degenerate, independent modes, we can use the Gram-Schmidt
orthogonalization procedure to obtain $N$ orthogonal modes if we so
desire.

\section {{{ Rectangular Waveguides}}}
\index{Rectangular waveguide}
\begin{figure}[htb]
\begin{center}
\hfil\includegraphics[width=3.6truein]{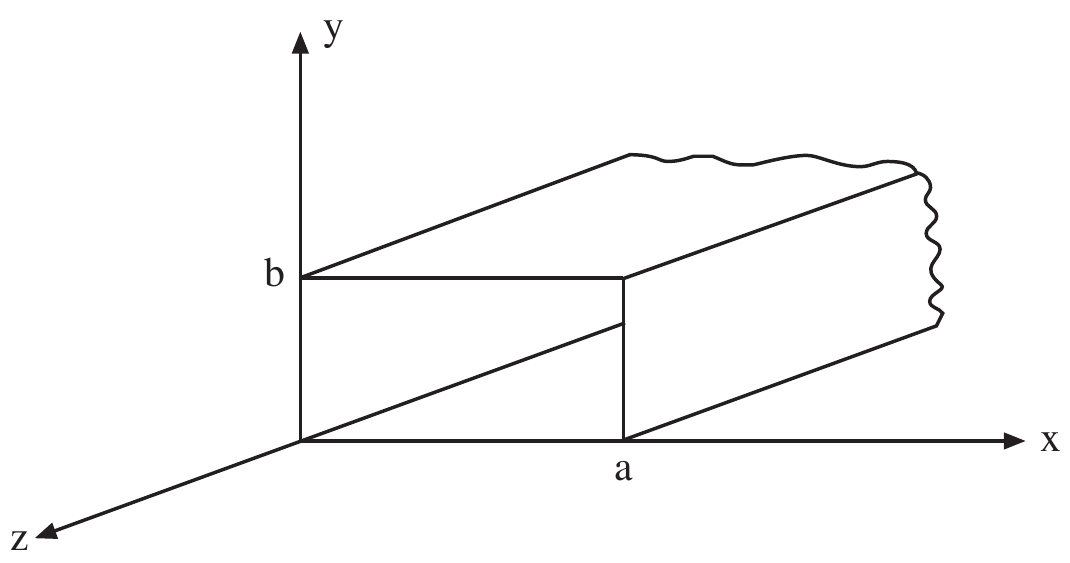}\hfil
\end{center}
\caption{A rectangular waveguide.}\label{fgRectangular}
\end{figure}


The rectangular waveguide is the most commonly used hollow
waveguide. By adjusting the aspect ratio, $a$ to $b$, of the
waveguide, one can obtain a good bandwidth for single mode
propagation. Furthermore, the analysis of this waveguide is simple.

\subsection { TE Modes (H Modes)}
\index{Rectangular waveguide!TE modes (H modes)}

A TE mode in a rectangular waveguide is characterized by $H_z$
satisfying Equation (\ref{eq23-1}) with the requisite Neumann
boundary condition. An $H_z$ that will satisfy (\ref{eq23-1}) with
$\frac {\partial }{\partial n}H_z=0$ on the waveguide wall is
\begin{equation}
H_z=H_0\cos \left(\frac {m\pi }ax\right)\cos \left( \frac {n\pi
}by\right) e^{ik_zz}, \label{261-1}
\end{equation}
where $k_s^2 = (\frac {m\pi }a)^2+(\frac {n\pi }b)^2$, $k_z=\sqrt
{k^2-k_s^2}$.  The fact that $k_s^2$ has to satisfy the prescribed
values is also known as the guidance condition.  The transverse
fields can be found using (\ref{eq27}) and (\ref{eq28}). The mode
becomes evanescent\index{Mode!evanescent} or non-propagating when $k_s^2 > k^2$, i.e., when
$k_z$ becomes imaginary. Since $k^2 = \omega ^2 \mu\epsilon$, the
cutoff frequency \index{Cutoff frequency} (the frequency below which the mode is evanescent)
is given by
\begin{equation}
\omega _{mnc}=\frac 1{\sqrt {\mu\epsilon }}\left[\left(\frac {m\pi
}a\right)^2+\left( \frac {n\pi }b\right)^2\right]^{\frac 1{2}}.
\label{eq261-2}
\end{equation}
The corresponding mode is usually labeled as TE$_{mn}$ (or $H_{mn}$)
mode. The wavelength of a wave at $\omega _c$ in the medium denoted
by $\mu$, $\epsilon $ is the cutoff wavelength. It is
\begin{equation}
\lambda _{mnc}=\frac 2{[(\frac m{a})^2+(\frac {n} b)^2]^{\frac
1{2}}}. \label{eq261-3}
\end{equation}

In waveguide conventions, $a$ is assumed larger than $b$. Then, the
\index{Mode!dominant}dominant mode (fundamental mode\index{Mode!fundamental}) with the \index{Mode!lowest TE}lowest cutoff frequency is
the $m = 1$, $n = 0$ mode, also known as the TE$_{10}$ mode\index{Mode!TE$_{10}$} (or
$H_{10}$ mode). The TE$_{00}$ mode does not exist because, in this
case, $k_s=0$, and $\v E_s$ and $\v H_s$ diverge from (\ref{eq27})
and (\ref{eq28}) when $k_s\rightarrow 0$.

\subsection { TM Modes (E Modes)}
\index{Rectangular waveguide!TM modes (E modes)}

A TM mode in a rectangular waveguide is characterized by $E_z$
satisfying Equation (\ref{eq23-2}) with the requisite Dirichlet
boundary condition. An $E_z$ that will satisfy (\ref{eq23-2}) with
$E_z = 0$ on the waveguide wall is
\begin{equation}
E_z=E_0\sin \left(\frac {m\pi }ax\right)\sin \left(\frac {n\pi
}by\right)e^{ik_zz}, \label{eq262-4}
\end{equation}
where $k_s^2 =(\frac {m\pi }a)^2+(\frac {n\pi }b)^2$, $k_z=\sqrt
{k^2-k_s^2}$. The TM$_{mn}$ mode has the same \index{Cutoff frequency}cutoff frequency as
the TE$_{mn}$ mode. However, when either $m = 0$, or $n = 0$, the
mode does not exist since $E_z = 0$ then. Therefore, the lowest TM
mode is the TM$_{11}$ mode with a cutoff frequency above that of the
TE$_{10}$ mode. Given the $z$ components of the fields, all other
field components of a waveguide can be derived. Figure {\ref{fgCut}
shows the field plots of some modes of a rectangular waveguide
\cite{LEE_LEE_CHUANG}.\footnote{The plots here are reproduced by A.
Greenwood according to this reference.}

\section {{{ Circular Waveguides}}}

Certain modes of a circular waveguide have less attenuation from
wall loss compared to a rectangular waveguide. Hence, it is
sometimes preferred over a rectangular waveguide.

\subsection { TE Modes (H Modes)}

\begin{figure}[htb]
\begin{center}
\hfil\includegraphics[width=3.0truein]{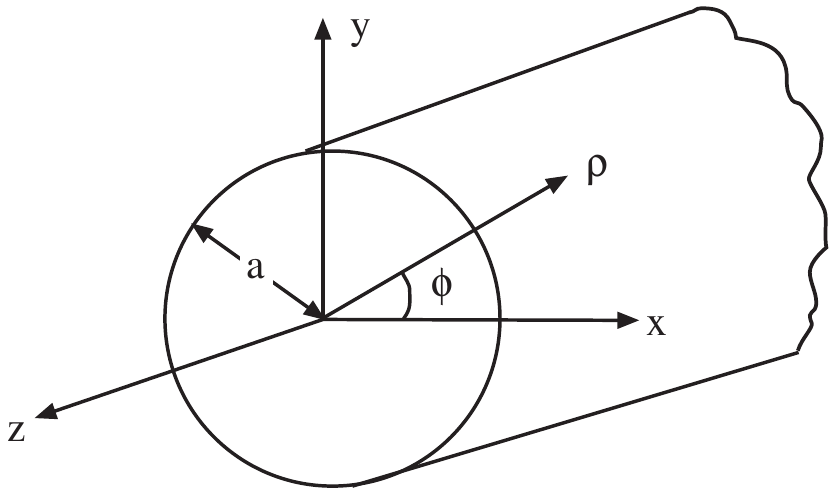}\hfil
\end{center}
\caption{A circular waveguide.}\label{fgAcircular}
\end{figure}


The $H_z$ component of a TE mode satisfies Equation (\ref{eq23-1})
in cylindrical coordinates, i.e.,
\begin{equation}
\left[\frac 1{\rho }\frac {\partial }{\partial\rho }\rho\frac
{\partial }{\partial\rho }+\frac 1{\rho ^2}\frac {\partial
^2}{\partial\phi ^2}+k_s^2\right]H_z=0, \label{eq27-1}
\end{equation}

\noindent with the boundary condition that $\frac {\partial
}{\partial\rho } H_z=0$, $\rho = a$, where $a$ is the radius of the
waveguide. If we assume that $H_z$ has $e^{\pm in\phi }$, $\sin
{n\phi }$ or $\cos {n\phi}$ dependence, where $n$ is an integer, we
can replace $\frac {\partial ^2}{\partial \phi ^2}$ by $-n^2$.
Equation (\ref{eq27-1}) then becomes
\begin{equation}
\left[\frac 1{\rho }\frac {\partial }{\partial\rho }\rho\frac
{\partial }{\partial\rho }-\frac {n^2}{\rho ^2}+k_s^2\right]H_z=0.
\label{eq27-2}
\end{equation}

Equation (\ref{eq27-2}) is the \index{Bessel equation} Bessel equation, whose solutions are
either $J_n(k_s\rho )$, $Y_n(k_s\rho )$, $H_n^{(1)}(k_s\rho )$ or
$H_n^{(2)} (k_s\rho )$. Of these four solutions, only two are
independent because of the relations
\begin{subequations}
\begin{equation}
H_n^{(1)}(k_s\rho )=J_n(k_s\rho )+iY_n(k_s\rho ), \label{eq27-3a}
\end{equation}
\begin{equation}
H_n^{(2)}(k_s\rho )=J_n(k_s\rho )-iY_n(k_s\rho ), \label{eq27-3b}
\end{equation}
\end{subequations}
$J_n(k_s\rho )$ is regular about the origin when $\rho\rightarrow
0$, but $Y_n(k_s\rho )$ is singular (so are $H_n^{(1)}(k_s\rho )$
and $H_n^{(2)}(k_s\rho )$). Since the field cannot be infinite at
the center of the waveguide due to the absence of sources, the
solution to (\ref{eq27-1}) is of the form
\begin{equation}
H_z=H_0J_n(k_s\rho )e^{\pm in\phi +ik_zz}. \label{eq27-4}
\end{equation}
We require that $\frac {\partial }{\partial\rho }H_z=0$ at $\rho =
a$, implying that
\begin{equation}
J_n'(k_sa)=0, \label{eq27-5}
\end{equation}
with $k_z =\sqrt {k^2-k_s^2}$. If the $m$-th zero of $J_n'(x)$ is
defined to be $\beta _{nm}$ such that $J_n'(\beta _{nm}) = 0$, the
values of possible $k_s$ are
\begin{equation}
k_s=\frac {\beta _{nm}}a. \label{eq27-6}
\end{equation}
The above is also the \index{Guidance condition}guidance condition for the waveguide mode. The
subscript $n$ denotes the orders of the Bessel function $J_n(x)$ and
the circular harmonic $e^{\pm in\phi }$. The subscript $m$ denotes
the $m$-th zero of $J_n'(x)$ discounting the zero at the origin. The
corresponding mode is usually denoted as the TE$_{nm}$ mode. Cutoff
occurs when $k = \omega \sqrt {\mu\epsilon } <k_s$. From Figure
\ref{fgPlots}, we see that the TE$_{11}$ mode corresponding to the
first zero of $J_1(x)$ has the lowest cutoff frequency. The cutoff
frequency for the TE$_{nm}$ mode is given by
\begin{equation}
\omega _{nmc}=\frac 1{\sqrt {\mu \epsilon }}\frac {\beta _{nm}}a,
\label{27-7}
\end{equation}
and the corresponding cutoff wavelength is
\begin{equation}
\lambda _{nmc}=\frac {2\pi }{\beta _{nm}}a. \label{eq27-8}
\end{equation}

\subsection { TM Modes (E Modes)}
\index{Mode!TM}

\begin{figure}[htb]
\begin{center}
\hfil\includegraphics[width=4.0truein]{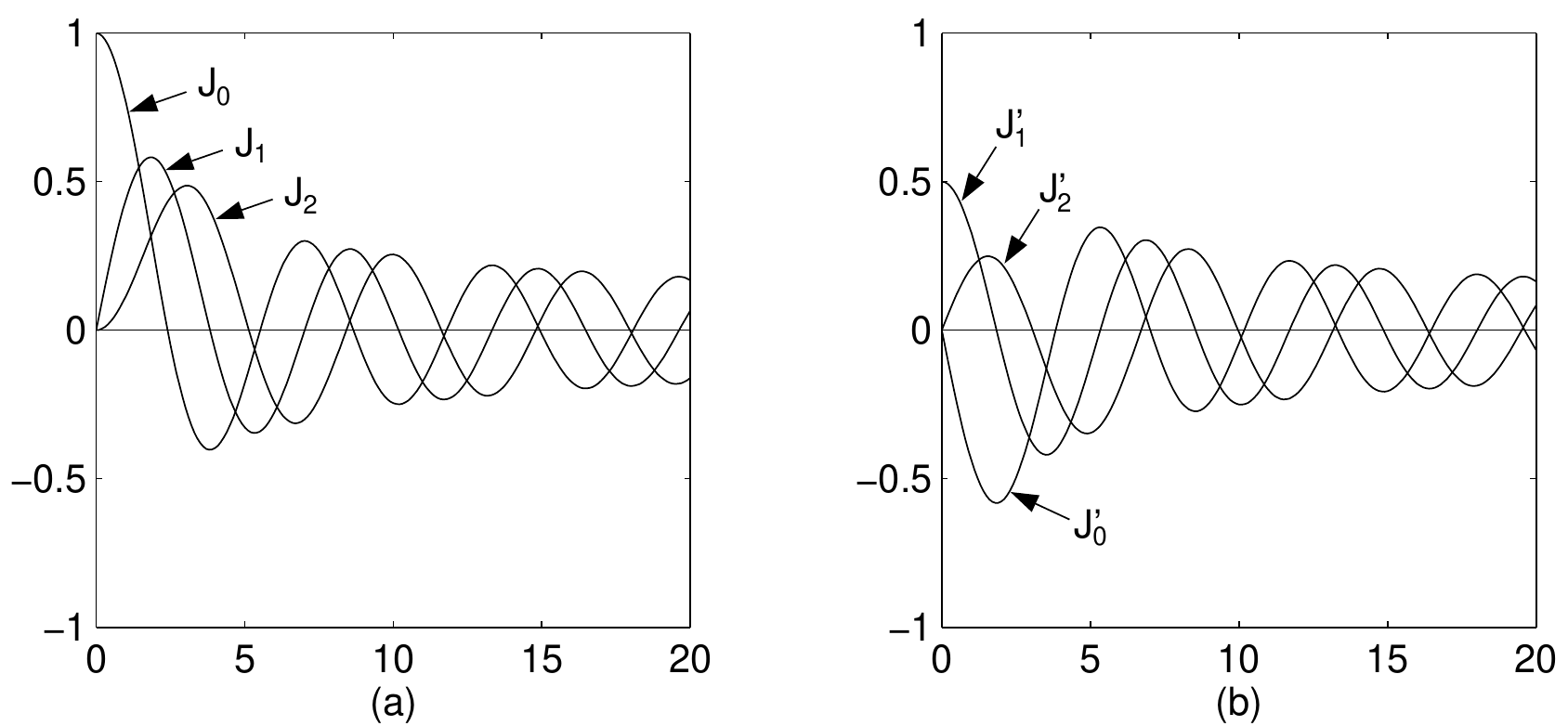}\hfil
\end{center}
\caption{Plots of Bessel functions and their
derivatives.}\label{fgPlots}
\end{figure}


Similar to a TE mode, the $E_z$ component of a TM mode has $e^{\pm
in\phi }$ dependence, satisfying the equation
\begin{equation}
\left[\frac 1{\rho }\frac {\partial }{\partial\rho }\rho\frac
{\partial }{\partial\rho }-\frac {n^2 }{\rho ^2}+k_s^2\right]E_z=0,
\label{eq27-9}
\end{equation}
with the boundary condition $E_z (\rho =a) =0$. Hence,
\begin{equation}
E_z=E_0J_n(k_s\rho )e^{\pm in\phi +ik_zz}, \label{eq27-10}
\end{equation}
with $J_n(k_sa) = 0$. If we denote the $m$-th zero of $J_n(x)$ by
$\alpha_{nm}$, then $k_s$ has possible values of
\begin{equation}
k_s=\frac {\alpha _{nm}}a, \label{eq27-11}
\end{equation}
where the subscript $n$ denotes the order of the Bessel function,
$J_n(x)$, and the subscript $m$ denotes the $m$-th zero of $J_n(x)$,
discounting the zero at the origin. It is also the guidance
condition, and the corresponding mode is known as the TM$_{nm}$
mode. The cutoff frequency of the TM$_{nm}$ mode is given by
\begin{equation}
\omega _{nmc}=\frac 1{\sqrt {\mu\epsilon }}\frac {\alpha _{nm}}a,
\label{eq27-12}
\end{equation}
and the corresponding cutoff wavelength is
\begin{equation}
\lambda _{nmc}=\frac {2\pi }{\alpha _{nm}}a. \label{eq27-13}
\end{equation}
Looking at Figure \ref{fgPlots}, we see that the \index{Mode!lowest TM}lowest TM mode is
the TM$_{01}$ mode\index{Mode!TM$_{01}$}, but it has a higher cutoff frequency compared to
the TE$_{11}$ mode.

\vskip 20pt

\centerline {Table 2.3.1. Roots of $J'_n(x) =0$.}
\nopagebreak \centerline { \vbox{\offinterlineskip \hrule
\halign{&\vrule#&
\strut\quad\hfil#\quad\cr
height2pt&\omit&&\omit&&\omit&&\omit&&\omit&\cr &n\hfil&&$\beta
_{n1}$&&$\beta_{n2}$&&$\beta_{n3}$&&$\beta_{n4}$&\cr
height2pt&\omit&&\omit&&\omit&&\omit&&\omit&\cr \noalign{\hrule}
height2pt&\omit&&\omit&&\omit&&\omit&&\omit&\cr
&0&&3.832&&7.016&&10.174&&13.324&\cr
&1&&1.841&&5.331&&8.536&&11.706&\cr
&2&&3.054&&6.706&&9.970&&13.170&\cr
&3&&4.201&&8.015&&11.346&&14.586&\cr
&4&&5.318&&9.282&&12.682&&15.964&\cr
&5&&6.416&&10.520&&13.987&&17.313&\cr
height2pt&\omit&&\omit&&\omit&&\omit&&\omit&\cr} \hrule}}

\vskip 20pt

\vskip 12pt\centerline{Table 2.3.2. Roots of $J_n(x)=0$.}
\nopagebreak \centerline { \vbox{\offinterlineskip \hrule
\halign{&\vrule#&
\strut\quad\hfil#\quad\cr
height2pt&\omit&&\omit&&\omit&&\omit&&\omit&\cr &n\hfil&&$\alpha
_{n1}$&&$\alpha_{n2}$&&$\alpha_{n3}$&&$\alpha_{n4}$&\cr
height2pt&\omit&&\omit&&\omit&&\omit&&\omit&\cr \noalign{\hrule}
height2pt&\omit&&\omit&&\omit&&\omit&&\omit&\cr
&0&&2.405&&5.520&&8.654&&11.792&\cr
&1&&3.832&&7.016&&10.174&&13.324&\cr
&2&&5.135&&8.417&&11.620&&14.796&\cr
&3&&6.380&&9.761&&13.015&&16.223&\cr
&4&&7.588&&11.065&&14.373&&17.616&\cr
&5&&8.771&&12.339&&15.700&&18.980&\cr
height2pt&\omit&&\omit&&\omit&&\omit&&\omit&\cr} \hrule}}

\vskip 20pt

It can be shown that the TE$_{01}$\index{Mode!TE$_{01}$} mode has the lowest loss at high
frequencies. The TE$_{01}$ mode is axially symmetric with $\v E = \^
{\phi }E_{\phi }$. Therefore, it can be enhanced by various means.
One way is to use a \index{Mode filter}mode filter as shown in Figure \ref{fgAmode}(a)
\cite{ARCHER}. The radial conducting wire will short out modes with
radial components of the electric field. However, the TE$_{01}$ mode
is oblivious to the presence of the radial conducting strips, and is
little affected. Another way to enhance the TE$_{01}$ mode in a
circular waveguide is to use a \index{Waveguide!ribbed}ribbed waveguide. For the TE$_{01}$
mode, the current is purely circumferential and is oblivious to the
presence of the ribs. However, the TM modes, which have axial
components of the current, will be affected by the ribbed wall of
the waveguide. In other words, the ribbed waveguide wall does not
support the axial current flow effectively. If the periodicity of
the waveguide corrugation is small compared to the wavelength, the
TE$_{01}$ mode will not be affected much.

\begin{figure}[htb]
\begin{center}
\hfil\includegraphics[height=1.25in,width=4.2in]{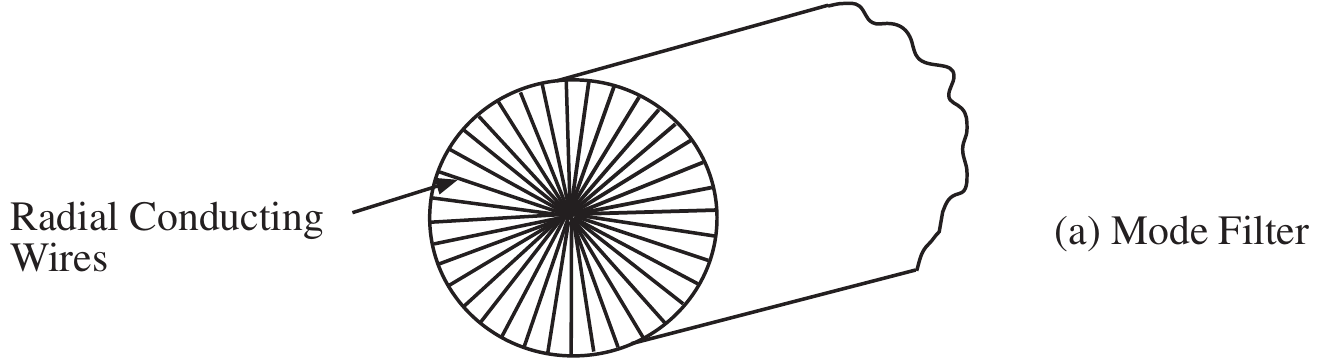}\hfil
\end{center}
\begin{center}
\hfil\includegraphics[width=3.0truein]{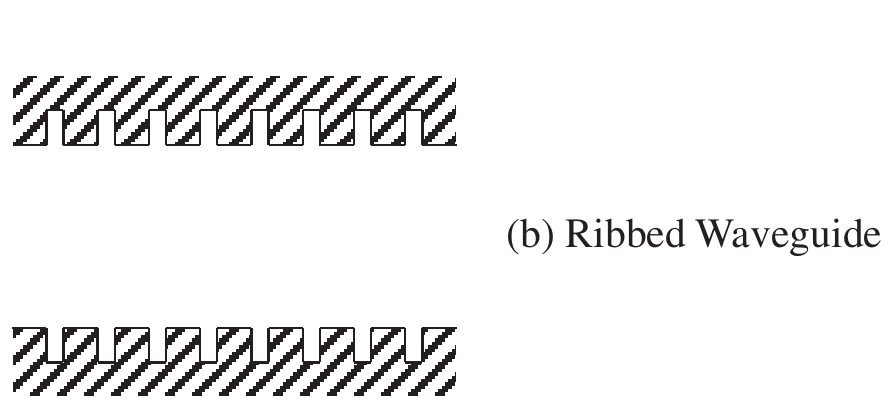}\hfil
\end{center}
\caption{(a) A mode filter used to filter out most other modes, but
it allows the TE$_{01}$ modes to pass through. (b) A ribbed
waveguide prevents axial current flow, hence attenuates the TM
modes, which has $E_z\neq 0$.} \label{fgAmode}
\end{figure}

The transverse field components of a circular waveguide are easily
obtained given the axial components. Figure \ref{fgCutoff} shows the
field plots of some modes in a circular waveguide
\cite{LEE_LEE_CHUANG}.



\section{{{ Power Flow in a Waveguide}}}
\index{Waveguide!power flow}

Because of the power orthogonality of the distinct modes in a
waveguide with perfectly impenetrable walls (perfect electric
conductors or perfect magnetic conductors), we can study the power
flow due to each mode and the total power flow is the sum of the
power flow from each mode.

\subsection { Power Flow and Group Velocity}
\index{Velocity!group}

The time average power flow in a waveguide is given by
\begin{equation}
P_f=\frac 1{2}\Re e\int\limits _S(\v E_s\times\v H_s^*)\cdot\^ z
dS. \label{eq241-1}
\end{equation}
For a \index{Mode!TE}TE mode with $e^{ik_zz}$ dependence, from (\ref{eq28}),
\begin{equation}
\v E_s=\frac {i\omega\mu }{k_s^2}\nabla _s\times\^ zH_z,
\label{eq241-2}
\end{equation}
\begin{equation}
\v H_s=\frac {ik_z}{k_s^2}\nabla _sH_z. \label{eq241-3}
\end{equation}
Substituting into (\ref{eq241-1}), we can show that
\begin{equation}
P_f=\frac 1{2}\Re e\left\{ \frac {\omega\mu k_z^*}{k_s^4}\^
z\int\limits _S (\nabla_sH_z)\cdot (\nabla _sH_z^*) dS\right\}.
\label{eq241-4}
\end{equation}
($k_s^2$ is always real in a homogeneously-filled waveguide).
Using the fact that $\nabla _s\cdot (H_z\nabla _sH_z^*)=(\nabla
_sH_z)\cdot (\nabla _sH_z^*)+H_z\nabla _s^2H_z^*$, the above
becomes
\begin{equation}
P_f=\frac 1{2}\Re e\left\{\frac{\omega\mu k_z^*}{k_s^4}\^ z\left[
\int \limits _C dl\^ n\cdot H_z\nabla _sH_z^*-\int\limits
_SdSH_z\nabla _s^2 H_z^*\right]\right\}. \label{eq241-5}
\end{equation}
The first integral vanishes by virtue of the boundary condition.
While using $\nabla _s^2H_z^*=-k_s^2H_z^*$, we have
\begin{equation}
P_f=\frac {\^ z}2\Re e\left\{\frac{\omega\mu
k_z^*}{k_s^2}\int\limits _SdS| H_z|^2\right\}. \label{eq241-6}
\end{equation}
For a \index{Mode!TM}TM mode, we can similarly show that
\begin{equation}
P_f=\frac {\^ z}2\Re e\left\{\frac {\omega\epsilon^*
k_z}{k_s^2}\int \limits _SdS|E_z|^2\right\}. \label{eq241-7}
\end{equation}
When the mode is cutoff, i.e., when $k^2 < k_s^2$, so that $k_z =
\sqrt {k^2-k_s^2}$ is pure imaginary, and the waveguide is
lossless, there is no time average power flow down the waveguide.
In such a case, the mode is evanescent.

The time average energy stored per unit length in the electric
field for the TE mode is given by
\begin{equation}
\begin{split}
\langle W_e\rangle  & =\frac 1{4}\epsilon \int\limits _S\v
E_s\cdot\v E_s^*dS\\ & =\frac {\omega ^2|\mu |^2\epsilon
}{4k_s^4}\int\limits _S(\nabla _sH_z) \cdot (\nabla_sH_z^*)dS.
\end{split}
\label{eq241-8}
\end{equation}
In the same manner as (\ref{eq241-4}), we can show that
\begin{equation}
\langle W_e\rangle =\frac {\omega ^2|\mu |^2\epsilon }{4k_s^2}\int
\limits _S|H_z|^2dS. \label{eq241-9}
\end{equation}
It can be shown that $\langle W_e\rangle  = \langle W_m\rangle $
for a lossless waveguide where $\langle W_m\rangle $ is the time
average energy stored in the magnetic field. Therefore, the total
time average energy stored per unit length is
\begin{equation}
\langle W_T\rangle =\langle W_e\rangle +\langle W_m\rangle =\frac
{\omega ^2 \mu ^2\epsilon }{2k_s^2}\int\limits _S |H_z|^2 dS.
\label{eq241-10}
\end{equation}
For a lossless waveguide with a propagating mode, $k_z$ is pure
real and (\ref{eq241-6}) for the TE mode becomes
\begin{equation}
P_f=\frac {\^ z}2\frac {\omega\mu k_z}{k_s^2}\int\limits _S
dS|H_z|^2. \label{eq241-11}
\end{equation}
Comparing (\ref{eq241-10}) and (\ref{eq241-11}), we note that for
a lossless waveguide,
\begin{equation}
P_f=\frac {k_z}{\omega\mu\epsilon }\langle W_T\rangle ,
\label{eq241-12}
\end{equation}
where ${k_z}/{\omega\mu\epsilon }$ has the dimension of velocity.
In a waveguide,
\begin{equation}
k_z^2=\omega ^2\mu\epsilon -k_s^2. \label{eq241-13}
\end{equation}
The group velocity in a waveguide is derived to be
\begin{equation}
v_g=\frac {d\omega }{dk_z}=\frac {k_z}{\omega\mu\epsilon }.
\label{eq241-14}
\end{equation}
Therefore, (\ref{eq241-12}) is just
\begin{equation}
P_f=v_g\langle W_T\rangle . \label{eq241-15}
\end{equation}
In other words, in a lossless waveguide, the time average energy
stored per unit length, moving at the group velocity $v_g$
contributes to the power flow.  The group velocity in a waveguide
is the velocity of energy propagation, and it is also the \index{Velocity!of signal} signal
velocity.

The \index{Velocity}\index{Velocity!phase}phase velocity of a wave in a waveguide is defined to be
\begin{equation}
v_{ph}=\frac {\omega }{k_z}=\frac {\omega }{\sqrt {k^2-k_s^2}}.
\label{eq241-16}
\end{equation}
It is the velocity of the phase of the wave. Since a signal does
not travel at the phase velocity, it could be larger than the
speed of light. This happens near cutoff when $k^2\rightarrow
k_s^2$ for a mode. Group velocity or signal velocity cannot be
larger than the \index{Speed of light}speed of light, a limit dictated by Einstein's
theory of special relativity. Note that $v_{ph} v_g=c^2$ where $c$
is the velocity of light in the medium.

\subsection { Pulse Propagation in a Waveguide}
\index{Pulse propagation}

Since the phase and the group velocities inside a hollow waveguide
are frequency dispersive, a pulse propagating inside a hollow
waveguide will be distorted due to \index{Waveguide!dispersion}frequency dispersion. Different
frequency components will travel with different velocities. Hence,
after a certain distance of propagation, different Fourier
components lose their phase coherence, causing pulse distortion.
Therefore, to minimize pulse distortion, the bandwidth of the
pulse should be narrow. It can be shown that for a narrow-band
pulse, the \index{Envelope of pulse} envelope of the pulse propagates with the group
velocity while the \index{Carrier signal} carrier signal propagates with the phase
velocity.

A narrow-band pulse can be written as:
\begin{equation}
\begin{split}
p (\v r, t) & =\frac{1}{2 \pi}\int^{\infty}_{-\infty}d \omega
\tilde{p}(\v r,\omega) e^{-i\omega t}\\ & = \frac{1}{\pi}\Re
e\int^{\infty}_{0}d\omega\tilde{p}(\v r,\omega)e^{-i\omega t}
\label{eq241_1}
\end{split}
\end{equation}
where $p(\v r,t)$ may represent a component (e.g., the $z$
component) of the electric field or magnetic field inside a
waveguide. Since
\begin{equation}
\left (\nabla^2-\frac{1}{c^2}\frac{\partial^2}{\partial
t^2}\right) p(\v r, t)=0, \label{eq241_2}
\end{equation}
by substituting (\ref{eq241_1}) into (\ref{eq241_2}), we require
that
\begin{equation}
\left(\nabla^2+\frac{\omega^2}{c^2}\right)\tilde{p}(\v
r,\omega)=0, \label{eq241_3}
\end{equation}
If $p(\v r,t)$ corresponds to a particular mode, for example, in
the case of a single mode propagation, then
\begin{equation}
\left(\nabla^2_s + k^2_s\right)\tilde{p}(\v r, \omega)=0,
\label{eq241_4}
\end{equation}
and subtracting (\ref{eq241_3}) from (\ref{eq241_4}), we have
\begin{equation}
\left(\frac {\partial^2}{\partial z^2}+{k^2_z}\right)\tilde{p}(\v
r, \omega)=0, \label{eq241_5}
\end{equation}
where $k^2_z=\frac{\omega^2}{c^2}-k^2_s$. Hence
\begin{equation}
\tilde{p}(\v r,\omega)=\tilde{p}_o(\v r_s,\omega)e^{i k_z z}.
\label{eq241_6}
\end{equation}
where $\v r_s=\hat x x+\hat y y$. Consequently, we can rewrite
(\ref{eq241_1}) as
\begin{equation}
p(\v r,t)=\frac{1}{\pi}\Re e \int^{\infty}_{0}
d\omega\tilde{p_o}(\v r_s,\omega)e^{i k_z z-i\omega t}.
\label{eq241_7}
\end{equation}
If $p(\v r,t)$ is a narrow-band pulse with a carrier frequency at
$\omega_o$, the above can be approximated by an integral
\begin{equation}
p(\v r,t)\simeq\frac{1}{\pi}\Re e
\int^{\omega_0+\Delta}_{\omega_0-\Delta} d \omega{\tilde p_o}(\v
r_s,\omega)e^{ik_z z-i\omega t}. \label{eq241_8}
\end{equation}
In the above, we can approximate $k_z(\omega)$ as
\begin{equation}
k_z(\omega)\simeq k_z(\omega_0)+(\omega-\omega_0)\frac{d
k_z(\omega_0)}{d\omega}=k_z(\omega_0)+\frac{(\omega-\omega_0)}{v_g}
\label{eq241_9}
\end{equation}
and obtain
\begin{equation}
\begin{split}
p(\v r,t) &\simeq \frac{1}{\pi}\Re e \left\{
e^{i\left[k_z(\omega_0) z -{\omega_0} t\right]}
\int^{\omega_0+\Delta}_{\omega_0-\Delta} d \omega\,\tilde{p}_o(\v
r_s,\omega)
e^{\frac{i(\omega-\omega_0)(z-v_g t)}{v_g}}\right\} \\
& \simeq 2\Re e \left\{ e^{i\left[k_z(\omega_0)z-\omega_0
t\right]} F(\v r_s,z-v_g t)\right\}
\end{split}
\label{eq241_10}
\end{equation}
where
\begin{equation}
F(\v
r_s,z)=\frac{1}{2\pi}\int^{\omega_0+\Delta}_{\omega_0-\Delta}d\omega\tilde{p}_o(\v
r_s,\omega)e^{\frac{i(\omega-\omega_0)z}{v_g}} \label{eq241_11}
\end{equation}
Since $\omega-\omega_0$ is small, $F(\v r_s,z)$ is a slowly
varying function of $z$. It represents an envelope function, which
in (\ref{eq241_10}), propagates at the group velocity $v_g$. The
envelope function \index{Modulation} modulates a rapidly varying function
\begin{equation}
e^{i(k_z z-\omega_0 t)} \label{eq241_12}
\end{equation}
which travels at the phase velocity $v_{ph}=\frac{\omega_0}{k_z}$.
Hence for a narrow-band signal, a shape-retaining envelope pulse
can propagate in a dispersive waveguide.

\subsection { Attenuation in a Waveguide}
\index{Waveguide!attenuation}

When we have a lossy dielectric medium inside a perfectly
conducting waveguide, the attenuation due to the lossy dielectric
can be ascertained from the formula
\begin{equation}
k_z=\sqrt {\omega ^2\mu\epsilon -k_s^2}={k_z}'+i{k_z}'',
\label{eq241-17}
\end{equation}
where $\epsilon =\epsilon' +i\epsilon''$ is complex. This is
because the mathematical boundary value problem has not changed
when $\epsilon $ becomes complex. Hence, (\ref{eq23-1}) and
(\ref{eq23-2}) hold true even for lossy dielectric.

When the attenuation is due to wall losses because of the finite
conductivity of the metallic wall, the calculation is more
involved. In this case, we can use the following formula derived
from energy conservation
\begin{equation}
{k_z}''=\frac {P_d}{2P_f}, \label{eq241-18}
\end{equation}
where $P_d$ is the time average power dissipated per unit length
while $P_f$ is the total time average power flow in the waveguide.
Since $\^ n \times\v E$ is not identically zero on the waveguide
wall now, $\^ n\cdot(\v E\times\v H^*)$ is not zero on the
waveguide wall.  We can calculate the time average power
dissipated per unit length by integrating $\^ n\cdot(\v E\times \v
H^*)$ over the circumference of the waveguide wall, i.e.,
\begin{equation}
P_d=\frac 1{2}\Re e\oint _C dl\^ n\cdot(\v E\times\v H^*),
\label{eq241-19}
\end{equation}
where $C$ is a contour defining the cross-section of the
waveguide. Since $\^n\times\v E$ is not zero, it can be
approximated by
\begin{equation}
\hat n\times\v E=\sqrt {\frac {\mu\omega }{i\sigma }}\v
H=(1-i)\frac 1{\sigma\delta }\v H, \label{eq241-20}
\end{equation}
where $\delta =\sqrt {\frac 2{\omega\mu\sigma }}$ is the \index{Skin depth}skin
depth in a metallic conductor. The above follows from that in the
metal of the waveguide, the $\v E$ and $\v H$ are related by the
intrinsic impedance of metal which is $\sqrt {{\mu\omega }/{i
\sigma}}$ [see Problem 2-2]. Therefore, (\ref{eq241-19}) becomes
\begin{equation}
P_d=\frac 1{2\sigma\delta }\oint _C dl|\v H|^2, \label{eq241-21}
\end{equation}
where $\v H$ is purely tangential on the waveguide wall, if the
wall is a perfect conductor.  For a perturbation calculation, we
can use the $\v H$ field of a perfectly conducting waveguide to
estimate (\ref{eq241-21}). Note that $\sigma\delta$ is also the
surface conductance of the waveguide.

For a TE mode,
\begin{equation}
\v H=\^ zH_z+\frac {ik_z}{k_s^2}\nabla _sH_z. \label{eq241-22}
\end{equation}
Therefore, from (\ref{eq241-21}), $P_d$ becomes
\begin{equation}
P_d=\frac 1{2\sigma\delta }\oint _C dl\left\{ |H_z|^2+\frac
{|k_z|^2}{k_s^4}|\nabla _sH_z|^2\right\}. \label{eq241-23}
\end{equation}
Using (\ref{eq241-11}), (\ref{eq241-18}), and (\ref{eq241-23}), we
obtain that
\begin{equation}
{k_z}''=\frac {k_s^2\delta }{4k_z}\frac {\oint _C dl\left\{
|H_z|^2+\frac {|k_z|^2}{k_s^4}|\nabla _sH_z|^2\right\}
}{\int\limits _S dS|H_z|^2}. \label{eq241-24}
\end{equation}

For a TM mode,
\begin{equation}
\v H=-\frac {i\omega\epsilon }{k_s^2}\nabla _s\times\^ zE_z.
\label{eq241-25}
\end{equation}
Then, $P_d$ becomes
\begin{equation}
P_d=\frac {\omega ^2\epsilon ^2}{2\sigma\delta k_s^4}\oint _C
dl|\nabla _sE_z|^2, \label{eq241-26}
\end{equation}
and
\begin{equation}
{k_z}''=\frac {k^2\delta }{4k_zk_s^2}\frac {\oint _C dl|\nabla
_sE_z|^2} {\int\limits _S dS|E_z|^2}. \label{eq241-27}
\end{equation}
The above method of computing the attenuation of a waveguide is
also known as the \index{Power-loss method}power-loss method. It is inadequate when the
modes of the waveguides are degenerate.

For the TE case, $k_z\rightarrow k$ when
$\omega\rightarrow\infty$, while $k_s^2$ remains a constant
independent of frequencies. Therefore, from (\ref{eq241-24}), we
have
\begin{equation}
{k_z}''\sim\frac {\delta k}{4k_s^2}\frac {\oint _C dl|\nabla
_sH_z|^2} {\int\limits _S dS|H_z|^2}, \label{eq241-28}
\end{equation}
which increases as the frequency increases since $\delta k =
{\sqrt {2\omega \epsilon / \sigma}}$.

For the TM case, when $\omega\rightarrow \infty$, from
(\ref{eq241-27})
\begin{equation}
{k_z}''\sim\frac {\delta k}{4k_s^2}\frac {\oint _C dl|\nabla
_sE_z|^2} {\int\limits _S dS|E_z|^2}, \label{241-29}
\end{equation}
which also increases as the frequency increases. Therefore, a
metallic waveguide becomes more inefficient at high frequencies.

From (\ref{eq241-24}) and (\ref{eq241-27}), we see that ${k_z}''$
diverges when $k_z \rightarrow 0$, i.e., when the wave tends to
cutoff. This is because there is no real power flow $P_f$ at
cutoff, and hence $P_f\rightarrow 0$ in (\ref{eq241-18}). Since
Equation (\ref{eq241-18}) embodies a perturbation concept, it is
only valid when ${k_z}'' \ll {k_z}'$ .  Therefore, Equations
(\ref{eq241-24}) and (\ref{eq241-27}) are invalid near the cutoff
of the wave. But still, the trend is that $k''_z$ becomes larger
close to cutoff.

\subsubsection{The Magic Modes}
\index{Magic modes}

For some special modes of a waveguide, due to symmetry, the second
term in \eqref{eq241-22} vanishes.  This can happen to some modes of
a highly symmetrical waveguide such as a parallel plate waveguide or
a circular waveguide.  In this case, \eqref{eq241-24} becomes
\begin{equation}
{k_z}''=\frac {k_s^2\delta }{4k_z}\frac {\oint _C dl\left\{
|H_z|^2\right\} }{\int\limits _S dS|H_z|^2}. \label{eq241-24a}
\end{equation}
In this case, $k_z''$ becomes smaller as the frequency increases.
For these modes, the electric field is tangential to the waveguide
wall, with magnetic field normal to the $k$ vector and the electric
field. As the frequency increases, the $k$ vector becomes almost
parallel to the $z$ axis.  The magnetic field becomes almost
vertical to the waveguide wall with a small tangential component.
Hence, the induced surface current on the waveguide wall actually
becomes smaller. Consequently, the attenuation of the waveguide mode
actually decreases with increasing frequency.   Some TE modes of the
circular waveguide are such a mode, and they are known as ``magic
modes'' (see Problems 2.9 and 2.12).

\begin{figure}[htb]
\begin{center}
\hfil\includegraphics[width=2.5truein]{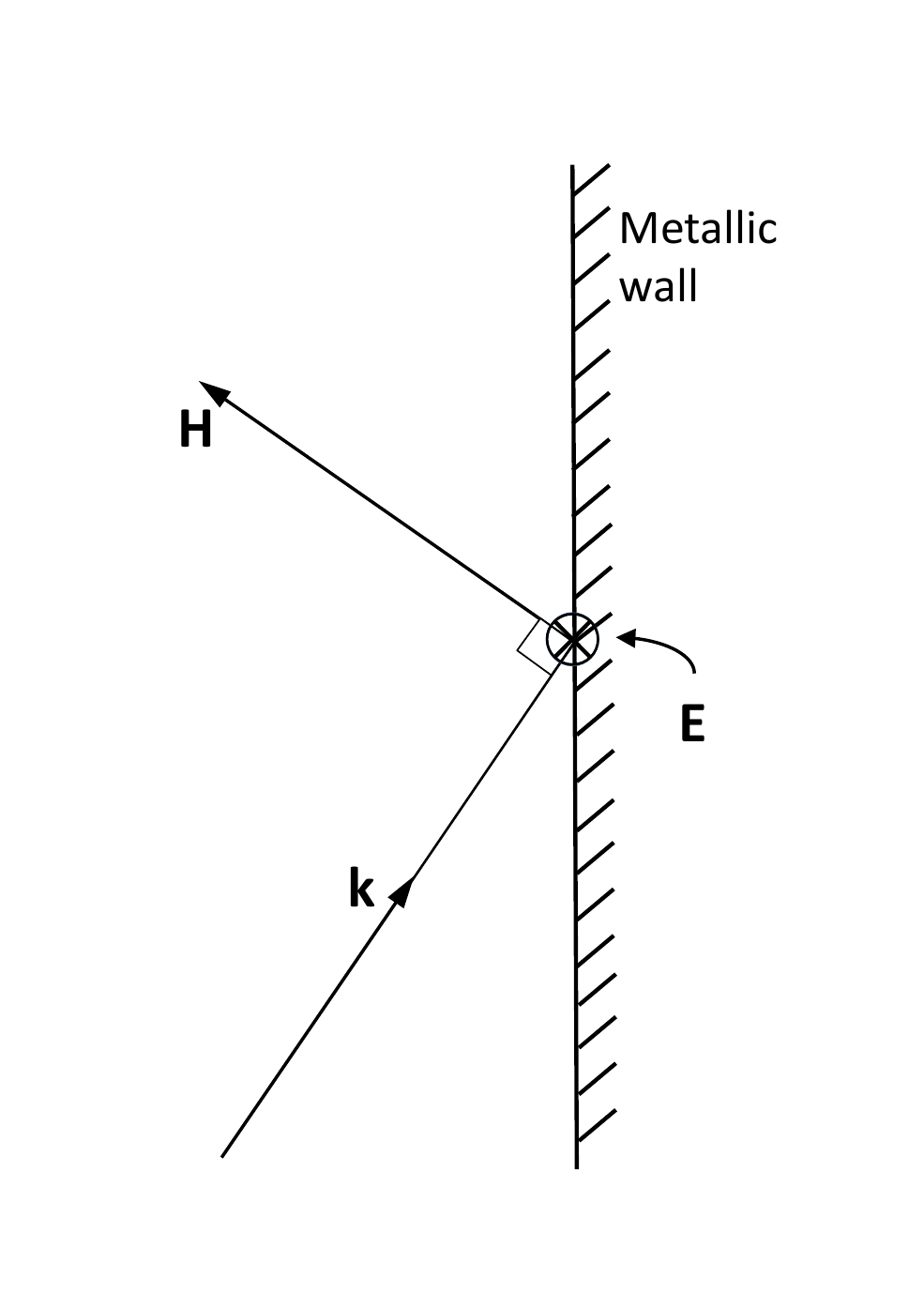}\hfil
\end{center}
\caption{The field configuration of a magic mode on the wall of a
waveguide.  As the frequency increases, the $k$ vector becomes more
parallel to the waveguide wall, and tangential $\v H$ becomes
smaller, since $\v H$ is orthogonal to both $\v E$ and $\v k$.  This
reduces the surface current, and hence, the wall
loss.}\label{MagicModeFields}
\end{figure}

Figure {\ref{fgLoss} shows some typical losses of different modes in
a rectangular and a circular waveguide including a ``magic mode''.
Such low loss is desirable in radio astronomy where the frequency is
high and the signal low.  Hence, circular waveguides with a
corrugated wall to ``discourage'' other modes, but promote the
propagation of this magic mode, is actually used in the design of
the VLA (very large array) of NRAO (National Radio Astronomy
Observatory) in New Mexico.

\begin{figure}[htb]
\begin{center}
\hfil\includegraphics[width=3.5truein]{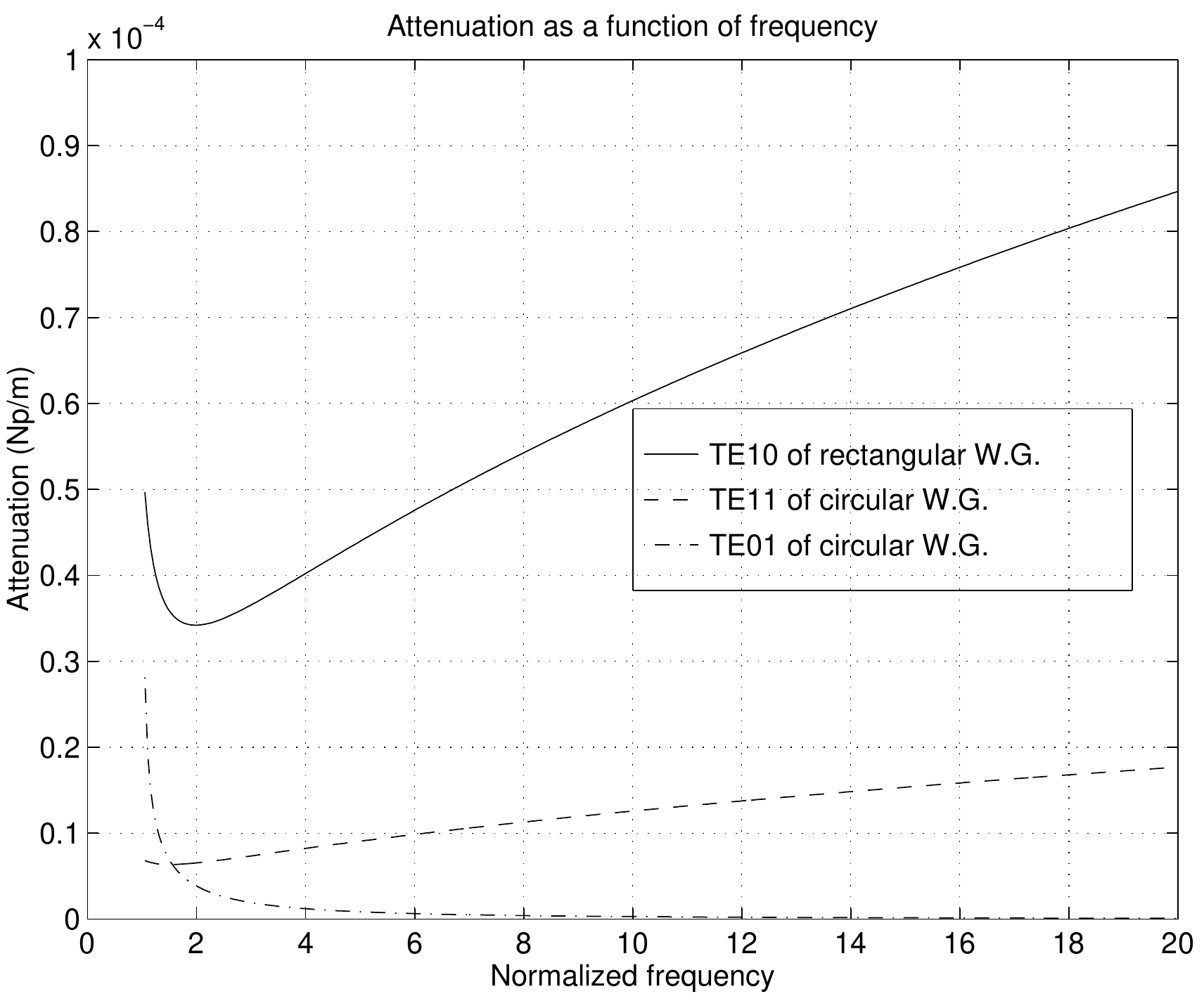}\hfil
\end{center}
\caption{Loss as a function of frequency for different waveguide
modes.}\label{fgLoss}
\end{figure}

%

\section {{{ Excitation of Modes in a Waveguide}}}
\index{Waveguide!excitation of modes}

The modes of a waveguide are excited by putting sources inside a
waveguide.  This is usually in terms of \index{Waveguide!probe}waveguide probe carrying a
current or a charge.  So the probe can be dipole-like, producing
mainly an electric field, or it can be loop-like producing mainly a
magnetic field.  The location of the probe is imperative if certain
desirable modes are to be excited.  We will study the relation
between the \index{Waveguide!probe!location}probe location, the source type, and the modes

\subsubsection{Periodic Boundary Condition}
\index{Boundary conditions!periodic}

The excitation of modes in a waveguide can be made very similar to
the excitation of modes in a cavity by the use of periodic boundary
condition.  Take the example of a \index{Rectangular cavity}rectangular cavity.  The modes are
countably infinite in all three directions and we have
$k^2=\left(\frac{m\pi}{a}\right)^2+\left(\frac{n\pi}{b}\right)^2+\left(\frac{p\pi}{d}\right)^2$
where $m$, $n$, and $p$ are the indices for counting the modes in
the $x$, $y$, and $z$ directions respectively.  In a rectangular
waveguide of infinite length, then
$k^2=\left(\frac{m\pi}{a}\right)^2+\left(\frac{n\pi}{b}\right)^2+k_z^2$
where $k_z^2$ now becomes uncountably infinite as it becomes a
continuum variable rather than a discrete variable in the cavity
case.

However, we can use periodic boundary condition to discretize the
waveguide wavenumber associated with the $z$ axis.  If we have a
traveling wave in the $z$ direction indicated by $\exp(ik_zz)$ with
the requirement that this function repeats itself after distance
$d$, then
\begin{align}\label{PBC1}
e^{ik_zz}|_{z=0}=e^{ik_zz}|_{z=d}
\end{align}
The above implies that
\begin{align}\label{PBC2}
e^{ik_zd}=1
\end{align}
implying that
\begin{align}\label{PBC3}
k_z=\frac{2p\pi}{d},\quad \forall\quad \text{integer}\quad p
\end{align}
The traveling wave then becomes a Fourier mode
\begin{align}\label{PBC4}
e^{ik_zz}=e^{i\frac{2p\pi}{d}z}
\end{align}
When $d\rightarrow\infty$, $k_z$ assumes a continuum of modes as in
Fourier transform.  When $d$ is finite, the modes in the $z$
direction is countably infinite just as the modes in a cavity.

\subsubsection{Generalized Eigenfunction Expansion for Vector Wave Equation}
\index{Wave equation!eigenfunction expansion}
\index{Wave equation!vector}

Given an electric field that satisfies
\begin{align}\label{GEE1}
\nabla\times\nabla\times\v E(\v r)-k_0^2\v E(\v r)=i\omega\mu\v J(\v
r)
\end{align}
we can expand the field in terms of the eigenfunctions of the
following equation
\begin{align}\label{GEE2}
\nabla\times\nabla\times\v F_m(\v r)-k_m^2\v F_m(\v r)=0
\end{align}
assuming that $\v F_m$ satisfies the same boundary condition as the
electric field $\v E$.   First, we let
\begin{align}\label{GEE3}
\v E(\v r)=\sum_m a_m\v F_m(\v r)
\end{align}
On substituting \eqref{GEE3} into \eqref{GEE1}, we have
\begin{align}\label{GEE4}
\sum_m a_m (k_m^2-k_0^2)\v F_m(\v r)=i\omega\mu\v J(\v r)
\end{align}
Assuming an orthonormal relationship for the eigenfunctions such
that\footnote{Such an orthogonality relation with a conjugation is
needed since the operator $\nabla\times\nabla\times$ is a Hermitian
(self-adjoint) operator.}
\begin{align}\label{GEE5}
\int_V d\v r \v F_{m'}^*(\v r) \v F_m(\v r)=\delta_{m'm}
\end{align}
we can deduce that
\begin{align}\label{GEE6}
a_m =i\omega\mu\frac{\langle \v F_m^*, \v J\rangle}{k_m^2-k_0^2}
\end{align}
Consequently, we get
\begin{align}\label{GEE7}
\v E(\v r)=i\omega\mu\sum_m \v F_m(\v r) \frac{\langle\v F_m^*, \v
J\rangle}{k_m^2-k_0^2}
\end{align}
The above is a general eigenfunction expansion formula if we
orthonormalize the eigenfunctions. It says physically that the field
due to a source in a waveguide can be expanded in terms of the
eigenfunctions of the waveguide.  The same expression also holds for
cavity mode expansion.  The \index{Excitation coefficients}excitation coefficients of the
eigenfunctions are given by \eqref{GEE6}.  These coefficients are
proportional to $\langle \v F_m^*, \v J\rangle$ which is the inner
product between the eigenfunction and the source. Hence, it is
important that we learn how to find these eigenfunctions.

Modes are often excited by \index{Current probe}current probes in the waveguide or
cavity. The above expression tells us if we want a certain mode $\v
F_m$ to be strongly excited, we need the inner product $\langle\v
F_m^*, \v J\rangle$ to be large. Hence, the current on the probe
should be located at where the field of the mode $\v F_m$ is strong.
If the probe is a short wire, it can be approximated by an electric
dipole with strong charge accumulation that produces a strong
electric field.  This electric dipole should be placed close to the
maxima of the mode in order to excite it.

On the other hand, if the current source $\v J$ consists of a
\index{Current loop}current loop with constant current,\footnote{This is possible if the
loop size is much smaller than the wavelength.} then the inner
product
\begin{align}
\langle \v F_m^*, \v J\rangle=\int_V d\v r \v F_m^*(\v r)\cdot\v
J(\v r)=\oint_C d\v l\cdot\v F_m^*(\v r)=\int_A dS \hat n\cdot
\nabla\times\v F_m^*(\v r)
\end{align}
where $C$ is the contour of the loop, and $A$ is the cross section
area of the loop.  If $\v F_m$ represents electric field, then
$\nabla\times\v F_m$ represents the magnetic field. Hence, the
current loop has to be placed in location where the magnetic field
of the mode is strong in order to excite it.  A small current loop
behaves like a magnetic dipole and hence, it needs to be placed near
strong magnetic field in order to excite the mode strongly.

Also, due to the $1/(k_m^2-k_0^2)$ dependence of the excitation
coefficient, if the operating frequency of the source is closed to
the resonant frequency of the mode, that mode will be strongly
excited.  This is the phenomenon of \index{Resonance coupling}resonance coupling. Energy can
be coupled to a mode if we operate close to the resonant frequency
of the mode.

\subsection { Vector Wave Functions in a Waveguide}
\index{Waveguide!vector wave functions}

In scalar function theory, we know that an arbitrary function can be
expanded as $f(x)=\sum _{n=0}^{\infty } a_n \cos\left(\frac {n\pi x
}a \right)$ by the completeness of the Fourier cosine basis for
$0<x<a$.  It turns out that the waveguide modes that we have
previously studied do not constitute a \index{Waveguide!complete set}complete set.  First, they
satisfy source-free Maxwell's equations and hence, they are
divergence free.  Hence, they cannot constitute a non-divergence
free field in a waveguide. Second, they propagate in either plus or
minus $z$ directions.  If a source is present in a waveguide, it is
not clear what direction the field is propagating in the source
region. To obtain a complete set, we need to derive the vector wave
functions.

An arbitrary vector function in a waveguide can be expanded in terms
of vector wave functions of the waveguide which is the analogue of
Fourier basis.  However, drawing such an analogy has its pitfall.
The expansion of modes in a cavity excited by a current source has a
colorful history. There are basically two kinds of vector wave
functions in a waveguide or a cavity: the divergence-free type
(solenoidal) and the curl-free type (irrotational). It was believed
by some for a while that the usual divergence free functions are
complete. But in fact, both kinds of functions are needed for
completeness \cite{HANSEN,STRATTON,COLLINF}. More references on this
topic can be found in \cite{WFIME}.

We shall discuss the derivation of such vector wave functions for
a uniform hollow waveguide. Both the electric field and the
magnetic field in a waveguide satisfy the equation
\begin{equation}
\nabla\times\nabla\times\v F-k^2\v F=0. \label{eq28-9}
\end{equation}
It is to be noted that the above is an eigenvalue problem where
$k^2$ is the eigenvalue, and $\v F$ is the eigenfunction. It can be
shown that with the appropriate boundary conditions, the
$\nabla\times\nabla\times$ operator is Hermitian, and that the
eigenvalues $k^2$ are always real.

The above is not actually the equation satisfied by the fields of
the waveguide since $k^2$ is arbitrary and not fixed to be
$\omega^2\mu\epsilon$.  However, solving the above equation for all
possible $k^2$ or eigenvalues generates enough eigenfunctions that
form a complete set.  This includes the cases where $k^2=0$,
implying the functions that belong to the null space of the
$\nabla\times\nabla\times$ operator.  Since $k^2$ is equivalent to
frequency, we can think of the above as solving for all the
resonance solutions (eigensolutions) of a very long cavity. As we
shall see next, if the cavity is of finite length, then $k^2$ will
assume discrete values. But if the cavity is of infinite length,
then $k^2$ assumes a continuum of values.

\subsubsection{Divergence-Free Eigenfunctions}
\index{Divergence-free eigenfunctions}

We can convert the \eqref{eq28-9} into a scalar wave equation via
the transformation
\begin{equation}
\v F=\nabla\times\v c\psi. \label{eq28-10}
\end{equation}
The above vector function is clearly divergence free.
Substituting
(\ref{eq28-10}) into (\ref{eq28-9}), and if $\psi$ satisfies
\begin{equation}
(\nabla ^2+k^2)\psi=0, \label{eq28-11}
\end{equation}
then $\v F$ in (\ref{eq28-10}) will satisfy (\ref{eq28-9}). It is
clear that $\v F$ is transverse to $\v c$. Since in a waveguide, as
well as an infinitely long cavity, we can decompose the field into
TE and TM types with respect to $z$, it is natural to choose $\v c
=\^ z$. For example, if $\psi _h$ satisfies (\ref{eq28-11}) with the
Neumann boundary condition, i.e.,
\begin{equation}
(\nabla ^2+k^2)\psi _h=0, \qquad\frac {\partial\psi _h}{\partial
n}=0, \qquad \text {on}\qquad C \label{eq28-12}
\end{equation}
then the vector wave function
\begin{equation}
\v M_e=\nabla\times (\^ z\psi _h), \label{eq28-13}
\end{equation}
is analogous to the $\v E$-field of a TE mode.  Again, it is to be
emphasized that (\ref{eq28-12}) defines an eigenvalue problem.

Since $\nabla\times\v M_e$ is also a solution to (\ref{eq28-9}),
we can define another vector wave function
\begin{equation}
\v N_h=\frac 1{k}\nabla\times\v M_e=\frac
{1}k\nabla\times\nabla\times\^ z\psi _h. \label{28-14}
\end{equation}
$\v N_h$ is a vector wave function analogous to the magnetic field
of a TE mode.

Similarly, we define a $\psi _e$ satisfying
\begin{equation}
(\nabla ^2+k^2)\psi _e=0, \qquad\psi _e=0, \qquad \text {on}\quad
C. \label{eq28-15}
\end{equation}
Then,
\begin{equation}
\v M_h=\nabla\times\^ z\psi _e, \label{eq28-16}
\end{equation}
is analogous to the $\v H$-field of the TM mode in a waveguide.
Similarly, we have
\begin{equation}
\v N_e=\frac {1}k\nabla\times\v M_h=\frac
{1}k\nabla\times\nabla\times\^ z\psi _e, \label{eq28-17}
\end{equation}
which is analogous to the $\v E$ field of a TM mode.
The eigenfunctions obtained from solving \eqref{eq28-12} and
\eqref{eq28-15} are complete, and hence they can be used to generate
all possible $\v M$ and $\v N$ functions above.

In the previous discussion, the subscript $e$ denotes a quantity
that is related to the $\v E$ field while the subscript $h$
denotes a quantity that is related to the $\v H$-field in terms of
their boundary conditions. $\psi _h$ and $\psi_e$ are sometimes
known as the \index{Magnetic Hertzian potential}magnetic Hertzian potential and the \index{Electric Hertzian potential}
electric Hertzian potential respectively. By looking back at Equations (\ref{eq27})
and (\ref{eq28}), one notes that they are analogous to the $H_z$
and $E_z$ components of the field respectively.

As an example, for a rectangular waveguide,
\begin{subequations}
\begin{equation}
\psi _{hmn}(k_z,\v r)=\cos \left( \frac {m\pi x}a\right)\cos \left(
\frac {n\pi y}b\right)e^{ik_zz}, \label{eq28-18a}
\end{equation}
\begin{equation}
\psi _{emn}(k_z,\v r)=\sin \left( \frac {m\pi x}a\right)\sin \left(
\frac {n\pi y}b\right)e^{ik_zz}. \label{eq28-18b}
\end{equation}
\end{subequations}
We can generate the vector wave function from the above by
\begin{subequations}
\begin{equation}
\v M_{emn}(k_z,\v r)=\nabla\times\^ z\cos\left( \frac {m\pi x
}a\right)\cos \left( \frac {n\pi y}b\right)e^{ik_zz}, \label{28-19a}
\end{equation}
\begin{equation}
\v N_{hmn}(k_z,\v r)=\frac 1{k}\nabla\times\nabla\times\^
z\cos\left( \frac {m\pi x }a\right)\cos \left( \frac {n\pi
y}b\right)e^{ik_zz}, \label{eq28-19b}
\end{equation}
\end{subequations}
and
\begin{subequations}
\begin{equation}
\v M_{hmn}(k_z,\v r)=\nabla\times\^ z\sin \left(\frac {m\pi x
}a\right)\sin \left( \frac {n\pi y}b\right) e^{ik_zz},
\label{eq28-20a}
\end{equation}
\begin{equation}
\v N_{emn}(k_z,\v r)=\frac 1{k}\nabla\times\nabla\times\^ z\sin
\left( \frac {m\pi x }a\right)\sin \left( \frac {n\pi y}b\right)
e^{ik_zz}. \label{eq28-20b}
\end{equation}
\end{subequations}
We allow $k_z$ to be a free variable (arbitrary variable) in
equations (\ref{eq28-18a}) and (\ref{eq28-18b}). This would render
$\psi_{hmn}$ and $\psi_{emn}$ to be complete scalar functions in a
hollow waveguide.  At least, we know that an arbitrary function of
$z$ can be Fourier expanded in terms of $e^{ik_z z}$.


The above is analogous to finding the eigenvalues of a rectangular
cavity of dimension $a\times b\times d$ with periodic boundary
condition in the $z$ direction. In this case, the eigenvalue is
given by $k^2=\left( \frac {m\pi }a\right) ^2+\left (\frac {n\pi
}b\right)^2+\left (\frac {2p\pi }d\right)^2$, where $m$, $n$, and
$p$ are all integers.  We can identify $k_z=\frac {2p\pi }d$ . As we
let $d\rightarrow\infty$, $k_z$ which previously takes on discrete
values, becomes a continuum variable. Since $k_z$ can be any
continuum real variable, the eigenvalue $k^2=k_s^2+k_z^2$ can also
take on any continuum real variable. In the above $k_s$ is the
transverse eigenvalue of the waveguide problem similar to
\eqref{eq23-1} and \eqref{eq23-1}.  We shall denote it as $k_{is}$
subsequently, where the index $i$ implies an ordered pair $(m,n)$ in
the case of a rectangular waveguide.

As of this point, these vector wave functions are not physical
modes of a waveguide.  In order for them to be \index{Waveguide!physical modes}physical modes,
$k_z^2$ has to satisfy the dispersion relation $k_0^2=k_s^2+k_z^2$
where $k_0$ is the wavenumber inside the waveguide.  Since the
dispersion relation describes an equation of a sphere in the $k$
space, this sphere is known as the \index{Ewald sphere}Ewald sphere or the \index{Energy shell}energy
shell in physics.  The eigenfunctions derived so far do not
satisfy the dispersion relation, and they are said to be off the
energy shell.  When they are forced to satisfy the dispersion
relation, they are said to be on the energy shell.

Given the above information, we can easily show that
\begin{equation}
\begin{split} \int dV\psi _{hmn}(k_z,\v r)\psi^*
_{hm'n'}({k_z}',\v r)&=\\
&(1+\delta _{0m})(1+\delta _{0n})\frac {\pi ab}2\delta
(k_z-{k_z}')\delta _{mn, m'n'}, \label{eq28-21}
\end{split}
\end{equation}
where $\delta (x)$ is the Dirac delta function and $\delta _{ij}$,
the Kronecker delta function. In general, for a waveguide of
arbitrary cross-section, \index{Waveguide!orthogonality relations}orthogonality relations exist for the
$\psi _h$'s and $\psi _e$'s as [analogous to (\ref{eq23-3}) and
(\ref{eq23-4})] \footnote {See \cite{WFIME} p. 388.}
\begin{subequations}
\begin{equation}
\int dV\psi _{hi}(k_z,\v r)\psi^* _{hj}({k_z}',\v r)=A_{hi}\delta
(k_z-{k_z}')\delta _{ij}, \label{28-22a}
\end{equation}
\begin{equation}
\int dV\psi _{ei}(k_z,\v r)\psi^* _{ej}({k_z}',\v r)=A_{ei}\delta
(k_z-{k_z}')\delta _{ij}, \label{eq28-22b}
\end{equation}
\end{subequations}
where $A_{hi}$ and $A_{ei}$ are the appropriate normalization
constants.\footnote{Notice that when $k_z$ is real, $\psi
_{ej}(-{k_z})=\psi^* _{ej}({k_z})$, but since $k_z$ will become
complex in Cauchy integration technique applied on the complex
plane, a technique that we will use later, we will retain this
notation in the above.}

For the vector wave functions, the orthogonality relationships are
\begin{subequations}
\begin{equation}
\int dV\v M_{{\scriptstyle e\atop \scriptstyle h }i}(k_z,\v
r)\cdot\v M^*_{{\scriptstyle e\atop \scriptstyle h }j}({k_z}',\v
r)=k_{is{\scriptstyle h\atop \scriptstyle e}}^2A_{{\scriptstyle
h\atop\scriptstyle e}i}\delta (k_z-{k_z}')\delta _{ij},
\label{eq28-23a}
\end{equation}
\begin{equation}
\int dV\v N_{{\scriptstyle e\atop \scriptstyle h }i}(k_z,\v
r)\cdot\v N^*_{{\scriptstyle e\atop \scriptstyle h }j}({k_z}',\v
r)=k_{is{\scriptstyle e\atop \scriptstyle h }}^2A_{{\scriptstyle
e\atop\scriptstyle h}i}\delta (k_z-{k_z}')\delta _{ij},
\label{eq28-23b}
\end{equation}
\begin{equation}
\int dV\v M_{{\scriptstyle e\atop \scriptstyle h }i}(k_z,\v
r)\cdot\v N^*_{{\scriptstyle e\atop \scriptstyle h }j}({k_z}',\v
r)=0, \qquad \text {for all}\quad i,\ j, \ k_z, \ {k_z}'.
\label{eq28-23c}
\end{equation}
\end{subequations}

Because the vector functions $\v M$ and $\v N$ are divergence free,
and that their curls are not zero, these functions are also termed
the \index{Solenoidal vector wave functions}solenoidal vector wave functions. They can be used to expand
divergence-free fields.

\subsubsection{Curl-Free Eigenfunctions}
\index{Curl-free eigenfunctions}

The expressions (\ref{eq28-16}) and (\ref{eq28-18a}) and
(\ref{eq28-18b}) are clearly divergence free. However, $\v M$ and
$\v N$ vector wave functions are not, in general, complete.  They
cannot be used to represent fields whose divergence is non zero. To
remedy this, the functions in the null-space of the
$\nabla\times\nabla\times$ operator need to be considered. We need
the $\v L$ functions, which are defined as\footnote{It is to be
noted that \index{Theorem!Helmholtz}Helmholtz theorem says that an arbitrary vector field can
be decomposed into the sum of divergence-free field and curl-free
field.  In other words, $\v F=\nabla\times A+\nabla\psi$. We expect
to see such decomposition here.}
\begin{equation}
\v L_h=\nabla\psi _h, \qquad\qquad\v L_e=\nabla\psi _e,
\label{eq28-27}
\end{equation}
where $\psi _h$ and $\psi _e$ are as defined in (\ref{eq28-12}) and
(\ref{eq28-15}). $\v L_h$ satisfies the magnetic field boundary
condition while $\v L_e$ satisfies the electric field boundary
condition on the waveguide wall. The vector wave functions have zero
curl and non-zero divergence. Hence, they are also known as the
\index{Irrotational vector wave functions}irrotational vector wave functions. They are solutions to Equation
(\ref{eq28-9}) corresponding to when $k^2=0$.  Hence, they are the
null-space solution of the $\nabla\times\nabla\times$ operator.

Their divergence is
\begin{equation}
\nabla\cdot\v L_h=-k^2\psi _h, \qquad\qquad \nabla\cdot\v
L_e=-k^2\psi _e. \label{eq28-28}
\end{equation}
Since the divergence of field is proportional to charge, the right
hand sides of the above represent charges. But since $\psi_h$ and
$\psi_e$ form complete sets, they can be used to expand fields due
to arbitrary sources inside the waveguide.

It can be shown that
\begin{subequations}
\begin{equation}
\int dV\v L_{{\scriptstyle e\atop \scriptstyle h }i}(k_z,\v
r)\cdot\v L^*_{{\scriptstyle e\atop \scriptstyle h }j}({k_z}',\v
r)=k^2A_{{\scriptstyle e\atop \scriptstyle h}i}\delta
(k_z-{k_z'})\delta _{ij}, \label{eq28-29a}
\end{equation}
\begin{equation}
\int dV\v M_{{\scriptstyle e\atop \scriptstyle h }i}(k_z,\v
r)\cdot\v L^*_{{\scriptstyle e\atop \scriptstyle h }j}({k_z}',\v
r)=0, \qquad \text {all}\quad i, \ \ j, \ \ k_z, \ \ {k_z}',
\label{eq28-29b}
\end{equation}
\begin{equation}
\int dV\v N_{{\scriptstyle e\atop \scriptstyle h }i}(k_z,\v
r)\cdot\v L^*_{{\scriptstyle e\atop \scriptstyle h }j}({k_z}',\v
r)=0, \qquad \text {all}\quad i, \ \ j, \ \ k_z, \ \ {k_z}',
\label{eq28-29c}
\end{equation}
\end{subequations}

\subsubsection{Eigenfunction Expansion of Arbitrary Fields}

An arbitrary field in a waveguide can, in general, be expanded as
\begin{subequations}
\begin{equation}
\v E(\v r)=\int^\infty_{-\infty} dk_z\sum _i[a_{ei}(k_z)\v
M_{ei}(k_z,\v r)+b_{ei}(k_z)\v N_{ei}(k_z,\v r)+c_{ei}(k_z)\v
L_{ei}(k_z,\v r)], \label{eq28-30a}
\end{equation}
\begin{equation}
\v H(\v r)=\int^\infty_{-\infty} dk_z\sum _i[a_{hi}(k_z,\v r)\v
M_{hi}(k_z,\v r)+b_{hi}(k_z)\v N_{hi}(k_z,\v r)+c_{hi}(k_z)\v
L_{hi}(k_z,\v r)]. \label{eq28-30b}
\end{equation}
\end{subequations}
The coefficients can be found from the orthogonality
relationships.

\subsection { Dyadic Green's Function}
\index{Green's function!dyadic}

The dyadic Green's function in a waveguide is the solution to the
equation
\begin{equation}
\nabla\times\nabla\times\dyad G(\v r, \v r')-k_0^2\dyad G(\v r, \v
r')=\dyad I\delta (\v r-\v r'), \label{eq28-31}
\end{equation}
satisfying the boundary condition $\^ n\times\dyad G(\v r, \v
r')=0$, for $\v r$ on the waveguide wall.  Once this Green's
function is known, the field due to an arbitrary distributed
source in a waveguide can be written as
\begin{equation}
\v E=i\omega\mu\int \dyad G(\v r, \v r')\cdot\v J(\v r')d\v r'.
\label{eq28-32}
\end{equation}

\subsubsection{A General Dyadic Green's Function}
\index{Green's function!general dyadic}

From the definition of the dyadic Green's function as given by
\eqref{eq28-32}, and from the generalized formula for mode expansion
due to a current source, as given in \eqref{GEE7}, we can deduce
that the general dyadic Green's function is of the form
\begin{align}
\dyad G(\v r,\v r')=\sum_m \frac{\v F_m(\v r) \v F_m^*(\v
r')}{k_m^2-k_0^2}
\end{align}
The above is a succinct way to express the dyadic Green's function
in terms of the modes of the cavity, where the modes are assumed to
be orthonormal.

\subsubsection{Dyadic Green's Function for Hollow Waveguide}
\index{Green's function!for hollow waveguide}

The derivation of the dyadic Green's function has a colorful
history as the mode expansion in a cavity
\cite{COLLINF,TAI,TAI2,YAGHJIAN}.  The controversy comes from the
incompleteness of the divergence-free modes in a waveguide or a
cavity.  More references can be found in \cite{WFIME}.

To solve Equation (\ref{eq28-31}), we expand $\dyad G(\v r, \v
r')$ in terms of the vector wave functions $\v M_e$, $\v N_e$, and
$\v L_e$. In other words,
\begin{equation}
\begin{split}
\dyad G(\v r, \v r')=\int^\infty_{-\infty} dk_z\sum _i & [\v
M_{ei}(k_z, \v r)\v a_{ei}(k_z,
\v r')+\v N_{ei}(k_z, \v r)\v b_{ei}(k_z, \v r')\\
& +\v L_{ei}(k_z, \v r)\v c_{ei}(k_z, \v r')]. \label{eq28-33}
\end{split}
\end{equation}
Substituting $\dyad G(\v r, \v r')$ into (\ref{eq28-31}), and
using Equation (\ref{eq28-1}), we have
\begin{equation}
\begin{split}
\int^\infty_{-\infty} dk_z\sum _i\{ (k^2- & k_0^2)[\v M_{ei}(k_z,
\v r)\v a_{ei}(k_z,
\v r')+\v N_{ei}(k_z, \v r)\v b_{ei}(k_z, \v r')]\\
& -k_0^2\v L_{ei}(k_z, \v r)\v c_{ei}(k_z, \v r')\} =\dyad I\delta
(\v r -\v r'). \label{eq28-34}
\end{split}
\end{equation}
From the orthogonality relations for $\v M$, $\v N$, and $\v L$
functions, we have
\begin{subequations}
\begin{equation}
\v a_{ei}(k_z, \v r')  =\frac 1{k_{ish}^2A_{hi}(k_z^2-k_{izh}^2)}\v
M^*_{ei} (k_z, \v r'), \label{eq28-35a}
\end{equation}
\begin{equation}
\v b_{ei}(k_z, \v r')  =\frac 1{k_{ise}^2A_{ei}(k_z^2-k_{ize}^2)}\v
N_{ei}^*(k_z, \v r'), \label{eq28-35b}
\end{equation}
\begin{equation}
\v c_{ei}(k_z, \v r')  =\frac {-1}{k^2A_{ei}k_0^2}\v L_{ei}^*(k_z,
\v r'). \label{eq28-35c}
\end{equation}
\end{subequations}
In the above, we have replaced $k^2-k_0^2$ with
$k_z^2-k_{iz{\scriptstyle e\atop\scriptstyle h}}^2$, because $k^2
= k_{is{\scriptstyle e\atop\scriptstyle h}}^2 + k_z^2$, $k_0^2 =
k_{is{\scriptstyle e\atop\scriptstyle h}}^2 +k_{iz{\scriptstyle
e\atop\scriptstyle h}}^2$.  Here, $k_{iz}$ denotes values on the
energy shell, while $k_z$ denotes values off the energy shell. The
subscripts $e$ and $h$ on the transverse eigenvalue $k_{is}$ and
longitudinal wavenumber $k_{iz}$ denote the association of these
values with either the Dirichlet or the Neumann problem
respectively.

Therefore, the dyadic Green's function is
\begin{equation}
\begin{split}
\dyad G(\v r, \v r')=\int^\infty_{-\infty} dk_z & \sum _i \left[
\frac {\v M_{ei}(k_z, \v r)\v M_{ei}^*(k_z, \v
r')}{k_{ish}^2A_{hi}(k_z^2-k_{izh}^2)}\right. \\ & \left. +\frac {\v
N_{ei}(k_z, \v r)\v N_{ei}^*(k_z, \v
r')}{k_{ise}^2A_{ei}(k_z^2-k_{ize}^2)} -\frac {\v L_{ei}(k_z, \v
r)\v L_{ei}^*(k_z, \v r')}{k^2A_{ei}k_0^2}\right]. \label{eq28-36}
\end{split}
\end{equation}

\subsubsection{Cauchy Integration Technique}
\index{Cauchy integration}

The first two integrals are of the form
\begin{equation}
I=\int^\infty_{-\infty} dk_z\frac
{e^{ik_z(z-z')}f(k_z)}{k_z^2-k_{iz}^2}. \label{eq28-37}
\end{equation}
There are \index{Poles} poles at $k_z =\pm k_{iz}$.  If we assume a small loss
in the medium, then the poles are off the real axis and the
integral (\ref{eq28-37}) is well-defined. If
${f(k_z)}/{k_z^2}\rightarrow 0$ when $k_z \rightarrow\infty$, for
$z>z'$, we can deform the contour of integration from the real
axis to the contour $C$. By virtue of \index{Jordan's lemma} Jordan's lemma, the integral
over $C$ vanishes and the integral (\ref{eq28-37}) is then equal
to the residue of the pole at $k_z = k_{iz}$. When $z < z'$, we
can deform the path of integration downward, and equate the
integral (\ref{eq28-37}) to the residue of the pole at $k_z =
-k_{iz}$. Therefore, it follows that
\begin{equation}
I=\pi i \frac {e^{\pm ik_{iz}(z-z')}}{k_{iz}} f(\pm k_{iz}),
\qquad
\begin{aligned}
\ &z > 0\\
\ &z <0.
\end{aligned}
\label{eq28-38}
\end{equation}
Note that the process of Cauchy integration technique forces $k_z$
to be on the energy shell or on the Ewald sphere.

We can apply Cauchy integration technique to the first two integrals
since $\v M_{ei}$ and $\v N_{ei}$ tend to be constants when
$k_z\rightarrow\infty$. However, because
\begin{equation}
\v N_{ei}=\frac {1}{k}\nabla\times\v M_{hi}=\frac
{1}k\nabla\times\nabla\times\^z\psi _{ei}, \label{eq28-39}
\end{equation}
\begin{figure}[htb]
\begin{center}
\hfil\includegraphics[width=4.0truein]{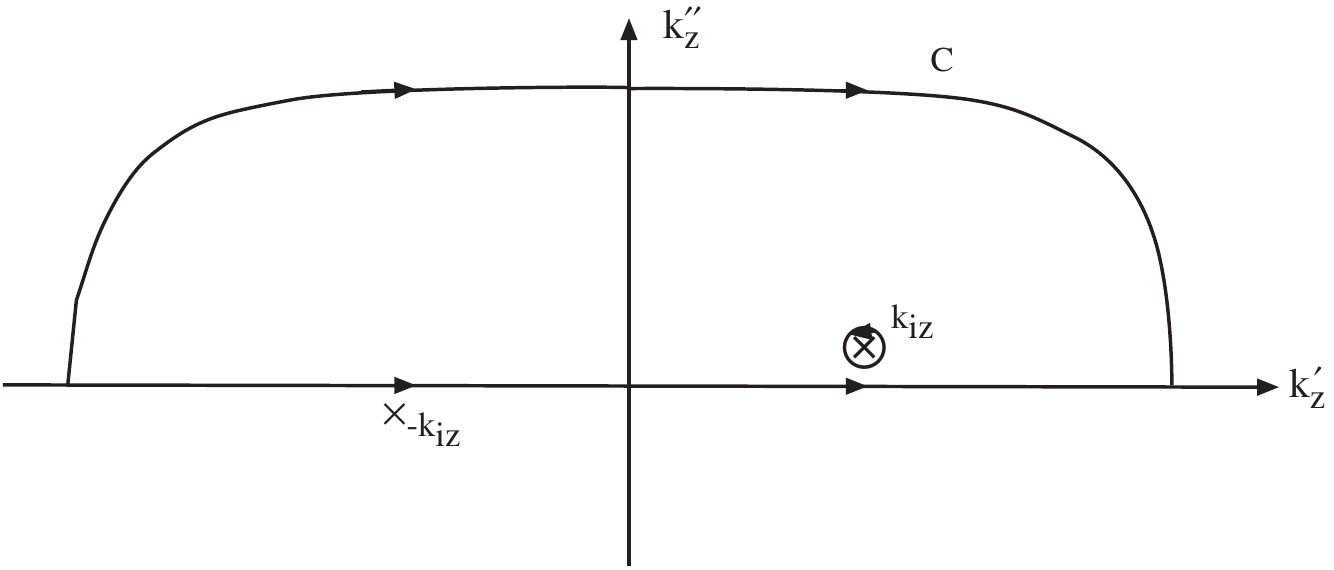}\hfil
\end{center}
\caption{Contour integration on the complex $k_z$ plane.}\label{fgContour}
\end{figure}
%
%
%
$\v N_{ei}(\v k_z, \v r)\v N_{ei}(-k_z,\v r')$ is proportional to
$1/{k^2}$. But $k^2=k_{ise}^2 + k_z^2$, implying that there are
additional poles at $k_z = \pm ik_{ise}$ for the second term in
(\ref{eq28-36}). Similarly, the third term in (\ref{eq28-36}) also
has a $1/{k^2}$ dependence with poles at $k_z = \pm ik_{ise}$. It
can be shown that the pole contributions at $k_z =\pm ik_{ise}$
for the second and the third terms cancel each other.

Since
\begin{equation}
\v L_{ei}(k_z, \v r)\sim\^ zik_z\psi _{ei}, \qquad
k_z\rightarrow\infty, \label{eq28-40}
\end{equation}
the third term tends to be a constant when $k_z\rightarrow\infty$.
Therefore, contour integration cannot be applied to the third
term, since Jordan's lemma is not satisfied. To remedy this, we
write (\ref{eq28-36}) as
\begin{equation}
\begin{split}
\dyad G(\v r, \v r') &=\int^\infty_{-\infty} d k_z\sum _i\left[
\frac {\v M_{ei}(k_z, \v r)\v M_{ei}^*(k_z, \v
r')}{k_{ish}^2A_{hi}(k_z^2-k_{izh}^2)}\right. \\
&\hskip1in +\frac {\v
N_{ei}(k_z, \v r)\v N_{ei}^*(k_z, \v
r')}{k_{ise}^2A_{ei}(k_z^2-k_{ize}^2)}
\\
&\left. - \left( \frac {\v L_{ei}(k_z, \v r)\v L_{ei}^*(k_z, \v
r')}{k^2A_{ei}k_0^2}-\frac {\^ z\^ z\psi _{ei}(k_z, \v r)\psi
_{ei}^*(k_z, \v
r')}{k_0^2A_{ei}}\right) \right] \\
& -\int^\infty_{-\infty} dk_z\sum _i \frac {\^ z\^ z\psi _{ei}(k_z,
\v r)\psi _{ei}^*(k_z, \v r')}{k_0^2A_{ei}}.
\end{split}
\label{eq28-41}
\end{equation}
Contour integrations can now be performed on the third term,
giving rise to a pole contribution that cancels a similar
contribution from the second term. The second term contains a pole
at $k=0$ becuse the $\v N$ function is proportional to $1/k$ as
evident from (\ref{28-14}).  Hence, the second term is
proportional to $1/k^2$ just as the third term. Since $k=0$ for
this pole contribution, for the sake of discussion, we will call
this the static pole.

It is to be noted that if the $\v L$ functions are not used (which
is erroneous) in the expansion, a modal contribution from the
second term due to this static pole will exist.  This mode will
satisfy the dispersion relation $k_x^2+k_y^2+k_z^2=0$, making it
an nonphysical mode in the presence of a time-harmonic excitation.
Fortunately, it is cancelled by a similar contribution from the
third term, the curl-free term.  On first sight, it may seem
strange that a contribution from the divergence-free term should
cancel one from the curl-free term.  However, a closer examination
shows that this static pole contributes both to a divergence-free
and curl-free field.  The role of the curl-free modes outside the
source region is also discussed in \cite{JOHNSON_HOWARD_DUDLEY}.

Consequently, we have
\begin{equation}
\begin{split}
\dyad G(\v r,\v r')=\sum _i & \pi i \left[ \frac {\v M_{ei}(\pm
k_{izh}, \v r)\v M_{ei}^*(\pm k_{izh}, \v
r')}{k_{izh}k_{ish}^2A_{hi}} \right.\\ &\left. + \frac {\v
N_{ei}(\pm k_{ize}, \v r)\v N_{ei}^*(\pm k_{ize}, \v
r')}{k_{ize}k_{ise}^2A_{ei}} \right] -\frac {\^ z\^ z}{k_0^2}\delta
(\v r-\v r'),
\begin{aligned}
\ &z > z'\\
\ &z <z'.
\end{aligned}
\end{split}
\label{eq28-42}
\end{equation}
The upper sign is chosen when $z > z'$ and the lower sign is
chosen when $z < z'$. The identity
\begin{equation}
\delta (\v r-\v r')=\int_{-\infty}^{\infty} dk_z\sum _i\frac {\psi
_{ei}(k_z, \v r)\psi _{ei}^*(k_z, \v r')}{A_{ei}}, \label{eq28-43}
\end{equation}
has been used to simplify the last integral. All the vector wave
functions in (\ref{eq28-42}) evaluated on the Ewald sphere. Hence,
they are now physical wave functions which are solutions to
Maxwell's equations.

The Dirac delta function part of the Green's function in
(\ref{eq28-42}) has a $\^z\^ z$ component. This is because we have
performed the $dk_z$ integration first, letting $k_z
\rightarrow\infty$, before letting the index $i$ go to infinity.
When $k_z\rightarrow\infty$, it also implies that we are looking
at length scales in the $\^ z$ direction with infinite resolution
before the length scales in the transverse direction. Hence, the
singularity in (\ref{eq28-42}) is exactly the one obtained if one
were to use a disk-shaped pill box in performing the principal
volume integral. Therefore, in order to obtain a unique, correct
solution when applying (\ref{eq28-42}), one has to use a
disk-shaped principal volume integral.\footnote {See \cite{WFIME},
Chapter 7.}

Once the dyadic Green's function of a hollow waveguide is known,
the \index{Excitation of modes}excitation of modes due to an arbitrary current source in a
waveguide can be found using (\ref{eq28-32}). By substituting
Equation (\ref{eq28-42}) into (\ref{eq28-32}), we have
\begin{equation}
\begin{split}
\v E(\v r)&=\\
&-\omega \mu \pi \sum _i \left[ \frac {1}{k_{izh} k^2_{ish} A_{hi}}
\int \v M_{ei}(\pm k_{izh}, \v r) \v M_{ei}^*(\pm k_{izh}, \v r')
\cdot \v J(\v r')d\v r' \right. \\
& \left. +\frac {1}{k_{ize} k^2_{ise} A_{ei}}\int \v N_{ei}(\pm
k_{ize}, \v r)
\v N_{ei}^* (\pm k_{ize}, \v r')\cdot\v J(\v r')d\v r' \right] \\
& - \frac {i}{\omega \epsilon} \^ z J_z(\v r).
\end{split}
\label{eq28-44}
\end{equation}
The above integral is over the support of the current $\v J(\v
r').$ If $z > z'_{max}$ or $z < z'_{min}$, where $z'_{max}$ and
$z'_{min}$ define the range of the support of  $\v J(\v r')$ in
the $z$-direction, i.e., $ \v J(\v r')=0$ for  $z > z'_{max}$ and
$z < z'_{min}$, then the above can be written as
\begin{equation}
\begin{split}
\v E(\v r) &= \\
&- \omega \mu \pi  \sum _i \left[ \frac {1} {k_{izh} k^2_{ish}
A_{hi}} \v M_{ei}(\pm k_{izh}, \v r) \int \v M_{ei}^*(\pm k_{izh},
\v r')\cdot
\v J(\v r')d\v r' \right. \\
& \left. +\frac {1} {k_{ize} k_{ise} A_{ei}} \v N_{ei}(\pm k_{ize},
\v r)\int \v N_{ei}^* (\pm k_{ize}, \v r')\cdot \v J(\v r')d\v r'
\right],
\begin{aligned}
\ &z > z'_{max}\\
\ &z <z'_{nmn}.
\end{aligned}
\label{eq28-45}
\end{split}
\end{equation}
Notice that $\v M_{ei}$ and $\v N_{ei}$ denote the \index{Mode!TE}TE and \index{Mode!TE}TM modes
of a waveguide, respectively. The integrals are proportional to
the \index{Waveguide!excitation coefficients}excitation coefficients of the waveguide modes. From the
above, we can see that the excitation coefficient of a waveguide
mode $\v E_j$ is proportional to
\begin{equation}
{\text{Excitation Coefficient}} \sim \int \v E_j(\v r')\cdot\v
J(\v r')d\v r'. \label{eq28-46}
\end{equation}

In other words, to excite a certain mode $\v E_j$ in a waveguide
maximally, there should be as much projection of $\v J$  onto $\v
E_j$. To avoid the excitation of a mode $\v E_j$, the current $\v
J$ should be chosen to be orthogonal to the mode $\v E_j$.

\subsection { Excitation of Modes by a Filamental Current}
\index{Filamental current}

Consider a probe in a waveguide as shown in Figure
\ref{fgExcitation} \cite{KONG}. A current in the \index{Waveguide!probe}probe will produce
an electromagnetic field that couples to the modes of the waveguide.
We shall discuss how to calculate the amplitudes of the excited
waveguide modes. Let us assume that the current on the probe is
described by a current sheet
\begin{equation}
\v J_s=\^ y I_0\delta (x-d). \label{eq28-1}
\end{equation}
The above is a current sheet in the $xy$ plane, and has a dimension
of amperes per meter.  In this simplified case, it is a current
flowing in the $y$ direction, and is a function of $x$ only. This
probe current does not produce an $E_z$ component of the electric
field. Hence, we do not expect the TM modes to be excited. However,
the \index{Mode!TE}TE modes will be excited because the probe current will produce
an $H_z$ component of the magnetic field. If, on the other hand, the
current on the probe is not a constant, there will be charge build
up on the probe from $\nabla\cdot\v J-i\omega\rho =0$. This charge
will induce an $E_z$ component of the electric field, coupling to
the TM modes [see Problem 2-11].

\begin{figure}[htb]
\begin{center}
\hfil\includegraphics[width=3.5truein]{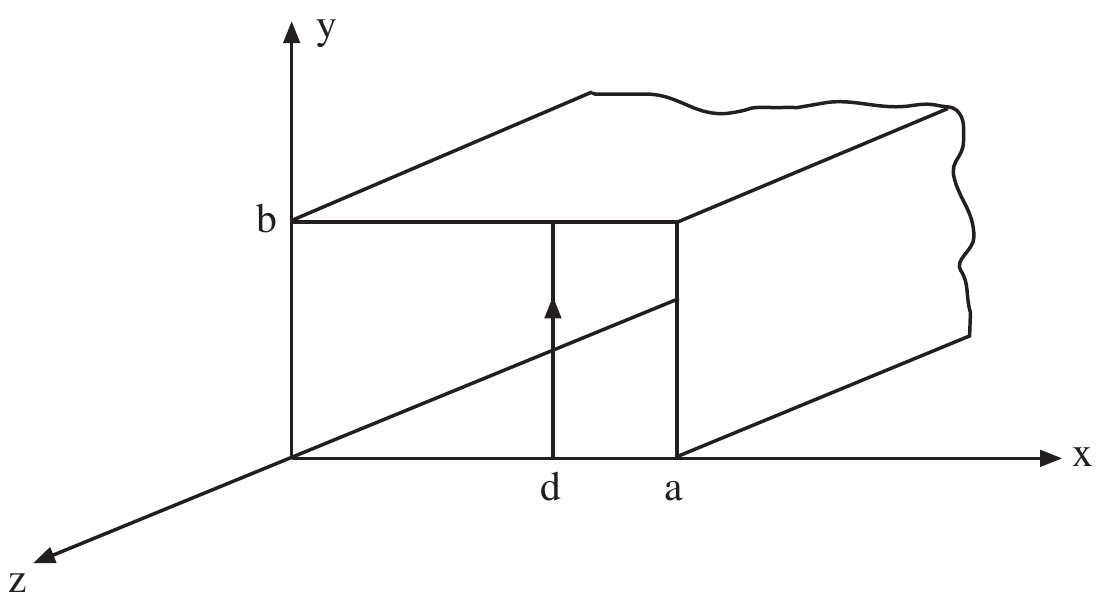}\hfil
\end{center}
\caption{Excitation of a rectangular waveguide by a current
probe.}\label{fgExcitation}
\end{figure}


For the current probe described by (\ref{eq28-1}), only TE modes
will be excited and we can write the field as
\begin{equation}
H_z=\left\{
\begin{aligned}
&\sum \limits _{m, n} H_{mn}^+ \cos \left(\frac {m\pi x}a\right)\cos \left(\frac {n\pi y}b\right) e^{ik_zz}, \hskip30pt z>0\\
&\sum \limits _{m, n} H_{mn}^- \cos \left(\frac {m\pi x}a\right)\cos \left(\frac
{n\pi y}b\right) e^{-ik_zz}, \hskip25pt z<0.
\end{aligned}
\right. \label{eq28-2}
\end{equation}
where $k_z=\sqrt{k^2_0-\left(\frac {m\pi }{a}\right)^2-\left(\frac
{n\pi }{b}\right)^2}$. The boundary condition across a current sheet
is that
\begin{equation}
\^ z\times [\v H(z+)-\v H(z-)]=\v J_s, \qquad\^ z\times[\v E(z+)-\v
E(z-)]=0, \label{eq28-3}
\end{equation}
where $\^ z\times\v H$ gives the transverse field component of the
magnetic field. Since $\v J_s$ has no $y$ variation, we expect $\hat
z \times \v H$ not to have any $y$ variation. Hence, we can assume
that $H_{mn}^{\pm} = 0$, $n\ne 0$. Assuming this, we can derive the
transverse field components from (\ref{eq28-2}) using (\ref{eq27})
and (\ref{eq28}), giving
\begin{subequations}
\begin{equation}
\v H_s=-\sum _{m}H_{m0}^+\frac {ik_z}{k_s^2}\^ x\frac {m\pi }a\sin
\left (\frac {m\pi x}a\right)e^{ik_zz},\qquad z>0, \label{eq28-4a}
\end{equation}
\begin{equation}
\v H_s=\sum _{m}H_{m0}^-\frac {ik_z}{k_s^2}  \^ x\frac {m\pi }a\sin
\left ( \frac {m\pi x}a\right)e^{-ik_zz},\qquad z<0, \label{eq28-4b}
\end{equation}
\end{subequations}
where $k_s^2 =(\frac {m\pi }a)^2+(\frac {n\pi }b)^2$. Applying the
first boundary condition, we get
\begin{equation}
-\sum _{m}\frac {ik_z}{k_s^2}\^ y\frac {m\pi }a\sin \left( \frac
{m\pi x}a\right) ( H_{m0}^++H_{m0}^-) =\^ yI_0\delta (x-d).
\label{eq28-5}
\end{equation}
Therefore, from the above boundary condition, only the TE$_{m0}$
modes will be excited. Equation (\ref{eq28-5}) alone is not
sufficient to determine the unknowns $H_{m0}^+$ and $H_{m0}^-$.
Therefore, we need to apply the second boundary condition in
(\ref{eq28-3}). However, if one were to note that $H_z(z+) =
H_z(z-)$ (a boundary condition which is the subset of the second
boundary condition in (\ref{eq28-3}), we note that
\begin{equation}
H_{m0}^+=H_{m0}^-. \label{eq28-6}
\end{equation}
Using (\ref{eq28-6}) in (\ref{eq28-5}) will uniquely determine the
solution of (\ref{eq28-5}). Consequently, (\ref{eq28-5}) becomes
\begin{equation}
\sum _{m=1}^{\infty }-\frac {2ik_z}{k_s^2}H_{m0}^+\frac {m\pi }a\sin
\left (\frac {m\pi x}a\right) =I_0\delta (x-d). \label{eq28-7}
\end{equation}
A simple Fourier series analysis shows that
\begin{equation}
H_{m0}^+=\frac {-k_s^2I_0}{ik_zm\pi }\sin \left(\frac {m\pi
d}a\right). \label{eq28-8}
\end{equation}
where $k^2_s=(\frac {m\pi}{a})^2$, and $k_z=\sqrt{k^2_0-k^2_s}$.
This analysis shows that only TE modes with no $\^ y$-variation of
the field are excited. This is because the probe current in
(\ref{eq28-1}) generates a field with no $y$-variation. $H_{m0}^+$
is sometimes known as the excitation coefficient of the mode. For a
particular TE$_{m0}$ mode, we can change $d$ to alter the amplitude
of the excitation coefficient. For example, letting $d/a = 1/2$ will
maximize the excitation coefficient of the TE$_{10}$ mode, while the
TE$_{m0}$ mode with $m$ even will not be excited.

\section {{{ Modes of a Hollow Waveguide of Arbitrary Cross-Section}}}
\index{Waveguide!hollow}
\index{Waveguide!arbitrary cross-section}
\index{Mode!arbitrary hollow waveguide}

If a metallic waveguide has an arbitrary cross-section, whose
shape does not fall on any of the curvilinear coordinates, we will
have to find the modes numerically.  We have learned that the TE
and TM modes are characterized by solutions of the following
equations:
\begin{equation}
\text {TE} \qquad\qquad (\nabla _s^2+k_s^2)H_z(\v r_s)=0, \qquad
\frac {\partial H_z}{\partial n}=0 \quad \text {on}\quad C,
\label{eq29-1}
\end{equation}
\begin{equation}
\text {TM} \qquad\qquad (\nabla _s^2+k_s^2)E_z(\v r_s)=0, \qquad
E_z=0 \quad \text {on}\quad C. \label{eq29-2}
\end{equation}

\subsection{{Differential Equation Method}}
\index{Differential equation method}

The above are eigenvalue problems. Eigenvalue problems can be converted into variational problems by
defining \index{Rayleigh quotient}Rayleigh quotient
\begin{equation}\label{eqn:1}
k_{s}^{2} = -\frac{\langle\phi,\nabla_s^{2}\phi\rangle}{\langle\phi,\phi\rangle}
\end{equation}
\noindent The numerator can be simplified by integration by parts.
\begin{align}\label{eqn:2}
\nonumber \langle\phi,\nabla_s^{2}\phi\rangle &= \int_{S}dS\phi\nabla_s^{2}\phi \\
&= -\int_{S}dS\nabla_s\phi\cdot\nabla_s\phi + \oint_{C}dl\hat{n}\cdot(\phi\nabla_s\phi)
\end{align}
\noindent With the choice of appropriate boundary conditions, the last term can be made to vanish. Hence,
the Rayleigh quotient becomes
\begin{equation}\label{eqn:3}
k_{s}^2=\frac{\langle\nabla_s\phi,\nabla_s\phi\rangle}{\langle\phi,\phi\rangle}
\end{equation}
\noindent The above can be shown to be variational, meaning that a first order error in $\phi$ gives rise to
a second-order error in $k_{s}^{2}$. By letting
\begin{equation}\label{eqn:4}
\phi = \phi_{e}+\delta\phi
\end{equation}
\begin{equation}\label{eqn:5}
k_{s}^{2} = k_{se}^2+\delta k_{s}^{2}
\end{equation}
where $\phi_e$ and $k_{se}^2$ are exact values for the function and
the eigenvalue, respectively. Then, after cross-multiplying, and
taking the first variation, we have
\begin{equation}\label{eqn:6}
k_{se}^{2}2\langle\phi_{e},\delta\phi\rangle + \langle\phi_{e},\phi_{e}\rangle\delta k_{s}^2
= -2\langle\nabla_s\phi_{e},\nabla_s\delta\phi\rangle
\end{equation}
\noindent After doing integration by parts on the right-hand side, the term involving $\delta\phi$ cancel each other, and hence
$\delta k_{s}^{2}=0$.  In other words, the exact eigenfunctions and eigenvalues of
the problems \eqref{eq29-1} and \eqref{eq29-2} are at the
stationary values or stationary points of the Rayleigh
quotient \eqref{eqn:3}.

When complex function $\phi$ is allowed, a Rayleigh quotient
\begin{equation}\label{eqn:3a}
k_{s}^2=\frac{\langle\nabla_s\phi^*,\nabla_s\phi\rangle}{\langle\phi^*,\phi\rangle}
\end{equation}
The above ensures that $k_s^2$ is always real for all $\phi$.

\subsubsection{{Rayleigh-Ritz Method}}
\index{Rayleigh-Ritz method}

In this method, we let
\begin{equation}\label{eqn:7}
\phi=\sum_{n=1}^{N}a_{n}\phi_{n}
\end{equation}
\noindent and pick $a_{n}$ to make \eqref{eqn:3} stationary. By using \eqref{eqn:7} in \eqref{eqn:3}, we arrive at
\begin{equation}\label{eqn:8}
k_{s}^{2}=\frac{\sum_{n,n'}a_{n}a_{n'}M_{nn'}}{\sum_{n,n'}a_{n}a_{n'}B_{nn'}}
= \frac{{\bf a}^{t}\cdot{\dyad M}\cdot {\bf a}}{{\bf a}^t \cdot \bar{{\bf B}} \cdot {\bf a}}
\end{equation}
\noindent where ${\dyad M}$ and ${\dyad B}$ are symmetric matrices, $M_{nn'}=
\langle\nabla_s\phi_{n},\nabla_s\phi_{n'}\rangle, B_{nn'}=\langle\phi_{n},\phi_{n'}\rangle$.
Here, $B_{nn'}$ is also called the \index{Gram matrix}Gram matrix.  In
the equation above, \eqref{eqn:8}, has stationary points.
When we increase the number of unknowns in \eqref{eqn:7},
the stationary points of \eqref{eqn:8} will approach the
exact stationary points and hence, the exact answers.
Meanwhile, \eqref{eqn:7} will approach the exact
eigenfunction.
We assume that
\begin{equation}\label{eqn:9}
{\bf a}={\bf a}_{0}+\delta{\bf a},\qquad k_{s}^{2}=k_{s0}^{2}+\delta k_{s}^{2}
\end{equation}
\noindent where ${\bf a_{0}}$ is the value that will optimize \eqref{eqn:8},
and ${k_{s0}}^{2}$ is the optimal value of ${k_{s}}^{2}$.
In other words, we want ${\bf a_{0}}$ to be at the stationary
point of \eqref{eqn:8}.  In this case, $\delta k_s^2$ will
be zero.
Subsequently, after cross-multiplying, and taking the first variation, (\ref{eqn:8})
becomes
\begin{equation}\label{eqn:11}
{k_{s0}}^2 2 \delta {\bf {a}^{t}} \cdot{\dyad{B}}
\cdot
{{\bf{a_{0}}}} +  \delta {k_{s}}^{2} {\bf {a_{0}}^{t}} \cdot {{\dyad B}} \cdot {\bf a_{0}}   = 2\delta{\bf a^{t}}\cdot{{\dyad M}}\cdot{\bf
a_{0}}
\end{equation}
\noindent In order for the $\delta {k_{s}}^2$ to be zero, so that $\bf{a_{0}}$ represents the optimal solution, we require that
the $\delta \bf{a^{t}}$ terms cancel each other. Then it is
necessary
that
\begin{equation}\label{c2eq:12}
{\dyad M} \cdot {\bf a_{0}} = {{k_{s0}}^2} {\dyad {B} } \cdot {\bf a_{0}}
\end{equation}
\noindent The above is the matrix eigenvalue form which we can solve for $\bf a_{0}$ and ${k_{s0}}^2$.
Once $\bf a_{0}$ is found, the eigenfunction $\phi$ is found via (\ref{eqn:7}).

Myriads of methods can be used to choose
$\phi_{n}$ in \eqref{eqn:7}.
If the cross section of the waveguide is arbitrary, it is more
practical to triangulate the cross section
and pick subdomain basis functons such as pyrimidal functions. Such a method of solution is known as the
finite element method. \index{Finite element method (FEM)}Finite element method (FEM) is vastly popular in solving many differential equation
problems\cite{ZIENKIEWICZ,SILVESTER&FERRARI,JIN_JM,VOLAKISETAL}.

\begin{figure}[!htb]
\begin{center}
\hfil\includegraphics[width=2.1truein]{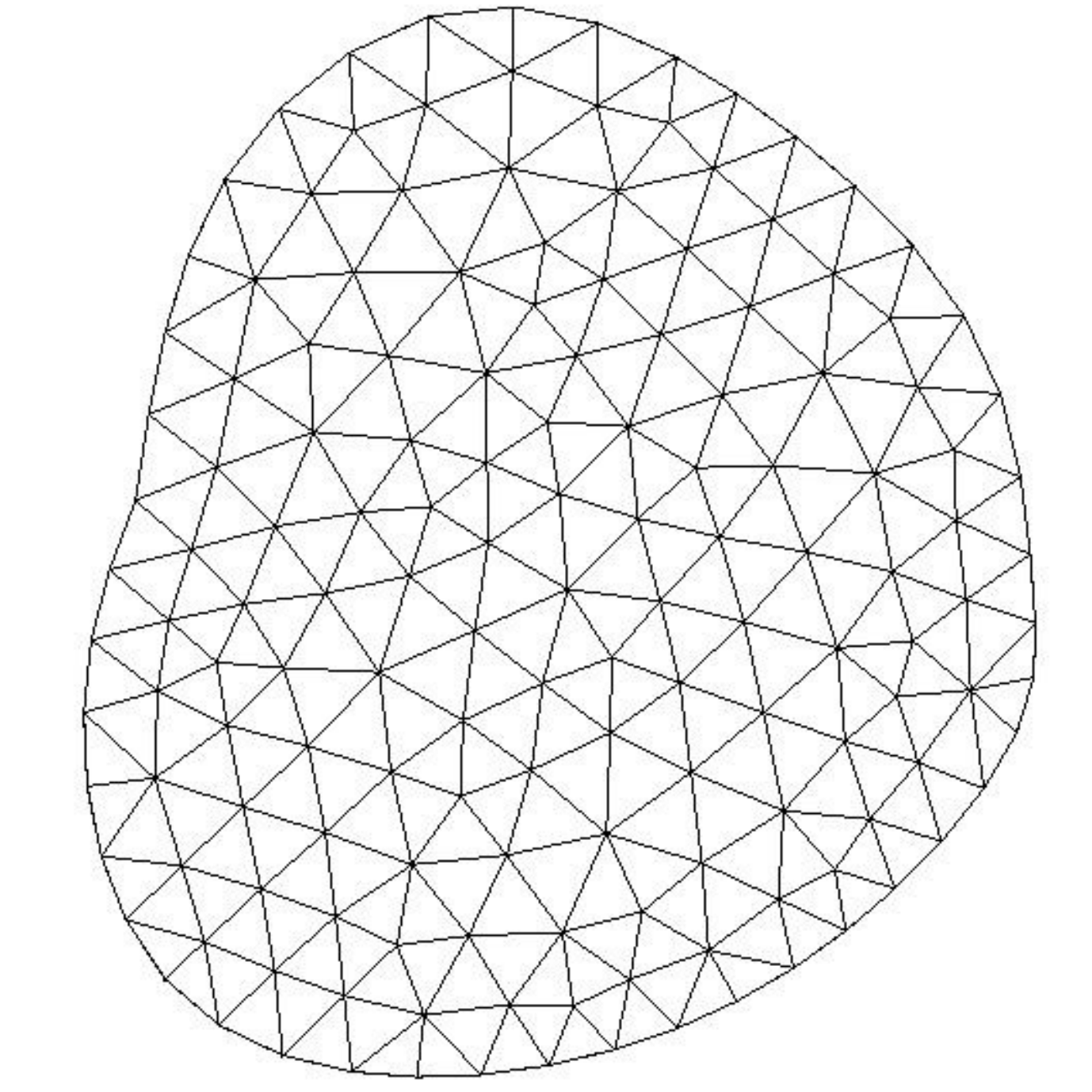}\hfil
\end{center}
\caption{Two-dimensional FEM mesh for the
waveguide.}\label{2dFEMMesh}
\end{figure}

In the choice of basis functions in (\ref{eqn:7}), we can pick the functions to satisfy the homogeneous Dirichlet or
Neumann boundary condition to make the last term in \eqref{eqn:2} vanish. If we pick basis functions whose values float at the contour $C$,
the solutions will satisfy the homogeneous Neumann boundary condition. Hence, the homogeneous boundary condition is also
known as the natural
boundary
condition\cite[p. 308]{WFIME}.

\subsection {Integral Equation Method}
\index{Integral equation method}

A more rigorous way of finding the modes of an arbitrarily shaped
waveguide is to use the integral equation
approach\cite{HARRINGTON,FAECEM,CHEW_TONG_HU}. For this, we need a
\index{Green's function}Green's function which is the solution to the equation
\begin{equation}\label{eq29-14}
(\nabla_s^2 + k_s^2)g(\v r_s, \v r_s' ) = - \delta ( \v r_s - \v
r_s'), \quad \v r_s \in S_1, \quad \v r_s' \in S.
\end{equation}
Multiplying (\ref{eq29-1}) by $g(\v r_s, \v r_s')$ and
(\ref{eq29-14}) by $H_z(\v r_s)$, subtracting the two resultant
equations, and integrating over $S$, we obtain
\begin{equation}
\int\limits_S [g(\v r_s, \v r_s') \nabla_s^2 H_z(\v r _s) - H_z(\v
r_s) \nabla_s^2 g(\v r_s, \v r_s')]dS = H_z(\v r_s'), \quad \v r_s'
\in S \label{eq29-15}
\end{equation}
But since $g{\nabla}_s^2 H_z-H_z {\nabla}_s^2 g =\nabla_s \cdot(g
\nabla_s H_z- H_z{\nabla_s}g)$, Gauss' theorem can be used to
convert (\ref{eq29-15}) into
\begin{equation}
\oint\limits_{C} dl \hat n \cdot [g(\v r_s, \v r_s')\nabla_s H_z(\v
r_s)-H_z (\v r_s) \nabla_s g(\v r_s, \v r'_s)] = H_z(\v r'_s),\quad
\v r'_s \in S. \label{eq29-16}
\end{equation}
Since $\hat n \cdot \nabla _sH_z(\v r_s) =\frac{\partial}{\partial
n} H_z(\v r_s) = 0$ for $\v r_s \in C$ from Equation
(\ref{eq29-1}) we have
\begin{equation}
-\oint dl H_z (\v r_s)\hat n \cdot \nabla_s g(\v r_s, \v r_s') =
H_z(\v r'_s)  , \quad \v r'_s \in C. \label{eq29-17}
\end{equation}
We have imposed $\v r'_s \in C$ in the above so that it is now an
integral equation with the unknown $H_z (\v r_s')$,  $\v r_s' \in
C$.  Since $\v r_s$ can be equal to $\v r_s'$ in the integral, the
singularity of the Green's function $\hat n \cdot \nabla g(\v r_s,
\v r _s')$ makes a straightforward evaluation of the above
integral divergent. A principal value integral has to be taken to
obtain a convergent integral \cite[p. 455]{WFIME}.

\begin{figure}[htb]
\begin{center}
\hfil\includegraphics[width=3.1truein]{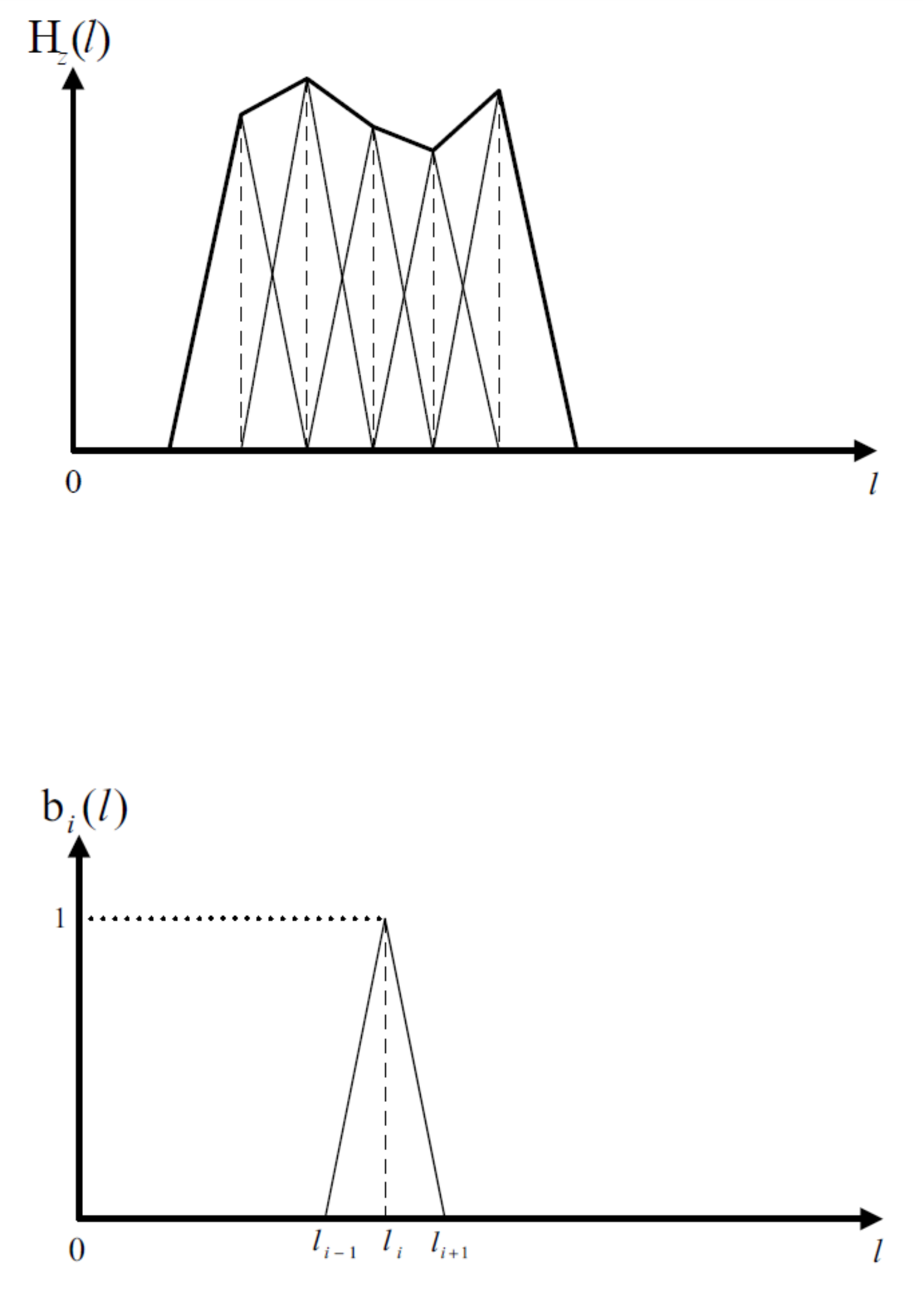}\hfil
\end{center}
\caption{Choice of basis functions for $H_z(l)$. $b_i(l)$ is a
triangular (hat or chapeau) basis function. It can approximate an
arbitrary function as a piecewise linear function.}\label{fgChoice}
\end{figure}


In the above, $H_z(\v r_s)$ can be parameterized as a function of
$l$ where $l$ is a variable defining the contour $C$ of the
waveguide. For simplicity, we can expand $H_z(l)$ in terms of
triangular (also known as hat or chapeau) basis functions\index{Triangular basis functions}, i.e.,
$H_z(l) = \sum\limits_{i=1}^{N} a_i b_i(l)$. Then, (\ref{eq29-17})
becomes
\begin{equation}
\sum \limits_{i=1}^{N} a_i \int \limits_{l_{i-1}}^{l_{i+1}} dl \hat
n \cdot \nabla_s g(l, l') p_i(l) = -\sum\limits_{i=1}^{N} a_i p_i
(l'). \label{eq29-18}
\end{equation}
After the $dl$ integration, the summand in the above equation is a
function of $l'$. A matrix equation can be obtained by point
matching and fixing the above equation at $l'=l_j$, $j=1,\dots, N,$
yielding
\begin{equation}
\sum\limits_{i=1}^{N} A_{ji} a_j = 0, \quad j=1,\dots, N,
\label{eq29-19}
\end{equation}
where
\begin{equation}
A_{ji} = \int \limits _{l_{i-1}}^{l_{i+1}} dl\hat n \cdot \nabla_s
g(l,l_j) p_i(l) + p_i (l_j). \label{eq29-20}
\end{equation}

Equation (\ref{eq29-19}), hence, reduces to a matrix equation
\begin{equation}
\dyad A(k_s)\cdot \v a=0. \label{eq29-20a}
\end{equation}
A nontrivial solution exists for $\v a$ only if
\begin{equation}
\det \left( \v A (k_s) \right) = 0. \label{eq29-21}
\end{equation}
Here, $\v A$ is a function of $k_s$ because the Green's function in
Equation (\ref{eq29-14}) is a function of $k_s$.  The above can be
satisfied only at certain values of $k_s$.  At these values,
\eqref{eq29-20a} has a non-trivial null space solution, and hence
$\v a$ is nonzero, and can be found.

The derivation so far requires $g(\v r_s, \v r_s')$ to be a
solution of (\ref{eq29-14}) for $\v r_s \in S$, and $\v r_s' \in
S$. A simple Green's function that satisfies this requirement is
\begin{equation}
g(\v r_s, \v r_s') = g(\v r_s - \v r_s') = \frac{i}{4}
H_0^{(1)}(k_s | \v r_s - \v r_s'|). \label{eq29-22}
\end{equation}

A similar integral equation can be derived for TM polarization
which is
\begin{equation}
\oint \limits_C dl g(\v r_s, \v r_s') \hat n \cdot \nabla_s E_z(\v
r_s) = 0 \quad \v r_s' \in C. \label{eq29-23}
\end{equation}
This can also be used to find the waveguide modes of an arbitrarily
shaped waveguide.  In the above, there is no derivative on the
Green's function.  Hence, this integral equation is less singular
compared to the one for TE polarization.

\subsection { Ad Hoc Method}

We describe an ad hoc method for solving for the waveguide modes,
but this method is not rigorous. It can only yield the modes
satisfactorily if the shape of the waveguide is not too oblong, or
if the wall of the waveguide does not have sharp corners.

One way to solve (\ref{eq29-1}) and (\ref{eq29-2}) is to expand the
scalar field $H_z$ or $E_z$ in terms of functions which are known to
be solutions of (\ref{eq29-1}) or (\ref{eq29-2}), but do meet the
specified boundary conditions. For example, we can let
\begin{equation}
H_z(\v r_s)=\sum _na_n\psi _n(\v r_s), \label{eq29-3}
\end{equation}
where $\psi _n(\v r_s)$ could be \footnote{ The basis given by
(\ref{eq29-4}) is actually not complete in an arbitrarily shaped
waveguide. This method is hence akin to the method of Rayleigh's
hypothesis \cite{WFIME}.}
\begin{equation}
\psi _n(\v r_s)=J_n(k_s\rho )e^{in\phi }. \label{eq29-4}
\end{equation}
$\psi _n(\v r_s)$ is clearly a solution to (\ref{eq29-1}), but the
boundary condition is not met. In order to satisfy the boundary
condition, we require that the normal derivative of $H_z=0$, or
\begin{equation}
\frac{\partial H_z}{\partial n}=\sum\limits _na_n\frac{\partial\psi
_n(\v r_s)}{\partial n}=0 \ \ \text{on} \quad C. \label{eq29-5}
\end{equation}

\begin{figure}[htb]
\begin{center}
\hfil\includegraphics[width=2.4truein]{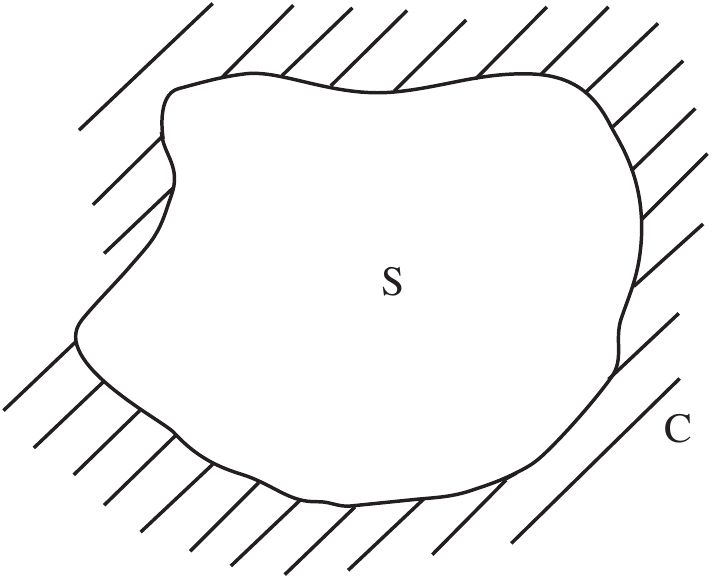}\hfil
\end{center}
\caption{A waveguide of arbitrary
cross-section.}\label{fgAwaveguide}
\end{figure}

%
The general method to solve (\ref{eq29-5}) is to weight it with a
test function and integrate over $C$, e.g.,
\begin{equation}
\sum\limits_n a_n\left\langle\phi_m,\frac{\partial \psi_n}{\partial
n}\right\rangle =0, \label{eq29-6}
\end{equation}
where
\begin{equation}
\left <\phi_m,\frac{\partial\psi_n}{\partial n}\right >
=\int\limits_C dl\phi_m(\v r_s)\frac{\partial\psi_n}{\partial
n}=A_{mn}(k_s) \label{eq29-7}
\end{equation}
is only a number dependent on $k_s$ and the indices $m$ and $n$.
Equation (\ref{eq29-6}), hence, reduces to a matrix equation
\begin{equation}
\dyad A(k_s)\cdot \v a=0. \label{eq29-8}
\end{equation}
A nontrivial solution exists for $\v a$ only if
\begin{equation}
\det \left( \dyad A(k_s)\right) =0. \label{eq29-9}
\end{equation}
We can search Equation (\ref{eq29-9}) numerically for the values of
$k_s$ that satisfy the equation. These are the eigenvalues of the
problem for the TE modes. The eigenvector $\v a$ can be found from
(\ref{eq29-8}) at these values of $k_s$.  With the knowledge of $\v
a$, we can construct $H_z$, the eigenfunction.

\begin{figure}[htb]
\begin{center}
\hfil\includegraphics[width=2.40truein]{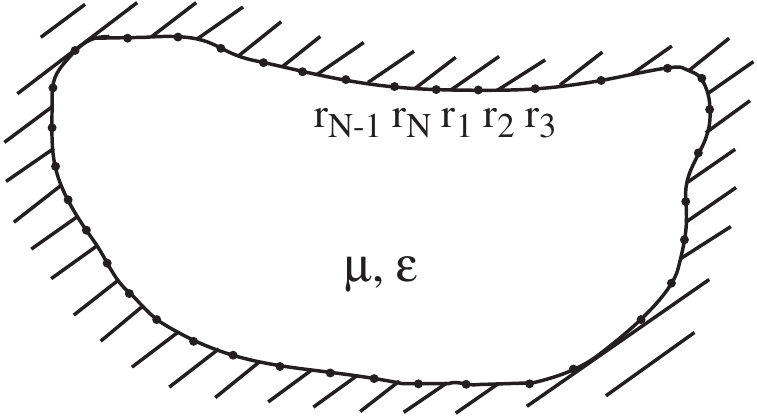}\hfil
\end{center}
\caption{The solution of a waveguide problem via the method of
point-matching.}\label{fgThesolution}
\end{figure}

%
For the TM problem, similar analysis gives rise to an $\v A$ matrix
given by
\begin{equation}
A_{mn}(k_s)=\int\limits _Cdl\phi _m(\v r_s)\psi _n(\v r_s).
\label{eq29-10}
\end{equation}

\index{Point-matching technique}
Another way of solving (\ref{eq29-5}) is to first assign $N$ points
on the perimeter of the wave guide $C$ and evaluate the Equation
(\ref{eq29-6}) at these location. Equation (\ref{eq29-5}) then
becomes
\begin{equation}
\sum\limits _na_n\frac {\partial \psi _n(\v r_m)}{\partial n}=0,
\quad m=1, \hdots, N. \label{eq29-11}
\end{equation}
The above is a matrix equation similar to (\ref{eq29-8}) where
\begin{equation}
A_{mn}(k_s)=\frac {\partial \psi _n(\v r_m)}{\partial n}.
\label{eq29-12}
\end{equation}
Similar analysis for the TE modes gives
\begin{equation}
A_{mn}(k_s)=\psi _n(\v r_m). \label{eq29-13}
\end{equation}
This particularly simple way of solving Equation (\ref{eq29-5}) is
known as the point matching technique.

\begin{figure}[htb]
\begin{center}
\includegraphics[height=2.50in]{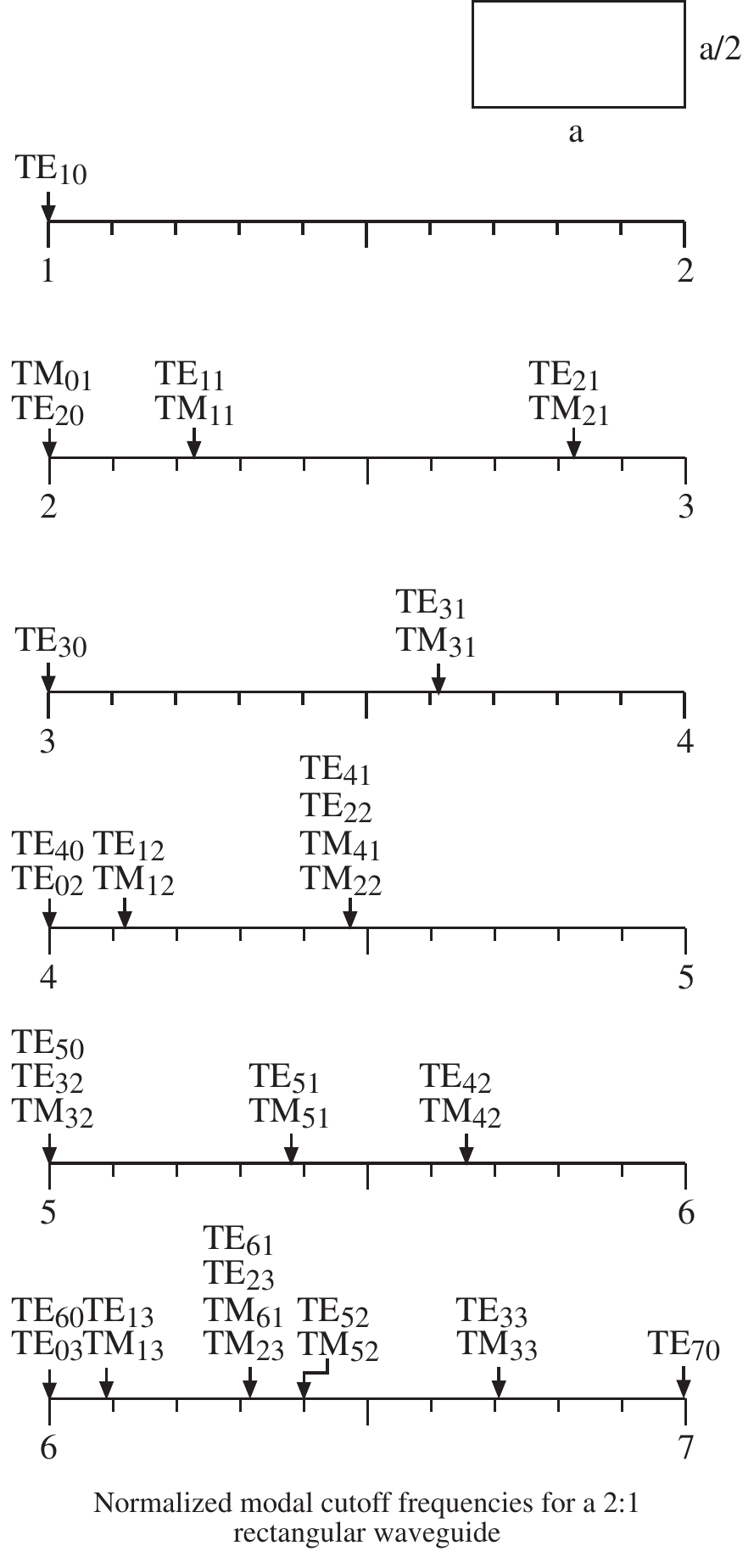}
\end{center}
\vskip 15pt
\begin{center}
\includegraphics[height=4.5truein]{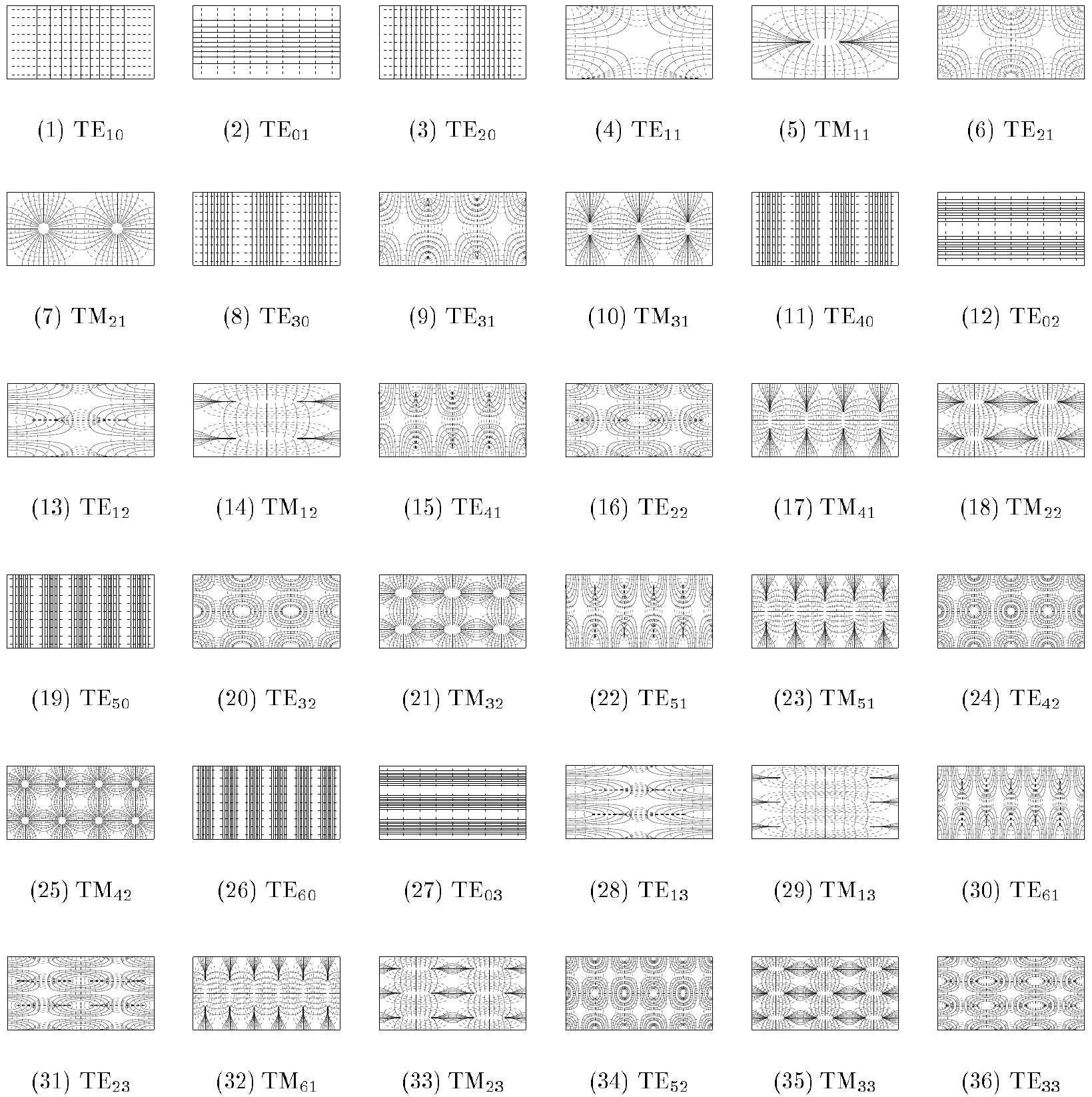}
\end{center}
\caption{Cutoff frequencies (normalized with $\omega _{10c}=1$) and
field plots of the first 36 modes of a rectangular waveguide
(courtesy of A. Greenwood).}\label{fgCut}
\end{figure}

\begin{figure}[htb]
\begin{center}
\includegraphics[height=2.5in]{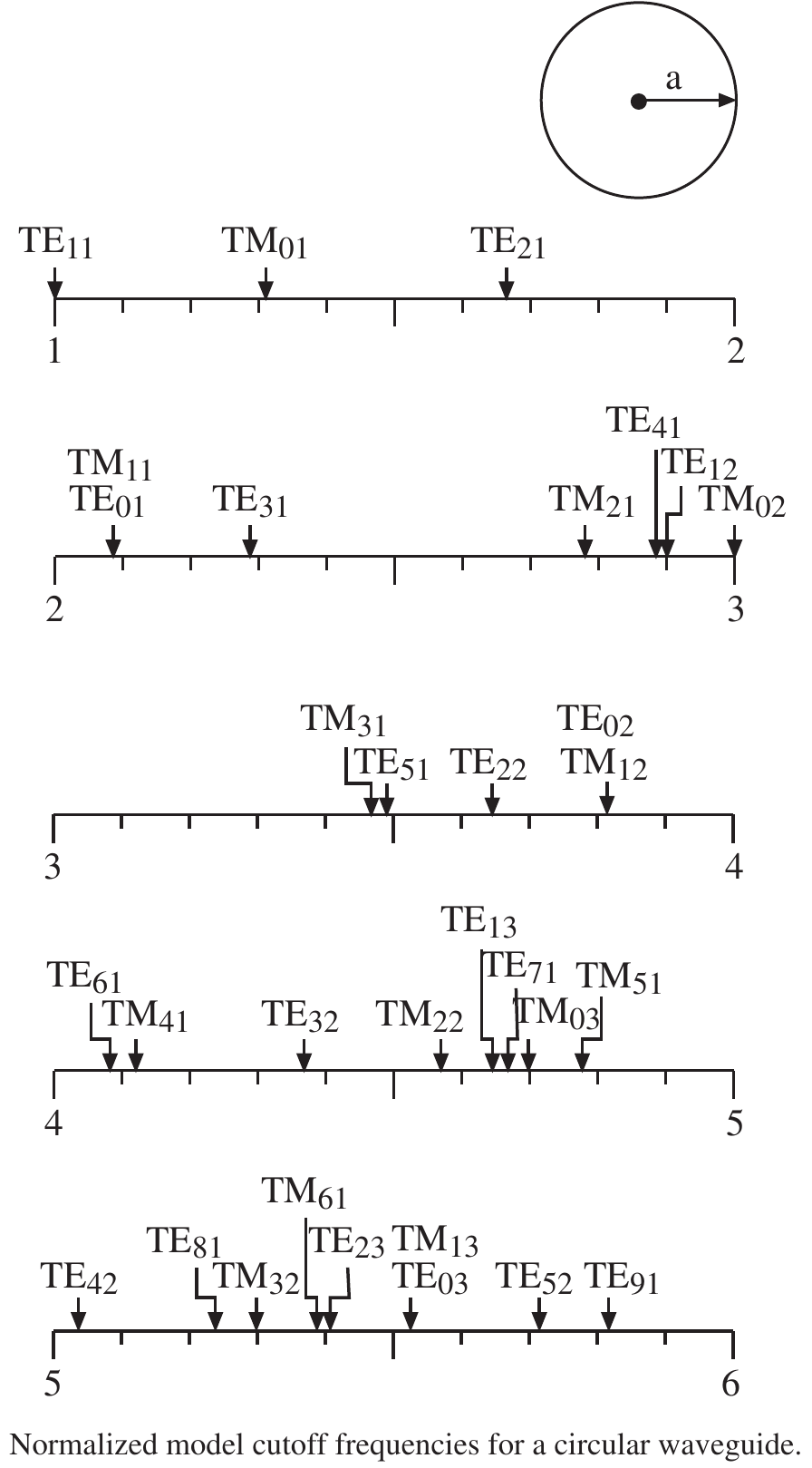}
\end{center}
\vskip 16pt
\begin{center}
\includegraphics[height=4.50in]{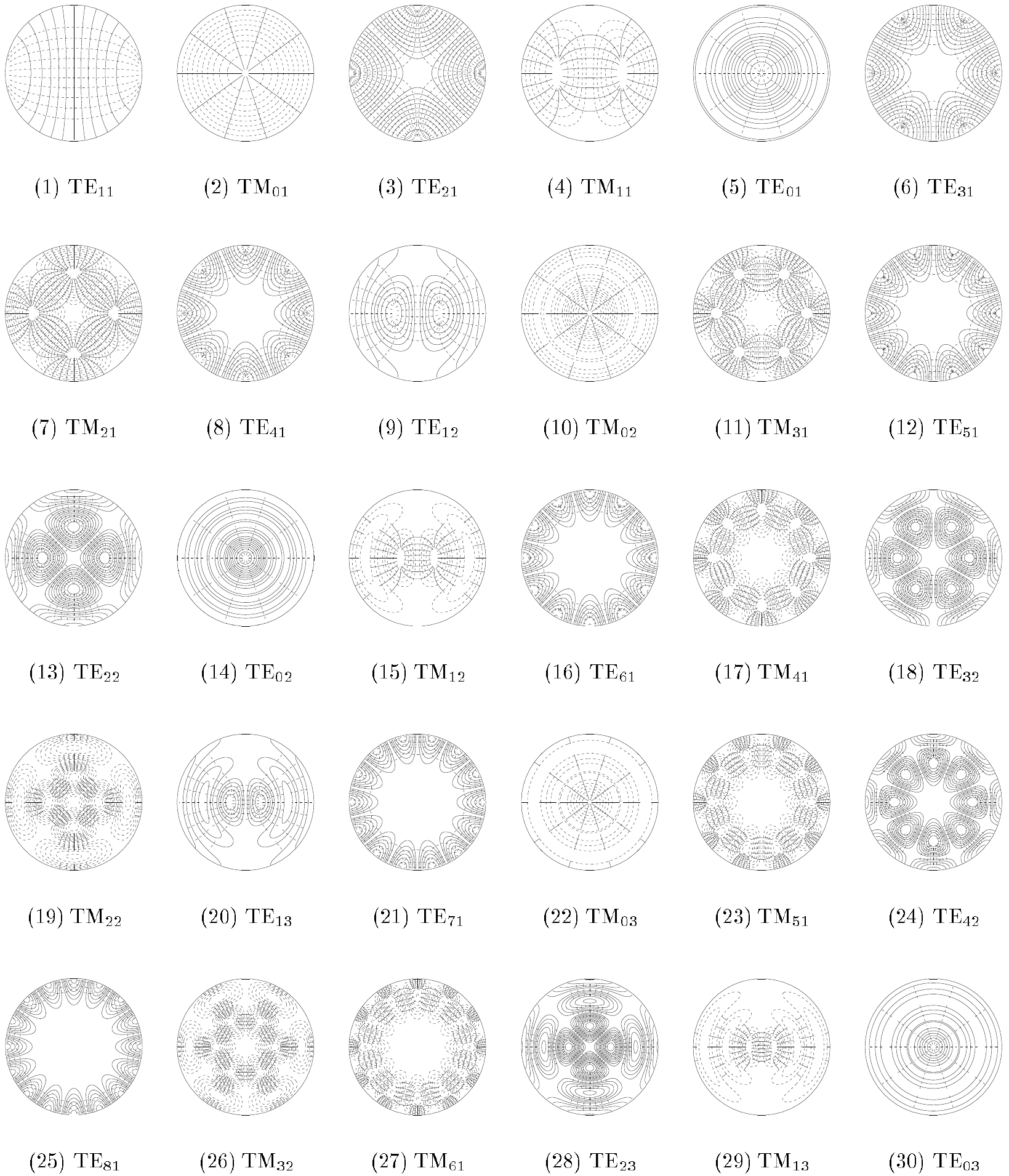}
\end{center}
\caption{Cutoff frequencies (normalized with $\omega _{11c}=1$) and
field plots of the first 30 modes of a circular waveguide (courtesy
of A. Greenwood).} \label{fgCutoff}
\end{figure}


\vskip 20pt

\centerline{\bf Exercises for Chapter 2 }

\vskip 12pt \vskip 8pt \noindent {\bf Problem 2-1:} \noindent Many
transmission line systems have more than two conductors, like your
home telephone line, computer bus etc. If you have an $N$
conductors (a general cylindrical waveguide with translational
invariance in $z$ direction) embedded in a homogeneous isotropic
medium, how many {\it independent} TEM (Transverse
Electromagnetic) modes can propagate down this line.

\begin{figure}[!htb]
\begin{center}
\hfil\includegraphics[width=2.10truein]{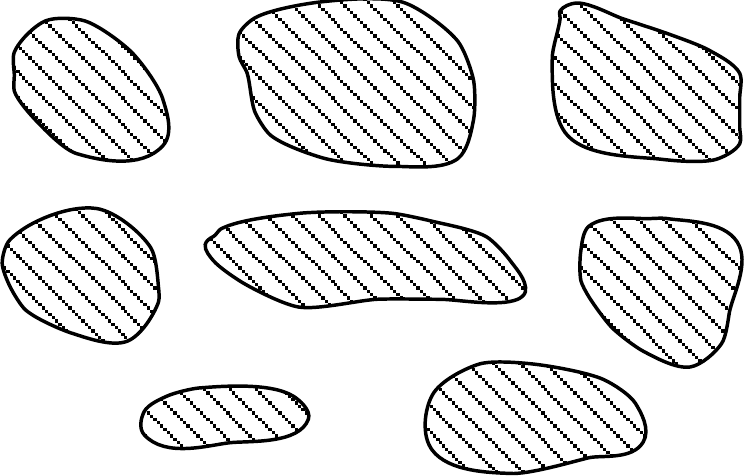}\hfil
\end{center}
\caption{Problem 2-1}
\end{figure}


\vskip 8pt \noindent {\bf Problem 2-2:} \noindent The reflection
of a plane wave by a half space has a closed form solution. Write
down the solutions of the reflection of plane TE and TM waves from
a half space. Now make the lower half space a metallic conductor
whose conductivity is high but not infinite. Deduce a relationship
between the tangential component of the electric field and the
magnetic field at the metallic surface, i.e.,
$$
E_t=Z_m H_t.
$$
$Z_m$ is also known as the surface impedance of the metallic
conductor. It is useful for calculating the power absorbed by the
metallic conductor.

\vskip 8pt \noindent {\bf Problem 2-3:} \noindent For a coaxial
transmission line with an inner conductor of radius $a$ and an
outer conductor of radius $b$, find the electric potential
$\phi_s$ and the magnetic potential $\psi_s$.  Is $\psi_s$ a
single value or a multi-value function for each point in space?
Comment on this.

There are two ways to calculate the attenuation in a transmission
line when the conductor is not perfect.  One way is to first
calculate the series resistance per unit length, $R$, and then
find the attenuation constant from the formula $(k'+ik'')^2=-YZ$
where $Y$ is the shunt admittance per unit length, and $Z$ is the
series impedance per unit length, and $k''$ is the attenuation
constant. Another way is to find  $P_d$, the power dissipated per
unit length by integrating the time average Poynting vector
$\langle\v E \times \v H\rangle$ over the surface of the
conductors. Since the tangential electric field is not zero
anymore on a non-perfectly conducting metallic surface, the time
average Poynting vector is not zero on the metallic surfaces.
After finding the $P_d$, one can calculate the attenuation
constant from
$$
k''= \frac{P_d}{2P_0}
$$
where $P_0$ is the time average total power flow down the
transmission line. You can use the surface impedance found from
Problem 2 to calculate the power absorbed by the metallic
conductor.
\begin{itemize}
\item[(a)] Find the attenuation constant for a coaxial line using these two methods, and show that they are the same.
\item[(b)] For a 50 ohm coaxial transmission line with an outer radius of .5 cm, filled with teflon with
$\epsilon=2\epsilon_0$, calculate the loss due to attenuation in terms of dB/Km at 20 KHz and 1 GHz, if we
assume that the conductor is made out of copper with $\sigma = 5.8\times 10^7$
mho/m.
\end{itemize}

\noindent {\bf Problem 2-4:} \noindent Can a single piece of
conductor embedded in a homogeneous medium support a TEM mode?
Discuss why and why not.

\vskip 8pt \noindent {\bf Problem 2-5:} \noindent
\begin{itemize}
\item[(a)] Explain why the line capacitance (farad/m) of the TEM mode
in a transmission line can be found by solving an electrostatic problem.
\item[(b)] Given the knowledge of the line capacitance, show that it has
to be of the form $C=\epsilon K$ where $K$ is a dimensionless, geometry dependent factor.
\item[(c)] Show that now if there is dielectric loss such that $\epsilon=\epsilon'+i\sigma/\omega$,
then the admittance per unit length to be used in the telegraphists equation (for lossy line)
is $Y=-i\omega\epsilon' K +\sigma K$.  Hence, $G=\sigma K$.
\end{itemize}

\noindent {\bf Problem 2-6:} \noindent The eigenvalue $k_{is}^2$
of the partial differential equation $(\nabla_s^2
+k_{is}^2)\psi_{iz} = 0$ is always real, and can be proven as
follows:
\begin{itemize}
\item[(a)] Assume that $k_{is}^2$ is complex so that
$\psi_{iz}^{\ast}$ is a solution of $(\nabla_s^2
+{k_{is}^\ast}^2)\psi_{iz}^{\ast} = 0$. Show that
$$
\psi_{iz}^{\ast}\nabla_s^2\psi_{iz}-\psi_{iz}\nabla_s^2\psi_{iz}^{\ast}
=({k_{is}^{\ast}}^2-k_{is}^2)|\psi_{iz}|^2 .
$$
\item[(b)] Assume that $\psi_{iz}$ is either $E_{iz}$ or $H_{iz}$,
integrate the equation in part (a) over the cross-section of a
metallic waveguide and show that
$$
({k_{is}^{\ast}}^2-k_{is}^2)\int_S|\psi_{iz}|^2 dS = 0.
$$
From the above, explain why $k_{is}^2$ has to be real.
\end{itemize}

\noindent {\bf Problem 2-7:} \noindent Prove that
$$
\int_S dS \v H_{is}\cdot\v H_{js} = 0, \qquad i\neq j
$$
for any two distinct modes for a homogeneously filled closed
waveguide. $S$ in the above refers to the cross-sectional area of
the waveguide that is permeated by the field.

\vskip 8pt \noindent {\bf Problem 2-8:} \noindent
\begin{itemize}
\item[(a)] Prove that $\v E_{is}$ and $\v E_{js}$ are always orthogonal if one field is coming from
a TM (TE) mode while the other field is coming from a TE (TM) mode.  Hint: Express these fields in terms
of the longitudinal
components of the fields.
\item[(b)]{Prove that
$$
\int_S (\v E_{is}\times \v H_{js}^\ast)\cdot\hat z dS = 0
$$
for $i\neq j$.}
\item[(c)]{From the above, explain how the concept of energy
orthogonality and power orthogonality follow.}
\end{itemize}

\noindent {\bf Problem 2-9:} \noindent For the TE$_{10}$ mode of a
rectangular waveguide, find its attenuation constant due to wall
loss. Do the same for the TE$_{11}$ and the TE$_{01}$ modes of a
circular waveguide. Sketch the attenuation constants as functions
of frequency. Which mode has the lowest loss at high frequencies?

\vskip 8pt \noindent {\bf Problem 2-10:} \noindent Derive the
complete expression for the fields of the TE$_{11}$ mode of a
rectangular waveguide and sketch the field patterns for both the
$\v E$ and $\v H$ fields in the cross-section of the waveguide. Do
the same for the TM$_{11}$ mode of the rectangular waveguide, and
the TE$_{12}$ mode of the circular waveguide. (A good feel for the
field pattern of the waveguide mode is essential in determining
how to excite it with a probe, see Equation (\ref{eq28-46}).)

\begin{figure}[!htb]
\begin{center}
\hfil\includegraphics[width=3.20truein]{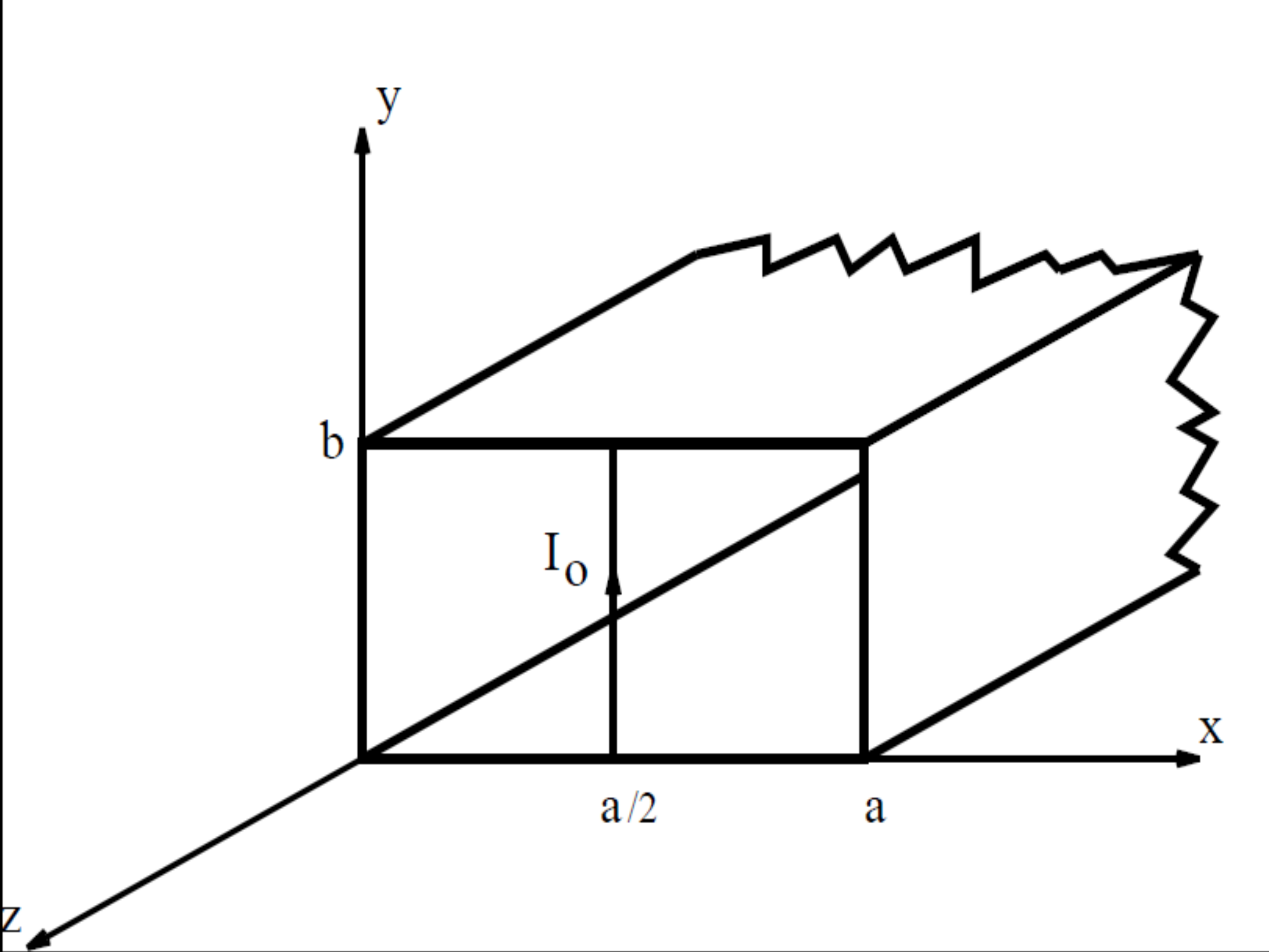}\hfil
\end{center}
\caption{Problem 2-10}
\end{figure}

\vskip 8pt \noindent {\bf Problem 2-11:} \noindent

A probe with a current distribution represented by a current sheet
given by
$$
\v J_s(\v r) = \hat y \delta (x-a/2) \sin (\frac{\pi y}{b})
$$
is placed in a waveguide.  Calculate the excitation coefficients
of all the modes excited by this probe.

\vskip 8pt \noindent {\bf Problem 2-12:} \noindent Certain modes
of metallic waveguides have decreasing loss for increasing
frequency.
\begin{itemize}
\item[(a)] For the TE$_{10}$ mode of a rectangular waveguide, find its attenuation constant due to wall loss.
Do the same for the
TE$_{11}$ and the TE$_{01}$ modes of a circular waveguide. Hint: Integral of product of Bessel functions
is needed in the solution. Equation (11.4.5) of Abramowitz and
Stegun comes in useful here.
\item[(b)] Sketch the attenuation constants as functions of frequency. Which mode has the lowest loss at high frequencies?
\item[(c)] For the TE$_1$ mode of a parallel plate waveguide, does its attenuation constant increase or
decrease with frequency?
\item[(d)] From the above, explain the feature of the ``magic" mode
that has decreasing loss for increasing frequency.
\end{itemize}

\noindent {\bf Problem 2-13:} \noindent
\begin{itemize}
\item[(a)] Give an intuitive explanation as to why the TE$_{11}$ mode of a circular waveguide has a lower
cutoff frequency than the TE$_{01}$ mode.
\item[(b)] Explain why the TE$_{00}$ mode of a rectangular waveguide cannot exist.
\end{itemize}

\noindent {\bf Problem 2-14:} \noindent Prove the mutual
orthogonality of the vector wave functions in a hollow waveguide,
that is
$$
\int dV \v M_{{e\atop h} i}(k_z,\v r)\cdot \v N_{{e\atop
h}j}(-k_z',\v r) = 0,
$$
$$
\int dV \v L_{{e\atop h}i}(k_z,\v r)\cdot \v N_{{e\atop
h}j}(-k_z',\v r) = 0,
$$
$$
\int dV \v L_{{e\atop h}i}(k_z,\v r)\cdot \v M_{{e\atop
h}j}(-k_z',\v r) = 0,
$$
for all $i$, $j$, $k_z$ and $k_z'$.

\vskip 8pt \noindent {\bf Problem 2-15:} \noindent

\begin{figure}[htb]
\begin{center}
\hfil\includegraphics[width=2.0truein]{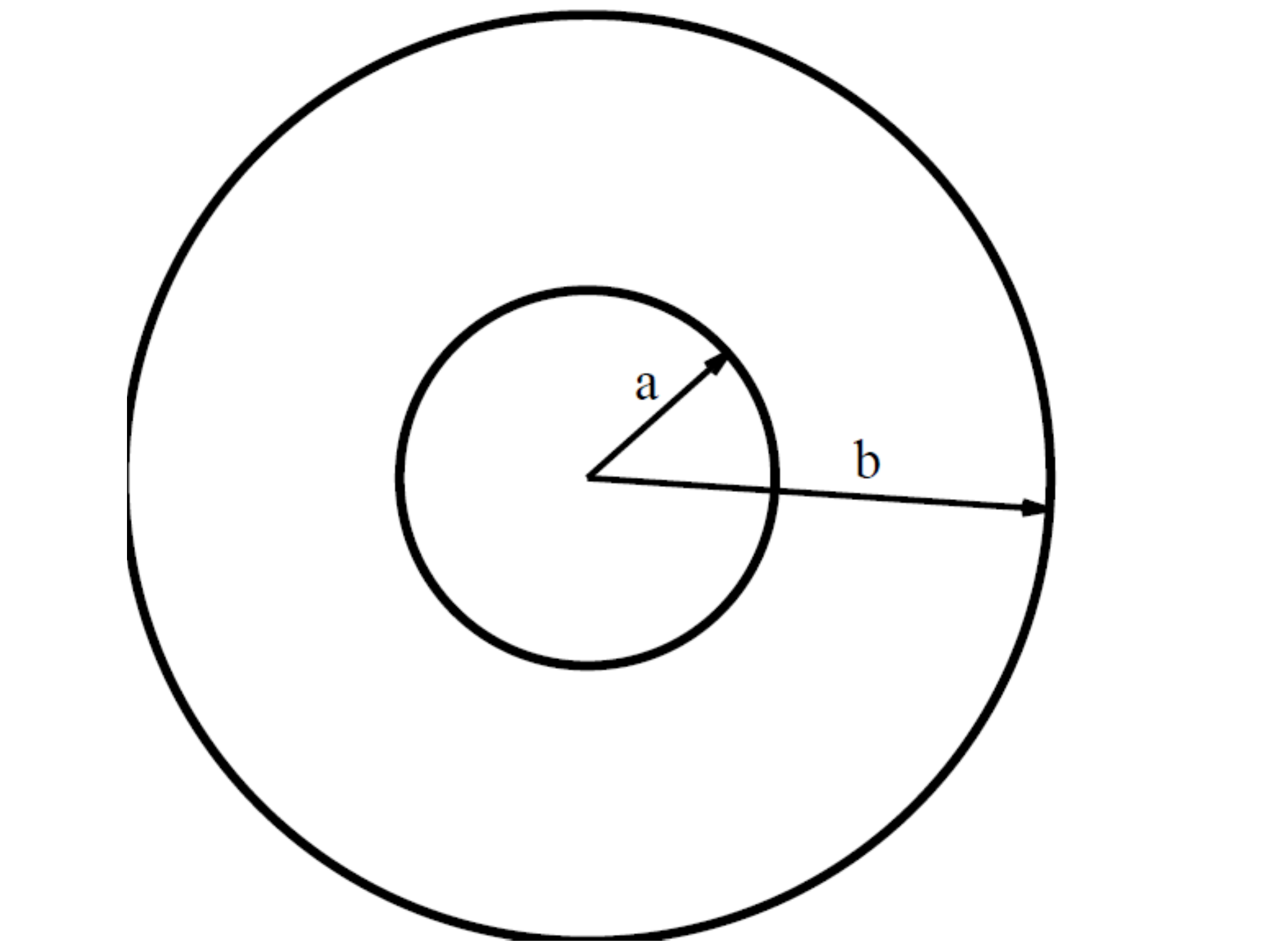}\hfil
\end{center}
\caption{Problem 2-12}
\end{figure}

For the coaxial waveguide shown:
\begin{itemize}
\item [(a)] Write down the guidance conditions for all the modes in the waveguide using Bessel and Neumann functions.
Is the TEM mode a special case of these guidance
conditions?
\item [(b)] Derive the $\v M$, $\v N$, and $\v L$
vector wave functions for the electric and magnetic fields of this
waveguide. (Hint: Use sine and cosine functions for the $\phi$
dependence.)
\item [(c)] Write down the orthogonality relationships for the vector wave functions with the $A_{ei}$ and $A_{hi}$
(defined in the text) derived
explicitly.
\item [(d)] For a source in the circular waveguide described by
$$\v J(\v r) = \hat z
I\ell \delta \left (\rho-{a+b\over 2}\right )\delta
(\phi)\delta(z)/\rho,$$ find the $\v E$ field in terms of the
vector wave function.
\end{itemize}

\noindent {\bf Problem 2-16:} \noindent Go through the exercise of
deriving the dyadic Green's function of a waveguide. Most of the
derivation is already outlined in the text. Fill in the details.

\vskip 8pt \noindent {\bf Problem 2-17:} \noindent For a
rectangular waveguide:
\begin{itemize}
\item [(a)] Write down the guidance conditions for all the modes in the waveguide.
\item [(b)] Derive the $\v M$, $\v N$ and $\v L$ vector wave functions for the electric
and magnetic fields of this waveguide.
\item [(c)] You are given three Hertzian electric dipoles. Describe how you
would place these dipoles in a rectangular waveguide and weight
their amplitudes so that the TE$_{10}$ is excited but the
TE$_{20}$ and TE$_{30}$ modes would not be excited (note: you may
excite all the other modes).
\end{itemize}

\noindent {\bf Problem 2-18:} \noindent
\begin{itemize}
\item[(a)] Derive the integral equation for finding the modes of the TM polarization of a hollow
waveguide:
$$
\oint_C dl g(\v r_s,\v r_s') \hat n\cdot \nabla_s E_z(\v r_s) =
0,\quad \v r_s'\in C.
$$
\item[(b)] Describe how you would solve the above equation numerically.
\end{itemize}

\bibliographystyle{plain}
\bibliography{emt,matrix,fmm,math,fastsum,parallel,cs,my}


\def\chaptitle{Inhomogeneously Filled Waveguides}

\chapter{\chaptitle}

\markboth{\smallbooktitle}{\chaptitle}




\def\v #1{{\bf #1}}
\def\vg #1{{\boldsymbol #1}}
\def\dyad#1{\overline {\bf #1}}
\def\dyadg#1{\overline {\vg #1}}
\def\beq{\begin{equation}}\def\eeq{\end{equation}}
\def\tinf{\text{\it inf\,}}\def\^{\hat}
\def\cal#1{\mathcal{#1}}
\def\ed{\end{document}}
\def\dalpha{\dyadg\alpha}
\def\dbeta{\dyadg\beta}
\def\Rg{\Re g}
\def\l{\label}
\def\sinc{\text{sinc}}

\index{Waveguide!inhomogeneously filled}
The theory of inhomogeneously filled waveguides covers a large
class of waveguides \cite{MARCUVITZ,COLLIN,COLLIN2,YARIV,OKOSHI}.
Waveguides are filled with inhomogeneous material to give the
waveguide a certain property. For instance, phase-shifters,
polarizers, and attenuators are made out of inhomogeneously filled
waveguides. Often, waveguides are filled inhomogenously with
ferrite material to make non-reciprocal waveguides.  To make the
fabrication of waveguides simpler, many waveguides are also filled
with inhomogeneous materials.  Optical waveguides almost
exclusively comprise inhomogeneously filled material.


\section{The Need for Hybrid modes}
\index{Mode!hybrid}

When a hollow waveguide is filled with inhomogeneity, the most
general case is the existence of the hybrid modes. In these modes,
the TE$_z$ and TM$_z$ modes are coupled\index{Coupled modes}. Only for cases with special
symmetry they are decoupled. The reason is that both TE$_z$ and
TM$_z$ fields are needed to match the boundary condition at the
dielectric interface.

Assume that we only have TE$_z$ field inside the waveguide. Then,
 \begin{align}\label{c3eq:1}
  {\v{H}}_s=\frac{1}{k_s^2}\left[\frac{\partial}{\partial z}\nabla_s {H}_z \right]=\frac{1}{k_s^2}ik_z\nabla_s {H}_z
 \end{align}
 The boundary condition for $H_z$ is that it is continuous, or $H_{1z}=H_{2z}$ at the interface between two dielectric regions.
 Furthermore, we require that,
 \begin{align}\label{c3eq:2}
 \hat{n}\times {\v{H}}_{1s}=\hat{n}\times {\v{H}}_{2s}
 \end{align}
 implying that at the dielectric interface,
 \begin{align}\label{c3eq:3}
 \frac{1}{k^2_{1s}} ik_z\hat{n}\times\nabla_s {H}_{1z}=
 \frac{1}{k^2_{2s}} ik_z\hat{n}\times\nabla_s {H}_{2z}
 \end{align}
 Due to \index{Phase matching}phase matching, $k_z$ is the same in all regions. If $H_{1z}$ is continuous at a dielectric interface,
 then $\hat{n}\times\nabla_s {H}_{1z}=\hat{n}\times\nabla_s
 {H}_{2z}$ also,
 since these are tangential derivatives. Therefore, (\ref{c3eq:3}) cannot be satisfied since $k_{1s}\ne k_{2s}$ in general.
 In order to satisfy the boundary condition, (\ref{c3eq:1}) has to be augmented with the contribution from TM$_z$
 field.

 However, under special circumstances, Equation (\ref{c3eq:3}) can be satisfied
 if:
 \begin{enumerate}[label={(\arabic*)}]
 \item $k_z=0$, implying that, $\v H_s=0$ and ${\v{H}}=\hat{z} {H}_z$ only.  In this case, it reduces to a two-dimensional problem;
 \item $\hat{n}\times\nabla_s {H}_{iz}=0$ at the interface, implying symmetry that exists for certain modes, for instance, in a dielectric slab or in an axi-symmetric
 geometry such as a circular optical fiber;
 \item The surface is a PMC surface so that ${H}_z=0$, $\hat{n}\times {\v{H}}_s=0$.  This boundary condition is sufficient to guarantee
 the uniqueness of the TM$_z$ mode alone
 in the waveguide;
 \item The surface is a PEC surface so that the boundary condition is for $\hat{n}\times {\v{E}}=0$, giving rise to $\hat{n}\cdot\nabla {H}_z=0$ on the surface.
Again, this boundary condition is sufficient to guarantee the
uniqueness of the solution, needless for the coupling of the TE$_z$
and TM$_z$ modes.
 \end{enumerate}
 Similarly, arguments above apply to the TM$_z$ field.  Under these
 special circumstances, the field is not depolarized at the interface. Namely,
 if the field is TE$_z$ or TM$_z$ before impinging on the interface,
 the scattered field off the interface remains the same as the original
 polarization.

\section{{{ Derivation of Pertinent Equation}}}

\setcounter{equation}{0} \setcounter{figure}{0}

When a uniform waveguide is filled with inhomogeneous materials,
the guided modes of the structure cannot be decomposed into TE and
TM modes, except for some very special cases.  In other words, the
$E_z$ and $H_z$ components of the fields are always coupled
together. Such modes are also called the \index{Waveguide!hybrid modes}\index{Mode!hybrid}hybrid modes. This
coupling can be shown from Maxwell's equations, which imply
\begin{equation}
\nabla\times\mu ^{-1}\nabla\times\v E-\omega ^2\epsilon\v E=0.
\label{eq1-1}
\end{equation}
The above governs the $\v E$-field propagating in an
inhomogeneously filled waveguide.  Assuming $e^{ik_zz}$
dependence, we can decompose
\begin{equation}
\v E=\v E_s+\v E_z, \qquad\nabla =\nabla _s +\^ zik_z,
\label{eq1-2}
\end{equation}
for a guided mode in the waveguide.  With the use of the above,
\begin{equation}
\begin{split}
\nabla\times\mu ^{-1} &\nabla\times\v E \\
&= \nabla _s\times\mu ^{-1}\nabla _s\times\v E_s+ik_z\^ z\times\mu
^{-1}\nabla _s\times\v E_z-k_z^2\mu ^{-1}\^z\times\^ z\times\v
E_s\\ & +\nabla _s\times\mu ^{-1}\nabla _s\times\v E_z+ik_z\nabla
_s\times\mu ^{-1}\^z\times\v E_s. \label{eq1-3}
\end{split}
\end{equation}

\begin{figure}
\begin{center}
\hfil\includegraphics[width=3.0truein]{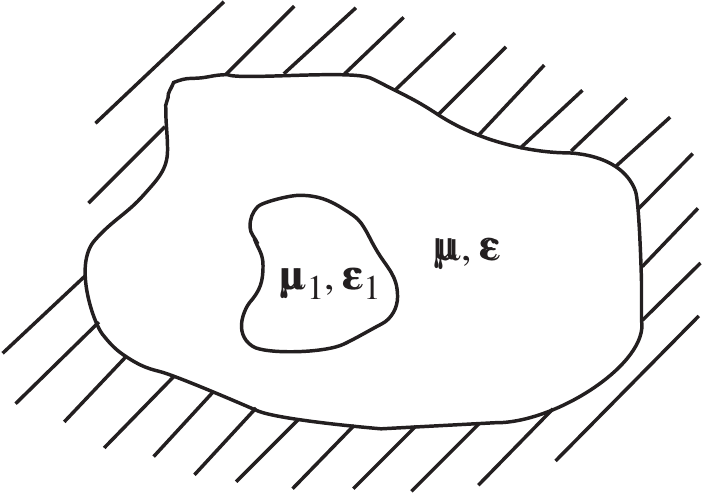}\hfil
\end{center}
\caption{An inhomogeneously filled waveguide.}\label{fg311}
\end{figure}


Because $\mu$ is a function of $\v r_s$, it does not commute with
$\nabla _s$.  The last two terms in (\ref{eq1-3}) are directed in
the $z$ direction. Therefore, by equating the $z$ components in
(\ref{eq1-1}), we have
\begin{equation}
\nabla _s\times\mu ^{-1}\nabla _s\times\v E_z+ik_z\nabla
_s\times\mu ^{-1}\^z\times\v E_s-\omega ^2\epsilon \v E_z=0.
\label{eq1-4}
\end{equation}

\subsection{$E_z$-$H_z$ Formulation}
\index{Waveguide!$E_z$-$H_z$ formulation}

It can be shown that
\begin{subequations}
\begin{equation}
\nabla _s\times\mu ^{-1}\nabla _s\times\v E_z=-\nabla _s\cdot\mu
^{-1}\nabla_s\v E_z, \label{eq1-5a}
\end{equation}
\begin{equation}
\nabla _s\times\mu ^{-1}\^ z\times\v E_s=\^ z\nabla _s\cdot\mu
^{-1}\v E_s. \label{eq1-5b}
\end{equation}
\end{subequations}
In \eqref{eq1-5a}, $\nabla _s\cdot\mu ^{-1}\nabla_s$ is a scalar
operator, and hence, the right-hand side is still a vector pointing
in the $\^z$ direction. Using\footnote {This equation is valid in an
inhomogeneously filled waveguide, because the only assumption made
in deriving it is that the field varies as $\exp(ik_zz)$. However,
$k_s$ is not constant anymore.}
\begin{equation}
\v E_s=\frac i{k_s^2}[ k_z\nabla _sE_z+\omega\mu\nabla _s\times\v
H_z] , \label{eq1-6}
\end{equation}
where now $k_s^2 = \sqrt {\omega ^2 \mu \epsilon - k_z^2}$ is a
function of $\v r$, we can rewrite (\ref{eq1-4}) as
\begin{equation}
\nabla _s\cdot\mu ^{-1}\nabla _sE_z-ik_z\nabla _s\cdot\left (
\frac {ik_z}{\mu k_s^2}\nabla _sE_z+\frac {i\omega }{k_s^2}\nabla
_s\times\v H_z\right) +\omega ^2\epsilon E_z=0. \label{eq1-7}
\end{equation}
In the above, $k_s = \sqrt {k^2-k_z^2}$. Since $k$ is a function of
$\v r_s$, so is $k_s$. Therefore, in general, the equation governing
$E_z$ is coupled to $H_z$ as well.  By duality principle, the
equation governing $H_z$, which is also coupled to $E_z$, is
\begin{equation}
\nabla _s\cdot\epsilon ^{-1}\nabla _sH_z-ik_z\nabla _s\cdot\left(
\frac {ik_z}{\epsilon k_s^2}\nabla _sH_z-\frac
{i\omega}{k_s^2}\nabla _s\times \v E_z\right) +\omega ^2\mu H_z=0.
\label{eq1-8}
\end{equation}
However, if $\mu$ and $\epsilon$ are constants, then $k_s^2$ is a
constant, and $\nabla _s\cdot\nabla _s\times\v E_z =\nabla
_s\cdot\nabla _s\times\v H_z=0$, the $E_z$ and $H_z$ equations are
decoupled again. Therefore, for general $\mu$ and $\epsilon$ which
are inhomogeneous, the TE and the TM fields in a waveguide are
coupled.  Exceptions are sometimes found, e.g., in the axial
symmetric modes of a circular optical fiber.

The above equations show that the TE and TM fields have to co-exist
in the waveguide, hence giving rise to hybrid modes.  However, they
are not cast in terms of eigenvalue problems.  We will derive
equations from which the eigenvalues and eigenfunctions of the
waveguide can be derived.

\subsection{Transverse Field Formulation}
\index{Waveguide!transverse field formulation}

 The transverse components of the field in
Equation (\ref{eq1-1}) can be isolated to obtain
\begin{equation}
\nabla _s\times\mu ^{-1}\nabla _s\times\v E_s+ik_z\mu ^{-1}\^
z\times\nabla_s\times\v E_z+k_z^2\mu ^{-1}\v E_s-\omega ^2\epsilon
\v E_s=0. \label{eq1-9}
\end{equation}
Using $\nabla \cdot\epsilon\v E=0$, we have
\begin{equation}
ik_zE_z=-\epsilon ^{-1}\nabla _s\cdot\epsilon\v E_s.
\label{eq1-10}
\end{equation}
Therefore
\begin{equation}
ik_z\^ z\times\nabla _s\times\v E_z=ik_z\nabla _sE_z=-\nabla
_s\epsilon^{-1}\nabla _s\cdot\epsilon\v E_s. \label{eq1-11}
\end{equation}
Consequently, an equation that governs the transverse electric
field for the $i$-th mode is
\begin{equation}
\mu\nabla _s\times\mu ^{-1}\nabla _s\times\v E_{is}-\nabla
_s\epsilon^{-1}\nabla _s\cdot\epsilon\v E_{is}-k^2\v
E_{is}+k_{iz}^2\v E_{is}=0, \label{eq1-12}
\end{equation}
where $k^2 = \omega ^2\mu\epsilon$. For reason to be explained
later, we can multiply the above by $z\times$ to get
\begin{equation}
z\times\mu\nabla _s\times\mu ^{-1}\nabla _s\times\v
E_{is}-z\times\nabla _s\epsilon^{-1}\nabla _s\cdot\epsilon\v
E_{is}-k^2 z\times\v E_{is}+k_{iz}^2 z\times\v E_{is}=0,
\label{eq1-12a}
\end{equation}
In addition, an equation for the transverse magnetic field for the
$j$-th mode is easily obtained by invoking duality, namely,
\begin{equation}
z\times\epsilon\nabla _s\times\epsilon ^{-1}\nabla _s\times\v
H_{js}-z\times\nabla _s\mu^{-1}\nabla _s\cdot\mu\v
H_{js}-k^2z\times\v H_{js}+k_{jz}^2z\times\v H_{js}=0.
\label{eq1-13a}
\end{equation}
In the above, $k_z^2$ are the eigenvalues since it is constant
throughout the uniform waveguide. Due to the phase matching
condition, $k_z$ is a constant everywhere in an inhomogeneous
waveguide. We need to solve either Equation (\ref{eq1-12a}) or
(\ref{eq1-13a}) since they can be shown to be transpose of each
other.

Notice that whether if we describe the modes in a waveguide with
Equation (\ref{eq1-7}) and (\ref{eq1-8}), or Equations
(\ref{eq1-12a}) and (\ref{eq1-13a}), only two components of the
electric field or the magnetic field are required to describe the
modes in an inhomogeneously filled waveguide. This is because
$\nabla \cdot \v B = 0$ and $\nabla \cdot  \v D = 0$ and hence, not
all three components of the $\v B $ field or the $\v D $ field are
independent of each other.

As a last note, as shall be shown later, \eqref{eq1-12a} and
\eqref{eq1-13a} are transpose of each other.  Hence, eigenfunctions
from these equations are mutually orthogonal.

\subsection { Physical Interpretation of the Depolarization Effect}

The fact that the $E_z$ and $H_z$ \index{Wave!coupled}waves are in general coupled at
a dielectric interface is also known as \index{Depolarization effect}depolarization effect. An
$H_z$ (TE to $z$) wave incident at a dielectric rod, in general,
generates both $H_z$ and $E_z$ (TM to $z$) waves, and hence,
causes the \index{Wave!depolarization}depolarization of the wave. This happens conversely for
$E_z$ wave incident on a dielectric rod. This depolarization
effect occurs only for $E_z$ or $H_z$ waves that vary as a
function of $z$ $(k_z \ne 0)$. The depolarization effect
disappears when the waves do not vary as a function of $z$  $
(k_z= 0)$, or when the scatterer is a cylindrical perfect electric
conductor (PEC) or a perfect magnetic conductor.

A guided mode can be thought of as a wave bouncing around in a
waveguide such that phase coherence or constructive interference
occurs. The condition for phase coherence or constructive
interference is precisely the guidance condition of the waveguide
modes. In the case of a hollow waveguide, the waves are bouncing
off a PEC cylindrical surface and hence, the polarization purity
can be maintained. Therefore, the mode can be either \index{Mode!pure $E_z$ or $H_z$}purely $E_z $
or $H_z$ type. When a waveguide is inhomogeneously filled, the
waves have to bounce off a dielectric rod, and in general, $E_z$
or $H_z$ mode purity cannot be maintained. These modes are termed
the hybrid modes.

\begin{figure}
\begin{center}
\hfil\includegraphics[width=4.0truein]{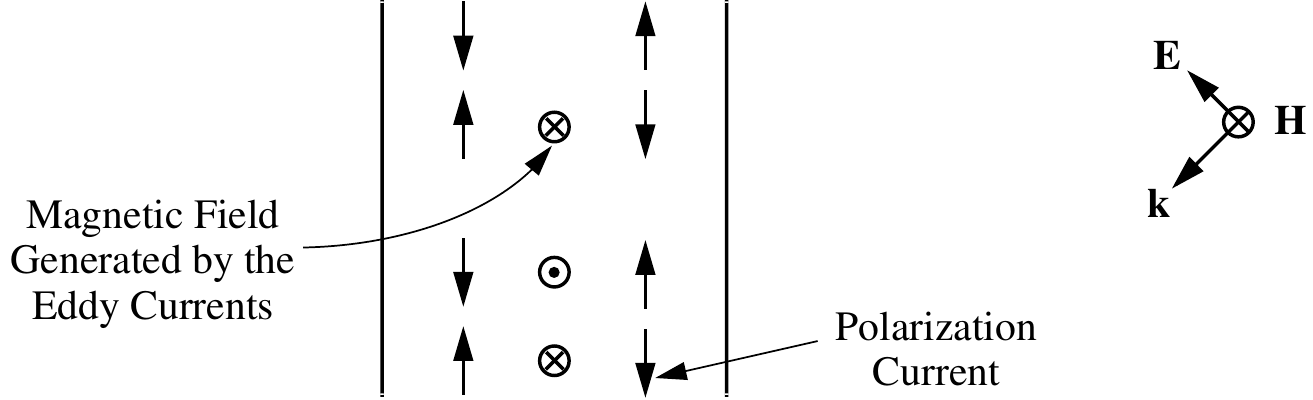}\hfil
\end{center}
\caption{Polarization current and magnetic field generated by the
induced current on a dielectric rod illuminated by an $E_z$ wave.}\label{fg312}
\end{figure}

We can argue by contradiction, except for special symmetric cases,
that only a hybrid mode consisting of TE and TM waves is possible.
When an $E_z$ polarized (TM to $z$) wave is obliquely incident on
a dielectric slab as shown in Figure \ref{fg312}, polarization
purity can be kept if the slab is infinitely wide coming out of
the paper.   The magnetic field will be horizontal in the slab
with the electric polarization current flowing around it.  The
magnetic field alternates in its polarity as one moves in the $z$
direction.  The $k$ vector lies in the plane of the paper both in
the air and in the slab.   Now assume that the slab is truncated
so that it extends in finitely into the paper as well as out of
the paper.   Assume that polarization purity is still preserved.
Then at the interface at the truncated surfaces, the $k$ vectors
are both parallel to the surface inside the dielectric as well as
in the air.  This is an impossibility since the phase velocity of
the waves on two sides of the interface are different and the
boundary condition can never be met since phase matching is
violated.  Hence, the $k$ vector has to ``tilt" in order to
satisfy the boundary condition, introducing the $z$ component of
the magnetic field.

Alternatively, we can consider an $H_z$ incident wave (TE to $z$).
Let us assume that the field remains TE inside the dielectric rod
and see that it will lead to a contradiction. If this is the case,
the transverse electric field will at least induce polarization
currents flowing in the $xy$ directions. When these currents meet
a dielectric interface, polarization charges are induced at the
interface. Because of the $z$-variation of the incident field,
these charges must be sign-changing in the $z$-direction.  The
electric field has to turn around due to the different phase
velocity it has in the air compared to the dielectric. Therefore,
an $E_z$ field must exist due to these charges.

\begin{figure}
\begin{center}
\hfil\includegraphics[width=4.0truein]{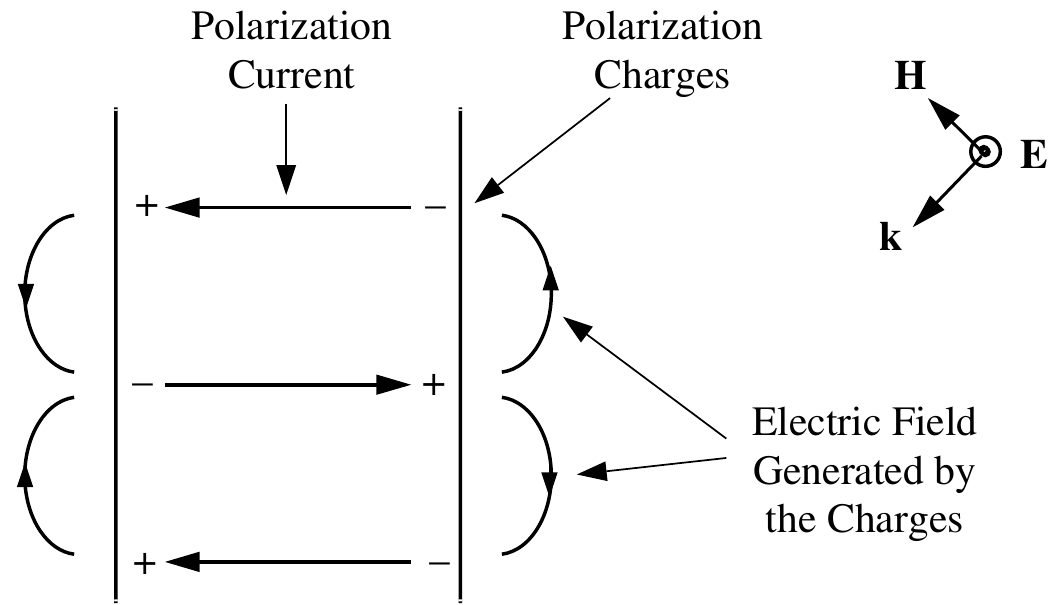}\hfil
\end{center}
\caption{Induced polarization currents and charges in a dielectric
rod if the field remains TE to $z$ when the rod is illuminated by
an $H_z$ wave.}\label{fg313}
\end{figure}


The above explains the general depolarization of the wave.
However, certain symmetrical rods may not depolarize certain
incident field. For instance, an infinite dielectric slab will not
depolarize a TE (or TM) wave where the $\v E$ (or $\v H$) field is
aligned parallel to the dielectric interface. A circular
dielectric rod will not depolarize an axially symmetric $E_{\phi}
$ or $H_{\phi }$ polarized TE or TM wave incident on it. In the
above discussion, we are referring to TE to $z$ or TM to $z$
waves.

\subsection { Mode Orthogonality}
\index{Mode!orthogonality}

In an inhomogeneous waveguide, the differential operators in
(\ref{eq1-12a}) and (\ref{eq1-13a}) are not self-transpose (or
symmetric) which is unlike a homogeneously filled cylindrical
waveguide. Therefore, general orthogonality between $\v E_{is}$ and
$\v E_{js}$, or $\v H_{is}$ and $\v H_{js}$ does not exist. However,
mode orthogonality exists between $\v E_{is}$ and $\v H_{js}$. We
can prove the orthogonality of $\v E_{is}$ and $\v H_{js}$ for a
general, inhomogeneous, anisotropic waveguide using the Lorentz
reciprocity theorem.

\begin{figure}
\begin{center}
\hfil\includegraphics[width=4.5truein]{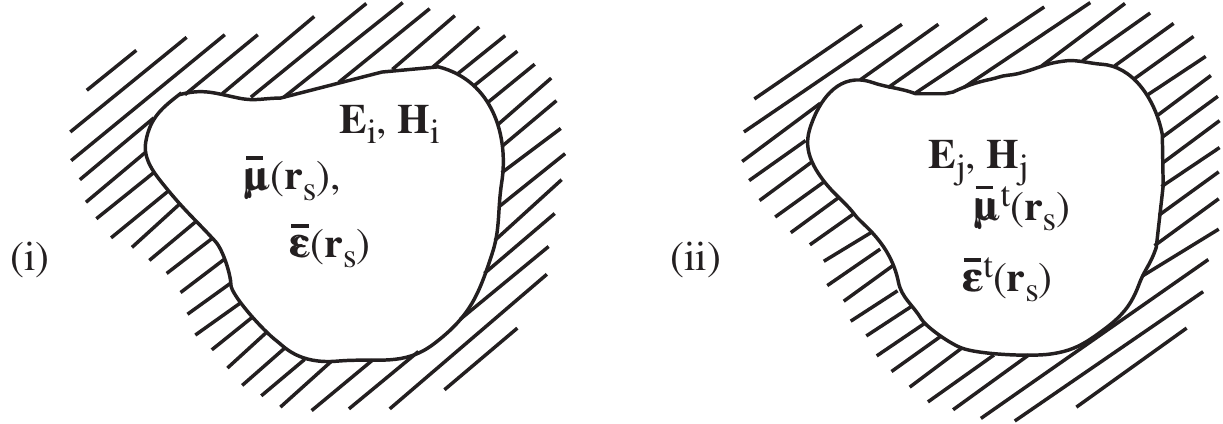}\hfil
\end{center}
\caption{Proof of mode orthogonality.  The second problem is the
auxiliary of the first where the waveguide is filled with transposed
material tensors.}\label{fg314}
\end{figure}

Consider two waveguides with identical walls but different
anisotropic, inhomogeneous media: one waveguide is filled with
$\dyadg\mu$, $\dyadg\epsilon$, while the other one is filled with
$\dyadg\mu^t(\v r_s)$, $\dyadg\epsilon^t(\v r_s)$ where the
superscript $t$ stands for transpose (see Figure \ref{fg314}, the
second problem is the auxiliary of the first one).  It can be shown
that
\begin{equation}
\int _S(\v E_{is}\times\v H_{js})\cdot\^ z dS=0, \qquad i\ne j,
\label{eq1-14}
\end{equation}
where $\v E_i$ is the $\v E$-field inside the first waveguide,
while $\v H_j$ is the $\v H$-field inside the second waveguide. In
the first waveguide,
\begin{equation}
\nabla \times\v E_i=i\omega\dyadg\mu\cdot\v H_i,
\qquad\nabla\times\v H_i=-i\omega\dyadg\epsilon\cdot\v E_i,
\label{eq1-15}
\end{equation}
while in the second waveguide,
\begin{equation}
\nabla\times\v E_j=i\omega\dyadg\mu^t\cdot\v H_j,
\qquad\nabla\times\v H_j=-i\omega\dyadg\epsilon^t\cdot\v E_j.
\label{eq1-16}
\end{equation}
We can show that
\begin{equation}
\begin{split}
\nabla\cdot (\v E_i\times\v H_j-\v E_j\times\v H_i)
& =i\omega (\v H_j\cdot\dyadg\mu\cdot\v H_i-\v H_i\cdot\dyadg\mu^t\cdot\v H_j)\\
& +i\omega (\v E_i\cdot\dyadg\epsilon^t\cdot\v E_j-\v
E_j\cdot\dyadg\epsilon\cdot\v E_i). \label{eq1-17}
\end{split}
\end{equation}
Since $\v A\cdot\dyad B^t\cdot\v C=\v C\cdot\dyad B\cdot\v A$,
\begin{equation}
\nabla\cdot (\v E_i\times\v H_j-\v E_j\times\v H_i)=0.
\label{eq1-18}
\end{equation}
The above is the generalized Lorentz reciprocity theorem. If $\v
E_i$, $\v H_i\sim e^{ik_{iz}z}$ while $\v E_j$, $\v H_j\sim
e^{ik_{jz}z}$ in their $z$-dependence, the above becomes
\begin{equation}
\nabla _s\cdot (\v E_i\times\v H_j-\v E_j\times\v
H_i)=-i(k_{iz}+k_{jz})\^z\cdot (\v E_{is}\times\v H_{js}-\v
E_{js}\times\v H_{is}) \label{eq1-19}
\end{equation}
Integrating (\ref{eq1-19}) over the cross-section of the
waveguide, we have

\begin{equation}
\oint _C dl\^ n\cdot (\v E_i\times\v H_j-\v E_j\times\v
H_i)=-i(k_{iz}+k_{jz})\int _S dS\^ z\cdot (\v E_{is}\times\v
H_{js}-\v E_{js}\times\v H_{is}). \label{eq1-20}
\end{equation}
The line integral vanishes by virtue of the boundary conditions.
Hence,
\begin{equation}
(k_{iz}+k_{jz})\int dS\^ z\cdot (\v E_{is}\times\v H_{js}-\v
E_{js}\times\v H_{is})=0. \label{eq1-21}
\end{equation}
If the waveguide has reflection symmetry, a mode propagating in
the $+z$ direction is almost the same as a mode propagating in the
$-z$ direction, except for a change of the sign of the field.  If
$\v E_{js}$, $\v H_{js}$ corresponding to a mode with $k_{jz}$
propagating in $+z$ direction, then by Poynting theorem, $\v
E_{js}$, $-\v H_{js}$ are the transverse fields of a mode with
$-k_{jz}$ propagating in the $-z$ direction. Then, Equation
(\ref{eq1-21}) can be rewritten as
\begin{equation}
(k_{iz}-k_{jz})\int dS\^ z\cdot (\v E_{is}\times\v H_{js}+\v
E_{js}\times\v H_{is})=0. \label{eq1-22}
\end{equation}
If $k_{iz}$ and $k_{jz}$ are both non-zero, then the integral in
(\ref{eq1-21}) must be zero. If $k_{iz}\ne k_{jz}$, then the
integral in (\ref{eq1-22}) must be zero. The combination of
(\ref{eq1-21}) and (\ref{eq1-22}) implies that
\begin{equation}
\int _S dS\^ z\cdot (\v E_{is}\times\v H_{js})=0, \qquad i\ne j.
\label{eq1-23}
\end{equation}

If the \index{Medium!reciprocal}medium is reciprocal, then $\dyadg\mu=\dyadg\mu ^t$,
$\dyadg\epsilon=\dyadg\epsilon^t$ and waveguides (i) and (ii) in
Figure \ref{fg311} are the same waveguide, and $\v E_{is}$ and $\v
H_{js}$ are the fields from the same waveguide.

Furthermore, if $\dyadg\mu$ and $\dyadg\epsilon$ are Hermitian
corresponding to a lossless medium with reflection symmetry, similar
proof shows that
\begin{equation}
\int _S dS\^ z\cdot (\v E_{is}\times\v H_{js}^*)=0, \qquad i\ne j.
\label{eq1-24}
\end{equation}
The above is the power orthogonality condition for two different
modes in an inhomogeneous, anisotropic, lossless, waveguide with
reflection symmetry.  Otherwise, the mode of the original waveguide
is orthogonal to another mode of another waveguide where the medium
is filled with a conjugate transpose medium.

\subsection { Reflection Symmetry and Conservation of Parity}
\index{Symmetry!reflection}
\index{Conservation of parity}

A commonly accepted law of physics is that the classical laws of
physics hold true in the mirrored world (the reflected
world)\cite{FEYNMAND}. For electromagnetics, this means that a
right-hand rule becomes a left-hand rule. This is also known as
the conservation of parity, and is found to be violated in modern
physics by some weak interactions.

If a waveguide has reflection symmetry, we say that it appears to
be the same waveguide in the mirrored world as it is in the real
world. This is certainly true of all uniform hollow waveguide. If
a mode propagates in the $+z$ direction in the real world, the
corresponding mode propagates in the mirrored $+z $ direction,
plus that all the field components are mirrored. We can rotate the
mirrored waveguide by $180^o$ about an axis perpendicular to the
$z$ axis, and we recover the original waveguide. The field of the
\index{Mode!mirrored, real}mirrored mode satisfies the left-hand rule rather than the
right-hand rule. However, we can convert the field of the mirrored
mode into a real-world mode by switching $\v H $ to $-\v H$, and
now the field will satisfy the right-hand rule as before.

When the waveguide is filled with an anisotropic material, the
problem is more tricky, because in the mirrored world, the
anisotropic material may not reflect to be the same material in
the real world. However, if the permeability and permittivity
tensors are of the form
\begin{equation}
\dyadg\mu =
\begin{pmatrix}
\dyadg\mu _s & 0 \\
 0 & \mu_{zz}
\end{pmatrix},
\qquad  \qquad \dyadg\epsilon =
\begin{pmatrix}
\dyadg\epsilon _s & 0  \\
 0 & \epsilon _{zz}
\end{pmatrix},
\label{eq1-25}
\end{equation}
then the waveguide has reflection symmetry, i.e., the waveguide in
the mirrored world is the same as the original waveguide. Again,
if we switch $\v H$ to $-\v H$, we can obtain a real world mode.

\begin{figure}
\begin{center}
\hfil\includegraphics[width=4.5truein]{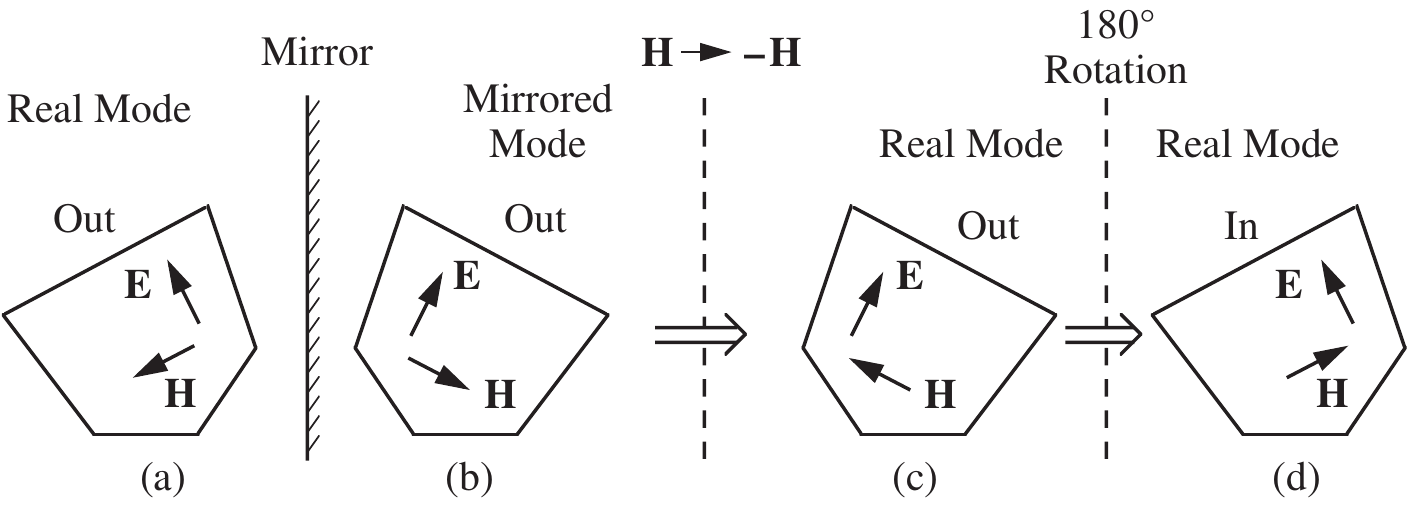}\hfil
\end{center}
\caption{Reflection and the conservation of parity of a
waveguide.}\label{fg315}
\end{figure}


We can convince ourselves more by looking at \index{Maxwell's equations}Maxwell's equations.
By separating it into axial and transverse components, Maxwell's
equations for a guided mode for the above tensors become
\begin{subequations}
\begin{equation}
\nabla _s \times \v H_s =-i\omega \epsilon _{zz} \v E_z,
\label{eq1-26a}
\end{equation}
\begin{equation}
ik_z \^z \times \v H_s + \nabla _s \times \v H_z =-i\omega
\dyadg\epsilon _s\cdot \v E_s , \label{eq1-26b}
\end{equation}
\begin{equation}
\nabla _s \times \v E _s =i \omega \mu  _{zz} \v H_z,
\label{eq1-26c}
\end{equation}
\begin{equation}
ik_z \hat z \times \v E_s +\nabla _s \times \v  E_z =i\omega
\dyadg\mu _s \cdot\v H _s . \label{eq1-26d}
\end{equation}
\end{subequations}
If the direction of propagation changes, which is what a mirror
reflection does, $k_z$ changes to $-k_z$. Notice that now if we
let
\begin{equation}
\v H_s \rightarrow -\v H_s, \quad \v H_z \rightarrow \v H_z ,
\quad \v E _s \rightarrow \v E_s, \quad\v E_z \rightarrow -\v E_z,
\label{eq1-27}
\end{equation}
the above equations remain invariant\index{Maxwell's equations!invariance}. Therefore if the waveguide
mode changes direction, the above transformation in the field is
needed to obtain another solution to Maxwell's equations.

To see how this is related to the conservation of parity, we
imagine a waveguide in (a) above with the $ \v E $ and $\v H$
fields as shown. They could have a $\^z$ component perpendicular
to the paper which is not shown. We assume that the mode is
propagating out of the paper. In the mirrored world, as shown in
(b), we have a mirrored mode and mirrored waveguide. The
electromagnetic field does not satisfy Maxwell's equations
according to the right-hand rule. To make the mode in (b) into a
real mode, we let $ \v H \rightarrow -\v H $ as shown in (c).
However, compared to (a), the waveguide is  not the original
waveguide. If the constitutive parameters are those  in
(\ref{eq1-25}), then (c) is just a $180^o$ rotation of the
waveguide in (a) about an axis perpendicular to the $z$ axis. By
rotating the waveguide by $180^o$, we obtain (d). However, $\v H_s
$ has changed sign while $\v H_z$ remains unchanged compared to
(a). This is precisely the field transformation prescribed by
(\ref{eq1-27}). A closer examination also indicates that $\v E_s$
does not change sign, while $E_z$ changes sign.

\section {{{ General Anisotropic Waveguide}}}\label{Section3.3}
\index{Waveguide!general anisotropic}

The vector wave equation governing the field inside an anisotropic
waveguide is
\begin{equation}
\nabla\times\dyadg\mu^{-1}\cdot\nabla\times\v E-\omega
^2\dyadg\epsilon\cdot\v E=0. \label{eq2-1}
\end{equation}
In general, the modes inside such a waveguide cannot be decomposed
into TE and TM modes. Hence, the problem is again vector,
requiring at least two components of the field. Assuming the
waveguide to have reflection symmetry in the $z$ direction, then
\begin{equation}
\dyadg\epsilon=
\begin{bmatrix}
\dyadg\epsilon_s & 0\\
 0 & \epsilon _{zz}
\end{bmatrix},
\qquad\qquad \dyadg\mu=
\begin{bmatrix}
\dyadg\mu_s & 0\\
 0 & \mu _{zz}
\end{bmatrix} ,
\label{eq2-2}
\end{equation}
where $\dyadg\epsilon_s$ and $\dyadg\mu_s$ are $2\times 2$ tensors
with components in the transverse direction. The transverse
component of (\ref{eq2-1}) can be extracted to obtain
\begin{equation}
\nabla _s\times\mu _{zz}^{-1}\nabla _s\times\v E_s-ik_z\^
z\times\dyadg\mu_s^{-1}\cdot\^ z\times\nabla _sE_z-k_z^2\^
z\times\dyadg\mu_s^{-1}\cdot\^z\times\v E_s-\omega
^2\dyadg\epsilon_s\cdot\v E_s=0. \label{eq2-3}
\end{equation}
With the use of the divergence condition,
\begin{equation}
ik_zE_z=-\epsilon _{zz}^{-1}\nabla _s\cdot\dyadg\epsilon_s\cdot\v
E_s. \label{eq2-4}
\end{equation}
Hence, the $z$ component of the field in (\ref{eq2-3}) can be
replaced to yield
\begin{equation}
\begin{split}
\nabla _s\times\mu _{zz}^{-1}\nabla _s\times\v E_s &+\^ z\times\dyadg
\mu_s^{-1}\cdot\^ z \times\nabla _s\vg\epsilon _{zz}^{-1}\nabla _s\cdot\dyadg\epsilon_s\v E_s\\
& -k_z^2\^ z\times\dyadg\mu_s^{-1}\cdot\^ z\times\v E_s
 -\omega ^2\dyadg\epsilon_s\cdot\v E_s=0.
\label{eq2-5}
\end{split}
\end{equation}

Consequently, the field in an anisotropic waveguide can be
characterized by $\v E_s$ alone. We can multiply the above by
$\dyadg\mu_s\cdot\^ z\times$ to get
\begin{equation}
\begin{split}
\dyadg\mu_s\cdot\^ z\times\nabla _s\times\mu _{zz}^{-1} &\nabla
_s\times\v E_s-\hat z\times\nabla _s\epsilon _{zz}^{-1}\nabla _s\cdot\dyadg\epsilon_s\cdot\v E_s\\
& -\omega ^2 \dyadg\mu_s\cdot\^ z\times\dyadg\epsilon_s\cdot\v
E_s+k_z^2\hat z \times \v E_s=0. \label{eq2-6}
\end{split}
\end{equation}
Note that the above equation is a function of $k_z^2$, implying
that if $\v E_s$ with $e^{ik_zz}$ dependence is a solution to
(\ref{eq2-6}), an $\v E_s$ with $e^{-ik_zz}$ dependence is also a
solution. This is a consequence of reflection symmetry, or the
assumptions about $\dyadg\epsilon$ and $\dyadg\mu$ in
(\ref{eq2-2}). The corresponding equation for the transverse
magnetic field is obtained by duality, yielding
\begin{equation}
\begin{split}
\dyadg\epsilon_s\cdot\^ z\times\nabla _s\times\epsilon_{zz}^{-1}
&\nabla _s\times\v H_s-\hat z\times\nabla _s\vg \mu _{zz}^{-1}\nabla _s\cdot\dyadg\mu_s\cdot\v H_s\\
& -\omega ^2\dyadg\epsilon_s\cdot\^ z\times\dyadg\mu_s\cdot\v
H_s+k_z^2\hat z\times\v H_s=0. \label{eq2-7}
\end{split}
\end{equation}
Since the solution to (\ref{eq2-6}) is orthogonal to the solution to
(\ref{eq2-7}), for two different modes, Equation (\ref{eq2-7}) is
also the transpose equation of (\ref{eq2-6}) \cite{CHEW&NASIR,WFIM}.
This shall be elaborated in the next section.

%

\section{{Proof of Transpose of Operators}}
\index{Transposed operators}

        For an anisotropic waveguide with reflection symmetry,
        \begin{align}\label{PTO1}
            \dyadg{\mu}_{s}\cdot\hat{z}\times\nabla_{s}\times\mu_{zz}^{-1}\nabla_{s}\times{\v E}_{s}
            -\hat{z}\times\nabla_{s}\epsilon_{zz}^{-1}
            \nabla_{s}\cdot\dyadg{{\epsilon}}_{s}\cdot\v{E}_{s}\nonumber \\
            -\omega^{2}{\dyadg\mu}_{s}\cdot\hat{z}\times{\dyadg\epsilon}_{s}\cdot{\v E}_{s}+k_{z}^{2}\hat{z}\times\v{E}_{s}=0
        \end{align}
        We shall show that the above equation is transpose to the equation
               \begin{align}\label{PTO2}
            {\dyadg\epsilon}_{s}^t\cdot\hat{z}\times\nabla_{s}\times\epsilon_{zz}^{-1}\nabla_{s}\times\v{H}_{s}-\hat{z}\times\nabla_{s}
            \mu_{zz}^{-1}\nabla_{s}\cdot{{\dyadg\mu}}_{s}^t\cdot {\v H}_{s}\nonumber \\
            -\omega^{2}{\dyadg\epsilon}_{s}^t\cdot\hat{z}\times {\dyadg\mu}_{s}^t\cdot{\v H}_{s}+k_{z}^{2}\hat{z}\times{\v H}_{s}=0
        \end{align}
                The definition of the transpose operator is\footnote{When applied to a matrix operator, this becomes
$\v u^t\cdot\dyad{\v L}\cdot \v v = \left(\v v^t\cdot\dyad{\v
L}^t\cdot \v u \right)$. Similar formula can be derived for defining
conjugate transpose or adjoint of an operator \cite{WFIM}.}
        \begin{equation}
        \langle{\v u,\mathcal L\v v}\rangle = \langle \v v,\mathcal L^{t}\v u\rangle
        \end{equation}
        where \index{Hilbert space}inner product between two vector fields in the infinite dimensional space (also called the Hilbert space)
        is $\langle{\v f,\v g}\rangle=\int dS ~\v f\cdot \v g$.  The
        integration in this case is taken over the cross-sectional
        area of the waveguide.  Furthermore, the $\v H_s$ field in
        (\ref{PTO2}) is the field of the auxiliary problem as
        indicated in Figure \ref{fg314}.

        To prove this, we start with the first term in \eqref{PTO1}, and calculate the expression
        \begin{equation}
            I =\langle\v{H}_{s},{\dyadg\mu}_{s}\cdot\hat{z}\times\nabla_{s}\times\mu_{zz}^{-1}\nabla_{s}\times{\v E}_{s}\rangle
        \end{equation}
        It can be easily shown that
        \begin{equation}
            I=\langle\hat{z}\times {\dyadg\mu}_{s}^t\cdot{\v H}_{s},\nabla_{s}\times\mu_{zz}^{-1}\nabla_{s}\times {\v E}_{s}\rangle
        \end{equation}

        Using Gauss theorem, or integration by parts, and that $\nabla_{s}\cdot(\v{A}\times\v{B}) = \v{B}\cdot \nabla_{s} \times \v{A} - \v{A} \cdot
         \nabla_{s} \times \v{B}
        $
        we have
        \begin{equation}
        I = -\int dS \nabla_{s} \cdot [(\hat{z}\times {\dyadg{\mu}}_{s}^{t}\cdot \v H_{s})\times(
        \bar{\bf{\mu}}_{zz}^{-1}\nabla_{s} \times \v{E}_{s})] + \int dS (\nabla_{s}\times (\hat{z}\times \dyadg{\mu}_{s}^{t}\cdot
        \v{H}_{s}))\cdot\mu_{zz}^{-1}\nabla_{s}\times \v{E}_{s})
        \end{equation}
        It can be shown that $\nabla_{s}\times(\hat{z}\times\dyadg{{\mu}}_{s}^{t}\cdot\v{H}_{s}) = \hat{z}\nabla_{s}
        \cdot(\dyadg{{\mu}}_{s}^{t}\cdot\v{H}_{s})$, and that $\hat{z}\cdot\nabla_{s}\times\v{E}_{s} = -\nabla_{s}
        \cdot(\hat{z}\times\v{E}_{s})$.  The first term above can be
        converted to a boundary integral over the waveguide wall
        using Gauss' divergence theorem.  It vanishes by virtue of
        the boundary condition on the waveguide wall that $\hat n \cdot \v B=0$.  Hence, only
        the second term remains.  Consequently,
\begin{equation}
I=-\int dS \nabla_s\cdot (\dyadg{\mu}_s^t\cdot\v
H_s)\mu_{zz}^{-1}\nabla_s\cdot  (\hat z\times \v E_s)
\end{equation}
        Using integration by parts one more time, and then the vector
        identity, that $\nabla_s\cdot (\phi\v A)=\phi\nabla_s\cdot\v
        A+\nabla_s\phi\cdot\v A$, where $\phi=\nabla_s\cdot (\dyadg{\mu}_s^t\cdot\v
H_s)\mu_{zz}^{-1}$ and $\v A=\hat z\times \v E_s$, yields
        \begin{equation}
        I = -\int dS \nabla_{s}\cdot\left[\nabla_s\cdot (\dyadg{\mu}_s^t\cdot\v
H_s)\mu_{zz}^{-1}  (\hat z\times \v E_s)\right]
          +  \int dS \hat{z} \times \nabla_{s} \mu_{zz}^{-1} \nabla_{s}\cdot(\dyadg{\mu}_{s}^{t}\cdot\v{H}_{s})\cdot\v{E}_{s}
        \end{equation}
        where the cyclic relation of cross-dot product has been
        used.
        The first integral can be converted to a boundary integral
        and vanishes by virtue of the boundary condition and that $\hat z \times \v E_s=0$ on the waveguide wall.
        Hence,
\begin{equation}
        I =  \int dS \hat{z} \times \nabla_{s} \mu_{zz}^{-1}
        \nabla_{s}\cdot(\dyadg{\mu}_{s}^{t}\cdot\v{H}_{s})\cdot\v{E}_{s}=-\langle\v
        E_s,\hat{z} \times \nabla_{s} \mu_{zz}^{-1}
        \nabla_{s}\cdot(\dyadg{\mu}_{s}^{t}\cdot\v{H}_{s})\rangle
        \end{equation}
        Hence, the first term in \eqref{PTO1} is the negative transpose of
        the second term in \eqref{PTO2}.  Similarly, the second term
        in \eqref{PTO1} is the transpose of the first term in
        \eqref{PTO2}.

        Furthermore, it can be shown that $(\hat{z}\times)^{t} = -\hat{z}\times $, $({\dyadg\epsilon \cdot \hat{z} \times \dyadg\mu_{s}})^{t} = -\dyadg{{\mu}}_s^{t}\cdot
        \hat{z}\times\dyadg{{\epsilon}}_{s}^{t}$ .
        Hence if,
        \begin{equation}
         \mathcal L(*) = \dyadg{{\mu}}_{s}\cdot\hat{z}\times\nabla_{s} \times \mu_{zz}^{-1} \nabla_{s} \times(*) - \hat{z} \times \nabla_{s}
         \epsilon_{zz}^{-1}\nabla_{s}\cdot\dyadg{{\epsilon}}_{s}(*)-\omega^{2}\dyadg{{\mu}}_{s}\cdot\hat{z}\times\dyadg{{\epsilon}}_{s}(*)+
         k_{z}^{2}\hat{z}\times(*)
        \end{equation}
        Then
        \begin{equation}
         -\mathcal L^{t}(*) = \dyadg{{\epsilon}}_{s}^t\cdot\hat{z}\times\nabla_{s}\times\epsilon_{zz}^{-1}\nabla_{s}\times(*)
         -\hat{z}\times\nabla_{s}\mu_{zz}^{-1}\nabla_{s}\cdot\dyadg{{\mu}}_{s}^t(*)
         -\omega^{2}\dyadg{{\epsilon}}_{s}^{t}\cdot\hat{z}\times\dyadg{{\mu}}_{s}^{t}(*)+k_{z}^{2}\hat{z}\times(*)
        \end{equation}
        where $(*)$ represents a vector function that these
        operators act on.
        In general, \eqref{PTO2} is the field equation for a waveguide
        filled with \index{Medium!transpose}\index{Waveguide!transpose medium}transpose medium compared to the original
        waveguide equation \eqref{PTO1}, and \eqref{PTO2} is the
        negative transpose of \eqref{PTO1}.
        However, if the waveguide is filled with reciprocal medium,
        then \eqref{PTO1} and \eqref{PTO2} are field equations for the same
        waveguide.

\section {{{ Dielectric-Slab-Loaded Rectangular Waveguides}}}
\index{Waveguide!dielectric-slab-loaded}

\begin{figure}
\begin{center}
\hfil\includegraphics[width=4.5truein]{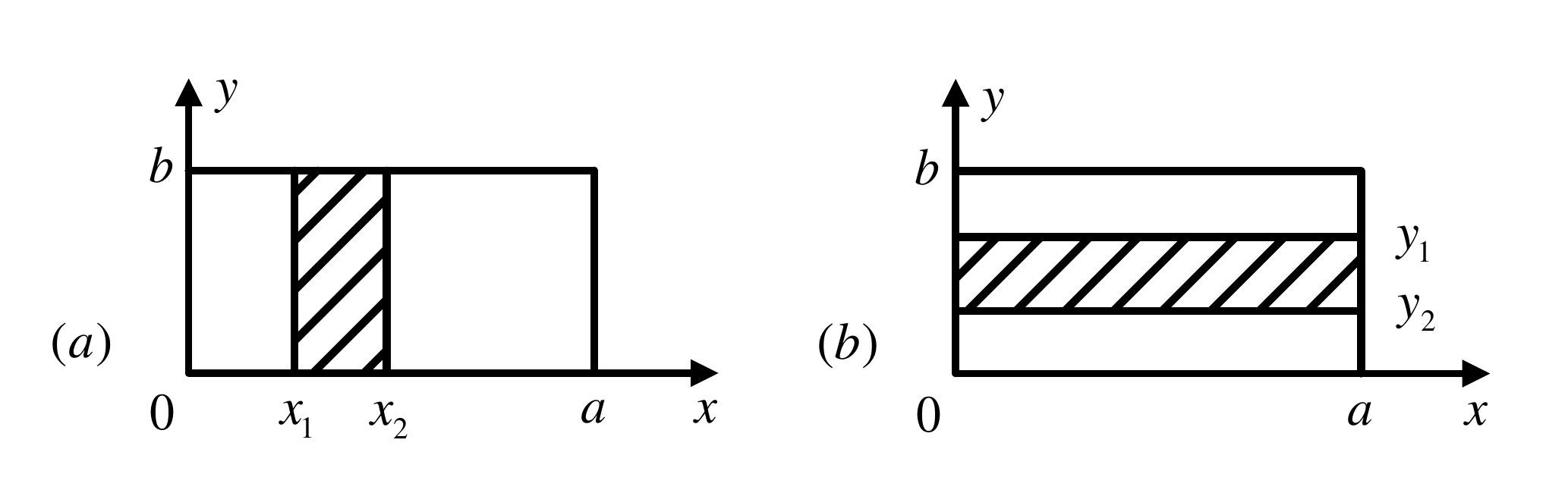}\hfil
\end{center}
\caption{Dielectric-slab-loaded waveguide.}\label{fg331}
\end{figure}

For the analysis of a general inhomogeneously filled waveguide, a
numerical method has to be sought \cite{CHEW&NASIR,JFLEEETAL,JIN}.
However, when the waveguide has certain symmetry such as a slab
loaded rectangular waveguide, analytic method for its analysis is
possible.

 Dielectric-slab-loaded waveguides find applications in a
number of microwave components, because the \index{Phase velocity}phase velocity of a
mode can be altered with dielectric loading
\cite{COLLIN,VARTANIAN_AYRES_HELGESSON}.  A waveguide can also be
loaded with a ferrite slab \cite{LAX_UTKM_ROTH,BRESLER}.  When a
ferrite slab is biased with a magnetic field, it becomes an
anisotropic, \index{Medium!gyrotropic}gyrotropic medium. Such a medium is non-reciprocal.
Hence, \index{Ferrite slab}ferrite slabs can be used to design non-reciprocal devices
such as isolators.

The analysis of a ferrite-slab-loaded waveguide is complicated. We
will focus our analysis on a dielectric-slab-loaded waveguide.
In such a waveguide, with the exception of special cases, the
modes cannot be decomposed into TE and TM modes with respect to
the $z$ direction. Hence, a general mode is \index{Mode!hybrid}hybrid. However, for
the slab-loaded waveguide as shown in Figure \ref{fg331}(a), the
modes in the waveguide can be decomposed into modes with $\v E$
field transverse to $x$ or modes with $\v H$ field transverse to
$x$. Similar decomposition exists for Figure \ref{fg331}(b) since
it is just a $90^o$ rotation of Figure \ref{fg331}(a).  The modes
with $\v E$ field transverse to $x$ are known as the \index{Mode!Longitudinal Section Electric}LSE
(Longitudinal Section Electric) modes, while the modes with $\v H$
field transverse to $x$ are known as the \index{Mode!Longitudinal Section Magnetic}LSM (Longitudinal Section
Magnetic) modes.  There exist closed form expressions for the
guidance condition of these modes, but the exact wave number $k_z$
has to be found numerically.

\begin{figure}
\begin{center}
\hfil\includegraphics[width=3.5truein]{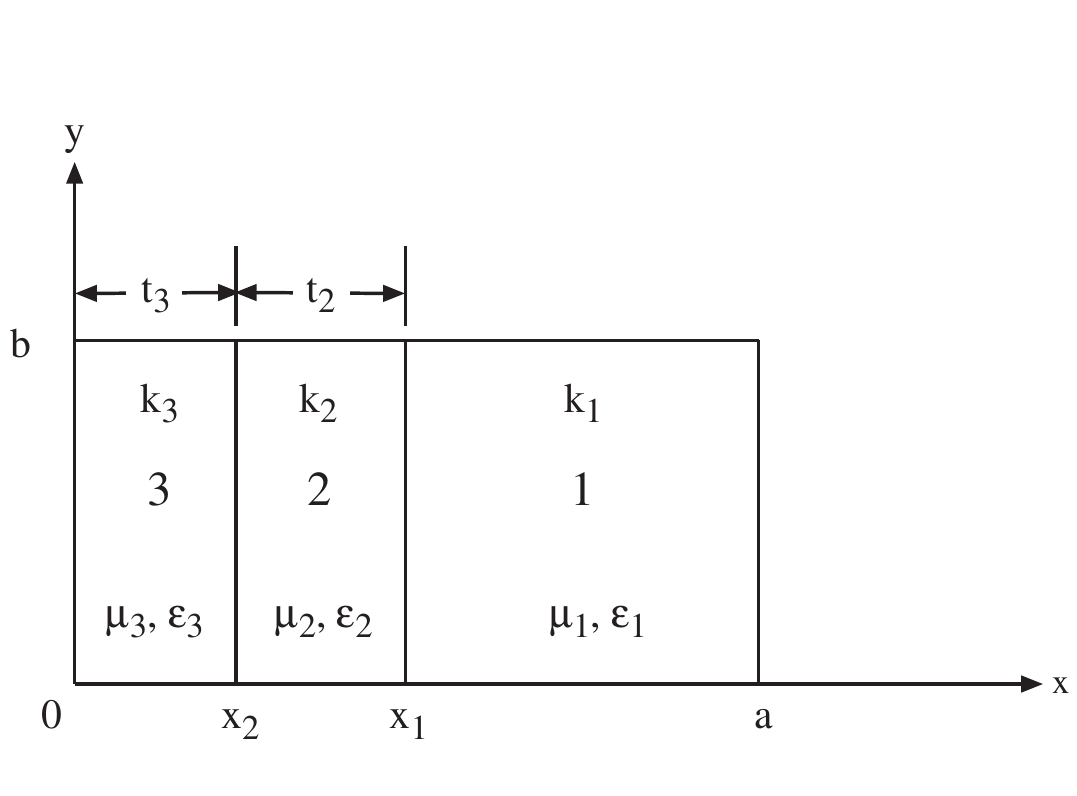}\hfil
\end{center}
\caption{A special case of a slab loaded waveguide.}\label{fg332}
\end{figure}


Since the LSE mode is transverse to $x$, it can be characterized
with $H_x$. Similarly, the LSM mode can be characterized with
$E_x$.  The field has to have $e^{ik_zz}$ dependence everywhere
inside the waveguide due to the phase matching condition.
Consequently, the equations satisfied by $H_x$ and $E_x$ are
\begin{equation}
(\nabla _s^2+k_i^2-k_z^2)H_{ix}=0, \qquad\text {LSE modes},
\label{eq3-1}
\end{equation}
\begin{equation}
(\nabla _s^2+k_i^2-k_z^2)E_{ix}=0, \qquad\text {LSM modes}.
\label{eq3-2}
\end{equation}
where subscript $i$ denotes the region $i$ and subscript $s$
denotes transverse to $x$. The solutions to (\ref{eq3-1}) and
(\ref{eq3-2}) are of the form
\begin{equation}
H_{ix}=H_0\{ e^{\pm ik_{ix}x}\} \{ e^{\pm ik_yy}\} e^{ik_zz},
\label{eq3-3}
\end{equation}
\begin{equation}
E_{ix}=E_0\{ e^{\pm ik_{ix}x}\} \{ e^{\pm ik_yy}\} e^{ik_zz},
\label{eq3-4}
\end{equation}
where the braces imply linear superpositions. Here, $k_y$ in each
region must be the same due to the phase matching condition. The
fields transverse to $x$ can be found from $H_{ix}$ and $E_{ix}$
in each region, i.e.,
\begin{subequations}
\begin{equation}
\v E_{is}=\frac 1{k_y^2+k_z^2}\left[ \frac {\partial }{\partial
x}\nabla _sE_{ix}-i\omega \mu _i\^ x\times\nabla _sH_{ix}\right],
\label{eq3-5a}
\end{equation}
\begin{equation}
\v H_{is}=\frac 1{k_y^2+k_z^2}\left[ \frac {\partial }{\partial
x}\nabla_sH_{ix}+i\omega \epsilon _i\^ x\times\nabla
_sE_{ix}\right], \label{eq3-5b}
\end{equation}
\end{subequations}
where the subscript $s$ implies transverse to $x$. It can be seen
from the above that the boundary conditions for $H_{ix}$ and
$E_{ix}$ are that
\begin{subequations}
\begin{equation}
\frac {\partial H_{ix}}{\partial y}=0, \qquad \text {at}\quad
y=0\quad \text {and} \quad y=b, \label{eq3-6a}
\end{equation}
\begin{equation}
E_{ix}=0, \qquad \text {at}\quad y=0\quad \text {and}\quad y=b.
\label{eq3-6b}
\end{equation}
\end{subequations}
Therefore, forms for $H_{ix}$ and $E_{ix}$ that satisfy the above
boundary conditions are
\begin{subequations}
\begin{equation}
H_{ix}=H_0\{ e^{\pm ik_{ix}x} \}\cos \left( {\frac {n\pi
y}b}\right) e^{ik_zz}, \label{eq3-7a}
\end{equation}
\begin{equation}
E_{ix}=E_0\{ e^{\pm ik_{ix}x} \}\sin \left( {\frac {n\pi
y}b}\right) e^{ik_zz}, \label{eq3-7b}
\end{equation}
\end{subequations}
where we have let $k_y = \frac {n\pi }b$, and $n$ is an integer.
In the above, $k_{ix} = \sqrt {k_i^2 - k_z^2 - (\frac {n\pi
}b)^2}$. It is now clear that $H_{ix}$ represents bouncing waves
that are TE to $x$ while $E_{ix}$ represents bouncing waves that
are TM to $x$. Hence, in region 1, we can write the solution as
\begin{subequations}
\begin{equation}
H_{1x}=H_0[ e^{-ik_{1x}(x-x_1)}+\tilde
R_{12}^{TE}e^{ik_{1x}(x-x_1)}] \cos {\left( \frac {n\pi y}b\right)
} e^{ik_zz}, \label{eq3-8a}
\end{equation}
\begin{equation}
E_{1x}=E_0[ e^{-ik_{1x}(x-x_1)}+\tilde
R_{12}^{TM}e^{ik_{1x}(x-x_1)}] \sin {\left( \frac {n\pi y}b\right)
} e^{ik_zz}, \label{eq3-8b}
\end{equation}
\end{subequations}
where $\tilde R_{12}^{TE, TM}$ is a generalized \index{Fresnel reflection coefficient}Fresnel reflection
coefficient for a TE or a TM wave incident from the left at $x =
x_1$.  It includes subsurface reflections. However, we need to
impose the boundary conditions that
\begin{equation}
H_{ix}(x=a)=0, \qquad \frac {\partial E_{ix}}{\partial x}(x=a)=0.
\label{eq3-9}
\end{equation}
The above boundary conditions can be derived from \eqref{eq3-5a} and
\eqref{eq3-5b} as the most general case.  They can also be derived
from $\hat n\cdot \v H=0$ and $\nabla\cdot\v E=0$, respectively.
They could only be satisfied if
\begin{subequations}
\begin{equation}
1+\tilde R_{12}^{TE}e^{2ik_{1x}(a-x_1)}=0, \qquad \text {LSE
modes}, \label{eq3-10a}
\end{equation}
\begin{equation}
1-\tilde R_{12}^{TM}e^{2ik_{1x}(a-x_1)}=0, \qquad \text {LSM
modes}. \label{eq3-10b}
\end{equation}
\end{subequations}
The above are the \index{Guidance condition}guidance conditions for the LSE modes and the
LSM modes in a slab loaded rectangular waveguide.

To find $\tilde R_{12}$, we note that a wave impinging on a slab
with one subsurface interface will consist of a surface reflection
from the top interface plus a sequence of subsurface reflections.
The single interface reflections are governed by the Fresnel
reflection coefficients.  Hence \cite{WFIM}
\begin{equation}
\begin{split}
\tilde R_{12}=R_{12} & +T_{12}R_{23}T_{21}e^{2ik_{2x}t_2}+T_{12}R_{23}^2R_{21}T_{21}e^{4ik_{2x}t_2}\\
& +T_{12}R_{23}^3R_{21}^2T_{21}e^{6ik_{2x}t_2}+\hdots,
\label{eq3-11}
\end{split}
\end{equation}
where $T_{ij} = 1 + R_{ij}$ is the \index{Fresnel transmission coefficient}Fresnel transmission
coefficient at the $ij$ interface. The above could be summed to
yield
\begin{equation}
\tilde R_{12}=R_{12}+\frac
{T_{12}R_{23}T_{21}e^{2ik_{2x}t_2}}{1-R_{23}R_{21}e^{2ik_{2x}t_2}}=\frac
{R_{12}+R_{23}e^{2ik_{2x}t_2}}{1-R_{23}R_{21}e^{2ik_{2x}t_2}}.
\label{eq3-12}
\end{equation}
In the above,
\begin{equation}
R_{ij}^{TE}=\frac {\mu _jk_{ix}-\mu _ik_{jx}}{\mu _jk_{ix}+\mu
_ik_{jx}} , \qquad R_{ij}^{TM}=\frac {\epsilon _jk_{ix}-\epsilon
_ik_{jx}}{\epsilon _jk_{ix}+\epsilon _ik_{jx}}, \label{eq3-13}
\end{equation}
depending on whether we are calculating $\tilde R_{12}$ for a TE
wave or a TM wave.

If there are subsurface layers below region 3, $R_{23}$ in
(\ref{eq3-12}) can be replaced with $\tilde R_{23}$, or
\begin{equation}
\tilde R_{12}=\frac {R_{12}+\tilde
R_{23}e^{2ik_{2x}t_2}}{1-R_{21}\tilde R_{23}e^{2ik_{2x}t_2}}.
\label{eq3-14}
\end{equation}

\begin{figure}
\begin{center}
\hfil\includegraphics[width=3.0truein]{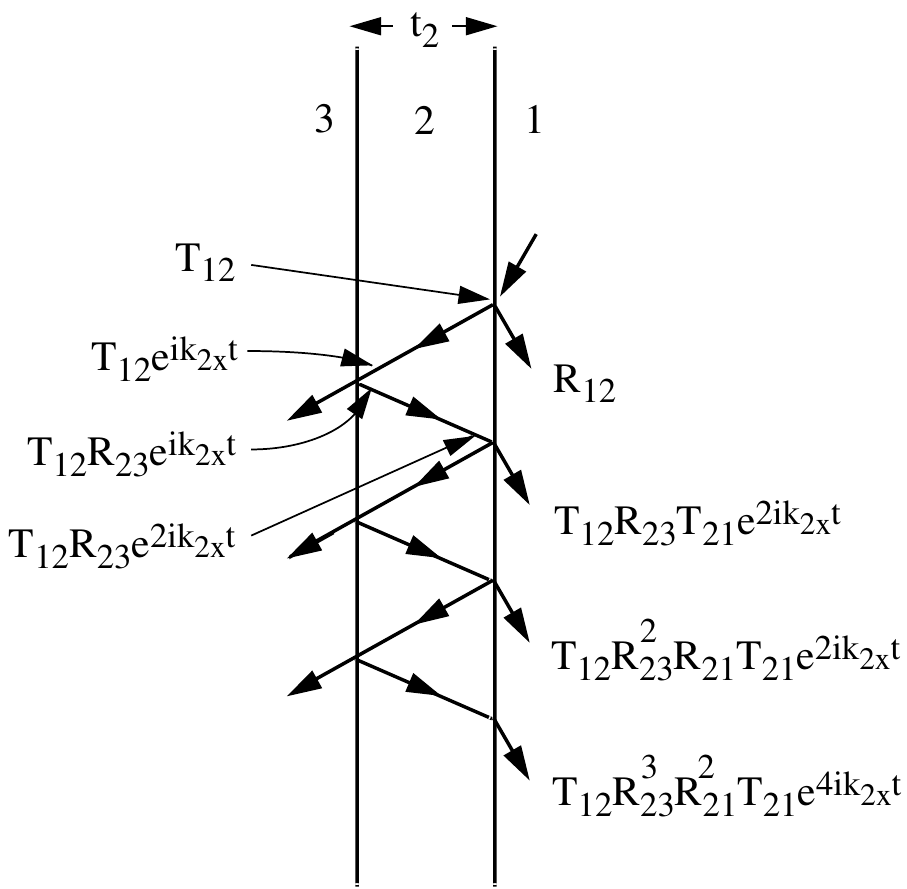}\hfil
\end{center}
\caption{Multiple reflections and transmissions through a
dielectric slab.}\label{fg333}
\end{figure}


The above is a recursive relation from which one can calculate
$\tilde R_{ij}$ for any number of layers.  For example, if there
is a metallic wall at $x = 0$, as in Figure \ref{fg332},
$R_{34}^{TE}=-1$, $R_{34}^{TM}=1$, then
\begin{equation}
\tilde R_{23}^{TE}=\frac
{R_{23}^{TE}-e^{2ik_{3x}t_3}}{1+R_{32}^{TE}e^{2ik_{3x}t_3}},
\qquad \tilde R_{23}^{TM}=\frac
{R_{23}^{TM}+e^{2ik_{3x}t_3}}{1-R_{32}^{TM}e^{2ik_{3x}t_3}}.
\label{eq3-15}
\end{equation}


\begin{figure}
\begin{center}
\includegraphics[height=2.5in,width=3.5truein]{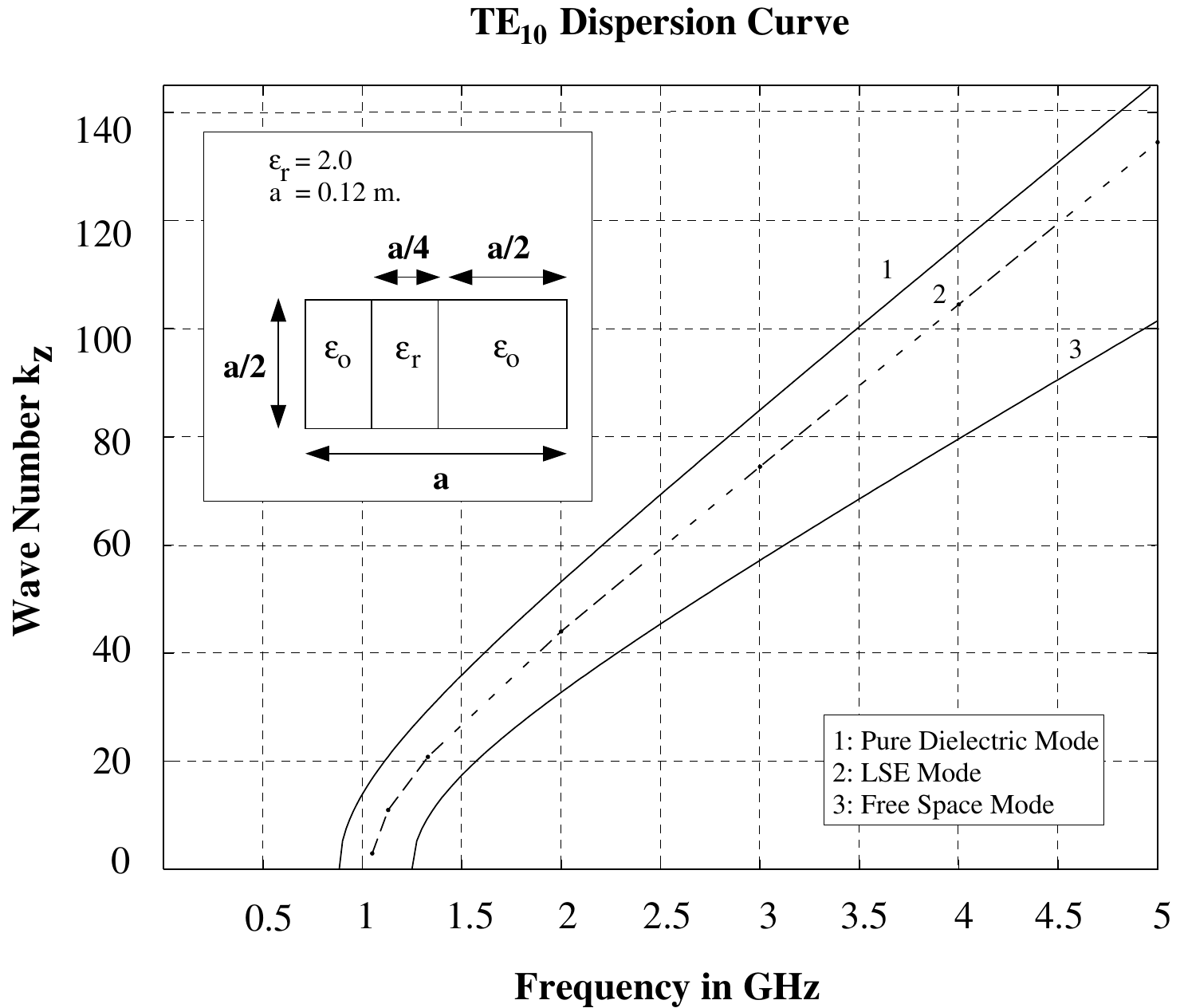}
\end{center}
\begin{center}
 (a)
\end{center}
\begin{center}
\includegraphics[height=2.5in,width=3.5truein]{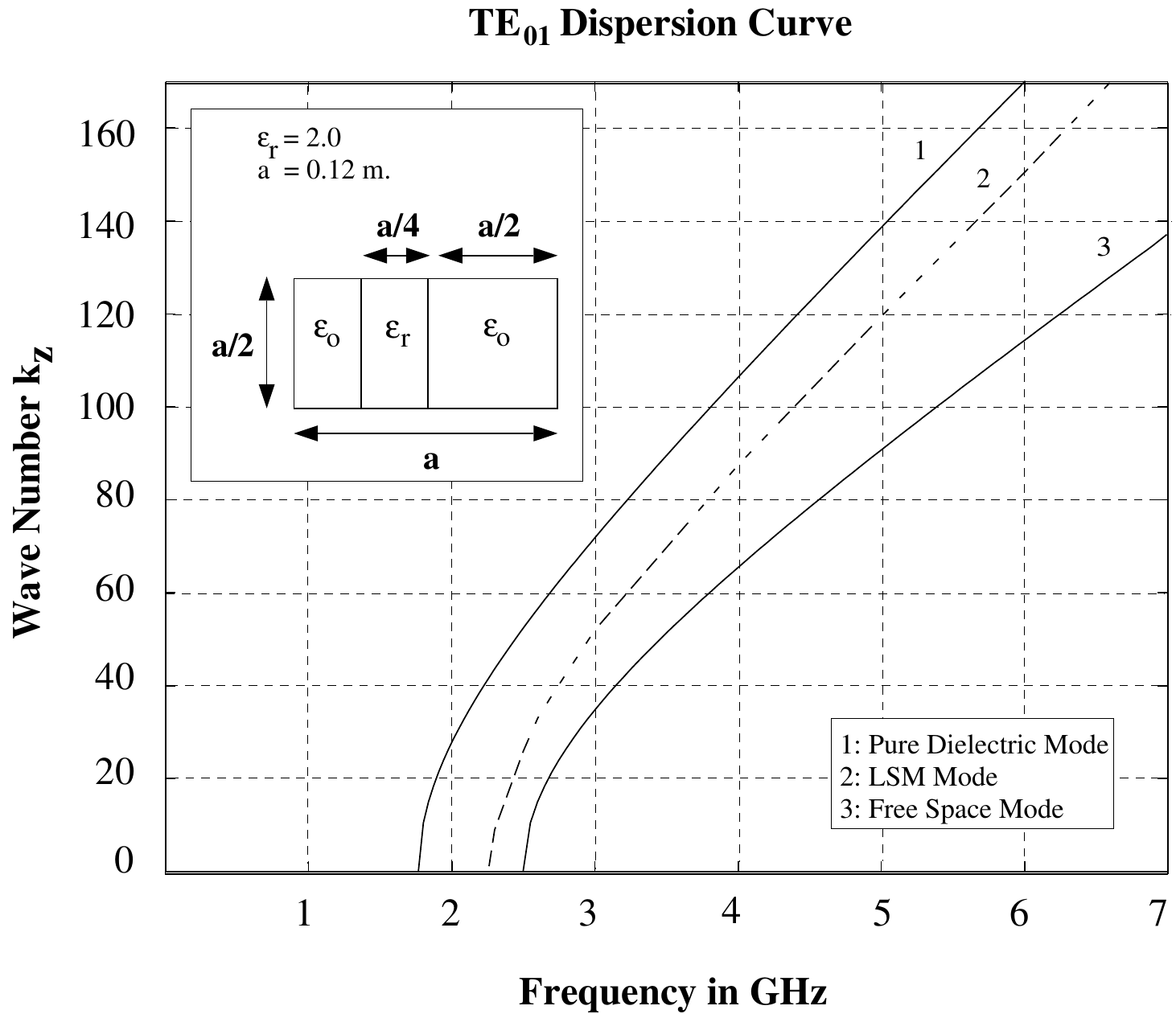}
\end{center}
\begin{center}
  (b)
\end{center}
\caption{Dispersion curves for various dielectric loading for (a)
TE$_{10}$ mode, and (b) TE$_{01}$ mode.}\label{fg334b}
\end{figure}


\begin{figure}
\begin{center}
\hfil\includegraphics[height=3.0in,width=4.0truein]{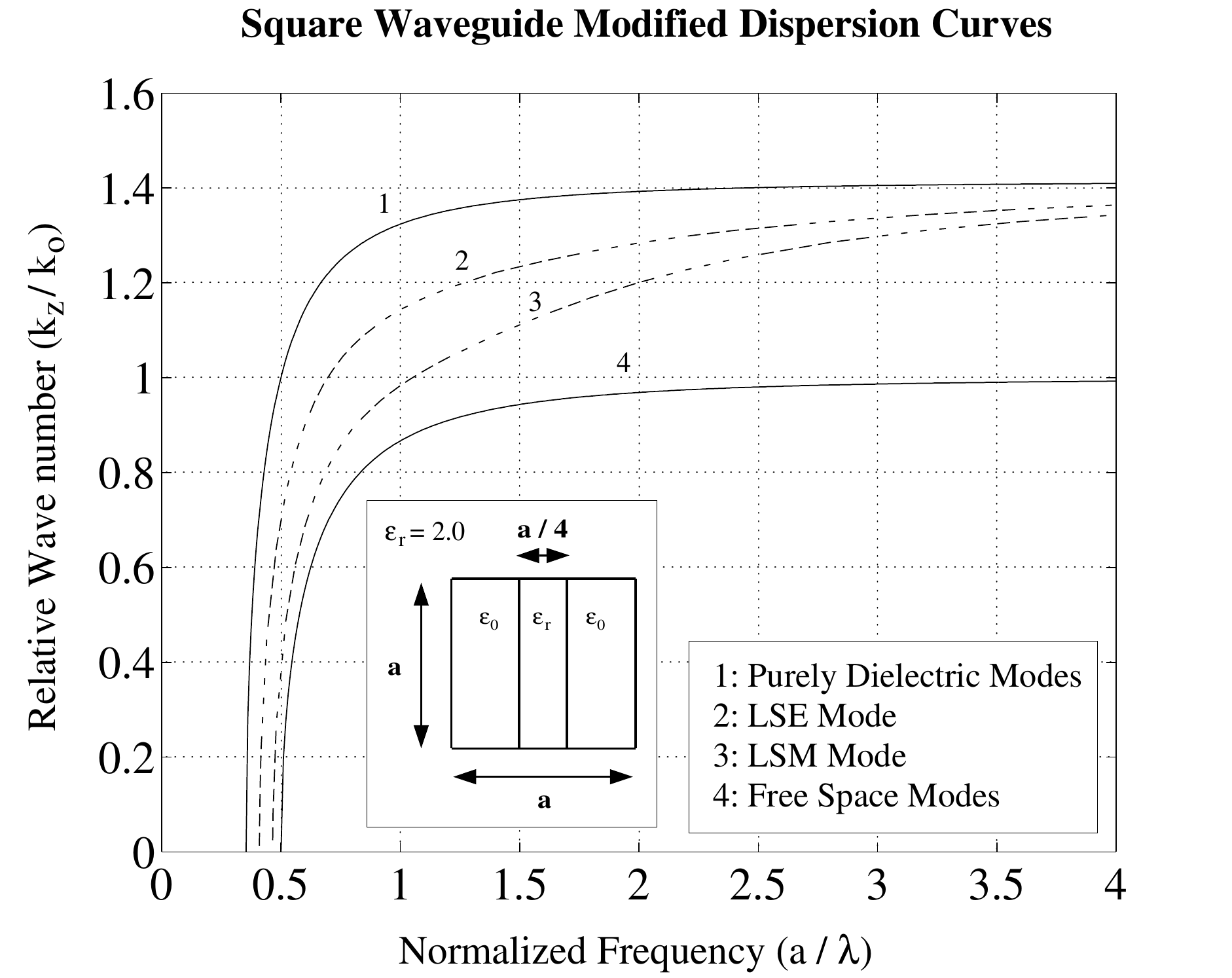}\hfil
\end{center}
\caption{The dispersion curves for a slab-loaded square waveguide
showing anisotropy.}\label{fg335}
\end{figure}


With $\tilde R_{12}$ defined by (\ref{eq3-14}) and (\ref{eq3-15}),
Equations (\ref{eq3-10a}) and (\ref{eq3-10b}), in general, have to
be solved numerically with a root solver like the Newton-Raphson
method, or the Muller's method. All the $R_{ij}$'s are defined in
terms of $k_{ix}$ in Equation (\ref{eq3-13}), where
\begin{equation}
k_{ix}=\sqrt {k_i^2-k_z^2-\left(\frac {n\pi }b\right)^2}.
\label{eq3-16}
\end{equation}
Hence, Equations (\ref{eq3-10a}) and (\ref{eq3-10b}) can be solved
for values of $k_z$. For simple geometry, graphical solutions to
Equations (\ref{eq3-10a}) and (\ref{eq3-10b}) may be found.

The guidance condition given by Equations (\ref{eq3-10a}) and
(\ref{eq3-10b}) are also known as the \index{Transverse resonance condition}transverse resonance
condition which will be discussed in greater detail in the next
section. The guidance conditions in (\ref{eq3-10a}) and
(\ref{eq3-10b}) are obtained by considering waves bouncing in the
air region. If a guided mode is trapped in the dielectric slab
region, the wave becomes evanescent in the air region and
Equations (\ref{eq3-10a}) and (\ref{eq3-10b}) can become
ill-posed. To remedy this, it is better to write down the guidance
condition in the slab region using the transverse resonance
condition described in the next section.

Figure \ref{fg334b} shows the dispersion curves of the TE$_{10}$
and TE$_{01}$ (with respect to $z$) modes of a slab loaded
dielectric waveguide. Case 1 is when the whole waveguide is filled
with dielectric material, and case 3 is when the waveguide is
empty. Notice that for the perturbed modes (case 2), the
dispersion curve is parallel to case 1, The reason is that the
mode is entrapped in the dielectric slab for high frequencies, and
the group velocity of the mode ($d\omega /dk_z$) approaches that
of the dielectric slab.

Also, notice that at lower frequencies, the TE$_{10}$ mode is
perturbed more by the slab than the TE$_{01}$ mode, because for
the TE$_{01}$ mode, the electric field is normal to the slab while
for the TE$_{10}$ mode, the electric field is parallel to the
slab. This fact can be used to create anisotropy in a
\index{Waveguide!symmetric}symmetric
waveguide like a circular waveguide or a square waveguide. Figure
\ref{fg335} shows the dispersion curves for the LSE mode, which is
the perturbed TE$_{10}$ mode, and the LSM mode, which is the
perturbed TE$_{01}$ mode. Notice that the LSE mode is affected
more by the dielectric slab than the LSM mode. This anisotropic
effect can be used to make a quarter-wave plate out of a
dielectric slab loaded waveguide. This will be discussed in
greater detail later.

\section {{{ Transverse Resonance Condition}}}
\index{Transverse resonance condition}

The guidance conditions given by (\ref{eq3-10a}) and
(\ref{eq3-10b}) can also be derived by the transverse resonance
condition. The transverse resonance condition is a powerful
condition that can be used to derive the \index{Layered medium!guidance condition}guidance condition of a
mode in a \index{Layered medium}layered medium.

\begin{figure}
\begin{center}
\hfil\includegraphics[width=4.5truein]{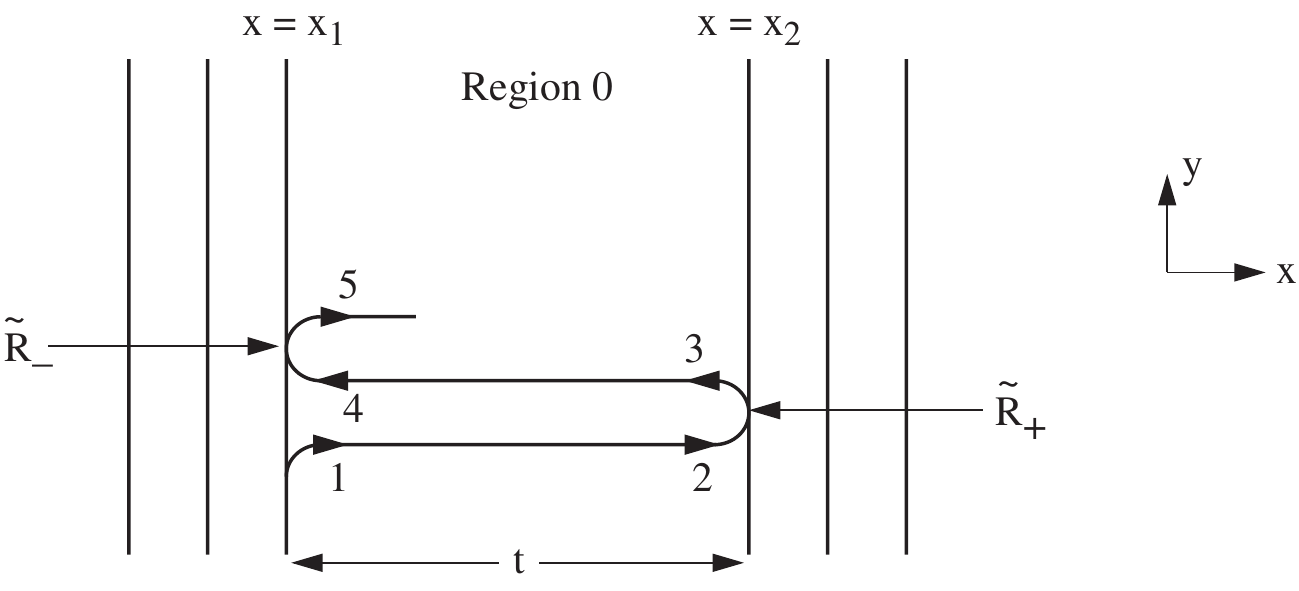}\hfil
\end{center}
\caption{The transverse resonance condition for a layered medium.
The phase of the wave at position 5 should be equal to the phase
at position 1.}\label{fg341}
\end{figure}


To derive this condition, we first have to realize that a guided
mode in a waveguide is due to the coherent or constructive
interference of the waves. This implies that if a plane wave
starts at position 1 and is multiply reflected as shown, it will
regain its original phase in the $x$ direction at position 5.
Since this mode progresses in the $y$ direction. it will gain a
phase in the $y$ direction. But, for it to coherently interfere in
the $x$ direction, the transverse phase at 5 must be the same as
1.

Assuming that the wave starts with amplitude 1 at position 1, it
will gain a transverse phase of $e^{ik_{0x}t} $ when it reaches
position 2. Upon reflection at $x=x_2$, at position 3, the wave
becomes $\tilde R_+ e ^{ik_{0x}t}$. Finally, at position 5, it
becomes $\tilde R_-\tilde R_+e ^{2ik_{0x}t }$. For constructive
interference to occur or for the mode to exist, we require that
\begin{equation}
\tilde R_-\tilde R_+e ^{2ik_{0x}t} =1  . \label{eq4-1000}
\end{equation}
The above is the transverse resonance condition. It is also the
\index{Guidance condition}guidance condition for a mode travelling in a layered medium.

In (\ref{eq3-10a}), a metallic wall has a reflection coefficient
of 1 for a TM wave, hence if $\tilde R_+ $ is 1, Equation
(\ref{eq4-1000}) becomes
\begin{equation}
1-\tilde R_-e ^{2ik_{0x}t} =0. \label{eq4-2000}
\end{equation}
In (\ref{eq3-10b}), a metallic wall has a reflection coefficient
of $-1$, and Equation (\ref{eq4-1000}) becomes
\begin{equation}
1+\tilde R_- e^{2ik_{0x}t} =0. \label{eq4-3000}
\end{equation}

\section{Fabry-Perot Etalon}
\index{Fabry-Perot etalon}
Since we have the machinery in place, it is convenient to study the
Fabry-Perot etalon which is often used as an optical filter
\cite{YARIV}. By tracing the plane wave or ray through the slabs, we
can show that the \index{Generalized transmission coefficient}generalized transmission coefficient from region 1
to region 3 is
 \begin{align}\label{fbeq:1}
 \tilde{T}_{13}=\frac{T_{12}T_{23}e^{ik_{2x}d}}{1-R_{23}R_{21}e^{2ik_{2x}d}}
 \end{align}
where $d$ is the thickness of the slab. In etalon application,
regions 1 and 3 are free space while region 2 is dielectric.
Specializing to the case when regions 1 and 3 have the same
parameters, we have
 \begin{align}\label{fbeq:2}
 \tilde{T}=\frac{T_{12}T_{21}e^{ik_{2x}d}}{1-R_{21}^2e^{2ik_{2x}d}}
 \end{align}
Since $T_{12}=1+R_{12}$, $T_{21}=1+R_{21}=1-R_{12}$, we have
 \begin{align}\label{fbeq:3}
 \tilde{T}=\frac{(1-R_{21}^2)e^{ik_{2x}d}}{1-R_{21}^2e^{2ik_{2x}d}}
 \end{align}

\begin{figure}
\begin{center}
\includegraphics[width=.7\textwidth]{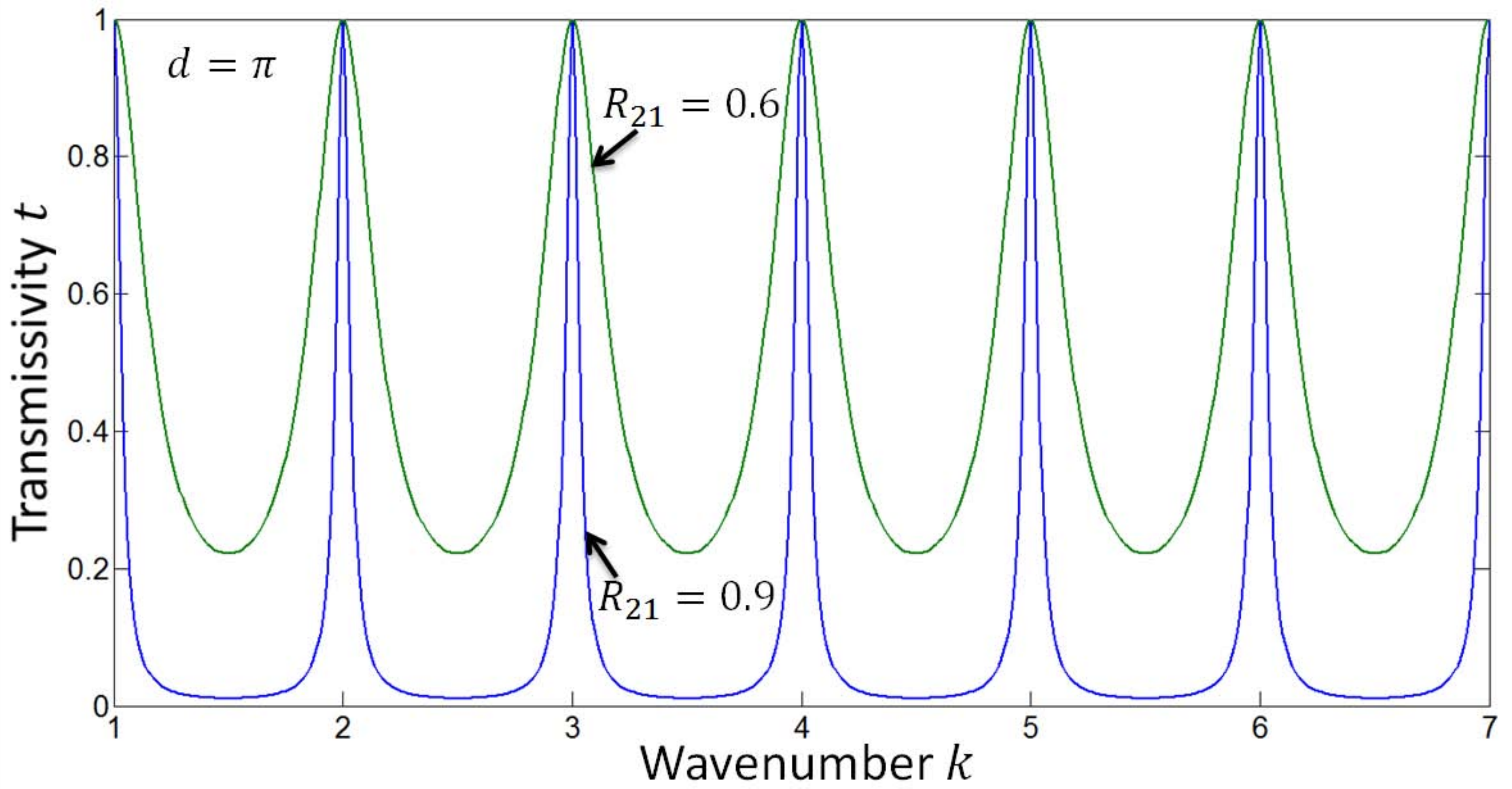}
\end{center}
\caption{The transmissivity of the etalon as a function of
frequency. }\label{EtalonTransmissivity}
\end{figure}

 For normally incident wave, $k_{2x}=k_2$, and the above becomes
 \begin{align}\label{fbeq:4}
 \tilde{T}=\frac{(1-R_{21}^2)e^{ik_{2}d}}{1-R_{21}^2e^{2ik_{2}d}}
 \end{align}
where
\begin{equation}
R_{21} = \frac{k_{2}-k_{1}}{k_{2}+k_{1}} = \frac{\sqrt{\epsilon_{2}}
- \sqrt{\epsilon_{1}}}{\sqrt{\epsilon_{2}} + \sqrt{\epsilon_{1}}}
\end{equation}
assuming that $ \mu_{1} = \mu_{2} = \mu$ . The transmissivity is
\begin{equation}
t = |{\tilde{T}}|^{2} =
\frac{|1-R_{21}^{2}|^{2}}{|1-R_{21}^{2}e^{2ik_{2}t}|^{2}}
\end{equation}
The above is maximum with $t=1$ when $e^{2ik_{2}d} =1$ or $2k_{2}d =
2 m \pi$ with integer $m$.  In other words, at such a frequency, the
etalon is transparent with no reflected wave.  Specializing
\eqref{eq3-14} to this case, and using the fact that
$R_{12}=-R_{21}$,
\begin{equation}
\tilde R=\frac {R_{12}\left(1-e^{2ik_{2}d}\right)}{1-R_{21}^2
e^{2ik_{2}d}}. \label{eq3-14a}
\end{equation}
It is seen that the above can be zero if $2k_{2}d = 2 m \pi$.  The
zero comes about because of the destructive interference of the
reflected waves.  The transmissivity of the etalon as a function of
frequency is shown in Figure \ref{EtalonTransmissivity}.

Next, we can perform the \index{Pole analysis}pole analysis of the etalon.  In the
vicinity of the maxima, $k_2 d-m\pi$ is small, and
\begin{align}\label{fbeq:7}
  e^{2ik_{2}d}=e^{2ik_{2}d-2im\pi}\approx 1+\frac{2id}{c_2}\left(\omega-\omega_m\right)=1+\frac{2id}{c_2}\Delta\omega
 \end{align}
where $\Delta\omega=\omega-\omega_m$,
 \begin{align}\label{fbeq:8}
  \omega_m=\frac{m\pi c_2}{d}
 \end{align}
and
 \begin{align}\label{fbeq:8a}
  c_2=\frac{1}{\sqrt{\mu \epsilon_2}}
 \end{align}
Hence,
\begin{equation}
\tilde{T} \approx
\frac{1-R_{21}^{2}}{1-R_{21}^{2}-R_{21}^{2}2id\frac{\Delta
\omega}{c_{2}}}
\end{equation}
near the maxima. The pole locations can be found to be when
\begin{equation}
\Delta \omega \cong
\frac{1-R_{21}^{2}}{R_{21}^{2}}\left(\frac{-ic_{2}}{2d}\right)
\end{equation}
Hence, the pole location for the $m$-th resonance mode in the
complex plane is
\begin{equation}
\omega_{pm} \simeq \omega_{m} - \frac{ic_{2}}{2d}\frac
{1-R_{21}^{2}}{R_{21}^{2}} = \frac{m \pi c_{2}}{d}-
\frac{ic_{2}}{2d}\frac{1-R_{21}^{2}}{R_{21}^{2}}
\end{equation}

\begin{figure}
\begin{center}
\includegraphics[width=5.50truein]{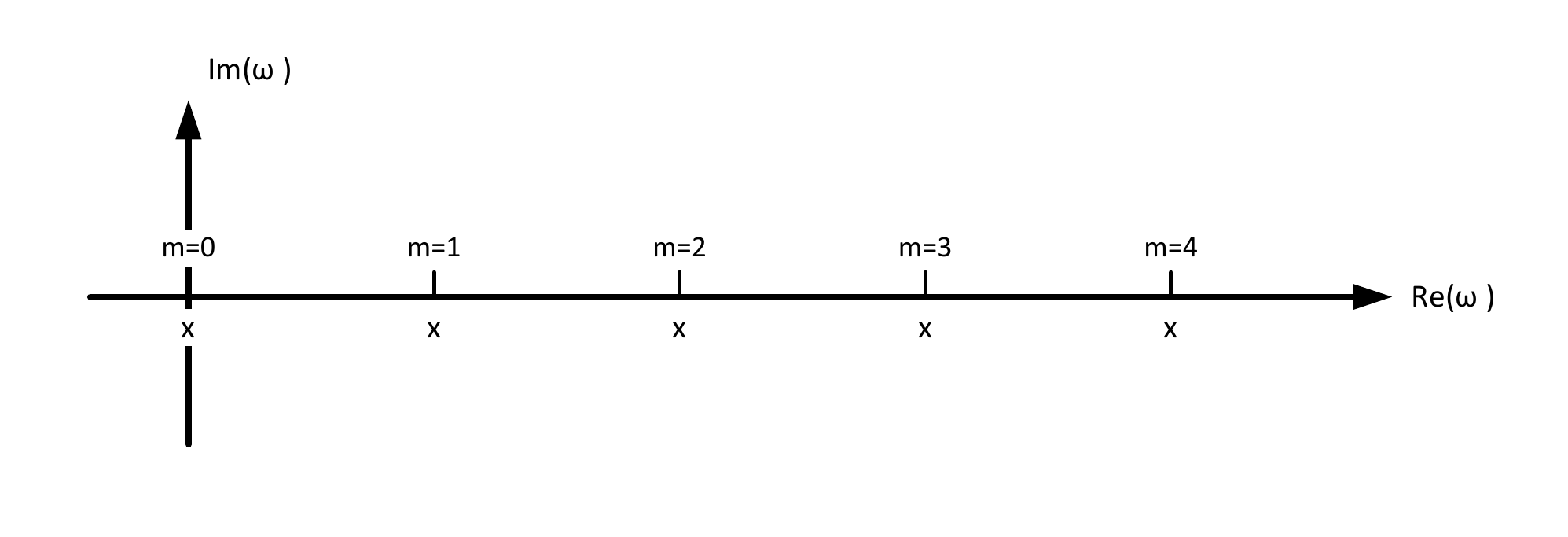}
\end{center}
\caption{The locations of the poles in the complex $\omega$ plane
for the Fabry-Perot modes.}\label{RealImagAxis}
\end{figure}

The above is accurate when $1- R_{21}^{2}$ is small. It is slightly
below the real $\omega$ axis which is in agreement with a lossy
resonant mode with $ e^{-i\omega t}$ time convention. Notice that
the imaginary part of the pole location is independent of frequency
(See Figure \ref{RealImagAxis}).


The mechanism for transmission here is \index{Resonance tunneling}resonance tunneling. The
Fabry-Perot etalon is used as an \index{Optical filter}optical filter. To obtain a narrow
band filter, the $Q$'s of these modes have to be high. The $Q$ of a
resonator is
\begin{equation}
    Q=\frac{\omega_{r}W_{T}}{P_{d}}
\end{equation}
where
$$\frac{W_{T}}{P_{d}}=T$$
is the time constant for which the energy store $W_{T}$ will
diminish to ${e}^{-1}$ of its original value. So $Q=\omega_{r}T$ is
also the number of cycles in radian for this decay to happen. $Q$ is
an asymptotic concept that has meaning only when $Q$ is large. When
the pole location is known,
$\omega_{p}=\omega_{p}^{'}+i\omega_{p}^{''}$, then
\begin{equation}
    Q=-\frac{\omega_{p}^{'}}{2\omega_{p}^{''}}
\end{equation}
Applying the above to the $m$-th mode in (11), we have
\begin{equation}
    Q_{m}=\frac{m\pi R_{21}^{2}}{(1-R_{21}^{2})}
\end{equation}


\section {{{ Rod-Loaded Circular Waveguide}}}
\index{Waveguide!rod-loaded circular}

When a circular waveguide is loaded with a concentric circular
dielectric rod, the equation for the guidance condition can be
found in closed form.  \index{Ferrite rod}Ferrite rods are usually used to load a
circular waveguide to engender \index{Faraday rotation}Faraday rotation.  Since ferrites
are anisotropic, the analysis of a ferrite loaded circular
waveguide is in general very complex. Hence, we will analyze the
case of a circular waveguide loaded with a circular dielectric
rod.   The general case of a waveguide periodically loaded with
dielectric rods can be used to make microwave filters
\cite{AMARI_VAHLDIECK_BORNEMANN}.

\subsection { Reflection off a Dielectric Rod}

Since $E_z$ and $H_z$ waves are in general coupled in uniform
dielectric rod, we have to consider both polarizations together in
this reflection problem.  We will follow an analysis presented in
\cite{WFIM}.   The $z$-components of the fields satisfy
\begin{equation}
\left(\nabla ^2 +k^2\right)
\begin{bmatrix}
 E_z  \\
 H_z
\end{bmatrix}
    =0   .
\label{eq5-1000}
\end{equation}
Assuming $e^{ik_z z+in\phi }$ dependence in the wave, the equation
becomes
\begin{equation}
\left( \frac {1} {\rho }\frac {\partial }{\partial \rho } \rho
\frac {\partial }{\partial  \rho } +\frac {n^2}{{\rho }^2}
+k^2-k^2_z \right)
\begin{bmatrix}
  E_z \\
  H_z
\end{bmatrix}
    =0  .
\label{eq5-2}
\end{equation}
The general solution to the above is of the form
\begin{equation}
\begin{bmatrix}
 E_z \\
 H_z
\end{bmatrix}
=[\v a_n J_n (k_{\rho }\rho)+\v b_n H^{(1)}_n (k_{\rho} \rho ) ]
 e^{ik_zz+in\phi },
\label{eq5-3}
\end{equation}
where $ k_{\rho}=\sqrt{ k^2-k^2_z}$ , and $ H^{(1)}_n (x)$ is the
Hankel function of the first kind.

We can assume an incident field on a dielectric rod in region 0 as
\begin{equation}
\v f ^i_{0z}=
\begin{bmatrix}
 E_{0z}^i \\
 H_{0z}^i
\end{bmatrix}
=\v a _{0n} J_n (k_{0\rho}\rho) \label{eq5-4}
\end{equation}
where $ k_{0 \rho}=\sqrt{k^2_0-k^2_z}$ , and the $e^{ik_z z+in\phi
}$ dependence is implied. The above is the incident wave in the
absence of the dielectric rod. Hence, it cannot have a \index{Hankel wave}Hankel wave
for it would be singular at the origin. When a circular dielectric
rod is put at  the origin, it will reflect the incident wave
generating an outgoing wave satisfying the radiation condition at
infinity. Therefore, The total solution in region 0 must be of the
form
\begin{equation}
\v f_{0z}=
\begin{bmatrix}
 E_{oz} \\
 H_{0z}
\end{bmatrix}
=\v a _{0n} J_n (k_{0\rho}\rho)+\v b _{0n}
H^{(1)}_n(k_{0\rho}\rho). \label{eq5-5000}
\end{equation}
Since $\v b_n$ must be linearly dependent on $\v a_n$ , we express
\begin{equation}
\v b _{0n}=\dyad R_{01}\cdot \v a_{0n} \label{eq5-6000}
\end{equation}
In region 1, the solution can only admit \index{Bessel wave}Bessel waves since a
Hankel wave is singular at the origin. Therefore, the general
solution is of the form

\begin{figure}
\begin{center}
\hfil\includegraphics[height=2.0truein,width=2.8truein]{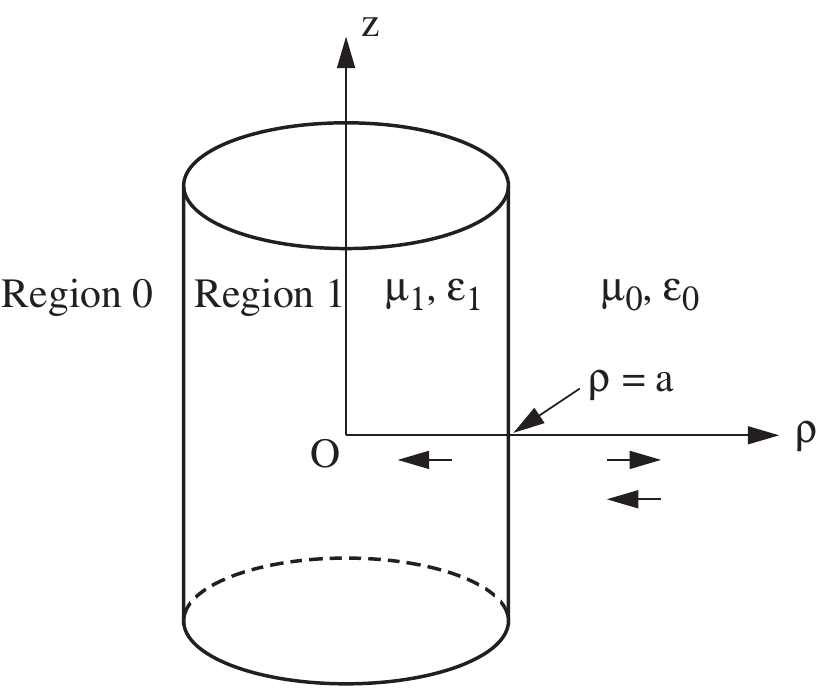}\hfil
\end{center}
\caption{Geometry for defining the reflection of a wave off a
circular dielectric rod.}\label{fg351}
\end{figure}


\begin{equation}
\v f_{1z}=
\begin{bmatrix}
 E_{1z} \\
 H_{1z}
\end{bmatrix}
=\v a _{1n} J_n (k_{1\rho }\rho ). \label{eq5-7000}
\end{equation}
Since $\v a_{n}$ is linearly related to $\v a_{0n}$, we let
\begin{equation}
\v a_{1n}=\dyad T_{01} \cdot \v a _{0n}. \label{eq5-8}
\end{equation}
The transverse to $z$ components of the field can be obtained from
the following equations in the $j$-th region
\begin{subequations}
\begin{equation}
\v E_{js} =\frac 1 {k^2_{j\rho}} [ik_z\nabla _s E_{jz} -i\omega
\mu \^z \times \nabla _s H_{jz}], \label{eq5-9a}
\end{equation}
\begin{equation}
\v H_{js} =\frac 1 {k^2_{j\rho}} [ik_z\nabla _s H_{jz} +i\omega
\epsilon \^z \times \nabla _s E_{jz}], \label{eq5-9b}
\end{equation}
\end{subequations}
where $\nabla _s =\^{\rho} \frac {\partial }{\partial \rho}
+\^{\phi} \frac{1}{\rho} \frac {\partial }{\partial \phi }
=\^{\rho} \frac {\partial }{\partial\rho } +\^{\phi} \frac
{in}{\rho}$, and $k_{j{\rho }}^2=k^2_j-k^2_z$. From the
phase-matching condition, $k_z$ is the same in all regions.

In order to find $\dyad R_{01} $ and $\dyad T_{01}$ in
(\ref{eq5-6000}) and (\ref{eq5-8}), we need to match the continuity
of the tangential components of $\v E$ and $\v H$ across the
interface. These tangential components are the $z$ and the $\phi$
components. The $z$ components are already given in (\ref{eq5-5000})
and (\ref{eq5-7000}). The corresponding $\phi$ components can be
derived using (\ref{eq5-9a}) and (\ref{eq5-9b}). From
(\ref{eq5-5000}) and (\ref{eq5-9a}) and (\ref{eq5-9b}), we deduce
that
\begin{equation}
\v f_{0\phi}=
\begin{bmatrix}
 H_{0\phi} \\
 E_{0\phi}
\end{bmatrix}
=\dyad J_n (k_{0\rho }\rho )\cdot \v a_{0n}+ \dyad H^{(1)}_n
(k_{0\rho } \rho )\cdot \dyad R_{01}\cdot \v a _{0n}
\label{eq5-10}
\end{equation}
\begin{equation}
\v f_{1\phi}=
\begin{bmatrix}
 H_{1\phi} \\
 E_{1\phi}
\end{bmatrix}
=\dyad J_n (k_{0\rho }\rho )\cdot \dyad T_{01} \cdot \v a _{0n} ,
\label{eq5-11}
\end{equation}
where
\begin{equation}
\dyad B_n (k_{j\rho} \rho)= \frac 1 {k^2_{j\rho}\rho}
\begin{bmatrix}
 i\omega\epsilon _j k_{j\rho} \rho B'_n (k_{j\rho}\rho)
 & -nk_z B_n (k_{j\rho}\rho)\\
  -nk_z B_n (k_{j\rho}\rho)
 & -i\omega\mu _j k_{j\rho} \rho B'_n (k_{j\rho}\rho)
\end{bmatrix}.
\label{eq5-12}
\end{equation}
In the above, $B_n$ is either $H^{(1)}_n $ or $J_n$ depending on
if we are defining $\dyad H^{(1)}_n $ or $\dyad J_n $ matrix. Note
that $\dyad B_n $ is diagonal when $n=0$. This also implies the
decoupling of $E_z$ and $H_z$ waves when $n=0$ .

Matching the boundary condition at the boundary where $\rho =a$,
we have
\begin{subequations}
\begin{equation}
[J_n (k_{0\rho }a ) +H^{(1)}_n (k_{0\rho} a)\dyad R_{01}]\cdot \v
a_{0n}= J_n (k_{1\rho }a ) \dyad T_{01}\cdot \v a_{0n},
\label{eq5-13a}
\end{equation}
\begin{equation}
[\dyad J_n (k_{0\rho }a ) +\dyad H^{(1)}_n (k_{0\rho} a)\cdot\dyad
R_{01}]\cdot \v a_{0n}=\dyad J_n (k_{1\rho }a ) \cdot \dyad
T_{01}\cdot \v a_{0n}, \label{eq5-13b}
\end{equation}
\end{subequations}
The above can be solved to yield
\begin{subequations}
\begin{equation}
\dyad R_{01}=\dyad D^{-1}\cdot [J_n (k_{1\rho }a ) \dyad J_n
(k_{0\rho } a)
 - J_n (k_{0\rho }a ) \dyad J_{n}(k_{1\rho} a)],
\label{eq5-14a}
\end{equation}
\begin{equation}
\dyad T _{01} =\frac {2\omega }{\pi k^2_{0\rho} a } \dyad D^{-1}
  \cdot
\begin{bmatrix}
 \epsilon _0 & 0\\
  0   & -\mu _0
\end{bmatrix},
\label{eq5-14b}
\end{equation}
where
\begin{equation}
\dyad D= \left[\dyad J_n (k_{1\rho }a ) H^{(1)}_n (k_{0\rho} a)-
\dyad H_n^{(1)} (k_{0\rho }a )  J_n (k_{1\rho} a) \right].
\label{eq5-14c}
\end{equation}
\end{subequations}
The \index{Wronskian for Hankel function}Wronskian for Hankel function, which is
\begin{equation}
H^{(1)}_n (x)J'_n(x)-J_n(x)H^{(1)'}_n (x)=- \frac {2i}{\pi x }
\label{eq5-15}
\end{equation}
has been used to simplify the above.

In general $\dyad R_{01}$ and $\dyad T_{01}$ are non-diagonal
implying the coupling of the $E_z$ and $H_z$ waves by the
dielectric rod.

\subsection { Reflection off a PEC Waveguide Wall}
\index{Reflection off PEC waveguide wall}

We have asserted that a PEC or a PMC cylindrical surface does not
depolarize an $E_z$ or $H_z$ wave, We can further confirm this
assertion by looking at Equation (\ref{eq5-9a}). For a PEC, we
require that $E_z=0$ and that $\^n \times \v E_s =0$ on the
cylindrical surface. It is seen that if we set $E_z=0$ on a
surface, then $\^n \times \nabla E_z =0$ or the first term in
(\ref{eq5-9a}) is zero without any help from the second term in
(\ref{eq5-9a}). Therefore, the $E_z$ wave alone can satisfy the
boundary condition on a PEC independently of the $H_z$ wave. In
order for $\^n \times \v E_s =0 $ for the $H_z$ wave, we require
that
\begin{subequations}
\begin{equation}
\^n \times \^z \times \nabla _s H_z =0, \label{eq5-16a}
\end{equation}
or that
\begin{equation}
\^n \cdot \nabla _s H_z =0 . \label{eq5-16b}
\end{equation}
\end{subequations}
In other words, if we impose the homogeneous Neumann boundary
condition (\ref{eq5-16b}) on a PEC cylindrical surface, the
tangential electric field that arises from the $H_z $ wave will
satisfy the requisite boundary  condition independently of the
$E_z$ wave.

As a result of the above discussion, when an outgoing Hankel Wave
impinges on a PEC waveguide wall which is circular, it reflects
back into a Bessel wave. Hence, the wave in region 0 can be
written as
\begin{equation}
\v f_{0z} =
\begin{bmatrix}
 E_{0z}\\
 H_{0z}
\end{bmatrix}
 = H^{(1)}_n (k_{0\rho }\rho ) \v b _{0n} +J_n(k_{0\rho }\rho ) \dyad R_{02}\cdot \v b _{0n}
\label{eq5-17}
\end{equation}
where $\dyad R_{02}$ is a diagonal $2 \times 2 $ matrix due to the
decoupling of the $E_z$ and $H_z$ waves. Matching the requisite
boundary condition on the waveguide wall at $\rho =b$, we obtain
that
\begin{equation}
\dyad R_{02}=
\begin{bmatrix}
 -H^{(1)}_n (k_{0\rho }b)/J_n(k_{0\rho }b) & 0\\
 0      & -H^{(1)'}_n (k_{0\rho }b)/J'_n(k_{0\rho }b)
\end{bmatrix}.
\label{eq5-18}
\end{equation}
Note that if region 2 is a dielectric region, the above will be a
non-diagonal matrix as shown in the next subsection.

\begin{figure}
\begin{center}
\hfil\includegraphics[height=1.7truein,width=2.0truein]{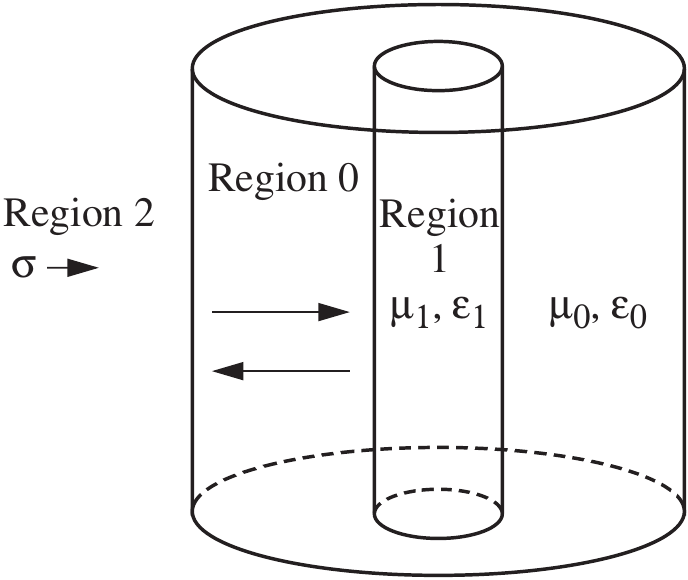}\hfil
\end{center}
\caption{Geometry of a dielectric-rod-loaded circular waveguide.}\label{fg352}
\end{figure}


\subsection { Reflection off an Outer Dielectric Wall}
\index{Outer dielectric wall}

If region 2 is a dielectric region rather than a perfectly
conducting region, a \index{Reflection matrix}reflection matrix can be similarly derived as
in Subsection 3.5.1. In this case, we can show that
\begin{subequations}
\begin{equation}
\dyad R_{02}=\dyad D^{-1} \cdot [H_n^{(1)}(k_{0\rho }a )\dyad
H_n^{(1)} (k_{2\rho }a ) - H_n^{(1)}(k_{2\rho }a )\dyad
H_n^{(1)}(k_{0\rho }a )] \label{eq5-19a}
\end{equation}
\begin{equation}
\dyad T_{02}=\frac {2\omega }{\pi k^2_{0\rho }a}\dyad D^{-1}\cdot
\begin{bmatrix}
 \epsilon _0 & 0\\
 0           & -\mu _0
\end{bmatrix}.
\label{eq5-19b}
\end{equation}
where
\begin{equation}
\dyad D=[\dyad J_n(k_{0\rho }a )H_n^{(1)} (k_{2\rho }a ) - \dyad
H_n^{(1)}(k_{2\rho }a )J_n(k_{0\rho }a )]. \label{eq5-19c}
\end{equation}
\end{subequations}

\subsection { The Guidance Condition}
\index{Guidance condition}

The guidance condition in a dielectric-rod-loaded circular
waveguide can be obtained by considering the solution in region 0
and matching boundary condition on the waveguide wall at $\rho =b
$. The solution in region 0 is given by (\ref{eq5-5000}) is rewritten
here as
\begin{equation}
\v f_{0z} =[J_n (k_{0\rho} \rho ) \dyad I+ H^{(1)}_n (k_{0p}\rho )
\dyad R _{01}]\cdot \v a _{0n}. \label{eq5-20}
\end{equation}

\begin{figure}
\begin{center}
\hfil\includegraphics[width=1.5truein]{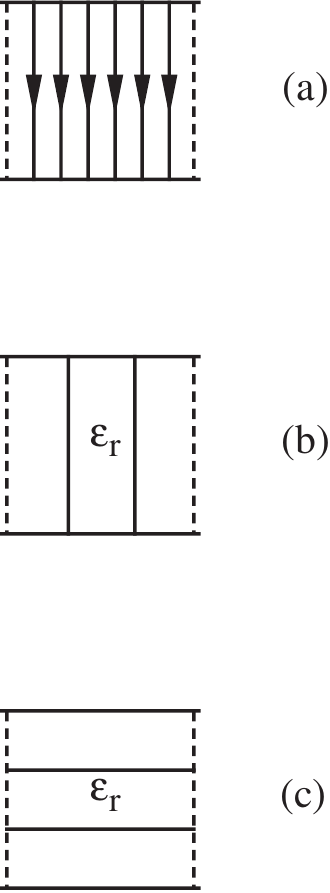}\hfil
\end{center}
\caption{Different dielectric-slab loading of a parallel plate waveguide.}\label{fg1.1}
\end{figure}

However, according to (\ref{eq5-17}), the field in region 0 can
also be written as
\begin{equation}
\v f_{0z} =[J_n (k_{0\rho} \rho )\dyad R_{02} + H^{(1)}_n
(k_{0\rho }\rho ) \dyad I]\cdot \v b _{0n}. \label{eq5-21}
\end{equation}
Hence, we conclude that
\begin{subequations}
\begin{equation}
\dyad R_{02} \cdot \v b_{0n} =\v a _{0n}, \label{eq5-22a}
\end{equation}
\begin{equation}
\dyad R_{01} \cdot \v a_{0n} =\v b _{0n}, \label{eq5-22b}
\end{equation}
\end{subequations}
or that
\begin{equation}
(\dyad R_{02} \cdot \dyad R_{01}-\dyad I ) \cdot \v a_{0n} =0 .
\label{eq5-23}
\end{equation}
In order for $\v a_{0n} \ne 0$, we require that
\begin{equation}
\rm {det}(\dyad R_{02} \cdot \dyad R_{01} -\dyad I)=0 .
\label{eq5-24}
\end{equation}
The above is the guidance condition or the transverse resonance
condition for a cylindrically layered circular waveguide.

\section {{{ Applications of Inhomogeneously Filled Waveguides}}}
\index{Waveguide!inhomogeneously filled}

Inhomogeneous filled waveguides can be used to make variable phase
shifters and attenuators \cite{COLLIN,COLLIN2}. When the
inhomogeneity filling the waveguide is nonreciprocal like ferrite,
isolators, gyrators, and \index{Attenuators}attenuators can be made.

\subsection { The Effect of Inhomogeneous Fillings on the Phase Velocity}
\index{Velocity!phase}

An intuitive  understanding of how inhomogeneous fillings affect
the phase velocity of a guided mode can be acquired by studying a
\index{Waveguide!parallel plate}parallel plate waveguide. Considering a square region of a
parallel plate waveguide away from  the edges so that fringing
field effect can be ignored. Hence, we assume that the field lines
for the TEM mode are perfectly vertical.

The capacitance of (a) can be thought of as three capacitances in
parallel or as three capacitances in series. When the waveguide is
filled as in (b), it affects one of the three capacitances in
parallel, yielding a resultant capacitance given by
\begin{equation}
C_b=\frac 2 3 C_a +\frac {\epsilon _r } {3} C_{a,} =\frac
{2+\epsilon _r }{3} C_a \label{eq6-1000}
\end{equation}
where $C_a $ is the capacitance of (a). When the waveguide is
filled as in (c), it affects one of the three capacitances in
series, yielding a resultant capacitance given by
\begin{equation}
C_c^{-1} =2(3C_a)^{-1} +(3 \epsilon _r C_a )^{-1} \label{eq6-2000}
\end{equation}
or
\begin{equation}
C_c= \frac {3\epsilon _r C_a }{2 \epsilon _r +1} . \label{eq6-3000}
\end{equation}
It is seen that
\begin{equation}
\frac {3\epsilon _r }{2 \epsilon _r +1 } < \frac {2+\epsilon _r
}3, \quad \epsilon _r >1. \label{eq6-4000}
\end{equation}
Therefore $C_b >C_c $  always for $ \epsilon _r > 1 $. The reason
is that enhancing a capacitor in parallel has more effect on the
total capacitance than enhancing a capacitor in parallel. When
$\epsilon _r \rightarrow \infty, C_c \rightarrow \frac 3 2 C_a$ or
saturates while $C_b \rightarrow \infty $. Since the phase
velocity of a \index{Mode!TEM}TEM mode in a parallel plate waveguide is given by
\begin{equation}
\upsilon  =\frac 1 {\sqrt{LC}} \label{eq6-5000}
\end{equation}
where $L$ and $C$ are line inductance and line capacitance
respectively, a dielectric loading in case (b) slows down the wave
more than case (c).

When the waveguide is a square waveguide, the \index{Mode!TE$_{10}$}TE$_{10}$ mode is
affected even more by a symmetrically located dielectric slab
because the electric field has a maximum at the center of the
waveguide. Hence, by dielectric-slab loading as in case (b), the
TE$_{10}$ mode propagates with a slower phase velocity than the
\index{Mode!TE$_{01}$}TE$_{01}$ mode. This gives rise to anisotropy in a waveguide.

\begin{figure}
\begin{center}
\hfil\includegraphics[width=4.5truein]{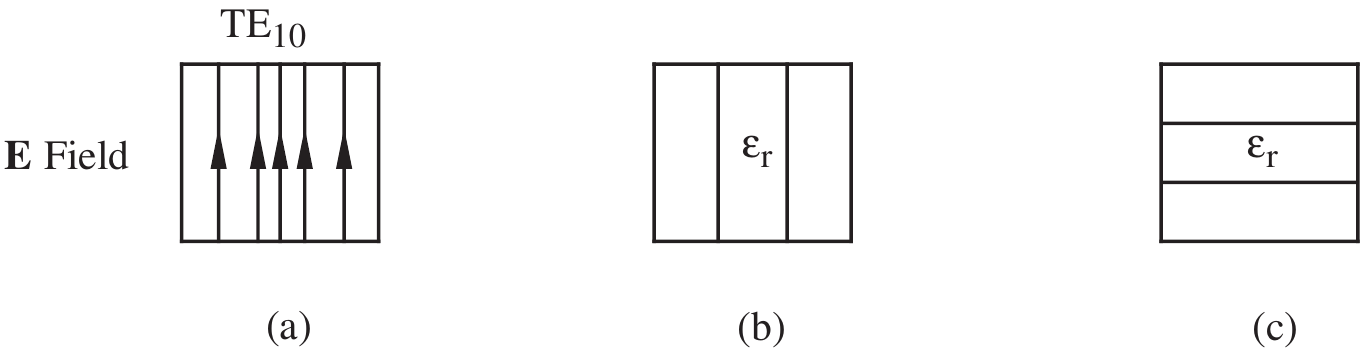}\hfil
\end{center}
\caption{Dielectric-slab loading of a square waveguide. The
TE$_{10}$ mode is affected more by case (b) than case (c).}\label{fg362}
\end{figure}


\subsection { Quarter-Wave Plate}
\index{Quarter-wave plate}

By slab loading a circular waveguide, one can make a quarter wave
plate. If a dielectric slab is oriented at $45^o$ with  respect to
the TE$_{11}$ mode of a circular waveguide, the mode can be
decomposed into two orthogonal modes, one with $\v E$ field
perpendicular to the dielectric slab, and another parallel to the
slab. The one with electric field parallel to the slab is going to
be slowed down more than the one with electric field perpendicular
to the slab. Hence after a certain distance, the phases of these
two modes are going to be out of phase. If the length of the
dielectric slab is chosen judiciously such that these two modes
are $90^o$ out of phase, then one obtains a circularly polarized
mode at the other end of the waveguide. Such an application of a
dielectric slab loading is similar to a quarter-wave plate in
optics and hence its name. A \index{Half-wave plate}half-wave plate will shift one
component of the wave by $180^o$ compared to the other orthogonal
component.

\begin{figure}
\begin{center}
\hfil\includegraphics[width=4.5truein]{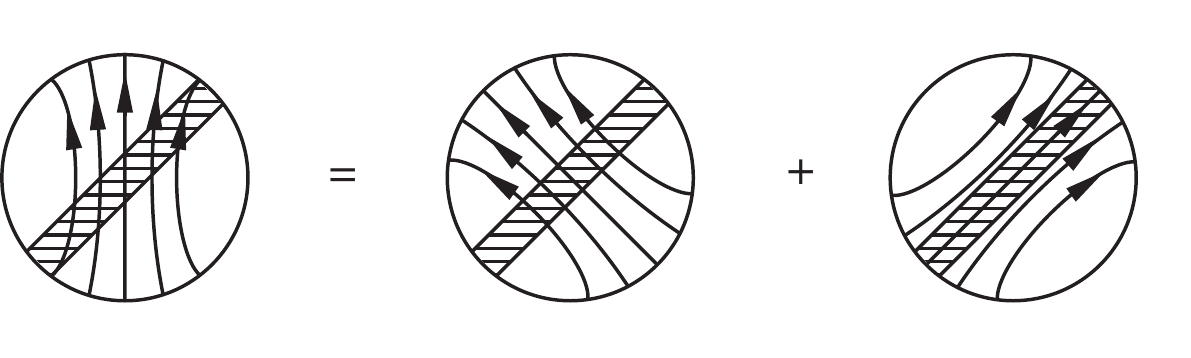}\hfil
\end{center}
\caption{Decomposition of a TE$_{11}$ mode of a circular waveguide
into two orthogonal modes. One perpendicular to the slab and
another parallel to the slab.}\label{fg363}
\end{figure}


\subsection { Variable Phase Shifter}
\index{Variable phase shifter}

\begin{figure}
\begin{center}
\hfil\includegraphics[width=4.5truein]{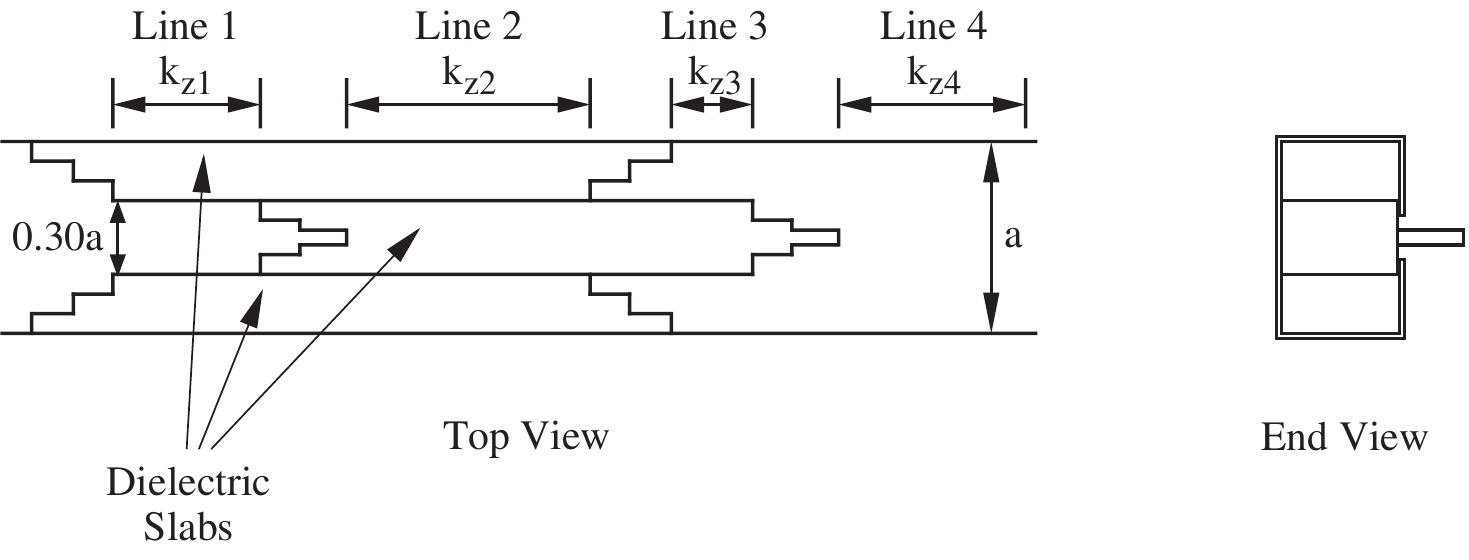}\hfil
\end{center}
\caption{A linear phase shifter using four different sections of
dielectric loading.}\label{fg365}
\end{figure}

Dielectric loading can be used to make variable phase shifters.
One way of achieving this is to vary the position of the slab
position in a rectangular waveguide. Since the \index{Mode!TE$_{10}$}TE$_{10}$ mode of a
rectangular waveguide has a weaker field near the side walls of
the waveguide, the phase velocity will be slowed down less when
the slab is near to the waveguide wall compared  to near the
center of the waveguide (See Figure \ref{fg365}).

A \index{Linear phase changer}linear phase changer can also be constructed by dielectric slab
loading as shown below. When the center slab is moved by a
distance $\Delta z $, line 1 and line 3 are increased by a length
$\Delta z $ while line 2 and line 4 are decreased by a length
$\Delta z $. Therefore, the total phase shift is
\begin{equation}
\Delta \phi =(k_{z1} +k_{z3}-k_{z2} -k_{z4}) \Delta z.
\label{eq6-6}
\end{equation}
Because of the sinusoidal distribution of the electric field of a
TE$_{10}$ mode of a rectangular waveguide, it is clear that
\begin{equation}
k_{z4}<k_{z1} <k_{z3} <k_{z2} . \label{eq6-7000}
\end{equation}
However, if the center section is chosen to be about $0.3a$, we
can have
\begin{equation}
k_{z1} +k_{z3}>k_{z2} +k_{z4}, \label{eq6-8000}
\end{equation}

\begin{figure}
\begin{center}
\hfil\includegraphics[width=3.0truein]{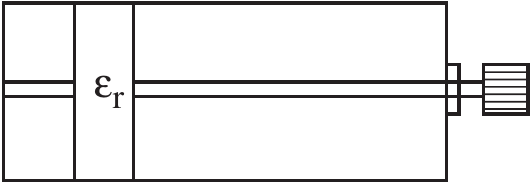}\hfil
\end{center}
\begin{center}
 (a)
\end{center}
\label{fg364a}
\begin{center}
\hfil\includegraphics[width=4.truein]{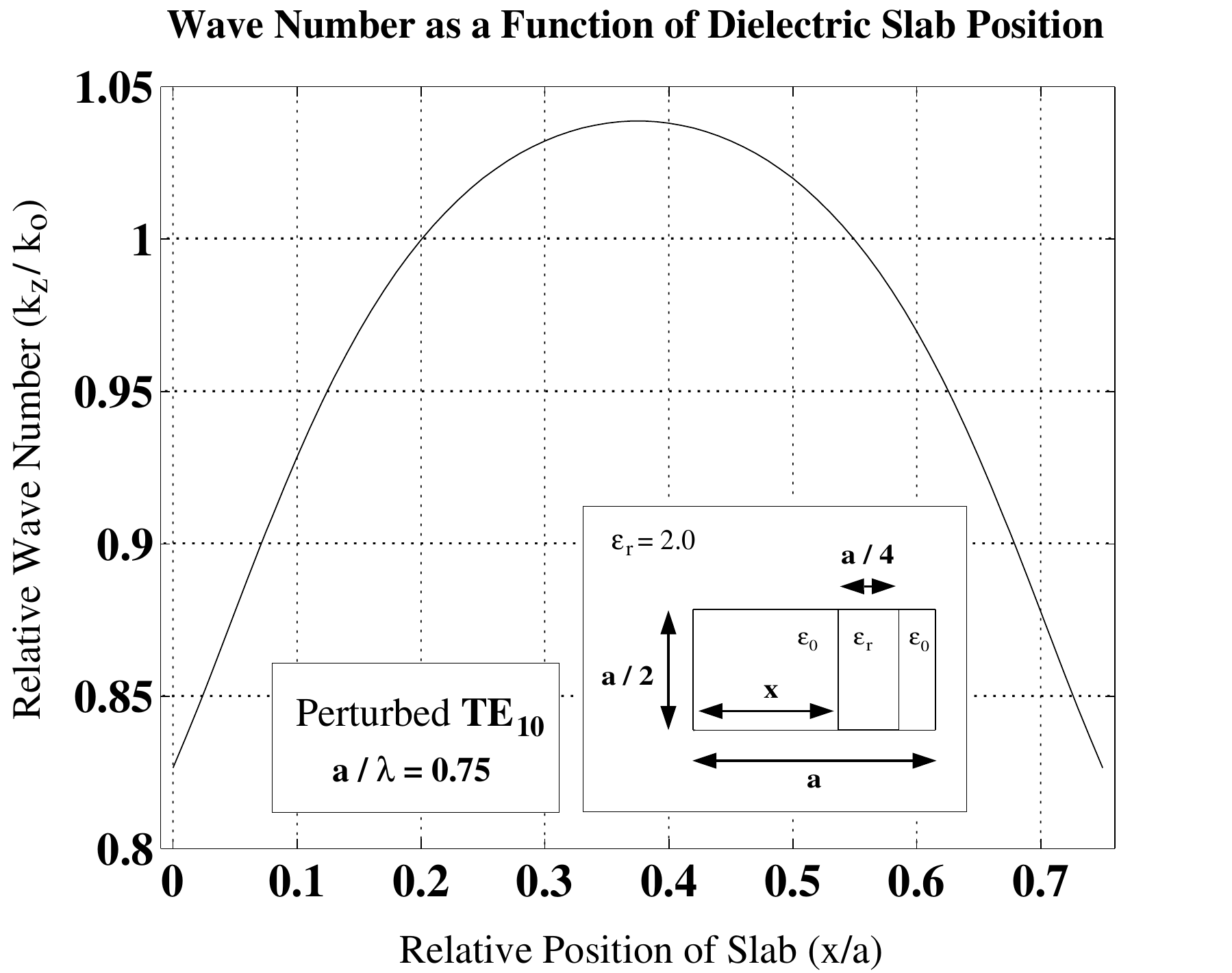}\hfil
\end{center}
\begin{center}
  (b)
\end{center}
\caption{(a) Variable dielectric slab locations can be used as a
\index{Variable phase shifter}variable phase shifter. (b) Variable dielectric slab locations
affects the phase velocity of the TE$_{10}$ mode.}
\label{fg364b}
\end{figure}

\noindent and a net positive phase shift linearly proportional to
$\Delta z $ becomes possible.

\begin{figure}
\begin{center}
\hfil\includegraphics[width=4.5truein]{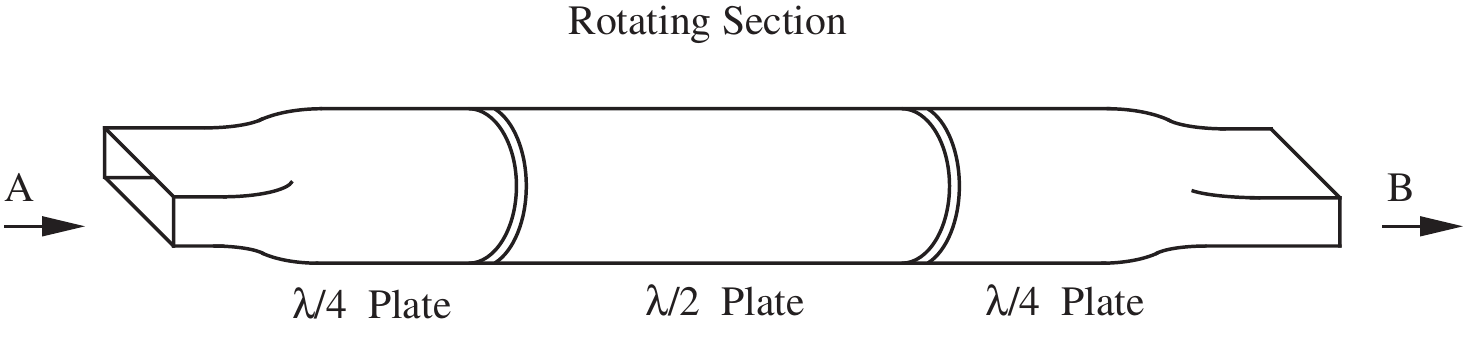}\hfil
\end{center}
\caption{A rotary phase shifter.}\label{fg366}
\end{figure}

A \index{Rotary phase shifter}rotary phase shifter can be made by sandwiching a section of
half-wave plate between two sections of quarter-wave plates as
shown in Figure \ref{fg366}. The quarter-wave plate and half-wave
plate sections are made by dielectric slab loading as described
previously. By rotating the middle section, one can vary the phase
of the wave going from port A to port B.

\begin{figure}
\begin{center}
\hfil\includegraphics[width=1.6truein]{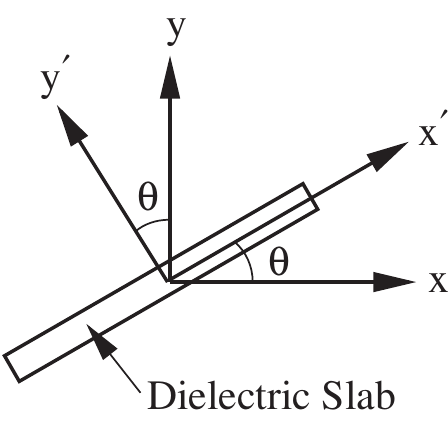}\hfil
\end{center}
\caption{Orientation of the dielectric slab in the half-wave plate
section compared to the original coordinates.}\label{fg367}
\end{figure}

To explain the operating principle of this device, we will use
linearly polarized waves in free space. The first quarter-wave
plate converts a linearly polarized wave into a circularly
polarized wave where the $\^x $ and $\^y$ components are $90^o$
out of phase, viz.,
\begin{equation}
\v E =(i\^x +\^y)E_0, \label{eq6-19}
\end{equation}
where $\^x$ direction is parallel to the dielectric slab in the
quarter-wave plate section, and $E_0$ is  a complex number. The
above represents a left-hand circularly polarized wave.  When this
wave impinges on the half-wave plate section, the dielectric slab
can be oriented at any angles. In Figure \ref{fg367}, we assume it
to be at an angle $\theta$ with  respect to the coordinates of the
dielectric slab in the first section. In this case, we have

\begin{equation}
\^x =\^x' \cos \theta -\^y' \sin \theta, \quad \^y=\^x' \sin
\theta +\^y' \cos \theta. \label{eq6-20}
\end{equation}
Using (\ref{eq6-20}) in (\ref{eq6-19}), we have
\begin{equation}
\v E =(i\^x' +\^y') e^{-i\theta} E_0 \label{eq6-21}
\end{equation}
The above is still a left-hand circularly polarized wave, and $
e^{-i\theta} $ comes about because we are rotating the coordinates
counterclockwise while the polarization is rotating clockwise.
After this wave has gone through the half-wave section, the $\^
y'$ component will be $180^o$  out of phase with respect to the
$\^x'$ component and we have
\begin{equation}
\v E =(-i\^x' +\^y') e^{-i\theta} E'_0 \label{eq6-22}
\end{equation}
which is a right-hand circularly polarized wave, and $E'_0$ is a
new complex number accounting for the additional phase delay the
wave has acquired in propagating through the middle section.
Projecting this back to the original coordinates by using
\begin{equation}
\^x'= \^x \cos \theta +\^y \sin \theta, \quad \^y'=-\^x \sin
\theta +\^y \cos \theta, \label{eq6-23}
\end{equation}
we have
\begin{equation}
\v E =(-i\^x +\^y) e^{-2i\theta} E'_0. \label{eq6-24}
\end{equation}
{If the wave had remained left-hand circularly polarized, the
projection back would have annulled the phase $e^{-i\theta}$. But
because it becomes a right-hand circularly polarized, it introduces
an additive phase instead.}

When this wave in \eqref{eq6-24} passes through the last section of
the quarter-wave plate, the $\^x $ component gains another $90^o$ of
phase with respect to the $\^y$ component, and it becomes
\begin{equation}
\v E =(\^x +\^y) e^{-2i\theta} E''_0 \label{eq6-25}
\end{equation}
where $E^{\prime\prime}_0$ is a new complex number. This wave is again linearly
polarized with respect to the  rectangular waveguide in the output
port. Note that there is a phase shift of $-2\theta$ which is
dependent on the orientation of the center rotating section .

\subsection { Variable Attenuator}
\index{Variable attenuator}

For the same reason that a dielectric slab parallel to the
electric field affects a mode more than a slab perpendicular to
the electric field, a resistive (lossy) dielectric sheet parallel
to the electric field will incur more loss on a mode than a
resistive sheet perpendicular to the electric field. Hence, a
variable attenuator can be made similar to a variable phase
shifter, except that in the first and the last sections, the
resistive sheets are loaded horizontally, and the middle section
the resistive sheet can be rotated.

\begin{figure}
\begin{center}
\hfil\includegraphics[width=4.5truein]{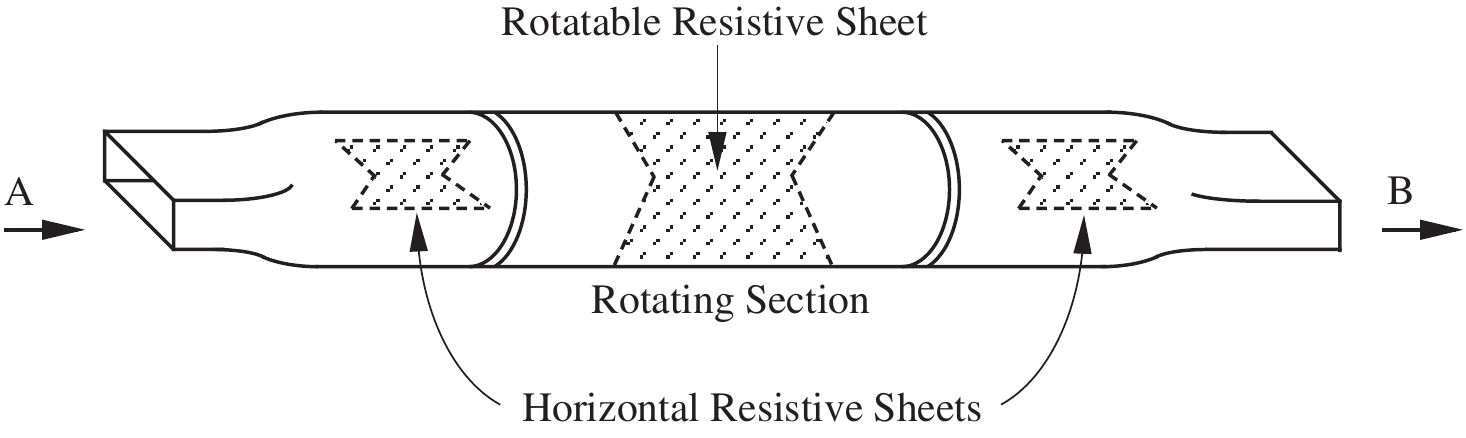}\hfil
\end{center}
\caption{A rotary variable attenuator.}\label{fg368}
\end{figure}


The horizontal resistive sheet in the first section ensures that
the mode is predominantly vertically polarization. The resistive
sheet in the middle section will attenuate the mode proportional
to the component of the electric field parallel to the resistive
sheet. Hence, the attenuation of the mode by the middle section
can be varied by rotating it. The last section ensures that any
horizontally polarized modes due to mode conversion be removed,
and only the vertically polarized component will exit from port B.

\section {{{Spin Dynamics and Ferrite Materials }}}
\index{Spin dynamics}
\index{Ferrite materials}

The understanding of the interaction of particle spins with
electromagnetic field is commonly encountered in the study of
nuclear magnetic resonance, and ferrite materials.  The first is
extremely useful in magnetic resonance imaging (MRI) which is an
important medical imaging modality.   A particle spin has angular
momentum as well as a magnetic moment.  The magnetic moment will
interact with an ambient magnetic field giving rise to the
precession of the spins.  The precession of the spins, in the case
of MRI, yields spin echoes that can be measured for imaging and
spectrocopic purposes.  In the case of ferrites, it gives rise to
\index{Medium!anisotropic}anisotropic, \index{Medium!gyrotropic}gyrotropic materials that exhibit \index{Faraday rotation}Faraday rotation.
Such effect can be used to design non-reciprocal microwave devices
such as isolators.

As mentioned, ferrite is commonly used as an anisotropic material in
a waveguide to make \index{Waveguide!nonreciprocal}nonreciprocal waveguides
\cite{COLLIN,COLLIN2,LAX,LAX_BUTTON,ZHOUetal}. Ferrite obtains its
anisotropy by having its \index{Spin of an electron}electron spins interact with a static
magnetic field. An electron spin has a \index{Magnetic dipole moment}magnetic dipole moment as
well as angular momentum. In the presence of a static magnetic
field, this dipole moment aligns itself with the magnetic field. If
a transverse force, via an electromagnetic field, is applied to tilt
the direction of the dipole moment, and hence, the direction of
angular momentum, the spin precesses about the static magnetic field
just as a spinning top precesses about a gravitational field.

The correct description of the motion of an electron in the presence
of a magnetic field requires quantum mechanics. However, when a
large number of electrons are considered, their average motion can
be described by a classical equation of motion similar to the
equation governing the motion of a spinning top.   From this
equation, we can understand the anisotropic nature of ferrites being
biased by a magnetic field.

For an electron with a magnetic dipole moment $\v m $ in the
presence of a magnetic field $\v B _0$, the torque exerted on the
electron is given by
\begin{equation}
\v T=\v m\times \v B_0. \label{eq7-1000}
\end{equation}
This torque is exerted on the angular momentum $\v P $ of the
electron, causing it to change. Therefore, we have
\begin{equation}
\frac {d\v P}{dt} =\v T =\v m \times \v B_0. \label{eq7-2000}
\end{equation}
But the magnetic dipole moment of an electron is antiparallel to
its angular momentum, i.e.,
\begin{equation}
\v m =-\gamma \v P \label{eq7-3000}
\end{equation}
where $\gamma $ is known as the \index{Gyromagnetic ratio}gyromagnetic ratio. Consequently,
the equation of motion for a spinning electron in a magnetic field
is
\begin{equation}
\frac{d \v P}{dt} =-\gamma \v P \times \v B _0=+\vg \omega _0
\times \v P \label{eq7-4000}
\end{equation}
where $\vg \omega _0=\gamma \v B _0$, and $\omega _0 =|\vg \omega
_0|=\gamma |\v B_0| $ is also known as the \index{Larmor frequency}Larmor frequency. For
electron spins in a reasonably strong magnetic field, the Larmor
frequency can be in the microwave regime.

Often time, the equation of motion is written as
\begin{equation}\label{eq7-4000a}
\frac{d \v P}{dt} =-\gamma \v P \times \v B _0-\lambda \v
P\times\left(\v P\times \v B_0\right)
\end{equation}
The last term above accounts for damping.  The above is known as the
\index{Landau-Lifshitz equation}Landau-Lifshitz equation.  Later, in 1955, T.L. Gilbert, starting from
first principles, replaced the last term with a different expression, depending on the time derivative of the angular momentum,
yielding
\begin{equation}\label{eq7-4000b}
\frac{d \v P}{dt} =-\gamma\left[ \v P \times \v B _0-\eta \v P\times
\left(\v P\times \frac{d \v P}{dt}\right)\right]
\end{equation}
The above is known as the \index{Landau-Lifshitz-Gilbert equation}\emph{Landau-Lifshitz-Gilbert equation}.  It can
be shown that the above reduces to the same form as the
Landau-Lifshitz equation
\begin{equation}\label{eq7-4000c}
\frac{d \v P}{dt} =-\gamma' \v P \times \v B _0-\lambda' \v
P\times\left(\v P\times \v B_0\right)
\end{equation}
but with different $\gamma'$ and $\lambda'$.

Even though we say that (\ref{eq7-4000}) to \eqref{eq7-4000c} are
the equations of motion for an electron, it actually governs the
average motion over an ensemble of a large number of electrons.
Hence, a classical picture applies here.  Other subatomic particles
like protons also possess a spin, but because of their larger mass,
the Larmor frequency is much lower. \index{Proton spin}Proton spins, prevalent in MRI,
has a Larmour frequency of 42.6 MHz per Tesla,\footnote{1 T=$10^4$
gauss=1 weber/m$^2$=1 Volt Second/m$^2$.} while electron spins have
Larmor frequencies in the GHz regime.

The Larmor frequency is the free precession frequency of a spin when
it is tipped from the vertical position in the presence of the
static biasing magnetic field. Equation (\ref{eq7-4000}) is
sometimes known as the \index{Bloch equation}Bloch equation.

Assuming that the field $\v B_0 $ is static, then it is easy to show
that the natural solution to (\ref{eq7-4000}) consists of a $\v P $
with a $\^ z $ component and a circulating $\^ x $ and $\^ y $
component. Let us assume that
\begin{equation}
\v P =\frac 1 {\sqrt 2} (\^ x +i\^ y) P_s e ^{-i\omega t } +\^ z
P_z . \label{eq7-5000}
\end{equation}
The above is not a pure time-harmonic signal as the second term can
have a different frequency from the first term.
 Note that $|\v P
|=\sqrt{P_s^2+P_z^2}=P_0 = const$ even though it is time varying as
angular momentum has to be conserved. Also, this angular momentum
comes from the intrinsic spin of the particle, which is a constant.
Also, $\v P(t) $ can be obtained in the real world by taking the
real part of Equation (\ref{eq7-5000}), or by adding a complex
conjugate term to the above. For convenience, we will leave
(\ref{eq7-5000}) with its phasors. Then, using (\ref{eq7-5000}) in
(\ref{eq7-4000}) yields
\begin{equation}
-i\omega (\^ x +i\^ y) P_s {\frac {e^{-i \omega t}} {\sqrt 2}}+\^
z \frac {dP_z}{dt}= (\omega _0 \^ y -i\omega _0 \^ x ) P_s {\frac
{e^{-i\omega t}} {\sqrt 2}}. \label{eq7-6000}
\end{equation}
It is seen that the above is satisfied when we have
\begin{equation}
\omega =\omega _0, \quad \frac {dP_z }{dt} =0. \label{eq7-7000}
\end{equation}
In other words, the \index{Spin precession}spin precesses at the Larmor frequency $\omega
_0 $ while the $\^ z $-component of its angular momentum remains
unchanged since the system is non-dissipative.  In a dissipative
system, the kinetic energy in the angular momentum will be lost to
the environment.  The precession of the spin will slow down, and
eventually, the spin will be completely aligned with the background
static magnetic field.

Now if we include an additional RF field $\v B_1$, in the
transverse direction which is circularly polarized such that
\begin{equation}
\v B_{1+} =\frac 1{\sqrt 2} (\^ x +i\^ y) B_{1+} e^{-i\omega t },
\label{eq7-8000}
\end{equation}
then the equation of motion becomes
\begin{equation}
\frac {d\v P}{dt} =\vg \omega _0 \times \v P +\gamma \v B _{1+}
\times \v P. \label{eq7-9000}
\end{equation}
The transverse RF field, which is the driving field, will tilt the
spin and force it to precess at the same frequency. Therefore, we
let
\begin{equation}
\v P =\frac 1 {\sqrt 2 } (\^ x +i\^ y) P_s e ^{-i\omega t } +\^ z
P_z. \label{eq7-10000}
\end{equation}
Using (\ref{eq7-10000}) in (\ref{eq7-9000}), we have
\begin{equation}
\begin{split}
-i\omega (\^ x +i\^ y ) P_s \frac {e^{-i\omega t}} {\sqrt 2} +\^ z
\frac {dP_z}{dt} &=-i\omega _0 (\^ x +i\^ y)P_s \frac {e^{-i\omega
t }}{\sqrt 2}\\
&+i\gamma (\^ x +i\^ y)B_{1+ }P_z \frac {e^{-i\omega t }}{\sqrt 2}
\label{eq7-11000}
\end{split}
\end{equation}
Again we have $dP_z /dt =0 $ and
\begin{equation}
P_s =\frac {\gamma B_{1+}} {\omega _0-\omega }P_z . \label{eq7-12000}
\end{equation}
The precession now is at the driving frequency $\omega$.  This
equation so far, has been derived with no approximation.  Note that
if the frequency $\omega =\omega _0$, a resonance occur, but $P_s $
does not go to infinity as $|\v P|=\sqrt{P_s^2+P_z^2}=P_0=const$
and $P_z \rightarrow 0$ when $P_s \rightarrow P_0$.  In other words,
as one drives the spin system closer to the resonance frequency,
$P_s\gg P_z$, but $P_s^2+P_z^2=const$.  Therefore, the vector $\v
P$ has to tilt, and at resonance, the vector $\v P$ is precessing
horizontally.

\begin{figure}
\begin{center}
\hfil\includegraphics[width=2.0truein]{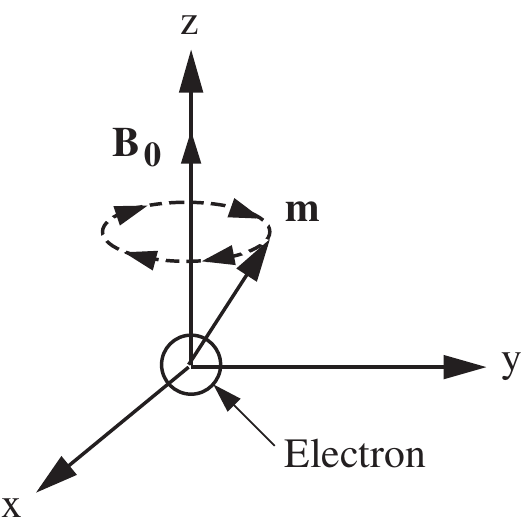}\hfil
\end{center}
\caption{The magnetic dipole moment of an electron precesses just
like a spinning top when an RF transverse to $z$ magnetic field is
applied.}\label{fg371}
\end{figure}


Next, we assume that the applied RF field has a magnetic field
component given by a circularly polarized field of opposite
polarity such that
\begin{equation}
\v B_{1-} =\frac 1{\sqrt 2} (\^ x -i\^ y) B_{1-} e^{-i\omega t }
\label{eq7-13000}
\end{equation}
and assuming then the spins are tipped from the vertical and
precess at the same frequency as the driving RF field, we let
\begin{equation}
\v P =\frac 1 {\sqrt 2 } (\^ x -i\^ y) P_s e ^{-i\omega t } +\^ z
P_z. \label{eq7-14000}
\end{equation}
Using (\ref{eq7-14000}) in (\ref{eq7-9000}) with $\v B_{1+} $ replaced
by $\v B_{1-} $, we have
\begin{equation}
-i\omega (\^ x -i\^ y ) P_s \frac {e^{-i\omega t}} {\sqrt 2} +\^ z
\frac {dP_z }{dt} =i\omega _0 (\^ x -i\^ y)P_s\frac {e^{-i\omega t
}}{\sqrt 2}-i\gamma (\^ x -i\^ y)B_{1- }P_z \frac {e^{-i\omega t
}}{\sqrt 2} . \label{eq7-15}
\end{equation}
The above implies that $dP_z/dt =0$, and that
\begin{equation}
P_s =\frac {\gamma B_{1-}} {\omega _0+\omega }P_z . \label{eq7-16}
\end{equation}

Assuming that the RF field is much smaller than the static field so
that $B_1\ll B_0$, then we can assume that $P_s$ is small or that
$P_z \simeq P_0.$ In this case, (\ref{eq7-12000}) and (\ref{eq7-16})
together become
\begin{equation}
P_{s\pm } \approx \frac {\gamma B_{1\pm }} {\omega _0\mp \omega
}P_0 , \label{eq7-17}
\end{equation}
and the transverse part of the spin angular momentum becomes
\begin{equation}
\v P_{s\pm } \approx \frac 1 {\sqrt 2} (\^ x \pm i\^ y )\frac
{\gamma B_{1\pm }} {\omega _0\mp \omega }P_0 e ^{-i\omega t } .
\label{eq7-18}
\end{equation}
By the above approximation, we have simplified the relationship
between $\v P_{s\pm }$ and the driving field $B_{1\pm }$.

The magnetic moment of a spin is given by $\v m =-\gamma \v P$.
Hence, the transverse part of the magnetic moment is
\begin{equation}
\v m_{s\pm } =\frac 1 {\sqrt 2} (\^ x \pm i\^ y )\frac {\gamma
B_{1\pm} m_0 } {\omega _0\mp \omega } e ^{-i\omega t } .
\label{eq7-19}
\end{equation}
where $m_0=-\gamma P_0$ is the component of the magnetic moment in
the $\^ z $-direction, where in the approximation here it is the
magnitude of the magnetic moment.

The magnetic dipole moment density or \index{Magnetization density}magnetization density is given
by $\v M=N\v m $ and we have
\begin{equation}
\v M_{s\pm } =\frac 1 {\sqrt 2} (\^ x \pm i\^ y )\frac {\gamma
B_{1\pm} M_0 } {\omega _0\mp \omega } e ^{-i\omega t } .
\label{eq7-20}
\end{equation}
where $\v M_s =N\v m_s $ and $M_0=Nm_0 $ are the dipole moment
density in the transverse and axial directions respectively.

The above is the RF response of the transverse magnetization
density in response to an applied RF magnetic field given by
\begin{equation}
\v B_{1\pm } =\frac 1 {\sqrt 2} (\^ x \pm i\^ y ) B_{1\pm} e
^{-i\omega t } . \label{eq7-21}
\end{equation}
In other words, we have
\begin{equation}
\v M_{s\pm } =\frac {\gamma M_0 } {\omega _0\mp \omega } \v B
_{1\pm } . \label{eq7-22}
\end{equation}
One can always make a linear polarization out of a linear
superposition of two circular polarizations via
\begin{equation}
\v B =\^ x B_x e^{-i\omega t}=\frac 1 2 \left[(\^ x +i\^ y )
B_x+(\^ x -i\^ y ) B_x \right] e^{-i\omega t }. \label{eq7-23}
\end{equation}
When this RF field is used to excite the spins, the RF magnetization
response is given by
\begin{equation}
\begin{split}
\v M_{s} &  =\left( \frac {\gamma M_0 }{\omega _0 - \omega
}\right) \frac 1 2 (\^ x +i\^ y ) B_x e^{-i\omega t }+\left(\frac
{\gamma M_0 } {\omega _0+ \omega }\right)
\frac 1 2 (\^ x -i\^ y )B_x  e^{-i\omega t }\\
         & =\^ x\gamma M_0 B_x \left(\frac {\omega _0 }{\omega ^2_0-
\omega ^2 } \right) e^{-i\omega t }+i\^ y \gamma M_0 B_x
\left(\frac
{\omega }{\omega ^2_0-\omega ^2 } \right)e^{-i\omega t }.\\
\label{eq7-24}
\end{split}
\end{equation}
Similarly, when we make the RF field $\^ y $ polarized such that
\begin{equation}
\v B =\^ y B_y e^{-i\omega t} =\frac 1 {2i} [(\^ x +i\^ y )B_y-(\^
x -i\^ y )B_y ] e^{-i\omega t }, \label{eq7-25}
\end{equation}
the RF magnetization response is
\begin{equation}
\begin{split}
\v M_{s} &=\left( \frac {\gamma M_0 } {\omega _0- \omega }\right)
\frac 1 {2i} (\^x +i\^y )B_y e^{-i\omega t }- \left( \frac {\gamma
M_0 }
{\omega _0+ \omega }\right) \frac 1 {2i} (\^ x -i\^ y )B_y e^{-i\omega t }\\
         &=-i\^x \gamma M_0 B_y \left(\frac {\omega }
{\omega ^2 _0-\omega ^2 } \right)
 e^{-i\omega t }+\^ y \gamma M_0 B_y \left(\frac {\omega _0}{\omega ^2 _0-
\omega ^2 } \right)e^{-i\omega t }.\\
\label{eq7-26}
\end{split}
\end{equation}
Consequently, $B_x =\mu _0 H_x $, and $B_y =\mu _0 H_y $, we have
in matrix form
\begin{equation}
\begin{bmatrix}
 M_x \\
 M_y
\end{bmatrix}
=
\begin{bmatrix}
\chi_{xx} &\chi _{xy}\\
\chi_{yx} & \chi _{yy}
\end{bmatrix}
\begin{bmatrix}
 H_x \\
 H_y
\end{bmatrix},  \quad \text {or}\quad \v M=\dyad\chi \cdot \v H
\label{eq7-27}
\end{equation}
where
\begin{subequations}
\begin{equation}
\chi _{xx} =\chi _{yy} =\frac {\mu _0\gamma  M_0 \omega_0}{\omega
^2_0 -\omega ^2 }, \label{eq7-28a}
\end{equation}
\begin{equation}
\chi _{xy} =-\chi _{yx} =\frac {-i\mu _0\gamma  M_0 \omega}{\omega
^2_0 - \omega ^2 }. \label{eq7-28b}
\end{equation}
\end{subequations}
The above is the RF magnetization response to an RF magnetic field
excitation in a ferrite medium when the RF signal is assumed small,
and we assume that $\^ z$ component of the magnetization response is
small and is of higher order.  The reason being that the $z$
component of the RF field has little effect on the spin momentum
when it is pointed primarily in the $z$ direction. In general,
\begin{equation}
\v B =\mu _0 (\v H+\v M )=\mu _0(\dyad I +\dyad\chi ) \cdot \v H
=\dyad\mu \cdot \v H \label{eq7-29}
\end{equation}
where
\begin{equation}
\dyad\mu =\mu _0
\begin{bmatrix}
{1+\chi _{xx}} & {\chi _{xy}}     & 0 \\
{\chi _{yx}}  & {1+\chi _{yy}} & 0 \\
 0      &      0       & 1
\end{bmatrix}.
\label{eq7-30}
\end{equation}
This is an example of an anisotropic magnetic medium, or a
gyrotropic magnetic medium.  The above permeability tensor is also
Hermitian implying that it represents a lossless medium.

\subsection { Natural Plane Wave Solutions in an Infinite Homogeneous Ferrite Medium}
\index{Medium!infinite homogeneous ferrite}

A ferrite medium,which is a gyrotropic material, admits \index{Wave!plane!circularly polarized}circularly
polarized plane waves as the natural plane waves propagating in a
homogeneous anisotropic ferrites \cite{BUDDEND}. We can see this by
looking at the solution of the vector wave equation. When
$\epsilon $ is homogeneous and isotropic, and $\dyad\mu $ is
anisotropic, the magnetic field satisfies the following vector
wave equation:
\begin{equation}
\nabla \times \nabla \times \v H -\omega ^2 \epsilon \dyadg\mu
\cdot \v H =0. \label{eq7-31}
\end{equation}
Assuming that a plane wave solution propagating in the $z$
direction exists such that
\begin{equation}
\v H =\v H_0 e^{ikz}, \label{eq7-32}
\end{equation}
then
\begin{equation}
-k^2 \^ z (\^ z \cdot \v H )+k^2 \v H -\omega ^2 \epsilon
\dyadg\mu \cdot \v H =0. \label{eq7-33}
\end{equation}
If $\dyadg\mu $ is of the form
\begin{equation}
\dyadg\mu =
\begin{bmatrix}
 {\dyadg\mu _s}  & 0 \\
   0         & \mu
\end{bmatrix}
\label{eq7-34}
\end{equation}
where $\dyadg\mu_s $ is a $2 \times 2$ tensor, then $H_z =0$ by
equating the $z$ component of (\ref{eq7-33}). Therefore $\v H $ has
only transverse to $z$ components, and Equation (\ref{eq7-33})
becomes
\begin{equation}
k^2 \v H_s -\omega ^2 \epsilon \dyadg \mu _s \cdot \v H_s =0
\label{eq7-35}
\end{equation}
where $\v H_s $ is a vector in the $xy $ plane. From Equation
(\ref{eq7-30}), it is clear that
\begin{equation}
\v B_{s\pm} = \dyadg\mu _s \cdot \v H_{s \pm }=\mu _{\pm} \v H
_{s\pm} \label{eq7-36}
\end{equation}
for a ferrite medium where $\chi_{xx}=\chi_{yy}$, and
\begin{equation}
\v H_{s \pm } =H_{0 \pm } (\^ x \pm i\^ y ). \label{eq7-37}
\end{equation}
Consequently, (\ref{eq7-35}) becomes
\begin{equation}
k^2 \v H_{s\pm } -\omega ^2 \epsilon \mu _{\pm }  \v H_{s\pm } =0
\label{eq7-38}
\end{equation}
where $k$ can have two possible values given by
\begin{equation}
k_{\pm} =\omega \sqrt{\mu _\pm \epsilon }. \label{eq7-39}
\end{equation}
Therefore, the natural solution in a ferrite medium is of the form
\begin{equation}
\v H_{\pm} =H_{0\pm } (\^ x \pm i\^ y) e^{ik_{\pm} z}.
\label{eq7-40}
\end{equation}
In essence, in a ferrite medium, a right-hand circularly polarized
wave ``feels'' a different permeability compared to a left-hand
circularly polarized wave. Therefore, the two polarizations
propagate with different velocities. When the wave is not
propagating in the $z$ direction, the propagation of the wave is
more complicated, and will not be discussed here.

\subsection { Faraday Rotation}
\index{Faraday rotation}

Faraday rotation occurs in a ferrite. It can also occur in the
earth ionosphere where the electron spins are biased by the earth
magnetic field. To understand Faraday rotation, we decompose a
linearly polarized wave into two circularly polarized waves, viz.,
\begin{equation}
\v E =\^ x E_0 =\frac 1 2 (\^ x +i\^ y ) E_0 +\frac 1 2 (\^ x -i\^
y) E_0, \label{eq7-41}
\end{equation}
where the first term is left-hand circularly polarized for a wave
propagating in the $\^ z $ direction, and likewise, the second
term is right-hand circularly polarized. In a ferrite medium,
these two polarizations will propagate with different phase
velocities. After a certain distance, there could be a phase
difference between them, and we have
\begin{equation}
\v E =\frac 1 2 (\^ x +i\^ y) E'_0e^{i\theta} +\frac 1 2 (\^ x
-i\^ y )E'_0. \label{eq7-42}
\end{equation}
Combining the terms in the above, we have
\begin{equation}
\begin{split}
\v E & =  \^ x e^{i\frac {\theta }{2}} \cos \left(\frac {\theta
}{2}\right)
E'_0 -\^ y e^{i\frac {\theta }{2}} \sin \left(\frac {\theta }{2}\right) E_0' \\
    & =\left[\^ x \cos \left(\frac {\theta }{2}\right ) -\^ y \sin \left(\frac{\theta }{2}\right) \right] E'_0 e^{i\theta /2}. \\
\label{eq7-43}
\end{split}
\end{equation}
Hence, the wave vector is now tilted by an angle $-\frac {\theta
}{2}$ or has rotated clockwise.

For a wave propagating in the negative $z$ direction, the
polarizations in the first and second term reverse roles. Now, the
second term in (\ref{eq7-42}) will have a phase gain over the first
term, and the wave vector is rotated counterclockwise after
propagating through a certain distance. Hence, the phenomenon is
nonreciprocal meaning that the waves propagating in the $\pm z$
directions are quite different in behavior.

\subsection { Applications of Faraday Rotation}
\index{Faraday rotation!applications}

Faraday rotation can be used to make a number of \index{Nonreciprocal microwave device}nonreciprocal
microwave devices. One simple example is a \index{Gyrator} gyrator, which is
defined as a device  whose transmission from port 1 to port 2 has
a $180^o$ phase shift compared to its transmission from port 2 to
port 1. The polarization rotation and the nonreciprocal natures of
ferrites can be used to make such a device.

Another device that can be made from ferrite loading is an
\index{Isolator}isolator. An isolator consists of first a $45^o$ mechanical twist
which is a reciprocal section. Then it is followed by another
section of $45^o$ Faraday rotation which is the nonreciprocal
section. Resistive cards are added to filter undesired modes other
than the TE$_{10}$ mode.

\begin{figure}
\begin{center}
\hfil\includegraphics[width=4.5truein]{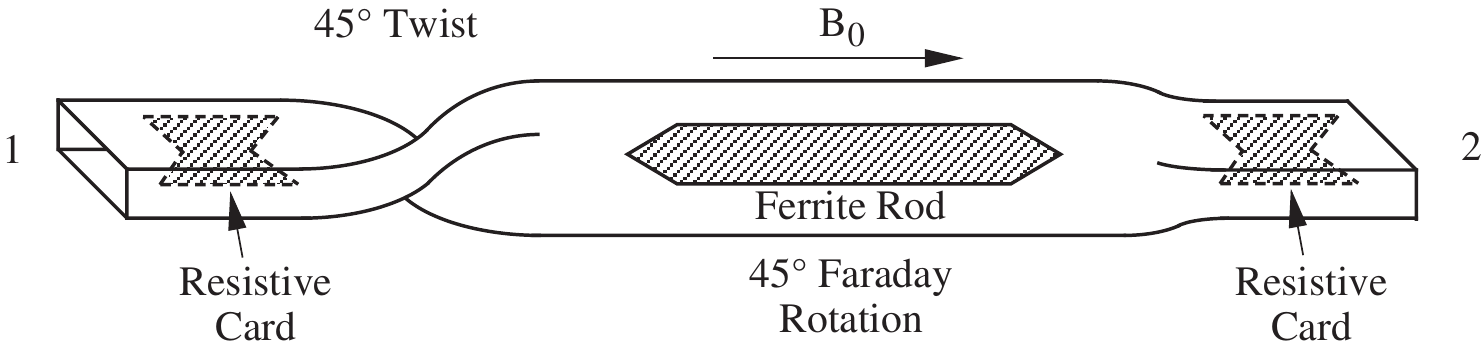}\hfil
\end{center}
\caption{A Faraday-rotation isolator.}\label{fg372}
\end{figure}


For a mode transmitting from port 1 to port 2, the mechanical
twist section rotates the polarization counterclockwise by $ 45
^o$ while the ferrite section undoes it by rotating the
polarization clockwise by $45^o$. Hence the transmission from port
1 to port 2 is little affected.

For transmission from port 2 to port 1, the rotation in the
ferrite section is counterclockwise, and hence is additive  with
respect to the rotation in the twisted section. Hence, when the
wave arrives at port 1, it has been rotated by $90^o$, and will
not be able to transmit as the TE$_{10}$ mode.

Another way to use ferrite as an isolator is to notice the
\index{Resonance behavior of ferrites}resonance behavior of ferrites. The precessing ferrite spins
acquire a larger component transverse to the static magnetic
fields when it is excited by one circular polarization compared to
the other. When there is a loss mechanism to dissipate the energy
of the spins, the polarization that excites a larger spin
amplitude will dissipate more energy into the ferrite than the
other polarization.

If we take the top view of the magnetic field distribution of the
\index{Mode!TE$_{10}$}TE$_{10}$ mode of a rectangular waveguide, the magnetic field is
actually circularly polarized away from the midsection of the
waveguide. The  polarization changes sign for a $+z$ or a $-z$
propagating wave. Therefore, if a waveguide is loaded with
ferrites away from the middle of the waveguide as shown, it will
attenuate a TE$_{10}$ mode by a different amount depending on the
direction of propagation of the mode. This can be used as an
isolator, and it is called the \index{Resonance isolator}resonance isolator.

\begin{figure}
\vspace{1.2 in}
\includegraphics[width=4.5truein]{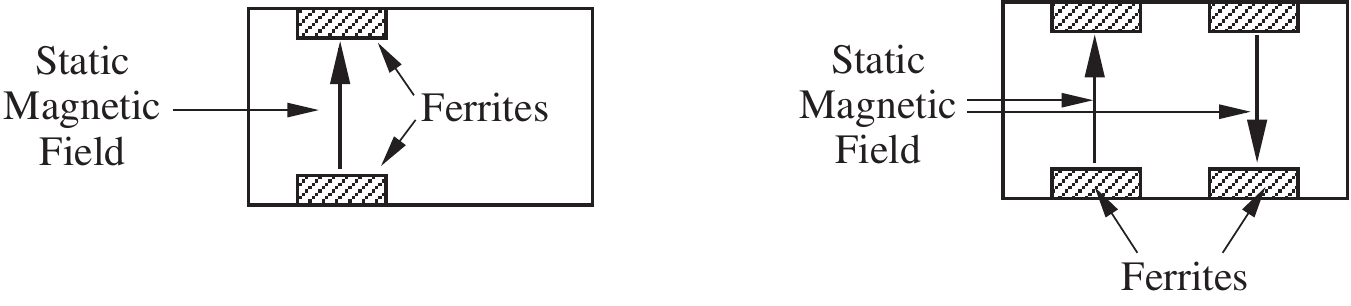}
\caption{Ferrite loading for resonance isolators.}\label{fg373}
\end{figure}



\subsection{Spintronics}
\index{Spintronics}

A fervent area of research related to spin dynamics is spintronics,
the art of making electronic devices by exploiting spin physics.
Some materials are highly magnetic because there are many unpaired
electrons in these materials. These unpaired electrons give these
materials magnetic dipole moments, such as ferromagnetic materials.
In \index{Ferromagnets}ferromagnets, the magnetic dipoles of the same orientation
cluster together to form microscopic domains that are random on a
macroscopic scale. These domains can be aligned macroscopically by
magnetization, making these materials into magnets.  For instance,
Fe, Co, Ni and their alloys have this property.

The conduction property of these materials can also be affected by
remnant magnetic field in the domain, or externally applied magnetic
field. In the presence of an external magnetic field, the energy
levels of the down spins are much higher than the energy levels of
the up spins. Hence, the down spins can be pushed into their
conduction band while the up spins are relegated to the valence
band. A flux of down-spin electrons can pass through such a medium,
but not the up-spin electrons.  This phenomenon can be used to
generate giant magneto-resistance (GMR), and it has been used to
generate magnetic storage devices.

\vskip 20pt

\centerline{ \bf Exercises for Chapter 3 }

\vskip 12pt \noindent {\bf Problem 3-1:} \noindent Prove the
identities in Equations (\ref{eq1-5a}) (\ref{eq1-5b}) of Section
3.1.

\vskip 8pt \noindent {\bf Problem 3-2:} Consider the scattering
problem involving a circular dielectric rod. The incident wave is
described by a TM wave given as
$$
E_z= E_0 J_0(k_\rho \rho) e^{ik_z z}.
$$
Match boundary condition at the surface of the dielectric rod, and
find the scattered field.  Does the scattered field involve both
TE and TM wave?  Explain why.

\vskip 8pt \noindent {\bf Problem 3-3:} If this world is dominated
by left-handed people, we may have used left-hand rule in cross
products rather than right-hand rule. How should Maxwell's
equations be rewritten if left-hand rule is used instead? Would
the law of electromagnetics be affected if left-hand rule is used
instead?

\vskip 8pt \noindent {\bf Problem 3-4:} Explain why if a general
uniform waveguide is filled with an anisotropic material such that
$\dyadg\epsilon$ and $\dyadg\mu$ are $3\times 3$ tensors, and that
$\epsilon_{xz}$, $\epsilon_{yz}$, $\epsilon_{zx}$,
$\epsilon_{zy}$, $\mu_{xz}$, $\mu_{yz}$, $\mu_{zx}$, and
$\mu_{zy}$ are all not zero, then the waveguide does not have
reflection symmetry. That is the waveguide in the mirrored world
is not the original waveguide anymore even after a 180$^o$
rotation.

\vskip 8pt \noindent {\bf Problem 3-5:}
\begin{itemize}
\item[(a)]
Prove that if two matrices are the transpose of each other, they
share the same eigenvalues. Do they share the same eigenvectors
also? Proof that their eigenvectors are orthogonal to each other.
\item[(b)] Prove that Equation (\ref{eq2-7}) is negative-transpose to Equation (\ref{eq2-6}) in Section \ref{Section3.3}.
\end{itemize}

\begin{figure}
\hfil\includegraphics[width=2.5truein]{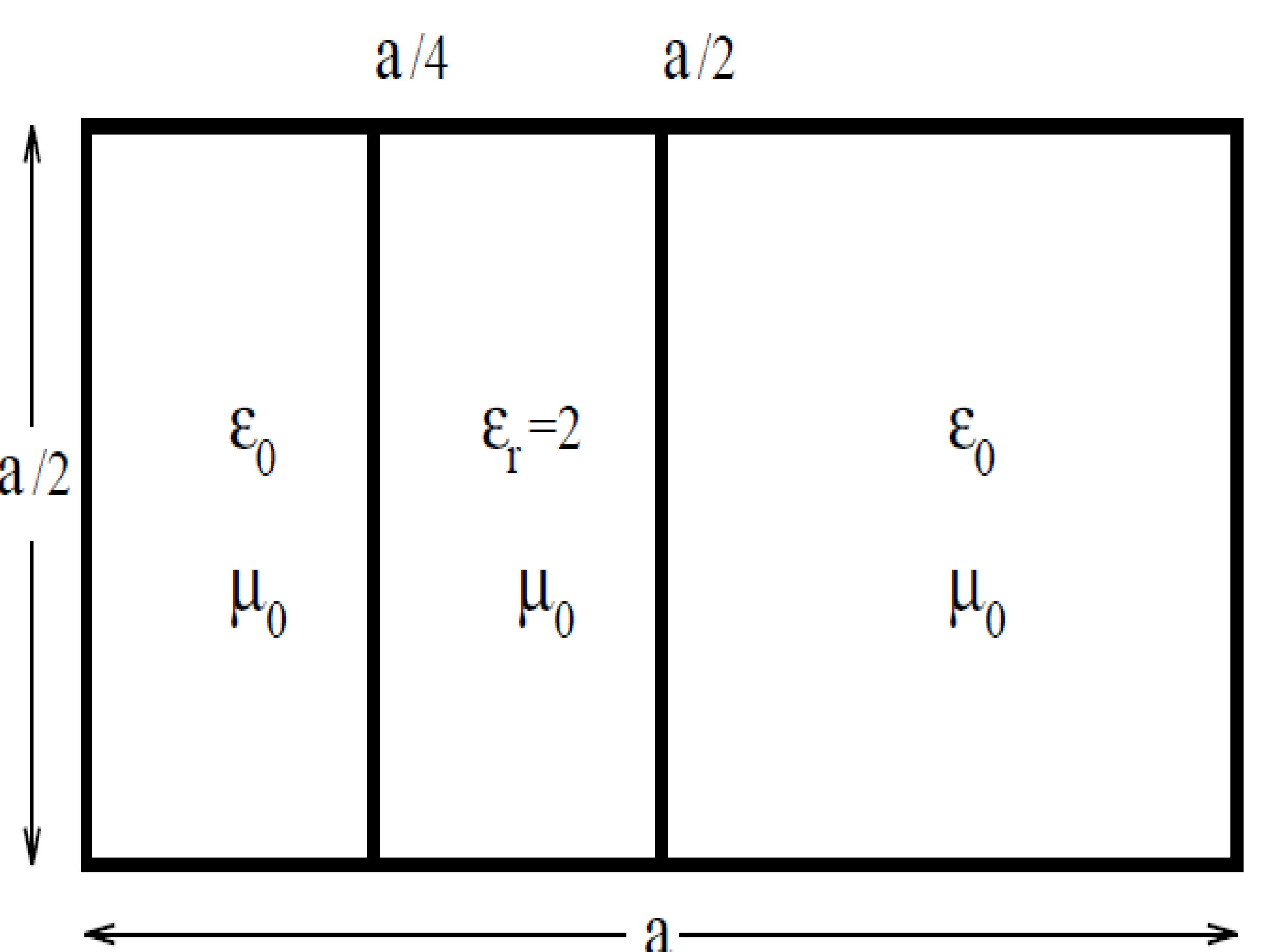}\hfil
\caption{Problem 3-5}
\end{figure}


\noindent {\bf Problem 3-6:}
 For the dielectric-slab-loaded
rectangular waveguide shown:
\begin{itemize}
\item [(a)]
Assume that the dielectric slab is absent, plot the dispersion
curve for the TE$_{10}$ and TE$_{01}$ modes above cut-off. That is
plot $k_z$ as a function of frequency.
\item[(b)] Now with the dielectric slab in place, the guidance properties of
the aforementioned two modes will be perturbed. Write down the
guidance conditions from which you can find the dispersion curves
of the perturbed TE$_{10}$ and TE$_{01}$ modes.  Define all the
variables in the guidance conditions so that if you need to
calculate these guidance conditions, you know how to.
\item[(c)]
Write a computer program to solve for the roots of the guidance
conditions and plot the dispersion curves for the perturbed modes.
(Note: You can use a Muller root solver which is available in
IMSL.)
\end{itemize}

\noindent {\bf Problem 3-7:} Simplify Equation (\ref{eq5-24}) of
Section 3.5 for the $n=0$ mode, or the axially symmetric mode.

\vskip 8pt \noindent {\bf Problem 3-8:} For an inhomogeneously
filled waveguide, the equations governing the $E_z$ and $H_z$
components of the fields are
$$
\mu \nabla_s\cdot\mu^{-1}\nabla_s E_z -
ik_z\mu\nabla_s\cdot\left[\frac{ik_z}{\mu k_s^2}\nabla_s E_z
+\frac{i\omega}{k_s^2} \nabla_s\times \v H_z\right] +
\omega^2\mu\epsilon E_z = 0,
$$
$$
\epsilon \nabla_s\cdot\epsilon^{-1}\nabla_s H_z -
ik_z\epsilon\nabla_s\cdot\left[\frac{ik_z}{\epsilon k_s^2}\nabla_s
H_z -\frac{i\omega}{k_s^2} \nabla_s\times \v E_z\right] +
\omega^2\mu\epsilon H_z = 0.
$$
\begin{itemize}
\item[(a)]
For inhomogeneities which are piecewise constant, show that these
two equations are coupled only at the discontinuities of the
piecewise constant inhomogeneity.
\item[(b)]
{For a homogeneously filled waveguide, show that the two equations
are decoupled from each other.  What are the equations in this
case?}
\end{itemize}

\begin{figure}
\hfil\includegraphics[width=2.5truein]{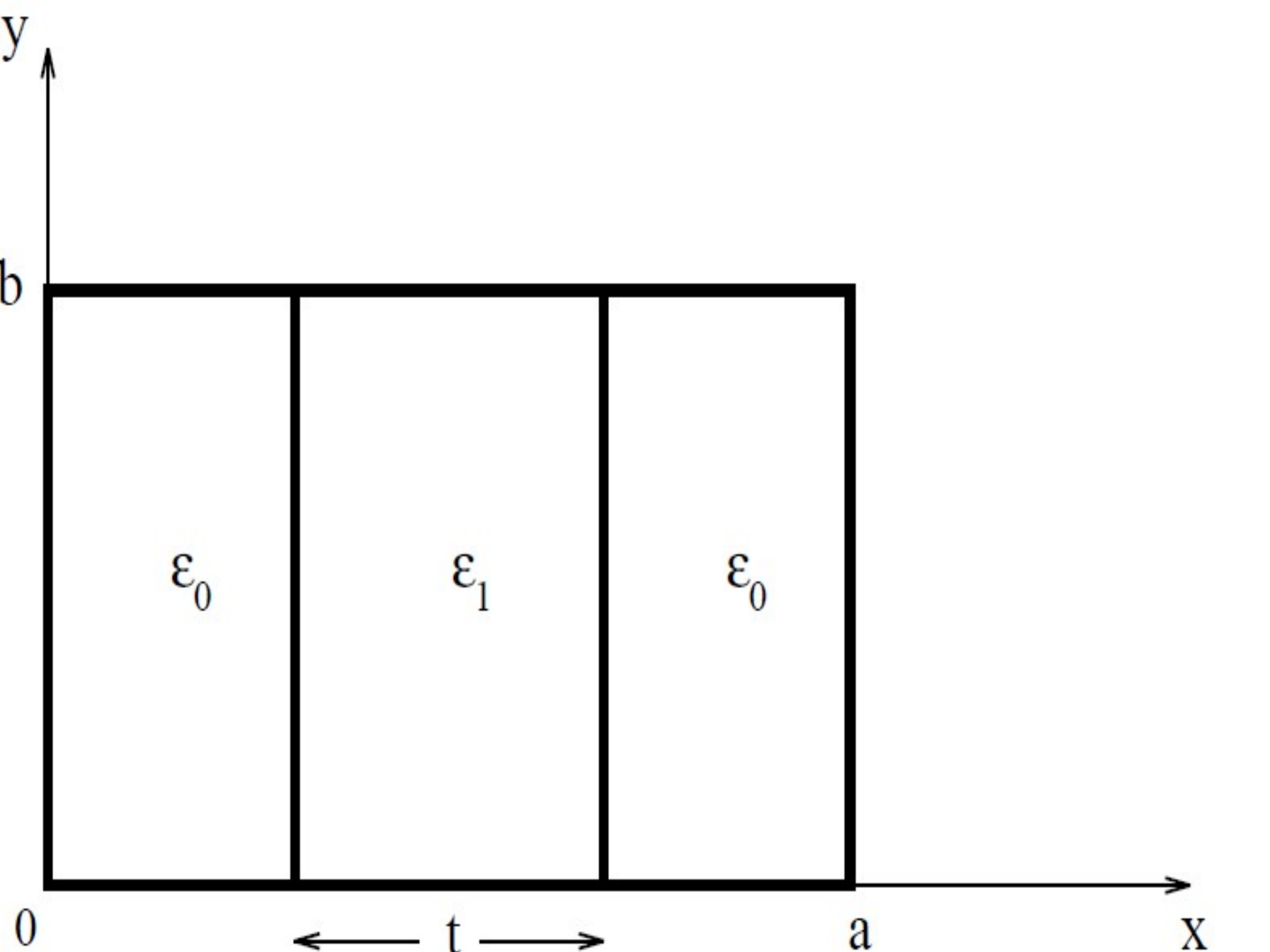}\hfil
\caption{Problem 3-9}
\end{figure}

\noindent {\bf Problem 3-9:}


\noindent Find the guidance conditions for the symmetrically
placed slab in the rectangular waveguide for the LSM and LSE
modes. Simplify the expressions for the guidance conditions as
much as possible.

\vskip 8pt \noindent {\bf Problem 3-10:} \noindent If the equation
for spin precession now has small loss terms, so that $P_z
\rightarrow P_{z0}$, and $P_s\rightarrow 0$, when $t\rightarrow
\infty$, namely,
$$
\frac{d \v P}{dt} = \vg \omega _0 \times \v P - \v P_s \frac{1}{T_2}
-\hat z \left(P_z-P_{z0}\right) \frac{1}{T_1},
$$
where $T_1$ is the relaxation time for $P_z$, and $T_2$ is the
relaxation for $\v P_s$. Find the solution to the above equation.
What is the steady solution if a time-harmonic RF $\v B_1$ field as
in \eqref{eq7-8000} is applied to excite the system?


\vskip 8pt \noindent {\bf Problem 3-11:} \noindent Analyze the
problem in Subsection 3.10.1 when the wave is propagating in the
negative $z$ direction for Faraday rotation.

\bibliographystyle{plain}
\bibliography{emt,matrix,fmm,math,fastsum,parallel,cs,my}


\def\chaptitle{Coupling of Waveguides and Cavities}
\index{Coupling of waveguides and cavities}

\chapter{\chaptitle}

\markboth{\smallbooktitle}{\chaptitle}



\def\v #1{{\bf #1}}
\def\dyad#1{\overline {\bf #1}}
\def\vg #1{\mbox{\boldmath$#1$}}
\def\beq{\begin{equation}}\def\eeq{\end{equation}}
\def\tinf{\text{\it inf\,}}\def\^{\hat}
\def\cal#1{\mathcal{#1}}



\noindent


Once we have a waveguide or a cavity, it is important to know how to
couple energy into it. Energy can be coupled into a waveguide by use
of a probe, or an aperture, or simply just by connecting one
waveguide to another.   We will first study the excitation of modes
in a waveguide by the use of a probe.  We will develop the integral
equation from which such a problem can be solved exactly, as well as
calculating the input impedance by a variational formula.  Such
method can also be applied as well to cavity coupling, and coupling
of electromagnetic energy into free space as in antennas.

To study aperture coupling, we will also discuss the pertinent
equivalent principle needed. Such problems have also been addressed
in  \cite{COLLINH,COLLINI}.

\section{{{ Excitation of Waveguides by a Probe}}}
\index{Waveguide!excitation by a probe}

\setcounter{equation}{0} \setcounter{figure}{0}

\begin{figure}[htb]
\begin{center}
\hfil\includegraphics[width=4.5truein]{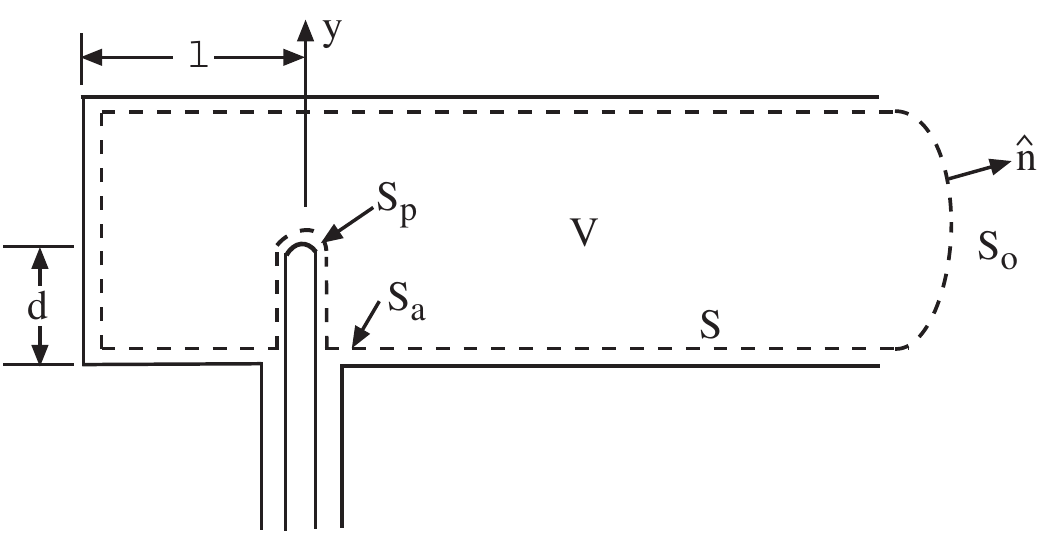}\hfil
\end{center}
\caption{Excitation of a cylindrical, hollow waveguide by a
probe.}\label{fg411}
\end{figure}

We will study the coupling of modes from a \index{Coaxial line}coaxial line to a
waveguide as shown in Figure \ref{fg411}. Coaxial line is quite
prevalent, and its characteristics can be easily understood via
transmission line theory.  This coupling problem is important in
understanding the transfer of power from a coaxial cable to a
waveguide system. By the proper adjustment of $d$ and $l$, we can
cause almost all the power from the coaxial cable to be
transferred to the waveguide.  From a transmission line theory
viewpoint, the coupling to the waveguide is reflected in the
transmission line being terminated with a load.  The load can be
changed by the proper adjustment of the dimension and location of
the probe until a matched load is arrived at.

\subsection { Derivation of the Equivalent Problem and the Integral Equation}

In the waveguide volume enclosed by the surface $S$, the electric
field satisfies the following vector wave equation
\begin{equation}
\nabla\times\nabla\times\v E(\v r)-k^2\v E(\v r)=0. \label{eq4-1}
\end{equation}
A \index{Green's function!dyadic}dyadic Green's function is defined to be a solution to the
following equation, i.e.,
\begin{equation}
\nabla\times\nabla\times\dyad G(\v r, \v r')-k^2\dyad G(\v r, \v
r')=\dyad I\delta (\v r-\v r'). \label{eq4-2}
\end{equation}
Dot-multiplying (\ref{eq4-1}) by $\dyad G(\v r,\v r')$ and
(\ref{eq4-2}) by $\v E(\v r)$, upon subtraction and integration
over $V$, we have
\begin{equation}
\int_V d\v r[ \v E(\v r)\cdot\nabla\times\nabla\times\dyad G(\v r,
\v r')-\nabla\times\nabla\times\v E(\v r)\cdot\dyad G(\v r, \v
r')]=\v E(\v r'). \label{eq4-3}
\end{equation}
Using
\begin{equation}
\begin{split}
& \nabla \cdot [-\v E(\v r)\times\nabla\times\dyad G (\v r, \v
r')-(\nabla\times\v E(\v r))\times\dyad G(\v r, \v r')]\\ & =\v
E(\v r)\cdot\nabla\times\nabla\times\dyad G(\v r. \v
r')-\nabla\times\nabla\times\v E(\v r)\cdot\dyad G(\v r, \v r'),
\label{eq4-4}
\end{split}
\end{equation}
and the divergence theorem, then
\begin{equation}
\v E(\v r')=-\int_S dS\^ n \cdot[\v E(\v r)\times\nabla\times\dyad
G(\v r, \v r')+(\nabla\times\v E(\v r))\times\dyad G(\v r, \v
r')],\quad\v r'\in V. \label{eq4-5}
\end{equation}
The above can also be written as
\begin{equation}
\v E(\v r)=-\int _S dS'[ \^ n'\times\v E(\v
r')\cdot\nabla'\times\dyad G(\v r', \v r)+i\omega\mu\^ n'\times\v
H(\v r')\cdot\dyad G(\v r', \v r)],\quad\v r\in V. \label{eq4-6}
\end{equation}
The above is actually a statement of \index{Huygens' principle}Huygens' principle for vector
electromagnetic field: Given the knowledge of tangential $\v E$ and
$\v H$ fields on the closed surface of a volume $V$, the field is
known everywhere inside $V$.\footnote{In \eqref{eq4-3}, if $\v r'$
is outside $V$, the right-hand side will evaluate to zero, leading
to the left-hand side of \eqref{eq4-6} to be zero.  This identity is
known as the extinction theorem.}

\begin{figure}[htb]
\vspace{-0.1 in}
\begin{center}
\hfil\includegraphics[width=4.0truein]{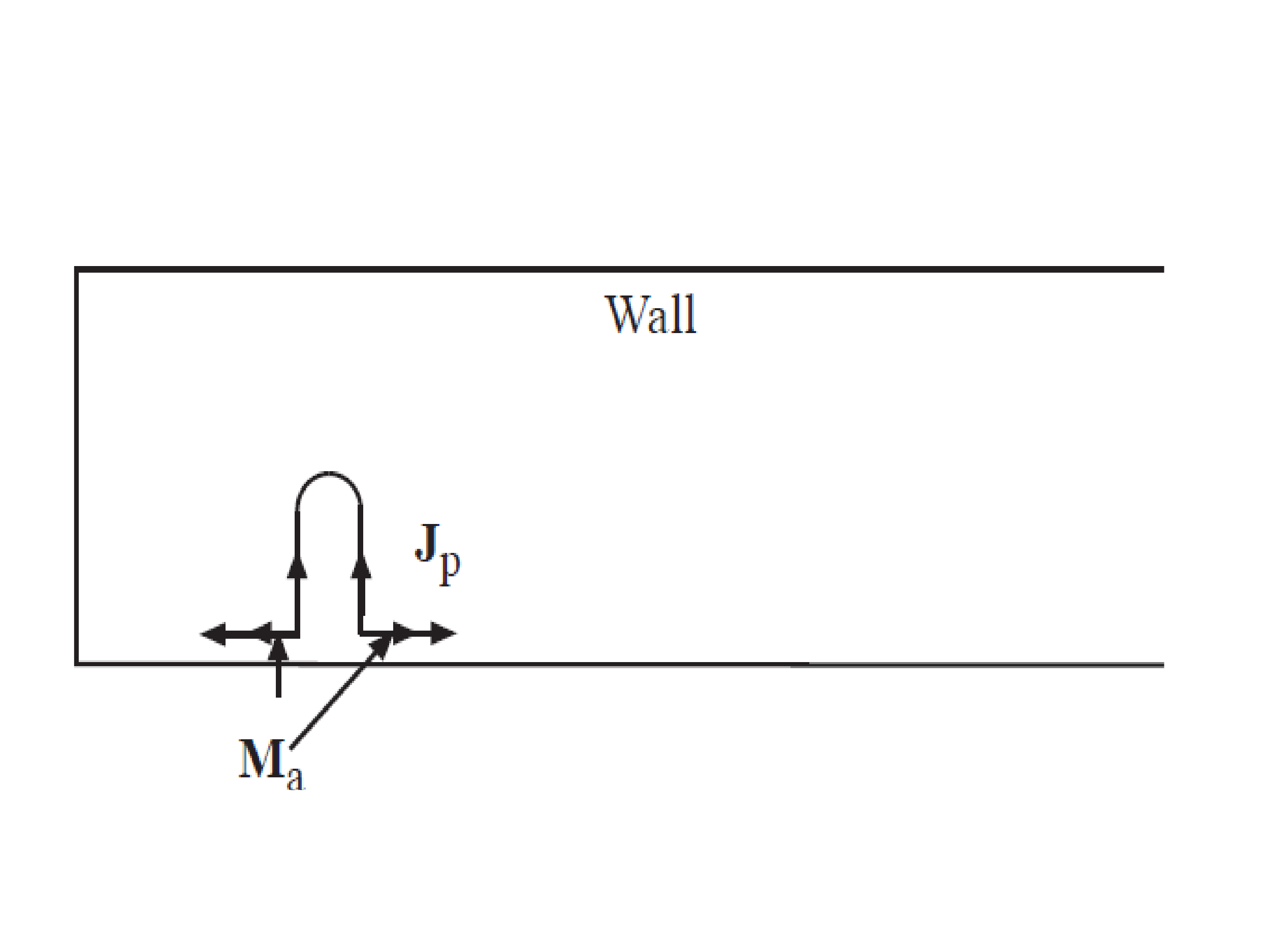}\hfil
\end{center}
\vspace{-0.5 in}
\caption{Equivalent sources for the probe excitation of a
waveguide.}\label{fg412}
\end{figure}

Furthermore, we can define the \index{Green's function!electric dyadic}electric dyadic Green's function that
will generate an electric field from an electric point current
source. We will label such a Green's function $\dyad G_e(\v r,\v
r')$ satisfying the requisite boundary condition on the waveguide
wall as shown in Figure \ref{fg412}. Similarly, we can define a
\index{Green's function!magnetic dyadic}magnetic dyadic Green's function that generates a magnetic field
from a magnetic point current source, and label it $\dyad G_m(\v
r,\v r')$.

Moreover, using the reciprocity relations, it can be shown that
\cite[p. 32]{WFIMF} [see also Problem 4-1]
\begin{subequations}
\begin{equation}
[\dyad G_e(\v r', \v r)]^t =\dyad G_e(\v r, \v r'), \label{eq4-11a}
\end{equation}
\begin{equation}
[\nabla ' \times \dyad G_e (\v r', \v r) ]^t = \nabla \times \dyad
G_m (\v r, \v r') \label{eq4-11b}
\end{equation}
\end{subequations}
Hence, we can rewrite the above as
\begin{equation}
\v E(\v r)=-\int _S dS'[ \nabla\times\dyad G_m(\v r, \v r')\cdot \^
n'\times\v E(\v r')+i\omega\mu\dyad G_e(\v r, \v r')\cdot \^
n'\times\v H(\v r')],\quad\v r\in V. \label{eq4-6a}
\end{equation}

At this point, we have not specified the boundary conditions to be
satisfied by the above dyadic Green's functions except that they are
solutions of (\ref{eq4-2}). A convenient choice for the electric
dyadic Green's function for the waveguide is that it satisfies the
boundary condition on the waveguide wall\index{Waveguide!boundary conditions} that
\begin{equation}
\^ n\times\dyad G_e(\v r, \v r')\cdot\v a=0, \qquad \v r\in \text
{wall}, \label{eq4-7}
\end{equation}
where $\v a$ is an arbitrary vector. In other words,
\begin{equation}
\^ n\times\dyad G_e(\v r, \v r')=0, \qquad \v r\in \text {wall},
\label{eq4-7a}
\end{equation}
 Also, $\dyad G_m(\v r,\v r')$ generates the magnetic field inside a waveguide due to
a magnetic current source. Its curl produces the electric field with
zero tangential component on the waveguide wall, or
\begin{equation}
\^ n\times\nabla\times\dyad G_m(\v r, \v r')=0, \qquad \v r\in \text
{wall}, \label{eq4-7b}
\end{equation}

Consequently, it is seen that the electric field thus generated in
\eqref{eq4-6a} has zero tangential component on the waveguide wall.
Hence, $\^ n'\times \v E(\v r')=0$ on the waveguide wall except for
$S_a$. Also, $\dyad G_e(\v r, \v r')\cdot \^ n'\times\v H(\v r')$ is
zero on the waveguide wall by reciprocity or by taking the transpose
of this expression and applying \eqref{eq4-7}.  In other words,
\begin{equation} \dyad G_e(\v r, \v r')\cdot\^ n'\times\v
H(\v r')= \^ n'\times\v H(\v r')\cdot\dyad G_e(\v r', \v r)=-\v H(\v
r')\cdot\^ n'\times\dyad G_e(\v r', \v r)=0, \qquad\v r'\in \text
{wall}. \label{eq4-8}
\end{equation}
Hence, the second integral in (\ref{eq4-6a}) is nonzero only on
$S_p$.   Therefore, (\ref{eq4-6a}) can be rewritten as
\begin{equation}
\v E(\v r)=-\int _{S_a} dS' \nabla\times\dyad G_m(\v r, \v r')\cdot
\^ n'\times\v E(\v r')-i\omega\mu \int _{S_p} dS'\dyad G_e(\v r, \v
r')\cdot \^ n'\times\v H(\v r'),\quad\v r\in V. \label{eq4-6b}
\end{equation}

The integral over $S_0$ in Figure \ref{fg411} can be made to vanish
by taking $S_0$ to infinity and introducing an infinitesimal amount
of loss. Also, $-\^ n'\times\v H(\v r')$ can be identified as $\v
J_p(\v r')$, the surface current on the probe, and $\^ n'\times\v
E(\v r')$ can be identified as $\v M_a(\v r')$, as equivalent
magnetic current impressed on $S_a$. Rewriting (\ref{eq4-6b}), it
becomes
\begin{equation}
\v E(\v r)=-\int _{S_a} dS'\nabla \times \dyad G_m (\v r, \v r')
\cdot \v M_a (\v r') + i\omega\mu\int _{S_p} dS'\dyad G_e(\v r, \v
r')\cdot\v J_p(\v r'). \label{eq4-12}
\end{equation}
If $\v M_a(\v r')$ is assumed known, then using $\^ n\times\v E(\v
r)=0$ on $S_p$ as the additional boundary condition, an integral
equation can be set up
\begin{equation}
i\omega\mu\^ n\times\int _{S_p} dS'\dyad G_e(\v r, \v r')\cdot\v
J_p(\v r')=\^ n\times\int _{S_a} dS'\nabla \times \dyad G_m (\v r,
\v r') \cdot \v M_a(\v r'), \label{eq4-13}
\end{equation}
from which $\v J_p(\v r')$ can be solved for. In Equation
(\ref{eq4-12}), the problem in Figure \ref{fg411} is replaced with
an equivalent problem consisting of impressed current $\v J_p$ on
$S_p$ and $\v M_a$ on $S_a$.  These are impressed currents because
they are sources impressed in the space $V$ radiating via the
waveguide dyadic Green's function.


\subsection{Generalization to Other Structures}
\begin{figure}[!h]
    \centering
    \includegraphics[height=5cm]{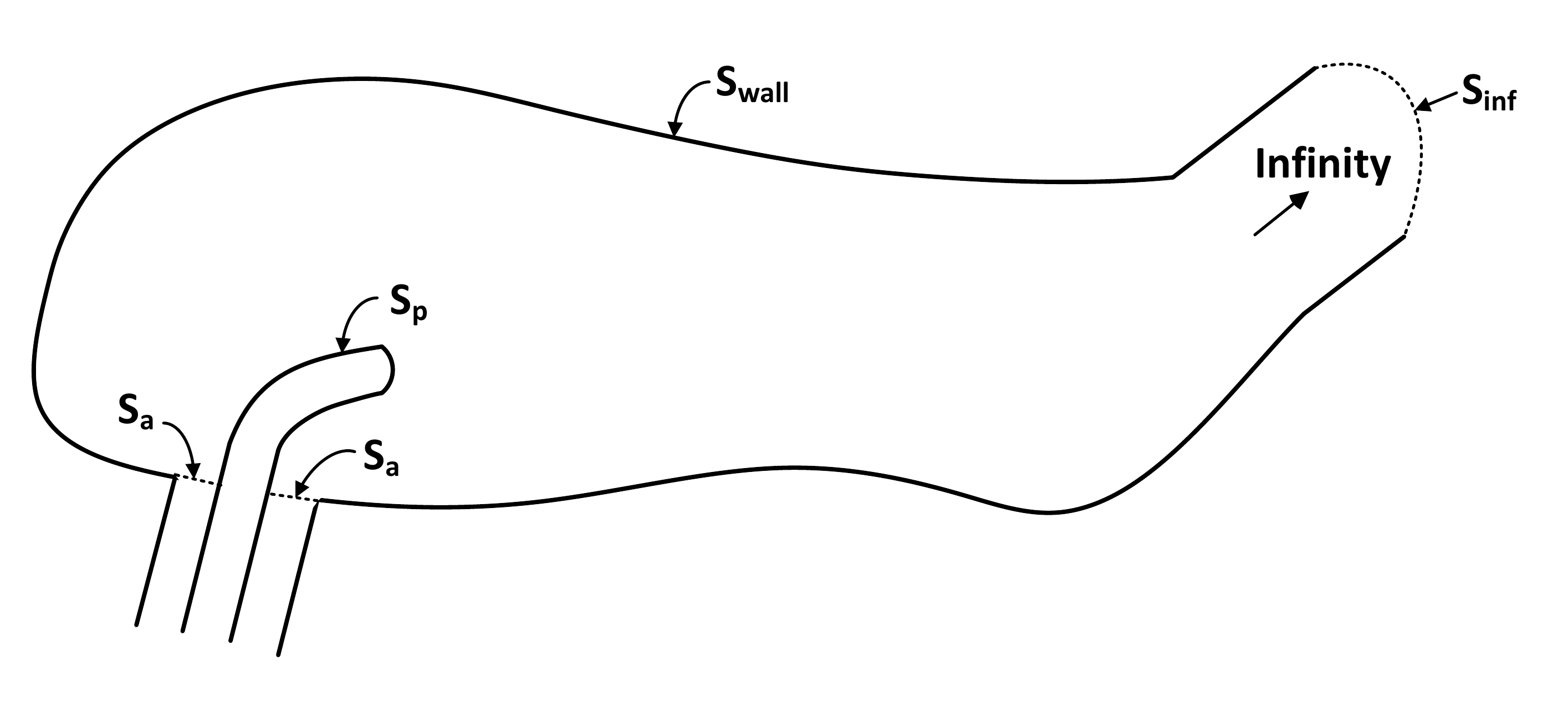}
    \includegraphics[height=5cm]{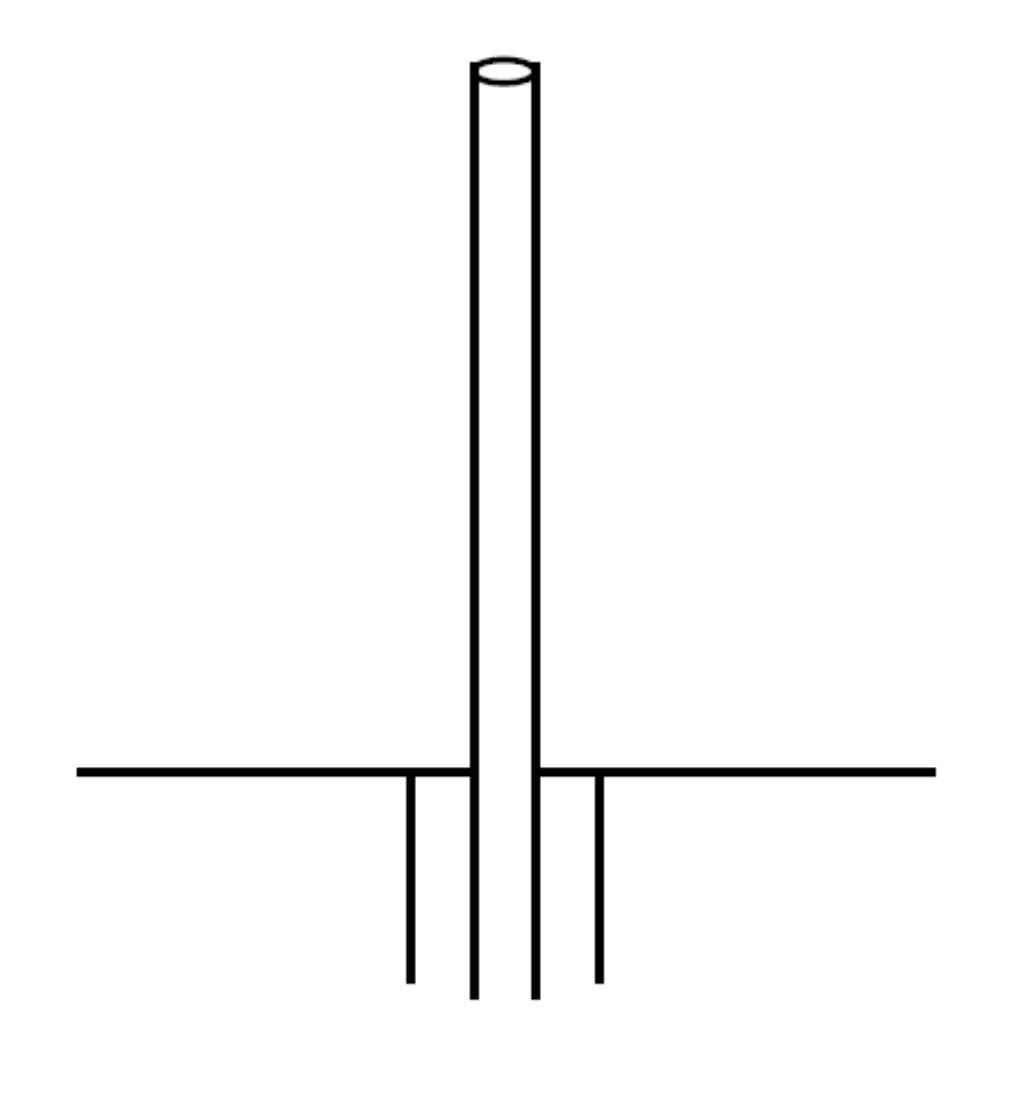}
    \includegraphics[height=5cm]{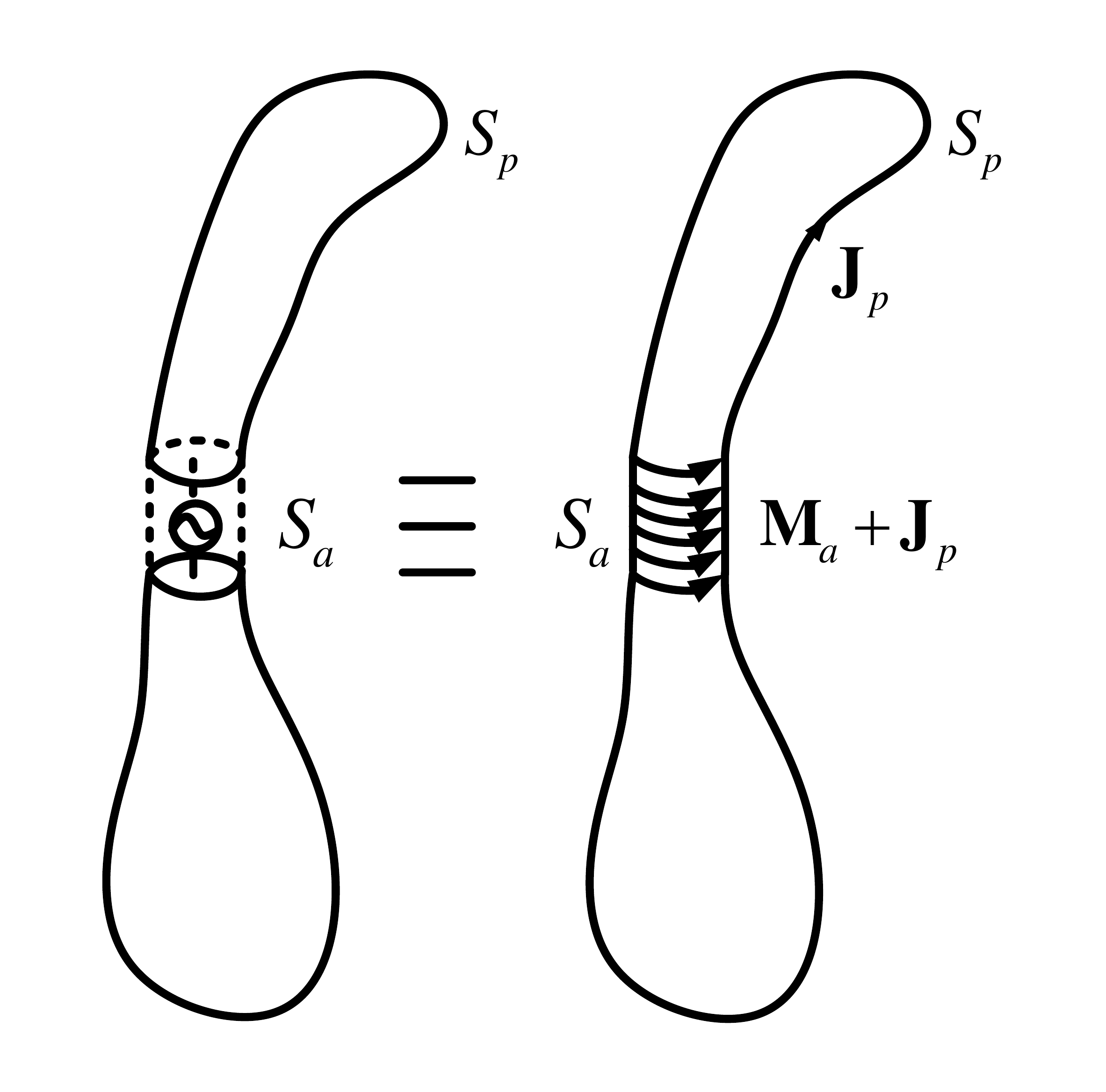}
    \caption{Excitation of a different structure with a magnetic current
    source.
    Top: A coaxial driven probe exciting an arbitrary waveguide.
    Bottom left: A monopole antenna driven by a coax via a ground plane.
    Bottom right: A general antenna driven by a voltage source
    represented by an equivalent magnetic current source.
    }
    \label{FigureGeneral}
    \end{figure}

 The above theory is quite general.  It obviously applies to
arbitrarily shaped waveguides or cavities.  When applied to an
arbitrarily shaped structure as shown in Figure \ref{FigureGeneral},
the onus is on finding the dyadic Green's function.  However, one
can also use free-space dyadic Green's function, but at the expense
of adding more unknowns to the integral equation.

The above formulation can  also be applied to a monopole antenna
mounted on a ground plane driven by a coaxial cable from below.  In
this case, we can use a dyadic Green's function that satisfies the
boundary condition of a metallic half space.  This Green's function
can be found easily using image theorem.

It can also be applied to a metallic antenna driven by a magnetic
current source in free space.  We can apply the free-space dyadic
Green's function in this case.  One can assume that $S_a$ and $S_p$
form a closed surface. In this case, there are impressed electric
current on the surface $S_p$.  But on the surface $S_a$, there would
be both impressed magnetic and electric current.

In the above cases, there is a bounding surface at infinity,
$S_{inf}$ that has to be included. By use of the radiation
condition, the contribution from this bounding surface can be shown
to vanish.

\section {{{ Input Impedance of the Probe}}}

The input impedance of a probe exciting a waveguide can be
calculated. This yields information on how the position and length
of the probe can be adjusted to arrive at the input impedance we
desire.  For maximum transfer of the power, the input impedance
should be matched to the characteristic impedance of the
transmission line to minimize reflections.

\subsection { Variational Expressions for Input Admittance}
\index{Variational expression}
\index{Input admittance}

Variational expressions for the input impedance of an antenna has
been discussed in  \cite{HARRINGTON4,JORDAN_BALMAIN,KONGD}. Such
expressions can also be used to find the input impedance of a probe
inside the waveguide, as the two problems are very similar. The
difference appears mainly in the Green's function.  In an antenna,
the source is radiating in free space, and hence, free-space Green's
function usually suffices.  However, in a waveguide, the free-space
Green's function has to be replaced by the Green's function of the
waveguide structure.

A variational expression for the input impedance or admittance of a
source driven complex structure can be derived. The source can be
either a voltage source or a current source.   When it is a voltage
source, it is a magnetic current ring, and when it is a current
source, it is an electric dipole such as a Hertzian dipole or its
equivalence.  The case of a current source has been adequately
discussed in  \cite{HARRINGTON}.   The voltage source is usually
modeled by a magnetic ring current, a ribbon current, or a magnetic
frill.  The case of a magnetic current exciting an antenna structure
has been discussed in  \cite{JORDAN_BALMAIN}.   However, our
discussion here is in accordance to  \cite{CHEW_NIE_LIU_LO}, which
is different from the previous treatment on the subject. However,
the variational formula for a magnetic current source driven antenna
seems to have been missed until presented in
\cite{CHEW_NIE_LIU_LO}.

In any case, the input impedance of the structure is predominantly
determined by the induced current on the structure.  A variational
expression has the advantage of yielding a second order error for
the input impedance when the error of the current on the structure
is first order.  The current on the complex structure can be
roughly estimated, or solved for from an integral equation.

If a probe current produces a magnetic field $\v H_p$, and the
aperture magnetic current produces an electric field $\v E_a$ and
a magnetic field $\v H_a$, then
\begin{equation}
-\langle \v M_a, \v H_T\rangle =-\int\limits _{S_a}dS'\v M_a(\v
r')\cdot \v H_T(\v r')=-\int\limits _{S_a}dS'\^ n'\cdot (\v
E_a\times \v H_T) \label{eq4-14}
\end{equation}
where $\v H_T=\v H_p+\v H_a$. On the aperture $S_a$, we can assume
that only the TEM mode of the coax is important. Therefore,
\begin{equation}
\v E_a=\^ \rho E_0, \quad \v H_T=\^ \phi H_0, \label{eq4-15}
\end{equation}
Substituting \eqref{eq4-15} into \eqref{eq4-14}, we have
\begin{equation}
-\int\limits _{S_a}dS'\^ n\cdot (\v E_a\times \v H_T)=\int\limits
_a^bd\rho' E_0\int\limits _0^{2\pi }\rho' d\phi' H_0=VI.
\label{eq4-16}
\end{equation}
In the above, the fields are represented in a local coordinate
system, and $\hat n'=-\hat z$.  Also,
 even though $\v H_T$ may depart from the \index{Mode!TEM}TEM mode
field of a waveguide, by the mode orthogonality theorem, only the
TEM-mode component of $\v H_T$ will contribute to the integral in
Equation (\ref{eq4-16}).

Does this mean that the higher-order modes in $\v H_T$ do not
contribute to the current $I$?  No, it does not. The higher-order
modes do contribute to the current $I$ near the aperture, but
their effect diminishes rapidly away from the aperture. Therefore,
the current $I$ in (\ref{eq4-16}) is only the TEM component of the
current, which can be easily related to the current in the rest of
the coaxial cable by transmission line theory.

Consequently,
\begin{equation}
VI=-\langle \v M_a, \v H_T\rangle. \label{eq4-17}
\end{equation}
By letting $I=Y_{in}V$, we deduce that
\begin{equation}
Y_{in}=-\frac {\langle \v M_a, \v H_T\rangle }{V^2}.
\label{eq4-18}
\end{equation}

The above is an expression for the input admittance of the probe
assuming only the TEM mode in the coax, but it is not variational.
To derive a variational expression, we write
 \cite{CHEW_NIE_LIU_LO}
\begin{equation}
Y_{in}=\frac {-\langle \v M_a, \v H_T\rangle +\langle \v J_p, \v
E_T\rangle }{V^2} \label{eq4-19}
\end{equation}
where $\v E_T=\v E_a+\v E_p$, the total electric field produced
both by $\v M_a$ and $\v J_p$. The above also falls under the
category of the reaction formula for the input impedance of
antennas. Notice that if $\v E_T$ is exact, then the tangential
component of $\v E_T$ is zero on the probe surface and $\langle \v
J_p, \v E_T\rangle $ would be zero. However, the second term in
the numerator of (\ref{eq4-19}) is required to make it a
variational expression. In other words, first order error $\v J_p$
will result in a second order error in $Y_{in}$. To prove that
(\ref{eq4-19}) is variational, we let $\v M_a$ and $V$ be known
and hence fixed, and let
\begin{equation}
\v J_p=\v J_{pe}+\delta \v J, \ \ \ Y_{in}=Y_{ine}+\delta Y
\label{eq4-20}
\end{equation}
where the subscript $e$ stands for ``exact''. Cross-multiplying
(\ref{eq4-19}), and taking the first variation, we have
\begin{equation}
\delta YV^2 = -\langle \v M_{a}, \delta \v H\rangle +\langle
\delta \v J, \v E_{Te}\rangle +\langle \v J_{pe}, \delta \v
E\rangle. \label{eq4-21}
\end{equation}
From reciprocity,
\begin{equation}
\langle \delta \v J, \v E_{Te}\rangle =-\langle \v M_{a}, \delta
\v H\rangle +\langle \v J_{pe}, \delta \v E\rangle. \label{eq4-22}
\end{equation}
Then
\begin{equation}
\delta YV^2=2\langle \delta \v J, \v E_{Te}\rangle. \label{eq4-23}
\end{equation}
Since
\begin{equation}
\langle \delta \v J, \v E_{Te}\rangle =0, \label{eq4-24}
\end{equation}
because $\v E_{TE}$ has no tangential components on the probe
surface and $\delta \v J$ is purely tangential on the probe
surface, (\ref{eq4-23}) implies that
\begin{equation}
\delta Y=0. \label{eq4-25}
\end{equation}
As a result, the first variation in the admittance about the exact
admittance $Y_{ine}$ is zero. Equation (\ref{eq4-19}) is a
variational expression for the input admittance. Given $\v M_a$, $\v
J_p$ and $V$ with first order errors, the errors incurred in
$Y_{in}$ is of second order.

The variational nature of Equation (\ref{eq4-19}) can be better
appreciated if its quadratic nature is written more explicitly.  To
this end, it can be written as
\begin{equation}
Y_{in}=-\frac {\langle \v M_a, \v H_a\rangle }{V^2}-\frac {\langle
\v M_a, \v H_p\rangle }{V^2}+\frac {\langle \v J_p, \v E_p\rangle
}{V^2}+\frac {\langle \v J_p, \v E_a\rangle }{V^2} \label{eq4-34}
\end{equation}
where we assume $\v M_a$, and hence, $V$, $\v H_a$, and $\v E_a$ are
fixed. When $\v J_p$ is varied, only the last three terms would
vary.  By reciprocity, $-\langle \v M_a, \v H_p\rangle=\langle \v
J_p, \v E_a\rangle$, and the above becomes
\begin{equation}
Y_{in}=-\frac {\langle \v M_a, \v H_a\rangle }{V^2}+2\frac {\langle
\v J_p, \v E_a\rangle }{V^2}+\frac {\langle \v J_p, \v E_p\rangle
}{V^2}. \label{eq4-35}
\end{equation}
Furthermore,
\begin{equation}
\begin{split}
\langle \v J_p, \v E_p\rangle & =i\omega \mu \langle \v J_p, \dyad
G_e, \v
J_p\rangle\\
& =i\omega\mu\int\limits _{S_p}dS\v J_p(\v r)\cdot \int\limits
_{S_p}dS'\dyad G_e(\v r, \v r')\cdot \v J_p(\v r'). \label{eq4-36}
\end{split}
\end{equation}
The double commas in the above implies that there is a double
integration in the inner product.  The above is analogous to $\v
a^t\cdot \dyad A\cdot \v a$ in linear algebra.  When it is used in
\eqref{eq4-35}, it becomes
\begin{equation}
Y_{in}=-\frac {\langle \v M_a, \v H_a\rangle }{V^2}+2\frac {\langle
\v J_p, \v E_a\rangle }{V^2}+i\omega\mu\frac {\langle \v J_p, \dyad
G_e, \v J_p\rangle }{V^2}. \label{eq4-35a}
\end{equation}
The above is clearly quadratic and has a stationary point about the
exact solution.

 For an
exact $\v E_T$, $\langle \v J_p, \v E_{Te}\rangle =0$ and
(\ref{eq4-19}) reduces to (\ref{eq4-18}) again.  Since $\v H_T=\v
H_a+\v H_p$, we have from (\ref{eq4-18})
\begin{equation}
Y_{in}=-\frac {\langle \v M_a, \v H_a\rangle}{V^2}-\frac {\langle
\v M_a, \v H_p\rangle }{V^2} \label{eq4-26}
\end{equation}
for exact solutions.

Furthermore, $-\langle \v M_a, \v H_p\rangle =\langle \v J_p, \v
E_a\rangle =-\langle \v J_p, \v E_p\rangle$ from reciprocity and
that $\langle \v J_p, \v E_T\rangle =0$ for exact solutions.
Hence, (\ref{eq4-26}) becomes
\begin{equation}
Y_{in}=-\frac {\langle \v M_a, \v H_a\rangle }{V^2}-\frac {\langle
\v J_p, \v E_p\rangle }{V^2}. \label{eq4-27}
\end{equation}
If $\v M_a $ is assumed real, the first term is the complex
conjugate of the complex power $\langle \v M_a, \v H_a^*\rangle$
due to $\v M_a$ alone and can be related to the gap capacitance at
the base of the probe. The second term in (\ref{eq4-27}) is the
complex conjugate of the complex power $\langle \v E_p, \v
J_p^*\rangle$ due to $\v J_p$ alone. Hence, it is due to the probe
admittance.  Since the terms in (\ref{eq4-27}) are additive, the
gap capacitor is in parallel connection with the \index{Probe admittance}probe admittance.

It is to be reminded that in the use of (\ref{eq4-19}), $\v H_T$
and $\v E_T$ are to be calculated from $\v M_a$ and $\v J_p$.  In
particular,
\begin{equation}
\v H_p(\v r)=\int\limits _{S_p}dS'\nabla\times\dyad G_e(\v r, \v
r')\cdot \v J_p(\v r'), \label{eq4-28}
\end{equation}
\begin{equation}
\v E_a(\v r)=-\int\limits _{S_a}dS'\nabla \times \dyad G_m (\v r,
\v r') \cdot \v M_a (\v r'). \label{eq4-29}
\end{equation}

\subsection { Rayleigh-Ritz Method}
\index{Rayleigh-Ritz method}

The Rayleigh-Ritz method is named after Lord Rayleigh
 \cite{RAYLEIGH}, a prodigious English scientist, and Walter Ritz
 \cite{RITZ}, a Swiss mathematician.  It is extremely useful in
solving complex physical problems.  It seems that many physical
phenomena are always described by an equation which corresponds to
the minimization of a certain quantity.  So instead of solving the
equation directly, one can attempt to minimize the corresponding
quantity instead.

This method is extremely useful if we have an expression for
$Y_{in}$ which will have $Y_{ine}$ (exact $Y_{in}$) as the lower
bound or the upper bound to all the approximate $Y_{in}$.
(However, $Y_{in}$ is actually a complex number, but for the sake
of the ease for discussion, we will assume that $Y_{in}$ is real.)
Without loss of generality, let us discuss the lower bound case.
For example, if
\begin{equation}
Y_{in}=f(\v J_p)\geq Y_{ine}, \label{eq4-30}
\end{equation}
and the equality is satisfied only if $\v J_p = \v J_{pe}$, an
optimal value of $Y_{in}$ can be obtained even with an approximate
$\v J_p$. We can let
\begin{equation}
\v J_p=\sum _{n=1}^N a_n\v J_n, \label{eq4-31}
\end{equation}
where $\v J_n$ is a set of basis functions with which an arbitrary
$\v J_p$ can be approximated fairly well. The coefficients $a_n$'s
are yet to be determined to give the best approximation to $\v
J_p$.  The Rayleigh-Ritz procedure provides a systematic way to
determine the optimal values of $a_n$'s so as to best determine
$Y_{in}$ from (\ref{eq4-30}). If we substitute (\ref{eq4-31}) into
(\ref{eq4-30}), then
\begin{equation}
Y_{ina}=f\left( \sum _{n=1}^N a_n\v J_n\right) >Y_{ine}.
\label{eq4-32}
\end{equation}

\begin{figure}[htb]
\begin{center}
\hfil\includegraphics[width=2.40truein]{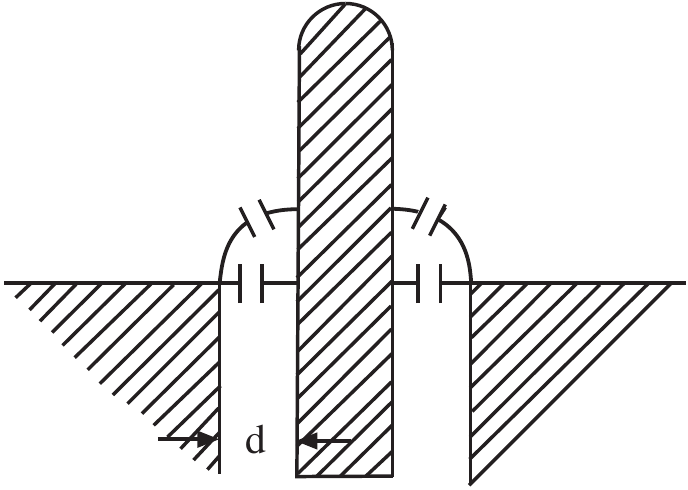}\hfil
\end{center}
\caption{Gap capacitances can be important in the input impedance of
a probe.}\label{fg421}
\end{figure}


\begin{figure}[htb]
\begin{center}
\hfil\includegraphics[width=3.5truein]{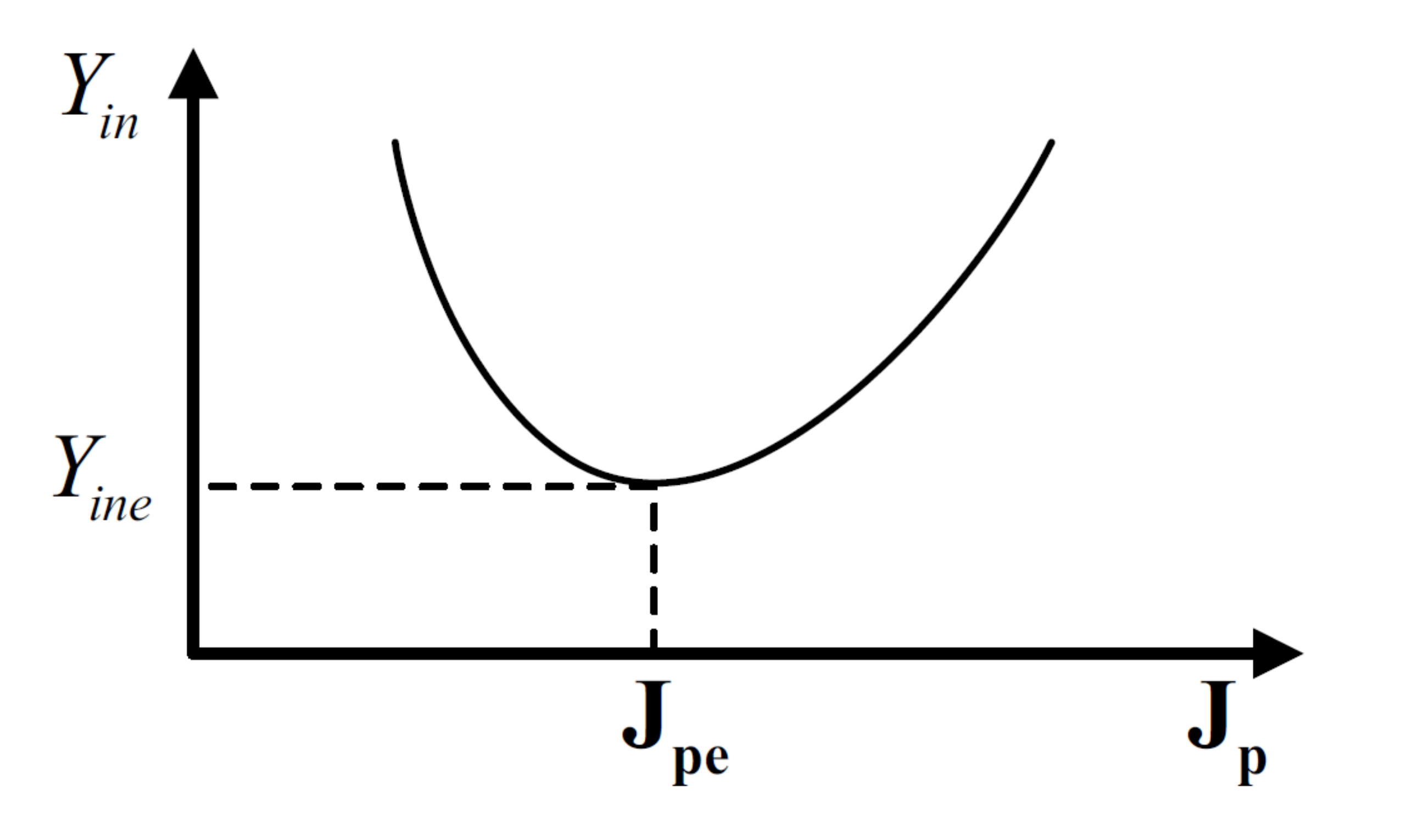}\hfil
\end{center}
\caption{A pictorial representation of a stationary point in a
multidimensional space.}\label{fg422}
\end{figure}

%

\begin{figure}[htb]
\begin{center}
\hfil\includegraphics[width=3.4truein]{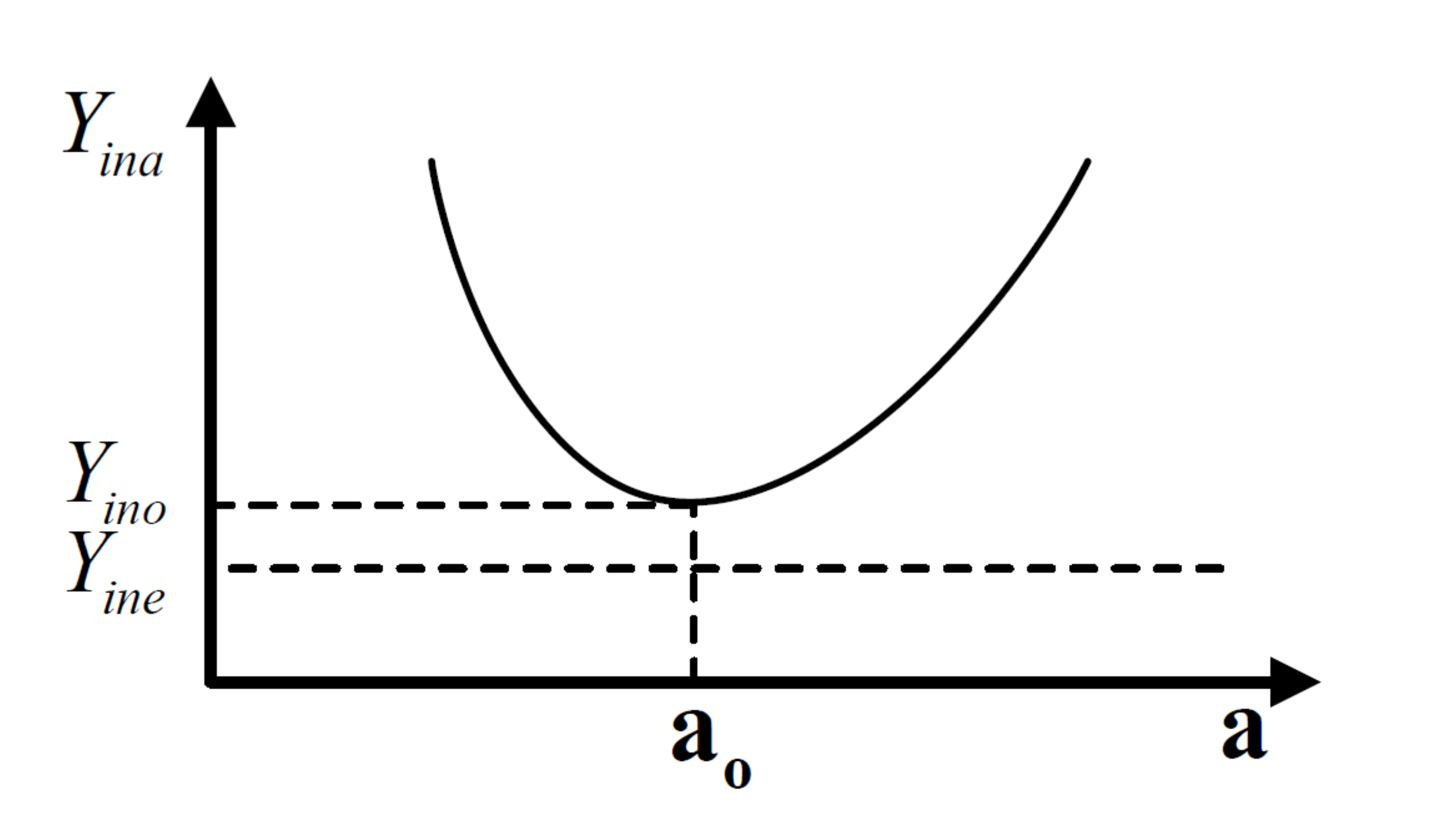}\hfil
\end{center}
\caption{The minimum $Y_{ino}$ achieved by the Rayleigh-Ritz
procedure compared to the exact minimum, $Y_{ine}$.}\label{fg423}
\end{figure}

The best choice of $a_n$'s will be one that minimizes the number
$Y_{ina}$. In other words, from Figure \ref{fg423}. the optimal
values of $a_n$'s are those that would make $Y_{ina}$ stationary.
Hence, there are $N$ equations
\begin{equation}
\frac {\partial f\left( \sum \limits _{n=1}^N a_n\v J_n\right)
}{\partial a_i}=0, \qquad i=1, \hdots , N, \label{eq4-33}
\end{equation}
from which we can solve for the optimal $a_i$'s, the $a_{io}$'s.

This concept may not work as well when $Y_{in}$ is a complex
function, or when the stationary point is not a global minimum or
maximum, but a saddle point instead. However, for these cases, the
Rayleigh-Ritz procedure converges despite as we shall explain
later.

Equation (\ref{eq4-35}) is a variational expression when the
solution is varied about $\v J_{pe}$. Hence, the Rayleigh-Ritz
procedure can be used to find the optimal $Y_{in}$. To do so, we
assume
\begin{equation}
\v J_p=\sum\limits _{n=1}^Na_n\v J_n. \label{eq4-37}
\end{equation}
Then (\ref{eq4-35a}) becomes
\begin{equation}
Y_{in}=-\frac {\langle \v M_a, \v H_a\rangle }{V^2}+2\frac
{\sum\limits _na_n \langle \v J_n, \v E_a\rangle }{V^2}+\frac
{\sum\limits _n\sum\limits _{n'}a_na_{n'} i \omega \mu \langle \v
J_n, \dyad G_e, \v J_{n'}\rangle }{V^2}.\label{eq4-38}
\end{equation}
The above is of the form
\begin{equation}
Y_{in}=-\frac {\langle\v M_a, \v H_a\rangle}{V^2}+2\frac {\v
a^t\cdot \v e}{V^2}+\frac {\v a^t\cdot \dyad A\cdot \v a}{V^2}
\label{eq4-39}
\end{equation}
where $[\v e]_n=\langle \v J_n, \v E_a\rangle$, $\left[\dyad
A\right]_{nn'}=i \omega \mu\langle \v J_n, \dyad G_e, \v
J_{n'}\rangle$, and $[a]_n=a_n$. The matrix $\dyad A$ is the matrix
representation of the dyadic Green's function $i\omega \mu\dyad
G_e$, while the vector $\v e$ is the vector representation of $\v
E$.

Taking the first variation of the above about $\v a_o$, the optimal
$\v a$, we have
\begin{equation}
\delta Y_{in}=2\frac {\delta \v a^t\cdot \v e}{V^2}+2\frac {\delta
\v a^t\cdot \dyad A\cdot \v a_o}{V^2}. \label{eq4-40}
\end{equation}
The first variation $\delta Y_{in}$ will vanish if
\begin{equation}
\dyad A\cdot \v a_o=-\v e. \label{eq4-41}
\end{equation}

The above could be solved to yield the optimal $\v a_o$ which can
in turn be used to find $\v J_p$ in (\ref{eq4-37}). $Y_{in0}$ can
also be found by the direct substitution of (\ref{eq4-41}) into
(\ref{eq4-39}) yielding
\begin{equation}
Y_{in0}=-\frac {\langle \v M_a, \v H_a\rangle}{V^2}+\frac {\v
a_o^t\cdot \v e}{V^2}. \label{eq4-42}
\end{equation}

Equation (\ref{eq4-41}) is also more directly obtained by solving
the following integral equation which follows from $-\^ n\times \v
E_a=\^ n\times \v E_p$ on the probe surface
\begin{equation}
-\^ n\times \v E_a=i \omega \mu \^ n\times \int\limits
_{S_p}dS'\dyad G_e(\v r, \v r')\cdot\v J_p(\v r'). \label{eq4-43}
\end{equation}
The use of Galerkin's method in solving (\ref{eq4-43}) yields
(\ref{eq4-41}).

The reason why the \index{Rayleigh-Ritz method!convergence}Rayleigh-Ritz procedure converges to the exact
solution is because most of the variational expressions we work
with are quadratic in nature. This quadratic nature is the
generalization of the quadratic expression in one dimension such
as $ax^2+bx+c$ to higher dimensions.  For example, (\ref{eq4-39})
is a quadratic expression in the variable $\v a$ while
(\ref{eq4-34}) is quadratic in the variable $\v J_p$.  The space
for the inner product in (\ref{eq4-39}) is an approximate finite
dimensional space, while the space for the inner product in
(\ref{eq4-34}) is infinite dimensional. All quadratic expressions
have only one stationary point. Consequently, as one increase the
dimension of the approximate finite dimensional space, the
stationary point will approach that of the infinite dimensional
space.

\subsection { Mode Matching Method---A Tour de Force Calculation}
\index{Mode matching method}

In the previous section, the equation for the input admittance,
Equation (\ref{eq4-18}) is valid only when the field at $S_a$ can
be approximated by only TEM modes. This is not true in general.  A
more accurate analysis of the input impedance of the probe
requires the use of the mode-matching method
 \cite{CHEW_NIE_LIU_LO}.  This will allow us to use higher order
modes in the \index{Waveguide!coaxial}coaxial waveguide.  It also allows us to control the
accuracy of the calculation as much as we want to.  When very high
accuracy solution is needed, we just need to add more modes in the
solution procedure.

Inside the coaxial waveguide, the field is assumed to be
\begin{equation}
\v E (\v r) = \v E_0 (\v r_s)e^{ikz} + \sum \limits
_{l=0}^{L-1}\Gamma _l \v E_l (\v r_s) e^{-ik_{lz}z},
\label{eq4-44}
\end{equation}
where $\v E_0 (\v r_s)$ is the field distribution of the TEM mode,
and $\v E_m(\v r _s)$, $m>0$ are the fields of the higher order
modes which are evanescent. The corresponding magnetic field is
\begin{equation}
\v H(\v r) = \v H_0 (\v r_s) e^{ikz} - \sum \limits _{l=0}^{L-1}
\Gamma _l \v H_l (\v r _s) e^{-ik_{lz}z} . \label{eq4-45}
\end{equation}
Hence, the total electric field at the aperture in accordance with
Equation (\ref{eq4-44}) is
\begin{equation}
\v E_a (\v r _s) = \v E_0 (\v r_s) + \sum \limits_{l=0}^{L-1}
\Gamma _l \v E_l (\v r_s). \label{eq4-46}
\end{equation}
The corresponding magnetic current is
\begin{equation}
\v M_a (\v r_s) =  \v E_a  \times \hat n = \v M_0 + \sum \limits
_{l=0}^{L-1} \Gamma _l \v M _l \label{eq4-47}
\end{equation}
where $ \v M_l = \v E_l \times \hat n$.

The electric field in the waveguide region in the volume $V$ due to
the magnetic current $\v M _a $ is given by
\begin{equation}
\begin{split}
\v E_a (\v r) &= \int \limits_{S_a} dS' \nabla \times \dyad G_m
(\v r, \v r')
\cdot \v M_a (\v r')\\
&= \int \limits _{S_a} dS' \nabla \times \dyad G_m (\v r, \v r')
\cdot \v M_0
(\v r' )\\
&+ \sum \limits _{l=0}^{L-1} \Gamma _l \int \limits_{S_a} dS'
\nabla \times \dyad G_m (\v r, \v r') \cdot \v M_l (\v r'), \quad
\v r \in V, \label{eq4-48}
\end{split}
\end{equation}
where $\dyad G_m (\v r, \v r')$ is the magnetic-type dyadic
Green's function.

$\v E_a$ from (\ref{eq4-48}) can be substituted into
(\ref{eq4-43}) to yield
\begin{equation}
\begin{split}
-\hat n \times \int \limits _{S_a} dS' \nabla &\times \dyad G_m
(\v r, \v r')
\cdot \v M_0 (\v r') \\
&- \sum \limits _{l=0}^{L-1} \Gamma _l \hat n \times \int\limits
_{S_a} dS' \nabla \times \dyad G_m (\v r, \v r') \cdot \v M_l (\v
r')\\ &= i\omega \mu \hat n \times \int \limits_{S_p} dS' \dyad
G_e (\v r, \v r') \cdot \v J_p (\v r'). \label{eq4-49}
\end{split}
\end{equation}
Expanding $\v J_p$ as in (\ref{eq4-37}), we obtain
\begin{equation}
\begin{split}
-\hat n \times \int \limits _{S_a} dS' \nabla &\times \dyad G_m
(\v r, \v r')
\cdot \v M_0 (\v r') \\
& - \sum \limits _{l=0}^{L-1} \Gamma _l \hat n \times \int\limits
_{S_a} dS' \nabla \times \dyad G_m (\v r, \v r') \cdot \v M_l (\v
r')\\ &= i \omega \mu\sum \limits_{n=1}^{N} a_n \hat n \times \int
\limits_{S_p} dS' \dyad G_e (\v r,\v r') \cdot \v J_n (\v r').
\label{eq4-50}
\end{split}
\end{equation}
Testing the above equation with $\hat n \times \v J_m (\v r)$,
$m=1, \dots, N,$ we have
\begin{equation}
\begin{split}
-\langle \v J_m (\v r ), \nabla &\times \dyad G_m (\v r, \v r'),
\v M_0 (\v
r')\rangle \\
&- \sum \limits _{l=0}^{L-1} \Gamma _l  \langle \v J_m (\v
r), \nabla\times \dyad G_m (\v r, \v r'), \v M_l (\v r') \rangle\\
&= i \omega \mu \sum \limits _{n=1}^{N}  a_n \langle \v J_m (\v
r), \dyad G _e (\v r, \v r'), \v J_n (\v r') \rangle , \quad m =
1, \dots , N. \label{eq4-51}
\end{split}
\end{equation}
In the above, the notation
\begin{equation}
\langle \v f(\v r), \dyad G(\v r, \v r') , \v g (\v r') \rangle =
\int \limits _{S_p} dS \v f(\v r) \cdot \int   \limits _{S_p} dS'
\dyad G(\v r, \v r') \cdot \v g(\v r'). \label{eq4-52}
\end{equation}

Notice that the total field inside the waveguide is given by
\begin{equation}
\v E_T = \v E_a + \v E_p \label{eq4-53}
\end{equation}
where $\v E_p$ is the field produced by the probe current, i.e.,
\begin{equation}
\v E_p = i \omega \mu \int \limits_{S_p}dS'\dyad G _e (\v r, \v
r') \cdot \v J_p (\v r'). \label{eq4-54}
\end{equation}
Because of the electric dyadic Green's function used here, $\hat n
\times \v E_p =0 $ on $S_a$. Therefore,
\begin{equation}
\hat n \times \v E_T = \hat n \times \v E_a \quad \text{ on } S_a.
\label{eq4-55}
\end{equation}
Consequently, the total field given by (\ref{eq4-53}) calculated
via the use of (\ref{eq4-48}), and (\ref{eq4-55}) satisfies the
boundary condition that $\hat n \times \v E_T$ is continuous at
$S_a$ from the coaxial waveguide to the main waveguide.

Next, we need to impose the boundary condition that the tangential
component of the magnetic field is continuous across $S_a$. To
this end, we find $\v H_T = \v H_a + \v H_p$ where
\begin{subequations}
\begin{equation}
\v H_a (\v r) = i \omega \epsilon \int \limits_{S_a} dS' \dyad G_m
(\v r, \v r') \cdot \v M_a (\v r'), \label{eq4-55a}
\end{equation}
\begin{equation}
\v H_p (\v r) = \int \limits _{S_p} dS' \nabla \times \dyad G_e
(\v r, \v r') \cdot \v J_p (\v r'). \label{eq4-55b}
\end{equation}
\end{subequations}
On expanding $\v J_p$ as in (\ref{eq4-37}), and $\v M_a$ as in
(\ref{eq4-47}), we have
\begin{equation}
\begin{split}
\v H_T(\v r) &= i \omega \epsilon \int \limits_{S_a} dS' \dyad G_m
(\v r, \v r')\cdot \v M_0 (\v r') \\
& + i \omega \epsilon \sum \limits_{l=0}^{L-1} \Gamma_l
\int\limits_{S_a} dS' \dyad G_m (\v r, \v r') \cdot \v M _l (\v
r')\\ &+ \sum \limits_{n=1}^{N} a_n \int \limits_{S_p} dS' \nabla
\times \dyad G_e (\v r, \v r') \cdot \v J_n (\v r').
\label{eq4-56}
\end{split}
\end{equation}
The boundary condition requires that the tangential components of
$\v H$ in (\ref{eq4-45}) and (\ref{eq4-56}) be continuous. Equating
(\ref{eq4-45}) and (\ref{eq4-56}), and testing the result with $\v
M_{l'} (\v r)$, $l'=0, \dots , L-1$, we have
\begin{equation}
\begin{split}
\delta_{0l'} \lambda_0 - \Gamma_{l'} \lambda_{l'} &= i
\omega\epsilon \langle \v M _{l'} (\v r),
\dyad G_m (\v r, \v r'), \v M_0 (\v r') \rangle\\
&+ i \omega \epsilon \sum \limits_{l =0}^{L-1} \Gamma_l \langle \v
M_{l'} (\v r),
\dyad G_m (\v r, \v r'), \v M_l (\v r') \rangle\\
&+ \sum \limits_{n=1}^{N} a_n \langle \v M_{l'} (\v r), \nabla
\times \dyad G_e (\v r, \v r') , \v J_n (\v r')\rangle , \quad
{l'}=0, \dots , L-1. \label{eq4-57}
\end{split}
\end{equation}
In the above, we have made use mode orthogonality to arrive at
\begin{equation}
\begin{split}
\langle \v M_{l'} , \v H_l \rangle &= \int \limits_{S_a} dS \v
M_{l'} \cdot \v H_l =
\int \limits _{S_a} dS \v E_{l'} \times \hat n \cdot \v H_l\\
& = - \int \limits_{S_a} dS \hat n \cdot \v E_{l'} \times \v H_l = -
\delta _{l'l} \lambda_{l'} \label{eq4-58}
\end{split}
\end{equation}
where
\begin{equation}
\lambda_{l'} = \int \limits_{S_a} dS \hat n \cdot \v E_{l'} \times
\v H_{l'} \label{eq4-59}
\end{equation}
Equations (\ref{eq4-51}) and (\ref{eq4-57}) constitute $N+L$
equations for $N+L$ unknowns, $\Gamma_l$, $l= 0, \dots , L-1$, and
$a_n$, $n=1, \dots , N$. They can be solved by using matrix
inversion. Once $\Gamma _0$, the reflection coefficient of the TEM
mode is found, and assuming that only the TEM mode propagates in
the coaxial waveguide region, the input impedance of the probe is
given by
\begin{equation}
Y_{in} = Y_0 \frac{1-\Gamma_0}{1+\Gamma_0}. \label{eq4-60}
\end{equation}
The advantage of this approach is that the input admittance can be
found to any desired numerical accuracy by increasing the number
of terms in (\ref{eq4-37}) and (\ref{eq4-47}). When only one mode
is used in the coaxial region, it can be shown that this method,
with the input impedance given by (\ref{eq4-60}), yields the same
answer as the method using the variational expression.


    \section{Excitation of a Microstrip Patch Antenna}
\index{Antenna}
\index{Antenna!microstrip patch}

A microstrip patch antenna is made by etching a patch on top of a
dielectric substrate backed by a ground plane. It was first proposed
by Deschamps in 1953  \cite{DESCHAMPS_MICROSTRIP}, and put into
practice by Munson in 1972  \cite{MUNSON}. Microstrip patch antenna
is a very popular antenna because of its ease of fabrication, light
weight, and conformal nature. Because of the proximity of the
radiation source to a ground plane, and cancellation of the radiation
field due to a negative image current on the ground plane, the
current on a microstrip patch is a poor radiator ordinarily.
However, it can be made to radiate well if resonant modes exist on
the patch. At the resonant frequency of the patch, the current
amplitude can be greatly enlarged, enhancing the radiation field
despite negative image current cancelation. Hence, it radiates by
\index{Resonance coupling}resonance coupling
 \cite{LO_RICHARDS,RICHARDS_LO_HARRISON,CHEW_KONG_SHEN,POZAR4}.

A microstrip patch antenna can be thought of as a \index{Antenna!cavity-backed slot}cavity-backed slot
antenna.  The radiation is actually from the side walls or slots of
the antenna, and will not radiate well unless it is backed by a
resonant structure.  Because of the cavity nature of the antenna, it
is generally narrow band, but much ingenious design has made these
antennas operate with a broader bandwidth.

\subsection{Magnetic Wall Model}
\index{Magnetic wall model}

A microstrip patch can be approximated by a magnetic wall model where the side walls of
the patch are replaced with magnetic walls and the top and bottom patches remain metallic.
The magnetic walls have $ \hat{n}\times{\bf{H}}=0$ boundary condition while that
on the metallic walls is
$\hat n\times\v E=0$.  In short, they are perfectly conducting walls. We can assume that the
substrate is thin so that $\frac{\partial}{\partial z}=0$ for the field. Only very high order modes will
have $\frac{\partial}{\partial z}\neq0$.  These modes will be
far away from the operating frequency of the patch, so that
they are weakly excited.

We assume that only \index{Mode!TM}TM$_z$ modes are important since TE$_z$ modes will be shorted out.
So the field inside the cavity can be written
as
\begin{equation}
\label{c4eq:1}
{\bf{E}}_{mn}=\hat{z}E_{mn}\cos\biggl(\frac{m\pi x}{a}\biggr)\cos\biggl(\frac{n\pi y}{b}\biggr)
\end{equation}
This is called the \index{Mode!TM$_{\text{mn0}}$} TM$_{mn0}$ mode, but we will call this the
 TM$_{mn}$ mode for short.
Notice that we have chosen the solution to satisfy the Neumann
boundary condition on the magnetic wall so that tangential magnetic
field is zero there.\footnote{We assume that the probe used is such
that there is no charge accumulation on the probe and hence, only
divergence-free modes need to be considered in the mode expansion.
Such is the case if the current on the probe has constant current
such that $\nabla\cdot\v J=0$.}

    \begin{figure}[!htb]
    \begin{center}
    \includegraphics[height=5cm]{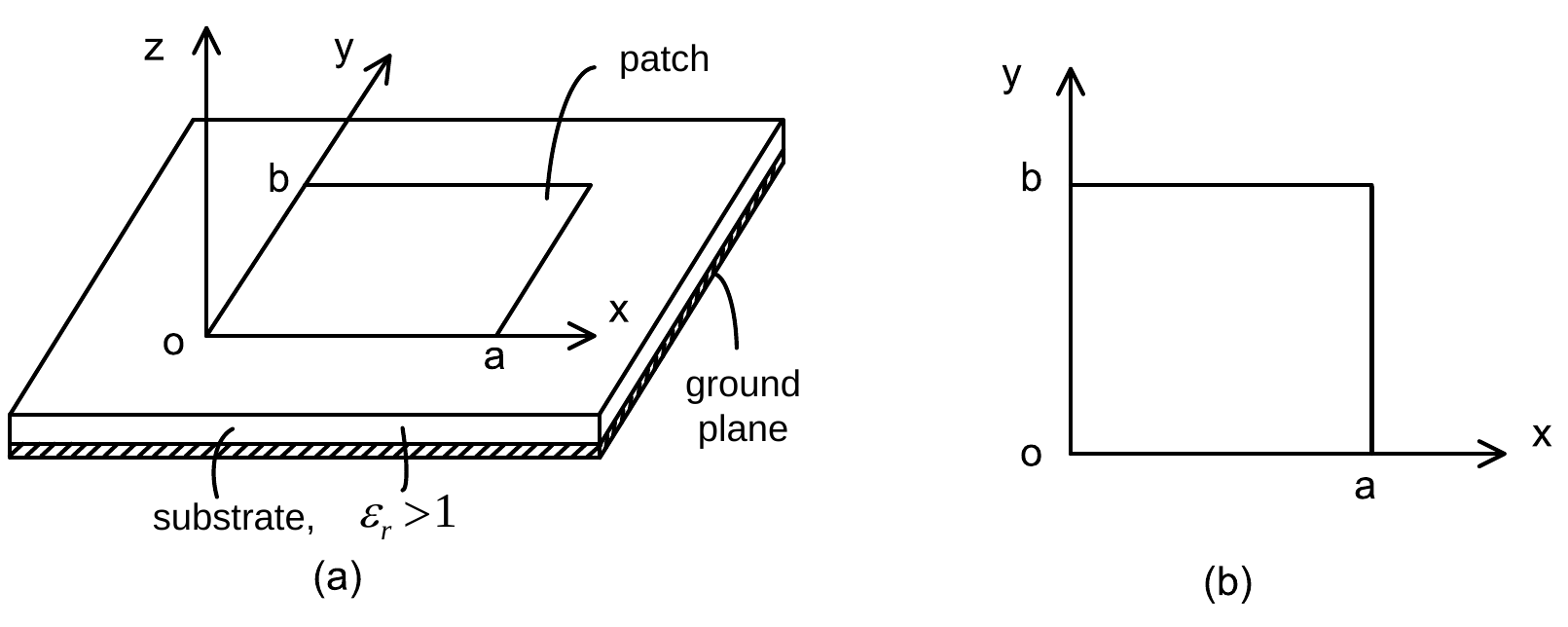}
    \caption{Microstrip patch antenna: (a) Side view. (b) Top view.}
    \label{MPAfg1}
    \end{center}
    \end{figure}

    The resonant frequency of the TM$_{mn}$ mode is given by
        \begin{equation}
        \label{c4eq:2}
         k_{mn}^2={\biggl(\frac{m\pi}{a}\biggr)}^2+{\biggl(\frac{n\pi}{b}\biggr)}^2
        \end{equation}
    The normalization constant is
        \begin{equation}
        \label{c4eq:3}
         E_{mn}=\left[\frac{4}{ab(1+\delta_{0m})(1+\delta_{0n})}\right]^{\frac{1}{2}}
        \end{equation}
        so that the modes are orthonormal.
    Using the fact that
        \begin{equation}
        \label{c4eq:4}
        \nabla\times\nabla\times{\bf{E}}-k^2{\bf{E}}=i\omega\mu{\bf{J}}
        \end{equation}
    and letting
        \begin{equation}
        \label{c4eq:5}
        {\bf{E}}=\sum_{m,n}{a_{mn}{\bf{E}}_{mn}(x,y)}
        \end{equation}
    and that
        \begin{equation}
        \label{c4eq:6}
        \nabla\times\nabla\times{\bf{E}}_{mn}-k_{mn}^2{\bf{E}}_{mn}=0
        \end{equation}
    we obtain that
        \begin{equation}
        \label{c4eq:7}
        a_{mn}=i\omega\mu\frac{\langle{{\bf{E}}^*_{mn},{\bf{J}}}\rangle}{k_{mn}^2-k^2}
        \end{equation}
    If we assume that
    \begin{equation} \label{c4eq:8}
        {\bf J}(x,y)=\hat{z}I_0\,\delta(y-y')B(x-x')
    \end{equation}
    where
    \begin{equation} \label{c4eq:9}
        B(x)=\frac{1}{w}
        \begin{cases}
            1, & \lvert x \rvert \leq w/2 \\
            0, & \lvert x \rvert > w/2
        \end{cases}
    \end{equation}
    is a box function.  Such a current is chosen so that it is not
    of zero thickness and width.   Otherwise, it will have infinite
    inductance.

    Then
    \begin{equation} \label{c4eq:11}
        \langle{\bf E}^*_{mn},{\bf J}\rangle=E_{mn}\cos\biggl(\frac{n\pi y'}{b}\biggr)\frac{1}{w}\int\limits_{x'-\frac{w}{2}}^{x'+\frac{w}{2}}
        \cos\biggl(\frac{m\pi x}{a}\biggr)\,dx
    \end{equation}
    By letting
    \begin{align} \label{c4eq:12}
        I=\frac{1}{w}\int\limits_{x'-\frac{w}{2}}^{x'+\frac{w}{2}}dx\,\cos\biggl(\frac{m\pi x}{a}\biggr)&=\frac{1}{w}
        \biggl(\frac{m\pi}{a}\biggr)^{-1}\sin\biggl(\frac{m\pi x}{a}\biggr)\bigg|_{x'-\frac{w}{2}}^{x'+\frac{w}{2}} \nonumber\\
        &=\frac{1}{w}\biggl(\frac{m\pi}{a}\biggr)^{-1}\biggl[\sin\biggl(\frac{m\pi(x'+\frac{w}{2})}{a}\biggr)-\sin\biggl(\frac{m\pi(x'-\frac{w}{2})}
        {a}\biggr)\biggr]
    \end{align}
    and by using
    \begin{align}\label{c4eq:13}
        \sin(A+B)-\sin(A-B)=2\cos A\sin B
        \end{align}
then
\begin{align} \label{c4eq:14}
        \qquad I=\frac{2a}{m\pi w}\cos\biggl(\frac{m\pi x'}{a}\biggr)\sin\biggl(\frac{m\pi w}{2a}\biggr)=
        \cos\biggl(\frac{m\pi x'}{a}\biggr)\text{sinc}\biggl(\frac{m\pi w}{2a}\biggr)
    \end{align}
    where $\text{sinc}(x)=(\sin x)/x$.

    Hence
    \begin{gather}
        \langle{\bf E}^*_{mn},{\bf J}\rangle=E_{mn}\cos\biggl(\frac{m\pi x'}{a}\biggr)
        \cos\biggl(\frac{n\pi y'}{b}\biggr)\text{sinc}\biggl(\frac{m\pi w}{2a}\biggr)  \label{c4eq:15}
        \\
        a_{mn}=i\omega\mu\frac{\phi_{mn}(x',y')\text{sinc}{\biggl(\frac{m\pi
        w}{2a}\biggr)}}{k_{mn}^{2}-k^{2}}\label{c4eq:15a}\\
        {\bf E}=i\omega\mu\hat{z}\sum_{m,n}\frac{1}{k_{mn}^{2}-k^{2}}\phi_{mn}(x,y)\phi_{mn}(x',y')
        \text{sinc}\biggl(\frac{m\pi w}{2a}\biggr) \label{c4eq:16}
    \end{gather}
    where
    \begin{equation} \label{c4eq:17}
        \phi_{mn}(x,y)=E_{mn}\cos\biggl(\frac{m\pi x}{a}\biggr)\cos\biggl(\frac{m\pi y}{b}\biggr)
    \end{equation}

The above is the \index{Magnetic wall cavity model}magnetic wall cavity model for the \index{Antenna!patch}patch antenna.
However, it has no loss and the resonant frequencies of the modes of
the cavity are purely real.  When the operating frequency coincides
with the resonant frequency, from \eqref{c4eq:7}, it is seen that
the excitation coefficient of the mode becomes infinite.  This is
unphysical, as the resonant frequency of the cavity is never real in
practice: the resonant modes of the patch antenna are radiationally
damped. Hence, its resonant frequencies are complex rather than
real.  In addition, there are material loss and copper loss of the
antenna that causes the modes to have complex resonant frequencies,
giving rise to damped resonances.   The dielectric loss can be
easily incorporated by using a complex dielectric.  The radiation
damping can be modeled by a \index{Lossy magnetic wall} lossy magnetic wall while the copper
loss can be modeled by a \index{Lossy electric wall} lossy electric wall.\footnote{The
excitation of the cavity modes by a source is elaborately dealt with
in  \cite{KUROKAWA,COLLINI}.  An application to microstrip antenna
is given in  \cite{CHEW_KONG_SHEN}.}

  \begin{figure}[!htb]
    \begin{center}
    \includegraphics[width=4in]{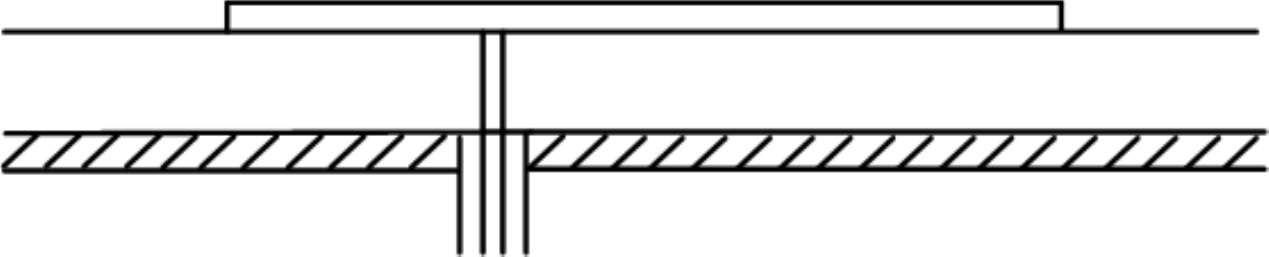}
    \caption{Excitation of the patch antenna with a probe (cross-section view).}
    \label{MPAEXfg1}
    \end{center}
    \end{figure}

For a real resonant frequency, the fields of the mode will have a
$e^{-i\omega_{mn}t}$ time dependence.  If the resonant frequency is
complex with $\tilde\omega_{mn}=\omega_{mn}-i\alpha_{mn}$, then the
time dependence of the fields is
$e^{-i\omega_{mn}t}e^{-\alpha_{mn}t}$.  The energy density of the
mode is proportional to $|\v E|^2$ and $|\v H|^2$, and hence the
stored energy $W_{mn,T}\sim e^{-2\alpha_{mn}t}$.  By energy
conservation, the power radiated by the mode is proportional to the
negative time rate of change of stored energy, or
$$
P_{mn,rad}=-\frac{d}{d t} W_{mn,T}=2\alpha_{mn} W_{mn,T}
$$
Hence, if the stored energy and the power radiated is known, the
imaginary part of the resonant frequency $\alpha_{mn}$ can be found.
This is equivalent to finding the Q of the resonant modes of the
antenna.

Actually, the \index{Antenna!reactive power leakage}reactive power leakage from the antenna also gives
rise to a real resonant frequency shift.  The estimate of this shift
is more difficult and method of estimating this shift is given in
\cite{CHEW_KONG1,CHEW_KONG2,CHEW_KONG3} using perturbation approach
and asymptotic approach. In retrospect, the fringing field at the
open edge of the patch makes the patch effectively larger, lowering
the real resonant frequency compared to that predicated by the
magnetic wall model.

\subsection{The Q of the Modes}
\index{Quality of resonant modes}

Due to radiation damping, and other losses in the cavity, the resonant frequency of
each mode is not purely real.  We shall discuss how to estimate the damping rate due to radiation by using a
\index{Perturbation approach} perturbation approach.
In this approach, we assume that the current distribution on the
patch is not changed a lot when the magnetic walls are removed to
allow for radiation of the patch current.

    To estimate the power radiated by the patch,
    we need to find the current on the patch. To this end, we derive the magnetic field in the cavity:
    \begin{equation} \label{c4eq:18}
        \nabla\times \mathbf{E}_{mn}=i\omega\mu\mathbf{H}_{mn}
    \end{equation}
    or that
    \begin{equation} \label{c4eq:19}
        \mathbf{H}_{mn}=-\frac{1}{E_{mn}\omega\mu}
        \left[\hat{x}\left(\frac{n\pi}{b}\right)\cos\left(\frac{m\pi x}{a}\right)\sin\left(\frac{n\pi
        y}{b}\right)-
        \hat{y}\left(\frac{m\pi}{a}\right)\sin\left(\frac{m\pi x}{a}\right)\cos\left(\frac{n\pi y}{b}\right)\right]
    \end{equation}
    The corresponding current on the top patch is
    \begin{equation} \label{c4eq:20}
        \mathbf{J}_{mn}=-\hat{z}\times\mathbf{H}_{mn}
        =\frac{1}{i\omega\mu}\left[\hat{x}\left(\frac{m\pi}{a}\right)
        \sin\left(\frac{m\pi x}{a}\right)\cos\left(\frac{n\pi y}{b}\right)
        +\hat{y}\left(\frac{n\pi}{b}\right)
        \cos\left(\frac{m\pi x}{a}\right)\sin\left(\frac{n\pi y}{b}\right) \right]
    \end{equation}
    If this current is radiating in free space, its field is given
    by the free space dyadic Green's function acting on the above
    current, namely,
    \begin{equation} \label{c4eq:21}
        \mathbf{E}_{mn,R}=i\omega\mu\int d\v{r}^{\prime} \dyad G\left(\v{r},\v{r}^{\prime}\right)\cdot\mathbf{J}_{mn}\left(\v{r}^{\prime}
        \right)
        =i\omega\mu\left(\mathbf{\bar{I}}+\frac{\nabla\nabla}{k^2}\right)\cdot \int d\v{r}^{\prime}\frac{e^{ik|\v{r}-\v{r}^{\prime}|}}
        {4\pi|\v{r}-\v{r}^{\prime}|}\mathbf{J}_{mn}\left(\v{r}^{\prime}\right)
    \end{equation}
    By letting letting $|\v r|=r \rightarrow \infty$,
    \begin{equation} \label{c4eq:22}
        \frac{e^{ik|\v{r}-\v{r}^{\prime}|}}{4\pi|\v{r}-\v{r}^{\prime}|} \sim
        \frac{e^{ikr}}{4\pi r}e^{-ik\hat{r}\cdot\v{r}^{\prime}}
    \end{equation}
    Hence,
    \begin{equation} \label{c4eq:23}
        \mathbf{E}_{mn,R}\approx i\omega\mu\left(\mathbf{\bar{I}}+\frac{\nabla\nabla}{k^2}\right)\cdot \frac{e^{ikr}}{4\pi r}\int d\v{r}^{\prime}e^{-ik\hat{r}\cdot\v{r}^{\prime}}\mathbf{J}_{mn}\left(\v{r}^{\prime}\right)
    \end{equation}
    Furthermore, when $r\rightarrow\infty$, the spherical wave becomes like a plane wave.  In other words, $e^{ikr}=e^{i\v k\cdot\v r}$, where $\v k=k\^k$ and $\v r=r\^r$.  Clearly,
    $\^r=\^k$.  Then, we let $\nabla\approx i\v k=ik\^ k$, and we have
    \begin{equation} \label{c4eq:24}
    \left(\mathbf{\bar{I}}+\frac{\nabla\nabla}{k^2}\right)\frac{e^{ikr}}{r}\approx
    \left(\mathbf{\bar{I}}-\hat{k}\hat{k}\right)\frac{e^{ikr}}{r}
    =\left(\hat{\theta}\hat{\theta}+\hat{\phi}\hat{\phi}\right)\frac{e^{ikr}}{r}
    \end{equation}
In the above, we have made use of the fact that $\hat k\hat k=\^ r\^
r$, and that $\dyad I=\^r\^r+\^\theta\^\theta+\^\phi\^\phi$.
Consequently,
    \begin{align}\label{c4eq:25}
     \mathbf{E}_{mn,R}& \approx i\omega\mu \frac{e^{ikr}}{4\pi r}\left( \hat{\theta}\hat{\theta}+\hat{\phi}\hat{\phi}\right)\cdot
     \int d\v{r}'e^{-ik\hat{r}\cdot \v{r}^{\prime}}\mathbf{J}(\v{r}') \nonumber\\
     &=i\omega\mu\frac{e^{ikr}}{4\pi r}\left( \hat{\theta}\hat{\theta}+\hat{\phi}\hat{\phi}\right)\cdot\tilde{\mathbf{J}}(k\hat{r})
     \end{align}
     The last integral corresponds to a Fourier integral with the
     Fourier spectral variable evaluated on the energy shell or Ewald sphere where $|\v k|=k=\omega\sqrt{\mu\epsilon}$.
     Hence, $\tilde{\mathbf{J}}(k\hat{r})$ is the Fourier transform
     of $\mathbf{J}(\v{r}')$ with $\v k=k\^ r$, or on the Ewald sphere.
Since
     $\v J(\v r)$ consists of sinusoidal functions, their Fourier
     transforms can be evaluated in closed form.  Hence, $\tilde{\v
     J}(k\hat r)$ can be found.  Also, the physical meaning is that
     only this Fourier component will radiate coherently in the
     $k\^r$ direction.

The above is the electric field radiated via the free-space dyadic
Green's function. In order to account for the fact that this current
is radiating on a dielectric substance backed by a ground plane, we
need only to add the reflected wave term. The reflected wave can be
added using ray physics since the observation point is in the far
field where ray physics applies  \cite{WFIMF}. Consequently,
    \begin{align}\label{c4eq:27}
     \mathbf{E}_{mn,R}\approx i\omega\mu\frac{e^{ikr}}{4\pi r}\left\{\hat{\theta}\tilde{{J}}_\theta\left(k\hat r\right)
     \left[ 1-\tilde{R}^{TM}(\hat{r})\right]+\hat{\phi}\tilde{{J}}_\phi\left(k\hat
     r\right)
     \left[ 1+\tilde{R}^{TE}\left(\hat{r}\right)\right]\right\}
    \end{align}
    where $\tilde{R}^{TM}$ and $\tilde{R}^{TE}$ are the generalized reflection coefficient for the \index{Layered medium}layered
    medium representing the substrate with a ground plane. The minus sign in front of
    $\tilde{R}^{TM}$ is because $J_\theta$ produces TM
    fields of opposite polarities above and below the source which is assumed to be an infinitely thin
    sheet.  The $J_\theta$ current resembles a \index{Hertzian dipole}Hertzian dipole
    radiating in endfire direction, and hence produces a TM field
    that is odd symmetric about $z=0$ plane.  But the $J_\phi$
    current resembles a Hertzian dipole radiating in the broadside
    diection produces a TE field that is even symmetric about $z=0$
    plane.

    The power density radiated by this mode is then
    \begin{align}\label{c4eq:28}
    S_{mn}(r,\theta,\phi)&=\frac{1}{2\eta}\left|\mathbf{E}_{mn,R}\right|^2 \\
    &=\frac{\omega^2\mu^2}{2\eta}\frac{1}{\left|4\pi r \right|^2}
\left\{ \left|
\tilde{{J}}_\theta(k\hat{r})\left[1-\tilde{R}^{TM}(\hat{r})\right]
\right|^2+\left|\tilde{{J}}_\phi(k\hat{r})\left[1+\tilde{R}^{TE}(\hat{r})\right]
\right|^2 \right\}
    \end{align}
where $\hat{r}$ is a function of $(\theta,\phi)$.

   \begin{figure}[!h]
    \centering
    \includegraphics[height=6cm]{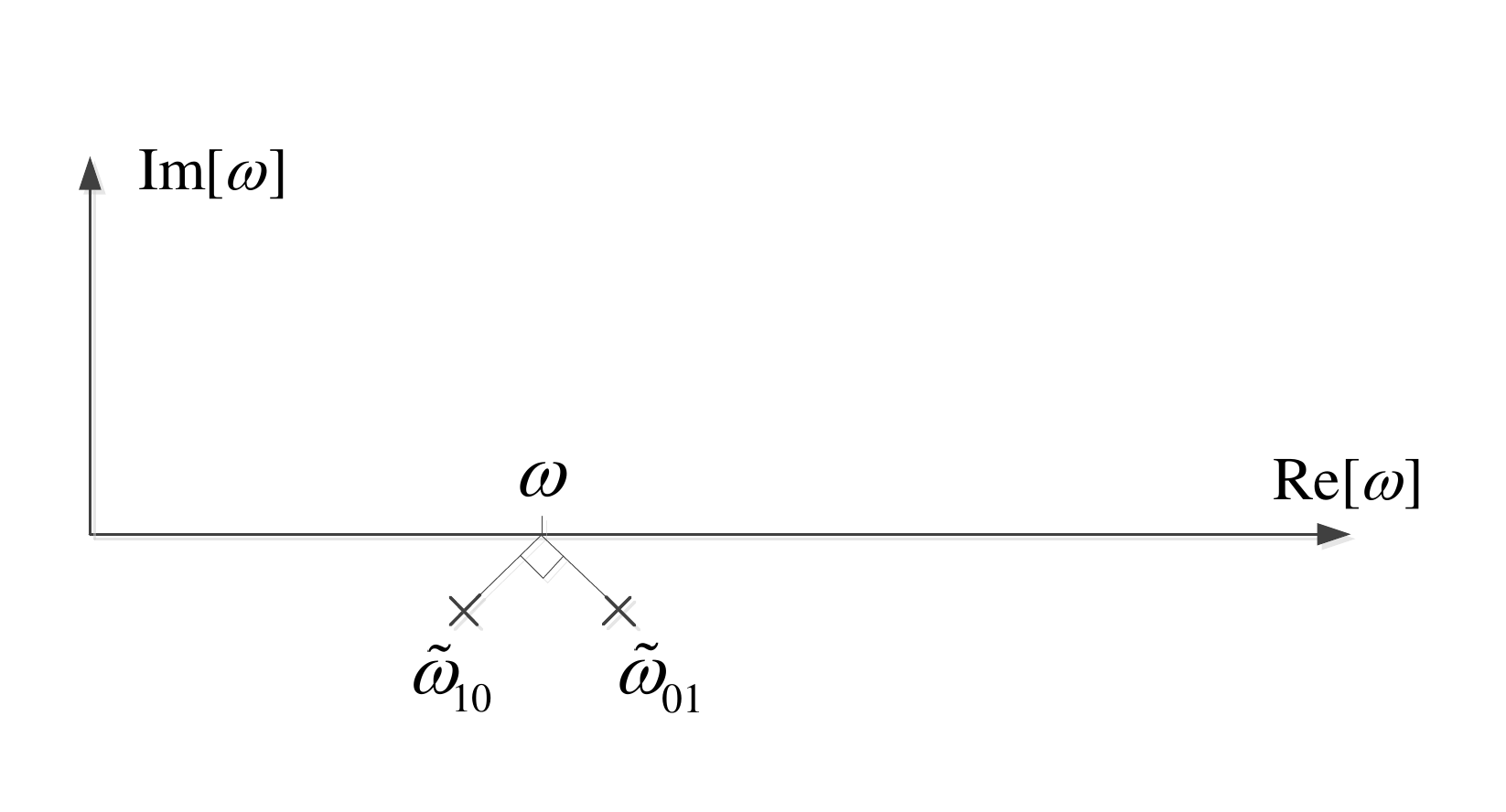}
    \caption{Splitting of the degenerate TM$_{10}$ and TM$_{01}$ modes.}
    \label{fg4p9}
    \end{figure}

    The total radiated power can be found by performing the following integral over a
    hemisphere,
    namely,\footnote{In addition, the patch current can excite a
    surface wave mode in the dielectric substrate layer causing further loss, and damping of the mode.  This is not included
    in this integral.}
    \begin{equation}\label{c4eq:29}
    P_{mn,rad} = \int^{2\pi}_{0} \int^{\frac{\pi}{2}}_{0} S_{mn}(r,\theta,\phi) r^{2} \sin\theta d\theta d\phi
    \end{equation}
    The stored energy of the mode is
    \begin{equation}\label{c4eq:29a}
    W_{mn,T} = \frac{1}{4}\int_{V} \left[ \epsilon |\mathbf{E}_{mn}|^{2} + \mu |\mathbf{H}_{mn}|^{2}\right] dV =\frac{1}{2} \int_{V}\epsilon |\mathbf{E}_{mn}|^{2} dV
    \end{equation}
    in the cavity between the patches. Therefore, for normalized
    modes,
    \begin{equation}\label{c4eq:30}
    W_{mn,T} = \frac{1}{2}\epsilon\int_{V} |\mathbf{E}_{mn}|^{2}dV =\frac{1}{2} \epsilon
    \end{equation}
    Consequently, the Q of the $mn$ mode is
    \begin{equation}\label{c4eq:31}
    Q_{mn} = \frac{\omega_{mn}W_{mn,T}}{P_{mn,rad}} = \frac{\omega_{mn}}{2\alpha_{mn}} = \omega_{mn}\tau_{mn}
    \end{equation}
    where $\tau_{mn}$ is the decay time constant of the $mn$ modes, and the complex resonant frequency of the mode is
    \begin{equation}\label{c4eq:32}
    \tilde \omega_{mn} = \omega_{mn} - i\alpha_{mn}
    \end{equation}
    From \eqref{c4eq:31},
    \begin{equation}\label{c4eq:33}
    \alpha_{mn} = \frac{\omega_{mn}}{2Q_{mn}}
    \end{equation}
    The Q can be estimated using the integral \eqref{c4eq:29} and expression
    \eqref{c4eq:31}.

    \subsection{Circular Polarization Excitation}
    \index{Circular polarization excitation}

    In a square \index{Antenna!microstrip patch}microstrip patch, the \index{Mode!TM$_{10}$} TM$_{10}$ and
    \index{Mode!TM$_{01}$} TM$_{01}$ modes are degenerate. If the operating frequency is chosen close to
    that of these modes, they will be dominant. Hence, \eqref{c4eq:5} can be approximated
    by only two modes, namely,
    \begin{equation} \label{c4eq:34}
        {\bf E}\cong a_{10}{\bf E}_{10}(x,y)+a_{01}{\bf E}_{01}(x,y)
    \end{equation}
    where
    \begin{equation} \label{c4eq:35}
        a_{10}=i{\omega}{\mu}\frac{\langle{\bf E}^*_{10},{\bf J}\rangle}{\tilde{k}_{10}^2-k^2},\qquad
        a_{01}=i{\omega}{\mu}\frac{\langle{\bf E}^*_{01},{\bf J}\rangle}{\tilde{k}_{01}^2-k^2}
    \end{equation}
    where $\tilde{k}_{10}$ and $\tilde{k}_{01}$ are the complex
    resonant frequencies of the modes.
    If the probe is located such that $x^\prime=y^\prime$ or
    along diagonal of the square patch, and
    $\text{sinc}(\frac{m{\pi}{w}}{2a})\approx 1 $
    for both TM$_{01}$ and TM$_{10}$ modes, then $a_{10}\approx a_{01}$,
    as can be seen from
    \eqref{c4eq:16}.
    The mode currents of these modes are orthogonal
    to each other in space,
    but the field produced is not circularly polarization.

    However, circular polarization can be obtained by making these modes non-degenerate by destroying the symmetry.
    We can let $a=b+\triangle$, so that the
    TM$_{10}$ mode has a slightly lower resonant frequency compared to the
    TM$_{01}$ mode as shown in Figure \ref{fg4p9}.
     Consequently, we have
    \begin{equation} \label{c4eq:36}
        a_{10}\simeq i{\omega}{\mu}\frac{\langle{\bf E}^*_{10},{\bf J}\rangle}{(\tilde{k}_{10}-k)2\tilde{k}_{10}},
        \qquad a_{01}\simeq i{\omega}{\mu}\frac{\langle{\bf E}^*_{01},{\bf J}\rangle}{(\tilde{k}_{01}-k)2\tilde{k}_{01}}
    \end{equation}
    If we split the modes appropriately, and get $\tilde{k}_{10}-k$ to be
    $90^o$ out of phase with $\tilde{k}_{01}-k$. Then $a_{10}$ and $a_{01}$
    will be $90^o$ out of phase, and $\bf E$ in \eqref{c4eq:34} will become circularly
    polarized.


\subsection{Perturbation Formula for Resonant Frequency Shift}
\index{Resonant frequency shift}

\begin{figure}[htb]
 \begin{center}
  \includegraphics[totalheight=0.3\textheight,width=1\textwidth]{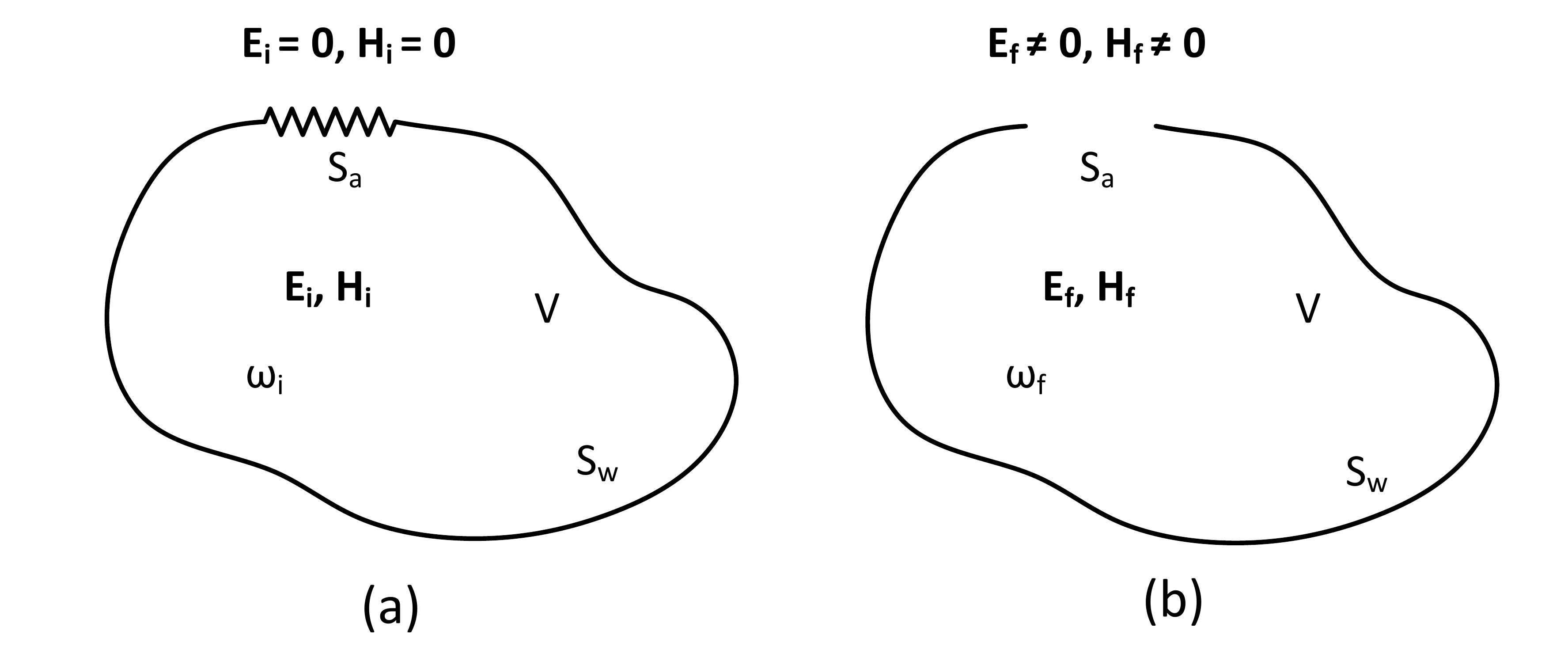}
  \end{center}
  \caption{Derivation of the perturbation formula: (a) Geometry before perturbation. (b) Geometry after perturbation.}
  \label{fg4p10}
\end{figure}

\newcommand\nv{\mathbf{n}}

The above derivation for resonant frequency shift due to radiation
damping is intuitive.  But a more formal procedure for ascertaining
the resonant frequency shift can be obtained by using \index{Perturbation approach}perturbation
concepts  \cite{HARRINGTON4}.  These concepts have been applied to
derive the resonant frequency shift for microstrip antennas
 \cite{CHEW_KONG1,CHEW_KONG2}.  To start, we take the divergence of
the following quantity:
\begin{align}
\nabla \cdot \v E_i^{*} &\times \v H_f+\v E_f \times \v H_i^{*})\nonumber\\
&=\v H_f\cdot\nabla\times \v E_i^{*}-\v E_i^{*}\cdot\nabla\times \v
H_f + \v H_i^{*}\cdot\nabla\times \v E_f-\v E_f\cdot\nabla\times
\v H_i^{*}\nonumber\\
&=-i\omega_i\v H_f\cdot\dyadg\mu^{*}\cdot\v H_i^{*}+i\omega_f\v
E_i^{*}\cdot \epsd\cdot\v E_f +i\omega_f\v H_i^{*}\cdot\mud\cdot\v
H_f-i\omega_i\v E_f\cdot\epsd^{*}
\cdot\v E_i^{*}\nonumber\\
\label{PEq1}
\end{align}
Integrating the above over the original volume of the cavity as
shown in Figure \ref{fg4p10}, we arrive at
\begin{align}
\int_{S}dS \hat{\nv}\cdot (\v E_i^{*} \times \v H_f+\v E_f \times \v
H_i^{*})= i(\omega_f-\omega_i)\int_V dV\left( \v
H_i^{*}\cdot\mud\cdot\v H_f+\v E_i^{*}\cdot \epsd\cdot\v E_f\right)
\end{align}
where we have assumed that the medium is lossless, and hence
$\dyadg\epsilon^\dagger=\dyadg\epsilon$ and
$\dyadg\mu^\dagger=\dyadg\mu$.
 Since
this is a perturbation, the final fields and the initial fields are
similar to each other.  Hence, we can approximate the right-hand
side with the initial field entirely.  Consequently,
\begin{align}
\int_{S}dS \hat{\nv}\cdot (\v E_i^{*} \times \v H_f+\v E_f \times \v
H_i^{*}) &\cong i(\omega_f-\omega_i) \int_{V}dV [\v
H_i^{*}\cdot\mud\cdot\v H_i+\v E_i^{*}\cdot\epsd\cdot\v E_i]
\label{PEq2}
\end{align}
The integral on the right-hand side is purely real now, and it can
be written in terms of the initial time-average stored energy
$\langle W_{T_i} \rangle$.
\begin{align}
\int_{S_A}dS \hat{\nv}\cdot (\v E_i^{*} \times \v H_f)+ \int_{S_W}dS
\hat{\nv}\cdot (\v E_f \times \v H_i^{*}) &\cong
i(\omega_f-\omega_i) 4\langle W_{T_i} \rangle \label{PEq3}
\end{align}
In the above, the first integral on the left can be approximated as
the complex power radiated by the hole.
\begin{align}
P^{*}_{rad}\cong\int_{S_A}dS \hat{\nv}\cdot (\v E_f^{*} \times \v
H_f)= 2 \langle P_{rad} \rangle-iR_{rad} \label{PEq4}
\end{align}
where $R_{rad}$ is the reactive power leaked by the cavity to the
outside.  The second integral on the left of \eqref{PEq3} can be
related approximately to the wall loss on the interior of the
cavity.  Hence, it can be approximated by
\begin{align}
P_{wall}&\cong\int_{S_W}dS \hat{\nv}\cdot (\v E_f \times \v H_i^{*})
= \int_{S_W}dS (\hat{\nv}\times \v E_f )\cdot \v H_i^{*}\nonumber\\
&=\frac{1}{\eta}\int_{S_W}dS |\v H_i^{*}|^2 \label{PEq5}
\end{align}
where
\begin{align}
\eta
&=\sqrt{\frac{\mu}{\epsilon}}=\sqrt{\frac{\omega\mu}{i\sigma}}=(1-i)\sqrt{\frac{\omega\mu}{2\sigma}}
\notag
\end{align}
Consequently, the wall loss becomes
\begin{align}
P_{wall}=(1-i)\sqrt{\frac{\omega\mu}{2\sigma}} \int_{S_W}dS |\v
H_i|^2=2\langle P_{wall}\rangle-iR_{wall} \label{PEq6}
\end{align}
Hence, the resonant frequency shift is given by
\begin{align}
\omega_f - \omega_i \approx -i \frac{2\langle P_{rad}\rangle +
2\langle P_{wall}\rangle  - i R_{rad} - i R_{wall} }{4 \langle
W_{Ti}\rangle }
\end{align}
Therefore,
\begin{align}
\Im m(\omega_f - \omega_i) \approx -i \frac{\langle
P_{rad}+P_{wall}\rangle }{2\langle W_{Ti} \rangle }
\end{align}
\begin{align}
\Re e(\omega_f - \omega_i) \approx - \frac{ R_{rad} + R_{wall} }{4
\langle W_{Ti}\rangle}
\end{align}
The dissipative loss causes the cavity to have a negative imaginary
part of the resonant frequency giving rise to \index{Cavity resonator!damping} damping. The reactive
power leakage and absorption by the wall make the cavity appear
larger and lower the resonant frequency.

The fringing field effect at the edge of a microstrip patch can also
be solved in closed form using Wiener-Hopf technique
 \cite{WIENER_HOPF,CHEW_KONG_PHILO}.  It has been used to ascertain
resonant frequency shift of microstrip antennas in
 \cite{CHEW_KONG3}.

   \subsection{Variational Impedance Formula for a Current Source}
    \index{Variational impedance formula}

    \begin{figure}[!htb]
    \begin{center}
    \includegraphics[width=8cm]{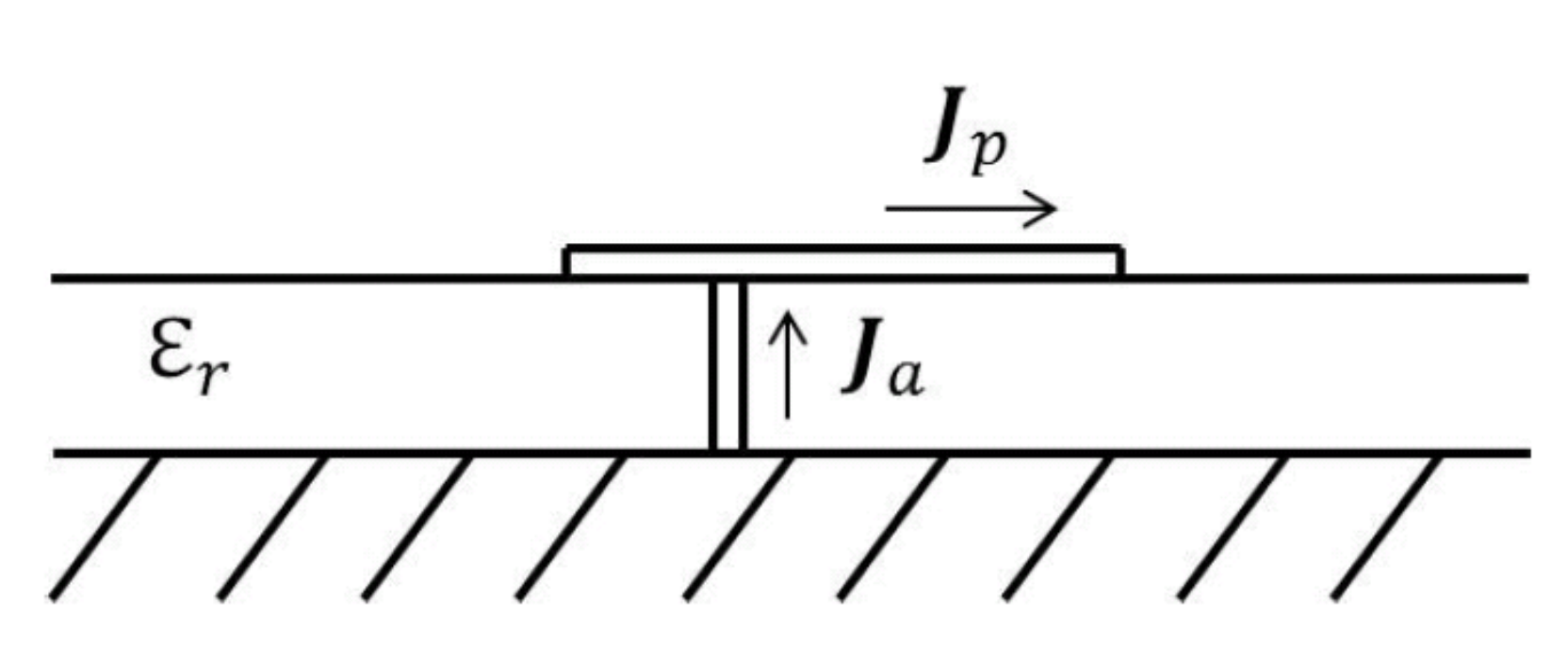}
    \caption{Excitation of a microstrip antenna by a current source.}
    \label{MPAEXCJfg1}
    \end{center}
    \end{figure}

    \begin{figure}[!htb]
    \begin{center}
    \includegraphics[width=8cm]{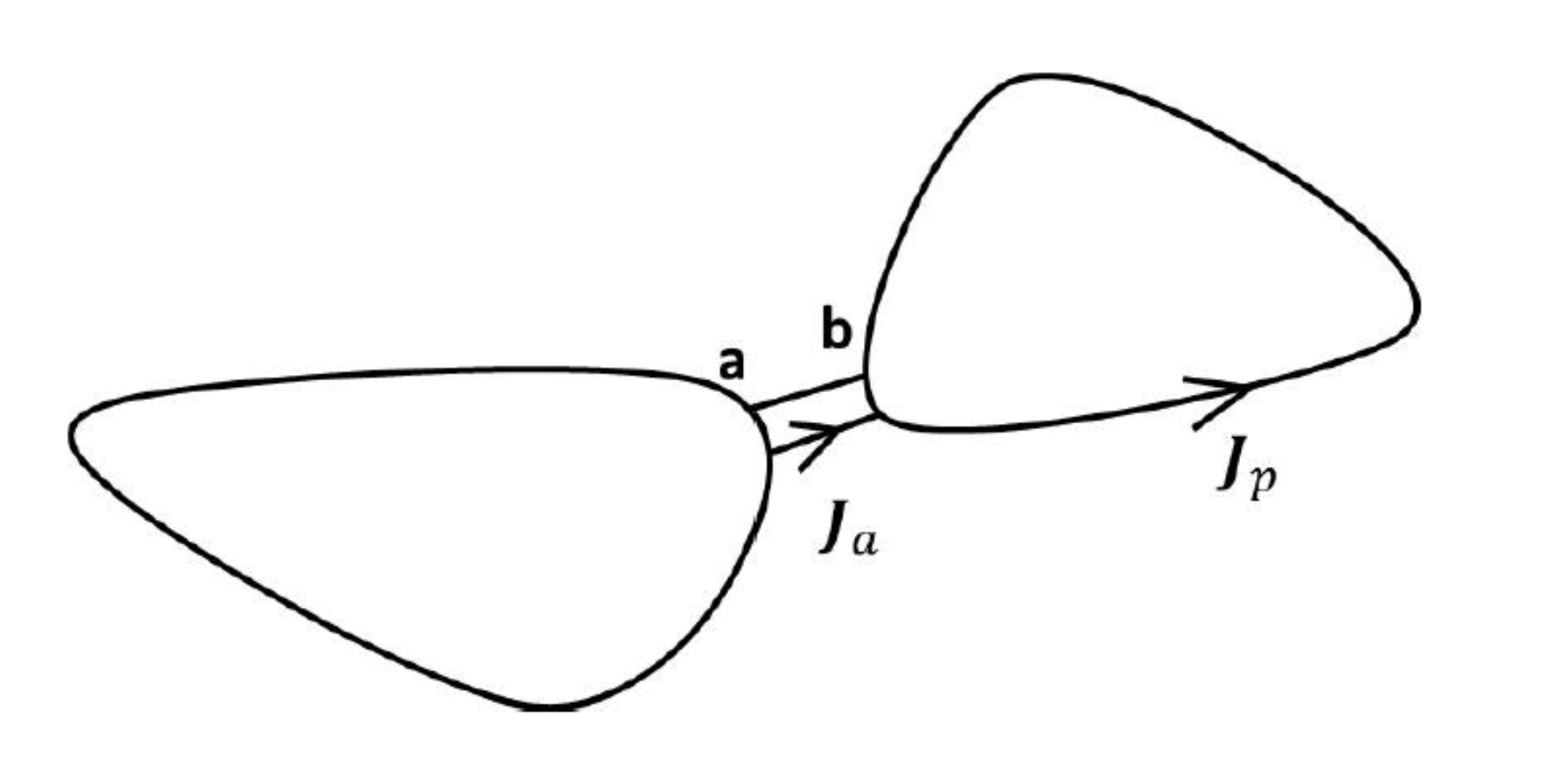}
    \caption{Excitation of a general antenna by a current source.}
    \label{fg2}
    \end{center}
    \end{figure}

In the previous input impedance formula, the magnetic current $\v
M_a$ is assumed known and immutable by its environment.  The
magnetic current is related to the tangential electric field; hence,
it is equivalent to a voltage source in circuit theory. Another
formula that is of importance is when the structure is driven by an
immutable applied (impressed) electric current $\v J_a$.  Hence, it
is equivalent to a current source in circuit theory.  The microstrip
antenna driven by such a current source is shown in Figure
\ref{MPAEXCJfg1} and a general structure case is shown in Figure
\ref{fg2}.  In this case, a variational formula for the input
impedance based on reaction is given by
 \cite{HARRINGTON4,CARTER}\footnote{This formula was used for input
impedance calculation before its variational nature was known.}
    \begin{equation}
    \label{c4veq:1}
    Z_{in}=-\frac{\langle\mathbf{E}_T,\mathbf{J}_T\rangle}{I^2}
\end{equation}
where $\v E_T=\v E_a+\v E_p$, and $\v E_a$ and $\v E_p$ are the
fields produced by applied (or impressed) current $\v J_a$ and the
induced current $\v J_p$, respectively, and $\v J_T=\v J_a+\v J_p$.
As shall be shown later, the above formula can admit approximate
solution for $\v J_p$ with second order error in the input impedance
$Z_{in}$.

In the exact limit, $\left\langle\v E_T,\v J_p\right\rangle=0$ and the above
reduces to
\begin{equation}
\label{c4veq:1a}
    Z_{in} =-\frac{\langle\mathbf{E}_T,\mathbf{J}_a+\mathbf{J}_p\rangle}{I^2}
    =-\frac{\langle\mathbf{E}_T,\mathbf{J}_a\rangle}{I^2}
    \end{equation}
To prove the above impedance formula, we assume that the exciting
source is small and hence, $\mathbf{J}_a$ is constant over space.
Then
    \begin{equation}
    \label{c4veq:2}
    -\langle\mathbf{E}_T,\mathbf{J}_a\rangle
    =-\int_V dV\mathbf{E}_T\cdot \mathbf{J}_a
    =-I\int_a^b d\mathbf{l}\cdot \mathbf{E}_a=V_{ab}I=VI=Z_{in}I^2
    \end{equation}
    asserting the correctness of the impedance formula in the exact
    limit.

To prove the variational form of the above formula, we express the
above in a quadratic form.  To this end, we have
    \begin{align}
    \label{c4veq:3}
    Z_{in}=-\frac{
    \langle\mathbf{E}_a+\mathbf{E}_p,\mathbf{J}_a+\mathbf{J}_p\rangle}{I^2}
    &=-\frac{\langle\mathbf{E}_a,\mathbf{J}_a\rangle
    +\langle\mathbf{E}_p,\mathbf{J}_a\rangle
    +\langle\mathbf{E}_a,\mathbf{J}_p\rangle
    +\langle\mathbf{E}_p,\mathbf{J}_p\rangle}{I^2}\nonumber\\
    &=-\frac{\langle\mathbf{E}_a,\mathbf{J}_a\rangle
    +2\langle\mathbf{E}_a,\mathbf{J}_p\rangle
    +\langle\mathbf{E}_p,\mathbf{J}_p\rangle}{I^2}\nonumber\\
    &=-\frac{\langle\mathbf{E}_a,\mathbf{J}_a\rangle
    +2\langle\mathbf{E}_a,\mathbf{J}_p\rangle
    +i\omega\mu\langle\mathbf{J}_p,\dyad{G}_e,\mathbf{J}_p\rangle}{I^2}
    \end{align}
It is straightforward to prove that the stationary point of the
above expression is at the exact solution.  In the above, the choice
of $\dyad G_e(\v r,\v r')$ is important.  It can be a free-space
Green's function, or Green's function that satisfies specific
boundary conditions.  For instance, on a PEC surface where $\hat
n\times \dyad G_e(\v r,\v r')=0$, reciprocity implies immediately
that an impressed electric current on such a surface does not
radiate, and its contribution can be ignored in the above
calculation.

Another popular formula for input impedance is the \index{Power formula for impedance}power formula
where the input impedance is given by
    \begin{equation}
    \label{c4veq:1ab}
    Z_{in}=-\frac{\langle\mathbf{E}_T,\mathbf{J}_T^*\rangle}{|I|^2}
\end{equation}
The above is based on power conservation, but its variational nature
cannot be proved.  However, in the limit when $\v E_T$ is exact, the
above reduces only to integration over the current $\v J_a$, and it
becomes
    \begin{equation}
    \label{c4veq:2a}
    Z_{in}=-\frac{\langle\mathbf{E}_T,\mathbf{J}_a^*\rangle}{|I|^2}
\end{equation}
If the current source $\v J_a$ is electrically small and constant
phase as is the case for a circuit component in circuit theory, the
above formula reduces to the variational formula based on reaction.

The names for these formulas have been rather confusing in the
literature.  The above power formula has been called the induced EMF
formula in  \cite{HARRINGTON4}, while the reaction formula has been
called the induced EMF formula in  \cite{JORDAN_BALMAIN}.  Much of
the controversy between the induced EMF formula and power formula
has also been discussed in  \cite{JORDAN_BALMAIN}.

\begin{figure}[!htb]
    \begin{center}
    \includegraphics[width=120mm]{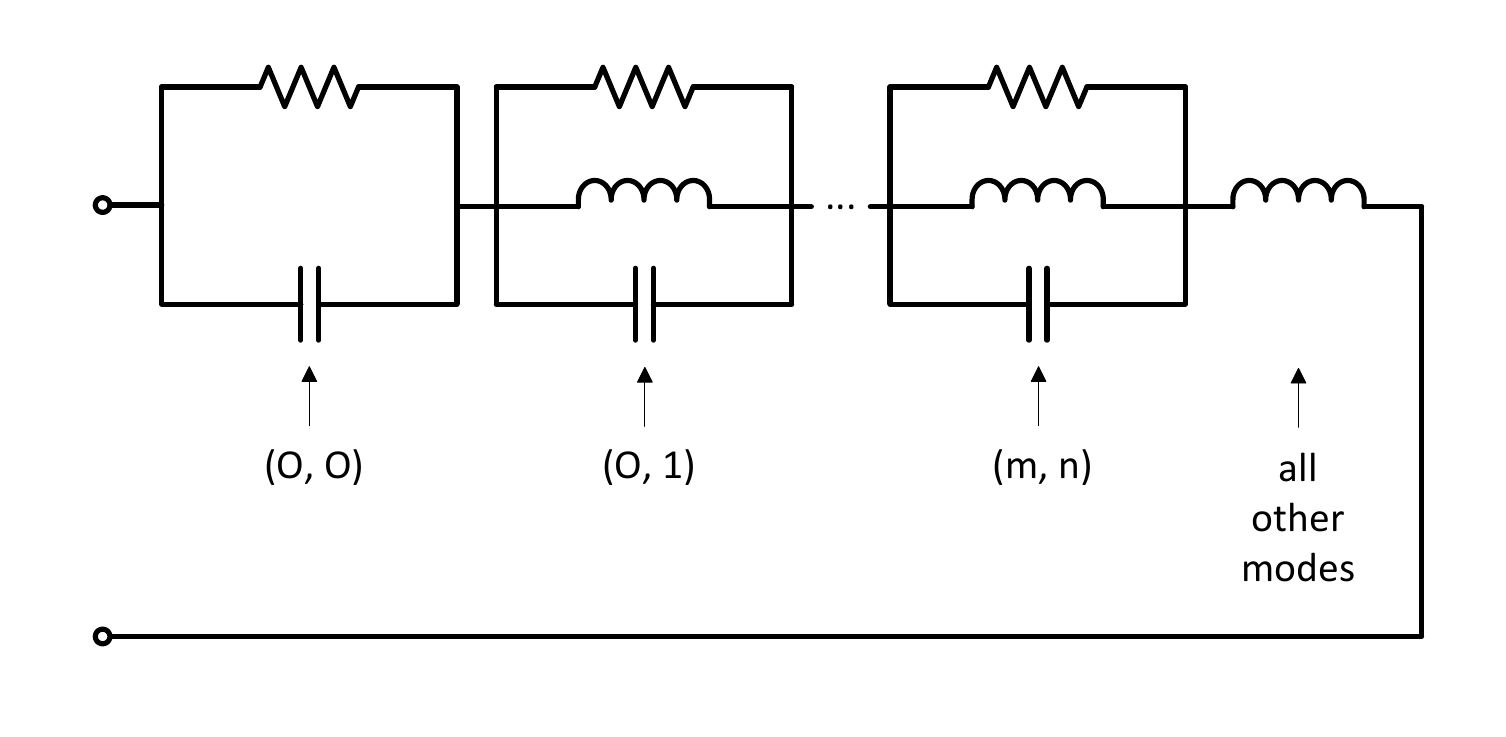}
    \caption{Circuit model for the input impedance of a microstrip patch antenna.}
    \label{circuitModelAntenna}
    \end{center}
    \end{figure}

\subsubsection{Application to Patch Antenna}
\index{Antenna!patch!application}

We will next show how to apply the above result to a microstrip
patch antenna.
    In the exact limit, as mentioned before,
    \begin{equation}
    \label{c4veq:4}
    Z_{in}
    =-\frac{\langle\mathbf{E}_T,\mathbf{J}_a\rangle}{I^2}
    \end{equation}
In the above, $\v E_T$ is the total field in the structure produced
by the applied current $\v J_a$ and the induced current $\v J_p$.
For the microstrip patch antenna, we have learned how to find the
total field in the cavity using a magnetic wall model when a current
source is inserted into it. Hence, the total field $\v E_T$ is
approximately given by\footnote{More elaborate model is given in
 \cite{KUROKAWA,COLLINI,CHEW_KONG_SHEN}.}
    \begin{equation}
    \label{c4veq:5}
    \mathbf{E}_T=i\omega\mu\sum_i\frac{\langle\mathbf{E}^*_i,\mathbf{J}_a\rangle}
    {\tilde{k}_i^2-k^2}\mathbf{E}_i\left(\mathbf{r}\right)
    \end{equation}
In the above, we have assumed that the \index{Green's function!dyadic!microstrip cavity}dyadic Green's function of
the microstrip cavity can be approximated by
   \begin{equation}
    \label{c4veq:5a}
    \dyad G(\v r,\v r') = \sum_i\frac{\v E_i(\v r)\mathbf{E}^*_i(\v r') }
    {\tilde{k}_i^2-k^2}
    \end{equation}
Hence, this Green's function automatically accounts for the field
generated by the induced current on the wall of the cavity.  This is
because we have chosen the eigenmodes to satisfy the requisite
boundary conditions on the wall of the cavity.

    Therefore,
    \begin{equation}
    \label{c4veq:6}
        \langle \v E_T,\v J_a\rangle=i\omega\mu\sum_i
        \frac{\langle \v E_i^*, \v J_a\rangle}{\tilde{k}_i^2-k^2}
        \langle \v E_i, \v
        J_a\rangle
    \end{equation}
    Hence, the input impedance, derivable from \eqref{c4veq:4}, is
    \begin{equation}
    \label{c4veq:7}
        Z_{in}=Z_a+\sum_i^M Z_i
    \end{equation}
    where $Z_a$ is higher-order mode contributions, and the second
    summation comes from the dominant mode contributions.
    \begin{equation}
    \label{c4veq:8}
        Z_i=-\frac{i\omega\mu}{I^2}
        \frac{\langle\v E_i^*,\v J_a\rangle\langle\v E_i,\v J_a\rangle}
        {\tilde{k}_i^2-k^2}
    \end{equation}
is the contribution to the input impedance from individual modes of
the cavity.  Notice that the above frequency dependence can be
fitted with a simple GLC model of a lossy tank circuit resonator.
Hence, for the $i$-th mode, we can pick $G_i$, $L_i$, and $C_i$
appropriately to fit the mathematical formula.  A circuit
approximation of the microstrip patch antenna hence can be expressed
as in Figure \ref{circuitModelAntenna}  \cite{RICHARDS_LO_HARRISON}.
For the microstrip patch, there is a static mode TM$_{00}$ mode with
zero resonant frequency. This mode represents the static capacitor
in the microstrip patch.  It is denoted by the lossy capacitor
model. Also, the probe produces a singular field, which can only be
constituted by a linear superposition of many high order modes.
Hence, the probe inductance comes from the higher order modes in the
cavity.  Notice that in this model, unlike the magnetic frill model,
the gap capacitance at the base of the probe is ignored.

\section {{{ Aperture Coupling in Waveguide}}}
\index{Aperture coupling in a waveguide}

\begin{figure}[htb]
\begin{center}
\hfil\includegraphics[width=3.0truein]{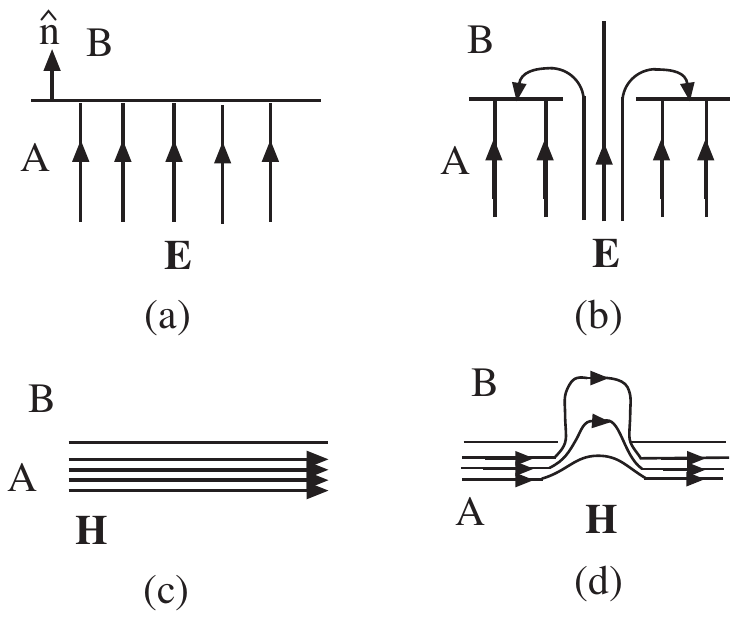}\hfil
\end{center}
\caption{Fields in the vicinity of an aperture in a waveguide.
Case (b) resembles an electric dipole, while case (d) resembles a
magnetic dipole.}\label{fg431}
\end{figure}


In addition to using probes to couple energy into a waveguide, a
simple way is to drill holes on the walls of the waveguide, and let
energy flow naturally from one waveguide to another. The simple
solution of aperture coupling was first derived by Hans Bethe
 \cite{BETHE,COLLINI,COHN}, who eventually received a Nobel prize,
not for one given contribution, but for his numerous contributions
in physics.

\subsection{Bethe Coupling}
\index{Bethe coupling}

An $\v E$ field in the vicinity of a waveguide wall, is
predominantly normal to the waveguide wall. If now, an aperture is
opened at the waveguide wall, the electric field in the vicinity of
the waveguide wall will be as shown in Figure \ref{fg431}(b). The
field looks like that of a vertical electric dipole in region $B$.
It has been shown by Bethe  \cite{BETHE,COLLINI} that the dipole
moment of the vertical electric dipole is proportional to the normal
component of the electric field. For a circular aperture of radius
$a_0$, it is
\begin{equation}
\v p=\frac 2{3} a_0^3\^ n (\^ n\cdot\epsilon _0\v E)=\alpha_e \^ n
(\^ n\cdot\epsilon _0\v E). \label{eq4-61}
\end{equation}
where $\alpha_e=\frac 2{3} a_0^3$, and $a_0$ is the radius of the
circular aperture.

\begin{figure}[htb]
\begin{center}
\hfil\includegraphics[width=5.0truein]{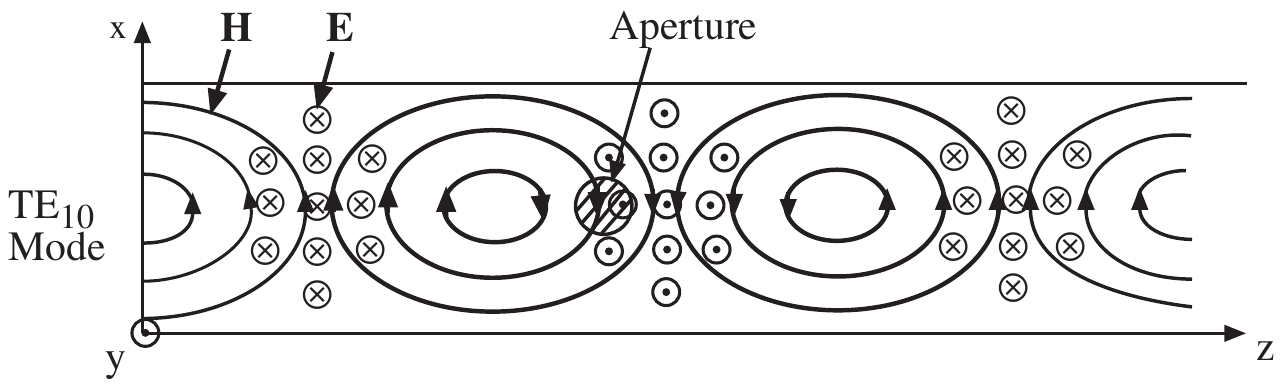}\hfil
\end{center}
\caption{Top view of the field in the neighborhood of an aperture
in a rectangular waveguide.}\label{fg432}
\end{figure}


A magnetic field in the vicinity of the waveguide wall, is
predominantly tangential. Now, if an aperture is present, the
magnetic field will leak into region $B$ as shown in Figure
\ref{fg431}(d). It looks like the field due to a horizontal magnetic
dipole in region $B$. Similarly, the dipole moment of the horizontal
magnetic dipole is  \cite{BETHE,COLLINI}
\begin{equation}
\v m=-\frac 4{3}a_0^3\v H_t==-\alpha_m\v H_t  . \label{eq4-62}
\end{equation}
where $\alpha_m=\frac 4{3} a_0^3$.

\begin{figure}[htb]
\begin{center}
\hfil\includegraphics[width=5.0truein]{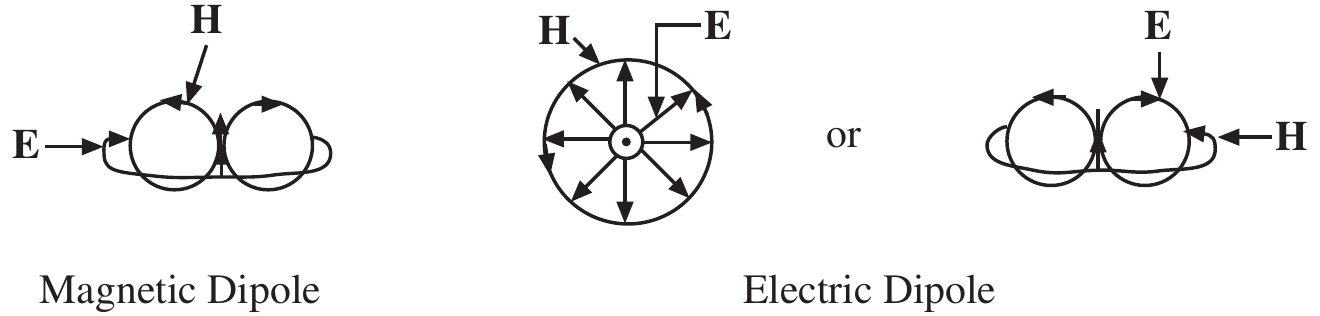}\hfil
\end{center}
\caption{Equivalent sources at the aperture of a waveguide.}\label{fg433}
\end{figure}


\begin{figure}[htb]
\begin{center}
\hfil\includegraphics[width=4.5truein]{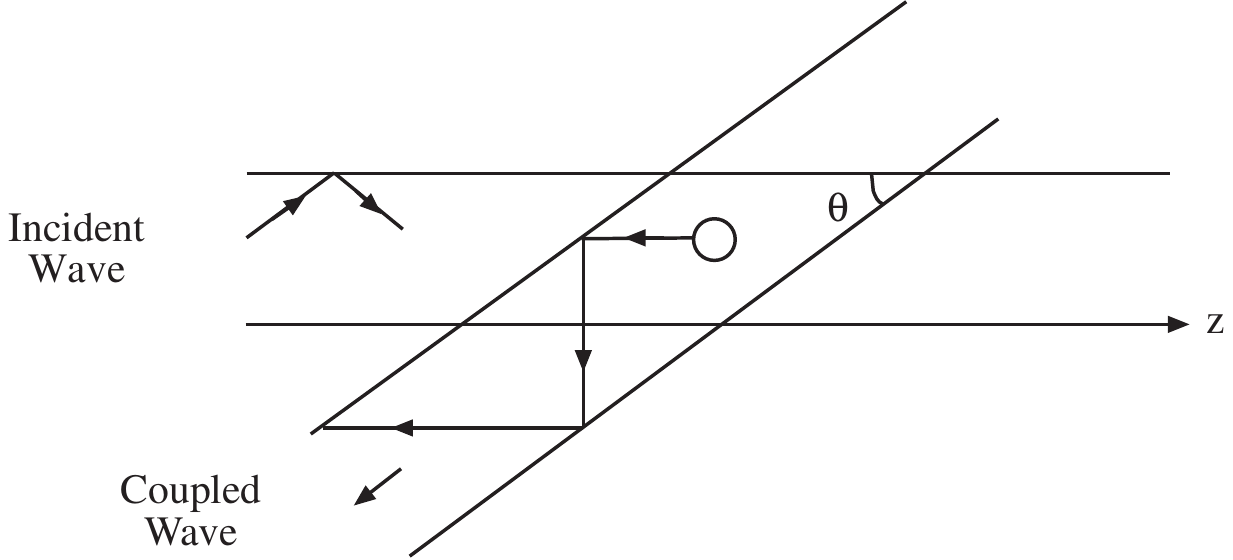}\hfil
\end{center}
\caption{A directional coupler using two rectangular waveguides,
one on top of another.}\label{fg434}
\end{figure}


If a TE$_{10}$ mode is propagating in a rectangular waveguide, and
we have a small aperture on top of the waveguide, then a time
harmonic electric and magnetic dipoles will be generated at the
aperture. In this case, the electric dipole and magnetic dipole are
in phase with respect to each other. For the case of Figure
\ref{fg432}, the magnetic dipole is pointing in the $x$ direction,
and the electric dipole is pointing in the $y$ direction. The
superposition of the vertical electric dipole and horizontal
magnetic dipole gives rise to the cancellation of fields in the $+z$
direction (see Figure \ref{fg433}). Hence, together, they radiate
predominantly in the $-z$ direction.\footnote{Antennas made by a
superposition of an electric dipole and a loop to increase their
directivity are known as \index{Antenna!Huygens}Huygens antenna.}

Now, if we lay another waveguide on top of the first waveguide, the
radiating electric and magnetic dipoles couple most efficiently into
the TE$_{10}$ mode of the top waveguide if the top guide is oriented
at an angle $\theta$ with respect to the bottom guide as shown in
Figure \ref{fg434}. This is because a TE$_{10}$ mode is actually a
bouncing plane wave in a rectangular waveguide.

\begin{figure}[htb]
\begin{center}
\hfil\includegraphics[width=4.0truein]{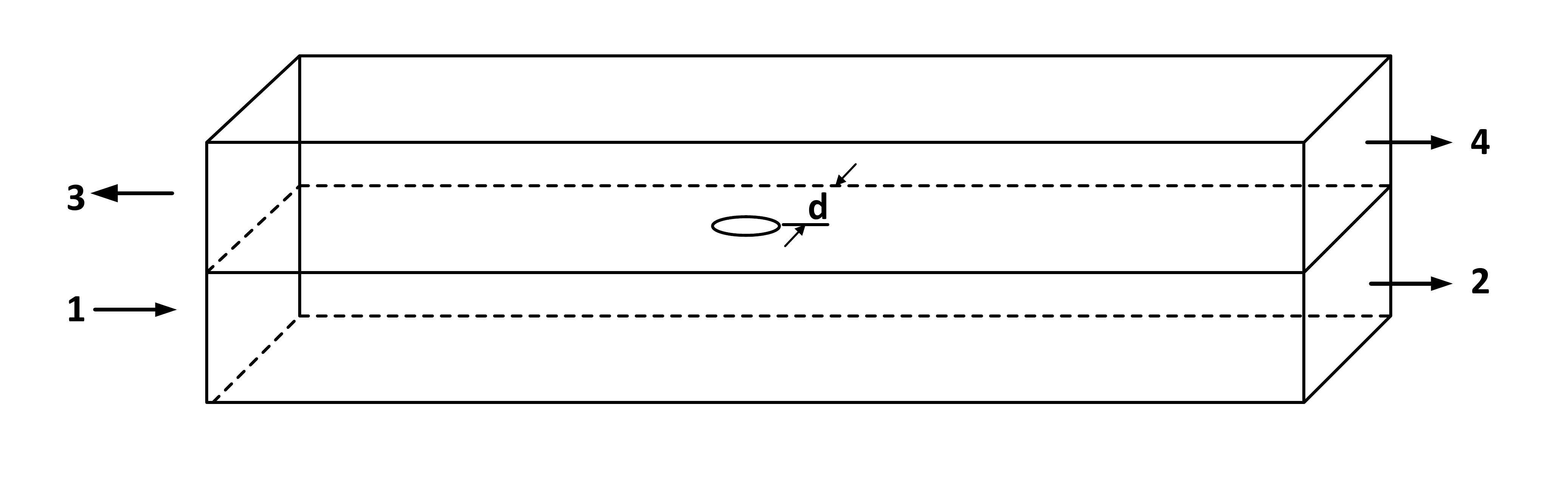}\hfil
\end{center}
\caption{An asymmetrically located aperture can also be used to
make a directional coupler where $d\ne a/2$.}\label{fg435}
\end{figure}


When the aperture is located at the \index{Waveguide!middle aperture}middle of the waveguide, the
magnetic dipole is pointing in the $x$ direction, requiring the top
waveguide to be tilted. We can generate a dipole pointing away from
the $x$ direction by placing the aperture away from the center of
the waveguide, or by using an \index{Elliptically shaped aperture} elliptically shaped aperture. In this
case, we can couple efficiently into the TE$_{10}$ mode without
having to tilt the top waveguide. Hence, a \index{Directional couplers}directional coupler can
also be made with $\theta =0$ if the aperture is not symmetrically
located as shown in Figure \ref{fg435} When an aperture is
asymmetrically located, the magnetic field that excites it is
elliptically polarized. It can be shown that the exciting field  of
this aperture due to an incident TE$_{10}$ in port 1 of the bottom
waveguide is given by


\begin{equation}
\v E=\hat y E_y =\hat y E_0 \sin \left(\frac {\pi d }{a}\right),
\label{eq4-63}
\end{equation}
\begin{equation}
\v H =-E_0 \frac {k_z}{\omega \mu _0} \left[ \^ x \sin \left(
\frac {\pi d}{a} \right)  +i \^ z \frac {\pi }{k_z a } \cos \left(
\frac {\pi d }{a} \right)\right]. \label{eq4-64}
\end{equation}
The equivalent electric dipole moment for radiation into the upper
guide is
\begin{equation}
\v p=\^ y \epsilon _0 \alpha_e E_0\sin \left(\frac {\pi d
}{a}\right), \label{eq4-65}
\end{equation}
The equivalent magnetic dipole moment is
\begin{equation}
\v m =\alpha_m  E_0 \frac {k_z }{\omega \mu _0}\left[ \^ x \sin
\left(\frac {\pi d }{a }\right) +i \^ z \frac {\pi }{k_z a } \cos
\left(\frac {\pi d }{a } \right)\right] , \label{eq4-66}
\end{equation}

\subsubsection{Excitation of Modes by Electric and Magnetic Current
Sources}
\index{Mode!excitation by current sources}

Since there will be an electric dipole source and a magnetic dipole
source induced in a waveguide due to \index{Mode!coupling via hole} coupling via a hole, it is
prudent to study how they would couple to other modes in a
waveguide. We have learned previously that an electric current
source excites modes in a waveguide or cavity as
\begin{align}\label{Betheeq:1}
  \v{E}(\v{r})=i\omega\mu\sum\limits_{i}\frac{\left<\v{E}_i,\v{J}_e\right>}{k_i^2-k^2}\v{E}_i(\v{r})
 \end{align}
We define
 \begin{align}\label{Betheeq:2}
  \v{H}_i=\frac{1}{k_i}\nabla\times\v{E}_i
 \end{align}
If $\v{E}_i$ is normalized, $\v{H}_i$ is also normalized. In fact
one can easily show that
 \begin{align}\label{Betheeq:3}
  \left<\v{H}_i^\ast,\v{H}_i\right> &=\frac{1}{k_i^2}\left<\nabla\times\v{E}_i^\ast,\nabla\times\v{E}_i\right>
  \notag\\ &=\frac{1}{k_i^2}\left<\v{E}_i^\ast,\nabla\times\nabla\times\v{E}_i\right>
  \notag\\ &=\left<\v{E}_i^\ast,\v{E}_i\right>=1
 \end{align}
Also, it can be easily shown that
 \begin{align}\label{Betheeq:4}
  \nabla\times\v{H}_i=\frac{1}{k_i^2}\nabla\times\nabla\times\v{E}_i=k_i\v{E}_i
 \end{align}
Hence, if we have a vector wave equation given by\footnote{We will
use $\v J_m$ to denote magnetic current and reserve $\v M$ to denote
magnetization density in this section.}
 \begin{align}\label{Betheeq:5}
  \nabla\times\nabla\times\v{H}-k^2\v{H}=i\omega\epsilon\v{J}_m
 \end{align}
 the orthonormal eigenmode expansion gives
 \begin{align}\label{Betheeq:6}
  \v{H}=i\omega\epsilon\sum\limits_{i}
  \frac{\left<\v{H}_i^\ast,\v{J}_m\right>}{\left(k_i^2-k^2\right)}\v{H}_i(\v{r})
 \end{align}
where $\v{H}_i$ is normalized. The corresponding $\v{E}$ field, via
the use of \eqref{Betheeq:4} and Maxwell's equations, is
 \begin{align}\label{Betheeq:7}
  \v{E}=\sum\frac{\left<\v{H}_i^\ast,\v{J}_m\right>}{\left(k_i^2-k^2\right)}k_i\v{E}_i
 \end{align}

The \index{Polarization density}polarization density can be expressed as
\begin{align}\label{Betheeq:8}
\v{P} = \hat{y} \epsilon \alpha_{e} E_{0} \sin\left(\frac{\pi
d}{a}\right)\delta(\v r-\hat x d)
\end{align}
to imply that the hole is located at $(x,y,z)=(d,0,0)$.  The
corresponding electric current density arising from time-varying
polarization density is
\begin{align}
\v{J}_{p} = -i \omega \v{P}
\end{align}

Similarly, the \index{Magnetization density}magnetization density is
\begin{align}\label{Betheeq:9}
\v{M} = \alpha_{m} \frac{k_{z} E_{0}}{\omega \mu} \left[\hat{x}
\sin\left(\frac{\pi d}{a}\right) + i\hat{z} \frac{\pi}{k_{z} a}
\cos\left({\pi d}{a}\right)\right]\delta(\v r-\hat x d )
\end{align}
The corresponding magnetic current density arising from a
time-varying magnetization density is given as
\begin{align}
\v{J_{m}} = i \omega \mu \v{M}
\end{align}

The corresponding TE$_{10}$ mode in the upper waveguide is
\begin{align}\label{Betheeq:10}
\v{E}_{10} = \hat{y} E_{10} \sin\left(\frac{\pi x}{a}\right) e^{\pm
i k_{z} z}
\end{align}
where $E_{10}$ is for normalization. The corresponding $\v{H}_{10}$
field is
\begin{align}\label{Betheeq:11}
\v{H}_{10} = \frac{E_{10}}{k_{10}} \left[\pm
\hat{x}ik_{z}\sin\left(\frac{\pi x}{a}\right)+ \hat{z} \frac{\pi}{a}
\cos\left(\frac{\pi x}{a}\right)\right]e^{\pm i k_{z} z}
\end{align}
which is normalized.  The $\pm$ sign implies $\pm z$ propagating
waves.

To see if the mode excited by the electric dipole will cancel the
one excited by the magnetic dipole, we need to compare
\eqref{Betheeq:1} and \eqref{Betheeq:7}.  Hence, we need to sum the
coefficients
$$
k_{10} \langle \v H_{10}^{\star} , \v J_{m} \rangle
$$
and
$$
i \omega \mu \langle \v E_{10}^{\star},\v{J}_{p} \rangle
$$
In details,
\begin{align}\label{Betheeq:12}
k_{10} \langle \v H_{10}^{\star} , \v{J}_{m} \rangle= \alpha_{m} i
k_{z} E_{0} E_{10} \left[\pm i k_{z} \sin^{2}\left(\frac{\pi
d}{a}\right) + i \frac{\pi^{2}}{k_{z} a^{2}} \cos^{2}\left(\frac{\pi
d}{a}\right)\right]
\end{align}
Furthermore,
\begin{align}\label{Betheeq:13}
i \omega \mu \langle \v E_{10}^{\star}, \v {J}_{p} \rangle=
\omega^{2} \mu \epsilon \alpha_{e}E_{0} E_{10} \sin^{2}(\frac{\pi
d}{a})
\end{align}
Hence, the TE$_{10}$ mode that is excited is proportional to
\begin{align}\label{Betheeq:14}
B_{\pm} \propto & \quad k_{10} \langle \v{H}_{10}^{\star}, \v{J}_{m}
\rangle
+ i \omega \mu \langle \v{E}_{10}^{\star}, \v{J}_{p} \rangle\nonumber\\
 &=
E_{0} E_{10}\left[\pm k_{z}^{2} \alpha_{m} \sin^{2}\left(\frac{\pi
d}{a}\right) + \left(\frac{\pi}{a}\right)^{2} \alpha_{m}
\cos^{2}\left(\frac{\pi d}{a}\right) + k^{2} \alpha_{e}
\sin^{2}\left(\frac{\pi d}{a}\right)\right] \nonumber\\&= 0, \quad
z<0.
\end{align}
The solution of the above is
\begin{equation}
\frac d a =\frac 1 \pi \sin ^{-1} \left(\frac 1 {\sqrt {6} } \frac
{\lambda _0}{a} \right) \label{eq4-71}
\end{equation}
where $\lambda_0$ is the free-space wavelength. The above coupler
using one hole to achieve directional coupling is known as the \index{Bethe hole coupler}Bethe
hole coupler.

\index{Coupling of waveguides}

When a hole is dug in a waveguide wall, the induced dipoles also
give rise to back action in the original waveguide, altering the
field.  This, in turn, gives rise to an alteration of the coupled
field from the first waveguide to the second waveguide.  This effect
is taken into account by Collin  \cite{COLLINI}.  The resulting
hierarchy of equations is rather complicated but the above
condition still holds true.  This can be thought of as a multiple
scattering or coupling effect, as in the Fabry-Perot etalon. It is
the cancellation of the leading order term that is important, for the
cancellation of the higher-order terms will follow suit.

\begin{figure}[htb]
\begin{center}
\hfil\includegraphics[width=4.0truein]{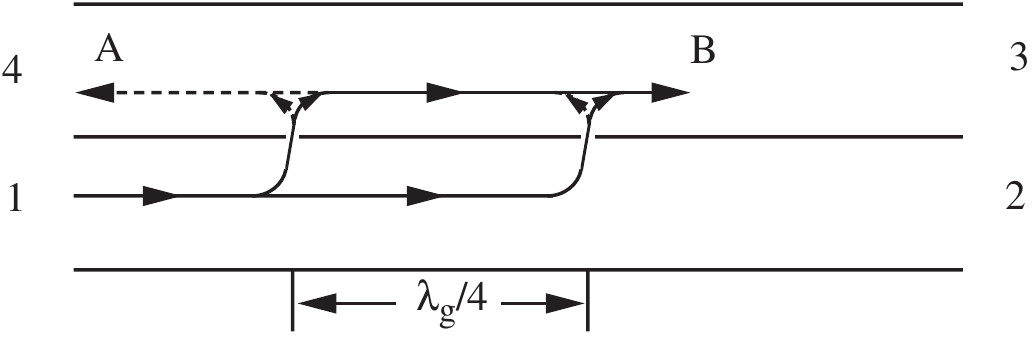}\hfil
\end{center}
\caption{A two-hole directional coupler.}\label{fg436}
\end{figure}

Other kinds of \index{Directional couplers} directional couplers are also possible. For instance,
one can make a \index{Two-hole coupler} two-hole coupler with the holes spaced $\lambda_g/4$
apart as shown in Figure \ref{fg436}. The apertures need not have a
preferred directional coupling. If the aperture coupling is weak,
then the wave reaching the second hole is essentially the same as
the wave that reaches the first hole. Hence, the two different waves
that reach $B$ via coupling through the two different holes are in
phase and will interfere constructively. Because of the $\lambda_g/4
$ separation of the two holes, the waves from the two different
holes that reach $A$ will be $\lambda_g /2$ or $180^o$ out of phase.
Therefore, at $A$, the waves interfere destructively, and there is
little energy coupled to port 4.


Since this coupler uses constructive and destructive interferences
to enhance its directivity, the directivity is frequency sensitive.
One remedy is to use more holes so as to broaden its bandwidth, or
to use the Schwinger reversed-phase coupler. \cite{COLLINI,ANDERSON}

\begin{figure}[htb]
\begin{center}
\includegraphics[width=1.5truein]{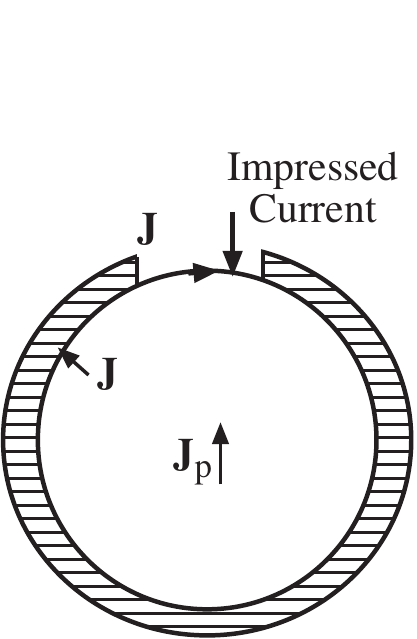}
\end{center}
\caption{Equivalent problem of a waveguide with an
aperture---nonradiating case.}\label{fg438}
\end{figure}


\subsection { Equivalence Principles in Aperture Coupling}
\index{Aperture coupling}


Equivalence principles have been discussed in
 \cite{COLLINI,HARRINGTON4}.  In the actual calculation of the
electric and magnetic dipole moments, certain equivalence principles
have to be invoked. We shall discuss them as follows.

There will be two kinds of currents in the following discussion:
induced currents and impressed currents.  \index{Induced current} Induced currents are
currents flowing in a conductor due to the presence of an incident
or exciting field next to the conductor.  A perfect conductor, for
instance, cannot have a non-zero field in it, and hence, current
flows on its surface to prevent the fields from penetrating it.  On
the other hand, \index{Impressed current}impressed currents are currents we assume as sources
in Maxwell's equations.  They are currents assumed to exist in free
space.  They are the driving source terms in Maxwell's equations,
that are immutable as we seek the solutions.  On the other hand,
induced currents follow from the solutions of Maxwell's equations.
They are due to currents flowing in conductors as we seek solutions
to Maxwell's equations.


\subsubsection{Equivalence Principle I}
\index{Equivalence principle}

\begin{figure}[htb]
\begin{center}
\hfil\includegraphics[width=5.5truein]{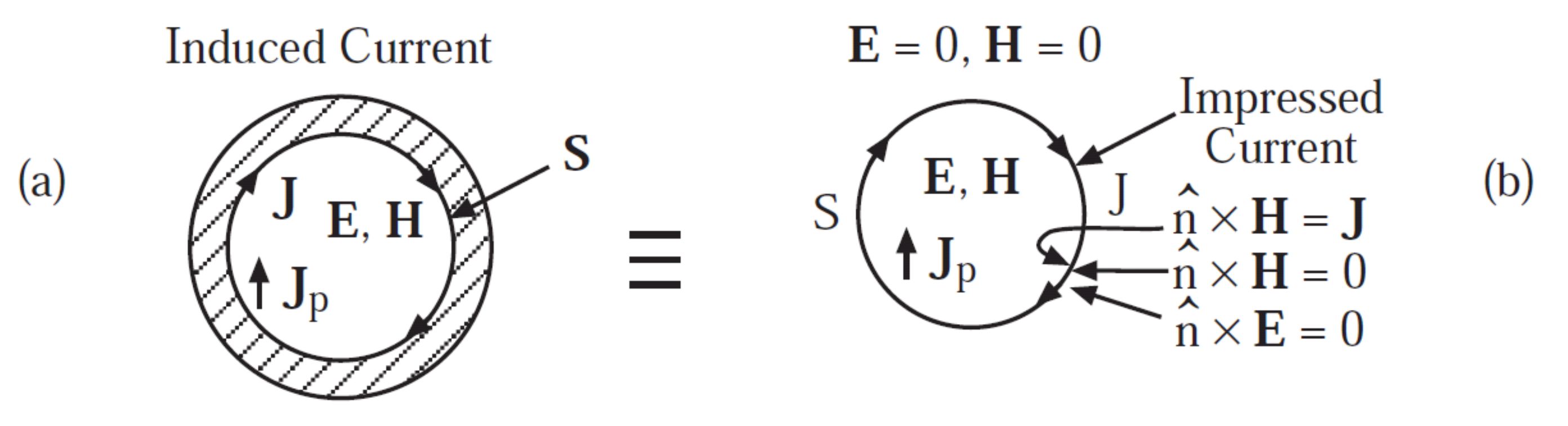}\hfil
\end{center}
\caption{Equivalent problem of a hollow waveguide---equivalence
principle I.}\label{fg437}
\end{figure}

If we have a source in a metallic waveguide, the source will
induce a current $\v J$ on the inner surface of the waveguide.
The current $\v J$ is known as the induced current.  On the
surface of the waveguide, we have
\begin{equation}
\^ n\times\v E=0, \qquad\^ n\times\v H=\v J. \label{eq4-72}
\end{equation}
This current is responsible for expelling the electromagnetic field
away from the \index{Perfect electric conductor (PEC)} perfect conductor.  It is also responsible for the
jump discontinuity for the magnetic field outside and inside the
perfect conductor.

Now if we remove the metallic wall that supports the induced current
and replace the induced current with an {\it impressed current} in
vacuum, the field inside the surface $S$ is identical to before.
Furthermore, the field outside $S$ is identically zero.  This is
because the impressed current $\v J$ supports a discontinuity in the
magnetic field.  Hence,  $\^ n\times \v H=0$ and $\^ n\times\v E=0$
just outside the impressed current $\v J$. By Huygens' principle,
the field must be zero everywhere outside $S$. We can check if the
solution satisfies all the requisite boundary conditions. If it
does, it is the only unique solution.

The equivalence principle can be proved by three means:
\index{Equivalence principle!proof}

\begin{itemize}
\item By performing a Gedanken experiment whereby the conductive
material with zero field inside is been chiseled away until the
induced current is replaced by impressed current in free space.  The
induced current has held the internal field of the cavity in place,
and the impressed current will still hold the internal fields in
place.

\item By using uniqueness principle argument,  when the induced
current is replaced by impressed current, the boundary conditions
for the fields remain the same.  By uniqueness principle, they must
be the same;

\item The equivalence principle can also be proved mathematically by
the use of Huygens principle.
\end{itemize}


\begin{figure}[htb]
\begin{center}
\hfil\includegraphics[width=5.0truein]{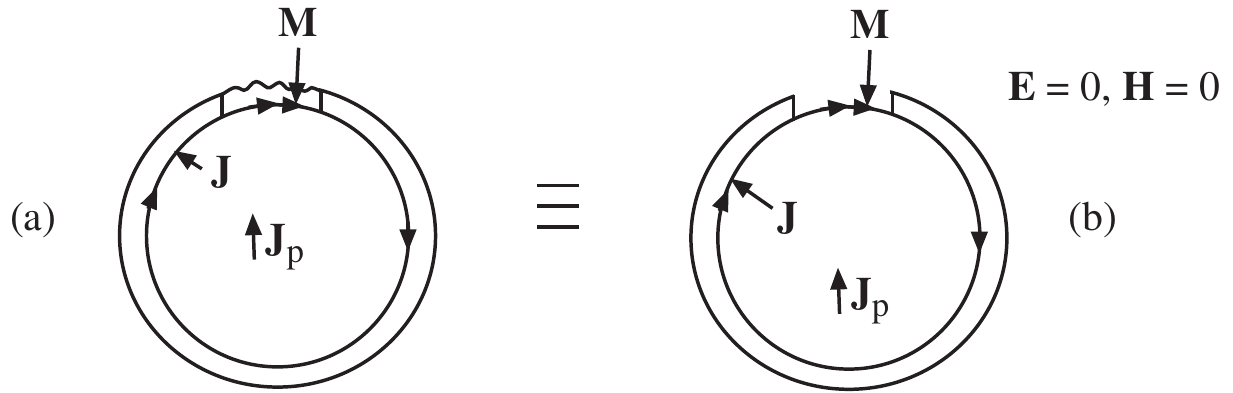}\hfil
\end{center}
\caption{Equivalent problem of a waveguide which is partly covered
with a magnetic wall-nonradiating case.}\label{fg4310}
\end{figure}


\subsubsection{Equivalence Principle II}

\begin{figure}[htb]
\begin{center}
\hfil\includegraphics[width=4.5truein]{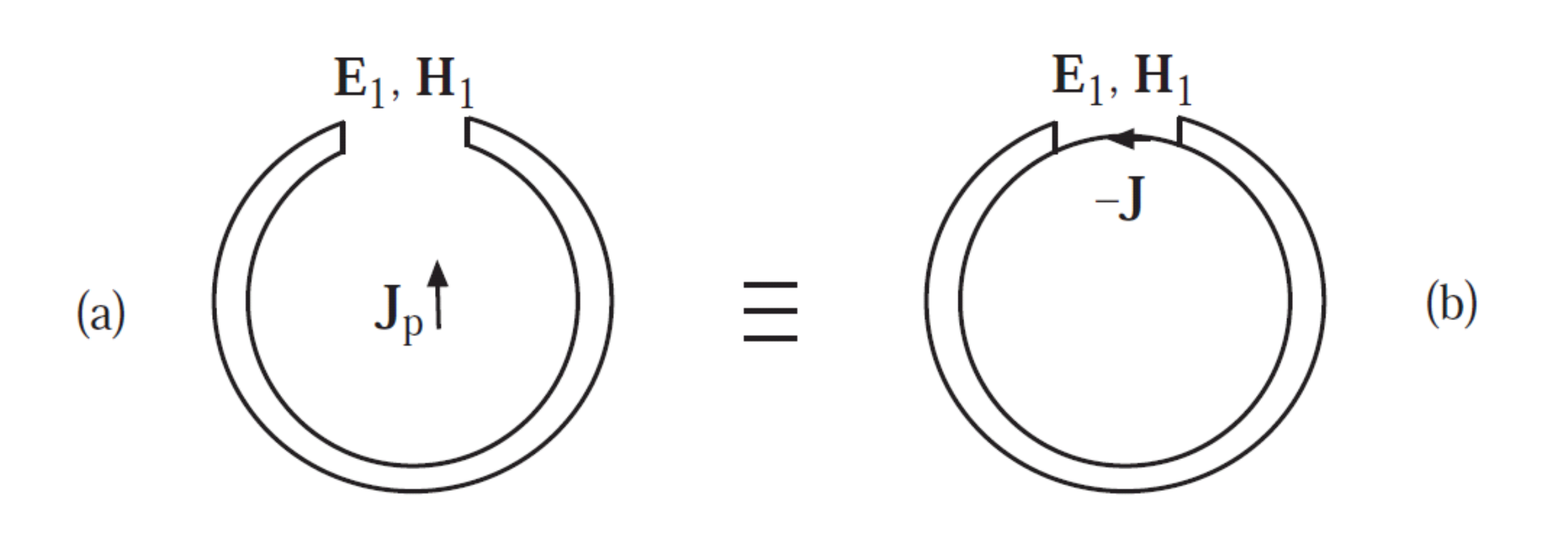}\hfil
\end{center}
\caption{Equivalent problem of a waveguide with an
aperture---radiating case---equivalence principle II.}\label{fg439}
\end{figure}

From equivalence principle I, if we have a small aperture in the
cavity now, and at the aperture, the original induced current from
Figure \ref{fg437}a is impressed, then the field will still be
identically zero outside the waveguide.  Hence, $\v J_p$ and its own
induced current on the PEC wall must have generated equal and
opposite field to that produced by impressed $\v J$ at the aperture
and its own induced current on the PEC wall.  Because of this, the
two currents in Figure \ref{fg438} generate zero field outside the
waveguide.  Also, because of this, the cases in Figure \ref{fg439}a
and Figure \ref{fg439}b generate equivalent field outside the
waveguide.  Note that the current in Figure \ref{fg439}b is exactly
opposite to that in Figure \ref{fg438}.


\subsubsection{Equivalence Principle III}

\begin{figure}[htb]
\begin{center}
\hfil\includegraphics[width=5.0truein]{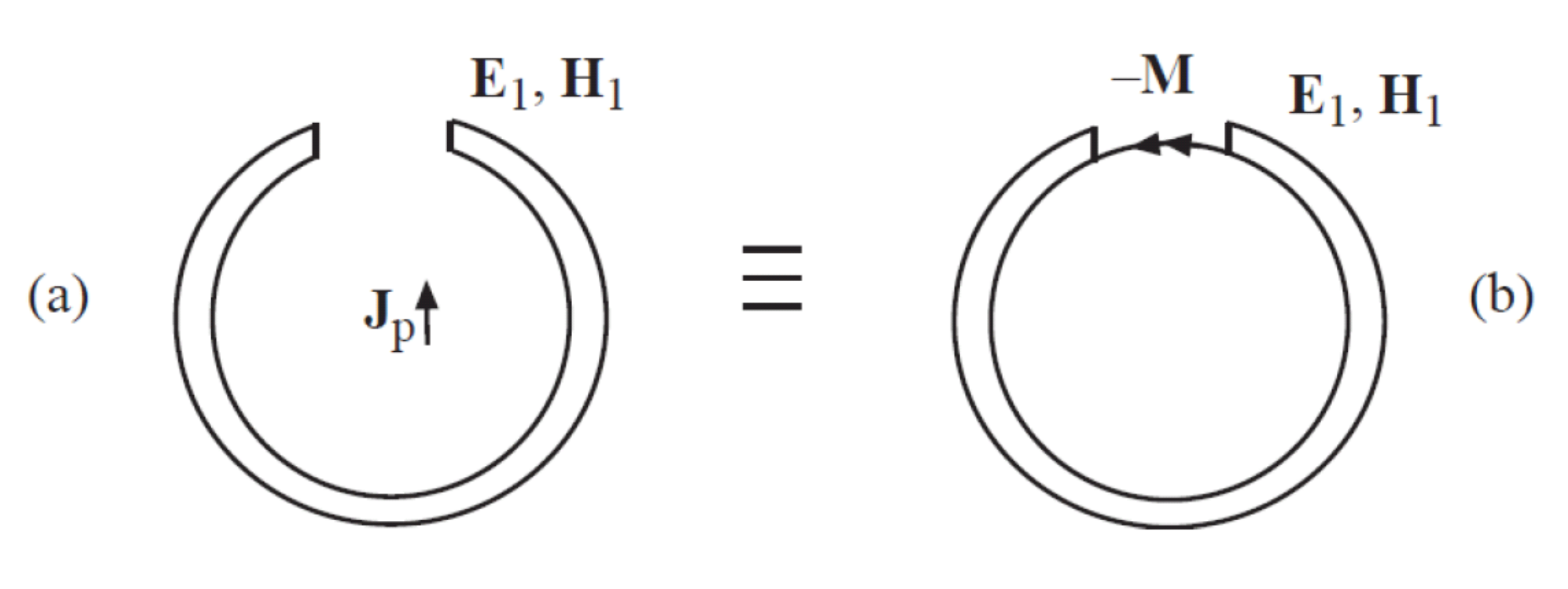}\hfil
\end{center}
\caption{Equivalent problem of a waveguide with an
aperture---radiating case--equivalence principle III.}\label{fg4311}
\end{figure}


\begin{figure}[htb]
\begin{center}
\hfil\includegraphics[width=5.0truein]{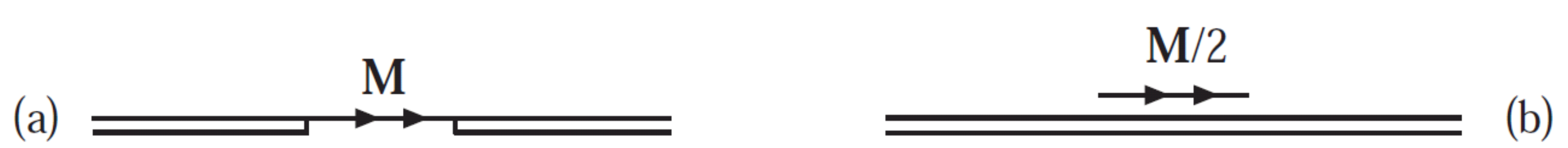}\hfil
\end{center}
\caption{Equivalent problem of a magnetic current radiating in the
vicinity of an aperture.}\label{fg4312}
\end{figure}


By a similar argument, if we have a closed cavity with a magnetic
wall over part of the cavity as in Figure \ref{fg4310}a, it is
completely equivalent to the case of Figure \ref{fg4310}b, with zero
field outside the waveguide.  Consequently, a waveguide with a small
aperture in Figure \ref{fg4311}a generates equivalent field outside
the waveguide as in the case of Figure \ref{fg4311}b.

Therefore, for radiation due to an aperture in a waveguide, there
are two equivalent problems denoted by Figure \ref{fg439} and
Figure \ref{fg4311}. For the case of Figure \ref{fg439}, to find
$\v J$, we have to first solve the closed waveguide problem
denoted by Figure \ref{fg437}a. For the case of Figure
\ref{fg4311}, we have to first solve the problem denoted by Figure
\ref{fg4310}a to find $\v M$. It turns out that equivalence
principle III is preferred over equivalence principle II because
the radiation of a magnetic current in a small aperture is easier
to calculate than the radiation of an electric current in a small
aperture. For example, if the surface of the waveguide is flat
enough, or that the aperture is small enough, we can replace the
problem in Figure \ref{fg4312}a with that in Figure \ref{fg4312}b.

When the PEC surface is flat, a horizontal \index{Magnetic dipole}magnetic dipole radiating
in the aperture is the same as the dipole radiating in free space,
as the dipole produces only horizontal magnetic field that satisfies
the boundary condition on the flat PEC surface.  So the source is
oblivious of the presence of the flat PEC surface.  By image
theorem, a horizontal magnetic dipole radiating in free space is
equivalent to one with half its original strength radiating on top
of a PEC ground plane.

\begin{figure}[htb]
\begin{center}
\hfil\includegraphics[width=5.0truein]{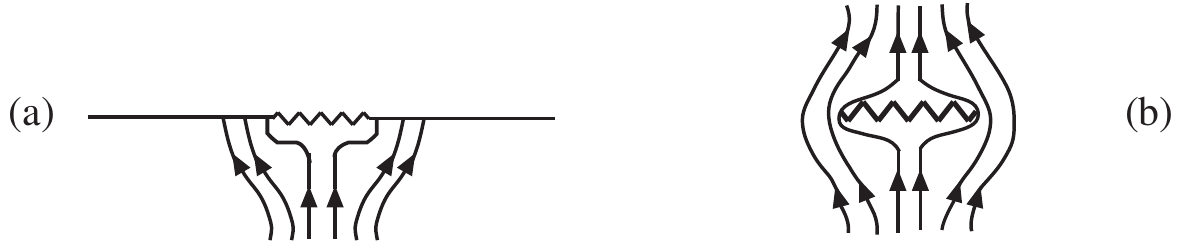}\hfil
\end{center}
\caption{Equivalent problem of the $\v E$ field around a waveguide
wall partially covered by a small magnetic wall.}\label{fg4313}
\end{figure}

\begin{figure}[htb]
\begin{center}
\hfil\includegraphics[width=5.0truein]{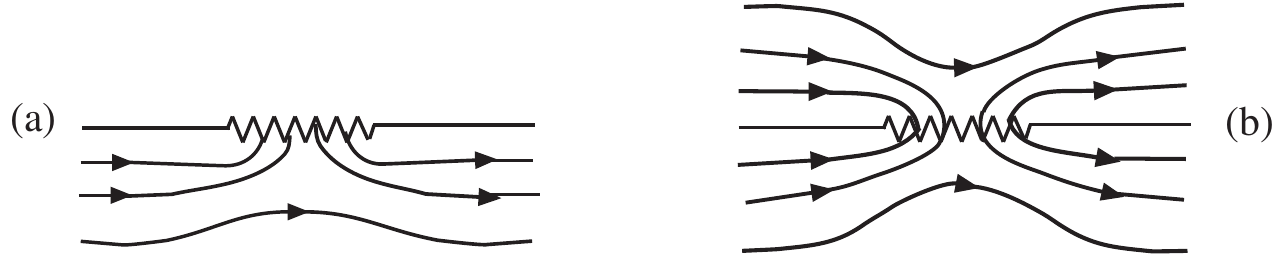}\hfil
\end{center}
\caption{Equivalent problem of the $\v H$ field around a waveguide
wall partially covered by a small magnetic wall.}\label{fg4314}
\end{figure}

As mentioned before, in order to find $\v M$, we need to solve the
closed problem with a magnetic wall patch as shown in Figure
\ref{fg4310}. If the magnetic wall patch is small enough, the
field around the magnetic wall may be approximated with a static
field solution. For example, if the field is predominantly
electric, the field in the vicinity of the magnetic wall patch
(assuming that the wall is reasonably flat) resembles that of
Figure \ref{fg4313}a. Due to the symmetry of the problem, it is
equivalent to that of Figure \ref{fg4313}b. What happens is that
the electric field induces circulating magnetic current on the
magnetic wall patch that expels the electric field. The
circulating magnetic current generates a vertical electric dipole
moment that expels the electric field from the magnetic disk.  If
the field is predominantly magnetic, the magnetic field around the
magnetic wall patch looks like that in Figure \ref{fg4314}a, which
is equivalent to that in Figure \ref{fg4314}b. The horizontal
magnetic field induces a horizontal magnetic dipole moment on the
magnetic wall patch.

\begin{figure}[htb]
\begin{center}
\hfil\includegraphics[width=2.0truein]{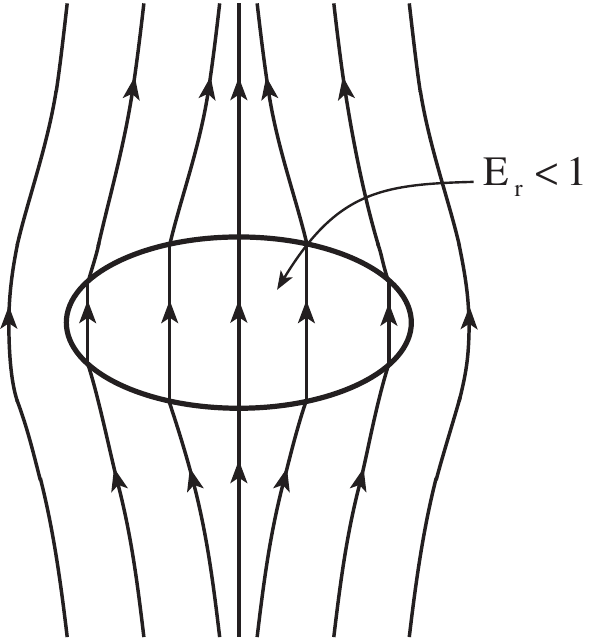}\hfil
\end{center}
\caption{Equivalent problem of a static electric field around an
ellipsoid.   When the relative permittivity becomes zero, and the
ellipsoid becomes a disk, the electric field around the disk is
the same as the electric around a PMC disk.}\label{fg4315}
\end{figure}

To obtain the solution for Figure \ref{fg4313},  we need to solve
for the solution of a static electric field in the vicinity of an
ellipsoid\index{Electric field of an ellipsoid}.  Consider the problem shown in Figure \ref{fg4315}. When
the relative permittivity, $\epsilon_r$ of the ellipsoid is less
than one, the electric field avoids the ellipsoid by skirting around
it.  The solution of the static electric field around the ellipsoid
can be obtained in ellipsoidal coordinates.  By letting one of the
axes of the ellipsoid shrink to zero, the ellipsoid becomes a disk.
Moreover, if we let $\epsilon_r=0$, no field can penetrate the disk
and the requisite solution for Figure \ref{fg4313} is obtained.

\begin{figure}[htb]
\begin{center}
\hfil\includegraphics[width=3.0truein]{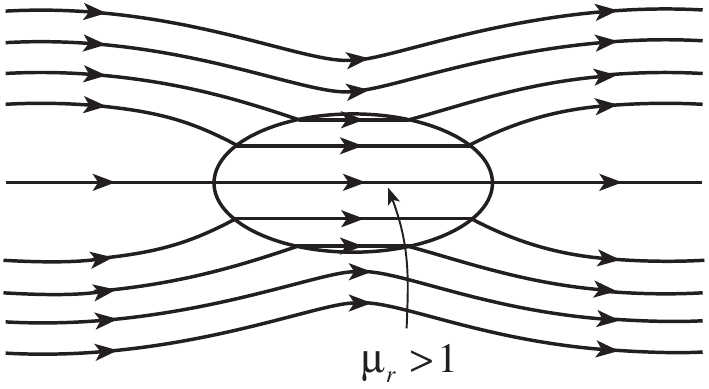}\hfil
\end{center}
\caption{ Equivalent problem of a static magnetic field around an
ellipsoid.   When the relative permeability becomes infinite, and
the ellipsoid becomes a disk, the magnetic field around the disk
is the same as the magnetic around a PMC disk.}\label{fg4316}
\end{figure}

By the same token, the solution needed for Figure \ref{fg4314} can
be obtained by studying the \index{Magnetic field of an ellipsoid} static magnetic field around an
ellipsoid with relative permeability $\mu_r$ larger than one.  In
this case, the magnetic field in the vicinity of the ellipsoid is
attracted to the ellipsoid as shown in Figure \ref{fg4316}.  When we
let the ellipsoid become a disk, and let $\mu_r\rightarrow\infty$,
we obtain the requisite solution for Figure \ref{fg4316}.  These
closed-form solutions can be used to obtain the coupling
coefficients for small aperture coupling, as in Bethe coupling, in a
waveguide.  This work illustrates the genius of Hans Bethe.

\vfil\eject

\centerline{\bf Exercises for Chapter 4 }

\vskip 12pt \noindent {\bf Problem 4-1:} \noindent By making use
of reciprocity, prove the validity of Equation (\ref{eq4-11a}) and
(\ref{eq4-11b}) of the text.

\vskip 8pt \noindent {\bf Problem 4-2:} Given the Rayleigh
quotient
$$
\frac{\v a^t \cdot \dyad A\cdot \v a}{\v a^t\cdot\dyad B\cdot \v
a},
$$
find the solution that will minimize it.  What equation does the
variational solution solve?

\vskip 8pt \noindent {\bf Problem 4-3:}

\begin{figure}[htb]
\begin{center}
\hfil\includegraphics[width=2.0truein]{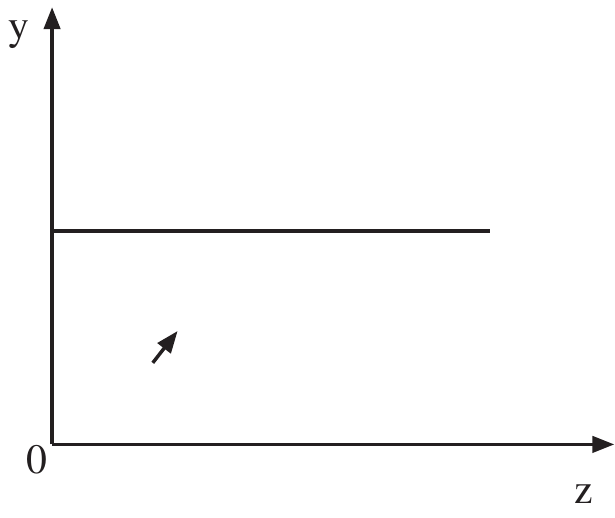}\hfil
\end{center}
\vspace{0.3 in}
\caption{Problem 4-3}
\end{figure}


The dyadic Green's function of a infinitely long waveguide is
given by
\begin{equation*}
\begin{split}
\dyad G (\v r,\v r') & = \sum_i \pi i \left[\frac{\v M_{ei}(\pm
k_{hiz},\v r)
\v M_{ei}(\mp k_{hiz},\v r')}{k_{hiz}k^2_{his}A_{hi}} \right.\\
&\left. +\frac{\v N_{ei}(\pm k_{eiz},\v r)\v N_{ei}(\mp k_{eiz},
\v r')}{k_{eiz}k^2_{eis}A_{ei}}\right]-\frac{\hat z \hat
z}{k^2_o}\delta (\v r-\v r'),
\begin{aligned}
\ &z>z' \\
\ &z<z'.
\end{aligned}
\end{split}
\end{equation*}
Using image theorem, find the dyadic Green's function of a
waveguide with a shorting plane at $z=0$.

\vskip 8pt \noindent {\bf Problem 4-4:}

\begin{figure}[htb]
\begin{center}
\hfil\includegraphics[width=1.0truein]{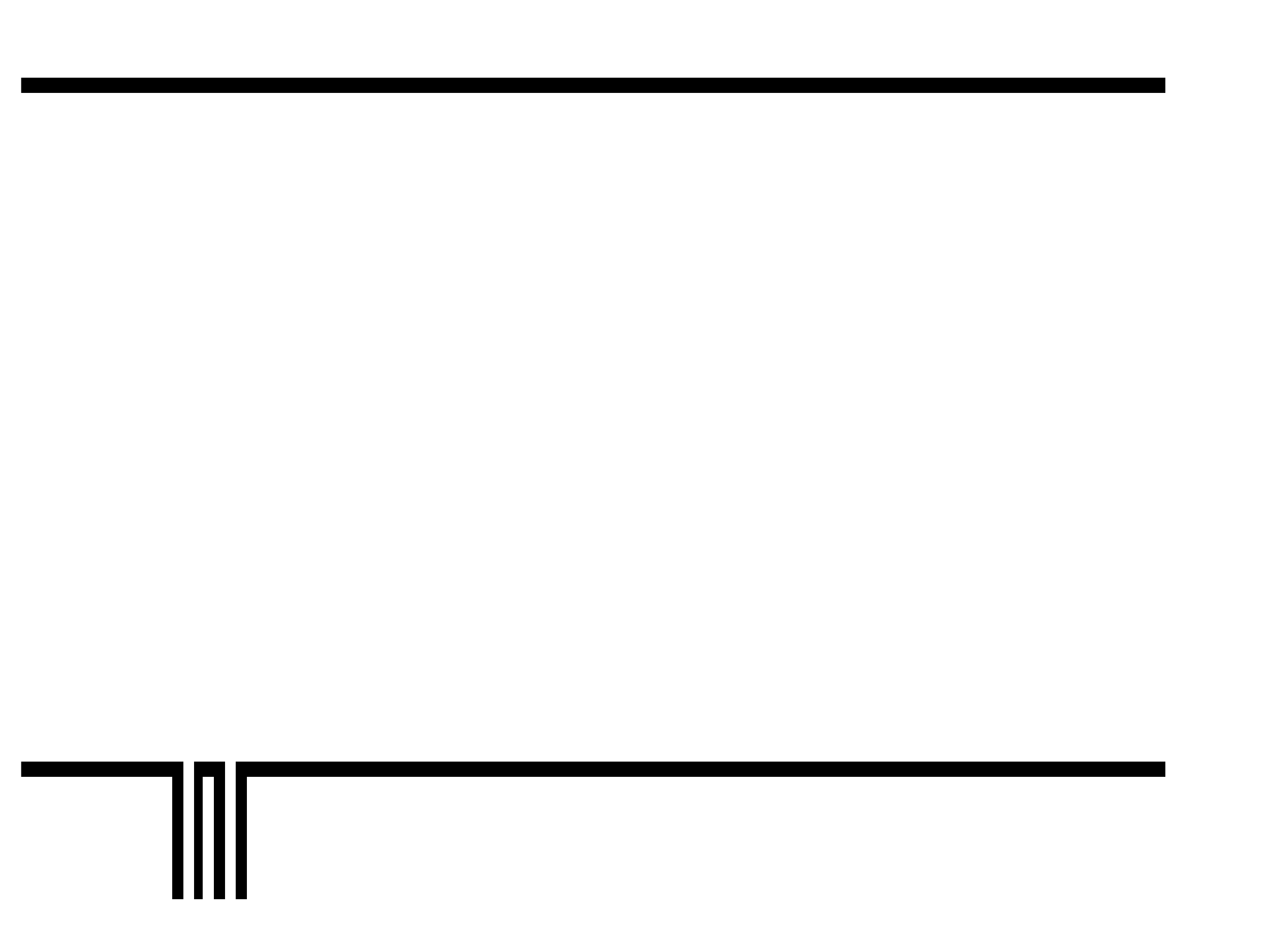}\hfil
\end{center}
\caption{Problem 4-4}
\vspace{0.3 in}
\end{figure}


\begin{itemize}
\item[(a)] For the geometry shown where a waveguide is fed by a
coaxial probe flushed with the waveguide wall, write down the
expression for the input admittance using the variational formula
derived in the text assuming that the field at the aperture is
that of the coaxial mode.
\item[(b)] Now, using mode-matching method, and assuming only a
TEM mode in the coaxial waveguide, derive an expression for the input
admittance using mode-matching method. Show that this result is the
same as that in part (a). (Hint: Do not write out the dyadic Green's
function explicitly. Leave it as a symbolic operator.)
\end{itemize}

\noindent {\bf Problem 4-5:}

\begin{figure}[htb]
\begin{center}
\hfil\includegraphics[width=2.5truein]{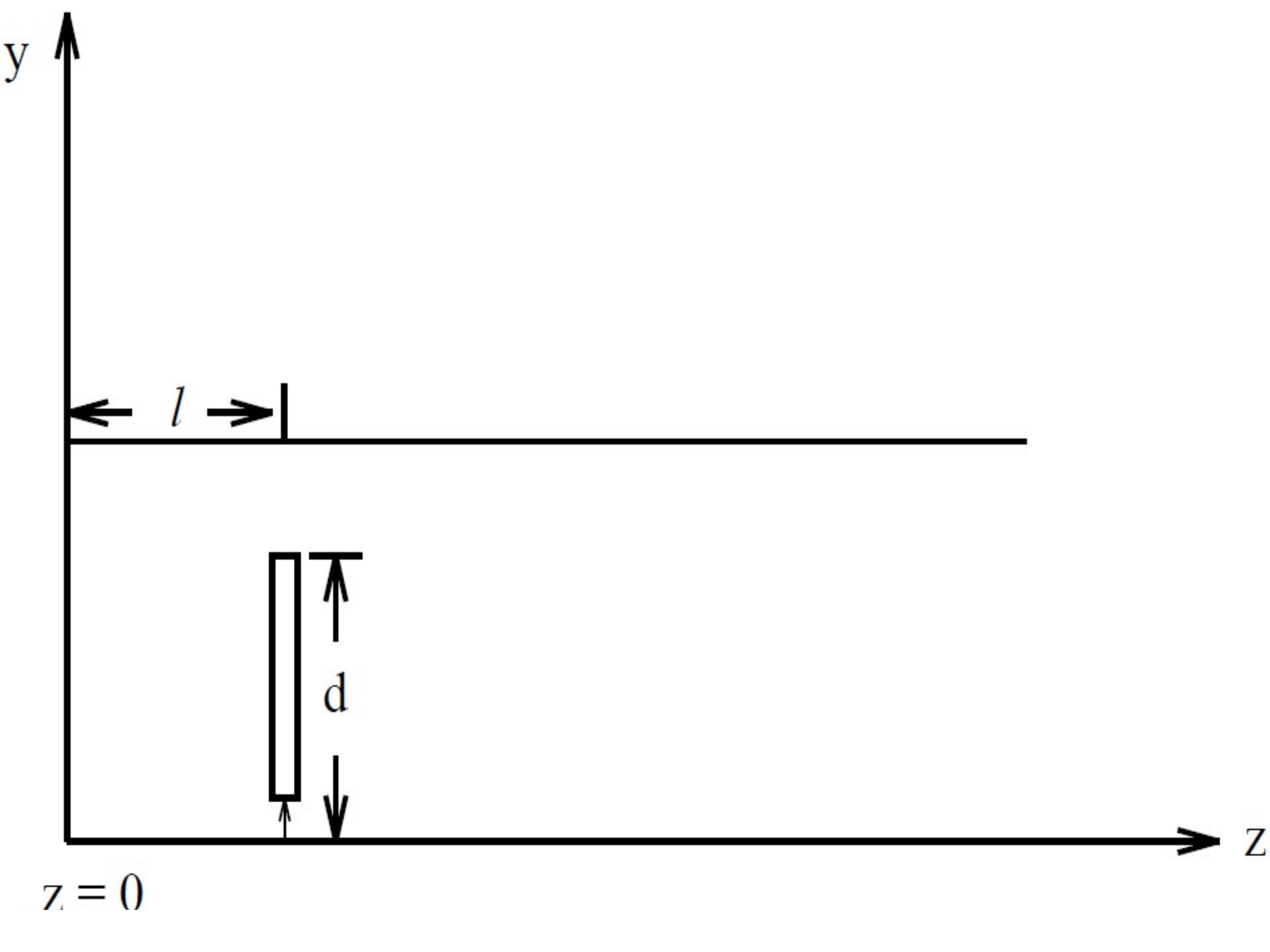}\hfil
\end{center}
\caption{Problem 4-5}
\end{figure}


\begin{itemize}
\item[(a)]
For a probe in a waveguide excited by a dipole as shown, show that
a variational expression for the input impedance at the base of
the probe is [see e.g., Harrington]
$$
Z_{in}=-\frac{i\omega\mu}{I^2}\langle\v J_p,\dyad G,\v J_p\rangle.
$$
\item[(b)]
If only the TE$_{10}$ mode is propagating in a rectangular
waveguide, we can approximate the dyadic Green's function only
with the term associated with the TE$_{10}$ mode. Assume that the
probe current is of the form $\v J_p=\hat y \sin [k(y-d)]$ so that
the probe current is zero at the tip of the probe. Use the
variational formula above to find an approximation to the probe
input impedance with the simplifying approximation on the dyadic
Green's function. (Note that this approximation is only good for
calculating the real part of the probe impedance. This is because
the real part of the probe impedance is related to the real power
radiated to infinity which is carried by the TE$_{10}$ mode. The
inductance of the probe will not be well approximated by this
method because the probe inductance is associated with the
singular field near the probe which can be well approximated only
if we include the higher order evanescent modes in the Green's
function.)
\end{itemize}

\vskip 8pt \noindent {\bf Problem 4-6:}

\begin{figure}[htb]
\begin{center}
\hfil\includegraphics[width=3.6truein]{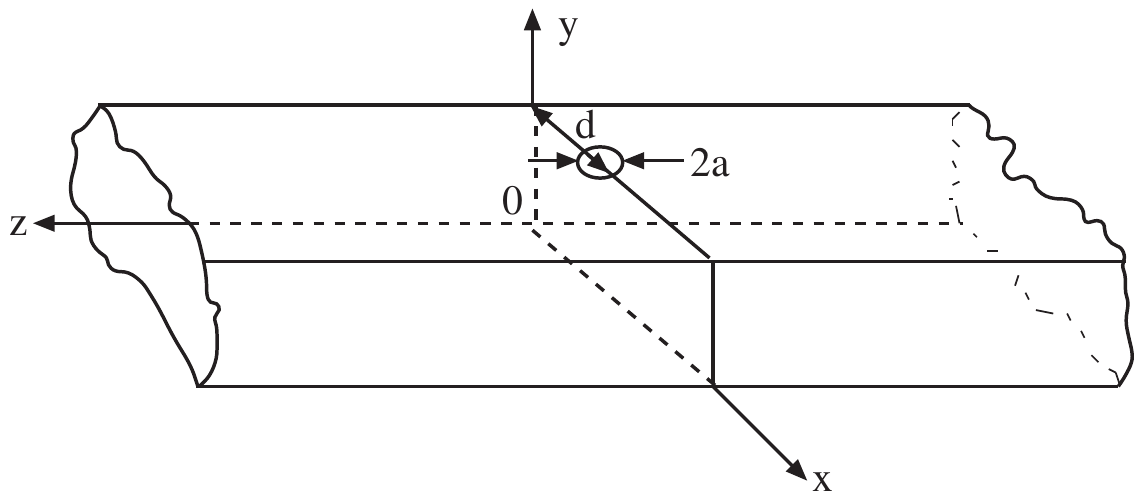}\hfil
\end{center}
\caption{Problem 5-2}
\end{figure}


For a TE$_{10}$ mode propagating in a rectangular waveguide with a
small circular aperture of radius $a$ on top of the waveguide, find
the polarization, amplitude and phase of the induced electric and
magnetic dipole moments at the aperture due to the field in the
waveguide.

\noindent {\bf Problem 4-7:} Derive the expressions for the
equations for Bethe coupling in \eqref{Betheeq:14}, and show that
\eqref{eq4-71} does solve the equation above it.


\vskip 8pt \noindent {\bf Problem 4-8:} Establish the equivalence
principles I, II, and III using Green's theorem or Huygens'
principle for vector electromagnetic fields.  In some of the cases,
it will be useful to assume that the dyadic Green's function
satisfies certain boundary conditions to reduce the size of the
surface integral that needs to be performed.

\vskip 8pt \noindent {\bf Problem 4-9:} Write down the electrostatic
solution of the potential due to an ellipsoid in the presence of a
constant electric field. Let one of its axes of the ellipsoid shrink
to zero. What should the permittivity of the ellipsoid be in order
to get the solution of Figures \ref{fg4313} and \ref{fg4314}

\bibliographystyle{plain}
\bibliography{emt,matrix,fmm,math,fastsum,parallel,cs,my}


\def\chaptitle{Discontinuities in Waveguides}
\index{Waveguide!discontinuities}

\chapter{\chaptitle}

\markboth{\smallbooktitle}{\chaptitle}



\def\v #1{{\bf #1}}
\def\vg #1{{\boldsymbol #1}}
\def\dyad#1{\overline {\bf #1}}
\def\dyadg#1{\overline {\vg #1}}
\def\beq{\begin{equation}}\def\eeq{\end{equation}}
\def\tinf{\text{\it inf\,}}\def\^{\hat}
\def\cal#1{\mathcal{#1}}
\def\ed{\end{document}}
\def\dalpha{\dyadg\alpha}
\def\dbeta{\dyadg\beta}
\def\Rg{\Re g}
\def\l{\label}
\def\sinc{\text{sinc}}



\noindent


Discontinuities in waveguides cannot be avoided.  When two
waveguides of different sizes are connected together, the junction
does not form a smooth transition introducing a discontinuity.
Such discontinuities will reflect the waveguide mode.  Moreover,
infinitely many modes are needed at the discontinuity in order to
match the boundary condition.  But most of the modes ``excited" by
the presence of discontinuities are evanescent modes.  They do not
convect energy away from a junction.  Hence, the higher-order
modes serve to store energy.  When most of the energy stored is in
the magnetic field, the junction discontinuity effect is
inductive, while if most of the stored energy is in the electric
field, the effect is capacitive.  Hence, simple equivalent models
for the junction can be either a capacitor or an inductor.
\index{Waveguide!junction capacitive effect}
\index{Waveguide!junction inductive effect}
However, the calculation of these inductive and capacitive effect
requires the use of some {\it tour de force} calculations, a
subject that we will discuss in this Chapter.

However, waveguide junctions appear in more complex forms when
applied to circulators and T junctions  \cite{HELSZAJN}.  The
analysis of these waveguide junctions is not amenable to analytic
methods, and hence, much numerical methods have been invoked in
analyzing them as is seen from the reference list in this chapter.

\section{{{ Transmission Line Equivalence of Waveguide}}}
\index{Waveguide!transmission line equivalence}

A wealth of engineering knowledge is built on circuit theory and
transmission line theory. Due to our familiarity with circuit
theory and transmission line theory, it is useful to relate the
propagation of modes in a waveguide to transmission line theory
 \cite{COLLINJ,COLLINK,MATTHEIetal}.

For a hollow waveguide of arbitrary shape, it is easy to show that
the ratio of the transverse components of the electric and
magnetic fields are
\begin{equation}
Z^{TE}=\frac {\^ z \times \v E_s }{\v H_s } =\frac {\omega \mu _0
}{k_z}, \quad Z^{TM}=\frac {\^ z \times \v E_s }{\v H_s } =\frac
{k _z }{\omega \epsilon _0}. \label{eq5-1}
\end{equation}
These are called the \index{Waveguide!wave impedance}wave impedances of a waveguide. Notice that
they are mode dependent. If they are to be likened to the
characteristic impedance of a transmission line, that it is
natural to define a voltage which is proportional to $\v E_s $ and
a current which is proportional to $\v H_s $ for a transmission
line equivalence. However, this proportionality constant should be
chosen so that the time average power given by $\Re e[VI^*/2]$, is
the same as the power flowing down the waveguide.

As an example, a mode propagating down a waveguide in the positive
$z$ direction may be expressed as \footnote {We shall use
$e^{j\omega t } $ time convention for agreement with the time
convention of circuit theory.}
\begin{subequations}
\begin{equation}
\v E_s =C_+\v e _s e ^{-jk_z z} \label{eq5-2a}
\end{equation}
\begin{equation}
\v H_s =C_+\v h _s e ^{-jk_z z} \label{eq5-2b}
\end{equation}
\end{subequations}
where $\^z \times \v e _s =Z_w \v h _s $ where $Z_w$ is wave
impedance of the particular mode under discussion. For a mode
propagating in the negative $z$ direction, the fields are given by
\begin{subequations}
\begin{equation}
\v E_s =C_-\v e _s e ^{jk_z z}, \label{eq5-3a}
\end{equation}
\begin{equation}
\v H_s =-C_-\v h _s e ^{jk_z z}. \label{eq5-3b}
\end{equation}
\end{subequations}
The expression for the field can be replaced by equivalent voltage
and current waves given by
\begin{subequations}
\begin{equation}
V=V_+ e^{-jk_z z}+ V_- e^{jk_z z}, \label{eq5-4a}
\end{equation}
\begin{equation}
I=I_+e^{-jk_z z} -I_- e^{jk_z z}, \label{eq5-4b}
\end{equation}
\end{subequations}
where $ V_+=K_1 C_+ $, $ V_-=K_1 C_- $, $I_+=K_2C_+$, and
$I_-=K_2C_-.$

To ensure that the equivalent circuit carries the same power, we
require that
\begin{equation}
\frac 1 2 V_+ I^*_+ =\frac {\vert C_+\vert ^2}{2} \int _s(\v e _s
\times \v h_s ^*)\cdot \^ z d S \label{eq5-5}
\end{equation}
or that
\begin{equation}
K_1 K_2^ *=\int _s(\v e _s \times \v h_s ^ *)\cdot \^ z d S.
\label{eq5-6}
\end{equation}
\index{Transmission line!characteristic impedance}
The characteristic impedance of this equivalent transmission line
is
\begin{equation}
Z_c=V_+/I_+=K_1/K_2. \label{eq5-7}
\end{equation}
Notice that there exists no unique way of choosing $K_1 $ and
$K_2$. Therefore, $\v e _s$ and $\v h_s $ in (\ref{eq5-6}) can be
normalized such that the right hand side of (\ref{eq5-6}) is 1. In
this case, we require $K_1K_2^*=1$. Also, the characteristic
impedance of the transmission line equivalence in (\ref{eq5-7})
can be made equal to 1 since all Smith charts are given for
normalized impedance values. Alternatively, one can choose
$K_1/K_2=Z_w$.

Armed with the transmission line model of a waveguide, much of the
tools that are found in transmission line theory like Smith chart,
and impedance matching techniques can be used. However, connecting
two waveguides of different wave impedances and hence sizes gives
rise to waveguide discontinuities.  These discontinuities scatter
a propagating mode into higher modes which are evanescent. Hence,
they give rise to localized stored energy in their vicinity.
Depending on if the stored energy is of electric or magnetic type,
these discontinuities are often modelled by shunt capacitances and
inductances which can be calculated. At other times, one may
deliberately introduce discontinuities in a waveguide call
diaphragms to give rise to a shunt capacitance or inductance for
the purpose of matching to the load or designing filters.

\section {{{ Waveguide Junction}}}
\index{Waveguide!junction}

Waveguide discontinuities have been studied by a number of workers
 \cite{MITTRA&LEED,KUHN}. Before the advent of digital computers,
analytic and variational methods were used to study these
discontinuities.  With the advent of high speed digital computers,
these discontinuities are routinely studied with numerical method,
requiring the solution of a large system of linear algebraic
equation (see reference list).

Here, we propose to characterize the scattering by a waveguide
discontinuity by reflection and transmission operators using mode
matching. Then, the generalization to an arbitrary number of
discontinuities (see Figure \ref{fg523}) becomes routine. The
waveguide discontinuity problem has also been studied previously,
but a more generalized formulation, valid for arbitrarily-shaped
waveguides, is presented here. The formulation is then related to
the integral equation formulation, and a new criterion for
convergence of the method is given and corroborated by numerical
simulation.

\begin{figure}
\begin{center}
\hfil\includegraphics[width=4.3truein]{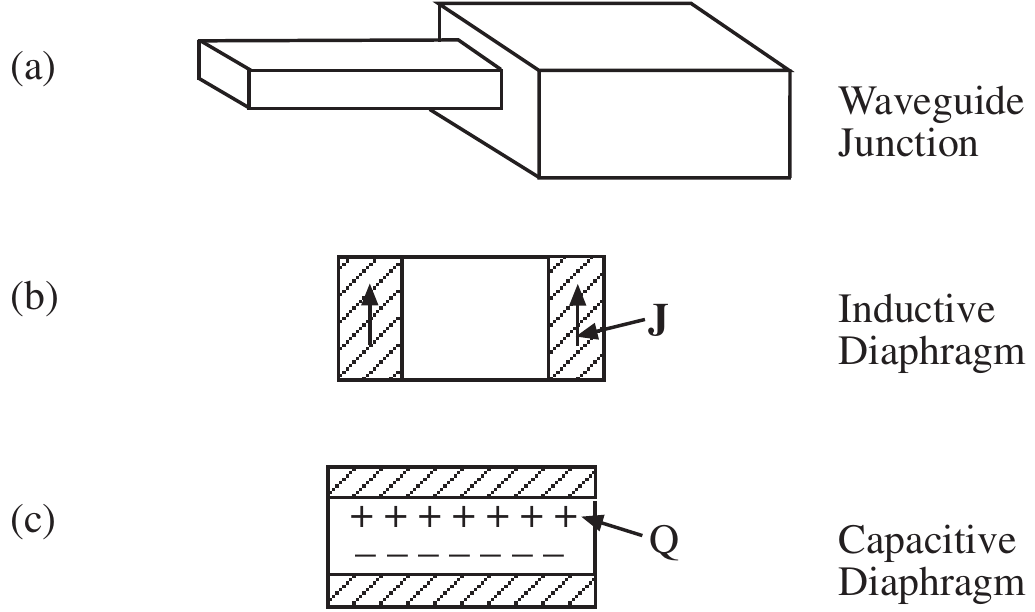}\hfil
\end{center}
\caption{Different kinds of waveguide junctions.  The inductive and
capacitive diaphragms are for TE$_{10}$ mode.}\label{fg521}
\end{figure}

Waveguide discontinuities arise in many occasions, for example, at
a waveguide junction (Figure \ref{fg521}(a)), or when \index{Waveguide!diaphragms}diaphragms
are added to change the phase and amplitude of a wave propagating
through a waveguide. A diaphragm can be of the inductive type as
shown in Figure 1(b), or of the capacitive type as shown in Figure
1(c).  An inductive diaphragm induces stored magnetic energy while
a capacitive diaphragm induces stored electric energy.   Moreover,
a capacitive diaphragm changes the phase of a mode differently
from an inductive diaphragm.  Hence, they can be used as tuning
elements for matching purposes in waveguides.

\subsection { Mode Matching--Eigenmode Expansion Method}
\index{Mode matching}
\index{Eigenmode expansion method}

\begin{figure}
\begin{center}
\hfil\includegraphics[width=4.3truein]{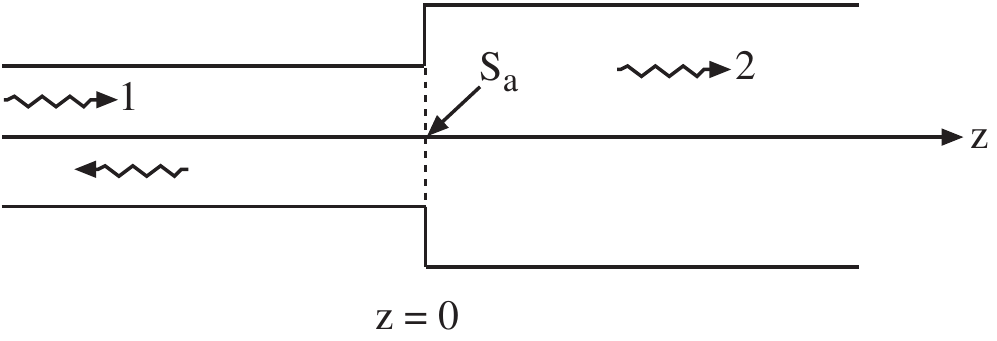}\hfil
\end{center}
\caption{Mode matching at a general waveguide junction.}\label{fg522}
\end{figure}

\begin{figure}
\begin{center}
\hfil\includegraphics[width=5.0truein]{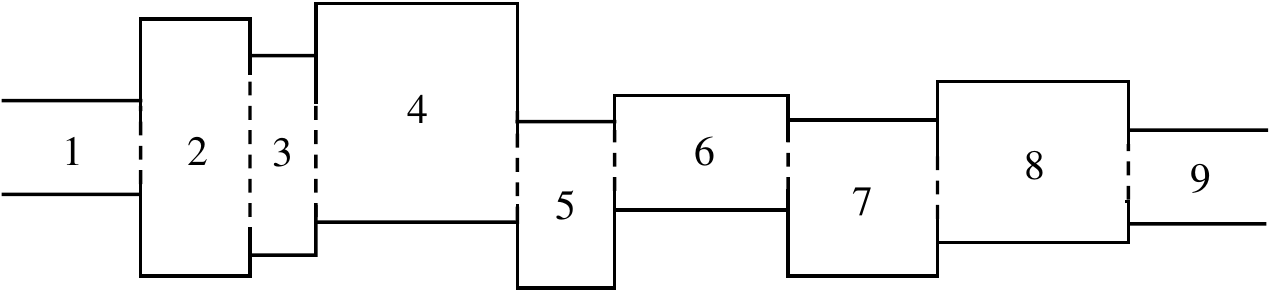}\hfil
\end{center}
\caption{A waveguide with many junction discontinuities.}\label{fg523}
\end{figure}

Waveguide discontinuity problems are generally solved by the method
of mode matching.\footnote {This section follows the development in
 \cite{CHEWetal}.} In this method, the waves in waveguide 1 and
waveguide 2 (see Figure \ref{fg522}) are expanded in terms of the
modes of the waveguide. The amplitudes of the modes are found by
matching the boundary conditions at the discontinuity.    The field
structure is quite complicated at the waveguide junction as the
field has to bend to adjust to the boundary conditions. A multitude
of modes is needed for matching the boundary condition. However,
most of these modes are evanescent giving rise to localized field at
the junction discontinuity.  The local field can store electric
field energy giving rise to a capacitive effect. If it stores
magnetic field energy, it gives rise to an inductive effect.

Another important physics that happens at a waveguide junction is
that if there is a pure waveguide mode impinging on the junction,
due to the need to match boundary conditions, all higher order modes
are excited at the junction giving rise to infinitely many reflected
modes and transmitted modes.  This is the physics of \index{Waveguide!mode conversion}mode
conversion.

In order to facilitate the ease for mode matching, we first develop
a succinct and compact notation that can encompass all modes in a
waveguide.  If we have a superposition of modes in an
arbitrarily-shaped, hollow waveguide, the $z$ components of the
fields can be written as
\begin{equation}
H_z=\sum _iH_i\psi _{hi}(\v r_s)e^{ik_{hiz}z}, \qquad (\text {TE
modes}) \label{eq5-2-1}
\end{equation}
\begin{equation}
E_z=\sum _iE_i\psi _{ei}(\v r_s)e^{ik_{eiz}z}. \qquad (\text {TM
modes}) \label{eq5-2-2}
\end{equation}
Here, $\psi _{hi}(\v r_s)$ and $\psi _{ei}(\v r_s)$ are solutions to
the equation $(\nabla _s^2+k_{is}^2)\psi _i(\v r_s)=0$ with Neumann
and Dirichlet boundary conditions, respectively, on the waveguide
wall. $\nabla_s^2=\frac {\partial ^2}{\partial x^2}+\frac {\partial
^2}{\partial y^2}$, and $k_{iz}^2$ is the eigenvalue corresponding
to the eigenmode $\psi _i(\v r_s)$. The subscripts $h$ and $e$ are
used to denote the Neumann problem (TE) and the Dirichlet problem
(TM), respectively.  For each mode, the transverse components of the
fields can be found via the equations derivable from Maxwell's
equations similar to those in Chapter 3:
\begin{equation}
\v e_{is}=\frac 1{k_{eis}^2}\frac {\partial }{\partial z}\nabla
_se_{iz}-\frac {i\omega\mu }{k_{his}^2}\^ z\times\nabla _sh_{iz},
\label{eq5-2-3}
\end{equation}
\begin{equation}
\v h_{is}=\frac 1{k_{his}^2}\frac {\partial }{\partial z}\nabla
_sh_{iz}+\frac {i\omega\epsilon }{k_{eis}^2}\^ z\times\nabla
_se_{iz}, \label{eq5-2-4}
\end{equation}
where $e_{iz}$ and $h_{iz}$ are the $z$ components of the electric
field and magnetic field respectively for each individual mode.
Consequently, $\v e_{is}$ and $\v h_{is}$ are the transverse
components of the electric field and magnetic field respectively
for each individual mode. Furthermore, $k_{eis}^2=k^2-k_{eiz}^2$
and $k_{his}^2=k^2-k_{hiz}^2$.

Using the above, we deduce that
\begin{equation}
\begin{split}
\v E_s&=\sum _i\v e_{is} \\
&=\sum _i\left[ \frac {E_i}{k_{eis}^2}ik_{eiz}\nabla_s\psi
_{ei}(\v r_s)e^{ik_{eiz}z} -\frac {i\omega\mu H_i}{k_{his}^2} \^z
\times \nabla_s\psi_{hi}(\v r_s)e^{ik_{hiz}z} \right],
\label{eq5-2-5}
\end{split}
\end{equation}

\begin{equation}
\begin{split}
\v H_s &=\sum _i\v h_{is}\\
& =\sum _i\left[ \frac {i\omega\epsilon E_i}{k_{eis}^2}\^
z\times\nabla _s\psi _{ei}(\v r_s)e^{ik_{eiz}z}+\frac
{H_i}{k_{his}^2}ik_{hiz} \nabla _s\psi _{hi}(\v
r_s)e^{ik_{hiz}z}\right]. \label{eq5-2-6}
\end{split}
\end{equation}
We see that $\^ z\times\v H_s$ is closely related to $\v E_s$,
i.e.,
\begin{equation}
\^ z\times\v H_s=\sum _i\left[ -\frac {i\omega\epsilon
E_i}{k_{eis}^2}\nabla _s\psi _{ei}(\v r_s)e^{ik_{eiz}z}+\frac
{H_i}{k_{his}^2} ik_{hiz}\^ z\times\nabla _s\psi _{hi}(\v
r_s)e^{ik_{hiz}z}\right]. \label{eq5-2-7}
\end{equation}

Using vector notation, we can write $\v E_s$ more compactly as
\begin{subequations}
\begin{equation}
\begin{split}
\v E_s & =\sum _i
\begin{bmatrix}
   \frac {\nabla _s\psi _{ei}(\v r_s)}{k_{eis}^2}\\
   \frac {\^ z\times\nabla _s\psi _{hi}(\v r_s)}{k_{his}^2}
\end{bmatrix}^t
\begin{bmatrix}
   e^{ik_{eiz}z} & 0\\
   0 & e^{ik_{hiz}z}
\end{bmatrix}
\begin{bmatrix}
   ik_{eiz}E_i\\
   -i\omega\mu H_i
\end{bmatrix} \\
& =\sum _i\underbrace{\dyad{\psi} _i^{\, t}(\v r_s)}_{2\times2}\cdot
\underbrace{e^{i\dyad k_{iz}z}}_{2\times2}\cdot\underbrace{\v
e_i}_{2\times1}, \label{eq5-2-8a}
\end{split}
\end{equation}
where
\begin{equation}
\dyad {\psi }_i(\v r_s)=
\begin{bmatrix}
   {\nabla _s\psi _{ei}(\v r_s)}/{k_{eis}^2}\\
   {\^ z\times\nabla _s\psi _{hi} (\v r_s)}/{k_{his}^2}
\end{bmatrix} , \qquad e^{i\dyad k_{iz}z}=
\begin{bmatrix}
   e^{ik_{eiz}z} & 0\\
   0 & e^{ik_{hiz}z}
\end{bmatrix},
\label{eq5-2-8b}
\end{equation}
and
\begin{equation}
\v e_i=
\begin{bmatrix}
   ik_{eiz}E_i\\
   -i\omega\mu H_i
\end{bmatrix}.
\label{eq5-2-8c}
\end{equation}
\end{subequations}
Similarly, we can write
\begin{equation}
\^ z\times\v H_s=-\sum _i\underbrace{\dyad {\psi }_i^{\,
t}}_{2\times2}(\v r_s)\cdot\underbrace{\dyad
g_i}_{2\times2}\cdot\underbrace{ e^{i\dyad
k_{iz}z}}_{2\times2}\cdot\underbrace{\v e_i}_{2\times1} ,
\label{eq5-2-9}
\end{equation}
where
\begin{equation*}
\dyad g_i =
\begin{bmatrix}
   {\omega\epsilon }/{k_{eiz}} & 0\\
   0 & {k_{{hiz}/{\omega\mu }}}
\end{bmatrix} .
\end{equation*}

Using the fact that one can write
\begin{subequations}
\begin{equation}
\sum _na_n\lambda _nb_n=\v a^t\cdot\dyad {\lambda }\cdot\v b,
\label{eq5-2-10a}
\end{equation}
where
\begin{equation}
\v a^t=[a_1, a_2, a_3, \hdots], \qquad\v b^t=[b_1, b_2, b_3,
\hdots ], \label{eq5-2-10b}
\end{equation}
\begin{equation}
\dyad {\lambda }=
\begin{bmatrix}
\lambda _1 & \qquad & \qquad & \qquad\\
\qquad & \lambda _2 & \qquad & \qquad\\
\qquad & \qquad & \lambda _3 & \qquad\\
\qquad & \qquad & \qquad & \ddots
\end{bmatrix} ,
\label{eq5-2-10c}
\end{equation}
\end{subequations}
we can rewrite (\ref{eq5-2-8a}) and (\ref{eq5-2-9}) compactly as
\begin{subequations}
\begin{equation}
\v E_s=\underbrace{\dyad {\Psi }^{\, t}(\v
r_s)}_{2\times\infty}\cdot\underbrace{ e^{i\dyad
Kz}}_{\infty\times\infty}\cdot\underbrace{\v e}_{\infty\times 1},
\label{eq5-2-11a}
\end{equation}
\begin{equation}
\^ z\times\v H_s=-\underbrace{\dyad {\Psi }^{\, t}(\v
r_s)}_{2\times\infty}\cdot\underbrace{\dyad
G}_{\infty\times\infty}\cdot \underbrace{e^{i\dyad
Kz}}_{\infty\times\infty}\cdot\underbrace{\v e}_{\infty\times 1},
\label{eq5-2-11b}
\end{equation}
\end{subequations}
where
\begin{subequations}
\begin{equation}
\dyad {\Psi }^{\, t}(\v r_s)=\left[\dyad {\psi }_1^{\, t}(\v r_s),
\dyad {\psi }_2^{\, t}(\v r_s), \dyad {\psi}_3^{\, t}(\v r_s),
\hdots\right], \label{eq5-2-12a}
\end{equation}
\begin{equation}
e^{i\dyad Kz}=
\begin{bmatrix}
e^{i\dyad k_{1z}z} & \qquad & \qquad & \qquad\\
\qquad & e^{i\dyad k_{2z}z} & \qquad & \qquad\\
\qquad & \qquad & e^{i\dyad k_{3z}z} & \qquad\\
\qquad & \qquad & \qquad & \ddots
\end{bmatrix},
\label{eq5-2-12b}
\end{equation}
\begin{equation}
\dyad G=
\begin{bmatrix}
\dyad g_1 & \qquad & \qquad & \qquad\\
\qquad & \dyad g_2 & \qquad &\qquad\\
\qquad & \qquad & \dyad g_3 & \qquad \\
\qquad & \qquad & \qquad & \ddots
\end{bmatrix},
\label{eq5-2-12c}
\end{equation}
\begin{equation}
\v e^t=[ \v e_1, \v e_2, \v e_3, \hdots ]. \label{eq5-2-12d}
\end{equation}
\end{subequations}

The above order of the elements of the matrices may be rearranged
for bookkeeping purposes. For instance, one may prefer to group
the TE modes and the TM modes separately rather than as a couplet
in the above. In this case, for example,
\begin{subequations}
\begin{equation}
\dyad G=
\begin{bmatrix}
g_1^{TM} &\qquad &\qquad &\qquad &\qquad &\qquad\\
\qquad & g_2^{TM} &\qquad &\qquad &\qquad &\qquad\\
\qquad &\qquad &\ddots &\qquad &\qquad &\qquad\\
\qquad &\qquad &\qquad & g_1^{TE} &\qquad &\qquad\\
\qquad &\qquad &\qquad &\qquad & g_2^{TE} &\qquad\\
\qquad &\qquad &\qquad &\qquad &\qquad &\ddots
\end{bmatrix}.
\label{eq5-2-13a}
\end{equation}
where,
\begin{equation}
g_i^{TM}=\omega\epsilon/k_{eiz},\qquad g_i^{TE}=k_{hiz}/\omega\mu.
\label{eq5-2-13b}
\end{equation}
\end{subequations}
Similar rearrangement need be done for (\ref{eq5-2-12a}),
(\ref{eq5-2-12b}) and (\ref{eq5-2-12d}) if this is in fact
preferred.

With the compact way to write the transverse components of the
fields as in (\ref{eq5-2-11a}), we can write down with physical
intuition the general solution for waveguides 1 and 2. In
waveguide 1, there will be an incident as well as a reflected
wave. Hence, we have
\begin{subequations}
\begin{equation}
\v E_{1s}=\dyad {\Psi }_1^{\, t}(\v r_s)\cdot \left( e^{i\dyad
K_1z}\cdot\v e+e^{-i\dyad K_1z}\cdot\v e_R\right),
\label{eq5-2-14a}
\end{equation}
\begin{equation}
-\^ z\times\v H_{1s}=\dyad {\Psi }_1^{\, t}(\v r_s)\cdot\dyad
G_1\cdot \left( e^{i\dyad K_1z}\cdot\v e-e^{-i\dyad K_1z}\cdot\v
e_R\right). \label{eq5-2-14b}
\end{equation}
In waveguide 2, we can write the fields as
\begin{equation}
\v E_{2s}=\dyad {\Psi }_2^{\, t}(\v r_s)\cdot e^{i\dyad
K_2z}\cdot\v e_T, \label{eq5-2-14c}
\end{equation}
\begin{equation}
-\^ z\times\v H_{2s}=\dyad {\Psi }_2^{\, t}(\v r_s)\cdot\dyad
G_2\cdot e^{i\dyad K_2z}\cdot\v e_T. \label{eq5-2-14d}
\end{equation}
\end{subequations}
We can define $\dyad R$ and $\dyad T$ which are reflection and
transmission operators such that
\begin{equation}
\v e_R=\dyad R\cdot\v e, \qquad\v e_T=\dyad T\cdot\v e.
\label{eq5-2-15}
\end{equation}
These operators entail the physics of mode conversion at a waveguide
junction.  They are infinite dimensional matrices, or operators.
 Then
\begin{subequations}
\begin{equation}
\v E_{1s}=\dyad {\Psi }_1^{\, t}(\v r_s)\cdot \left(e^{i\dyad
K_1z}+e^{-i\dyad K_1z}\cdot\dyad R\right)\cdot\v e,
\label{eq5-2-16a}
\end{equation}
\begin{equation}
-\^ z\times\v H_{1s}=\dyad {\Psi }_1^{\, t}(\v r_s)\cdot\dyad
G_1\cdot \left(e^{i\dyad K_1z}-e^{-i\dyad K_1z}\cdot\dyad R\right)
\cdot\v e, \label{eq5-2-16b}
\end{equation}
\end{subequations}
and
\begin{subequations}
\begin{equation}
\v E_{2s}=\dyad {\Psi }_2^{\, t}(\v r_s)\cdot e^{i\dyad
K_2z}\cdot\dyad T\cdot\v e, \label{eq5-2-17a}
\end{equation}
\begin{equation}
-\^ z\times\v H_{2s}=\dyad {\Psi }_2^{\, t}(\v r_s)\cdot\dyad
G_2\cdot e^{i\dyad K_2z}\cdot\dyad T\cdot\v e. \label{eq5-2-17b}
\end{equation}
\end{subequations}

From the continuity of the tangential electric field, we have
\begin{equation}
\dyad {\Psi }_1^{\, t}(\v r_s)\cdot \left(\dyad I+\dyad R\right)
\cdot\v e=\dyad {\Psi }_2^{\, t}(\v r_s)\cdot\dyad T\cdot\v e,
\quad \v r_s\in S_a, \label{eq5-2-18}
\end{equation}
Furthermore, we require that tangential electric fields equal zero
for $\v r_s\notin S_a$ at the plane of the discontinuity.

The continuity of the magnetic field implies
\begin{equation}
\dyad {\Psi }_1^{\, t}(\v r_s)\cdot\dyad G_1\cdot \left(\dyad
I-\dyad R\right) \cdot\v e=\dyad {\Psi}_2^{\, t}(\v r_s)\cdot\dyad
G_2\cdot\dyad T\cdot\v e, \quad \v r_s\in S_a. \label{eq5-2-19}
\end{equation}
The unknowns to be sought here are $\dyad R$ and $\dyad T$.

The aperture $S_a$ can be thought of as the cross-section of a
waveguide with modes $\dyad {\Psi }_a^{\, t}(\v r_s)$. We can then
equate (\ref{eq5-2-18}) to
\begin{equation}
\underbrace{\dyad {\Psi }_1^{\, t}(\v
r_s)}_{2\times\infty}\cdot\underbrace{\left( \dyad I+\dyad
R\right)}_{\infty\times\infty} \cdot\underbrace{\v e}_{\infty\times
1}=\underbrace{\dyad {\Psi }_2^{\, t}(\v r_s)}_{2
\times\infty}\cdot\underbrace{\dyad
T}_{\infty\times\infty}\cdot\underbrace{\v e}_{\infty\times
1}=\underbrace{\dyad {\Psi }_a^{\, t}(\v r_s)}_{\infty\times
N}\cdot\underbrace{\v a}_{N\times 1}, \quad \v r_s \in S_a,
\label{eq5-2-20}
\end{equation}
where $\dyad {\Psi }_a^{\, t}(\v r_s)=0$, $\v r_s \notin S_a$. In
this way, (\ref{eq5-2-18}) plus the auxiliary condition after
(\ref{eq5-2-18}) are satisfied. For practical purposes, we choose
$\v a$ to be a vector of length $N$ even though in theory, the
summation should be infinite. This is obviated by the integral
equation formulation in the next section. Hence, the dot product
for the last term in (\ref{eq5-2-20}) implies $N$-term summation
while the rest of the dot products imply infinite summation. From
this point onward, we shall denote inner products with infinite
summations with double-dot products, and leave the single-dot
product for inner product with a finite summation. Note that in
general, the length of the vector $\v e$ is determined by the
number of incident modes. We shall assume that $\v e$ is of
infinite length, but the following formulation is also valid for
$\v e$ of finite length.  It is to be noted that in
(\ref{eq5-2-20}), any complete set of basis functions instead of
waveguide modes that can be used to expand the aperture field for
$\v r_s \in S_a$ will also suffice.

Multiplying equation (\ref{eq5-2-20}) by $\dyad {\Psi }_1(\v r_s)$
and integrate over $\v r_s$, we get
\begin{subequations}
\begin{equation}
\underbrace{\dyad D_1}_{\infty\times\infty}:\underbrace{\left(\dyad
I+\dyad R\right)}_{\infty\times\infty}:\underbrace{\v
e}_{\infty\times 1}=\underbrace{\dyad L_{1a}}_{\infty\times
N}\cdot\underbrace{\v a}_{N\times 1}, \label{eq5-2-21a}
\end{equation}
where we define
\begin{equation}
\underbrace{\dyad D_i}_{\infty\times\infty}=\left< \underbrace{\dyad
{\Psi }_i}_{\infty\times 2}, \underbrace{\dyad {\Psi }_i^{\,
t}}_{2\times\infty}\right>, \qquad \underbrace{\dyad
L_{ia}}_{\infty\times N}=\left< \underbrace{\dyad {\Psi
}_i}_{\infty\times 2}, \underbrace{\dyad {\Psi}_a^{\, t}}_{2\times
N}\right> . \label{eq5-2-21b}
\end{equation}
\end{subequations}
$\dyad D_i$ is diagonal due to mode orthogonality while $\dyad
L_{ia}$ is in general non-square, infinite dimensional by $N$
dimensional matrix. $\dyad D_i$ can be made into an identity matrix
if the modes are made orthonormal. Similarly, multiplying
(\ref{eq5-2-20}) by $\dyad {\Psi }_2(\v r_s)$ and integrating over
$\v r_s$, we have
\begin{equation}
\dyad D_2:\dyad T:\v e=\dyad L_{2a}\cdot\v a. \label{eq5-2-22}
\end{equation}

From (\ref{eq5-2-21a}) and (\ref{eq5-2-22}), we deduce that
\begin{subequations}
\begin{equation}
\dyad R:\v e=\dyad D_1^{-1}:\dyad L_{1a}\cdot\v a-\v e,
\label{eq5-2-23a}
\end{equation}
\begin{equation}
\dyad T:\v e=\dyad D_2^{-1}:\dyad L_{2a}\cdot\v a.
\label{eq5-2-23b}
\end{equation}
\end{subequations}
The above allows us to express the unknowns $\dyad R$ and $\dyad T$
in terms of the new unknown $\v a$.   As $\dyad D_i$ is diagonal,
$\dyad D_i^{-1}$ is easily found even though it is infinite
dimensional.  In the above, \eqref{eq5-2-19} ensures the continuity
of tangential magnetic field.  To this end, upon substituting
(\ref{eq5-2-23a}) into (\ref{eq5-2-19}), and rearranging terms, we
have
\begin{equation}
\begin{split}
2\dyad {\Psi }_1^{\, t}(\v r_s):\dyad G_1:\v e= & \dyad {\Psi }_1^{\, t}(\v r_s):\dyad G_1:\dyad D_1^{-1}:\dyad L_{1a}\cdot\v a\\
& +\dyad {\Psi }_2^{\, t}(\v r_s):\dyad G_2:\dyad D_2^{-1}:\dyad
L_{2a}\cdot\v a, \qquad\v r_s{\in }{S_a}. \label{eq5-2-24}
\end{split}
\end{equation}
Remember that the above is derived from the continuity of the
magnetic field which needs to be imposed only on $S_a$. Hence, we
should weight the above equation with functions whose support is
over $S_a$. Weighting the above equation by $\dyad {\Psi }_a(\v
r_s)$ where $\dyad {\Psi}_a(\v r_s)$ is a vector of length $N$, we
have, after using the definition for $\dyad L_{ia}$ given in
(\ref{eq5-2-21b}),
\begin{equation}
2\dyad L_{1a}^{\, t}:\dyad G_1:\v e =\left(\dyad L_{1a}^{\,
t}:\dyad G_1:\dyad D_1^{-1}:\dyad L_{1a}+\dyad L_{2a}^{\, t}:\dyad
G_2:\dyad D_2^{-1}:\dyad L_{2a}\right)\cdot\v a. \label{eq5-2-25}
\end{equation}
Note that in general, $\dyad L_{ia}$ is nonsquare, but $\dyad
L_{ia}^{\, t}:\dyad G_i:\dyad D_i^{-1}:\dyad L_{ia}$ is an
$N\times N$ square matrix. Now, we can solve (\ref{eq5-2-25})
(which is unlike (\ref{eq5-2-24})) for $\v a$ giving
\begin{equation}
\v a=\left(\dyad L_{1a}^{\, t}:\dyad G_1:\dyad D_1^{-1}:\dyad
L_{1a}+\dyad L_{2a}^{\, t}:\dyad G_2:\dyad D_2^{-1}:\dyad
L_{2a}\right)^{-1}\cdot 2\left(\dyad L_{1a}^{\, t}:\dyad
G_1\right):\v e. \label{eq5-2-26}
\end{equation}
Substituting (\ref{eq5-2-26}) in (\ref{eq5-2-23a}), we can solve
for $\dyad R$ and $\dyad T$ giving
\begin{subequations}
\begin{equation}
\begin{split}
\dyad R=\dyad D_1^{-1}:\dyad L_{1a}\cdot &\left( \dyad L_{1a}^{\,
t}:\dyad G_1:\dyad D_1^{-1}:\dyad L_{1a}\right.\\ &\left.+\dyad
L_{2a}^{\, t}:\dyad G_2:\dyad D_2^{-1}:\dyad
L_{2a}\right)^{-1}\cdot 2\dyad L_{1a}^{\, t}:\dyad G_1-\dyad I,
\end{split}
\label{eq5-2-27a}
\end{equation}
\begin{equation}
\begin{split}
\dyad T=\dyad D_2^{-1}:\dyad L_{2a}\cdot &\left(\dyad L_{1a}^{\,
t}:\dyad G_1:\dyad D_1^{-1}:\dyad L_{1a}\right.\\ &\left.+\dyad
L_{2a}^{\, t}:\dyad G_2:\dyad D_2^{-1}:\dyad
L_{2a}\right)^{-1}\cdot 2\dyad L_{1a}^{\, t}: \dyad G_1.
\end{split}
\label{eq5-2-27b}
\end{equation}
\end{subequations}

The above expressions simplify further with the following
assumptions. If $S_a$ is as large as the smaller waveguide, then
$\dyadg {\psi }_a(\v r_s)=\dyadg {\psi }_1(\v r_s)$, $\dyad
L_{ia}=\dyad L_{i1}$ and $\dyad L_{11} = \dyad D_1$.
Consequently, the above simplifies to
\begin{subequations}
\begin{equation}
\begin{split}
\dyad R & =\left( \dyad D_1\cdot\dyad G_1+\dyad L_{21}^{\, t}:\dyad G_2:\dyad D_2^{-1}:\dyad L_{21}\right)^{-1}\cdot 2\dyad D_1\cdot\dyad G_1-\dyad I\\
& =\left(\! \dyad D_1\cdot\dyad G_1+\dyad L_{21}^{\, t}:\dyad G_2:
\dyad D_2^{-1}:\dyad L_{21}\!\right)^{-1}\\
& \cdot \left(\!\dyad D_1\cdot\dyad G_1-\dyad L_{21}^{\, t}: \dyad
G_2:\dyad D_2^{-1}:\dyad L_{21}\!\right), \phantom {\dyad L_2\,}
\label{eq5-2-28a}
\end{split}
\end{equation}
\begin{equation}
\dyad T=\dyad D_2^{-1}:\dyad L_{21}\cdot \left(\dyad D_1\cdot\dyad
G_1+\dyad L_{21}^{\, t}:\dyad G_2:\dyad D_2^{-1}:\dyad
L_{21}\right)^{-1}\cdot 2\dyad D_1\cdot\dyad G_1.
\label{eq5-2-28b}
\end{equation}
\end{subequations}
This is the case for the absence of the diaphragm. Note that
$\dyad R$ is now $N\times N$ and $\dyad T$ is $\infty\times N$.
This follows from that the basis functions used at the aperture is
orthogonal to the higher order modes in waveguide 1, and hence,
would not excite them.

If waveguides 1 and 2 are identical, but a diaphragm is present,
(\ref{eq5-2-27a}) simplifies to
\begin{subequations}
\begin{equation}
\dyad R=\dyad D_1^{-1}:\dyad L_{1a}\cdot \left(\dyad L_{1a}^{\,
t}:\dyad G_1:\dyad D_1^{-1}:\dyad L_{1a}\right)^{-1}\cdot\dyad
L_{1a}^{\, t}:\dyad G_1-\dyad I, \label{eq5-2-29a}
\end{equation}
\begin{equation}
\dyad T=\dyad D_1^{-1}:\dyad L_{1a}\cdot \left(\dyad L_{1a}^{\,
t}:\dyad G_1:\dyad D_1^{-1}:\dyad L_{1a}\right)^{-1}\cdot\dyad
L_{1a}^{\, t}:\dyad G_1. \label{eq5-2-29b}
\end{equation}
\end{subequations}
In this case, $\dyad I+\dyad R = \dyad T$.  Here, $\dyad R$ and
$\dyad T$ are again $\infty\times\infty$ matrices.

In general, $\dyad R$ and $\dyad T$ are nondiagonal matrices.
Physically, this means that a mode coming into the discontinuity
will excite many modes, giving rise to numerous reflected and
transmitted modes.  As mentioned before, this entails the physics of
mode conversion. However, if in both waveguides, only the dominant
modes are propagating, e.g., the TE$_{10}$ mode in a rectangular
waveguide, the excited higher order modes will be evanescent, and
hence, localized around the discontinuity. These evanescent fields
store electromagnetic energy, making the discontinuity either
capacitive or inductive depending on if the energy is stored in the
electric field or magnetic field.

Even though the double-dot products in the above should involve
infinite summations, a practical implementation necessitates the
truncation of the infinite summations. If the infinite summation
is truncated at $P$ terms, as shall be shown in the next section,
$P \gg N$ to ensure the accuracy of the double-dot products (see
also Problem 5-2).

If only a few elements of $\dyad R$ and $\dyad T$ are needed, to
save on computer memory, the matrix $\dyad L_{ia}$, which is
$P\times N$, need not be calculated and stored.  Explicit
expressions can usually be derived for the elements of $\dyad
L_{ia}^t:\dyad G_i:\dyad D_i^{-1}:\dyad L_{ia}$, which is $N\times
N$, so that only these matrices need be stored.  Then, the
corresponding elements of $\dyad L_{ia}$ are calculated to form
the desired elements of $\dyad R$ and $\dyad T$.

Note that the above formulation is general enough that it
encompasses the case of even if the waveguides are of different
shapes.

\begin{figure}
\begin{center}
\hfil\includegraphics[width=2.40truein]{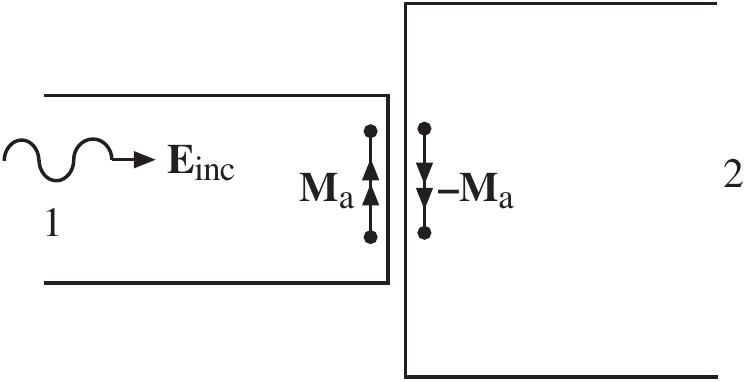}\hfil
\end{center}
\caption{Equivalent problem for a waveguide
discontinuity.}\label{fg524}
\end{figure}

\subsection {{{ Equivalence Principle and Integral Equation
Formulation}}}
\index{Equivalence principle}
\index{Integral equation formulation}

The above formulation shows a derivation of the reflection and
transmission operators of a waveguide discontinuity using a mode
matching procedure. Their calculation involves $\infty \times N$
dimensional matrices. From the above equations, we see that the
reflection and transmission operators could be calculated even if
we do not truncate the infinite dimension of the matrices. The
matrix that needs to be inverted is always $N\times N$. However,
to save computation resources, it is necessary to \index{Truncation of infinite summation}truncate the
infinite summations at some finite values. Many authors have
suggested using $P_1$ modes only in waveguide 1 and $P_2$ modes
only in waveguide 2 in the mode matching procedure
 \cite{MITTRA&LEED}. Consequently, the resultant matrices in the
above formulation, which are $\infty\times N$, become $P_1\times
N$ or $P_2\times N$. A rule is often given for the ratio between
$P_1$, $P_2$ and $N$ for a convergent result. This issue is
generally regarded as relative convergence. We shall give a new
rule for the choice of $P_1$ and $P_2$ for this problem.

Alternatively, the problem on the choice of $P_1$, $P_2$, and $N$
can be further enlightened by looking at an equivalent formulation
of the above problem. This equivalent formulation can be achieved
with the use of equivalence principle first, and later, deriving
an integral equation for the problem.

In the equivalent problem, the waveguide discontinuity problem is
divided into two waveguides with shorting planes at the location of
the discontinuity (see Figure \ref{fg524}). Moreover, magnetic
current $\v M_a=\v E\times \^ n$ is impressed at the location of
the aperture. By the equivalence principle, if $\v E\times\^ n$ is
known at the aperture, the reflected and the transmitted fields are
the same in the original problem and the equivalent problem.  This
equivalence problem can be easily derived using vector Green's
theorem and vector Huygens principle.

If the magnetic dyadic Green's functions are known for the shorted
waveguide 1 and waveguide 2, then the magnetic fields in waveguide
1 and waveguide 2 are
\begin{subequations}
\begin{equation}
\v H_1(\v r)=\v H_{inc}(\v r)+i\omega \epsilon _1\int\limits
_{S_a}dS'\dyad G_{1m}(\v r, \v r')\cdot \v M_a(\v r'),
\label{eq5-2-30a}
\end{equation}
\begin{equation}
\v H_2(\v r)=-i\omega\epsilon _2\int\limits _{S_a}dS'\dyad
G_{2m}(\v r, \v r')\cdot\v M_a(\v r'). \label{eq5-2-30b}
\end{equation}
\end{subequations}
Such a dyadic Green's function can be derived using the method
outlined in the previous chapter, which expands the Green's
function in terms of an infinite number of waveguide modes. Since
tangential magnetic field is continuous across the aperture, we
have
\begin{equation}
\^ n\times \v H_{inc}(\v r)=-i\omega\^n\times \int\limits
_{S_a}dS'[\epsilon _2\dyad G_{2m}(\v r, \v r')+\epsilon _1\dyad
G_{1m}(\v r, \v r')]\cdot \v M_a(\v r'), \quad\v r\in S_a.
\label{eq5-2-31}
\end{equation}
The above integral equation can be solved with the \index{Method of moments}method of moments
or Galerkin's method, and converted into an $N\times N$ matrix
equation, by using $N$ expansion functions and $N$ testing
functions. In this procedure, we let
\begin{align}
\v{M}_a(\v{r}^{\prime}) = \sum_{n=1}^{N} a_n \v{f}_n(\v{r}^{\prime})
\end{align}
By testing the above with $-\hat{n} \times \v{f}_m(\v{r}) $, we
obtain
\begin{align}
\langle \v{f}_m , \v{H}_{inc} \rangle = -i \omega
\sum_{n=1}^{\infty} \langle \v{f}_m, \epsilon_2
\overline{\v{G}}_{2m} + \epsilon_1 \overline{\v{G}}_{1m}, \v{f}_n
\rangle a_n,\quad m=1,\ldots,N
\end{align}
The above is a matrix equation of the form
\begin{align}
\v b=\dyad A\cdot \v a
\end{align}
where,
\begin{align}
[\v{b}]_m = \langle \v{f}_m,  \v{H}_{inc} \rangle ,  \hspace {2mm}
[\overline {\v{A}}]_{mn} = -i \omega  \langle \v{f}_m, \epsilon_2
\overline{\v{G}}_{2m} + \epsilon_1 \overline{\v{G}}_{1m}, \v{f}_n
\rangle ,\hspace {2mm} [\v{a}]_n = a_n
\end{align}
and
\begin{align}
\langle \v{f}_m,\overline{\v{G}},\v{f}_n \rangle &= \int _ {S_a} dS
\hspace {1mm} \v{f}_m(\v{r})\cdot \int_{S_a} dS^{\prime} \hspace
{1mm} \overline{\v{G}}(\v{r},\v{r}^{\prime}) \cdot
\v{f}_n(\v{r}^{\prime})\\
\langle \v{f}_m,\v H_{inc} \rangle &= \int _ {S_a} dS \hspace {1mm}
\v{f}_m(\v{r})\cdot  \v H_{inc}(\v r)
\end{align}

As has been seen, the method of moments yields the optimal solution
when $N$ basis functions are used to solve (\ref{eq5-2-31}).
Equation (\ref{eq5-2-31}) is an exact integral equation, and it is
clear that the numbers of waveguide modes needed to approximate
$\dyad G_{1m}$ and $\dyad G_{2m}$ should be infinite to render them
as accurate as possible. This is equivalent to choosing $P_1$ and
$P_2$ to be infinite in the previous formulation using mode
matching.

\subsection{Relative Convergence}
\index{Relative convergence}

If $S_a$ is made as large as waveguide 1, as is the case discussed
in Equation (\ref{eq5-2-28a}) in the previous section, and the
electric field of $N$ modes from waveguide 1 is used to
approximate $\v M_a$ via $\v E_1\times\hat n$, then the integral
\begin{equation}
i\omega\epsilon _1\int\limits _{S_a}dS'\dyad G_{1m}(\v r, \v
r')\cdot\v M_a(\v r') \label{eq5-2-32}
\end{equation}
would only yield $N$ waveguide modes for the reflected field in
waveguide 1 due to mode orthogonality.  Hence, $P_1=N$ in this
case. What then is the number of modes required in the expansion
of dyadic Green's function \index{Green's function!dyadic!expansion} in the following integral?
\begin{equation}
i\omega\epsilon _2\int\limits _{S_a}dS'\dyad G _{2m}(\v r, \v
r')\cdot \v M _a(\v r'). \label{eq5-2-33}
\end{equation}
This integral yields the transmitted field in waveguide 2. Since
mode orthogonality does not apply here, in theory, an infinite
number of modes is needed in $\dyad G_{2m}(\v r, \v r')$. But in
practice, we need only a finite number of modes $P_2$. How large
should $P_2$ be to accurately represent the field generated by
(\ref{eq5-2-33})?

Further insight can be obtained by studying a simpler problem---the
\index{Waveguide!parallel plate} parallel plate waveguide problem. The problem is scalar in this case
and a typical integral for (\ref{eq5-2-33}) looks like
\begin{subequations}
\begin{equation}
I=\int _0^{d_2}dx'\left [\sum\limits _{m=0}^{\infty }\frac
{e^{ik_{mz}|z|}} {k_{mz}}\cos\left (\frac {m\pi x}{d_2}\right
)\cos\left (\frac {m\pi x'}{d_2}\right )\right ]M_a(x'),
\label{eq5-2-34a}
\end{equation}
where we have expanded the scalar Green's function as
\begin{equation}
G_{2m}(\v r, \v r')\propto\sum\limits _{m=0}^{\infty }\frac
{e^{ik_{mz}|z-z'|}}{k_{mz}}\cos\left (\frac {m\pi x}{d_2}\right
)\cos\left (\frac {m\pi x'}{d_2}\right ). \label{eq5-2-34b}
\end{equation}
\end{subequations}
Here, $k_{mz}=\sqrt {k_2^2-\left (\frac {m\pi }{d_2}\right )^2}$
and $d_2$ is the separation of the parallel plate waveguide 2.
$M_a(x')$, the aperture field, following the above method, is
expanded in terms of the modes of waveguide 1, i.e.,
\begin{equation}
M_a(x')=\sum\limits _{n=0}^{N}a_n\cos\left (\frac{n\pi
x'}{d_1}\right), \label{eq5-2-35}
\end{equation}
where $d_1$ is the separation of parallel plate waveguide 1. Hence,
(\ref{eq5-2-34a}), after using (\ref{eq5-2-35}), letting $z=z'$, and
exchanging the order of summation and integration, becomes,
\begin{equation}
I=\sum\limits _{m=0}^{\infty }\frac {1}{k_{mx}}\cos\left (\frac
{m\pi x}{d_2}\right )\sum\limits _{n=0}^{N}a_n\int
_0^{d_2}dx'\cos\left (\frac {m\pi x'}{d_2}\right )\cos\left (\frac
{n\pi x'}{d_1}\right ). \label{eq5-2-36}
\end{equation}
The above integral could be integrated readily to yield
\begin{subequations}
\begin{equation}
I= \sum\limits _{n=0}^{N}a_n \sum\limits _{m=0}^{\infty }\frac
{1}{k_{mz}}\cos\left (\frac {m\pi x}{d_2}\right )f_{nm},
\label{eq5-2-37a}
\end{equation}
where
\begin{equation}
f_{nm}=\frac {1}{2}\left\{\frac{\sin\left [\left (\frac {m\pi
}{d_2} +\frac {n\pi }{d_1}\right )d_2\right ]}{\frac {m\pi
}{d_2}+\frac {n\pi }{d_1}}+ \frac {\sin\left [\left (\frac {m\pi
}{d_2}-\frac {n\pi }{d_1}\right )d_2 \right ]}{\frac {m\pi }
{d_2}-\frac {n\pi }{d_1}}\right\}. \label{eq5-2-37b}
\end{equation}
\end{subequations}
In the above, the integral $I$ is the result of the action of the
Green's function on the expansion or basis functions at the
aperture.  The integral in \eqref{eq5-2-36} is the \index{Cosine transform} cosine transform
of the basis functions in waveguide 1 using the cosine functions of
waveguide 2.  In other words, the basis function $\cos\left (\frac
{n\pi x'}{d_1}\right )$ is expanded in terms of the functions
$\cos\left (\frac {m\pi x'}{d_2}\right )$.  The coefficient of the
expansion peaks when $\frac {m\pi }{d_2}=\frac {n\pi }{d_1}$, or
when the wavelengths of the two Fourier harmonics match or about
equal.

\begin{figure}
\begin{center}
\hfil\includegraphics[width=3.5truein]{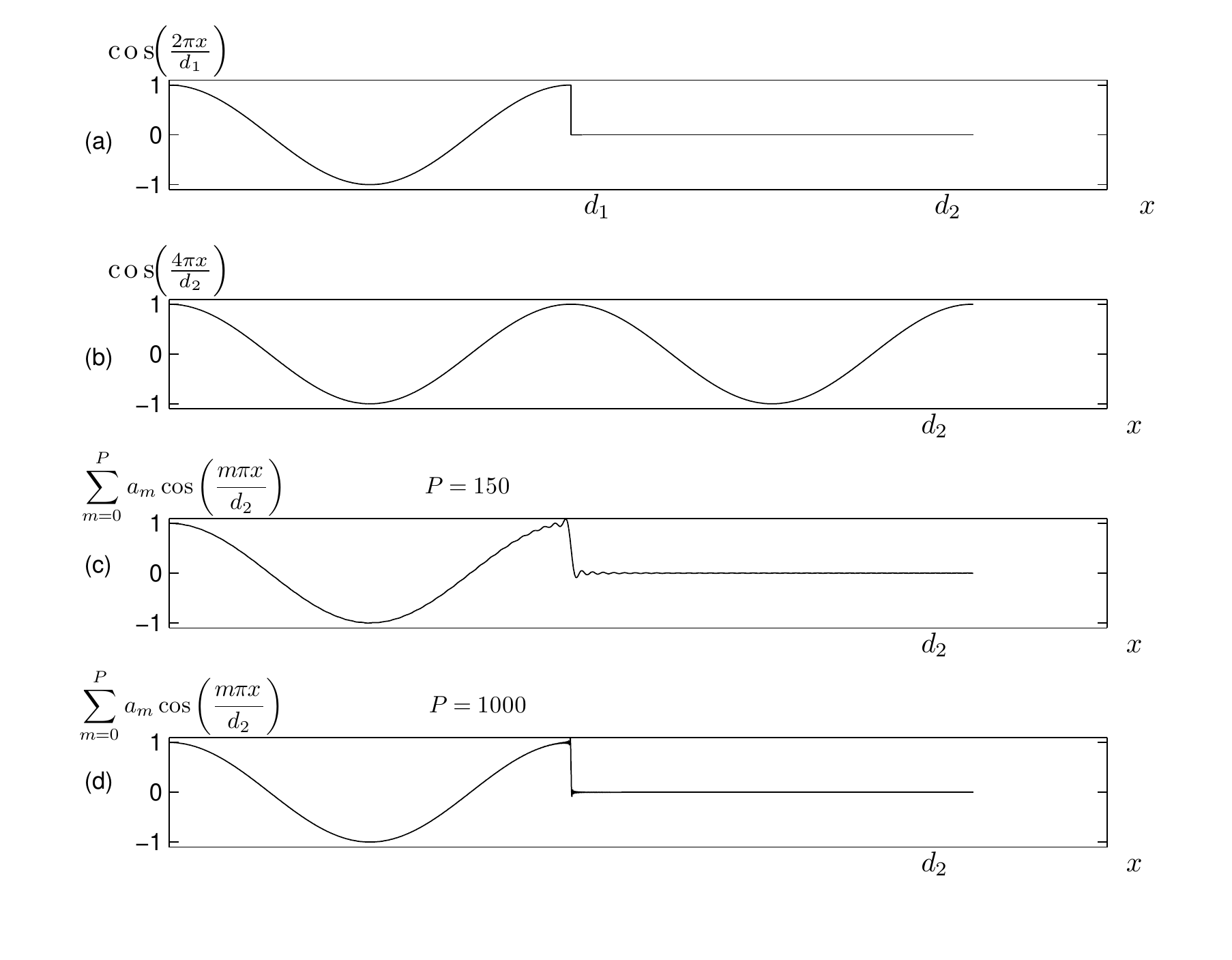}\hfil
\end{center}
\caption{Convergence of the cosine Fourier series expansion of the
mode of one waveguide with dimesion $d_1$ in terms of the modes of
the second waveguide with dimension
$d_2$.}\label{figure_Relative_Convergence}
\end{figure}

\begin{figure}
\begin{center}
\hfil\includegraphics[width=3.5truein]{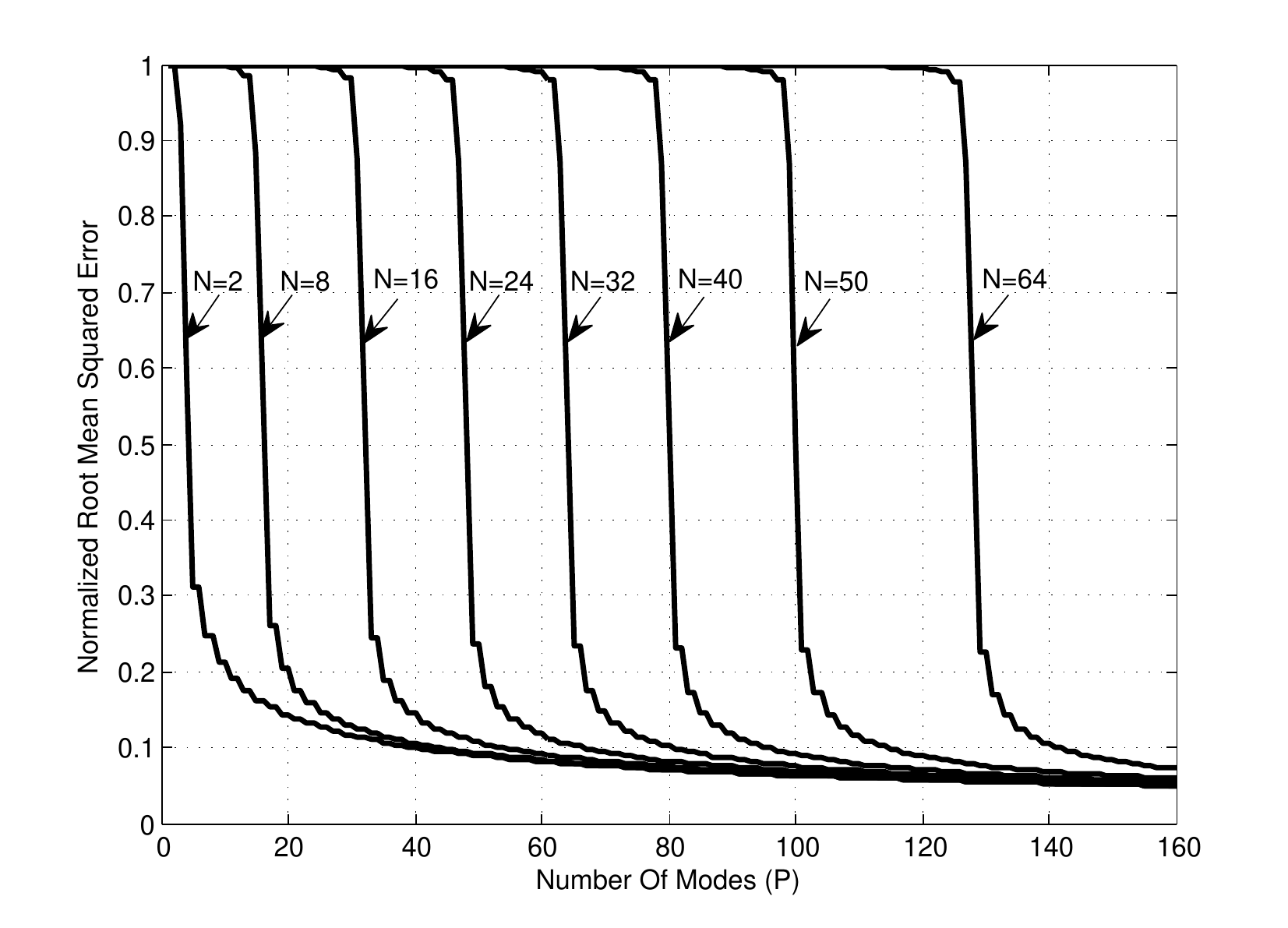}\hfil
\end{center}
\caption{Error convergence of the cosine Fourier series expansion
$\cos\left (\frac {N\pi x}{d_1}\right )=\sum\limits _{m=0}^{P }
a_m\cos\left (\frac {m\pi x}{d_2}\right )$ as a function of $P$ with
$d_2=2d_1$. Notice that as $P>N d_2/d_1$, the error drops rapidly.}
\end{figure}

When the integral (\ref{eq5-2-33}) is used in (\ref{eq5-2-31}) to
match boundary condition, $\v r\in S_a$, or $z=0$ in
(\ref{eq5-2-34a}) and (\ref{eq5-2-37a}) when used in
(\ref{eq5-2-31}). If we further test (or weight) (\ref{eq5-2-37a})
by $\cos\left(\frac {n'\pi x}{d_1}\right )$ and integrate over $x$,
as is required when the method of Galerkin is used to solve
(\ref{eq5-2-31}), then (\ref{eq5-2-37a}) can be transformed to (with
$z=0$)
\begin{equation}
\begin{split}
A &=\sum\limits _{n=0}^{N}a_n\sum\limits _{m=0}^{\infty
}f_{nm}\frac
{1}{k_{mz}}f_{n'm}\\
&\cong\sum\limits_{n=0}^{N}a_n\sum\limits _{m=0}^{P_2}f_{nm}\frac
{1}{k_{mz}}f_{n'm},\quad n'=0, 1, \cdots, N.
\end{split}
\label{eq5-2-38}
\end{equation}
The above is the matrix representation of the Green's function using
the eigenmodes from waveguide 1.  Note that in the above, $f_{nm}$
peaks at $m=\frac {d_2}{d_1}n$, and $f_{n'm}$ peaks at $m=\frac
{d_2}{d_1}n'$. Hence, in order to evaluate $A$ accurately in
(\ref{eq5-2-38}), the summation over $m$ in (\ref{eq5-2-38}) must at
least be large enough so that the contribution from the peaks of
$f_{nm}f_{n'm}$ are included. Since the largest $n$ and $n'$ are
$N$, for the approximate summation in (\ref{eq5-2-38}) to be
accurate when the infinite summation in (\ref{eq5-2-38}) is replaced
by a summation over $m$ from 0 to $P_2$, we require that
\begin{equation}
\frac {P_2}{d_2} \gg \frac {N}{d_1}. \label{eq5-2-39}
\end{equation}

From the above analysis, it is clear that

\vskip 8pt
\begin{itemize}
\item[(a)] if the dimension of the diaphragm region is $d_a$ such that $d_a<d_1$, $d_a<d_2$, and
\item[(b)] if $N$ basis functions are used to approximate the aperture field, and moreover,
\item[(c)] if truncated numbers of modes, $P_1$ and $P_2$ are used to represent the fields in waveguides 1 and 2, respectively,
\end{itemize}

\noindent then in order for this truncation to be accurate,
\begin{equation}
\frac {P_1}{d_1} \gg \frac{N}{d_a},\qquad\frac {P_2}{d_2}\gg \frac
{N}{d_a}. \label{eq5-2-40}
\end{equation}
This point has also been noted by Orta et al\footnote{R. Orta, R.
Tascone, and R. Zich, ``Multiple dielectric loaded perforated
screens as frequency selective surfaces," {\it IEE Proc}., vol.
135, pt. H, no. 2, pp. 75-82, 1988.} in the study of frequency
selective surfaces. Note that $P_1$ and $P_2$ need not be related
by a specific ratio, but if the same degree of accuracies is
required of the fields in both waveguides, then
\begin{equation}
\frac {P_1}{d_1}\simeq\frac{P_2}{d_2}. \label{eq5-2-41}
\end{equation}

The inequalities in (\ref{eq5-2-40}) can also be interpreted as
that $P_1$ and $P_2$ should be chosen large enough so that the
spectral components used in $\dyad G_{1m}$ and $\dyad G_{2m}$ are
large enough to capture (or accurately represent) the dominant
spectral components in $\v M_a$.

The above analysis can be extended to the rectangular waveguide
case. Since the $x$ and $y$ coordinates of a rectangular
waveguides are separable, the $x$-spectral components can be
considered separately from the $y$-spectral components.  Similar
inequalities as in (40) will hold separately for the $x$ and $y$
spectral components of the waveguide.  For the case of a circular
waveguide which is axially symmetric, a similar analysis will
yield the inequality as in (\ref{eq5-2-40}) but $d_1$, $d_2$, and
$d_a$ represent the diameter of the waveguides and aperture. This
is because Bessel functions behave like sinusoidal functions when
their arguments are large.

It is to be noted that other basis functions can be used to expand
the aperture field in (\ref{eq5-2-35}) other than the waveguide
modes. In this case, $f_{nm}$ will not be of the form given by
(\ref{eq5-2-37a}), but usually, a more complex form ensues. In
this case $P_1$ and $P_2$ should be chosen large enough to capture
the dominant spectral components in $\v M_a$. The rule for
choosing $P_1$ and $P_2$ will not be as simple as that given by
(\ref{eq5-2-40}). But for a fixed $N$, one should increase $P_1$
and $P_2$ so that they are large enough until the calculated
amplitudes of the reflected and transmitted modes stabilize.

\section {{{ Numerical Examples}}}

\begin{figure}
\begin{center}
\subfigure[]{\includegraphics[width=2.75truein]{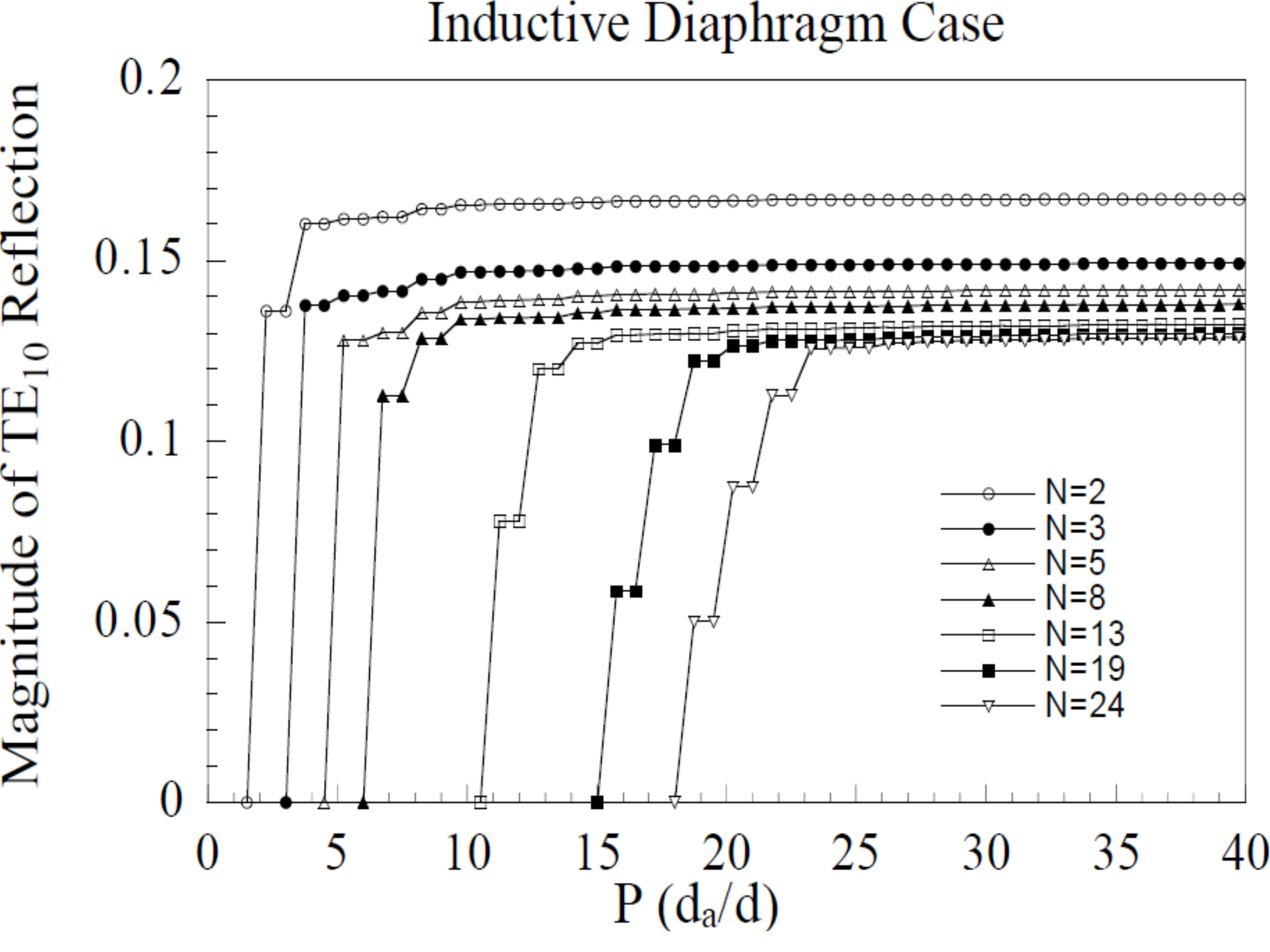}}
\subfigure[]{\includegraphics[width=2.75truein]{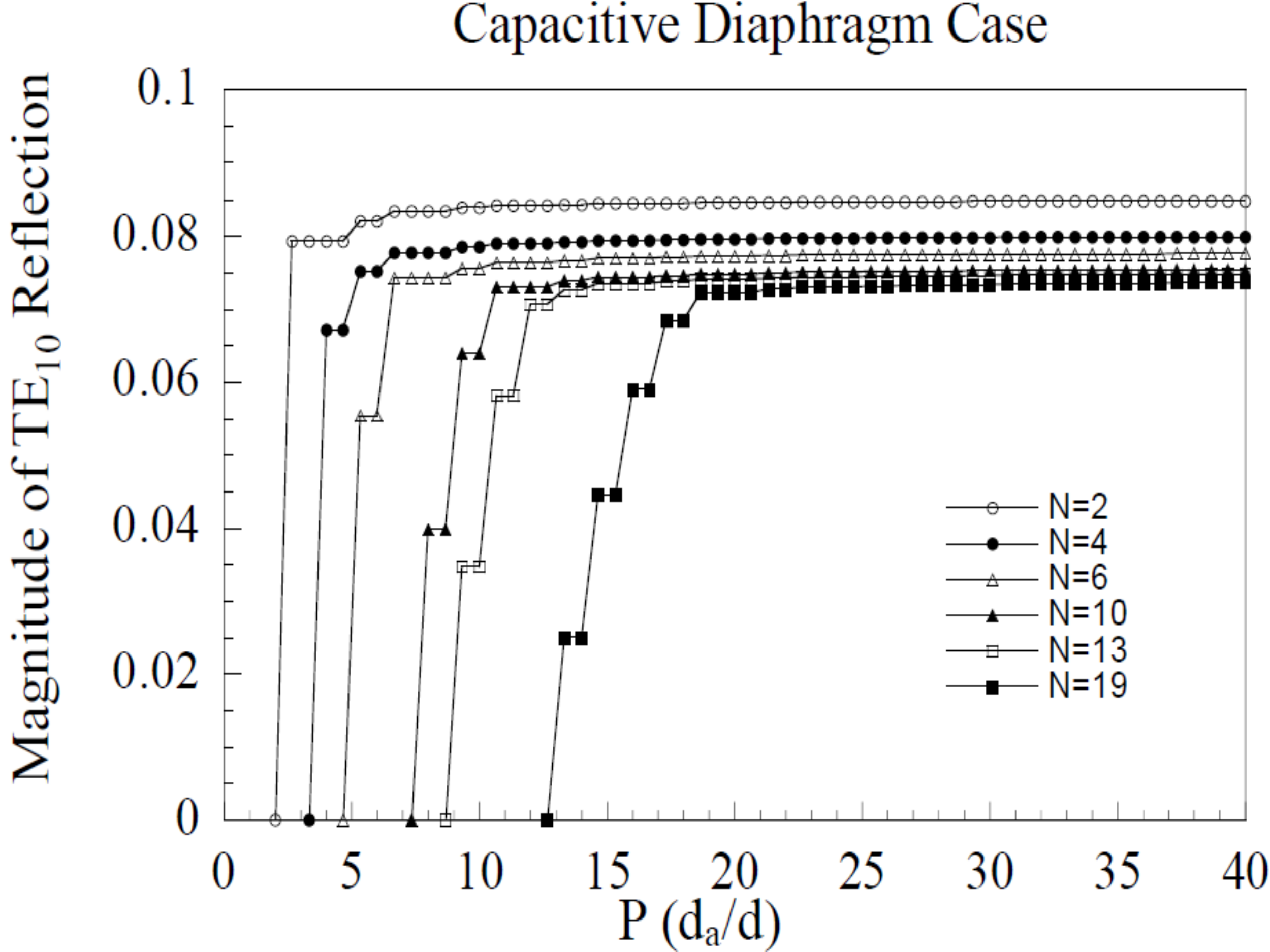}}
\end{center}
\caption{Plots showing the convergence of the reflection coefficient
amplitude of TE$_{10}$ mode of a rectangular waveguide when
(\ref{eq5-2-40}) is satisfied. The dimension of the waveguide is
$0.8 \times 0.6$ wavelength. (a) The inductive diaphragm case. The
dimension of the diaphragm is $0.6 \times 0.6$ wavelength. (b) The
capacitive diaphragm case. The dimension of the diaphragm is $0.8
\times 0.4$ wavelength. In these cases, $P_1 = P_2 = P$ (re-plotted
by F. Ling).}\label{fg545}
\end{figure}

A program has been developed using the formulation of Section 5.2.
This program yields the reflection and transmission operators due
to a junction discontinuity with a diaphragm of two rectangular
waveguides.
This program is used to study the convergence of the \index{Mode!TE$_{10}$!reflection coefficient}TE$_{10}$ mode
reflection coefficient when the two waveguides are identical but
separated by either an \index{Inductive diaphragm}inductive diaphragm or a capacitive
diaphragm. The dimension of the waveguide is $0.8\times0.6$
wavelength. The apertures are symmetrically located with dimension
$0.6\times0.6$ wavelength for the inductive diaphragm, and
$0.8\times 0.4$ wavelength for the capacitive diaphragm.

A TE$_{10}$ mode is assumed incident onto this junction
discontinuity. For the inductive diaphragm, due to symmetry, only
modes with $x$-variations will be excited. Figure \ref{fg545}(a)
shows the amplitude of the reflection coefficient as a function of
the number of waveguide modes when the number of basis function
$N$ in the aperture is kept fixed.  It is seen that for a fixed
$N$, the solution converges when (\ref{eq5-2-40}) is satisfied. It
is also seen that $N$ has to be sufficiently large before the
reflection coefficient amplitude is accurate.  Note that in this
case, $P\left (\frac {d_a}{d}\right )$ need not be very much
larger than $N$ for the calculation to stabilize. $P\left (\frac
{d_a}{d}\right )$ has to be only a little larger than $N$ to
capture the dominant components of $\v M_a$.

Figure \ref{fg545}(b) shows a similar convergence plot for the
case of a \index{Capacitive diaphragm}capacitive diaphragm. Due to symmetry, only higher order
modes with $y$-variation will be excited. Qualitatively, the
result is similar to that of Figure \ref{fg545}(a).


Note from Figures \ref{fg545}(a) and \ref{fg545}(b) that if the
conventional rule of choosing $\frac{P_1}{d_1} = \frac{N}{d_a} =
\frac{P_2}{d_2}$ is used, fairly good result could be obtained
when $N$ is large enough.  This is entirely an artifact of
choosing waveguide modes to expand the field at the aperture. The
conventional wisdom of choosing \index{Optimal number of modes}$P_1$ and $P_2$ for this case will
capture most of the \index{Spectral components}spectral components (the aperture basis
functions are expanded in terms of a Fourier series using the
waveguide modes, and we call the amplitude of a term of this
series a spectral component) this series of the basis functions at
the aperture. If, for instance, basis functions with edge conditions are used,
high spectral components will ensue and the
conventional wisdom will not apply.

As a final note, it can be shown that the mode matching method
always \index{Energy conservation} conserves energy irrespective of how many basis functions
are used at the aperture. Thus the conservation of energy can be
used to check the correctness of the implementation of the method,
but not the accuracy. As seen in the previous simulation, an
accurate solution can only be obtained by slowly increasing $N$
while ensuring that (\ref{eq5-2-40}) is satisfied.  An accurate
solution is obtained when the reflection and transmission
coefficients cease to change with increasing $N$.

\section {{{ Solution to the Multiple Waveguide Junction Problem}}}
\index{Waveguide!multiple junction}
\index{Waveguide!multiple junction!solution}

Once the solution to the one waveguide junction problem is known, a
two waveguide junction problem can be solved. Then, the
$N$-waveguide junction problem can be sought recursively in the
manner of  \cite{WFIMG}. The $N$-waveguide junction solution can be
used to model complex waveguide junctions. It can also be used to
model tapered waveguide junctions.

\subsection { A Two-Waveguide-Junction Problem}

\begin{figure}
\begin{center}
\hfil\includegraphics[width=3.5truein]{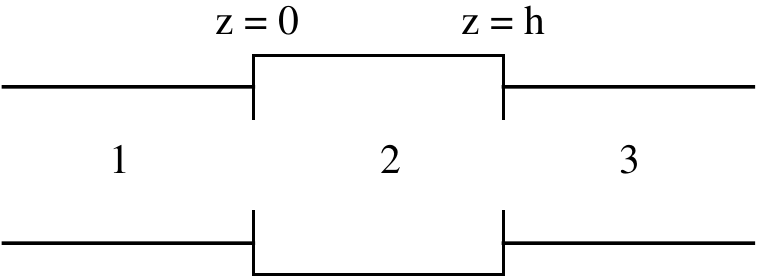}\hfil
\end{center}
\caption{A two-waveguide junction problem.}\label{fg551}
\end{figure}


Given a two-waveguide-junction geometry as shown, we can write
down the solution in waveguide 1 for the electric field as
consisting of
\begin{equation}
\v E_{1s} =\dyad \Psi ^t_1 (\v r_s)\cdot\left(e^{i\dyad K_1
z}+e^{-i\dyad K_1 z}\cdot\widetilde{\dyad R}_{12}\right)\cdot \v e
\label{eq5-5-1}
\end{equation}
where $\v e $ is the amplitude of the incident modes, and
$\widetilde{\dyad R}$ is the reflection operator including
multiple reflections between junctions 1 and 2. The first term
consists of modes travelling to the right while the second term
consists of modes travelling to the left.

In waveguide 2, we can write the solution for the electric field
as
\begin{equation}
\v E_{2s}=\dyad \Psi ^t_2(\v r _s)\cdot\left(e^{i\dyad K_2 z}\cdot
\v A_2 + e^{-i\dyad K_2 z}\cdot\v B_2\right). \label{eq5-5-2}
\end{equation}
In waveguide 3, the corresponding electric field is
\begin{equation}
\v E_{3s}=\dyad \Psi ^t_3(\v r _s)\cdot e^{i\dyad K_3 z}\cdot \v
A_3. \label{eq5-5-3}
\end{equation}
A relationship can be easily established between the amplitudes
$\v A_2 $ and $\v B _2$ in waveguide 2, because the left-going
wave in waveguide 2 is a consequence of the right-going wave in
the same waveguide. Therefore, at $z=h$, we must have
\begin{equation}
e^{-i\dyad K_2 h}\cdot \v B_2=\dyad R_{23}\cdot e^{i\dyad K_2
h}\cdot \v A_2, \label{eq5-5-4}
\end{equation}
where $\dyad R_{23} $ is the reflection operator for the single
waveguide junction. Consequently, we can rewrite (\ref{eq5-5-2})
as
\begin{equation}
\v E_{2s}=\dyad \Psi ^t_2(\v r _s)\cdot\left (e^{i\dyad K_2
z}+e^{-i\dyad K_2 (z-h)}\cdot \dyad R_{23}\cdot e^{i\dyad K_2
h}\right)\cdot \v A_2. \label{eq5-5-5}
\end{equation}
The right-going wave in waveguide 2 is a result of the
transmission of right-going wave in waveguide 1 plus a reflection
of the left-going wave in waveguide 2. Therefore, at $z=0$, we
must have
\begin{equation}
\v A_2=\dyad T_{12}\cdot \v e +\dyad R_{21}\cdot \v B_2.
\label{eq5-5-6}
\end{equation}
Using (\ref{eq5-5-4}) in (\ref{eq5-5-6}), we can solve for $\v
A_2$ to yield
\begin{equation}
\v A_2=\left(\dyad I-\dyad R_{21}\cdot e^{i\dyad K_2 h}\cdot \dyad
R_{23} \cdot e^{i\dyad K_2 h}\right)^{-1}\cdot \dyad T_{12}\cdot
\v e \label{eq5-5-7}
\end{equation}
Using (\ref{eq5-5-7}) in (\ref{eq5-5-4}), we can express $\v B_2$
in terms of $\v e $. Therefore, $\v A_2 $ and $\v B_2$ are found
in terms of $\v e$.

The right-going wave in waveguide 3 is a consequence of a
transmission of the right-going wave in waveguide 2. Therefore, at
$z=h$, we must have
\begin{equation}
e^{i\dyad K_3 h } \cdot \v A_3 =\dyad T_{23} \cdot e ^{i \dyad K_2
h}\cdot \v A_2 . \label{eq5-5-8}
\end{equation}
Consequently, $\v A_3 $ can be found in terms of $\v e $ via the
use of (\ref{eq5-5-7}) and (\ref{eq5-5-8}).

The left-going wave in waveguide 1 is a consequence of the
reflection of the right-going wave in waveguide 1 plus a
transmission of the left-going wave in waveguide 2. Therefore, we
can write
\begin{equation}
\widetilde{\dyad R}_{12} \cdot \v e =\dyad R_{12}\cdot \v e +\dyad
T_{21}\cdot \v B_2. \label{eq5-5-9}
\end{equation}
Using $\v B_2$ from (\ref{eq5-5-4}) and (\ref{eq5-5-7}), we have
\begin{equation}
\begin{split}
\widetilde{\dyad R}_{12} & =\dyad R_{12}+\dyad T_{21}\cdot e ^{i\dyad K_2 h}\cdot\dyad R_{23}\cdot e^{i\dyad K_2 h } \\
&\cdot\left(\dyad I -\dyad R_{21}\cdot e^{i\dyad K_2 h}\cdot\dyad
R_{23}\cdot e^{i\dyad K_2 h }\right)^{-1}\cdot \dyad T_{12}.
\label{eq5-5-10}
\end{split}
\end{equation}
Therefore, with $\v e $, the amplitude of the incident modes,
known, one can find the solution of the two waveguide junction
problem in all the three waveguides. We call $\widetilde{\dyad
R}_{12}$ the generalized reflection operator at the (1,2) junction
that accounts for multiple reflections.

\subsection { An N-Waveguide-Junction Problem}
\index{Waveguide!$N$-junction}

\begin{figure}
\begin{center}
\hfil\includegraphics[width=3.20truein]{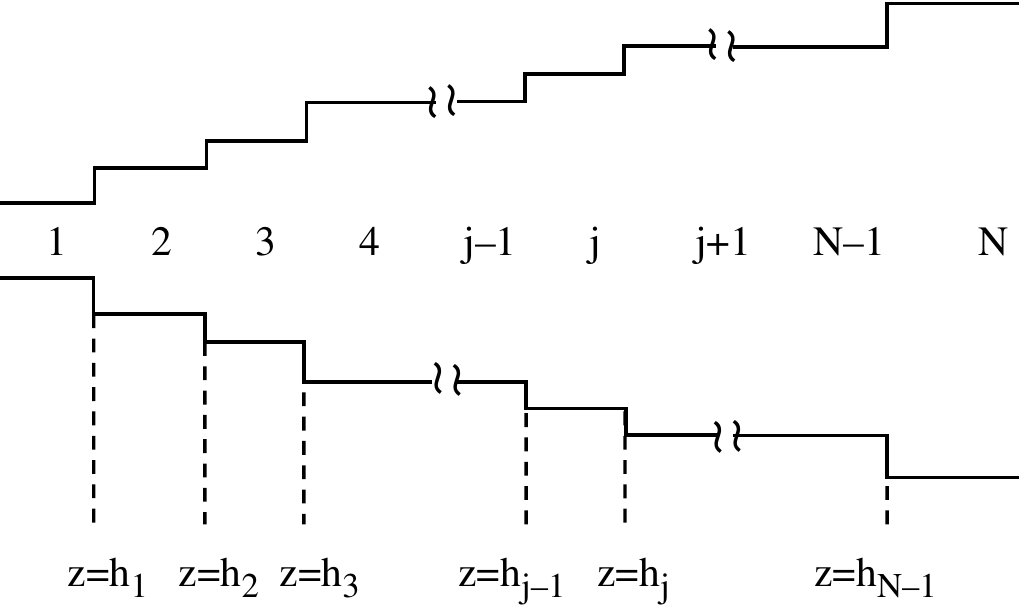}\hfil
\end{center}
\caption{A multiple-junction waveguide.}\label{fg552}
\end{figure}


Given an $N$-waveguide-junction problem where the discontinuities
of the waveguide are at $z=h_j$, $j=1,\cdots , N$, one can write
down the solution in region $j$ as
\begin{equation}
\v E_{js} ={\dyad \Psi }^t_j (\v r_s)\cdot \left(e^{i\dyad K_j z
}+e^{-i\dyad K_j(z-h_j)}\cdot \widetilde{\dyad R}_{j,j+1}\cdot
e^{i\dyad K_jh_j}\right)\cdot \v A_j  . \label{eq5-5-11}
\end{equation}
The amplitude of the left-going wave in the above is written in
such a way so that at $z=h_j$, the left-going wave amplitude is
just related to the right-going wave amplitude by
$\widetilde{\dyad R}_{j,j+1}$, the generalized reflection operator
at the $(j,j+1)$ junction. Motivated by (\ref{eq5-5-10}), we can
write down the expression for the \index{Generalized reflection operator}generalized reflection operator
as
\begin{equation}
\begin{split}
\widetilde{\dyad R}_{j,j+1} & =\dyad R _{j,j+1} +\dyad T_{j+1, j} \cdot  e^{i\dyad K_j (h_j-h_{j-1})}\cdot \widetilde {\dyad R}_{j+1, j+2} \cdot e^{i\dyad K_j(h_j-h_{j-1}) } \\
 & \cdot \left(\dyad I - \dyad R_{j,j+1} \cdot e ^{i\dyad K_j (h_j-h_{j-1} )}
\cdot \widetilde {\dyad R}_{j+1, j+2} \cdot e^{i\dyad
K_j(h_j-h_{j-1})} \right) ^{-1} \cdot \dyad T_{j,j+1}.
\label{eq5-5-12}
\end{split}
\end{equation}
Notice that we have replaced $\dyad R_{23} $ in (\ref{eq5-5-10})
with a generalized reflection operator $\widetilde{\dyad
R}_{j+1,j+2} $ at the $(j+1, j+2)$ junction because for a
multiple-junction waveguide, multiple reflections to the right of
the $(j,j+1)$ junction has to be accounted for.

Using (\ref{eq5-5-12}), starting at the right-most junction, the
generalized reflection operators in all the waveguides can be
found. Next, Using an equation similar to (\ref{eq5-5-7}), the
amplitude $\v A_j$ in waveguide $j$ can be related to $\v A_{j-1}$
in waveguide $j-1$ as
\begin{equation}
\begin{split}
e^{i\dyad K_j h_{j-1}}\cdot \v A_j & =\left(\dyad I - \dyad
R_{j,j-1} \cdot e^{i\dyad K_j (h_j-h_{j-1})}\cdot \widetilde
{\dyad R}_{j,j+1} \cdot
e^{i\dyad K_j(h_j-h_{j-1})} \right) ^{-1} \\
 & \cdot \dyad T_{j-1, j}\cdot e^{i\dyad K_{j-1} h_{j-1} }\cdot \v A_{j-1}.
\label{eq5-5-13}
\end{split}
\end{equation}
Starting with the left-most waveguide, the field solution for all
the waveguides can be obtained.


\subsection{Filter Design--A Resonance Tunneling Problem}
\index{Filter design}
\index{Resonance tunneling problem}

\begin{figure}
\vskip 0.1 in
\begin{center}
\hfil\includegraphics[width=2.20truein]{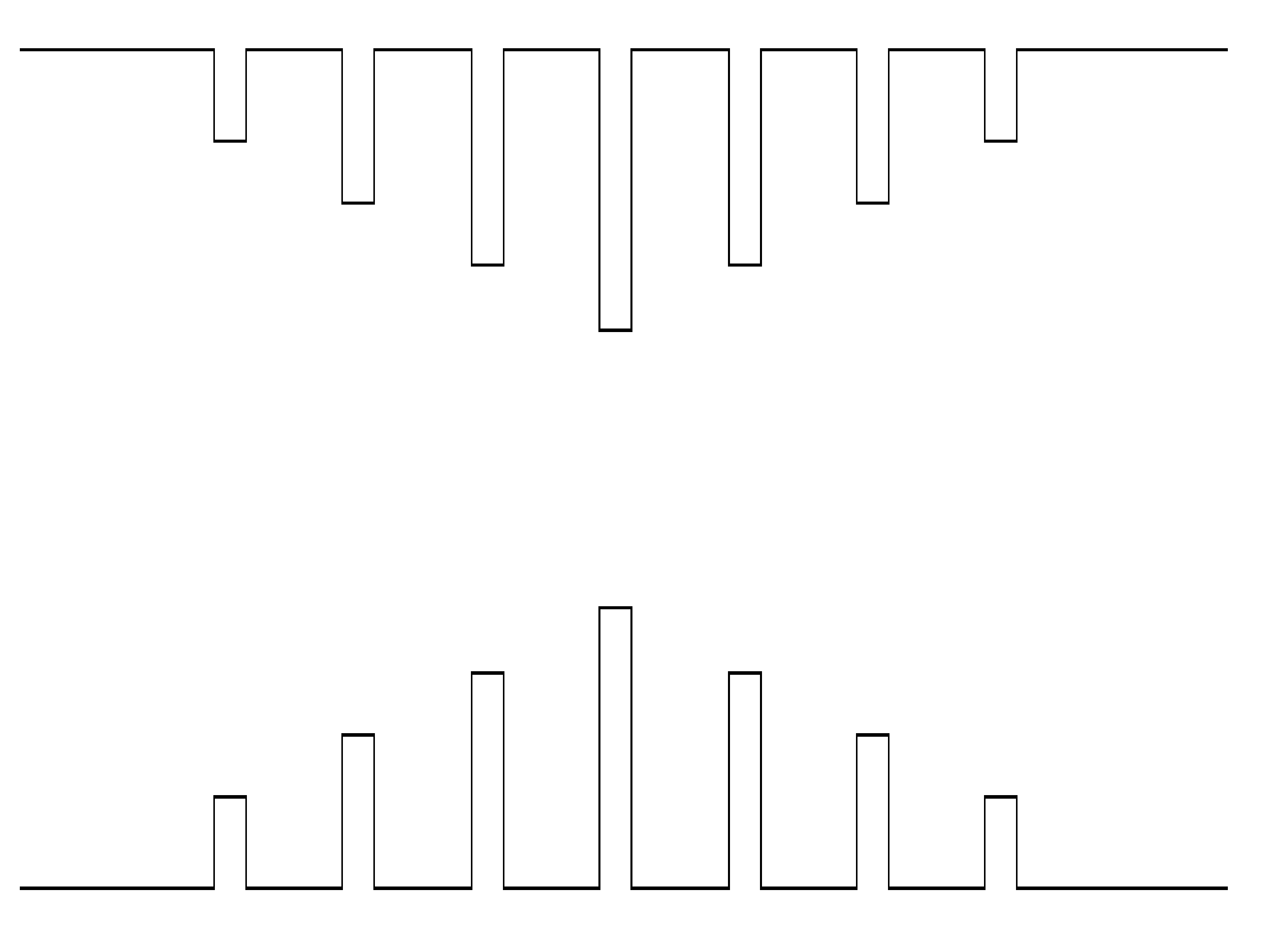}\hfil
\end{center}
\caption{A multiple-junction waveguide used to design a
filter.}\label{fg553}
\end{figure}


\begin{figure}
\vspace{-0.1 in}
\begin{center}
\hfil\includegraphics[width=2.60truein]{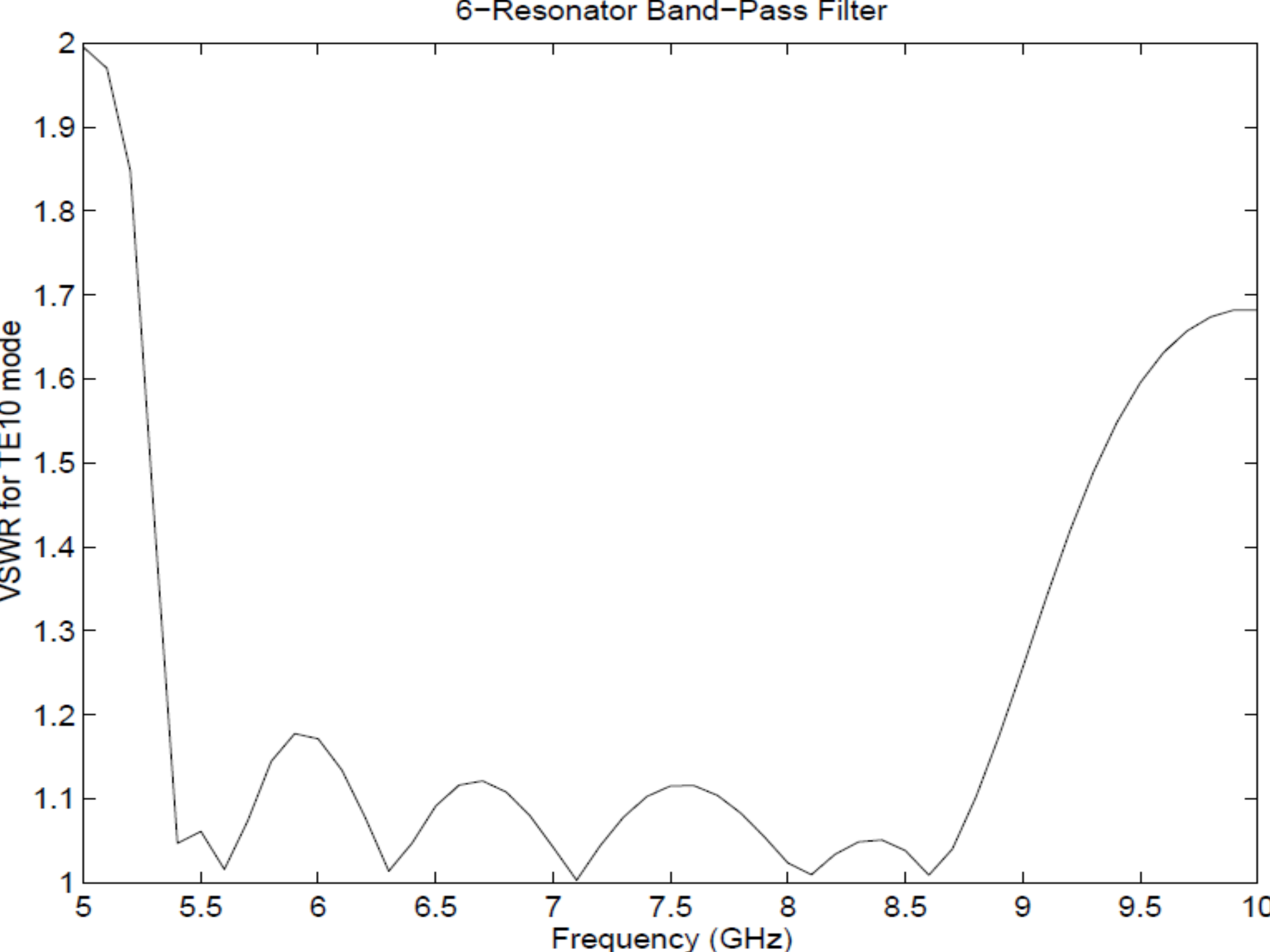}\hfil
\end{center}
\caption{The pass-band response of the filter in Figure \ref{fg553}
with six resonators. The waveguide dimensions are $13.7$ in by
$6.22$ in. The symmetric irises are $0.02$ in thick. The widths of
apertures are $1.026$, $0.958$, $0.898$, and $0.870$ in respectively
from one end to the center. The cavity lengths are $0.626$, $0.653$,
and $0.674$ in respectively from one end to the center (courtesy of
F. Ling).}\label{fg554}
\end{figure}

Multiple junctions in a waveguide can be used to design filters.
Junctions set up interference in a waveguide, giving rise to
resonant modes.  The \index{Waveguide!multiple junction!resonant mode}resonant mode of a multiple junction waveguide
is defined as one that a reflected mode can exist without the
presence of the incident mode.  This is the same as requiring that
\begin{equation}
\det\left(\widetilde{\dyad R}_{j,j+1} \right) =\infty
\end{equation}
The above is equivalent to
\begin{align}
\det\left(\dyad I - \dyad R_{j,j+1} \cdot e ^{i\dyad K_j
(h_j-h_{j-1} )} \cdot \widetilde {\dyad R}_{j+1, j+2} \cdot
e^{i\dyad K_j(h_j-h_{j-1})} \right) =0 \label{eq5-5-12a}
\end{align}
The above is the generalized transverse resonance condition.  One
can vary the frequency $\omega$ until the above equation is
satisfied.  It yields the poles of the system as is the case of the
Fabry-Perot etalon.  The locations of the poles can be used to guide
the design of filters as shown in Figures \ref{fg553} and
\ref{fg554}.

\section {{{ Hybrid Junctions}}}
\index{Waveguide!hybrid junction}

A hybrid-T junction is  a nifty device that when a TE$_{10}$ mode
is incident from port 1, the wave will couple to a mode in ports 2
and 3 but not 4. Similarly, a mode incident from port 4 will
couple to only ports 2 and 3 but not 1. A mode incident from port
2 couples to ports 1 and 4 but not 3, while a mode incident from
port 3 couples to ports 1 and 4 but not 2.

The understanding of the working of the hybrid-T, sometimes known
as a magic-T, can be derived from symmetry arguments. For a
TE$_{10}$ mode incident at port 1, the $\v E $ field is
symmetrical about the plane that bisects the center of the
waveguide of port 1. From the figure, the vertical component of
the electric field is even symmetric about the plane of symmetry,
while the horizontal component is odd symmetric. Hence, a
TE$_{10}$ mode cannot be excited in port 4.

\begin{figure}
\begin{center}
\hfil\includegraphics[width=5.0truein]{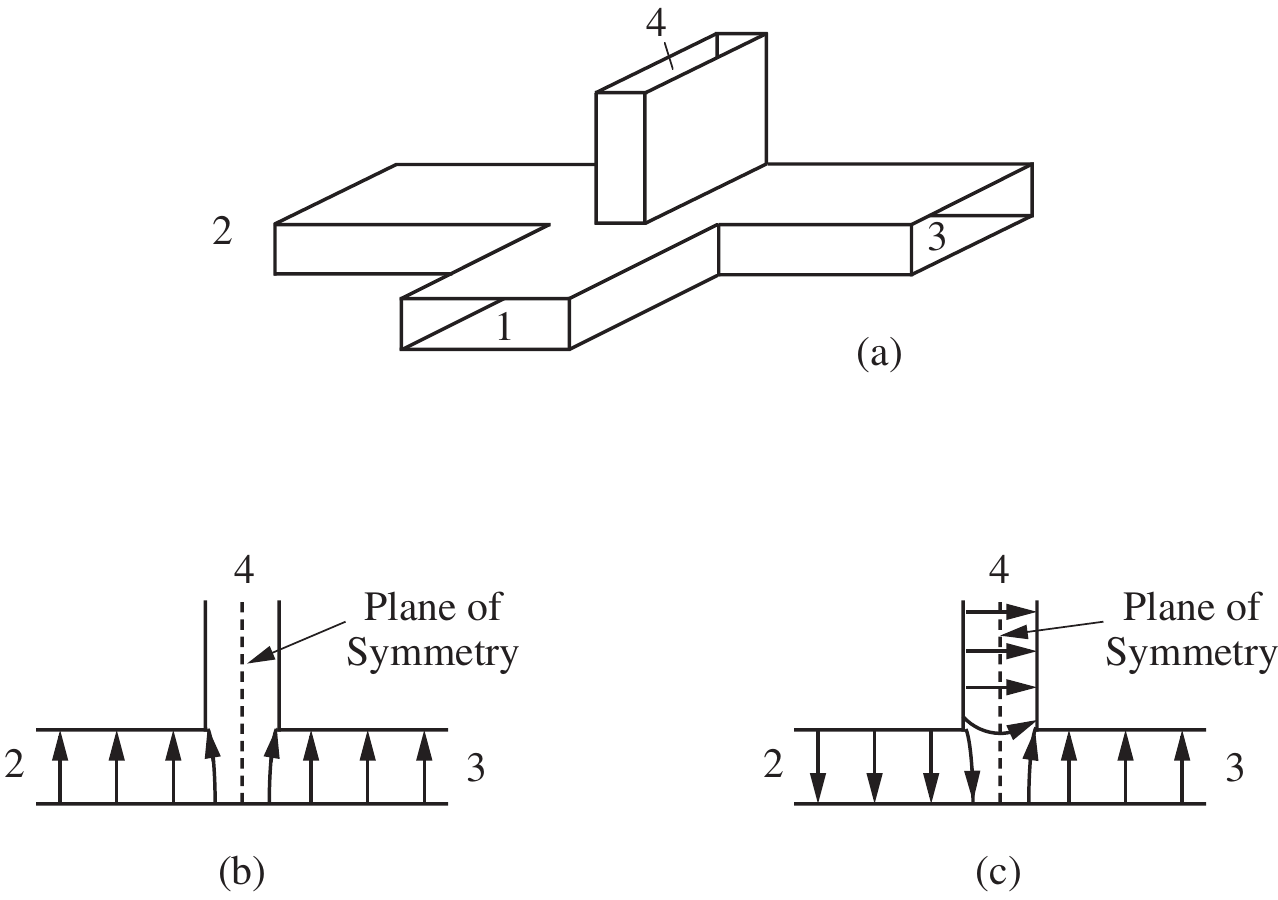}\hfil
\end{center}
\caption{(a) A hybrid-T junction. (b) Electric field pattern in port
2 and port 3 with a TE$_{10} $ mode incident from port 1. (c)
Electric field pattern in ports 2 and 3 with a TE$_{10}$ mode
incident from port 4.}\label{fg561}
\end{figure}


For a TE$_{10}$ mode incident from port 4, the horizontal
component of the electric field is even symmetric while the
vertical component is odd symmetric about the plane of symmetry.
Hence, the modes excited in ports 2 and 3 are of opposite
polarity. Furthermore, no TE$_{10} $ mode can be coupled to port 1
because of the symmetry condition.

When ports 2 and 3 are terminated in matched loads so that only
outgoing waves exist in them, ports 1 and 4 can be matched by
adding matching elements so that $S_{11}$ and $S_{44} $ are zero.
In this case, only incoming wave exists in port 1 or port 4 when
ports 2 and 3 are matched. The remaining property of the magic-T
can then be explained by time-reversal symmetry.

The \index{Waveguide!magic-T}magic-T is a linear and lossless device. By linearly
superposing a mode incident at port 1 and another mode incident at
port 4, one can obtain a mode exiting at port 3 but not a mode
exiting at port 2. Hence, this linear superposition is a solution
to Maxwell's equations. However, the time-reversed solution is
also a solution to Maxwell's equations. The time-reversed solution
corresponds to having ports 1 and 4 matched since there will only
be outgoing waves at these  ports. Furthermore, port 3 is matched
since only incoming wave exists at this port, and magically, there
is no coupling from port 3 to port 2! Similar argument leads to a
matched port 2 with matched loads at ports 1 and 4, and no
coupling from port 2 to port 3.

\begin{figure}
\begin{center}
\hfil\includegraphics[width=5.0truein]{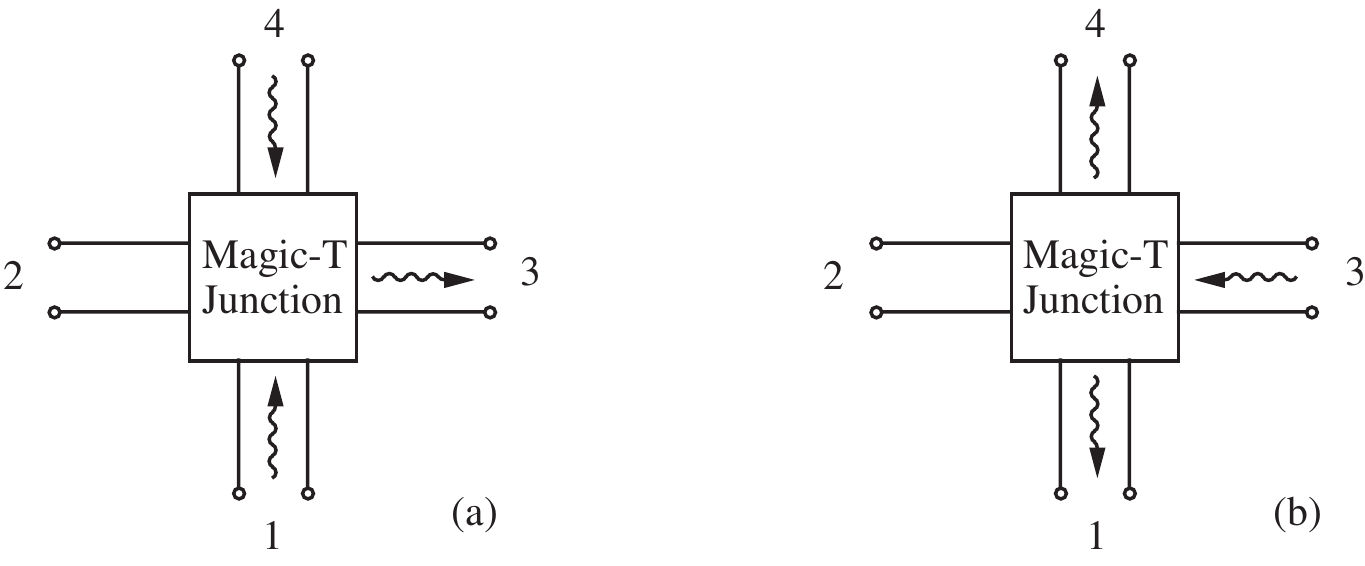}\hfil
\end{center}
\caption{(a) Normal-time operation of a magic-T. (b) Reversed-time
operation of a magic-T.}\label{fg562}
\end{figure}


Another useful hybrid junction is the \index{Rat-race ring circuit}rat-race ring circuit. When
a mode is incident in port 1, it will split evenly in two
directions into two waves. Due to the choice in the size of the
two ring, the waves will arrive out-of-phase at port 3. The
clockwise travelling wave will suffer a perturbation at port 2
while the counterclockwise travelling wave will suffer a
perturbation at port 4. Therefore, their amplitudes at port 3 are
equal and out of phase, and there is no coupling to port 3. The
phases of these two waves are equal at ports 2 and 4 even though
their amplitudes may not be equal because the clockwise travelling
wave and counterclockwise travelling wave are perturbed by
different amount. Therefore, there will be coupling to ports 2 and
4.

By the same argument, port 2 will not couple to port 4. Port 3
will not couple to port 1 and port 4 will not couple to port 2 by
reciprocity.

\begin{figure}
\begin{center}
\hfil\includegraphics[width=3.0truein]{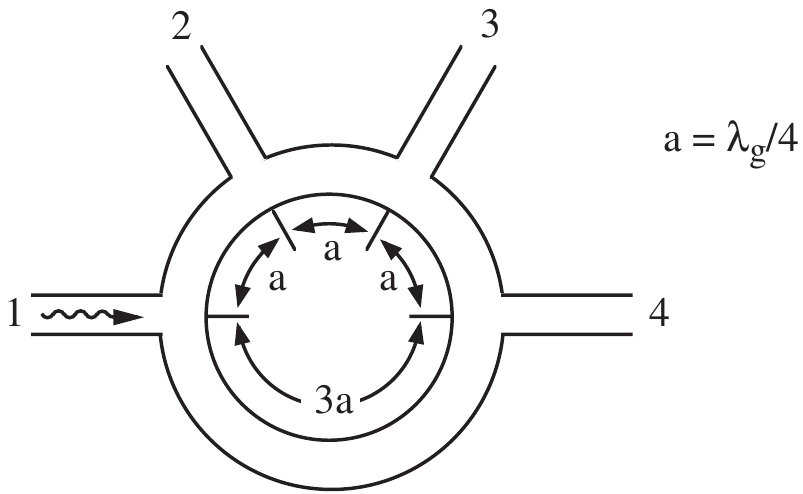}\hfil
\end{center}
\caption{A rat-race ring-type hybrid junction.}\label{fg563}
\end{figure}


Hybrid junctions is used to make microwave impedance bridges,
frequency discriminator circuits as in balanced mixers,
circulators as well as other applications.

\section {{{ Periodic Structures}}}
\index{Periodic structures}

A periodic structure is useful as a filter or a slow-wave structure.
The constructive and destructive interference phenomenon of a
periodic structure can generate passbands and stopbands: A \index{Passband} passband
is a frequency band that allows the propagation of a wave with no
attenuation while a \index{Stop band} stopband forbids the propagation of a wave.  A
periodic structure can be analyzed by the \index{Bloch-Floquet theorem} Bloch-Floquet theorem
\cite{COLLINJ,KITTEL,KONGE}.

\begin{figure}
\begin{center}
\hfil\includegraphics[width=4.0truein]{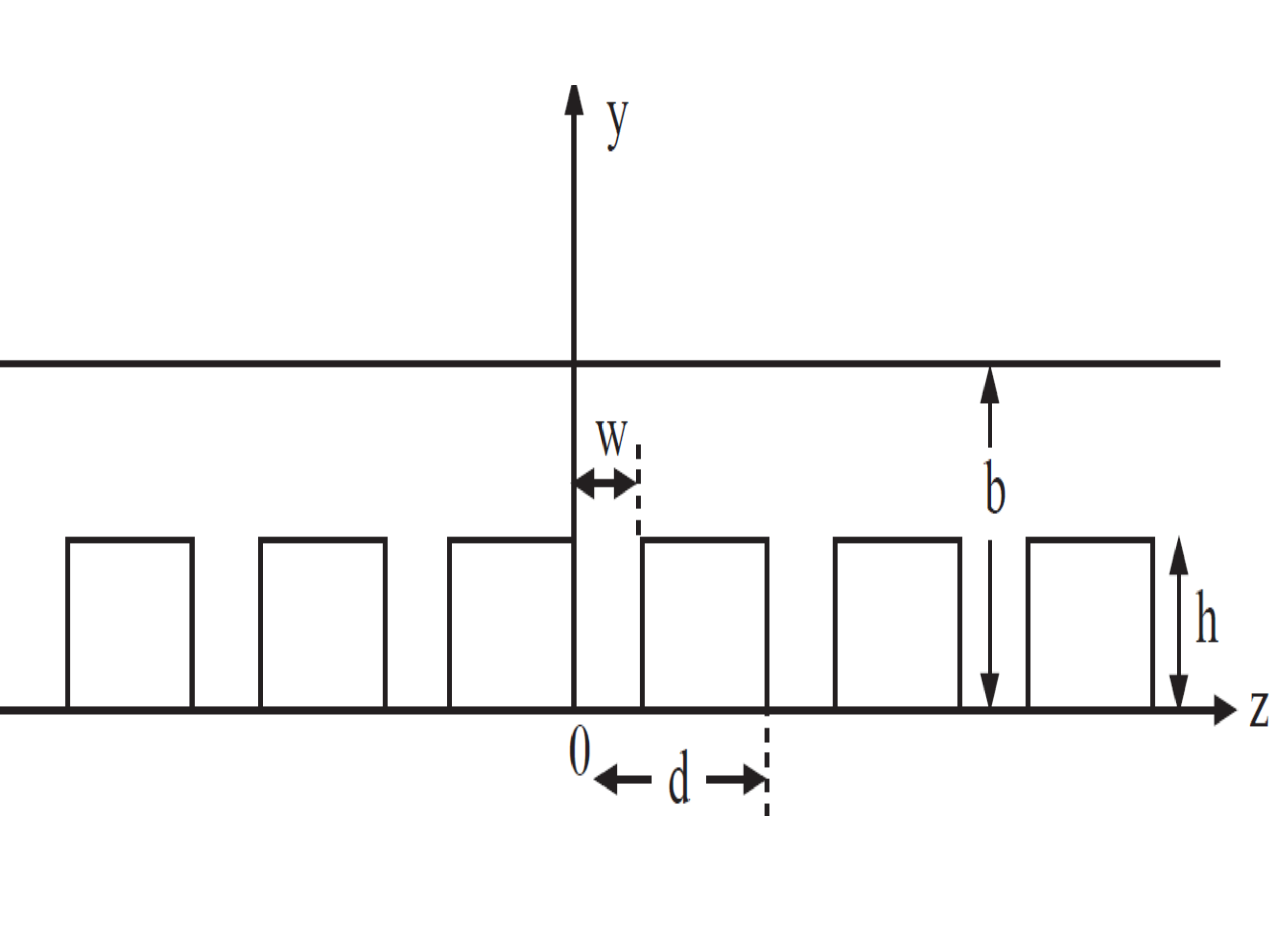}\hfil
\end{center}
\caption{A periodic structure in a parallel plate waveguide.}\label{fg571}
\end{figure}


If a wave is traveling inside a waveguide with a period $d$, it
implies that it has to be of the form
\begin{equation}
\v E(\v r)=e^{ik_zz}\v E_p(\v r), \label{eq5-7-2}
\end{equation}
\begin{equation}
\v H(\v r)=e^{ik_zz}\v H_p(\v r), \label{eq5-7-3}
\end{equation}
where $\v E_p(\v r)$ and $\v H_p(\v r)$ are periodic functions in
$z$ having the property
\begin{equation}
\v E_p(\v r+\hat znd)=\v E_p(\v r), \label{eq5-7-4}
\end{equation}
\begin{equation}
\v H_p(\v r+\hat znd)=\v H_p(\v r), \label{eq5-7-5}
\end{equation}
where $n$ is any positive or negative integer. The expression of a
field in (\ref{eq5-7-2}) and (\ref{eq5-7-3}) is known as the Floquet
theorem or the Bloch theorem. In other words, due to the periodicity
of this structure, a solution $\v E(\v r)$ when translated by a
distance $d$ in the $z$ direction, must be itself save an added
phase, viz.
\begin{equation}
\v E(\v r+\hat znd)=\v E(\v r)e^{ik_znd}, \quad \v H(\v r+\hat
znd)=\v H(\v r)e^{ik_znd}. \label{eq5-7-1}
\end{equation}

Any periodic function can be expanded in terms of a Fourier
series. Therefore, we can write
\begin{equation}
\v E_p(\v r)=\sum_{n=-\infty}^\infty \v E_{pn}(\v
r_s)e^{i{2n\pi\over d}z}, \label{eq5-7-6}
\end{equation}
\begin{equation}
\v H_p(\v r)=\sum_{n=-\infty}^\infty \v H_{pn}(\v
r_s)e^{i{2n\pi\over d}z}, \label{eq5-7-7}
\end{equation}
where $\v r_s=\hat xx+\hat yy$. Consequently, we can rewrite
(\ref{eq5-7-2}) and (\ref{eq5-7-3}) as
\begin{equation}
\v E(\v r)=\sum_{n=-\infty}^\infty \v E_{pn}(\v
r_s)e^{i(k_z+{2n\pi\over d})z} =\sum_{n=-\infty}^\infty \v
E_{pn}(\v r_s)e^{ik_{nz}z}, \label{eq5-7-8}
\end{equation}
\begin{equation}
\v H(\v r)=\sum_{n=-\infty}^\infty \v H_{pn}(\v
r_s)e^{i(k_z+{2n\pi\over d})z} =\sum_{n=-\infty}^\infty \v
H_{pn}(\v r_s)e^{ik_{nz}z}, \label{eq5-7-9}
\end{equation}
where $k_{nz}=k_z+{2n\pi\over d}$. Each of the terms above is
termed a \index{Floquet modes} Floquet mode, with a wavenumber $k_{nz}$. However, each
Floquet mode cannot exist by itself--they have to exist together
as a cluster.  The \index{Velocity!phase} phase velocity of the $n$-th Floquet mode is
\begin{equation}
v_{pn}={\omega\over k_{nz}}={\omega\over k_z+2n\pi /d},
\label{eq5-7-10}
\end{equation}
while the group velocity is given to be
\begin{equation}
v_{gn}={d\omega\over dk_{nz}}=\left ({dk_{nz}\over d\omega}\right
)^{-1}=v_g, \label{eq5-7-11}
\end{equation}
The \index{Velocity!group} group velocity is independent of the harmonics. Since $v_{pn}$
can be negative for some negative $n$, $v_{pn}$ can be opposite in
sign to $v_g$.

Let us assume that a TM mode is propagating in the parallel plate
waveguide such that $\v H=\hat xH_x$. There will be no
depolarization since $\partial /\partial x=0$, and this field
component is parallel to the surfaces involved. Consequently, we
have
\begin{equation}
H_x=\sum_{n=-\infty}^\infty h_n(y)e^{ik_{nz}z}, \label{eq5-7-12}
\end{equation}
Since $(\nabla ^2+{k_0}^2)H_x=0$, we have
\begin{equation}
{d^2h_n(y)\over dy^2}+k_{ny}^2h_n(y)=0, \label{eq5-7-13}
\end{equation}
where $k_{ny}^2=k_0^2-k_{nz}^2$.

For $y>h$, the coefficient $h_n(y)$ can be expanded as
\begin{equation}
h_n(y)=a_n \cos {\left [k_{ny}(y-b)\right ]}, \label{eq5-7-14}
\end{equation}
Since the Neumann boundary condition $\hat n\cdot\nabla H_x=0$ on
a metallic surface, we have chosen a cosine in (\ref{eq5-7-14}) so
that this condition is satisfied.

Inside the corrugation, we can expand $H_x$ in terms of the modes
of a shorted parallel waveguide, or
\begin{equation}
H_x=\sum_{m=0}^\infty g_m(y)\cos \left ({m\pi z\over w}\right ),
\label{eq5-7-15}
\end{equation}
By the same token, $g_m(y)$ satisfies
\begin{equation}
{d^2g_m(y)\over dy^2}+\gamma_{my}^2g_m(y)=0, \label{eq5-7-16}
\end{equation}
where $\gamma_{my}^2=k_0^2-\left ({m\pi\over w}\right )^2$.
Therefore, we derive that
\begin{equation}
g_m(y)=b_m\cos (\gamma_{my}y), \label{eq5-7-17}
\end{equation}
Consequently, we can write
\begin{equation}
H_x=\sum_{n=-\infty}^\infty a_n\cos
[k_{ny}(y-b)]e^{ik_{nz}z},\quad h<y<b, \label{eq5-7-18}
\end{equation}
\begin{equation}
H_x=\sum_{m=0}^\infty b_m\cos (\gamma_{my}y)\cos \left ({m\pi
z\over w}\right ),\quad 0<y<h. \label{eq5-7-19}
\end{equation}
The bottom equation is valid for $0<z<w$, and it replicates itself
with the correct phase shift for different slots.  To find the
guidance condition for nontrivial solutions to $a_n$ and $b_n$, we
match boundary condition across the $y=h$ interface. At this
interface, the tangential component of the magnetic field is
continuous, so is the tangential component of the electric field.
The latter is the same as ${\partial H_x\over\partial y}$ and is
continuous across this interface. Therefore, we arrive at
\begin{equation}
\sum_{n=-\infty}^\infty a_n\cos
[k_{ny}(h-b)]e^{ik_{nz}z}=\sum_{m=0}^\infty b_m\cos
\gamma_{my}h\cos \left ({m\pi z\over w}\right ), \quad 0<z<w
\label{eq5-7-20}
\end{equation}
\begin{equation}
\begin{split}
\sum_{n=-\infty}^\infty a_nk_{ny}\sin & [k_{ny}(h-b)]e^{ik_{nz}z}\\
 &= \left \{
\begin{aligned}
&\sum_{m=0}^\infty b_m\gamma_{my}\sin \left(\gamma_{my} h\right)\cos \left ({m\pi z\over w}\right ), 0<z<w,\\
& 0\hskip142pt\,w<z<d
\end{aligned}
\right .
\end{split}
\label{eq5-7-21}
\end{equation}
Equation (\ref{eq5-7-20}) is a cosine series expansion for
$0<z<w$.  We can solve for $b_m$ by multiplying (\ref{eq5-7-20})
by $\cos ({m\pi z\over w})$ and integrate from 0 to $w$ to obtain
\begin{equation}
b_m\cos (\gamma_{my}h){w\over
2}(1+\delta_{om})=\sum_{n=-\infty}^\infty a_n\cos [k_{ny}(h-b)]
\left< e^{ik_{nz}z}, \cos \left({m\pi z\over w}\right) \right>,
\label{eq5-7-22}
\end{equation}
where
\begin{equation}
\left< e^{ik_{nz}z}, \cos \left({m\pi z\over w}\right)\right> =
\int_0^w e^{ik_{nz}z}\cos \left( {m\pi z\over w}\right ) dz, \quad
m=0,..., \label{eq5-7-23}
\end{equation}
and can be evaluated in closed form if needed.

In (\ref{eq5-7-21}), by writing
\begin{equation}
e^{ik_{nz}z}=e^{ik_zz+i{2n\pi\over d}z}, \label{eq5-7-24}
\end{equation}
it is apparently a Fourier series expansion. We can find the
coefficients $a_n$ as
\begin{equation}
a_nk_{ny}\sin [k_{ny}(h-b)] d=\sum_{m=0}^\infty b_m\gamma_{my}\sin
(\gamma_{my}h) \left< \cos \left ({m\pi z\over w}\right ),
e^{-ik_{nz}z}\right>, \label{eq5-7-25}
\end{equation}
where
\begin{equation}
\left< \cos \left({m\pi z\over w}\right), e^{-ik_{nz}z}\right> =
\int_0^w \cos \left( {m\pi z\over w}\right) e^{-ik_{nz}z} dz,
\label{eq5-7-26}
\end{equation}
and $n$ varies from $-\infty$ to $+\infty$. Equations
(\ref{eq5-7-22}) and (\ref{eq5-7-25}) entail two infinite system
of equations.  Since $\gamma_{my}$ and $k_{ny}$ become large
imaginary numbers when $m$ and $n$ are large, $\cos
(\gamma_{my}h)$, $\sin (\gamma_{my}h)$, $\cos [k_{ny}(h-b)]$, and
$\sin [k_{ny}(h-b)]$ become exponentially large.  In this case, we
can define
\begin{equation}
\hat b_m= b_m\cos (\gamma_{my}h),
\end{equation}
\begin{equation}
\hat a_n= a_n\cos [k_{ny}(h-b)],
\end{equation}
to rewrite (\ref{eq5-7-22}) and (\ref{eq5-7-25}) as
\begin{equation}
\hat b_m {w\over 2}(1+\delta_{om})=\sum_{n=-\infty}^\infty \hat
a_n \left< e^{ik_{nz}z}, \cos \left({m\pi z\over w}\right)
\right>, \label{eq5-7-22a}
\end{equation}
\begin{equation}
\hat a_nk_{ny}\tan [k_{ny}(h-b)] d=\sum_{m=0}^\infty \hat
b_m\gamma_{my}\tan (\gamma_{my}h) \left< \cos \left ({m\pi z\over
w}\right ), e^{-ik_{nz}z}\right>, \label{eq5-7-25a}
\end{equation}

To solve the above, we need to truncate the infinite system. We
can rewrite (\ref{eq5-7-22a}) and (\ref{eq5-7-25a}) as
\begin{equation}
\hat b_m\lambda_m=\sum_{n=-N}^N \hat a_nA_{mn},\quad m=0,...,M,
\label{eq5-7-27}
\end{equation}
\begin{equation}
\hat a_n\beta_n=\sum_{m=0}^M \hat b_mB_{nm},\quad n=-N,...,N.
\label{eq5-7-28}
\end{equation}
The above can be written as matrix equations
\begin{equation}
\dyadg\lambda\cdot\hat {\v b}=\dyad A\cdot\hat{\v a},\quad\dyadg
\beta\cdot\hat{\v a}=\dyad B\cdot\hat{\v b}, \label{eq5-7-29}
\end{equation}
where $\dyadg \lambda$ and $\dyadg \beta$ are diagonal matrices,
$\dyad A$ is a $(M+1)\times(2N+1)$ matrix, $\dyad B$ is a
$(2N+1)\times(M+1)$ matrix, $\v a$ is a length $(2N+1)$ vector
while $\v b$ is a length $(M+1)$ vector. Equation (\ref{eq5-7-29})
can be rewritten as
\begin{equation}
\hat{\v b}-\dyadg \lambda ^{-1}\cdot\dyad A\cdot \dyadg
\beta^{-1}\cdot \dyad B\cdot \hat{\v b}=0. \label{eq5-7-30}
\end{equation}
Nontrivial solution for $\v b$ will exist if
\begin{equation}
\det (\dyad I-\dyadg \lambda^{-1}\cdot \dyad A\cdot
\dyadg\beta^{-1}\cdot \dyad B)=0. \label{eq5-7-31}
\end{equation}
The above is the guidance condition from which one can solve for $k$
given $k_z$.  It will be found that $k$ does not always exist for
all values for a given $k_z$.  Also, it could be that for a certain
window of $k$, no value exists for all $k_z$. Those are the stop
band of the periodic structure.

\begin{figure}
\begin{center}
\hfil\includegraphics[width=3.5truein]{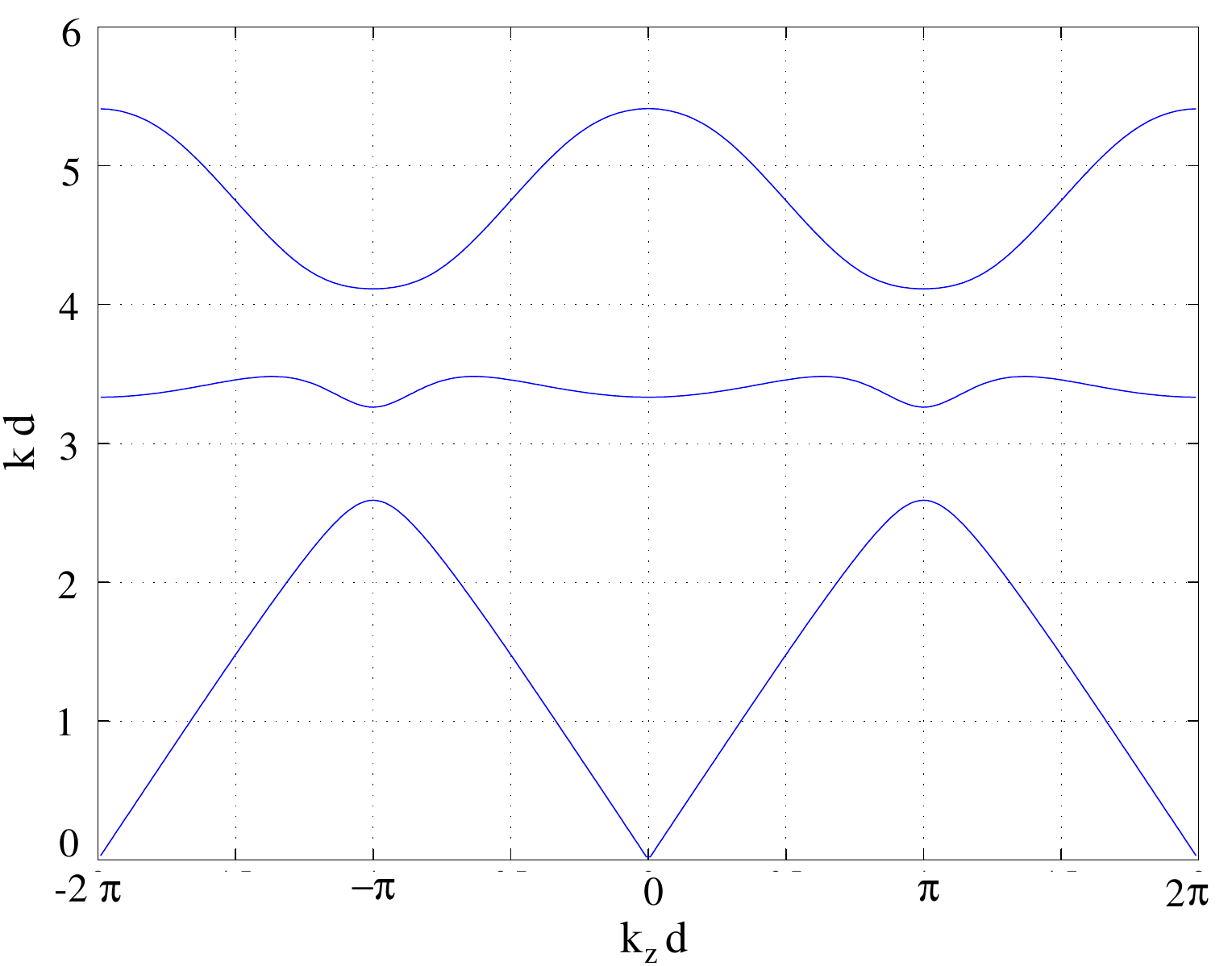}\hfil
\end{center}
\caption{Dispersion ($k-k_z$) diagram of a corrugated parallel
plate waveguide showing bandpass and bandstop sections ($w=d/2$,
$b=d$, $h=0.25b$) (courtesy of L.J. Jiang).}\label{fg572}
\end{figure}


\subsection{ Floquet Modes and Brillouin Zone}
\index{Floquet modes}
\index{Brillouin zone}

From the periodic structure theory, we notice that a wave
propagating in a periodic waveguide is of the form
\begin{equation}
\phi(\v r)=e^{ik_zz}\phi_p(\v r),\label{eq5-7-31a} \end{equation}
where $\phi_p(\v r)$ is a periodic function in $z$ with $\phi_p(\v
r+\hat z nd)=\phi_p(\v r)$, and
\begin{equation}
\phi_p(\v r)=\sum_{n=-\infty}^{\infty}\phi_{pn}(\v
r_s)e^{i\frac{2n\pi}{d}z}.\label{eq5-7-31b}
\end{equation}
The physical picture is that this wave is a cluster of Floquet
modes propagating in unison through the waveguide in order to
satisfy the boundary condition on the waveguide wall.  In other
words, all the Floquet modes of the form $\exp(i(k_z+2n\pi/d))$
move in lock step with respect to each other. We can express
(\ref{eq5-7-31a}) more explicitly in terms of the Floquet modes,
namely,
\begin{equation}
\phi(\v r)=\sum_{n=-\infty}^{\infty}\phi_{pn}(\v
r_s,k_z)e^{i\left(k_z+\frac{2n\pi}{d}\right)z}.\label{eq5-7-31c}
\end{equation}
Now, if we let $k_z\rightarrow k_z+\frac{2m\pi}{d}$, where $m$ is
an integer, the above becomes
\begin{equation}
\begin{split}
\phi(\v r)&=\sum_{n=-\infty}^{\infty}\phi_{pn}\left(\v
r_s,k_z+\frac{2m\pi}{d}\right)e^{i\left[k_z+\frac{2(n+m)\pi}{d}\right]z}\\
&=\sum_{n'=-\infty}^{\infty}\phi_{p,n'-m}\left(\v
r_s,k_z+\frac{2m\pi}{d}\right)e^{i\left(k_z+\frac{2n'\pi}{d}\right)z}.
\label{eq5-7-31d}
\end{split}
\end{equation}
It is clear that if (\ref{eq5-7-31c}) is an eigensolution that can
satisfy the boundary condition on the waveguide wall,
(\ref{eq5-7-31d}) can be easily made to satisfy the boundary
condition on the waveguide wall if
$$\phi_{p,n'-m}\left(\v r_s,k_z+\frac{2m\pi}{d}\right)
=c\phi_{pn'}(\v r_s,k_z)$$ where $c$ is a multiplicative constant.
 Therefore, the cluster of Floquet modes
is invariant with respect to the transform $k_z\rightarrow
k_z+\frac{2m\pi}{d}$. Consequently, the \index{Dispersion diagram} dispersion diagram for a
guided mode in a periodic structure is $\frac{2\pi}d$ periodic in
$k_z$ as shown in Figure \ref{fg572}.  Each of this periodic zone is
called the Brillouin zone.

Furthermore, we can show that if $\phi(\v r)$ is a solution, so is
$\phi^*(\v r)$.  Using this fact, one can show that the dispersion
diagram has to be even symmetric about the origin.

\section{{{ Stop Band and Coupled-Mode Theory}}}
\index{Stop band}
\index{Coupled-mode theory}

The existence of pass bands and stop bands in a \index{Waveguide!periodic} periodic waveguide
is due to the coupling, constructive and destructive interference of
the Floquet modes. To see how the interfering Floquet modes can
yield a stop band when they are coupled, we use the coupled modes
theory  \cite{HAUS5}.

\begin{figure}
\begin{center}
\hfil\includegraphics[width=3.5truein]{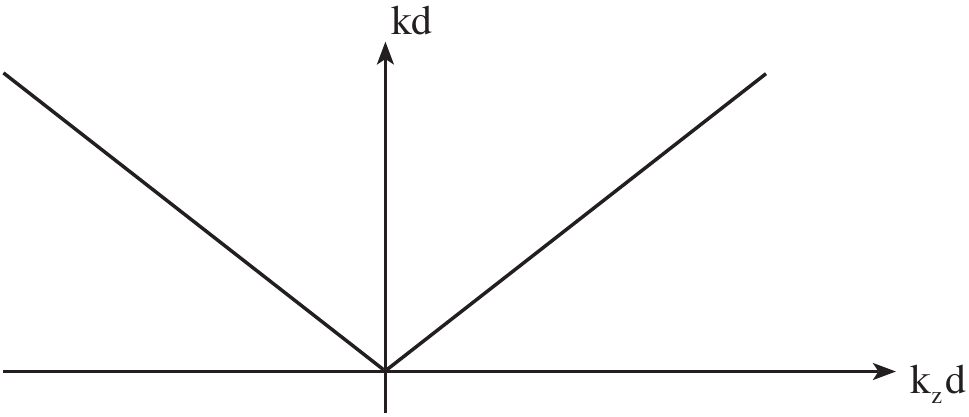}\hfil
\end{center}
\caption{Dispersion ($k-k_z$) diagram of a parallel plate
waveguide without periodic perturbation.}\label{fg581}
\end{figure}


\begin{figure}
\begin{center}
\hfil\includegraphics[width=3.5truein]{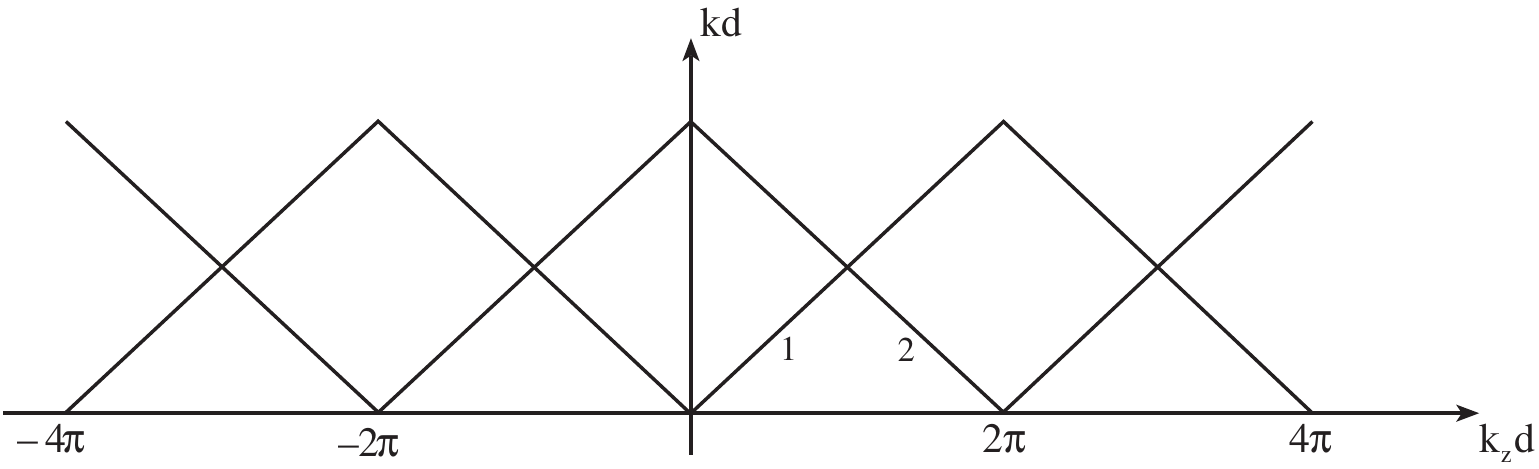}\hfil
\end{center}
\caption{Dispersion ($k-k_z$) diagram of a parallel plate waveguide
with weak or no periodic perturbation.  The Floquet modes are
generated, but they remain weakly coupled to each other.
}\label{fg21}
\end{figure}


To do this analysis, we think of the corrugation in a waveguide as a
perturbation. Before the \index{Corrugation perturbations} corrugation perturbations are introduced,
assume that we have only the forward and backward propagating TEM
modes within the parallel-plate waveguide. The dispersion diagram is
as shown in Figure \ref{fg581}. The moment a small periodic
perturbation is introduced, higher order Floquet modes emerge in the \index{Dispersion diagram}
waveguide and the \index{Dispersion diagram!periodic} dispersion diagram becomes periodic.  When this
happens, all the Brillouin zones are equivalent to each other, and
none is preferred over the others, as can be seen from the analysis
in the previous subsection.

If the perturbation is weak enough, little coupling occurs between
the forward Floquet modes and the backward Floquet modes. The
forward and backward Floquet modes carry energy in the forward and
backward directions respectively due to the sign of their \index{Velocity!group} group
velocities.   The dispersion curve in the case of vanishingly small
periodic perturbation is shown in Figure \ref{fg21}.

We can focus our attention on modes 1 and 2 and study the coupling
behavior between them. The couple-mode equation for describing
these two contra-propagating modes can be written as
\begin{equation}
\frac{d\phi_1}{dz} = ik_1\phi_1 + i\alpha_{12}\phi_2,
\label{eq11101}
\end{equation}
\begin{equation}
\frac{d\phi_2}{dz} = ik_2\phi_2 + i\alpha_{21}\phi_1,
\label{eq11102}
\end{equation}
$\alpha_{ij}$ is the coupling coefficient describing the coupling
of energy between the two modes. If $\alpha_{ij} = 0$, coupling
ceases to exist, and we have $k_z = k_1$ for mode 1, and $k_z =
k_2$ for mode 2 as expected. More specifically, $k_1 = k$, and
$k_2 = -k + \frac{2\pi}{d}$. We can rewrite (\ref{eq11101}) and
(\ref{eq11102}) using matrix notation as
\begin{equation}
\frac{d}{dz}\vg \phi = i \textsc{ }\dyad K \cdot \vg \phi,
\label{eq11103}
\end{equation}
where
\begin{equation}
\vg \phi = \bmatrix \phi_1\\ \phi_2\\ \endbmatrix,\quad \dyad K =
\bmatrix k_1 & \alpha_{12}\\ \alpha_{21} & k_2 \endbmatrix.
\label{eq11104}
\end{equation}
But letting the solution
\begin{equation}
\vg \phi = e^{ik_zz} \v {a_0}, \label{eq11105}
\end{equation}
we have from (\ref{eq11103}) that
\begin{equation}
k_z \v {a_0} = \dyad K \cdot \v {a_0} \label{eq11106},
\end{equation}
or that $\v {a_0}$ is the eigenvector of $\dyad K$, and then $k_z$
is its eigenvalue given by
\begin{equation}
\det \left( k_z \dyad I - \dyad K \right) = 0. \label{eq11107}
\end{equation}
Equation (\ref{eq11107}) yields
\begin{equation}
(k_1-k_z)(k_2-k_z) - \alpha_{12}\alpha_{21} = 0. \label{eq11108}
\end{equation}
Solving the above for $k_z$, we have
\begin{equation}
k_z = \frac{(k_1+k_2)}{2} \pm \frac{1}{2}\sqrt{(k_1 - k_2 )^2 +
\alpha_{12}\alpha_{21}}.\label{eq11109}
\end{equation}
From energy conservation, we rewrite that
\begin{equation}
\frac{d}{dz}\left(|\phi_1|^2 - |\phi_2|^2 \right) = 0,
\label{eq11110}
\end{equation}
since two modes propagate in the opposite directions, and hence,
their energy flow cancel each other. Equation (\ref{eq11110}) is
the same as
\begin{equation}
\frac{d}{dz} \vg \phi^\dag \cdot \dyad S \cdot \vg \phi = 0,
\label{eq11111}
\end{equation}
where $\dyad S = \bmatrix 1 & 0\\ 0 & -1\\ \endbmatrix$. From
(\ref{eq11111}), and using (\ref{eq11103}), we have
\begin{equation}
\begin{split}
\frac{d}{dz}\vg \phi^\dag \cdot \dyad S \cdot \vg \phi & = \left(
\frac{d}{dz}\vg \phi^\dag \right) \cdot \dyad S \cdot \vg \phi +
\vg
\phi^\dag \cdot \dyad S \cdot \frac{d\vg\phi}{dz}\\
& = -i \vg \phi^\dag \cdot \dyad K^\dag \cdot \dyad S \cdot \vg
\phi +
i\vg \phi^\dag \cdot \dyad S \cdot \dyad K \cdot \vg \phi\\
& = i \vg \phi^\dag \cdot \left(\dyad S \cdot \dyad K - \dyad
K^\dag \cdot \dyad S\right)\cdot \vg \phi \quad = 0.
\end{split}\label{eq11112}
\end{equation}
In order for the above to be zero, it is required that
$\alpha_{12} = -\alpha_{21}^*$. (If the two modes are
co-propagating, carrying energy the same direction, the condition
for energy conservation is $\alpha_{12} = \alpha_{21}^*$.)

For mode 1, $k_1 = k$, and for mode 2, $k_2 = -k+\frac{2\pi}{d}$.
From (\ref{eq11109}), we have that
\begin{equation}
k_z = \frac{\pi}{d} \pm \sqrt{\left(k - \frac{\pi}{d}\right)^2 -
|\alpha_{12}|^2}. \label{eq11113}
\end{equation}
But in the dispersion diagram, we usually fix $k_z$ and solve for
$k$.  Alternatively, we can invert Equation (\ref{eq11113}) to
obtain
\begin{equation}
k = \frac{\pi}{d} \pm \sqrt{\left(k_z - \frac{\pi}{d}\right)^2 -
|\alpha_{12}|^2} \label{eq11115}
\end{equation}
In the vicinity of $k_z = \frac{\pi}{d}$, $k = \frac{\pi}{d} \pm
|\alpha_{12}|$. In other words, $k$ splits into two distinct values
about $\frac{\pi}{d}$ due to forward and backward wave coupling.

\subsection {{{ Circuit Analysis of Periodic Structure}}}
\index{Periodic structures!circuit analysis}

The above numerical analysis solves the problem exactly within
numerical approximation. However, it offers little insight to the
problem. As an approximation to the problem, we can think of the
corrugated parallel plate waveguide problem as being a
\index{Transmission line} transmission line problem with shorted stubs connected in series.

\begin{figure}
\begin{center}
\hfil\includegraphics[width=4.0truein]{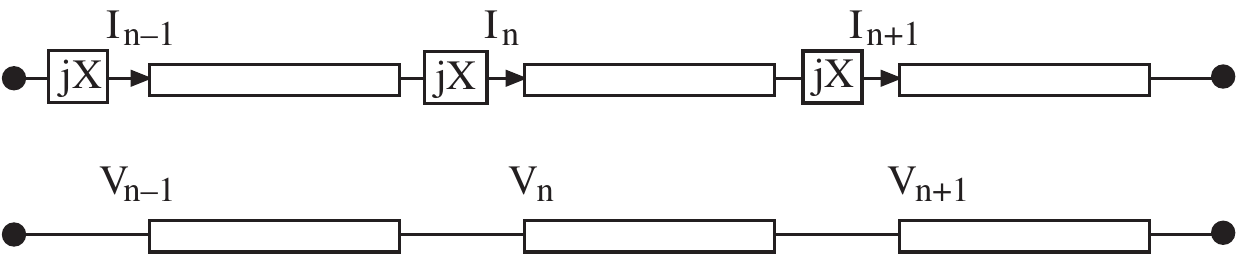}\hfil
\end{center}
\caption{Circuit equivalence of a periodic structure.}\label{fg581a}
\end{figure}


The structure in Figure \ref{fg581a} is best analyzed by the chain
matrix or the $ABCD$ transmission matrix. It can be shown that the
chain matrix connecting the voltages of a section of transmission
line of length $d$ is
\begin{equation}
\begin{bmatrix}
 V_1\\
 I_1
\end{bmatrix} =
\begin{bmatrix}
 \cos (kd)&jZ_c\sin (kd)\\
 jY_c\sin (kd)&\cos (kd)
\end{bmatrix}
\begin{bmatrix} V_2\\ I_2
\end{bmatrix} .
\label{eq5-7-32}
\end{equation}

\begin{figure}
\begin{center}
\hfil\includegraphics[width=2.0truein]{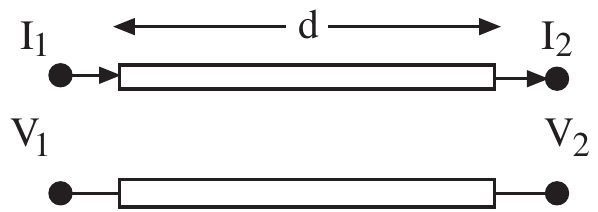}\hfil
\end{center}
\caption{A section of a transmission line.}\label{fg582}
\end{figure}


For a series reactance, the corresponding chain matrix is
\begin{equation}
\begin{bmatrix}
 V_1\\
 I_1
\end{bmatrix} =
\begin{bmatrix}
 1&jX\\
 0&1
\end{bmatrix}
\begin{bmatrix}
 V_2\\
 I_2
\end{bmatrix} .
\label{eq5-7-33}
\end{equation}
Therefore, the chain matrix connecting $V_n, I_n$ to $V_{n+1},
I_{n+1}$ is
\begin{equation}
\begin{bmatrix}
 V_n\\
 I_n
\end{bmatrix} =
\begin{bmatrix}
 \cos (kd)&jZ_c\sin (kd)\\
 jY_c\sin (kd)&\cos (kd)
\end{bmatrix}
\begin{bmatrix}
 1&j\\
 0&1
\end{bmatrix}
\begin{bmatrix}
 V_{n+1}\\
 I_{n+1}
\end{bmatrix} .
\label{eq5-7-34}
\end{equation}
But if a wave propagates on a periodic structure, then
\begin{equation}
V_n=V_{n+1}e^{jk_zd}, \quad I_n=I_{n+1}e^{jk_zd}. \label{eq5-7-35}
\end{equation}
Consequently, (\ref{eq5-7-34}) becomes
\begin{equation}
\left \{
\begin{bmatrix}
 \cos (kd)&jX\cos (kd)+jZ_c\sin (kd)\\
 jY_c\sin (kd)&-XY_c\sin (kd)+\cos (kd)
\end{bmatrix} -
\begin{bmatrix}
 e^{jk_zd}&0\\
 0&e^{jk_zd}
\end{bmatrix} \right \}
\begin{bmatrix}
 V_{n+1}\\
 I_{n+1}
\end{bmatrix} = 0 .
\label{eq5-7-36}
\end{equation}
The above will have a nontrivial solution only if
\begin{equation}
\det
\begin{pmatrix}
 A-e^{jk_zd}&B\\
 C&D-e^{jk_zd}
\end{pmatrix} = 0 ,
\label{eq5-7-37}
\end{equation}
The above is equivalent to
\begin{equation}
AD-BC-(A+D)e^{jk_zd}+e^{2jk_zd}=0, \label{eq5-7-38}
\end{equation}
Since $AD-BC=1$ for a reciprocal network, we have
\begin{equation}
\cos (k_zd)={A+D\over 2}, \label{eq5-7-39}
\end{equation}
The above gives the \index{Periodic structures!guidance condition} guidance condition for the wave number of a
wave propagating on a periodic structure. For the circuit of a
series reactance loaded periodic structure, this becomes
\begin{equation}
\cos (k_zd) = \cos (kd)-{XY_c\over 2}\sin (kd), \label{eq5-7-40}
\end{equation}
If $X$ is due to a shorted stub, then $X=Z_s\tan (kh)$, and we
have
\begin{equation}
\cos (k_zd) = \cos (kd)-{Z_sY_c\over 2}\tan (kh)\sin (kd),
\label{eq5-7-41}
\end{equation}

\begin{figure}
\begin{center}
\hfil\includegraphics[width=5.0truein]{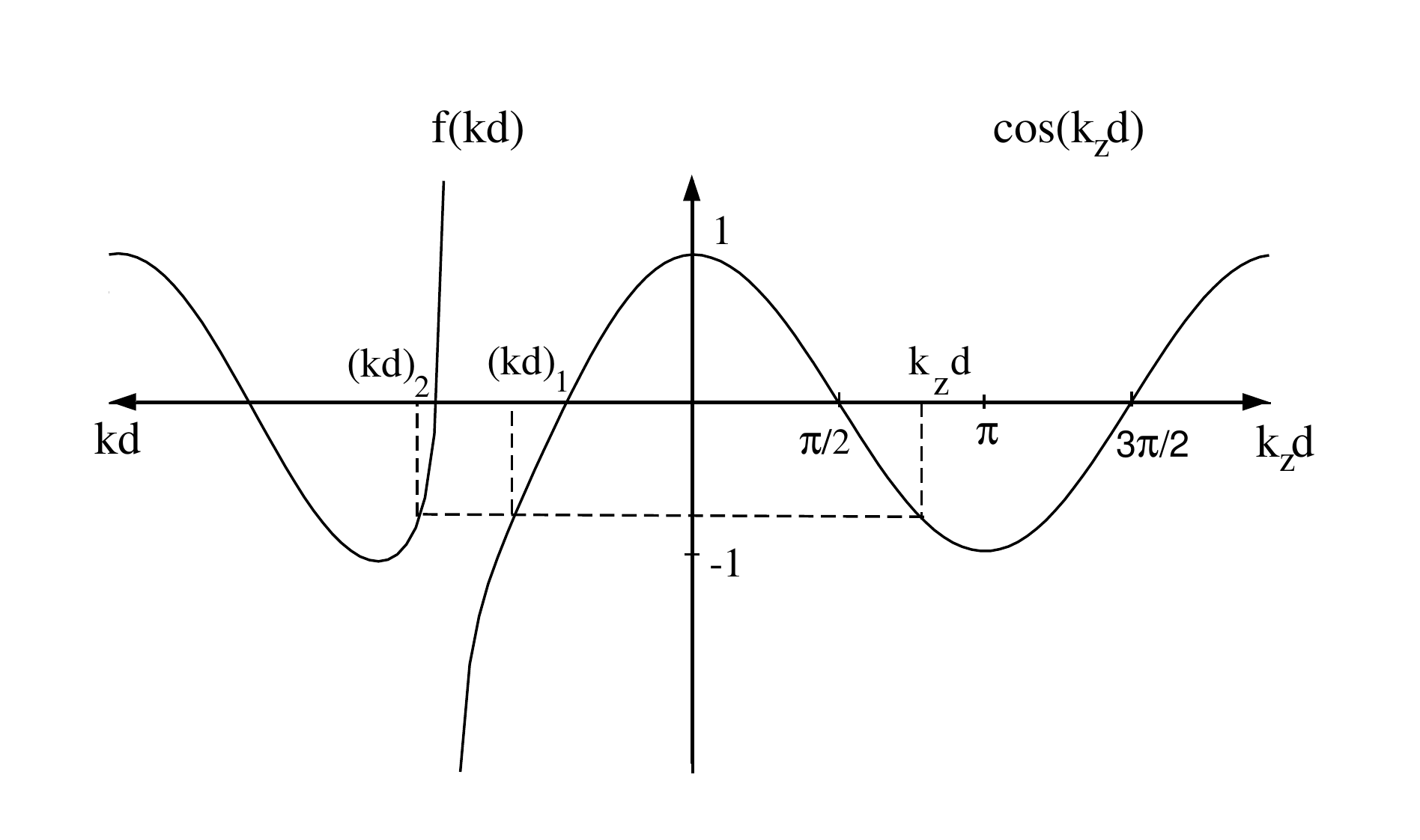}\hfil
\end{center}
\caption{Plot of $\cos (k_zd)$ versus $k_zd$ and $f(kd)$ versus
$kd$.}\label{fg583}
\end{figure}


Figure \ref{fg583} plots the right-hand side of (\ref{eq5-7-41})
which is $f(kd)$ as a function of $kd$, and the left-hand side of
(\ref{eq5-7-41}) which is just $\cos (k_zd)$. For every $k_zd$, we
can read off several values of $kd$ such that $\cos (k_zd) =
f(kd)$.  Hence we arrive at the following diagram for $kd$ versus
$k_zd$.

\begin{figure}
\begin{center}
\hfil\includegraphics[width=5.0truein]{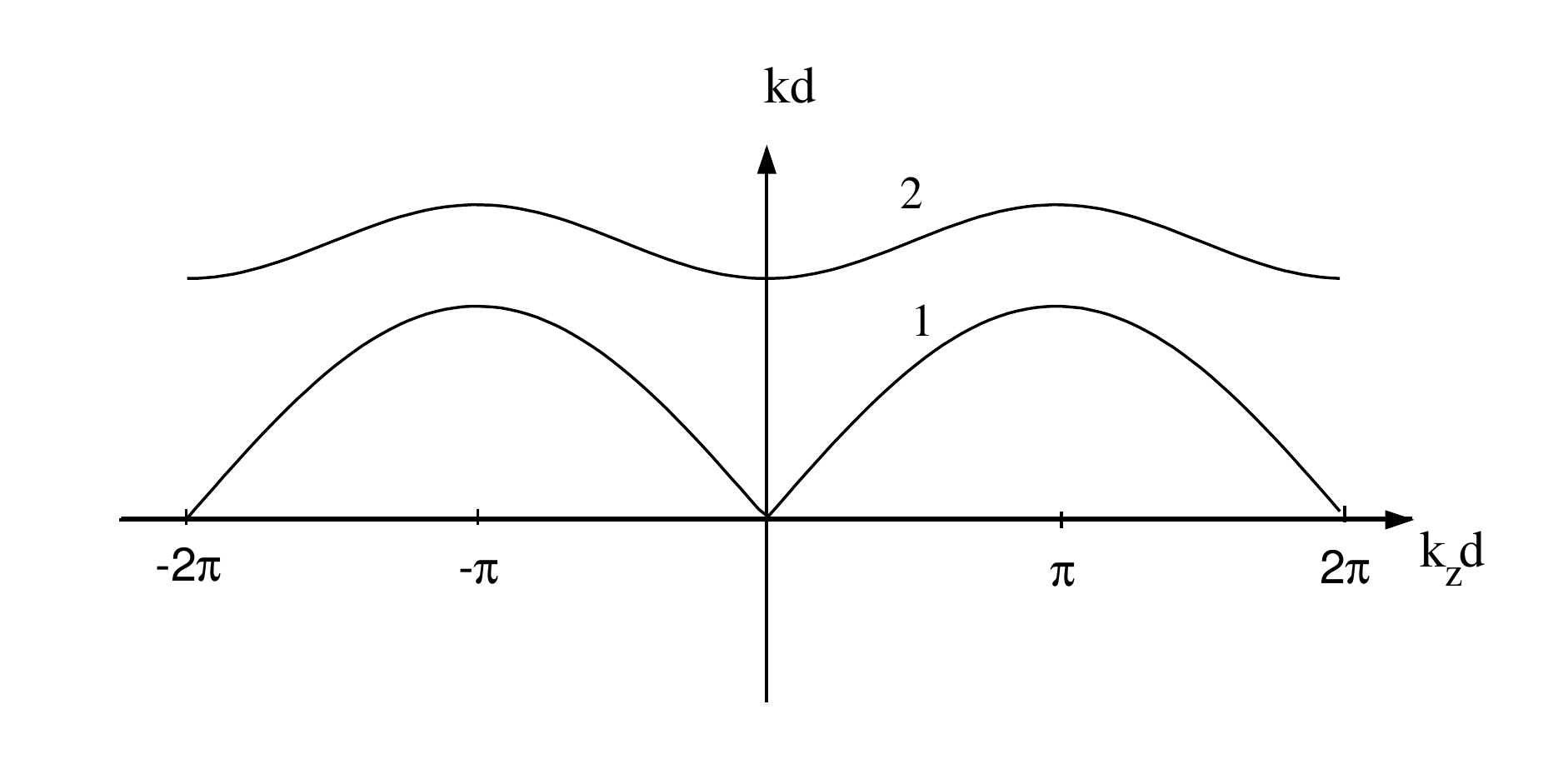}\hfil
\end{center}
\caption{The $kd-k_zd$ dispersion diagram of a periodically loaded
transmission line.}\label{fg584}
\end{figure}


As can be seen from Figure \ref{fg584}, there are frequency bands
at which real values of $k_zd$ could exist. These are the
\index{Passband} passbands. In Figure \ref{fg584}, we have only shown the first two
bands. The band of frequencies at which no real values of $k_zd$
can exist is the stopband. Because of this property, a periodic
structure can be used as a \index{Periodic structures!filter} filter.  Also, notice that the phase
velocity is smaller than that of just a TEM mode propagating in a
parallel plate waveguide, hence, the name slow-wave structure for
a periodic waveguide. It is also possible to have the group
velocity opposite in sign to the phase velocity as can be seen
from the above diagram. This is used in a backward wave oscillator
in microwave circuits.

\section{Metamaterials}
\index{Metamaterials}

The area of metamaterials has been inspired by the suggestion of
Vesalago \cite{VESALAGO} that if we have both $\epsilon$ negative
and $\mu$ negative material (also called a double-negative or DNG
material), then the wave will be a backward wave with the phase
velocity traveling in the opposite direction to the group velocity.
It was further suggested by Pendry \cite{PENDRY} that a DNG material
time reverses a field, and hence, can be used to make a perfect lens
(or superlens). As a result there have been a flurry of activities
in this field.

Unfortunately, many of the superlensing effect of DNG material
disappears with the slightest amount of loss or imperfection.
However, the excitement in search of the holy grail has inspired
many new ideas and structures that could be of interest.
Consequently, many other metamaterials have been proposed in recent
years. Examples of these are epsilon negative materials (ENG), mu
negative materials (MNG), zero index materials (ZIM), different
effective index materials.  In single negative materials (SNG), one
can show that a surface plasmon polariton can be excited at the
air-material interface. In ZIM, the wavelength of the field is
infinite, and hence, many concepts prevailing in circuit theory can
be applied in optical frequencies.
 Our ability to fabricate different
index materials also gives rise to the field of transformation
optics, which further yield the concept of cloaking. Since the
fabrication of 3D bulk metamaterials has been difficult, there has
been interest in meta-surfaces that can be fabricated easily by
epitaxial techniques. Also, advances in nano-fabrication technology
allow the fabrication of artificial atoms such as Cooper-pair boxes.
Suggestions of quantum metamaterials have emerged.

\begin{figure}
\begin{center}
\hfil\includegraphics[width=2.5truein]{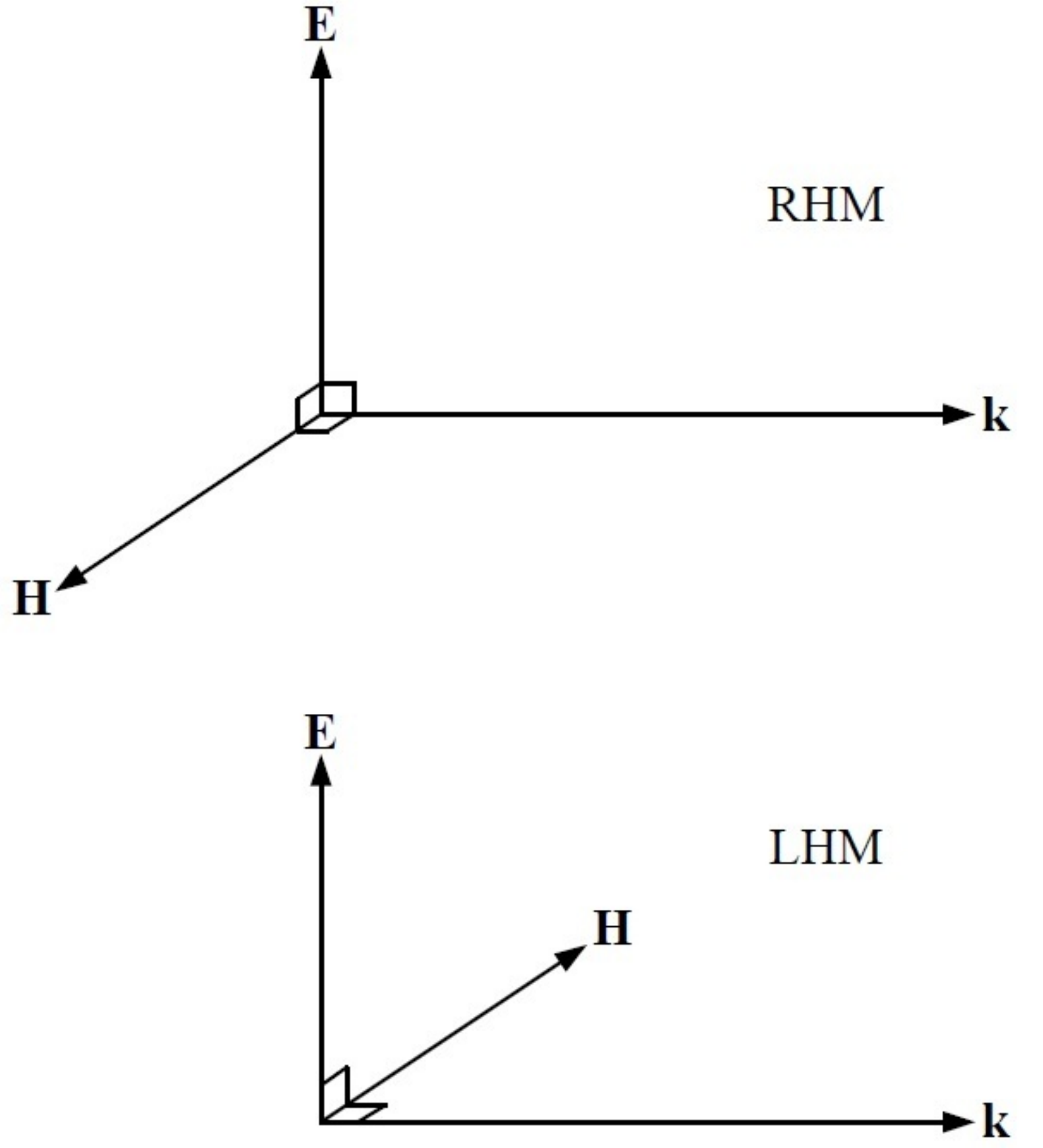}\hfil
\end{center}
\caption{An RHM plane wave in a regular medium versus an LHM plane
wave that propagates in a double negative medium.
}\label{LHMPlaneWave}
\end{figure}

Here, we will briefly review the physics of \index{Metamaterials!double-negative (DNG)}DNG materials.
We can start with Maxwell's equations for regular materials in the
frequency domain:
\begin{eqnarray}
\nabla \times \vec E&=&i\omega\mu \vec H\nonumber\\
\nabla \times \vec H&=&-i\omega\epsilon\vec E+\vec J\nonumber\\
\nabla\cdot\epsilon\vec E&=&\rho\nonumber\\
\nabla\cdot\mu\vec H&=&0\label{DNG1}
\end{eqnarray}
If we were to change the sign of $\epsilon$ and $\mu$ in Equations
\eqref{DNG1}, we arrive at:
\begin{eqnarray}
\nabla \times \vec E&=&-i\omega\mu\vec H\nonumber\\
\nabla \times \vec H&=&i\omega\epsilon \vec E +\vec J\nonumber\\
\nabla\cdot\epsilon\vec E&=&-\rho\nonumber\\
\nabla\cdot\mu\vec H&=&0\label{DNG2}
\end{eqnarray}
Notice that once the solution to \eqref{DNG1} is obtained, we can
obtain the solution to \eqref{DNG2} by changing the signs of $\bf
H$, $\bf J$, and $\rho$.  Hence power flow, which is defined to be
$\bf E\times\bf H^*$, points in the opposite direction for solutions
of Equations \eqref{DNG2} compared to that of solutions of Equations
\eqref{DNG1}.  Also, the $\v E$, $\v H$ and $\v k$ vectors form the
left-hand rule (see Figure \ref{LHMPlaneWave}).  Hence, a DNG medium
is also called a \index{Medium!left-handed} left-handed medium (LHM) as opposed to the regular
right-handed medium (RHM). As the plane wave has a $k$-vector that
points in the opposite direction to that of power flow, and that
$k=-\omega\sqrt{\epsilon\mu}$, a DNG material is also called a
\index{Negative index material} negative index material (NIM).

The fact that $k$-vector points in opposition to the power flow, it
implies that the \index{Velocity!phase} phase velocity is opposite in sign to the group
velocity.  This implies that the medium has to be \index{Medium!dispersive} dispersive as the
phase velocity $v_{ph}=\omega/k$ while the \index{Velocity!group} group velocity is
$v_{gr}=(dk/d\omega)^{-1}$.  They can only be opposite of each other
if $k=\omega\sqrt{\mu(\omega)\epsilon(\omega)}$.

We can also easily prove that LHM has to be \index{Medium!frequency dispersive} frequency dispersive by
{\it reductio ad absurdum} \cite{CHEW_DNG}. Assuming that the medium
is frequency independent or non-dispersive.  Then we can convert
Maxwell's equations back to the time domain.  Subsequently, a
conservation law for power flow can be easily derived for a
source-free region to be:
\begin{equation}
\nabla\cdot(\vec E\times\vec H)
=-\frac{1}{2}\frac{\partial}{\partial t} \left\{\mu|\vec
H|^2+\epsilon|\vec E|^2\right\} \nonumber
\end{equation}
or in integral form via the use of Gauss' divergence theorem:
\begin{equation}
\oint_S d\vec S\cdot(\vec E\times\vec H)
=-\frac{1}{2}\frac{\partial}{\partial t} \int_V dV \left\{\mu|\vec
H|^2+\epsilon|\vec E|^2\right\} \nonumber
\end{equation}
In the above, when LHM and a non-dispersive RHM, such as vacuum,
co-exist and mix with each other, $\epsilon$ and $\mu$ can have
different signs for different regions. Hence, with a proper choice
of surface and volume, the right-hand side of the above can be zero,
while the left-hand side is non-zero violating energy conservation!
However, if the medium is dispersive, as we have learned from
Chapter 1, the above is not the correct expression for energy
storage. Therefore, double negative materials can only be made over
a narrow bandwidth when it co-exists with regular materials.
Moreover, we still have to use $\vec E\times\vec H$ as the direction
of power flow if the DNG has been made from regular materials.
Consequently, the $\v k$ vector of a plane wave points in opposite
direction to that of power flow.

\begin{figure}
\begin{center}
\hfil\includegraphics[width=4.0truein]{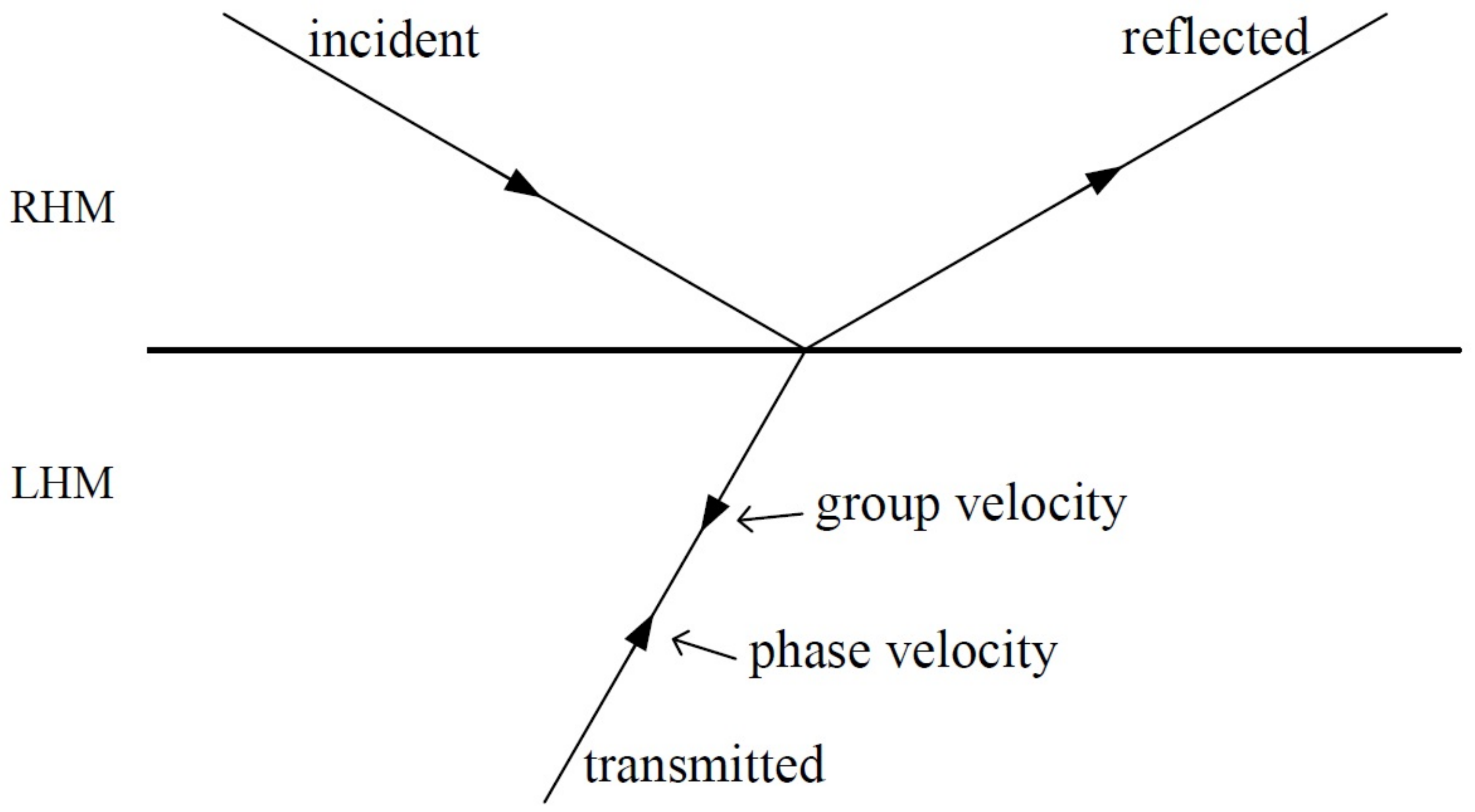}\hfil
\end{center}
\caption{Negative refraction happens at an interface because of the
need for phase matching and energy
conservation.}\label{NegativeRefraction}
\end{figure}

As a result, the group velocity of a DNG material has to be opposite
to that of the phase velocity. At a simple interface, due to phase
matching and energy conservation requirements, negative refraction
occurs as shown in Figure \ref{NegativeRefraction}. Phase matching
requires the $\v k$ vector to be aligned as shown, but energy
conservation requires that the group velocity be opposite to that of
the phase velocity.  Because of \index{Negative reflection} negative refraction, it has proposed
that negative refraction can be used for focusing \cite{VESALAGO}.

\subsection{Evanescent Amplification by a Matched DNG Slab}
\index{Matched DNG slab}
\index{Evanescent wave amplification}

A matched DNG slab has the capability of amplifying evanescent wave,
as shall be shown.  This has been suggested to use a DNG slab as a
super lens \cite{PENDRY}.

The transmission coefficient, as we have seen from the section on
Fabry-Perot etalon is given by
 \begin{align}\label{fbeq:2a}
 \tilde{T}=\frac{T_{12}T_{21}e^{ik_{2z}d}}{1-R_{21}^2e^{2ik_{2z}d}}
=\frac{(1-R_{21}^2)e^{ik_{2z}d}}{1-R_{21}^2e^{2ik_{2z}d}}
 \end{align}
with
\begin{equation}
R_{ij}^{TE}=\frac {\mu _jk_{iz}-\mu _ik_{jz}}{\mu _jk_{iz}+\mu
_ik_{jz}} , \qquad R_{ij}^{TM}=\frac {\epsilon _jk_{iz}-\epsilon
_ik_{jz}}{\epsilon_jk_{iz}+\epsilon _ik_{jz}}, \label{eq3-13a}
\end{equation}
depending on whether we are calculating $\tilde T$ for a TE wave or
a TM wave.  Also, $T_{ij}=1+R_{ij}$ in the above, and we have
assumed that the $z$ axis is normal to the slab.

For a matched medium, we pick $\mu_2=-\mu_1$ and
$\epsilon_2=-\epsilon_1$.  For propagating wave, we pick the phase
velocity in region 2 to be opposite to that of regions 1 and 3.  In
other words, $k_{2z}=-k_{1z}$.  Then it is quite clear that
$R_{12}=0$, and
\begin{align}\label{fbeq:2b}
 \tilde{T}=e^{-ik_{1z}h}
 \end{align}
In fact, it can be shown that the above result is independent of
whatever sign we pick for $k_{2z}$.  Because of the backward wave
nature of the wave in LHM, the above represents a phase advancement
instead of phase retardation in RHM.  The phase advancement
compensates for the phase retardation from the source to the first
slab interface, which can also be thought of as time reversal.

For the evanescent spectrum it can be shown that, irrespective of
the branch of square root for choose for $k_{2z}$, the transmission
coefficient is always
\begin{align}\label{fbeq:2c}
 \tilde{T}=e^{\alpha_{1z}h}
 \end{align}
where $\alpha_{1z}=\sqrt{k_1^2-k_x^2}$ where $k_x$ is the wavenumber
parallel to the slab. Hence, the evanescent wave has been amplified
after it has passed through the DNG slab. This can be used as a
super lens for \index{Super resolution phenomena} super resolution phenomena \cite{PENDRY}.

\begin{figure}
\begin{center}
\hfil\includegraphics[width=4.0truein]{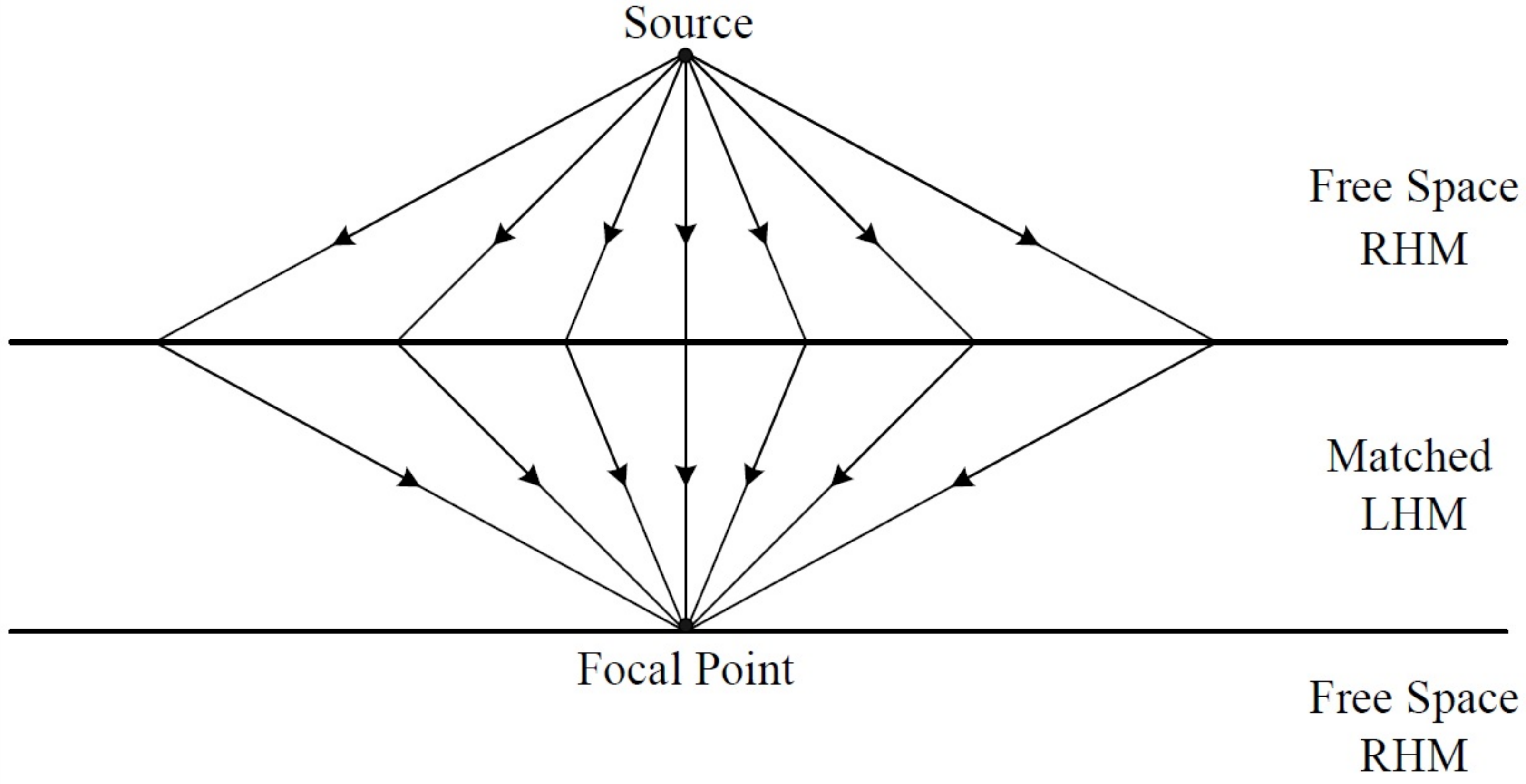}\hfil
\end{center}
\caption{Super lens or super resolution effect of a matched DNG
slab.  In the LHM material, the arrows show the direction of power
flow.  But the phase velocities are opposite to the directions of
the arrows.}\label{SuperLens}
\end{figure}

To understand this, we first look at the \index{Weyl identity} Weyl \cite{WFIM} identity
$$
\frac {e^{ik_0r}}r=\frac i{2\pi }\iint \limits _{-\infty
}^{\quad\infty }\,dk_xdk_y\,\frac {e^{ik_xx+ik_yy+ik_z|z|}}{k_z},
\label{Weyl}
$$
where $k_x^2+k_y^2+k_z^2=k_0^2$, or $k_z=(k_0^2-k_x^2-k_y^2)^{1/2}$.
The above says that the field produced by a point source consists of
both evanescent spectrum as well as the propagating spectrum.
However, it is the evanescent spectrum that contains the
high-resolution information of the point source.  As we move away
from the point source, the evanescent spectrum becomes smaller, and
therefore, the high-resolution information is lost.  However, if
this evanescent spectrum can be reconstituted by using the DNG slab,
then the high-resolution information can be regained.   This is
illustrated in Figure \ref{SuperLens}. However, this reconstitution
of the evanescent spectrum is a highly unstable undertaking and is
easily upset by loss or geometry imperfection
\cite{CHEW_DNG,ISHIMARU}.

\subsection {Composite Right-Left Handed Transmission Line}
\index{Transmission line!composite right-left handed}

     \begin{figure}[!htb]
     \centering
     \includegraphics[width=10cm]{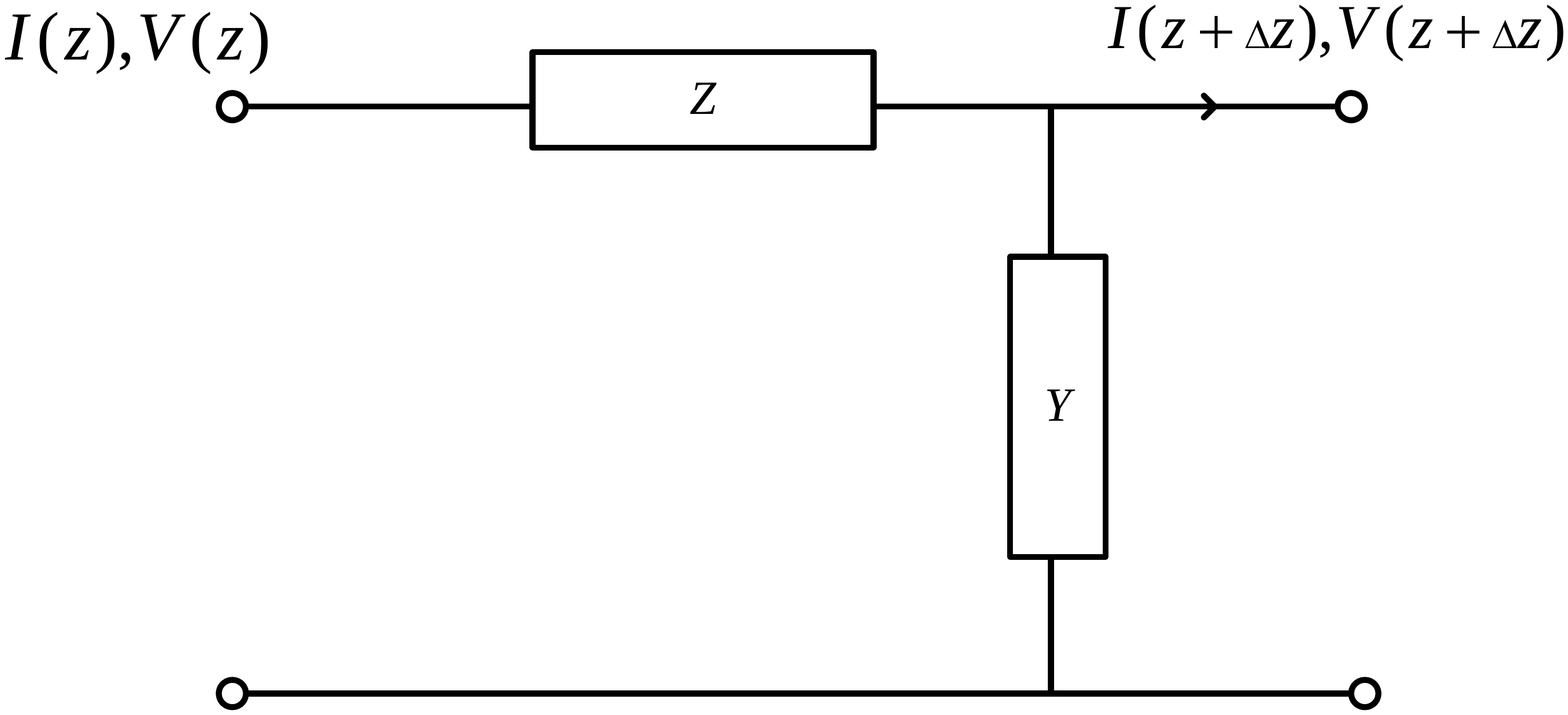}
     \caption{Lumped element model for deriving the CRLH transmission line
     equations.}
     \label{FigureP11}
     \end{figure}

One interesting technology that has emerged in this area is the
composite right-left handed (CRLH) transmission line
\cite{ITOH5,ITOH5a}. We will analyze this transmission line next
using circuit theory. For the case shown in Figure \ref{FigureP11},
we can write down Kirchoff voltage law to get
\begin{equation} \label{eq:1}
V(z+\Delta z)-V(z)=-I(z)Z\Delta z
\end{equation}
\begin{equation} \label{eq:2}
I(z+\Delta z)-I(z)=-V(z+\Delta z)Y\Delta z
\end{equation}
where $Z$ and $Y$ are per unit length impedance and admittance. The
above becomes the Telegraphers equations when $\Delta z\rightarrow
0$
\begin{equation} \label{eq:3}
\frac{\partial V(z)}{\partial z}=-ZI(z)
\end{equation}
\begin{equation} \label{eq:4}
\frac{\partial I(z)}{\partial z}=-YV(z)
\end{equation}
or
\begin{equation} \label{eq:5}
\frac{\partial^{2}V(z)}{\partial z^{2}}=ZYV(z)
\end{equation}

\begin{equation} \label{eq:6}
\frac{\partial^{2}I(z)}{\partial z^{2}}=ZYV(z)
\end{equation}
or
\begin{equation} \label{eq:7}
\beta^{2}=-YZ
\end{equation}
\begin{equation} \label{eq:8}
\beta=\pm j\sqrt{YZ}
\end{equation}
If
\begin{equation} \label{eq:9}
Z=\frac{1}{j\omega C_{L}}
\end{equation}
\begin{equation} \label{eq:10}
Y=\frac{1}{j\omega L_{L}}
\end{equation}
then above imitates a double negative material.  We can pick one
branch of the square root to get
\begin{equation} \label{eq:11}
\beta=-\frac{1}{\omega\sqrt{L_{L}C_{L}}}<0
\end{equation}
    \begin{equation} \label{eq:12}
         Z_{0}=\sqrt\frac{Z}{Y}=\sqrt\frac{L_{L}}{C_{L}}>0
    \end{equation}
    \begin{equation} \label{eq:13}
         v_{ph}=\frac{\omega}{\beta}=-{\omega}^2\sqrt{L_{L}C_{L}}<0
    \end{equation}
    \begin{equation} \label{eq:14}
         v_{g}=\frac{d\omega}{d\beta}=-\frac{\omega}{k}={\omega}^2\sqrt{L_{L}C_{L}}>0
    \end{equation}
    The above indicates that the group velocity is opposite to the phase
    velocity: it supports a backward wave.

     \begin{figure}[!htb]
     \centering
     \includegraphics[width=4in]{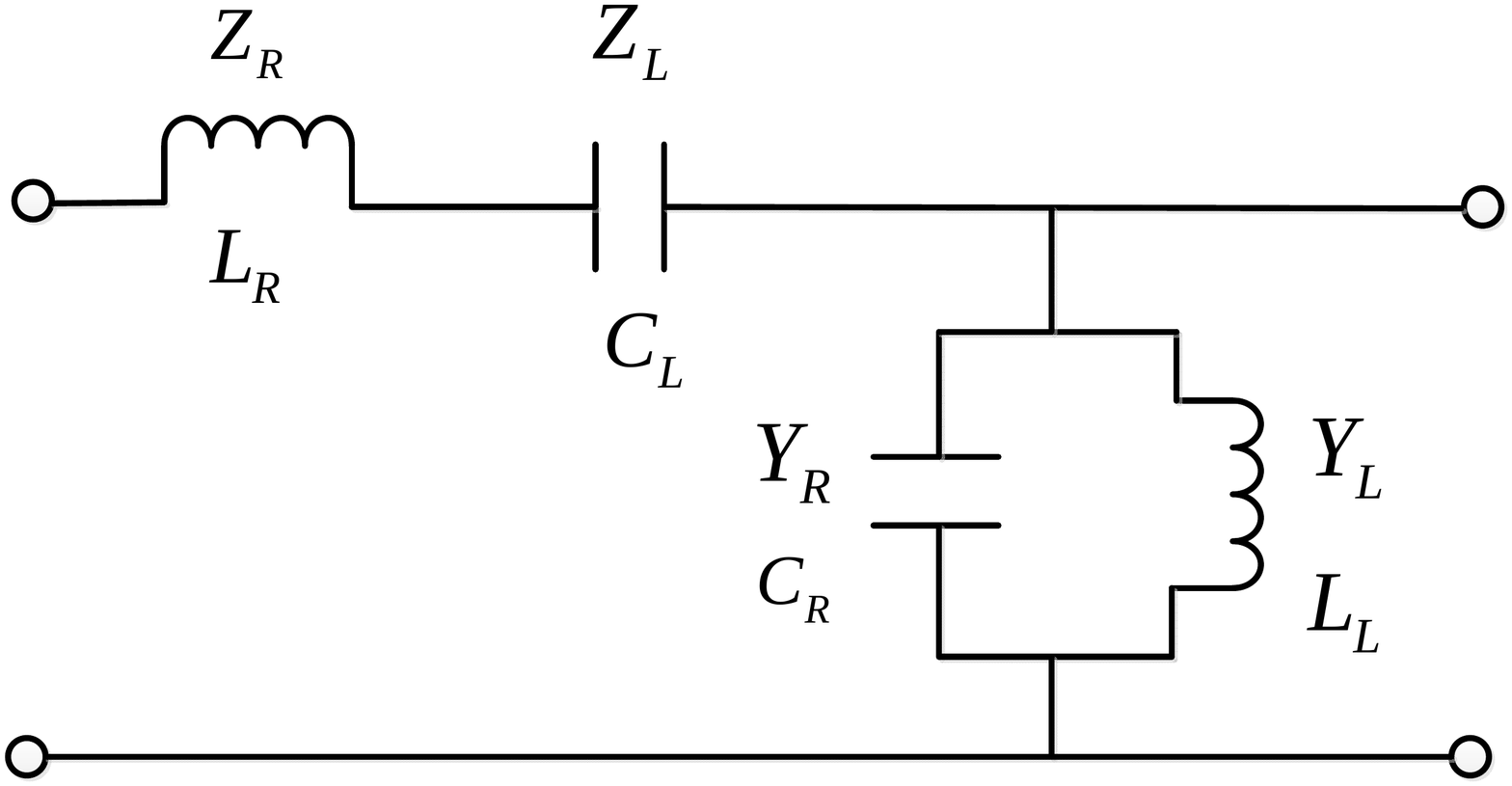}
     \caption{The lumped element model for a CRLH transmission
     line.}
     \label{P21}
     \end{figure}

For a CRLH transmission line, we have
    \begin{equation} \label{eq:15}
         Z=j\left({\omega}L_{R}-\frac{1}{{\omega}C_{L}}\right)
    \end{equation}
    \begin{equation} \label{eq:16}
         Y=j\left({\omega}C_{R}-\frac{1}{{\omega}L_{L}}\right)
    \end{equation}
    Then if ${\omega}>\omega_{se}=\frac{1}{\sqrt{L_{R}C_{L}}}$,
    ${\omega}>\omega_{sh}=\frac{1}{\sqrt{L_{L}C_{R}}}$, where
    $\omega_{se}$ and $\omega_{sh}$ are the resonant frequencies of
    the series and shunt impedances respectively, then
    \begin{equation} \label{eq:18}
         \beta=\sqrt{\left({\omega}L_{R}-\frac{1}{{\omega}C_{L}}\right)
         \left({\omega}C_{R}-\frac{1}{{\omega}L_{L}}\right)}>0
    \end{equation}
    If ${\omega}<\omega_{se}$, ${\omega}<\omega_{sh}$, then
\begin{equation} \label{eq:17}
         \beta=-\sqrt{\left(\frac{1}{{\omega}C_{L}}-{\omega}L_{R}\right)
         \left(\frac{1}{{\omega}L_{L}}-{\omega}C_{R}\right)}
    \end{equation}
and $\beta<0$.
       A band gap exists for $\min\left(\omega_{se},\omega_{sh}
       \right)<\omega<\max\left(\omega_{se},\omega_{sh}\right)$.
    If we make $\omega_{se}=\omega_{sh}$, the band gap disappears.

At $\omega=\omega_0>0$, there exists a wave with $\beta=0$, or a
constant phase wave. At this frequency, the series resonance becomes
a short and the shunt resonance becomes an open. This idea can be
used to design equi-phase loop antenna for RFID (radio frequency
identification) and MRI (magnetic resonance imaging) applications.
When $\beta<\omega/c$, the wave that propagates on the CRLH line
leaks energy to the space around it: the structure can be used to
make leaky-wave antennas.

    \begin{figure}
    \begin{center}
    \includegraphics[width=3.5in]{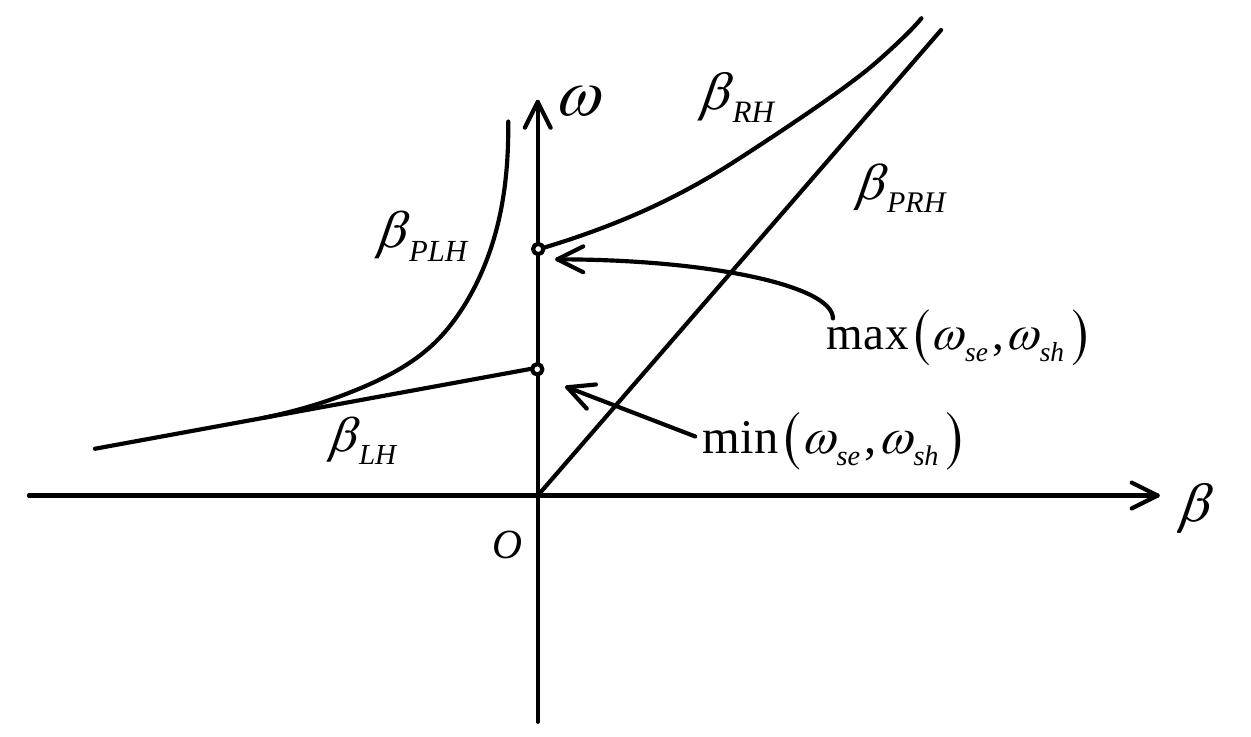}
    \label{P3_1}
    \end{center}
    \caption{The $\omega$-$\beta$ diagrams of the pure RH (PRH) line, the
    pure LH (PLH) line, and that of the CRLH line.  For CRLH, and band gap
    exists where the frequency $\omega$ has no solution for $\beta$
    or a propagating mode.}
    \end{figure}

    \begin{figure}
    \begin{center}
    \includegraphics[width=3.in]{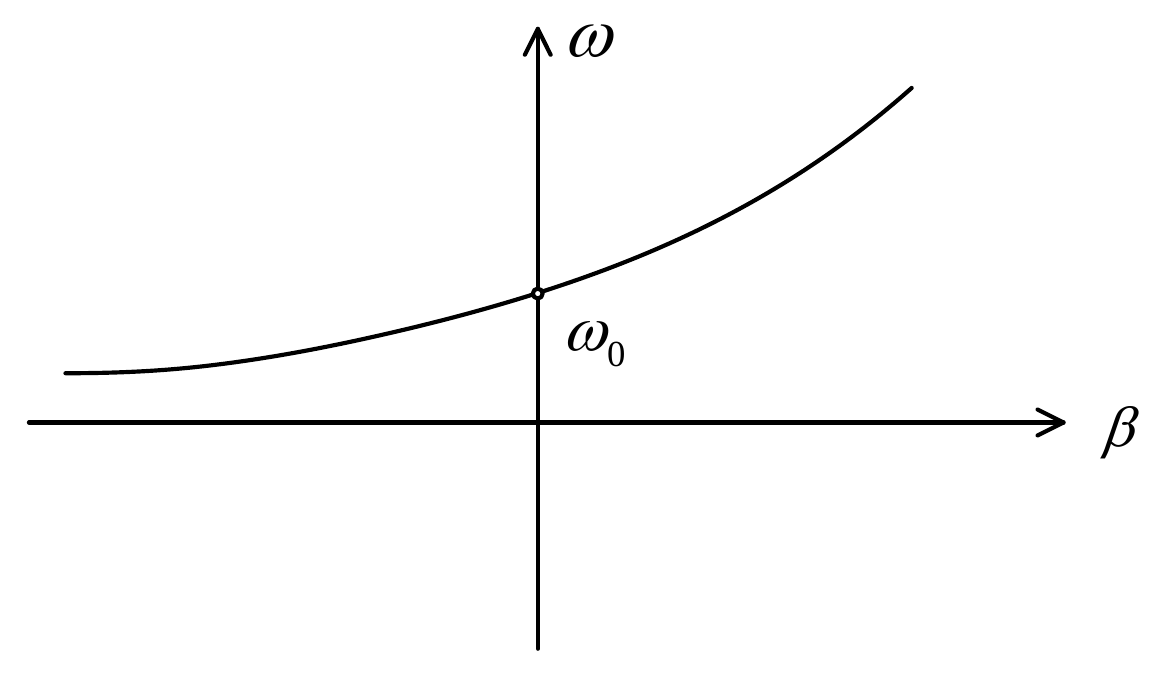}
    \caption{The choice of $L_R$, $L_L$, $C_R$, and $C_L$ where
    $\omega_{se} =  \omega_{sh}$, and the band gap disappears.
    }
    \label{P3_2}
    \end{center}
    \end{figure}


\vfil\eject

\centerline{\bf Exercises for Chapter 5 }

\vskip 12pt \noindent {\bf Problem 5-1:} \noindent A waveguide
junction is formed from two rectangular waveguides, one of which
is 2 cm by 4 cm, and the other is 2.5 cm by 4 cm. Assume that a 5
GHz TE$_{10}$ mode is incident on the waveguide junction from the
smaller waveguide. Use the transmission line model, ascertain the
approximate amplitudes of the reflection and transmission
coefficients for the TE$_{10}$ mode.


%
%
%
%
%
%

\vskip 8pt \noindent {\bf Problem 5-2:} To understand the number of
modes required for convergence in the mode-matching problem in a
waveguide, consider the matching of the following two Fourier
series:
\begin{align}
f_1(x)&=\sum_{n=-P_1}^{P_1}a_n e^{in\pi x/d_1},\qquad 0<x<d_1,\\
    &=0,\qquad \text{otherwise},\\
\hat f_1(x)&=\sum_{n'=-P_2}^{P_2}b_n' e^{in'\pi x/d_2}=
\begin{cases}
f_1(x), \qquad &0<x<d_1,\\
0, \qquad &d_1<x<d_2 ,
\end{cases}
\end{align}
where $d_2>d_1$.  Find the coefficents $b_n'$ in terms of $a_n$.
Show that in order for $\hat f_1(x)$ to approximate $f_1(x)$ well,
we require that $P_2/d_2>> P_1/d_1$.

\vskip 8pt \noindent {\bf Problem 5-3:}

\begin{figure}
\begin{center}
\hfil\includegraphics[width=4.5truein]{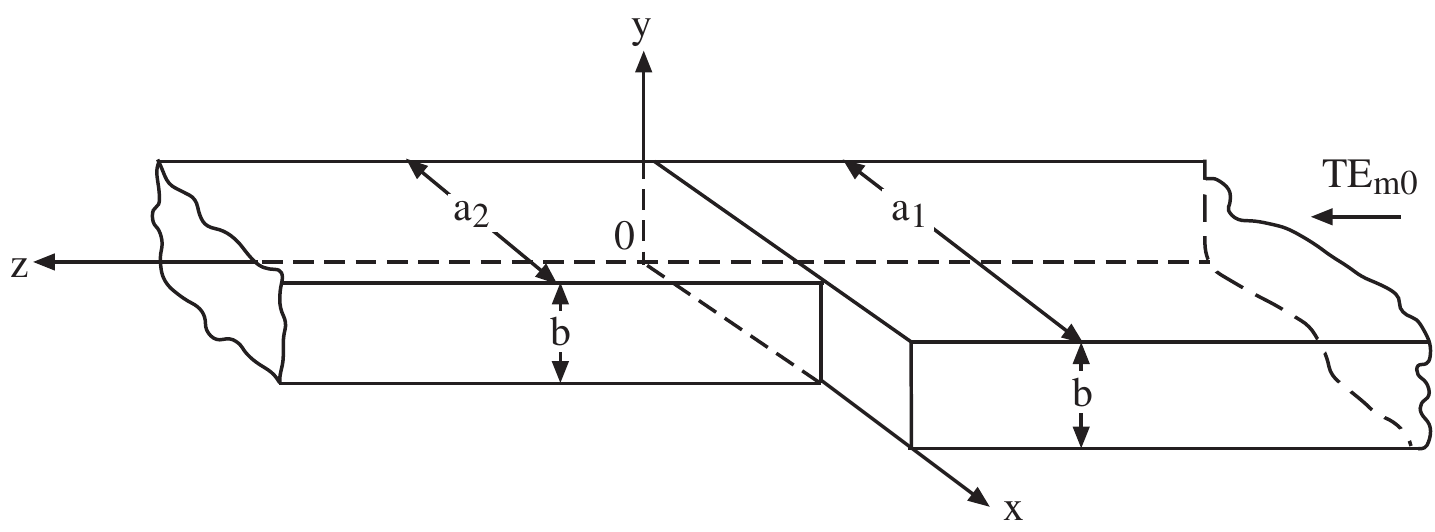}\hfil
\end{center}
\caption{Problem 5-3}\label{p5-3}
\end{figure}


If a sum of TE$_{m0}$ modes are incident at the discontinuity in the
 waveguide in Figure \ref{p5-3}, due to the symmetry of the problem, only TE$_{m0}$
modes are reflected and transmitted. Given that the incident modes
are described by
\begin{equation*}
H_z = \sum_m H_{m0} \cos\left(\frac{m\pi x}{a_1}\right)
e^{ik_{mz}z}
\end{equation*}
find the transmission and reflection operators that describe mode
conversions at the discontinuity.

\vskip 8pt \noindent {\bf Problem 5-4:}

\begin{figure}
\begin{center}
\hfil\includegraphics[width=4.0truein]{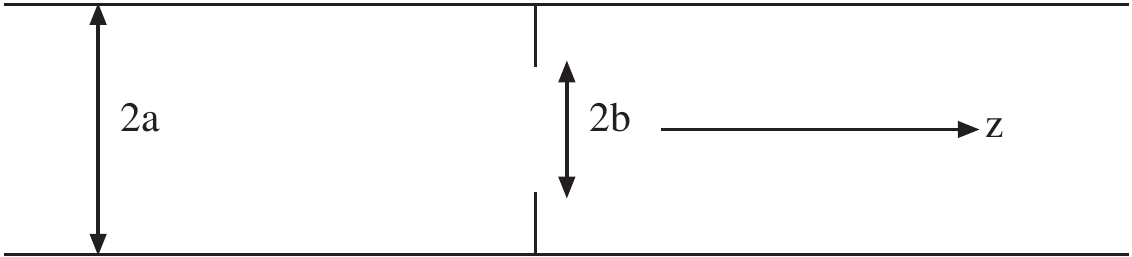}\hfil
\end{center}
\caption{Problem 5-4}\label{P5-4}
\end{figure}

%

A TE$_{11}$ mode is propagating in a circular waveguide with radius
$a$ as shown in Figure \ref{P5-4}. A circular diaphragm with radius
$b$ is placed at $z=0$.
\begin{itemize}
\item[(a)] Find the reflection and transmission operators due to the presence of this diaphragm.
\item[(b)] Give explicit expressions for the elements of the matrices involved in the description of the reflection and transmission operators.
\end{itemize}

\noindent {\bf Problem 5-5:}

\begin{figure}
\begin{center}
\hfil\includegraphics[width=4.0truein]{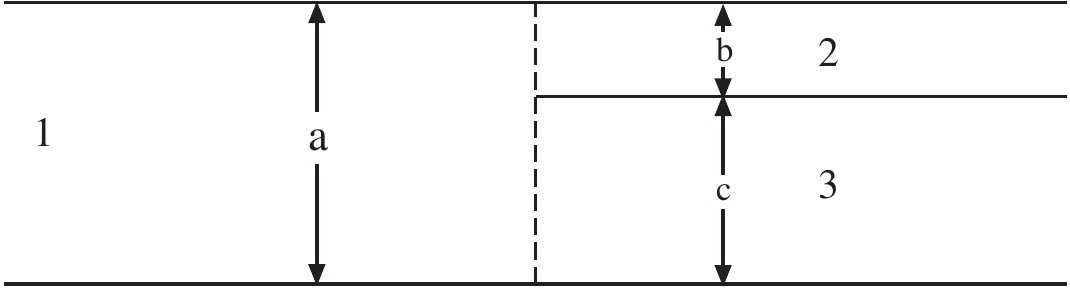}\hfil
\end{center}
\caption{Problem 5-5}\label{P5-5}
\end{figure}


A parallel plate waveguide is bifurcated as shown in Figure
\ref{P5-5}. Assume a bunch of TM$_n$ mode incident from the left of
the waveguide.
\begin{itemize}
\item[(a)] Using the formulation of this Chapter, find the reflection operator in waveguide 1, and transmission operators in waveguides 2 and 3.

The field at the waveguide discontinuity can either be
approximated by the modes of waveguide 1, or the modes of
waveguides 2 and 3.
\item[(b)] If the field at the discontinuity is approximated by $N$ modes of waveguide 1, how many reflected modes are there in waveguide 1, and how many transmitted modes are there in waveguides 2 and 3.
\item[(c)] If the field at the discontinuity is approximated by $M$ modes in waveguide 2 and $P$ modes in waveguide 3, how many reflected modes are there in waveguide 1, and how many transmitted modes are there in waveguides 2 and 3.  In this latter method of solving the problem, what should be the relative ratio of $M$ and $P$ in order to maintain the same accuracy for the transmitted field in waveguides 2 and 3. (Hint:  If only 1 mode is assumed in waveguide 2 while 10 modes are assumed in waveguide 3 at the discontinuity, the accuracy of the transmitted field in waveguide 1 will not be as accurate as that in waveguide 2.  So, a certain ratio needs be maintained to achieve the same order of accuracy. Think of this in terms of Fourier series expansions.)
\end{itemize}

\noindent {\bf Problem 5-6:}

\begin{figure}
\begin{center}
\hfil\includegraphics[width=4.5truein]{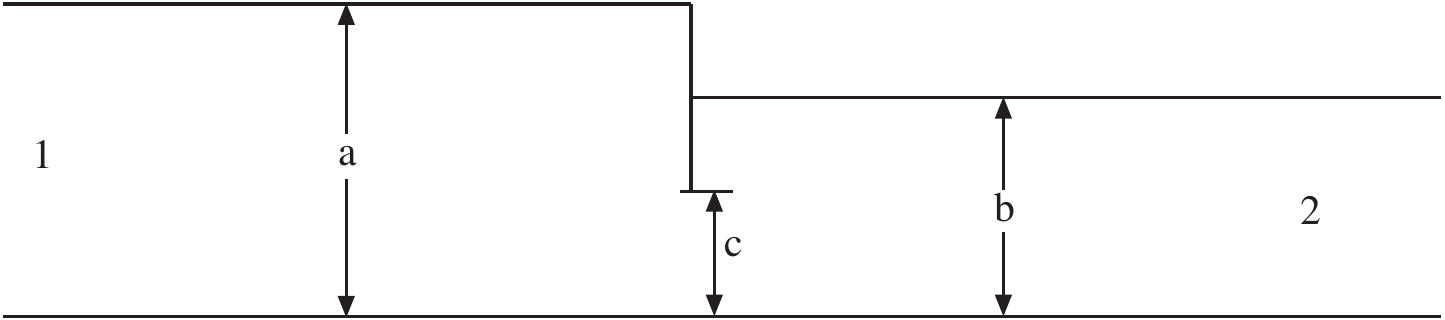}\hfil
\end{center}
\caption{Problem 5-6}\label{P5-6}
\end{figure}


A parallel plate waveguide is shown in Figure \ref{P5-6}. Assume a
TM$_1$ mode incident from the left of the waveguide.
\begin{itemize}
\item[(a)] Using the formulation of this Chapter, find the reflection operator in waveguide 1, and transmission operator in waveguide 2.
\item[(b)] If the field at the discontinuity is approximated by $N$ modes of a waveguide with the same dimension as the aperture, how many reflected modes are there in waveguide 1, and how many transmitted modes are there in waveguide 2 in theory, in order to obtain a numerically accurate solution.
\end{itemize}

\noindent {\bf Problem 5-7:} Show that a typical integral in
Equation (\ref{eq5-2-33}) looks like the expression in Equation
(\ref{eq5-2-34a}).

\vskip 8pt \noindent {\bf Problem 5-8:} Expand (\ref{eq5-5-10}) in a
geometrical series, and give a physical explanation of each term of
the geometrical series.  Would this series always converge? What can
you say about the norm of the reflection matrices?

\vskip 8pt \noindent {\bf Problem 5-9:} Explain how you would
construct a time reversed solution to Maxwell's equations once you
have found a time dependent solution. Is the time reversed
solution for a lossy waveguide a physical solution?

\vskip 8pt \noindent {\bf Problem 5-10:} Show that when the
Neumann boundary condition is satisfied for $H_x$ in the
corrugated periodic waveguide, then it is equivalent to tangential
electric field being zero on the surface of the waveguide wall.

\vskip 8pt \noindent {\bf Problem 5-11:} Evaluate Equation
(\ref{eq5-7-23}) and hence Equation (\ref{eq5-7-26}) in closeform.
Write a computer program to solve for different values of $k_z$
for different $k$.  Discuss how you would truncate the series
involved in the equations.

\vskip 8pt \noindent {\bf Problem 5-12:} Derive Equations
(\ref{eq5-7-32}) and (\ref{eq5-7-33}).

\vskip 8pt \noindent {\bf Problem 5-13:} Write a computer program
to plot $kd$ versus $k_zd$ according to Equation (\ref{eq5-7-41}).

\vskip 8pt \noindent {\bf Problem 5-14:} Derive the guidance
condition for a transmission line that is periodically loaded with
a shunt capacitance. Plot $kd$ versus $k_zd$ for this circuit.

\bibliographystyle{plain}
\bibliography{emt,matrix,fmm,math,fastsum,parallel,cs,my}

\def\chaptitle{Optical Waveguides}
\index{Waveguide!optical}

\chapter{\chaptitle}

\markboth{\smallbooktitle}{\chaptitle}


\def\v #1{{\bf #1}}
\def\dyad#1{\overline {\bf #1}}
\def\vg #1{\mbox{\boldmath$#1$}}
\def\beq{\begin{equation}}\def\eeq{\end{equation}}
\def\tinf{\text{\it inf\,}}\def\^{\hat}
\def\cal#1{\mathcal{#1}}







\setcounter{equation}{0} \setcounter{figure}{0}

Optical waveguides are some of the most important waveguides.
Their importance stems from the broad bandwidth and low loss
deliverable by optical communication systems
\cite{SENIOR,AGRAWAL}. As a result, optical fiber cables have
replaced transmission line cables as submarine cables throughout
the world for global communication \cite{RUNGE}.  The use of
optical fiber for communication was first proposed by Kao and
Hockham \cite{KAO&HOCKHAM}.  The success of Corning Glass Works on
making fiber of loss below 20 dB/km in the 1970s spurred
tremendous interest in the use of optical fiber.  An interesting
account of the history is given in Okoshi \cite{OKOSHID}.

\index{Optical fiber}

Optical fibers work by the physics of total internal reflection.
Waves are confined within a waveguide by total internal reflection
due to the presence of a dielectric interface. Hence, no metallic
part is needed in the construction of such a waveguide.  The
presence of metallic component is deleterious at high frequencies
due to the loss it incurs.   Optical fibers now can have a loss as
low as 0.2 dB/km \cite{AGRAWAL}.

In an optical waveguide, which usually is an open waveguide, a wave
is guided along a structure, but the field is not enclosed
completely in the structure --- the field extends to infinity.
However, outside the waveguide, the field is evanescent, an it
decays exponentially away from the guiding structure; hence, the
energy of the wave is still localized around the guiding structure.
The guiding structure is often filled with inhomogeneous medium.
Therefore, many properties of inhomogeneously filled waveguides are
also true in open, optical waveguides.

Due to the importance of optical waveguides, there has been a
tremendous amount of work on this subject.  Much of the work can
be found from the references for this chapter and the references
therein.  This chapter only serves to provide a sampling of some
topics available in the vast literature.  Optical waveguides are
still under intensive research by many workers (see reference
list).  A recent development is the use of photonic crystals for
optical waveguides.  Artificial photonic crystals can generate
\index{Stop band} stop bands (band gaps) in which the wave has to be evanescent in
the crystal. The band gap structures are then used to trap waves
inside the waveguide.  An excellent overview of planar lightwave
circuits is given by Okamoto \cite{OKAMOTO}.

\section {{{ Surface Waveguides--Dielectric Slab Wave\-guides}}}
\index{Waveguide!surface}
\index{Waveguide!dielectric slab}

An example of an open waveguide is a dielectric slab waveguide.
The waveguide is made with dielectric coating on a ground plane,
or in the case of optical thin film waveguides, it is a coating of
an optically more dense medium on top of an optically less dense
substrate
\cite{MARCUSE0,MARCUSE,HAUSE,SNYDER&LOVE,YARIVD,BELANGER,OKOSHID,CHUANG,OKAMOTO,KONGF}.

Due to the symmetry of the geometry, we can decompose the field
inside such a waveguide into TM and TE types. The mode is guided
by total internal reflection. This is only possible if $\epsilon
_1>\epsilon _0$ and $\epsilon _1>\epsilon _2$. At total internal
reflection, the fields in region 0 and 2 are evanescent, and hence
they decay exponentially away from the structure. Therefore, most
of the energy of the mode is still trapped and localized in the
vicinity of the structure.

\begin{figure}[ht]
\vspace{-1.2in}
\begin{center}
\hfil\includegraphics[width=3.6truein]{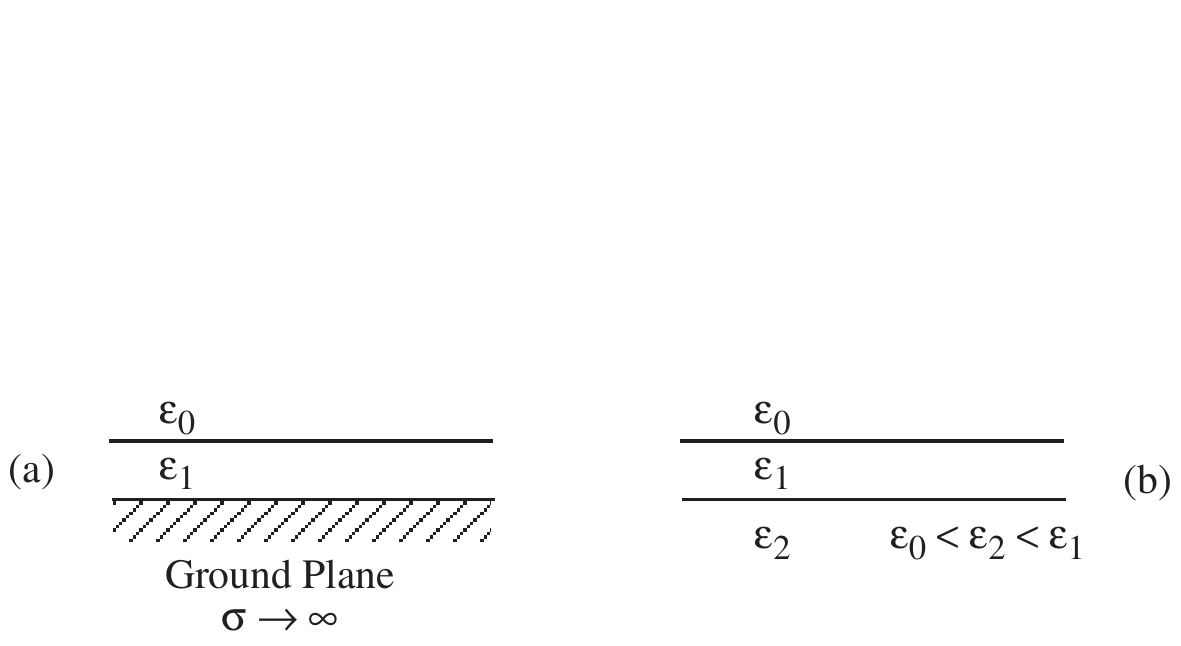}\hfil
\end{center}
\caption{Geometry for dielectric slab waveguides.}\label{fg611}
\end{figure}


\begin{figure}[ht]
\begin{center}
\hfil\includegraphics[width=4.5truein]{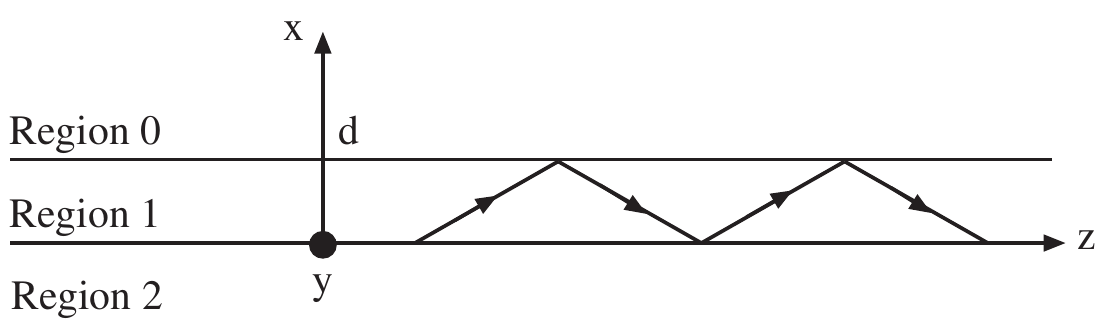}\hfil
\end{center}
\caption{Bouncing waves in a dielectric slab waveguide.}\label{fg612}
\end{figure}


If a TM wave is in a dielectric slab, we can write the field in
region 1 as
\begin{equation}
\v H_1=\^ y\left[A_1e^{ik_{1x}x}+B_1e^{-ik_{1x}x}\right]e^{ik_zz}.
\label{eq6-1}
\end{equation}
Equation (\ref{eq6-1}) has the physical meaning that the wave in
region 1 is representable as bouncing waves. At $x = 0$, the
upgoing wave is the reflection of the downgoing wave; hence, we
have
\begin{equation}
A_1=R_{12}^{TM}B_1, \label{eq6-2}
\end{equation}
where $R_{12}^{TM}$ is the TM reflection coefficient at the 1-2
interface. If there are subsurface layers, $R_{12}^{TM}$ could be
the generalized reflection coefficient that includes subsurface
reflections. Otherwise, it is just the single interface, Fresnel
reflection coefficient for a TM wave. At the upper interface at $x
= d$, we require that the downgoing wave is a reflection of the
upgoing wave, i.e.,
\begin{equation}
B_1e^{-ik_{1x}d}=R_{10}^{TM}e^{ik_{1x}d}A_1. \label{eq6-3}
\end{equation}
For non-trivial $A_1$ and $B_1$, Equations (\ref{eq6-2}) and
(\ref{eq6-3}) imply that
\begin{equation}
1-R_{10}^{TM}R_{12}^{TM}e^{2ik_{1x}d}=0. \label{eq6-4}
\end{equation}
The above is the guidance condition sometimes known as the
\index{Transverse resonance condition} transverse resonance condition for TM modes in a \index{Waveguide!dielectric slab} dielectric slab,
with
\begin{equation}
R_{ij}^{TM}=\frac {\epsilon _jk_{ix}-\epsilon _ik_{jx}}{\epsilon
_jk_{ix}+\epsilon _ik_{jx}}, \qquad k_{ix}=\sqrt {k_i^2-k_z^2}.
\label{eq6-5}
\end{equation}

\begin{figure}[ht]
\begin{center}
\hfil\includegraphics[width=4.truein]{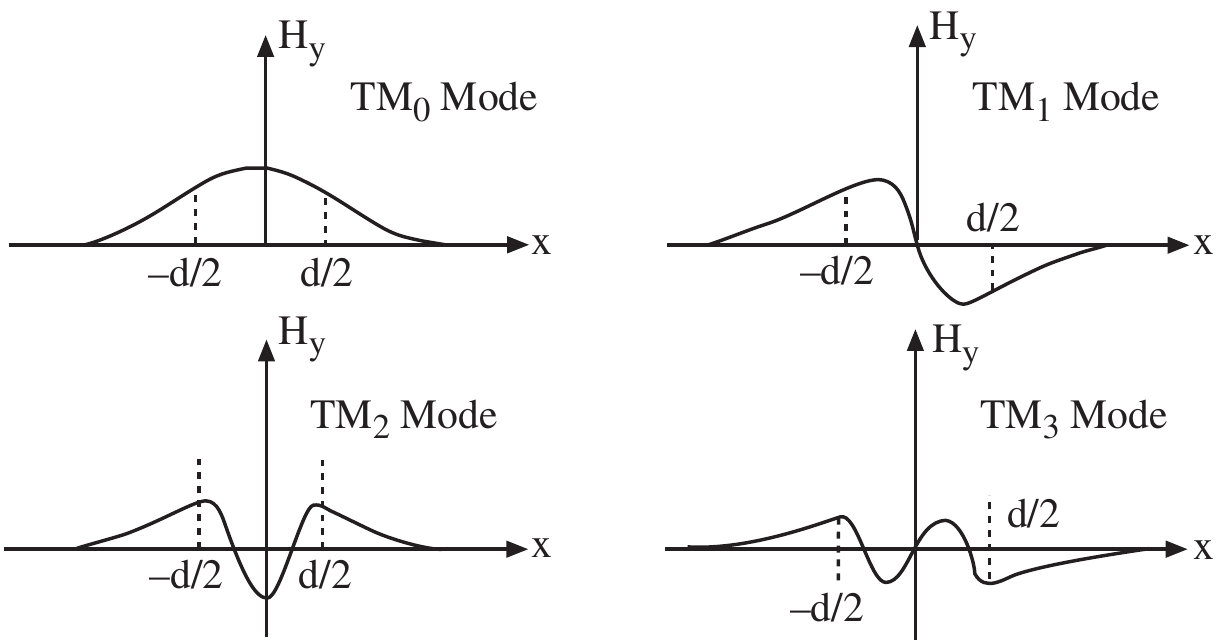}\hfil
\end{center}
\caption{Field distributions for different modes in a dielectric
slab waveguide.}\label{fg613}
\end{figure}


Due to Equation (\ref{eq6-5}), the guidance condition
(\ref{eq6-4}) can be expressed entirely as a function of $k_z$. We
can either solve Equation (\ref{eq6-4}) graphically or numerically
on a computer. Once the value of $k_z$ that satisfies
(\ref{eq6-4}) is found, it can be used in (\ref{eq6-2}) or
(\ref{eq6-3}) to find a relationship between $A_1$ and $B_1$. The
fields in region 0 and 2 can be found easily, i.e.,
\begin{subequations}
\begin{equation}
\v H_0=\^ yT_{10}^{TM}A_1e^{ik_{1x}d+ik_{0x}(x-d)+ik_zz},
\label{eq6-6a}
\end{equation}
\begin{equation}
\v H_2=\^ yT_{12}^{TM}B_1e^{-ik_{2x}x+ik_zz}. \label{eq6-6b}
\end{equation}
\end{subequations}
In other words, the field in region 0 is a consequence of the
transmission of the upgoing wave in region 1, while the field in
region 2 is a consequence of the transmission of the downgoing
wave in region 1. In the above, $T_{ij}$ is a transmission
coefficient with
\begin{equation}
T_{ij}^{TM}=1+R_{ij}^{TM}. \label{eq6-7}
\end{equation}
In order for a mode to be trapped, the field has to decay
exponentially in the $x$ direction. Therefore, $k_{0x}$ and
$k_{2x}$ have to be pure imaginary. In other words, we can find
the values of $k_z$ for guidance only in the range $k_z > k_0$ and
$k_z > k_2$.

\begin{figure}[ht]
\begin{center}
\hfil\includegraphics[width=3.5truein]{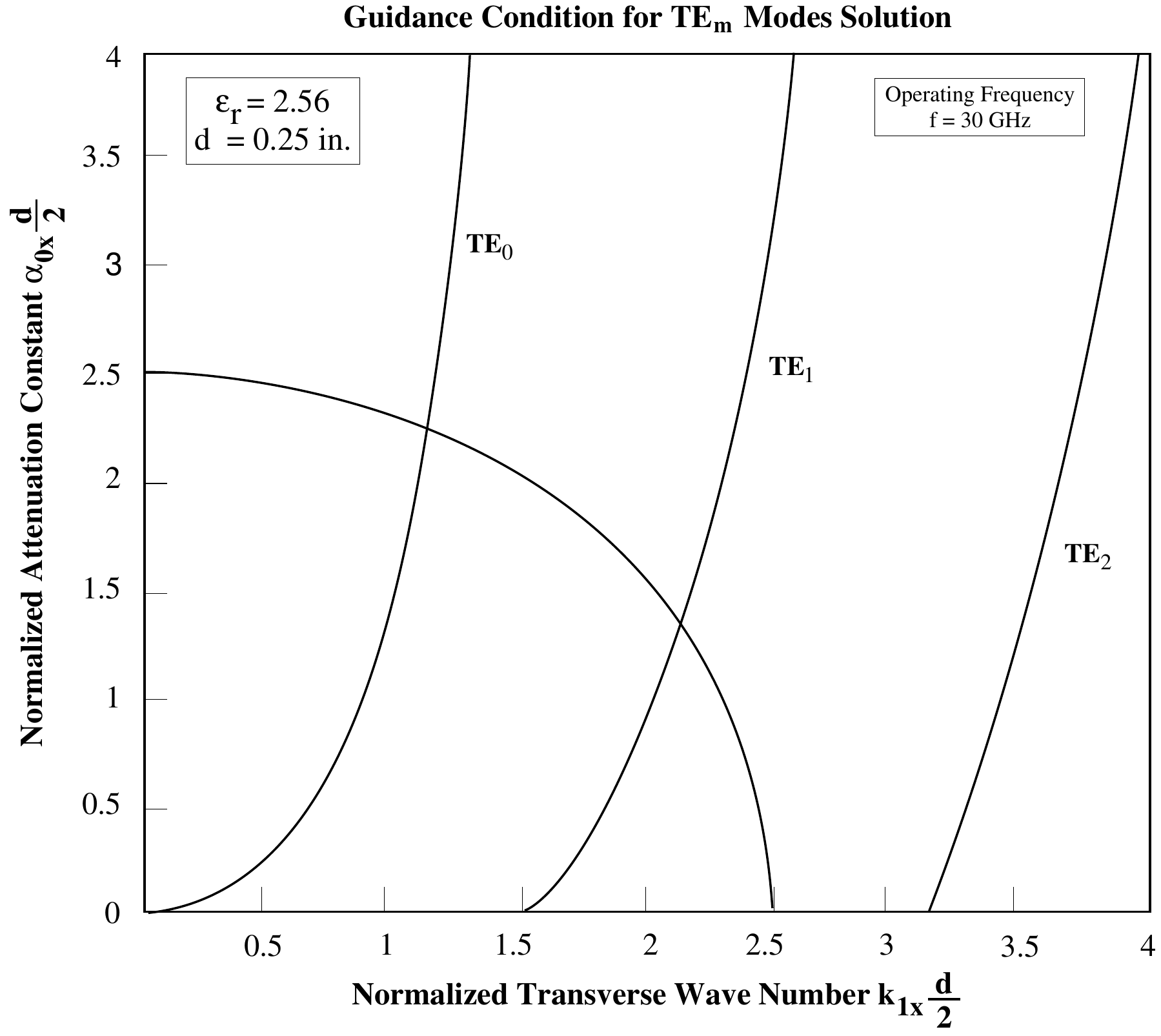}\hfil
\end{center}
\caption{Graphical solution of the guided TM modes of a dielectric
slab waveguide.}\label{fg614a}
\end{figure}

\begin{figure}[ht]
\begin{center}
\hfil\includegraphics[width=3.5truein]{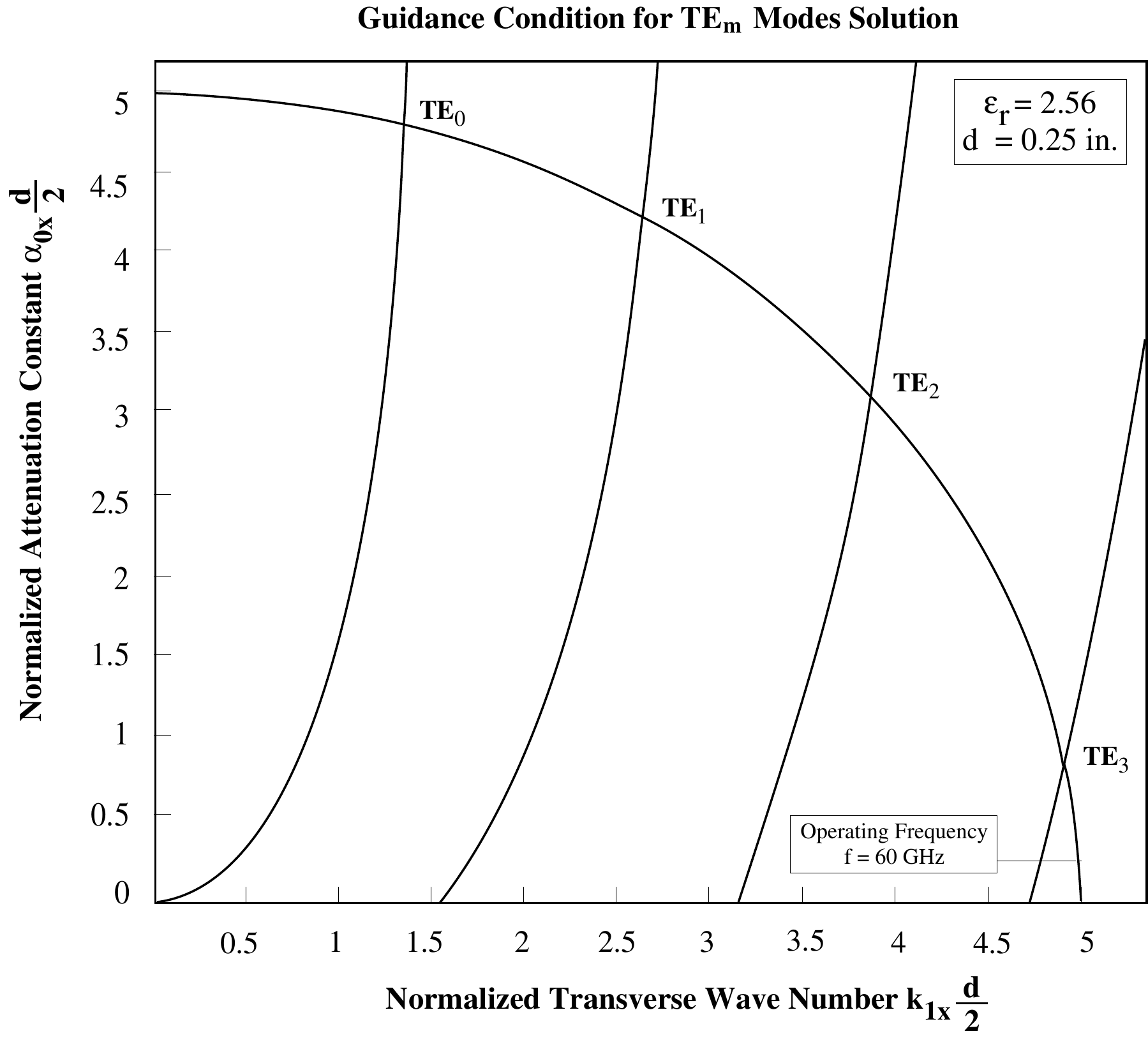}\hfil
\end{center}
\caption{Graphical solution of the guided TE modes of a dielectric
slab waveguide.}\label{fg614b}
\end{figure}


For a symmetric waveguide where regions 0 and 2 are the same, the
first few modes of the waveguides are as sketched in Figure
\ref{fg613}. These modes are, in general, more well-trapped when the
refractive index in region 1 is higher than those in regions 0 and
2, or when the frequency is high.
We can study the guidance of a TE mode which is the dual of a TM
mode in a similar fashion.

By a further manipulation of Equation (\ref{eq6-4}), and using the
definition of Fresnel reflection coefficients, Equation
(\ref{eq6-4}) can be written as
\begin{equation}
\alpha _{0x} \frac d 2 =\frac {\epsilon _0}{\epsilon _1 } k_{1x }
\frac d 2 \tan \left( \frac {k_{1x}d-m\pi }{2} \right),
\label{eq6-8}
\end{equation}
where $\alpha _{0x}=\sqrt {k^2_z -k^2_0}$,\, $k_{1x}= \sqrt {k^2_1
-k^2_z}$, and $m=1,2,3,4,\cdots $. The left-hand side of
(\ref{eq6-8}) can be expressed in terms of the $k_{1x} d $
variable, viz.,
\begin{equation}
\sqrt {k^2_z -k^2_0}\frac d 2= \sqrt {(k^2_1 -k^2_0) {\left( \frac
d 2 \right)} ^2- \left( \frac {k_{1x}d }{2} \right)^2 },
\label{eq6-9}
\end{equation}
which is the equation of a circle. Equation (\ref{eq6-8}) can hence
be solved graphically by plotting both sides of Equation
(\ref{eq6-8}) as a function of $k_{1z} d$.  The TE mode case can be
obtained by duality.  The graphical-solution plots are shown in
Figures \ref{fg614a} and \ref{fg614b}.

\begin{figure}[ht]
\begin{center}
\hfil\includegraphics[width=3.5truein]{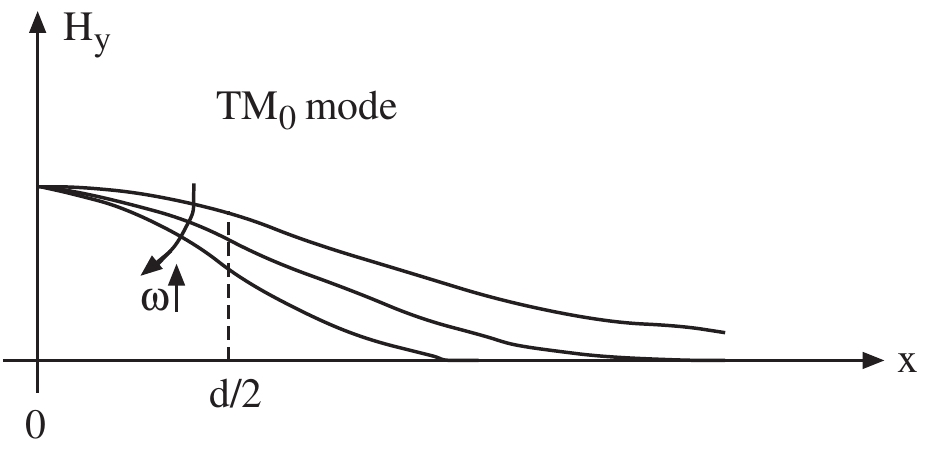}\hfil
\end{center}
\caption{For increasing frequency, a mode is better trapped inside
the dielectric slab. Only the field for $x>0$ is sketched in the
above.}\label{fg615}
\end{figure}

\begin{figure}[ht]
\begin{center}
\hfil\includegraphics[width=3.5truein]{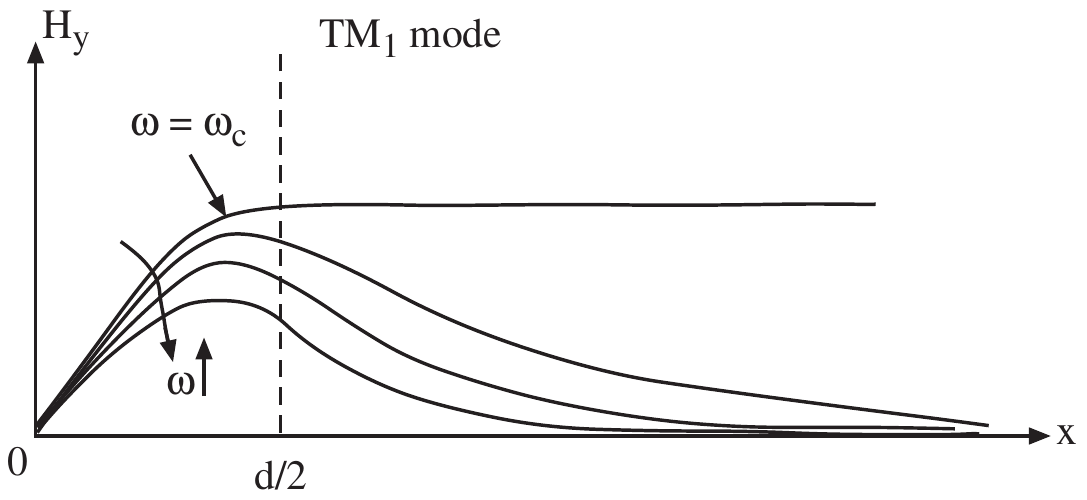}\hfil
\end{center}
\caption{The profile of the TM$_1$ mode for different frequencies.
When $\omega=\omega _c$, the cutoff frequency, the mode ceases to be
evanescent outside the waveguide.}\label{fg616}
\end{figure}


From Figure \ref{fg615}, it is clear that the TM$_0$ mode has no
cut-off, since continuity of the slope and field amplitude can be
satisfied for all frequencies and yet the field is evanescent
outside. This is not true of the higher order modes. Figure
\ref{fg616} shows the profile of the TM$_1$ mode for different
frequencies.

\begin{figure}[ht]
\begin{center}
\hfil\includegraphics[width=4.0truein]{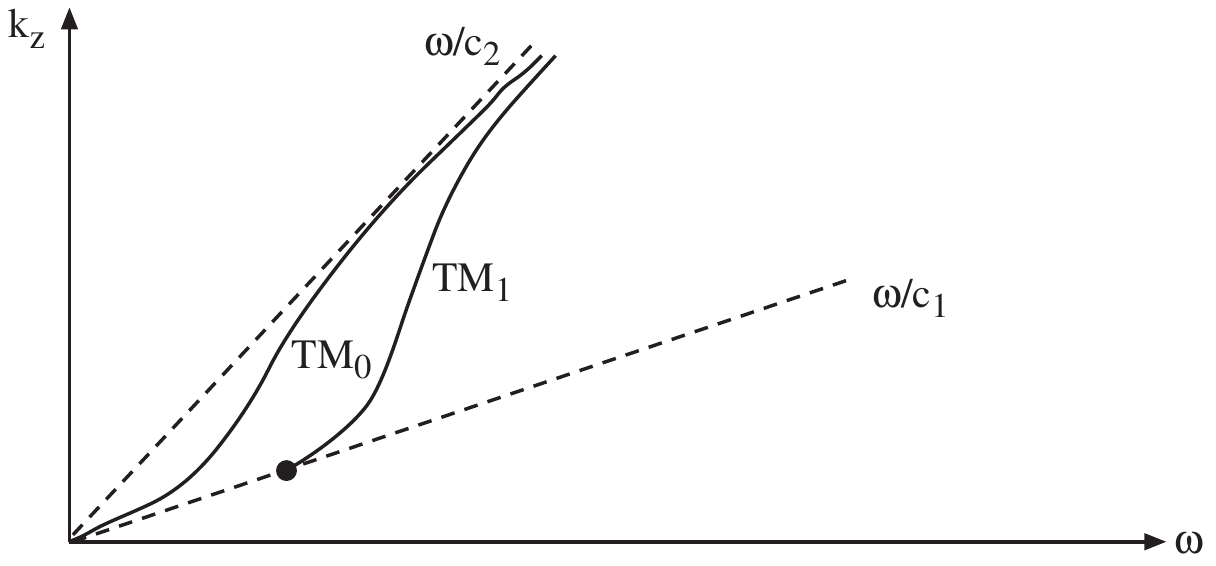}\hfil
\end{center}
\caption{Dispersion of the modes in a dielectric slab waveguide.}\label{fg617}
\end{figure}

A mode in a dielectric slab waveguide has part of its energy outside
the waveguide and inside the waveguide as shown in Figures
\ref{fg615} and \ref{fg616}. Hence, the \index{Velocity!group} group velocity of the mode
is between that of the slab region and the outer region.  For the
TM$_0$ mode, when the frequency is very low, the mode is weakly
evanescent outside the dielectric slab and the group velocity of a
mode is closer to that of the outer region because most of the
energy of the mode is outside the waveguide. When the frequency is
high, the mode is strongly evanescent outside and most of the energy
of the mode is trapped inside the slab. Hence, the group velocity of
the mode is close to that of the dielectric slab. Therefore, the
dispersion curve of the TM$_0$ mode is as shown in Figure
\ref{fg617}.

To elaborate on the TM$_1$ mode, for high frequencies, the mode is
\index{Mode!well-trapped} well-trapped inside the dielectric waveguide. When the frequency is
low such that $\lambda /4=d/2$, the field ceases to be evanescent
outside the dielectric slab. In this case, the field is a constant
outside, and this is the mode at the cut-off frequency precisely.
When $\lambda /4 > d/2$, the mode leaks energy to outside the slab,
it becomes a leaky mode and is not guided at all. Figure \ref{fg617}
shows the dispersion curve of the TM$_1$ mode in a dielectric slab
waveguide. Close to cut-off, the group velocity of the mode should
be close to that of the outside region, and vice-versa for high
frequencies. Figure \ref{fg616} allows us to determine the cut-off
frequency of the TM$_1$ mode quite readily.


\section {{{ Circular Dielectric Waveguide}}}

\begin{figure}[ht]
\begin{center}
\hfil\includegraphics[width=3.5truein]{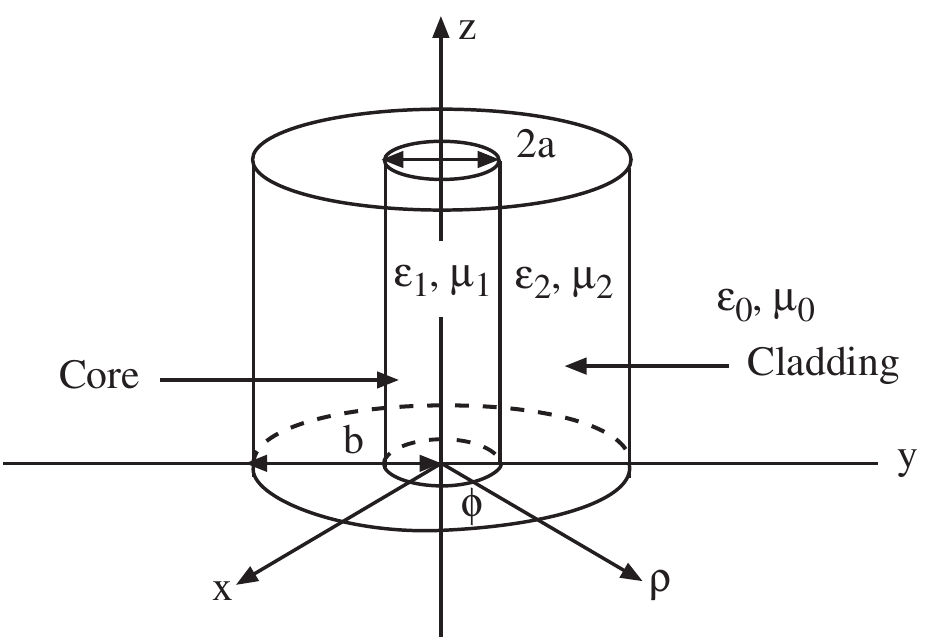}\hfil
\end{center}
\caption{Geometry of an optical fiber---a circular dielectric
waveguide.}\label{fg621}
\end{figure}

An optical fiber is a \index{Waveguide!circular dielectric} circular dielectric waveguide as shown in
Figure \ref{fg621}. Usually, an optical fiber has a protective
cladding as well. In the ensuing analysis, we will assume that the
protective cladding is infinitely thick, i.e., letting
$b\rightarrow\infty$ \cite{SNITZER,SNYDER,DIL&BLOK,YARIVD}.

In order for a mode to be guided, we require that $\epsilon _1\mu
_1>\epsilon _2\mu _2$. In other words, the light velocity in the
core region has to be slower than the light velocity in the
cladding. The field outside the core region is evanescent for a
guided mode. Therefore, the value of $b$ affects the guided mode
little when $b$ is large, especially if the mode is tightly bound to
the core; hence, $b\rightarrow\infty$ is a good approximation.


In the optical fiber modes, except for the axisymmetric modes, the
TE and TM fields are coupled to each other by the boundary
conditions as in an inhomogeneously filled waveguide. The $z$
components of the field characterize the TE and TM fields in the
fiber. Moreover, they are decoupled in a homogeneous region, and
 are solutions to the wave equation in
cylindrical coordinates for each of the homogeneous regions, i.e.,
\begin{equation}
\left[ \frac 1{\rho }\frac {\partial }{\partial\rho }\rho\frac
{\partial }{\partial \rho}+\frac 1{\rho ^2}\frac {\partial
^2}{\partial \phi ^2}+\frac {\partial ^2}{\partial z^2}+\omega ^2\mu
_j\epsilon _j\right]
\begin{bmatrix}
E_{jz}\\
H_{jz}
\end{bmatrix} =0.
\label{eq6-2-1}
\end{equation}
The solution to (\ref{eq6-2-1}) in region $j$ is of the general form
\begin{equation}
\left\{
\begin{matrix}
J_n  (k_{j\rho }\rho)\\
H_n^{(1)} (k_{j\rho }\rho)
\end{matrix}
\right\} e^{\pm in\phi +ik_zz}, \label{eq6-2-2}
\end{equation}
where $k_{j\rho }^2+k_z^2 =k_j^2$ and $k_j$ is the wave number in
region $j$.  Here, $H_n^{(1)}(k_{\rho }\rho )$ is singular when
$\rho\rightarrow 0$ while $J_n(k_{\rho }\rho )$ is regular.
Therefore, for the $e^{in\phi }$ harmonic, the solution in the core
region is
\begin{subequations}
\begin{equation}
E_{1z}=E_1J_n(k_{1\rho }\rho )e^{ik_zz+in\phi }, \label{eq6-2-3a}
\end{equation}
\begin{equation}
H_{1z}=H_1J_n(k_{1\rho }\rho )e^{ik_zz+in\phi }. \label{eq6-2-3b}
\end{equation}
\end{subequations}
In the cladding region, when $b\rightarrow\infty$, we should only
have outgoing waves. Therefore,
\begin{subequations}
\begin{equation}
E_{2z}=E_2H_n^{(1)}(k_{2\rho }\rho )e^{ik_zz+in\phi },
\label{eq6-2-4a}
\end{equation}
\begin{equation}
H_{2z}=H_2H_n^{(1)}(k_{2\rho }\rho )e^{ik_zz+in\phi }.
\label{eq6-2-4b}
\end{equation}
\end{subequations}
By the phase matching condition, $k_z$ is the same in both regions.
The choice of Hankel functions of the first kind in (\ref{eq6-2-4a})
and (\ref{eq6-2-4b}) stems from the fact that Hankel functions
represent outgoing waves.  The asymptotic expansions of the Hankel
and Bessel functions are
\begin{subequations}
\begin{equation}
H_n^{(1)}(k_{\rho }\rho )\sim\sqrt {\frac 2{\pi k_{\rho }\rho
}}e^{ik_{\rho }\rho -in\frac {\pi }2-i\frac {\pi }4}, \qquad
\rho\rightarrow\infty, \label{eq6-2-5a}
\end{equation}
\begin{equation}
J_n(k_{\rho }\rho )\sim\sqrt {\frac 2{\pi k_{\rho }\rho }}\cos
(k_{\rho }\rho -\frac {n\pi }2-\frac {\pi }4), \qquad
\rho\rightarrow\infty. \label{eq6-2-5b}
\end{equation}
\end{subequations}
Equation (\ref{eq6-2-5a}) corresponds to an outgoing wave while
(\ref{eq6-2-5b}) corresponds to a standing wave.  When a mode is
confined in the core, $k_{2\rho } = \sqrt {k_2^2- k_z^2} = i\alpha
_2$ is always positive imaginary.  Therefore, guidance is only
possible if $k_z>k_2$.

At this point, we introduce the modified Bessel function $K_n(x)$
such that
\begin{equation}
H_n^{(1)}(i\alpha _2\rho )=\frac 2{\pi i}e^{-i\frac {\pi
n}2}K_n(\alpha _2\rho ). \label{eq6-2-6}
\end{equation}
As such, Equations (\ref{eq6-2-4a}) and (\ref{eq6-2-4b}) can be
written as
\begin{subequations}
\begin{equation}
E_{2z}=E'_2K_n(\alpha _2\rho )e^{ik_zz+in\phi }, \label{eq6-2-7a}
\end{equation}
\begin{equation}
H_{2z}=H'_2K_n(\alpha _2\rho )e^{ik_zz+in\phi }, \label{eq6-2-7b}
\end{equation}
\end{subequations}
where $\alpha _2=\sqrt {k_z^2-k_2^2}$.  When
$\rho\rightarrow\infty$,
\begin{equation}
K_n(\alpha _2\rho )\sim\sqrt {\frac {\pi }{2\alpha _2\rho
}}e^{-\alpha _2\rho }, \label{eq6-2-8}
\end{equation}
corresponding to an evanescent wave. A guided mode must satisfy
the boundary conditions at the core-cladding interface. Therefore,
we need to find $E_{\phi }$ and $H_{\phi }$, the other tangential
components of the field.

In the core region, they are
\begin{subequations}
\begin{equation}
E_{1\phi }=\frac 1{k_{1\rho }^2}\left[ -\frac {nk_z}{\rho
}E_1J_n(k_{1\rho }\rho )-i\omega\mu _1H_1k_{1\rho }J'_n(k_{1\rho
}\rho )\right] e^{in\phi +ik_zz}, \label{eq6-2-9a}
\end{equation}
\begin{equation}
H_{1\phi }=\frac 1{k_{1\rho }^2}\left[ -\frac {nk_z}{\rho
}H_1J_n(k_{1\rho }\rho )+i\omega\epsilon _1E_1k_{1\rho
}J'_n(k_{1\rho }\rho )\right] e^{in\phi +ik_zz}. \label{eq6-2-9b}
\end{equation}
\end{subequations}
In the cladding region, they are
\begin{subequations}
\begin{equation}
E_{2\phi }=-\frac 1{\alpha _2^2}\left[ -\frac {nk_z}{\rho
}E'_2K_n(\alpha _2\rho)-i\omega\mu _2H'_2\alpha _2K'_n(\alpha
_2\rho )\right] e^{in\phi +ik_zz}, \label{eq6-2-10a}
\end{equation}
\begin{equation}
H_{2\phi }=-\frac 1{\alpha _2^2}\left[ -\frac {nk_z}{\rho
}H'_2K_n(\alpha _2\rho)+i\omega\epsilon _2E'_2\alpha _2K'_n(\alpha
_2\rho )\right] e^{in\phi +ik_zz}. \label{eq6-2-10b}
\end{equation}
\end{subequations}
The continuity of the tangential components of the fields from
(\ref{eq6-2-3a}), (\ref{eq6-2-3b}), (\ref{eq6-2-7a}),
(\ref{eq6-2-7b}), (\ref{eq6-2-9a}), (\ref{eq6-2-9b}), and
(\ref{eq6-2-10a}), (\ref{eq6-2-10b}) implies that
\begin{subequations}
\begin{equation}
E_1J_n(k_{1\rho }a)=E'_2K_n(\alpha _2a), \label{eq6-2-11a}
\end{equation}
\begin{equation}
H_1J_n(k_{1\rho }a)=H'_2K_n(\alpha _2a), \label{eq6-2-11b}
\end{equation}
\begin{equation}
\begin{split}
\frac 1{k_{1\rho }^2} & \left[ -\frac {nk_z}a E_1J_n(k_{1\rho
}a)-i\omega\mu
_1H_1k_{1\rho }J'_n(k_{1\rho }a)\right] \\
& =-\frac 1{\alpha _2^2}\left[ -\frac {nk_z}aE'_2K_n(\alpha
_2a)-i\omega\mu _2H'_2\alpha _2K'_n(\alpha _2a)\right],
\label{eq6-2-11c}
\end{split}
\end{equation}
\begin{equation}
\begin{split}
\frac 1{k_{1\rho }^2} & \left[ -\frac {nk_z}a H_1J_n(k_{1\rho }a)+i\omega\epsilon_1E_1k_{1\rho }J'_n(k_{1\rho }a)\right] \\
& =-\frac 1{\alpha _2^2}\left[ -\frac {nk_z}aH'_2K_n(\alpha
_2a)+i\omega\epsilon_2E'_2\alpha _2K'_n(\alpha _2a)\right].
\label{eq6-2-11d}
\end{split}
\end{equation}
\end{subequations}
Substituting (\ref{eq6-2-11a}) and (\ref{eq6-2-11b}) into
(\ref{eq6-2-11c}) and (\ref{eq6-2-11d}) to eliminate the unknowns
$E_1$ and $H_1$, there are only two remaining unknowns $E_2'$ and
$H_2'$. We can form a $2\times 2$ matrix equation with no driving
term.  For non-trivial solution, we set the determinant of the
resultant matrix to zero to arrive at
\begin{equation}
\begin{split}
k_z^2n^2 & \left( \frac 1{k_{1\rho }^2a^2}+\frac 1{\alpha
_2^2a^2}\right) ^2\\ & =\omega ^2\left[ \frac {\mu _1J'_n(k_{1\rho
}a)}{k_{1\rho }aJ_n(k_{1\rho
}a)}+ \frac {\mu _2K'_n(\alpha _2a)}{\alpha _2aK_n(\alpha _2a)}\right]\\
&\qquad \left[ \frac {\epsilon _1J'_n(k_{1\rho }a)}{k_{1\rho
}aJ_n(k_{1\rho }a)}+\frac {\epsilon _2K'_n(\alpha _2a)}{\alpha
_2aK_n(\alpha _2 a)}\right]. \label{eq6-2-12}
\end{split}
\end{equation}
Since $k_{1\rho }=\sqrt {k_1^2-k_z^2}$, $\alpha _2=\sqrt
{k_z^2-k_2^2}$, we can solve the above transcendental equation for
$k_z$.  Once the values of $k_z$ that satisfies (\ref{eq6-2-12})
are found, we can find the ratios of ${E_1}/{H_1}$ from
(\ref{eq6-2-11a})--(\ref{eq6-2-11d}). In particular,
\begin{equation}
\frac {E_1}{H_1}=\frac {nk_z}{i\omega }\left( \frac 1{k_{1\rho
}^2a^2}+\frac 1{\alpha _2^2a^2}\right)\left( \frac {\epsilon _1
J'_n(k_{1\rho }a)}{k_{1\rho }aJ_n(k_{1\rho }a)}+\frac {\epsilon
_2K'_n(\alpha _2a)}{\alpha_2aK_n(\alpha _2a)}\right) ^{-1}.
\label{eq6-2-13}
\end{equation}
This is the ratio of the \index{Mode!TM vs TE amplitude ratio} TM wave amplitude to the TE wave
amplitude inside the core.

In Equation (\ref{eq6-2-12}), $\frac {K'_n(x)}{xK_n(x)}$ is not
rapidly oscillating while $\frac {J'_n(x)}{xJ_n(x)}$ is rapidly
oscillating.  Moreover, it is a quadratic equation in terms of
$\frac {J'_n(x)}{xJ_n(x)}$.
We can solve Equation (\ref{eq6-2-12}) for $\frac {J'_n(k_{1\rho
}a)}{k_{1\rho }aJ_n(k_{1\rho }a)}$ giving
\begin{equation}
\begin{split}
\frac {J'_n(k_{1\rho }a)}{k_{1\rho }aJ_n(k_{1\rho }a)}= & -\frac
1{2}\left( \frac{\mu _2}{\mu _1}+\frac {\epsilon _2}{\epsilon
_1}\right) \frac {K'_n(\alpha
_2a)}{\alpha _2aK_n(\alpha _2a)}\\
& \pm\left[ \frac 1{4}\left(\frac {\mu _2}{\mu _1}-\frac {\epsilon
_2}{\epsilon _1}\right) ^2 \left( \frac {K'_n(\alpha
_2a)}{\alpha _2aK_n(\alpha _2a)}\right) ^2\right.\\
& \left. +\frac {n^2k_z^2}{k_1^2}\left( \frac 1{k_{1\rho
}^2a^2}+\frac 1{\alpha _2^2a^2}\right) ^2\right] ^{\frac 1{2}}.
\label{eq6-2-14}
\end{split}
\end{equation}
The plus and minus signs give rise to two classes of solutions. We
next make use of the recurrence relationship of Bessel functions
\begin{subequations}
\begin{equation}
 {J'_n(x)}=-J_{n+1}(x)+\frac n{x}J_n(x),
\label{eq6-2-15a}
\end{equation}
\begin{equation}
 {J'_n(x)}=J_{n-1}(x)-\frac n{x}J_n(x),
\label{eq6-2-15b}
\end{equation}
\end{subequations}
to get
\begin{subequations}
\begin{equation}
\begin{split}
\frac {J_{n+1}(k_{1\rho }a)}{k_{1\rho }aJ_n(k_{1\rho }a)} & =\frac
1{2}\left( \frac {\mu _2}{\mu _1}+\frac {\epsilon _2}{\epsilon
_1}\right) \frac
{K'_n(\alpha _2a)}{\alpha _2aK_n(\alpha _2a)}\\
& +\left( \frac n{(k_{1\rho }a)^2}-R\right) ,\qquad \text
{EH}\quad\text{modes}, \label{eq6-2-16a}
\end{split}
\end{equation}
\begin{equation}
\begin{split}
\frac {J_{n-1}(k_{1\rho }a)}{k_{1\rho }aJ_n(k_{1\rho }a)}= &
-\frac 1{2}\left( \frac {\mu _2}{\mu _1}+\frac {\epsilon
_2}{\epsilon _1}\right) \frac
{K'_n(\alpha _2a)}{\alpha _2aK_n(\alpha _2a)}\\
& +\left( \frac n{(k_{1\rho }a)^2}-R\right) ,\qquad \text
{HE}\quad\text{modes}, \label{eq6-2-16b}
\end{split}
\end{equation}
where
\begin{equation}
R=\left[ \frac 1{4}\left( \frac {\mu _2}{\mu _1}-\frac {\epsilon
_2}{\epsilon _1}\right) ^2\left( \frac {K'_n(\alpha _2a)}{\alpha
_2aK_n (\alpha _2a)}\right) ^2+\frac {n^2k_z^2}{k_1^2}\left( \frac
1{k_{1\rho }^2a^2}+\frac 1{\alpha _2^2a^2}\right) ^2\right]
^{\frac 1{2}}. \label{eq6-2-16c}
\end{equation}
\end{subequations}

\begin{figure}[ht]
\begin{center}
\hfil\includegraphics[width=3.5truein]{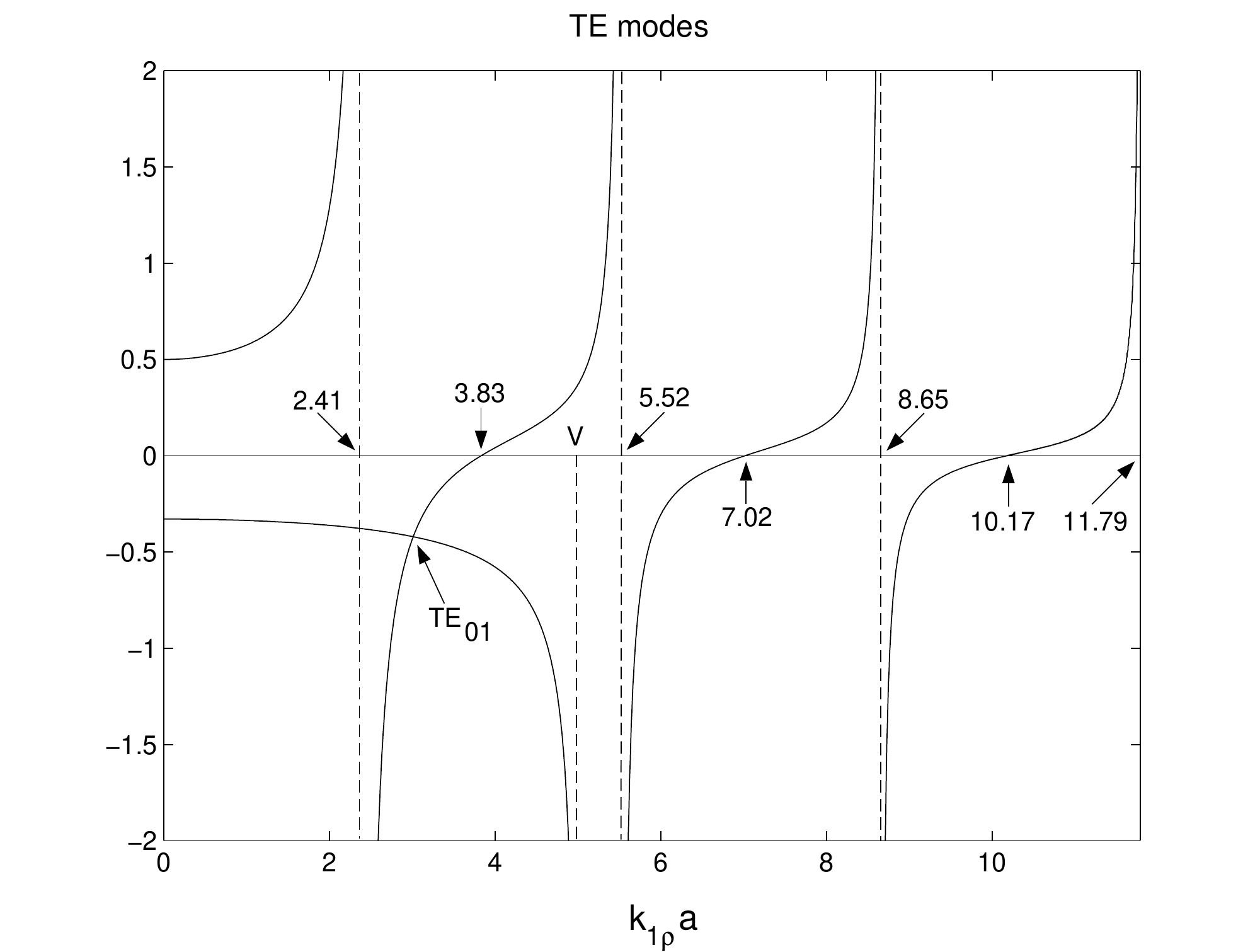}\hfil
\end{center}
\caption{Graphical solution of the axially symmetric TE modes.}\label{fg622}
\end{figure}


For the $n = 0$ case, or the axisymmetric case, the TE and TM
fields are decoupled, and the TE and TM modes are the dual of each
other. The guidance conditions are
\begin{subequations}
\begin{equation}
\frac {J_1 (k_{1\rho }a)}{k_{1\rho }aJ_0(k_{1\rho }a)}=- \frac
{\epsilon _2}{\epsilon _1}\frac {K_1(\alpha _2a)}{\alpha
_2aK_0(\alpha _2a)}, \qquad \text {TM}, \label{eq6-2-17a}
\end{equation}
\begin{equation}
\frac {J_1 (k_{1\rho }a)}{k_{1\rho }aJ_0(k_{1\rho }a)}=- \frac
{\mu _2}{\mu _1}\frac {K_1(\alpha _2a)}{\alpha _2aK_0(\alpha
_2a)}, \qquad \text {TE}. \label{eq6-2-17b}
\end{equation}
\end{subequations}
In the above, we have used $K'_0(x) = - K_1(x)$. Note that TE and TM
waves are decoupled from (\ref{eq6-2-11a})--(\ref{eq6-2-11c}), since
the TE and TM waves can satisfy the boundary conditions separately.
By writing
\begin{equation}
(\alpha _2a)^2=(k_1a)^2-(k_2a)^2-(k_{1\rho }a)^2=V^2-(k_{1\rho
}a)^2, \label{eq6-2-18}
\end{equation}
(\ref{eq6-2-17a}) can be solved in terms of $k_{1\rho }a$. To get
a feeling for the behavior of the solution, it can be solved
graphically. Since $\alpha _2a > 0$ for a guided mode, from
(\ref{eq6-2-18}), note that we need only to consider the case
where
\begin{equation}
0<k_{1\rho }^2 a^2<(k_1^2-k_2^2)a^2=V^2. \label{eq6-2-19}
\end{equation}
The right-hand side of (\ref{eq6-2-17a}) is always negative. For
the TE case, it has a value of ${-\mu _2K_1(V)}/{\mu _1VK_0(V)}$
at $k_{1\rho }a = 0$, and it has a value of $-\infty$, when
$k_{1\rho }a\rightarrow V$. More precisely,
\begin{equation}
-\frac {\mu _2K_1(\alpha _2a)}{\mu _1\alpha _2aK_0(\alpha
_2a)}\sim\frac {2\mu _2}{\mu _1(V^2-k_{1\rho }^2a^2)\ln
(V^2-k_{1\rho }^2a^2)}, \qquad k_{1\rho }a\rightarrow V.
\label{eq6-2-20}
\end{equation}

The left-hand side of (\ref{eq6-2-17a}) starts from $1/2$ at
$k_{1\rho }a= 0$ and goes to infinity at the zeros of
$J_0(k_{1\rho }a)$, and goes to zero at the zeros of $J_1(k_{1\rho
})$. A sketch of the left-hand side and the right-hand side of
(\ref{eq6-2-17a}) is shown in Figure \ref{fg622}. The number of
guided modes depends on $V = \sqrt {(k_1^2-k_2^2)}a$, the
normalized frequency. $V$ can be increased by increasing the
contrast between $\epsilon _1$ and $\epsilon _2$, by raising the
frequency, or by increasing $a$.  For $V < 2.405$, there could be
no possible guided modes. Hence, all axisymmetric modes have a
finite cut-off frequency.
\index{Graphical solution!for EH modes}

\begin{figure}[ht]
\vspace{-0.5 in}
\begin{center}
\hfil\includegraphics[width=4.0truein]{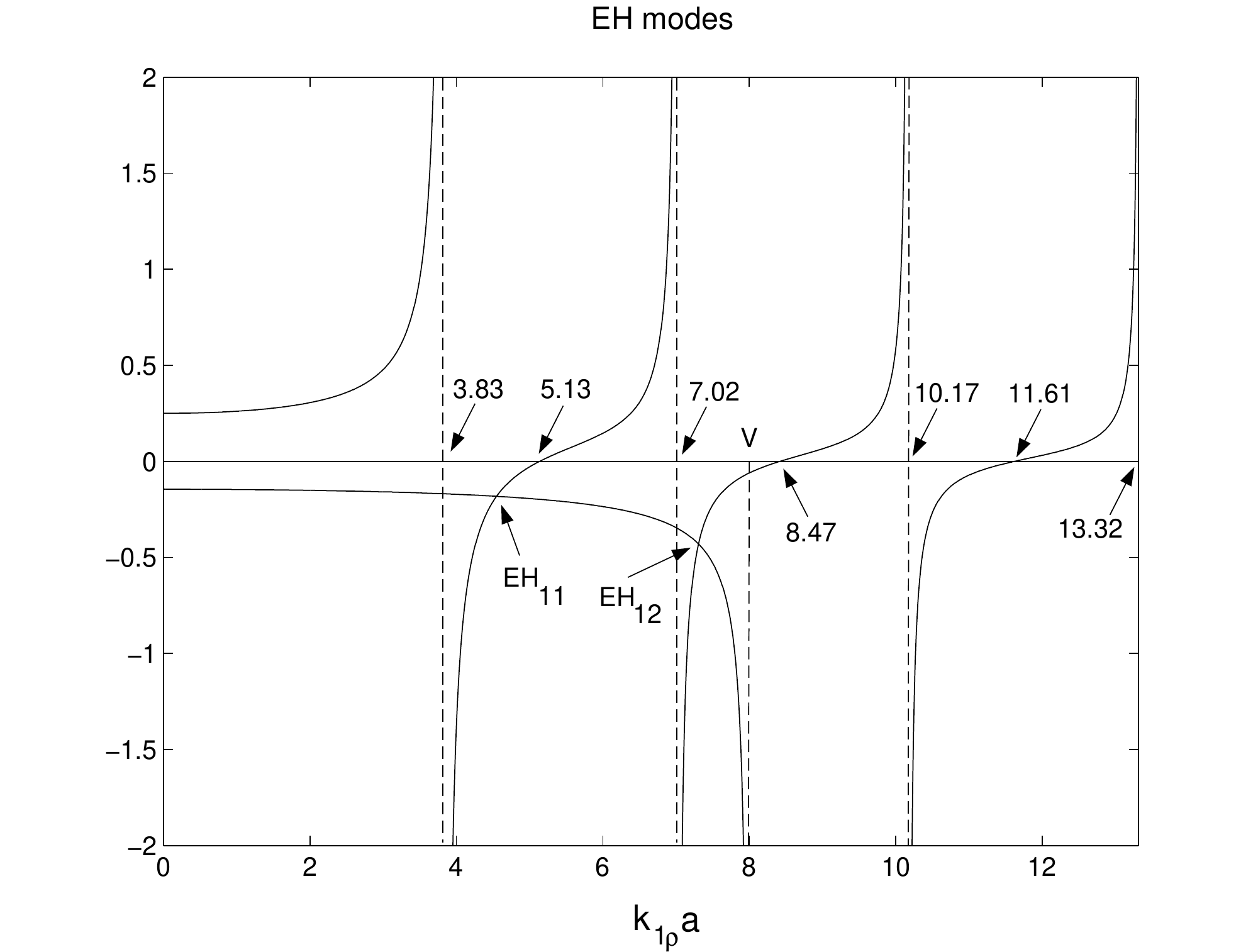}\hfil
\end{center}
\caption{Graphical solution of the EH modes.}\label{fg623}
\end{figure}


When $n = 1$, Equations (\ref{eq6-2-16a}), (\ref{eq6-2-16b}),
become
\begin{subequations}
\begin{equation}
\frac {J_2(k_{1\rho }a)}{k_{1\rho }aJ_1(k_{1\rho }a)}=\frac
1{2}\left( \frac {\mu _2}{\mu _1}+\frac {\epsilon _2}{\epsilon
_1}\right)\frac {K'_1(\alpha _2a)}{\alpha _2aK_1(\alpha
_2a)}+\left( \frac 1{(k_{1\rho }a)^2}-R\right),\quad \text {EH}
\label{eq6-2-21a}
\end{equation}
\begin{equation}
\frac {J_0(k_{1\rho }a)}{k_{1\rho }aJ_1(k_{1\rho }a)}=-\frac
1{2}\left( \frac {\mu _2}{\mu _1}+\frac {\epsilon _2}{\epsilon
_1}\right)\frac {K'_1(\alpha _2a)}{\alpha _2aK_1(\alpha
_2a)}+\left( \frac 1{(k_{1\rho }a)^2}-R\right), \quad \text {HE}.
\label{eq6-2-21b}
\end{equation}
\end{subequations}

\index{Graphical solution!for HE modes}

\begin{figure}[ht]
\begin{center}
\hfil\includegraphics[width=5.0truein]{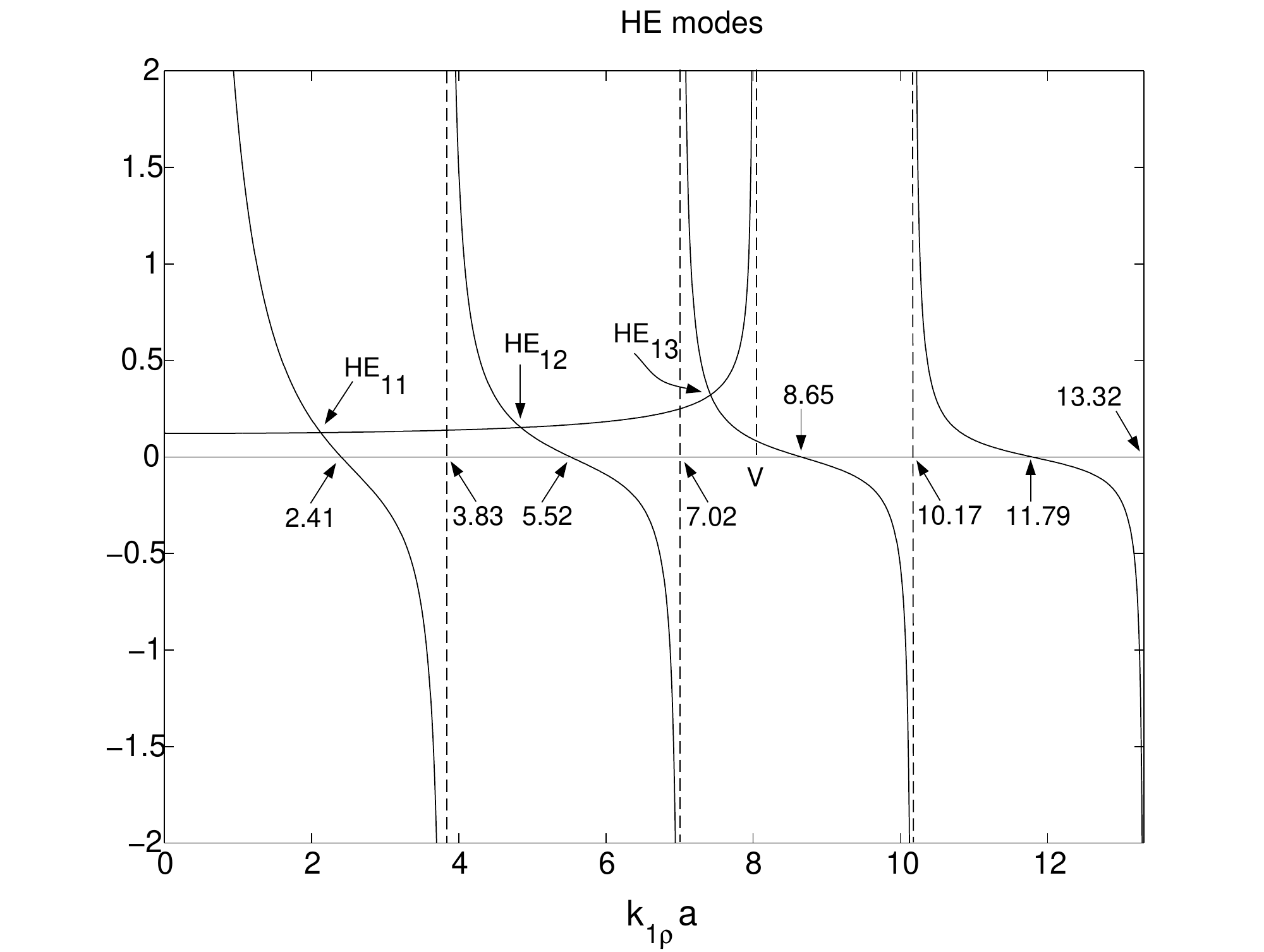}\hfil
\end{center}
\caption{Graphical solution of the HE modes.}\label{fg624}
\end{figure}


\index{Cutoff frequency}

\noindent The left and the right-hand side of (\ref{eq6-2-21a}) is
shown in Figures \ref{fg623} and \ref{fg624}. The right-hand side of
(\ref{eq6-2-21a}) goes to infinity at $k_{1\rho }a\rightarrow V$,
while the left-hand side goes to zero and infinity as before.  Note
that all the EH modes have finite cut-off frequencies while the
HE$_{11}$ mode does not have a cut-off frequency.  Therefore, the
HE$_{11}$ mode is the dominant mode in an optical fiber. If (i) the
optical fiber is small enough, or (ii) the frequency is low enough,
or (iii) when the contrast is very low, it is the only mode
propagating for the single mode operation of the optical fiber.

Usually, in a waveguide, an EH notation is used to denote a mode
where $E_z$ dominates over $H_z$, or the TM component dominates over
the TE component. However, due to a quirk in the history of optical
fibers, the EH notation is used to denote a mode whose TE component
dominates over its TM component, and vice versa for the HE
notation.


\begin{figure}[ht]
\begin{center}
\hfil\includegraphics[width=4.0truein]{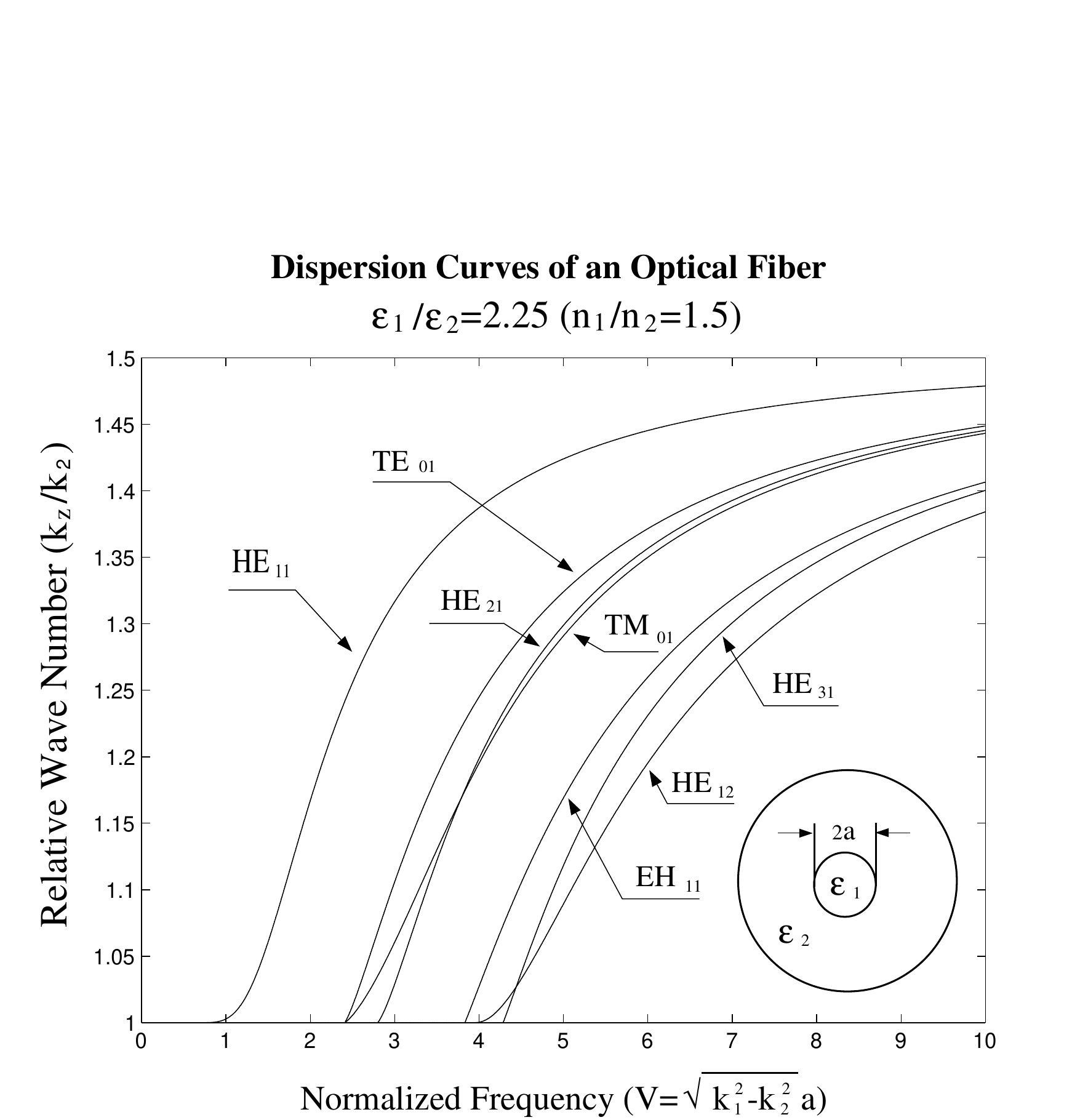}\hfil
\end{center}
\caption{Dispersion curves for various modes of an optical fiber for
the high contrast case.} \label{fg625b}
\end{figure}

\begin{figure}[ht]
\begin{center}
\hfil\includegraphics[width=4.0truein]{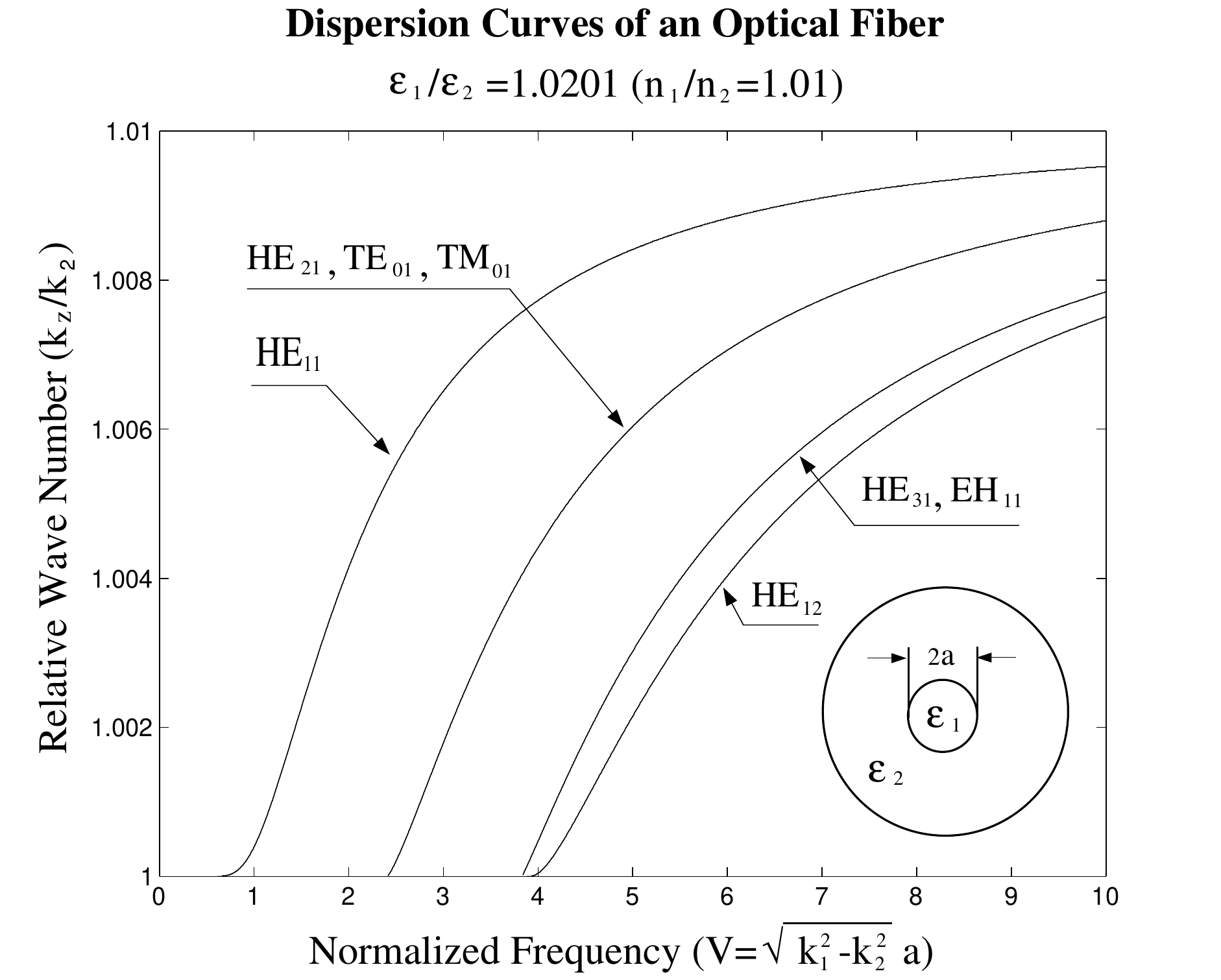}\hfil
\end{center}
\caption{Dispersion curves for various modes of an optical fiber for
the low contrast case.} \label{fg625a}
\end{figure}


A parameter of interest is the axial wave number $k_z$ as a
function of the \index{Normalized frequency} normalized frequency. In Figure \ref{fg625b}, we
plot ${k_z}/{k_0}$ versus $V$.  At very low frequencies, or near
the cut-off of a mode, ${k_z}/{k_0}$ approaches the refractive
index $n_2$ of the cladding.  This is because the mode is not well
confined, and most of the energy of the mode is in medium 2.
Hence, the mode phase velocity is close to that of medium 2. At
higher normalized frequency, the mode is well confined, and it
propagates in medium 1. Hence, ${k_z}/{k_0}$ approaches $n_1$
which is the refractive index of medium 1.

From Equation (\ref{eq6-2-12}), it is clear that when we find a
solution, a dual solution also exists. If ${\mu _1}/{\mu _2}
={\epsilon _1}/{\epsilon _2}$, then a solution and its dual are
degenerate. However, usually, ${\mu _1}/{\mu _2} = 1$; therefore, a
solution and its dual are not degenerate. However, when ${\epsilon
_1}/{\epsilon _2}\rightarrow 1$ as in the case of a weak-contrast
optical fiber, a solution and its dual are near degenerate.  The
dispersion curve for the weak contrast case is shown in Figure
\ref{fg625a}.

\section {{{ Weak-Contrast Optical Fiber}}}
\index{Weak-contrast optical fiber}

When the contrast between the cladding and the core of the fiber
is small, the analysis of the guided mode inside a fiber can be
greatly simplified.  In this case, polarization effect at the
interface of the fiber is unimportant, and scalar wave theory can
be applied \cite{GLOGE,YARIVD,OKAMOTO}.

The vector wave equations governing electromagnetic
fields in an inhomogeneous waveguide are
\begin{subequations}
\begin{equation}
\nabla\times\mu _r^{-1}\nabla\times\v E-\omega ^2\mu _0\epsilon
_0\epsilon _r\v E=0, \label{eq6-3-1a}
\end{equation}
\begin{equation}
\nabla\times\epsilon _r^{-1}\nabla\times\v H-\omega ^2\mu
_0\epsilon _0\mu _r\v H=0, \label{eq6-3-1b}
\end{equation}
\end{subequations}
where $\mu _r={\mu (\v r)}/{\mu _0} = f_1(\v r)$, and $\epsilon
_r= {\epsilon (\v r)}/{\epsilon _0}= f_2(\v r)$. If we find a
solution to (\ref{eq6-3-1a}), the solution to the dual problem is
obtained by letting $\v E\rightarrow - \v H$, $\v H\rightarrow\v
E$, $\mu _r\rightarrow\epsilon _r$, $\epsilon _r\rightarrow\mu
_r$. However, if $f_1(\v r)\ne f_2(\v r)$, the dual problem
corresponds to a different waveguide with $\mu _r=f_2(\v r)$ and
$\epsilon _r=f_1(\v r)$. In order for a dual problem to be itself,
we require that $f_1(\v r)=f_2(\v r)$.  If this is the case, then
a mode and its dual are degenerate. Therefore, we can associate
every mode in a waveguide with a dual mode. However, $f_1(\v r) =
1$ usually, while $f_2(\v r)\ne 1$; therefore, a mode is not
degenerate with its dual.

However, in the case of a weak-contrast optical fiber, $\epsilon
_r\rightarrow 1$; hence, a mode is near degenerate with its dual.
Furthermore, we can show that the vector nature of the wave is
unimportant.  If $\mu _r=1$, we can rewrite (\ref{eq6-3-1a}) and
(\ref{eq6-3-1b}) as
\begin{subequations}
\begin{equation}
\nabla\times\nabla\times\v E-\omega ^2\mu _0\epsilon _0\epsilon
_r\v E=0, \label{eq6-3-2a}
\end{equation}
\begin{equation}
\nabla\times\nabla\times\v H-(\nabla\ln\epsilon
_r)\times\nabla\times\v H-\omega ^2\mu _0\epsilon _0\epsilon _r\v
H=0. \label{eq6-3-2b}
\end{equation}
\end{subequations}
In the above $\nabla \ln \epsilon _r=\frac {\nabla\epsilon
_r}{\epsilon _r}$ is the effect of the polarization charges at the
dielectric interface. If the transverse and longitudinal
components of (\ref{eq6-3-2a}) and (\ref{eq6-3-2b}) are extracted,
the following equations ensue:
\begin{subequations}
\begin{equation}
\nabla ^2\v E_s+\nabla _s[\nabla _s\ln \epsilon _r \cdot \v
E_s]+k^2\v E_s=0, \label{eq6-3-3a}
\end{equation}
\begin{equation}
\nabla ^2 E_z+ik_z(\nabla _s\ln \epsilon _r )\cdot \v
E_s+k^2E_z=0, \label{eq6-3-3b}
\end{equation}
\end{subequations}
and
\begin{subequations}
\begin{equation}
\nabla ^2\v H_s+(\nabla _s\ln \epsilon _r)\times (\nabla _s\times
\v H_s)+k^2\v H_s=0, \label{eq6-3-4a}
\end{equation}
\begin{equation}
\nabla ^2\v H_z+(\nabla _s\ln\epsilon _r)\times (\nabla _s\times
\v H_z)+(\nabla _s\ln\epsilon _r)\times (\^ zik_z\times \v
H_s)+k^2\v H_z=0. \label{eq6-3-4b}
\end{equation}
\end{subequations}
In the limit when $\epsilon _r\rightarrow 1$,
the polarization charge terms in (\ref{eq6-3-3a}), \eqref{eq6-3-3b},
(\ref{eq6-3-4a}), and \eqref{eq6-3-4b} can be ignored with respect
to the other terms, yielding
\begin{subequations}
\begin{equation}
\nabla ^2\v E_s+k^2\v E_s=0, \label{eq6-3-5a}
\end{equation}
\begin{equation}
\nabla ^2 E_z+k^2 E_z=0, \label{eq6-3-5b}
\end{equation}
\begin{equation}
\nabla ^2\v H_s+k^2\v H_s=0. \label{eq6-3-5c}
\end{equation}
\begin{equation}
\nabla ^2 H_z+k^2 H_z=0. \label{eq6-3-5d}
\end{equation}
\end{subequations}

In other words, the wave guidance problem by a fiber of weak
contrast reduces to a scalar problem. Also, from the above
equations, it is apparent that $\nabla^2 \sim -k^2$, and hence, when
$\omega\rightarrow\infty$, the polarization terms in
\eqref{eq6-3-3a} and \eqref{eq6-3-4a} are much smaller than the
first and the last terms.  From \eqref{eq6-3-3b} and
\eqref{eq6-3-4b}, it is seen that the $z$ components of the fields
are induced by their transverse components; hence they are much
smaller than the transverse components.
Moreover, when the contrast $\epsilon_r= 1$, the guided mode in the
fiber becomes a TEM mode with $k=k_z$, and $E_z=H_z=0$.  This
represents the leading order solution when $\epsilon_r=1$.

As mentioned above, when $\epsilon_r > 1$, we see from Equation
(\ref{eq6-3-3b}) that a nonzero $E_z$ is induced by the presence of
$\v E_s$. Moreover, by comparing terms, and assuming high frequency,
$E_z \sim |\nabla \ln \epsilon_r\cdot \v E_s|/k \approx \ln
\epsilon_r|\v E_s|/(ka) \ll |\v E_s| $, because $\nabla \ln
\epsilon_r\approx \ln \epsilon_r/a$.  Therefore, when $\omega
\rightarrow \infty$, $|E_z| \ll |\v E_s|$ (see Problem 6-8.).  By
the same token, $|H_z| \ll |\v H_s|$ when $\omega\rightarrow
\infty$. Therefore, when the contrast is very weak, and the
frequency is very high, the mode is quasi-TEM, namely,
$\nabla_s\cdot \epsilon_r\v E_s = -ik_z\epsilon_r E_z\approx 0$ and
$\nabla_s\cdot \v H_s=-ik_z H_z\approx 0$, since $k_zE_z\ll |\v
E_s|/a$ and $k_zH_z \ll |\v H_s|/a$.
From the aforementioned analysis, it is clear that (\ref{eq6-3-5a})
and (\ref{eq6-3-5c}) are the equations to solve when
$\omega\rightarrow\infty$ and the contrast small.

Equations (\ref{eq6-3-5a}) and (\ref{eq6-3-5c}) are equivalent to
\begin{equation}
(\nabla ^2+k^2)\phi =0 \label{eq6-3-6}
\end{equation}
where $\phi$ is either $E_x$, $E_y$, $H_x$, or $H_y$.
For example, we can let
\begin{equation}
\phi =\left\{
\begin{aligned}
& AJ_n(k_{1\rho }\rho )e^{in\phi +ik_zz}, \quad \rho <a ,\\
& BK_n(\alpha _2\rho )e^{in\phi +ik_zz},  \quad \rho >a ,
\end{aligned}
\right . \label{eq6-3-7}
\end{equation}
The boundary conditions for $\phi$ at the interface where
$k^2=\omega ^2\mu_0\epsilon _0\epsilon _r$ displays a step
discontinuity is
\begin{subequations}
\begin{equation}
\phi _1=\phi _2, \label{eq6-3-8a}
\end{equation}
\begin{equation}
\^ n\cdot \nabla \phi _1=\^ n\cdot \nabla \phi _2.
\label{eq6-3-8b}
\end{equation}
\end{subequations}
These boundary conditions are derivable from Equation
(\ref{eq6-3-6}) alone.

Imposing the above boundary conditions at $\rho =a$ for the weak
contrast optical fiber, whose field is given by (\ref{eq6-3-7}), we
have
\begin{subequations}
\begin{equation}
AJ_n(k_{1\rho }a)=BK_n(\alpha _2 a), \label{eq6-3-9a}
\end{equation}
\begin{equation}
Ak_{1\rho }J'_n(k_{1\rho }a)=B\alpha _2K'_n(\alpha _2a).
\label{eq6-3-9b}
\end{equation}
\end{subequations}
The above yields
\begin{equation}
\frac {k_{1\rho }J'_n(k_{1\rho }a)}{J_n(k_{1\rho }a)}=\frac
{\alpha _2K'_n(\alpha _2a)}{K_n(\alpha _2a)}. \label{eq6-3-10}
\end{equation}
Using the recurrence relationship that $J'_n(x) = - J_{n+1}(x)
+\frac{n}{x} J_n (x)$ , and that $K'_n(x) = - K_{n+1}(x) +\frac
n{x} K_n(x)$, we can transform the above to
\begin{equation}
\frac {k_{1\rho }J_{n+1}(k_{1\rho }a)}{J_n(k_{1\rho }a)}=\frac
{\alpha _2K_{n+1}(\alpha _2a)}{K_n(\alpha _2a)}. \label{eq6-3-11}
\end{equation}
Similarly, using the recurrence relationship that $J'_n(x) =
J_{n-1}(x) -\frac{n}{x} J_n(x)$, and $K'_n(x) = - K_{n-1}(x) -
\frac{n}{x} K_n(x)$, we have
\begin{equation}
\frac {k_{1\rho }J_{n-1}(k_{1\rho }a)}{J_n(k_{1\rho }a)}=-\frac
{\alpha _2K_{n-1}(\alpha _2a)}{K_n(\alpha _2a)}. \label{eq6-3-12}
\end{equation}

\begin{figure}[ht]
\begin{center}
\hfil\includegraphics[width=3.5truein]{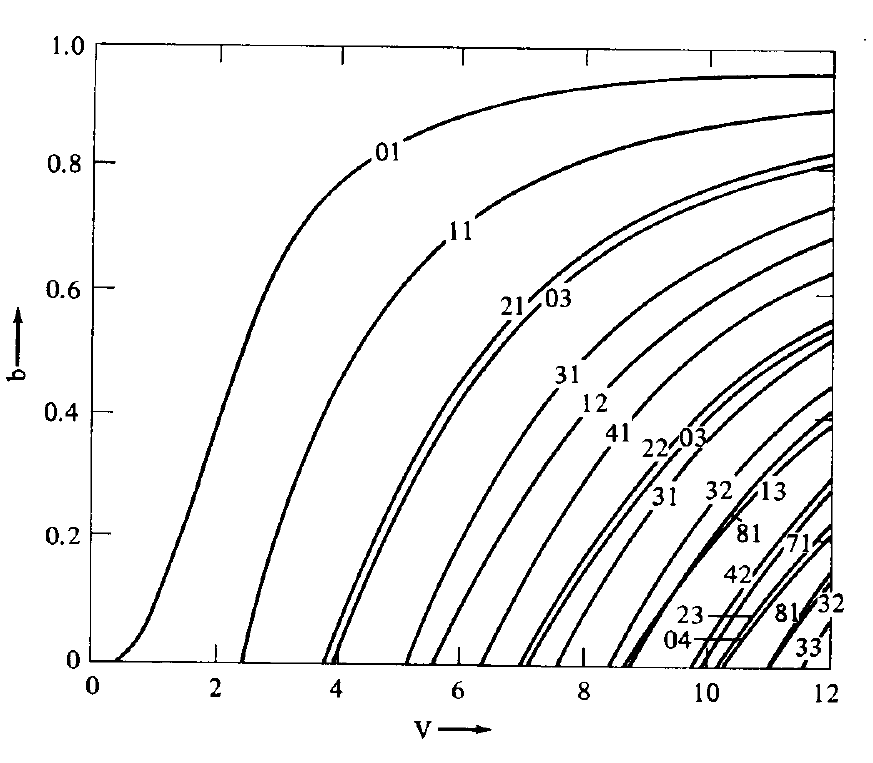}\hfil
\end{center}
\caption{Dispersion curves for a weak contrast optical fiber
\cite{GLOGE}.}\label{fg631}
\end{figure}

Comparing with Equations (\ref{eq6-2-16a}) and (\ref{eq6-2-16b}), we
note that now there are half as many solutions as before. This is
because when $\epsilon _r\rightarrow 1$, HE$_{n+1, m}$ and EH$_{n-1,
m}$ modes are degenerate.  We can solve (\ref{eq6-3-11}) graphically
as before. The modes thus found are designated the LP$_{nm}$ mode.
The lowest order mode is the \index{Mode!linearly polarized} LP$_{01}$ mode which is the degenerate
case of the HE$_{11}$ mode. LP here stands for ``linearly
polarized.''  {The LP modes are fragile, as they are actually the
linear superpositions of the degenerate HE and EH modes.  The moment
the contrast increases, this degeneracy splits into two modes
again.}\footnote{The author thanks Erhan Kudeki for this
discussion.} Figure \ref{fg631} shows the dispersion curves of the
LP$_{nm}$ modes. In the figure, $b = {\left(\frac
{k_z}{k_2}-n_2\right)}/{(n_1-n_2)}$, and $V$ is the normalized
frequency.

These modes are termed \index{Mode!weakly guided} ``weakly guided'' modes, but it is a
misnomer. When $V$ is large, a mode can still be tightly confined to
the waveguide as demonstrated by the phase velocity approaching that
of a core. Also, $E_x$ and $E_y$ are not independent of each other
since $\nabla_s\cdot\epsilon_r\v E_s\approx 0$. The same statement
applies to $H_x$ and $H_y$.  Hence, these modes are not linearly
polarized.

\section {{{ Perturbation Formula for Dielectric Waveguides}}}
\index{Waveguide!dielectric}
\index{Waveguide!perturbation formula}

If we know the solution to a waveguide geometry, and wish to
change the \index{Velocity!phase!adjusting}phase velocity of the waveguiding mode, dielectric
material can be added to achieve the purpose. A perturbative
approach can be used to analyze such problem
\cite{KONGF,HASHIMOTO,HAUSD}.

For the unperturbed problem, the electromagnetic field satisfies
\begin{equation}
\nabla\times\v E_0=i\omega\mu _0\v H_0, \label{eq6-4-1}
\end{equation}
\begin{equation}
\nabla\times\v H_0=-i\omega\epsilon _0\v E_0. \label{eq6-4-2}
\end{equation}
Now, let us change the permittivity and the permeability of the
waveguide. The new electromagnetic field satisfies
\begin{equation}
\nabla\times\v E=i\omega\mu\v H, \label{eq6-4-3}
\end{equation}
\begin{equation}
\nabla\times\v H=-i\omega\epsilon\v E, \label{eq6-4-4}
\end{equation}
where $\mu$ and $\epsilon$ are new. Taking the divergence of $\v
E_0^*\times\v H$ and $\v E\times\v H_0^*$, we have
\begin{subequations}
\begin{equation}
\nabla \cdot (\v E_0^*\times\v H)=-i\omega\mu _0\v H\cdot\v
H_0^*+i\omega\epsilon\v E_0^*\cdot\v E, \label{eq6-4-5a}
\end{equation}
\begin{equation}
\nabla \cdot (\v E\times\v H_0^*)=i\omega\mu\v H\cdot\v
H_0^*-i\omega\epsilon _0\v E_0^*\cdot\v E. \label{eq6-4-5b}
\end{equation}
\end{subequations}
Adding the above equations, we have
\begin{equation}
\nabla\cdot (\v E_0^*\times\v H+\v E\times\v
H_0^*)=i\omega\delta\mu\v H\cdot\v H_0^*+i\omega\delta\epsilon\v
E_0^*\cdot\v E, \label{eq6-4-6}
\end{equation}
where $\delta\mu =\mu -\mu _0$, and $\delta \epsilon
=\epsilon-\epsilon _0$. Since the unperturbed field has
$e^{ik_{0z}z}$ dependence, while the perturbed field has $e^{ik_zz}$
dependence, we have
\begin{subequations}
\begin{equation}
\begin{split}
\nabla\cdot (\v E_0^*\times\v H+\v E\times\v H_0^*)&=\nabla
_s\cdot (\v
E_0^*\times\v H+\v E\times\v H_0^*)\\
&+\^ zi(k_z-k_{0z})\cdot (\v E_0^*\times\v H+\v E\times\v H_0^*).
\end{split}
\label{eq6-4-7a}
\end{equation}
Integrating Equation (\ref{eq6-4-6}) over the cross-section of the
waveguide, making use of (\ref{eq6-4-7a}), we have
\begin{equation}
k_z-k_{0z}=\omega\frac {\iint \limits _S dS[\delta\mu\v H\cdot\v
H_0^*+\delta\epsilon\v E\cdot\v E_0^*]}{\iint \limits _S dS[\v
E_0^*\times\v H+\v E\times\v H_0^*]}. \label{eq6-4-7b}
\end{equation}
\end{subequations}

\index{Optical fiber!dispersion}
\begin{figure}[ht]
\vspace{0.05in}
\begin{center}
\hfil\includegraphics[width=2.5truein]{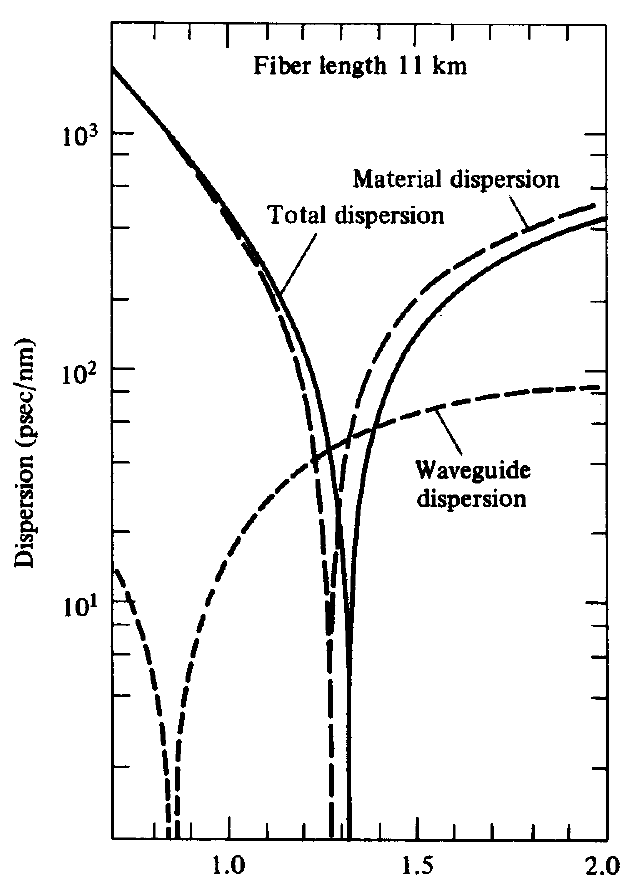}\hfil
\end{center}
\caption{Dispersion in a single-mode graded index optical fiber.
(From W.A. Gambling et al, ``Zero total dispersion in graded-index
single-mode fibers,'' {\it Electron. Lett.}, v. 15, p. 474,
1979.).}\label{fg641}
\end{figure}


\noindent Equation (\ref{eq6-4-7a}) is exact at this point, but is
not very useful because $\v E$ and $\v H$ are unknowns on the
right-hand side. However, when the perturbation is small, we can
approximate $\v E \simeq\v E_0$ and $\v H\simeq\v H_0$, and
(\ref{eq6-4-7a}) becomes
\begin{equation}
k_z-k_{0z}\simeq\omega\frac {\iint \limits _S dS[\delta\mu |\v
H_0|^2+\delta\epsilon |\v E_0|^2]}{2\Re e \iint \limits _S dS[\v
E_0\times\v H_0^*]}. \label{eq6-4-8}
\end{equation}

The time average power flow in the unperturbed waveguide is given
by
\begin{equation}
\langle P_f\rangle =\frac 1{2}\Re e \iint dS[\v E_0\times\v
H_0^*]. \label{eq6-4-9}
\end{equation}
The time average energy stored per unit length is
\begin{equation}
\langle W\rangle =\langle W_e\rangle +\langle W_m\rangle =2\langle
W_e\rangle =\frac 1{2}\iint \limits _S dS\epsilon |\v E|^2.
\label{eq6-4-10}
\end{equation}
In the case of a step-index optical fiber, where a perturbation is
in the dielectric constant, which is uniformly $\delta\epsilon _1$
in the core region and is uniformly $\delta\epsilon _2$ in the
cladding region, we can rewrite (\ref{eq6-4-8}) as
\begin{equation}
\delta k_z\simeq\omega\frac {\iint \limits _{S_1} dS\delta\epsilon
_1|\v E_1|^2+\iint \limits_{S_2} dS\delta \epsilon _2|\v
E_2|^2}{4\langle P_f\rangle }, \label{eq6-4-11}
\end{equation}
where $\v E_1$ is the original field in the core region, and $\v
E_2$ is the original field in the cladding region. Since
$\delta\epsilon _i$ is constant in region $i$, we can rewrite
(\ref{eq6-4-11}) as
\begin{equation}
\delta k_z\simeq \omega\frac {\delta\epsilon _1\epsilon
_1^{-1}\langle W_{e1}\rangle +\delta\epsilon _2\epsilon
_2^{-1}\langle W_{e2}\rangle }{\langle P_f\rangle },
\label{eq6-4-12}
\end{equation}
where $\langle W_{e1}\rangle $ is the time average energy stored
in the electric field in region 1 while $\langle W_{e2}\rangle $
is that for region 2. Since the total power flow in each region is
$2 v_g \langle W_{ei}\rangle $, we can rewrite (\ref{eq6-4-12}) as
\begin{equation}
\delta k_z\simeq \frac {\omega }{2v_g}[\delta\epsilon _1\epsilon
_1^{-1}\Gamma _1+\delta\epsilon _2\epsilon _2^{-1}\Gamma _2],
\label{eq6-4-13}
\end{equation}
where $\Gamma _i$ is the fraction of power flow in region $i$.

\begin{figure}[ht]
\vspace{-0.1in}
\begin{center}
\hfil\includegraphics[width=4.0truein]{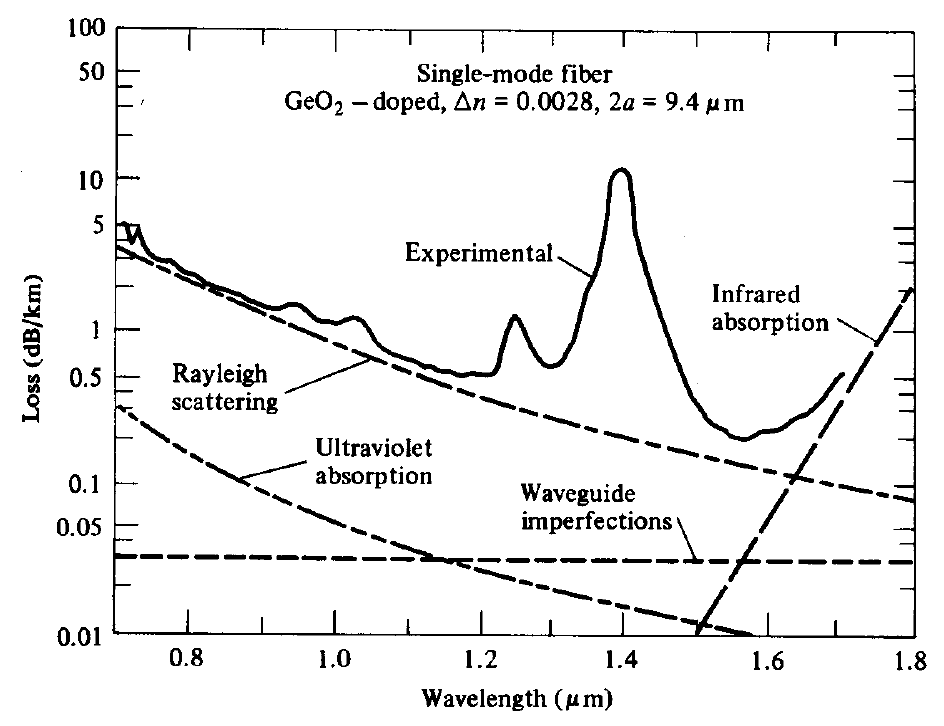}\hfil
\end{center}
\caption{Different loss mechanisms in a germanosilicate single-mode
optical fiber.  Updated figure can be found in Saleh and Teich
\cite{SALEH&TEICH}.}\label{fg642}
\end{figure}


For a weak contrast optical fiber, we can further approximate the
above with $\epsilon _1\simeq\epsilon _2$, $v_g\simeq c_1\simeq
c_2$,
\begin{equation}
\delta k_z\simeq\frac {k_1}{2\epsilon _1}[\delta\epsilon _1\Gamma
_1+\delta\epsilon _2\Gamma _2]. \label{eq6-4-14}
\end{equation}
Defining an effective refractive index $n={k_z}/{k_0}$, then
$\delta n={\delta k_z}/{k_0}$, and the above can be rewritten as
\begin{equation}
\delta n\simeq \frac 1{2n_1}[\delta n_1^2\Gamma _1+\delta
n_2^2\Gamma _2], \label{eq6-4-15}
\end{equation}
or that
\begin{equation}
\delta n\simeq [\delta n_1\Gamma _1+\delta n_2\Gamma _2],
\label{eq6-4-16}
\end{equation}
for a weak contrast optical fiber. Hence, the change in the
\index{Weak-contrast optical fiber!refractive index} effective refractive index is proportional to the change in the
refractive index in each region, weighted by the fraction of the
power flow in each region.

\section {{{ Mode Dispersion in an Optical Fiber}}}
\index{Optical fiber!mode dispersion}

Once the \index{Optical fiber!axial wave number} axial wave member $k_z$ of our optical fiber is found, we
can define an effective refractive index given by
\begin{equation}
n=\frac {k_z}{k_0}. \label{eq6-5-1}
\end{equation}
In other words,
\begin{equation}
k_z=n\frac {\omega }{c_0}, \label{eq6-5-2}
\end{equation}
where $n$ is a function of $n_1$, $n_2$ and $\omega $ or $n=n(n_1,
n_2, \omega)$. The signal in an optical fiber travels at the group
velocity.  Hence, we can study how the \index{Velocity!group} group velocity depends on
$n_1$, $n_2$  and $\omega $.
\begin{equation}
\frac 1{v_g}=\frac {d k_z}{d \omega }=\frac {\omega
}{c_0}\left[\frac {\partial n}{\partial n_1}\frac {d n_1}{d\omega
}+\frac {\partial n}{\partial n_2}\frac {d n_2}{d\omega }+\frac
{\partial n}{\partial\omega }\right] +\frac {n}{c_0}.
\label{eq6-5-3}
\end{equation}
This first two terms in the square bracket come from the materials
making up $n_1$ and $n_2$, and the last term in the square bracket
is a consequence of the waveguide geometry. From (\ref{eq6-4-16}),
we deduce that
\begin{equation}
\frac {\partial n}{\partial n_1}=\Gamma _1, \qquad\frac {\partial
n}{\partial n_2}=\Gamma _2, \label{eq6-5-4}
\end{equation}
This first two terms for a weak contrast optical fiber, and $\Gamma
_1$ and $\Gamma _2$ are the fractions of power flow in regions 1 and
2. With the approximation that
\begin{equation}
\frac {d n_1}{d \omega }\simeq \frac {d n_2}{d\omega }=\left(\frac
{\partial n}{\partial\omega }\right) _m, \label{eq6-5-5}
\end{equation}
where the subscript $m$ stands for dispersion from material
property, we can rewrite (\ref{eq6-5-3}) as
\begin{equation}
\frac 1{v_g}=\frac {dk_z}{d\omega }=\frac {\omega }{c_0}\left[
\left( \frac {\partial n}{\partial\omega }\right) _m +\left( \frac
{\partial n}{\partial\omega }\right) _w\right] +\frac {n}{c_0},
\label{eq6-5-6}
\end{equation}
where the subscript $w$ stands for dispersion from waveguide
geometry. With $\omega ={2\pi c_0}/{\lambda }$, we have ${\omega
}/{d\omega } = -{\lambda }/{d\lambda }$, and the above can be
rewritten as
\begin{equation}
\frac 1{v_g}=-\frac {\lambda }{c_0}\left[ \left(\frac {\partial
n}{\partial \lambda }\right) _m+\left( \frac {\partial n}{\partial
\lambda }\right) _w\right]+\frac n{c_0}. \label{eq6-5-7}
\end{equation}

A measure of dispersion along an optical fiber of length $L$ is
the \index{Velocity!group!dispersion} group velocity dispersion $D$ defined as
\begin{equation}
D=L^{-1}\frac {\partial T}{\partial \lambda }, \label{eq6-5-8}
\end{equation}
where $T$ is the travel time taken by a pulse to traverse the
length $L$ of the optical fiber. Since $T = {L}/{v_g}$, we have
\begin{equation}
D=\frac {\partial }{\partial\lambda }\frac 1{v_g}=-\frac {\lambda
}{c_0}\left[ \left( \frac {\partial ^2 n}{\partial \lambda
^2}\right) _m+\left( \frac {\partial ^2 n}{\partial \lambda
^2}\right) _w\right]. \label{eq6-5-9}
\end{equation}
In order to have least pulse distortion, we should operate at a
frequency where both the material dispersion, $\left( \frac
{\partial ^2 n}{\partial\lambda ^2}\right)_m$, and the waveguide
dispersion, $\left( \frac {\partial ^2n}{\partial\lambda ^2}\right)
_w$ are small. For GeO$_2$-doped silica, the \index{Material dispersion} material dispersion
passes through a minimum at $\lambda = 1.3 \mu m$.  The waveguide
dispersion can be altered by altering $a$ as well as $n_1$ and
$n_2$.  By choosing a core diameter between 4 and 5 $\mu$m, and
relative refractive index difference of ${(n_1 - n_2)}/{n_1} >
0.004$, the wavelength of the minimum group velocity dispersion can
be shifted to 1.5 to 1.6 $\mu$m region where the loss is lowest (see
Figure \ref{fg641} and Figure \ref{fg642}).   The unit of $D$ is
usually in picosecond per nanometer for a given fiber lenght $L$. In
the case of Figure \ref{fg641}, the fiber length is 11 km.

\section {{{ A Rectangular Dielectric Waveguide}}}
\index{Waveguide!rectangular dielectric}

When a dielectric waveguide is rectangular in shape, there is no
closed form solution to the problem.  The eigenmodes of the
waveguide have to be found numerically. We shall discuss two methods
of solving for the \index{Eigenmode problem} eigenmodes and eigenvalues of the rectangular
dielectric waveguide.

\begin{figure}[ht]
\begin{center}
\hfil\includegraphics[width=4.5truein]{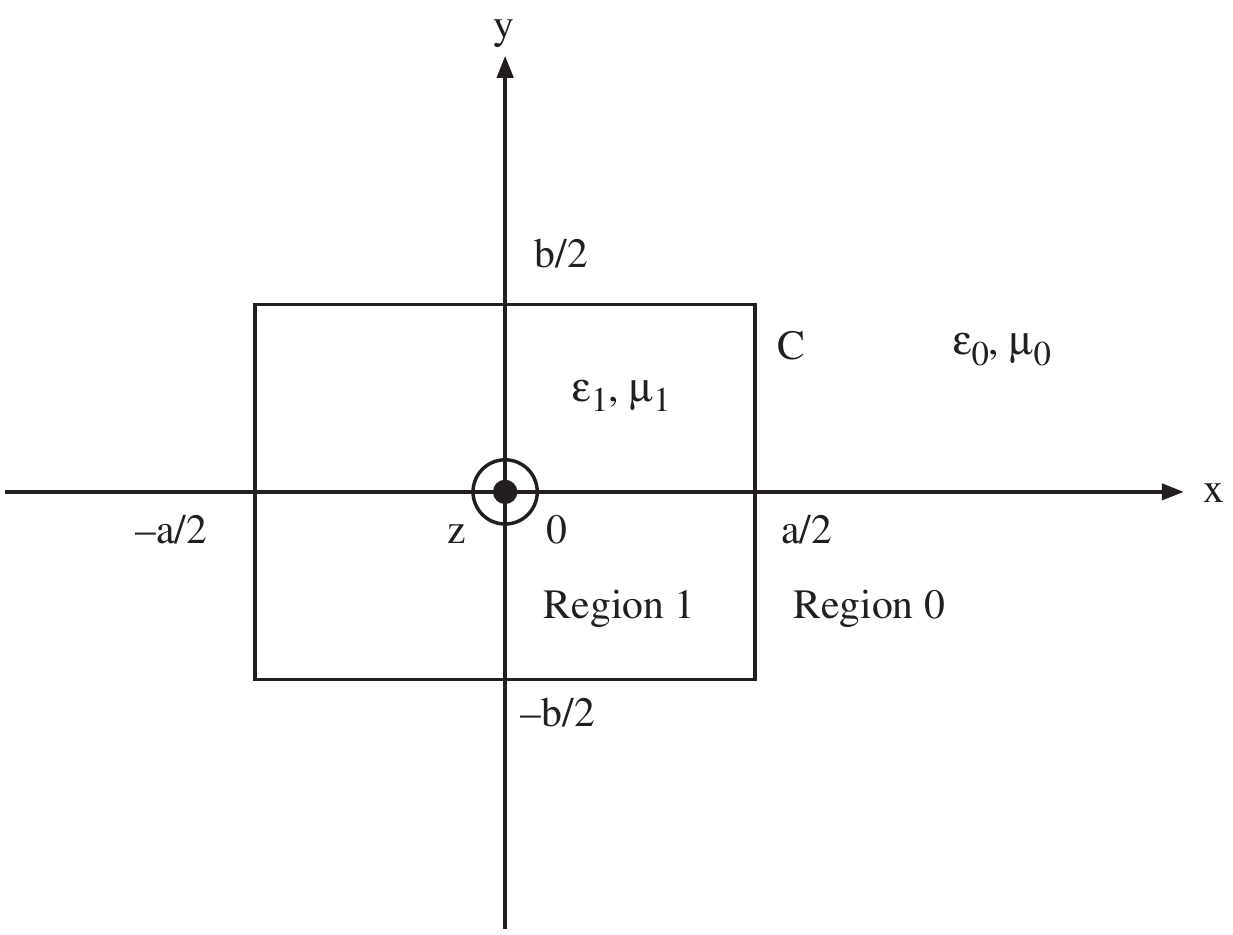}\hfil
\end{center}
\caption{The geometry of a rectangular dielectric waveguide.}\label{fg661}
\end{figure}


\subsection {{ Harmonic Expansion Method}}
\index{Eigenmode problem!harmonic expansion method}

\begin{figure}[ht]
\begin{center}
\hfil\includegraphics[width=5.0truein]{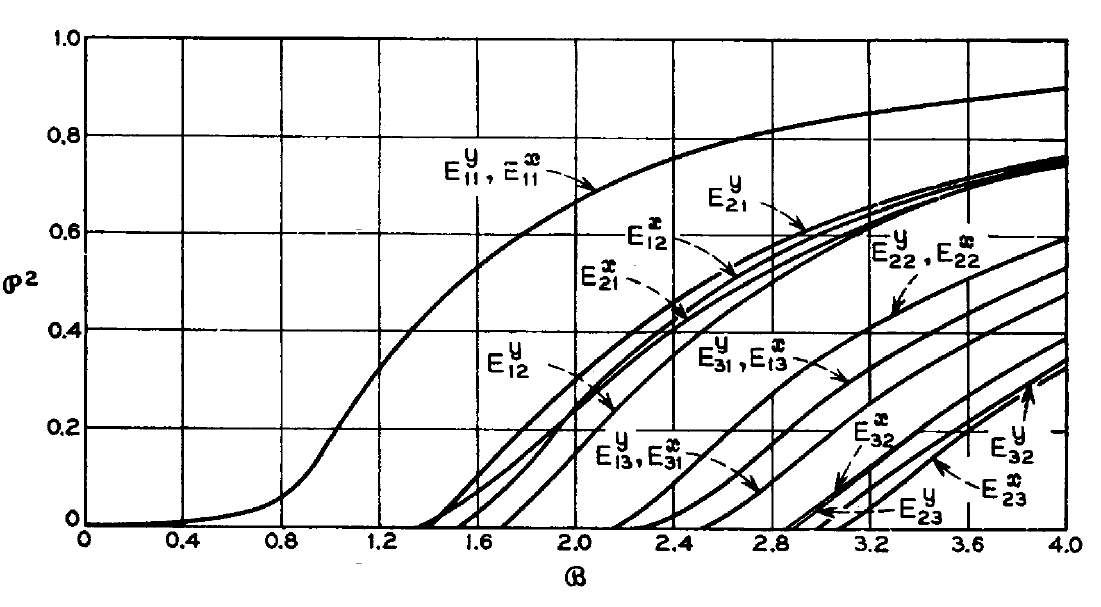}\hfil
\end{center}
\caption{Dispersion curves of different modes of a rectangular
dielectric waveguide where $({\epsilon _1}/{\epsilon
_0})^{1/2}=1.5$, and where the aspect ratio $a/b=1$ \cite{GOELL}.}
\label{fg662ac}
\end{figure}

\begin{figure}[ht]
\begin{center}
\hfil\includegraphics[width=5.0truein]{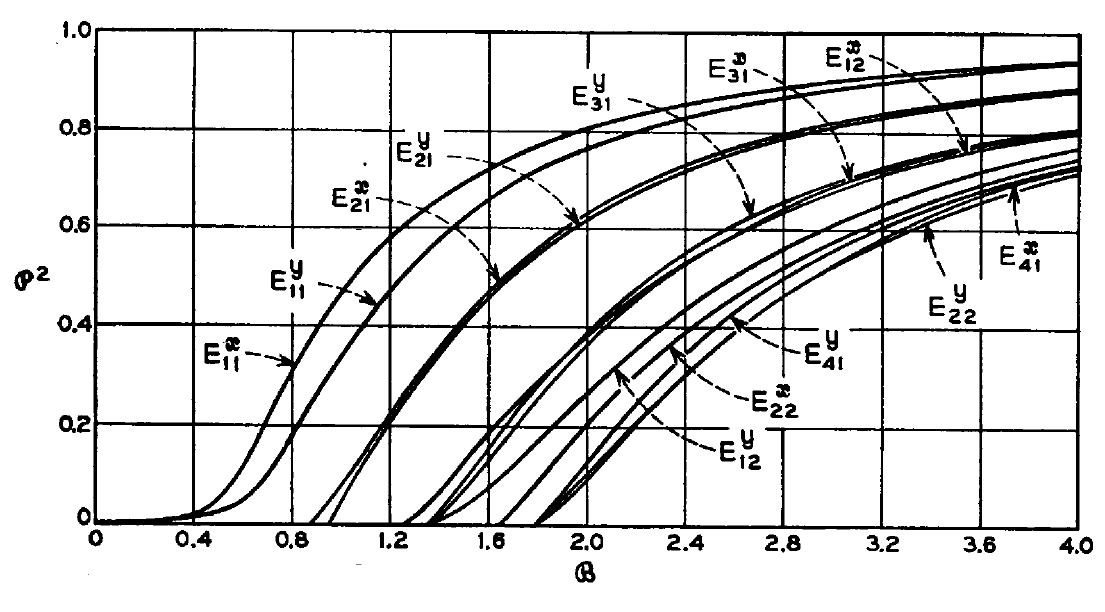}\hfil
\end{center}
\caption{Dispersion curves of different modes of a rectangular
dielectric waveguide where $({\epsilon _1}/{\epsilon
_0})^{1/2}=1.5$, and where the aspect ratio $a/b=2$ \cite{GOELL}.}
\label{fg662b}
\end{figure}


In this method \cite{GOELL} \footnote{This method is similar to
the method of Rayleigh hypothesis described in \cite{WFIMH}.  It
eventually will have ill-conditioning problem when the number of
unknowns is large. Mercatili \cite{MERCATILI} also analyzed this
problem approximately, although with a different approach.}, we
assume that the waveguide is piecewise homogeneous in each region.
Inside the waveguide, the field can be decomposed into TE and TM
to $z$ waves, each of which satisfies the following equations,
\begin{equation}
\text {TE}: \quad (\nabla _s^2+k_{1s}^2)H_{1z}(\v r)=0, \quad \v
r\in \text {region 1}, \label{eq6-6-1}
\end{equation}
\begin{equation}
\text {TM}: \quad (\nabla _s^2+k_{1s}^2)E_{1z}(\v r)=0, \quad \v
r\in \text {region 1} \label{eq6-6-2}
\end{equation}
where $k_{1s}^2=\omega ^2\mu _1\epsilon _1-k_z^2$ and $\nabla
_s^2={\partial ^2}/{\partial x^2}+{\partial ^2}/{\partial y^2}$.
We assume all the fields to have $e^{ik_zz}$ dependence due to the
phase matching condition.  In region 0, the fields satisfy similar
equations
\begin{equation}
\text {TE}: \quad (\nabla _s^2+k_{0s}^2)H_{0z}(\v r)=0, \quad \v
r\in \text {region 0}, \label{eq6-6-3}
\end{equation}
\begin{equation}
\text {TM}: \quad (\nabla _s^2+k_{0s}^2)E_{0z}(\v r)=0, \quad \v
r\in \text {region 0}. \label{eq6-6-4}
\end{equation}
The TE and TM waves will be coupled by the boundary conditions at
the dielectric boundary. In region 1, the general solution is
\begin{equation}
\begin{split}
\begin{bmatrix}
E_{1z}\\
H_{1z}
\end{bmatrix}
& =\sum\limits _{n=-\infty }^{\infty }
\begin{bmatrix}
a_n\\
b_n
\end{bmatrix}
J_n(k_{1s}\rho )e^{in\phi +ik_zz}\\
& =\sum\limits _{n=-\infty }^{\infty }
\begin{bmatrix}
a_n\\
b_n
\end{bmatrix}
\Re g\psi _n(k_{1s}, \v r_s) ,
\end{split}
\label{eq6-6-5}
\end{equation}
while in region 0, it is
\begin{equation}
\begin{split}
\begin{bmatrix}
E_{0z}\\
H_{0z}
\end{bmatrix} & =\sum\limits _{n=-\infty }^{\infty }
\begin{bmatrix}
c_n\\
d_n
\end{bmatrix} H_n^{(1)}(k_{0s}\rho )e^{in\phi +ik_zz}\\
& =\sum\limits _{n=-\infty }^{\infty }
\begin{bmatrix}
c_n\\
d_n
\end{bmatrix} \psi _n(k_{0s}, \v r_s) ,
\end{split}
\label{eq6-6-6}
\end{equation}
where $\Re g\psi _n(k_s, \v r_s)=J_n(k_s\rho )e^{in\phi }$, and
$\psi _n (k_s,\v r_s)=H_n^{(1)}(k_s\rho )e^{in\phi }$.  We need to
derive the transverse components of the fields in order to match
boundary conditions. Using the following equations,
\begin{subequations}
\begin{equation}
\v E_s=\frac i{k_s^2}[k_z\nabla _sE_z-\omega\mu\^ z\times\nabla
_sH_z], \label{eq6-6-7a}
\end{equation}
\begin{equation}
\v H_s=\frac i{k_s^2}[k_z\nabla _sH_z+\omega\epsilon\^
z\times\nabla _sE_z], \label{eq6-6-7b}
\end{equation}
\end{subequations}
we deduce that in region 1,
\begin{subequations}
\begin{equation}
\v E_{1s}=\sum\limits _{n=-\infty }^{\infty }\frac
i{k_{1s}^2}[k_za_n\nabla _s\Re g \psi _n(k_{1s}, \v r_s)-\omega\mu
_1b_n\^ z\times\nabla _s\Re g\psi _n(k_{1s}, \v r_s)],
\label{eq6-6-8a}
\end{equation}
\begin{equation}
\v H_{1s}=\sum\limits _{n=-\infty }^{\infty }\frac
i{k_{1s}^2}[k_zb_n\nabla _s\Re g\psi _n(k_{1s}, \v
r_s)+\omega\epsilon _1a_n\^ z\times\nabla _s\Re g\psi_n(k_{1s}, \v
r_s)], \label{eq6-6-8b}
\end{equation}
\end{subequations}
and in region 0,
\begin{subequations}
\begin{equation}
\v E_{0s}=\sum\limits _{n=-\infty }^{\infty }\frac
i{k_{0s}^2}[k_zc_n\nabla _s\psi _n(k_{0s}, \v r_s)-\omega\mu
_0d_n\^ z\times\nabla _s\psi _n(k_{0s}, \v r_s)], \label{eq6-6-9a}
\end{equation}
\begin{equation}
\v H_{0s}=\sum\limits _{n=-\infty }^{\infty }\frac
i{k_{0s}^2}[k_zd_n\nabla _s\psi _n(k_{0s}, \v r_s)+\omega\epsilon
_0c_n\^ z\times\nabla _s\psi _n(k_{0s}, \v r_s)]. \label{eq6-6-9b}
\end{equation}
\end{subequations}
The boundary condition requires that tangential $\v E$ and $\v H$ be
continuous across the dielectric interface. Matching the $z$
components of the fields, we have
\begin{subequations}
\begin{equation}
\sum\limits _{n=-\infty }^{\infty }a_n\Re g\psi _n(k_{1s}, \v
r_s)=\sum\limits _{n=-\infty }^{\infty }c_n\psi _n(k_{0s}, \v
r_s), \quad \v r_s\in C, \label{eq6-6-10a}
\end{equation}
\begin{equation}
\sum\limits _{n=-\infty }^{\infty }b_n\Re g\psi _n(k_{1s}, \v
r_s)=\sum\limits _{n=-\infty }^{\infty }d_n\psi _n(k_{0s}, \v
r_s), \quad \v r_s\in C. \label{eq6-6-10b}
\end{equation}
\end{subequations}
By defining a unit vector $\^ c$ to be pointing along the
circumference of the waveguide $C$, we can equate the tangential
components in (\ref{eq6-6-8a}) and (\ref{eq6-6-9a}) on $C$. By
doing so, we have
\begin{subequations}
\begin{equation}
\begin{split}
&\sum\limits _{n=-\infty }^{\infty }\frac
1{k_{1s}^2}\left[k_za_n\^c\cdot \nabla_s\Re g \psi _n(k_{1s}, \v
r_s)-\omega \mu _1b_n\^ c\cdot\^ z\times \nabla
_s\Re g\psi_n(k_{1s}, \v r_s)\right]\\
& =\sum\limits _{n=-\infty }^{\infty }\frac
1{k_{0s}^2}\left[k_zc_n\^ c\cdot\nabla_s\psi _n(k_{0s}, \v
r_s)-\omega\mu _0d_n\^ c\cdot\^ z\times\nabla _s\psi_n(k_{0s}, \v
r_s)\right], \label{eq6-6-11a}
\end{split}
\end{equation}
\begin{equation}
\begin{split}
&\sum\limits _{n=-\infty }^{\infty }\frac
1{k_{1s}^2}\left[k_zb_n\^c\cdot \nabla_s\Re g\psi _n(k_{1s}, \v
r_s)+\omega \epsilon _1a_n\^ c\cdot\^ z\times \nabla
_s\Re g\psi _n(k_{1s}, \v r_s)\right]\\
& =\sum\limits _{n=-\infty }^{\infty }\frac
1{k_{0s}^2}\left[k_zd_n\^ c\cdot\nabla_s\psi _n(k_{0s}, \v
r_s)+\omega\epsilon _0c_n\^ c\cdot\^ z\times\nabla
_s\psi_n(k_{0s}, \v r_s)\right]. \label{eq6-6-11b}
\end{split}
\end{equation}
\end{subequations}
Equations (\ref{eq6-6-10a}) and (\ref{eq6-6-11a}) can be written
as
\begin{subequations}
\begin{equation}
\sum\limits _{n=-\infty }^{\infty }\Re g\psi _n(k_{1s}, \v r_s)\v
a_n=\sum\limits_{n=-\infty }^{\infty }\psi _n(k_{0s}, \v r_s)\v
c_n, \quad \v r_s\in C, \label{eq6-6-12a}
\end{equation}
\begin{equation}
\begin{split}
\sum\limits _{n=-\infty }^{\infty } & \frac 1{k_{1s}^2}
\begin{bmatrix}
k_z\^ c\cdot\nabla _s\Re g\psi _n & -\omega\mu _1\^ c\cdot \^
z\times \nabla
_s\Re g\psi _n\\
{\omega\epsilon_1{\^c}\cdot\^ z\times \nabla _s\Re g\psi _n}i
 & {k_z\^ c\cdot\nabla_s\Re g\psi _n}
\end{bmatrix} \cdot \v a_n\\
 & = \sum\limits _{n=-\infty }^{\infty }\frac 1{k_{0s}^2}
\begin{bmatrix}
k_z\^ c\cdot\nabla_s\psi _n & -\omega\mu _0\^ c\cdot\^
z\times\nabla
_s\psi _n\\
\omega\epsilon _0\^ c\cdot\^ z\times \nabla _s\psi _n & k_z\^
c\cdot\nabla _s\psi _n
\end{bmatrix}
 \cdot\v b_n, \quad \v r_s\in C
\label{eq6-6-12b}
\end{split}
\end{equation}
\end{subequations}
where $\v a_n^t=[a_n, b_n]$, $\v c_n^t=[c_n, d_n]$.  We can truncate
the infinite summation to range from $-N$ to $+N$. In this case, the
$a_n$, $b_n$, $c_n$, and $d_n$ will constitute $4(2N+1)$ unknowns.
The method of \index{Point matching method} point matching can be used to convert
(\ref{eq6-6-12a}) and (\ref{eq6-6-12b}) into matrix equations,
\begin{subequations}
\begin{equation}
\sum\limits _{n=-N}^{N}\Re g\psi _n(k_{1s}, \v r_{ms})\v
a_n=\sum\limits _{n=-N}^N\psi _n(k_{0s}, \v r_{ms})\v c_n, \quad
m=-N, \hdots , +N, \label{eq6-6-13a}
\end{equation}
\begin{equation}
\begin{split}
\frac {k_{0s}^2}{k_{1s}^2}\sum\limits _{n=-N}^N &\Re g\dyad \psi
_n(k_{1s}, \v
r_{ms})\cdot\v a_n\\
&=\sum\limits _{n=-N}^N\dyad \psi _n(k_{0s}, \v r_{ms})\cdot\v
c_n,\quad m=-N, \hdots, N
\end{split}
\label{eq6-6-13b}
\end{equation}
\end{subequations}
where
\begin{equation}
\dyad \psi_n=
\begin{bmatrix}
{k_z\^c\cdot\nabla_s\psi_n} &
{-\omega\mu\^c\cdot\^ z\times\nabla _s\psi _n}\\
{\omega \epsilon\^c\cdot\^z\times\nabla_s\psi_n} &
{k_z\^c\cdot\nabla_s\psi_n}
\end{bmatrix}.
\label{eq6-6-14}
\end{equation}

Equations (\ref{eq6-6-13a}) are matrix equations of the form
\begin{subequations}
\begin{equation}
\dyad A\cdot \v a=\dyad A^{(1)}\cdot \v b, \label{eq6-6-15a}
\end{equation}
\begin{equation}
\dyad B\cdot\v a=\dyad B^{(1)}\cdot \v b. \label{eq6-6-15b}
\end{equation}
\end{subequations}
Eliminating $\v a$ from the above yields
\begin{equation}
[\dyad B^{-1}\cdot\dyad B^{(1)}-\dyad A^{-1}\cdot\dyad
A^{(1)}]\cdot\v b=\dyad M\cdot\v b=0. \label{eq6-6-16}
\end{equation}
Since $\dyad A$ and $\dyad B$ are functions of $k_{1s}=\sqrt
{k_1^2-k_z^2}$, and $\dyad A^{(1)}$ and $\dyad B^{(1)}$ are
functions of $k_{0s}=\sqrt {k_0^2-k_z^2}$, the matrix in
(\ref{eq6-6-16}) is a function of $k_z$. Nontrivial solutions
exist for $\v b$, and hence $\v a$, (i.e., the field) only if
\begin{equation}
\det\left(\dyad M(k_z)\right)=0. \label{eq6-6-17}
\end{equation}
Equation (\ref{eq6-6-17}) allows us to solve for the wavenumber
$k_z$ of a guided mode. The above method is in general, applicable
to waveguides of arbitrary shapes. For rectangular waveguides,
symmetry may be exploited to reduce the extend of the summation in
(\ref{eq6-6-12a}), and hence the number of unknowns.

In this method, since we are only assuming standing wave in region
1 and outgoing wave in region 0, it is not valid if the waveguide
is of very distorted shapes.  As mentioned earlier, this method is
similar to the Rayleigh's hypothesis  method of solving scattering
problem \cite{WFIMH}.

Because of the \index{Mode!hybrid} hybrid nature of the modes, and the variation in the
aspect ratio of a rectangular dielectric waveguide, the
classification of modes in a rectangular dielectric waveguide is a
complex subject. The EH$_{mn}$ and HE$_{mn}$ notations have been
adopted by some workers to denote the TM-like and TE-like nature,
respectively, of the modes. However, the EH$_{mn}$ and HE$_{mn}$
notations do not indicate if a mode is $x$-polarized or
$y$-polarized.  Hence, another notation is $E_{mn}^x$ or $E_{mn}^y$
to denote if the mode's electric field is predominantly $x$ or $y$
polarized. Yet, another notation, $H_{mn}^x$ or $H_{mn}^y$, is used
to denote if the magnetic field is predominantly $x$ or $y$
polarized, A combination of $E_{mn}^x$, $E_{mn}^y$, and $H_{mn}^x$,
$H_{mn}^y$ has also been suggested to denote the TM-like or TE-like
nature of a mode. The subscripts $mn$ in the above denote that the
mode has $m$ maxima in the $x$ direction and $n$ maxima in the $y$
direction.

\begin{figure}[ht]
\begin{center}
\hfil\includegraphics[width=4.0truein]{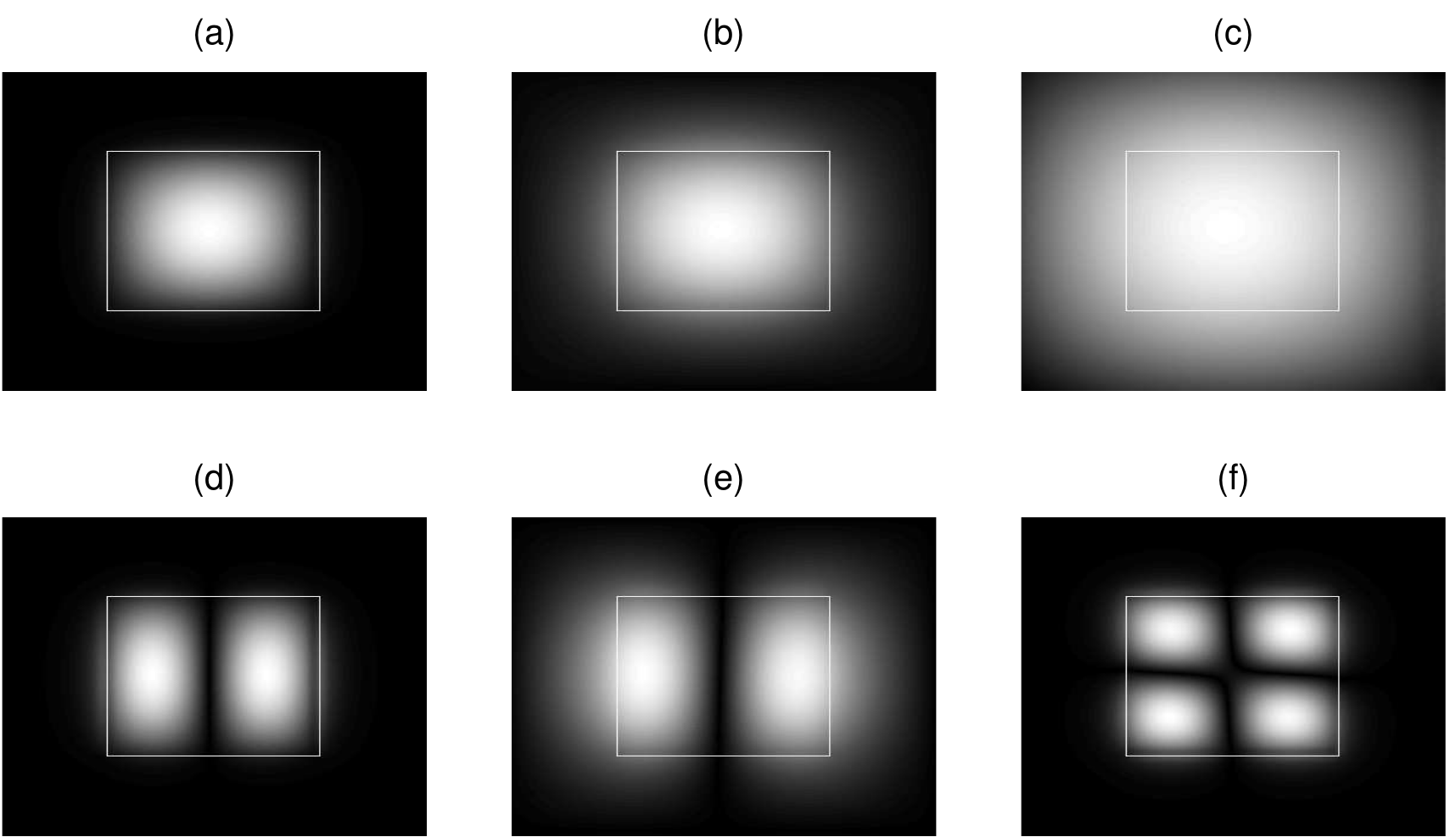}\hfil
\end{center}
\caption{(Top row) Field intensity plots of the $E_{11}^y$ for
different degrees of mode confinement: (a) $P^2=0.91$, (b)
$P^2=0.66$, and (c) $P^2=0.08$. (Bottom row) $E_{12}^y$ for (d)
$P^2=0.85$, (e) $P^2=0.56$, and $E_{22}^y$ mode for (f) $P^2=0.66$
(courtesy of K. Radhakrishnan \cite{RADHAKRISHNAN&CHEW}).}\label{fg663}
\end{figure}


Figure \ref{fg662ac} shows the dispersion curves for a rectangular
dielectric waveguide with \index{Waveguide!rectangular dielectric!dispersion curves}unity aspect ratio, $a/b=1$, and Figure
\ref{fg662b} shows the case when the aspect ratio, $a/b=2$. In the
figures,
$$P^2={[({k_z}/{k_0})^2-1]}/{[({k_1}/{k_0})^2-1]},$$ and
$$B={2b}/{\lambda _0}[({k_1}/{k_0})^2-1]^{1/2}.$$
Hence, $P^2$ is an indication of if the mode is trapped in the
cladding or the core.  If the mode is trapped in the core, then its
energy is mainly in the core, and $k_z$ is close to $k_1$, and
conversely, if its energy is in the cladding.  And $B$ is the
normalized dimension of the waveguide with respect to the
wavelength, and hence, is a normalized frequency.
The \index{Waveguide!rectangular dielectric!unity aspect ratio}unity aspect ratio causes some of the modes to be degenerate. In
Figure \ref{fg663}, the intensity field plot is displayed for
several modes of a rectangular dielectric waveguide.

\subsection {{ Variational Method}}
\index{Variational method}

The harmonic expansion method does not work if the rectangular
waveguide is a part of a substrate. In such a case, a more versatile
method like the variational method,\footnote{This section follows
the analysis in \cite{CHEW&NASIRD}.} or the finite element method
\cite{JIND} should be adopted for an inhomogeneous
waveguide.\footnote{Alternatively, it can be solved by the numerical
mode matching method \cite{CHEW}.} A variational expression for the
propagation constant of a waveguide mode can be derived from the
vector wave equation governing the fields of the waveguide. We have
shown in Chapter 3
that the equations governing the
electromagnetic field in an inhomogeneously filled waveguide are
\begin{equation}
\mu\nabla _s\times\mu ^{-1}\nabla _s\times \v E_s-\nabla
_s\epsilon ^{-1}\nabla _s\cdot\epsilon \v E_s-k^2\v E_s+k_z^2\v
E_s=0, \label{eq6-6-18}
\end{equation}
\begin{equation}
\epsilon\nabla _s\times\epsilon ^{-1}\nabla _s\times \v H_s-\nabla
_s\mu ^{-1}\nabla_s\cdot\mu\v H_s-k^2\v H_s+k_z^2\v H_s=0.
\label{eq6-6-19}
\end{equation}
We can dot multiply (\ref{eq6-6-18}) by $\^z \times \v H_s$ and
integrate the resultant equation over the cross-section of the
waveguide to yield
\begin{equation}
\begin{split}
\int \limits _SdS\^z \times\v H_s & \cdot\mu\nabla _s\times \mu
^{-1}\nabla _s\times \v E_s-\int \limits _SdS\^ z\times \v
H_s\cdot\nabla _s\epsilon
^{-1}\nabla _s\cdot\epsilon\v E_s\\
& -\int\limits _SdS k^2\^z\cdot(\v H_s\times\v
E_s)+k_z^2\int\limits _SdS\^ z\cdot(\v H_s\times\v E_s)=0.
\label{eq6-6-20}
\end{split}
\end{equation}
Using the identity that $\nabla _s\cdot (\v A\times \v B)=\v
B\cdot\nabla_s\times\v A-\v A\cdot\nabla _s\times \v B$, we have
\begin{equation}
\begin{split}
\^ z\times \v H_s\cdot\mu\nabla _s\times\mu ^{-1}\nabla _s\times\v
E_s &
=\nabla _s\cdot [(\mu^{-1}\nabla _s\times\v E_s)\times (\^ z\times\mu \v H_s)]\\
& \qquad\quad+\nabla
_s\times (\^ z\times\mu\v H_s)\cdot\mu ^{-1}\nabla _s\times\v E_s\\
& =\nabla _s\cdot [(\mu^{-1}\nabla _s\times \v E_s)\times
(\^ z\times\mu\v H_s)]\\
& \qquad\quad+(\nabla _s\cdot\mu\v H_s)\^ z\cdot \mu ^{-1}\nabla
_s\times \v E_s. \label{eq6-6-21}
\end{split}
\end{equation}
Using $\nabla _s\cdot\phi\v A=(\nabla _s\phi )\cdot \v
A+\phi\nabla _s\cdot\v A$, we have
\begin{equation}
\^ z\times \v H_s\cdot \nabla _s\epsilon ^{-1}\nabla
_s\cdot\epsilon \v E_s=\nabla _s\cdot [(\^ z\times \v H_s)\epsilon
^{-1}\nabla _s\cdot\epsilon\v E_s]-\nabla _s\cdot (\^ z\times \v
H_s)\epsilon ^{-1}\nabla _s\cdot\epsilon \v E_s. \label{eq6-6-22}
\end{equation}
Equations (\ref{eq6-6-21}) and (\ref{eq6-6-22}) can be used to
simplify the first two integrals in (\ref{eq6-6-20}). After
substituting them into (\ref{eq6-6-20}), the divergence can be
converted into a line integral on $C$, the outermost domain of the
waveguide. This integral becomes zero by virtue of the boundary
condition on $C$, or when $C\rightarrow \infty$. Consequently,
Equation (\ref{eq6-6-20}) becomes
\begin{equation}
\begin{split}
\int \limits _SdS(\nabla _s\cdot\mu\v H_s)\^ z & \cdot\mu
^{-1}\nabla _s\times\v E_s-\int \limits _SdS\^ z\cdot\nabla
_s\times\v H_s\epsilon ^{-1}\nabla
_s\cdot\epsilon\v E_s\\
& -\int \limits _SdSk^2\^ z\cdot (\v H_s\times\v E_s)+k_z^2\int
dS\^ z\cdot (\v H_s\times\v E_s)=0. \label{eq6-6-23}
\end{split}
\end{equation}
Applying the same operation to (\ref{eq6-6-19}), or by duality, we
have
\begin{equation}
\begin{split}
\int \limits _SdS & (\nabla _s\cdot\epsilon\v E_s)\^ z \cdot
\epsilon ^{-1}\nabla_s\times\v H_s -\int\limits _SdS\^ z\cdot
\nabla _s\times \v E_s\mu
^{-1}\nabla _s\cdot\mu\v H_s\\
& -\int\limits _SdSk^2\^ z\cdot (\v E_s\times \v H_s)+k_z^2\int
dS\^ z\cdot (\v E_s\times \v H_s)=0. \label{eq6-6-24}
\end{split}
\end{equation}

We note that Equations (\ref{eq6-6-23}) and (\ref{eq6-6-24}) are
identical. If we write Equations (\ref{eq6-6-18}) and
(\ref{eq6-6-19}) as
\begin{equation}
{\cal {L}}_e\cdot\v E_s+k_z^2\v E_s=0, \label{eq6-6-25}
\end{equation}
\begin{equation}
{\cal {L}}_h\cdot \v H_s+k_z^2\v H_s=0 \label{eq6-6-26}
\end{equation}
where ${\cal {L}}_e$ and ${\cal {L}}_h$ are the differential
operators in (\ref{eq6-6-18}) and (\ref{eq6-6-19}), then,
Equations (\ref{eq6-6-23}) and (\ref{eq6-6-24}) are the
consequences of
\begin{equation}
\langle\^ z\times \v H_s, {\cal {L}}_e\cdot\v E_s\rangle
+k_z^2\langle\^ z\times\v H_s, \v E_s\rangle =0, \label{eq6-6-27}
\end{equation}
\begin{equation}
\langle\^ z\times \v E_s, {\cal {L}}_h\cdot\v H_s\rangle
+k_z^2\langle\^ z\times\v E_s, \v H_s\rangle =0. \label{eq6-6-28}
\end{equation}
Hence (\ref{eq6-6-25}) and (\ref{eq6-6-26}) are transpose equation
of each other. A variational expression for $k_z^2$ is
\begin{equation}
k_z^2=-\frac {\langle \^ z\times \v H_s, {\cal {L}}_e\cdot \v
E_s\rangle }{\langle \^ z\times\v H_s, \v E_s\rangle }=-\frac
{\langle \^ z\times\v E_s, {\cal {L}}_h\cdot\v H_s\rangle
}{\langle \^ z\times\v E_s, \v H_s\rangle }. \label{eq6-6-29}
\end{equation}
In the above, $\langle \v A, \v B\rangle =\int\limits _SdS\v
A\cdot\v B$. The above also imply that the $(H_x, H_y)$ formulation
is the same as the $(E_x, E_y)$ formulation, if solved
variationally.

We can take the first variation of (\ref{eq6-6-29}) by letting $\v
E_s=\v E_{se}+\delta \v E_s$, $\v H_s=\v H_{se}+\delta \v H_s$,
where $\v E_{se}$ and $\v H_{se}$ are the exact solutions. Then,
after cross-multiplying the first equation in (\ref{eq6-6-29}) and
taking its first variation, we have
\begin{equation}
\begin{split}
\langle\^ z & \times\v H_{se}, {\cal {L}}_e \cdot\v E_{se}\rangle
+\langle\^ z\times\delta \v H_s, {\cal {L}}_e\cdot\v E_{se}\rangle
+\langle \^ z\times\v
H_{se}, {\cal {L}}_e\cdot\delta \v E_s\rangle\\
& +k_{ze}^2[\langle \^ z\times\v H_{se}, \v E_{se}\rangle +\langle
\^ z\times\delta\v H_s, \v E_{se}\rangle +\langle \^ z\times\v
H_{se}, \delta\v
E_s\rangle ]\\
& +\delta k_z^2\langle \^ z\times\v H_{se}, \v E_{se}\rangle
+\hdots =0. \label{eq6-6-30}
\end{split}
\end{equation}
The leading order terms cancel as a consequence of
(\ref{eq6-6-27}). Similarly, as a consequence of (\ref{eq6-6-25}),
\begin{equation}
\langle \^ z\times\delta\v H_s, {\cal {L}}_e\cdot\v E_{se}\rangle
+k_{ze}^2\langle\^ z\times \delta \v H_s, \v E_{se}\rangle =0.
\label{eq6-6-31}
\end{equation}
Because (\ref{eq6-6-23}) and (\ref{eq6-6-24}) are identical, we
can show that
\begin{equation}
\langle\^ z\times\v A, {\cal {L}}_h\cdot\v H_s\rangle =-\langle \^
z\times\v H_s, {\cal {L}}_e\cdot \v A\rangle \label{eq6-6-32}
\end{equation}
where $\v A$ is an arbitrary vector satisfying the boundary
conditions on $C$, or if it vanishes when $C\rightarrow \infty$.
As a result,
\begin{equation}
\langle\^ z\times \v H_{se}, {\cal {L}}_e\cdot \delta\v E_s\rangle
+k_{ze}^2\langle\^ z\times \v H_{se}, \delta \v E_s\rangle=0,
\label{eq6-6-33}
\end{equation}
and $\delta k_z^2=0$. Therefore, the first order variation in
$k_z^2$ vanishes, implying the stationarity of (\ref{eq6-6-29}).
The same thing can be shown for the other equation in
(\ref{eq6-6-29}).

Since (\ref{eq6-6-29}) is variational, a Rayleigh-Ritz procedure
can be adopted to obtain an optimal solution for it. We let
\begin{equation}
\v E_s=\sum\limits _{n=1}^Na_n\v E_{ns}, \quad \v H_s=\sum\limits
_{m=1}^Nb_m\v H_{ms}. \label{eq6-6-34}
\end{equation}
Substituting into (\ref{eq6-6-29}), we have
\begin{equation}
k_z^2=-\frac {\sum\limits _{n=1}^N\sum\limits _{m=1}^N
a_nb_m\langle \^ z\times\v H_{ms}, {\cal {L}}_e\cdot\v
E_{ns}\rangle }{\sum\limits _{n=1}^N\sum\limits _{m=1}^N
a_nb_m\langle \^ z\times\v H_{ms}, \v E_{ns}\rangle },
\label{eq6-6-35}
\end{equation}
or
\begin{equation}
k_z^2=-\frac {\v b^t\cdot\dyad A\cdot\v a}{\v b^t\cdot \dyad
M\cdot\v a} \label{eq6-6-36}
\end{equation}
where the $mn$ element of the matrices $\dyad A$ and $\dyad M$ are
\begin{subequations}
\begin{equation}
A_{mn}=\langle \^ z\times\v H_{ms}, {\cal {L}}_e\cdot\v
E_{ns}\rangle, \label{eq6-6-37a}
\end{equation}
\begin{equation}
M_{mn}=\langle \^ z\times\v H_{ms}, \v E_{ns}\rangle.
\label{eq6-6-37b}
\end{equation}
\end{subequations}
The optimal values of $\v a$ and $\v b$ in (\ref{eq6-6-36}) are
obtained by requiring the first variation of $k_z^2$ to vanish, or
that
\begin{equation}
\begin{split}
\v b_0^t\cdot\dyad A\cdot\v a_0 & +k_{z0}^2\v b_0^t\cdot\dyad
M\cdot\v a_0
+\delta\v b^t\cdot\dyad A\cdot\v a_0+k_{z0}^2\delta\v b^t\cdot\dyad M\cdot \v a_0\\
& +\v b_0^t\cdot \dyad A\cdot \delta\v a+k_{z0}^2\v
b_0^t\cdot\dyad M\cdot\delta\v a=0. \label{eq6-6-38}
\end{split}
\end{equation}
The leading order terms cancel each other by virtue of
(\ref{eq6-6-36}). The first order term vanishes if
\begin{subequations}
\begin{equation}
\dyad A\cdot\v a_0+k_{z0}^2\dyad M\cdot \v a_0=0,
\label{eq6-6-39a}
\end{equation}
\begin{equation}
\dyad A^t\cdot\v b_0+k_{z0}^2\dyad M^t\cdot\v b_0=0.
\label{eq6-6-39b}
\end{equation}
\end{subequations}
Equations (\ref{eq6-6-39a}) and (\ref{eq6-6-39b}) are matrix
eigenvalue problems. They have the same set of eigenvalues.
Hence, we need only to solve one of them. If $\dyad A$ and $\dyad
M$ are $N\times N$ matrices, in general, there will be $N$
eigenvalues and $N$ eigenvectors $\v a$ and $\v b$.  It can be
shown easily that for two eigenvectors $\v a_i$ and $\v b_j$
corresponding to two distinct eigenvalues,
\begin{equation}
\v b_j^t\cdot\dyad M\cdot\v a_i=D_i\delta _{ij}. \label{eq6-6-40}
\end{equation}
In other words, they are $\dyad M$ orthogonal.

\begin{figure}[ht]
\begin{center}
\hfil\includegraphics[width=3.truein]{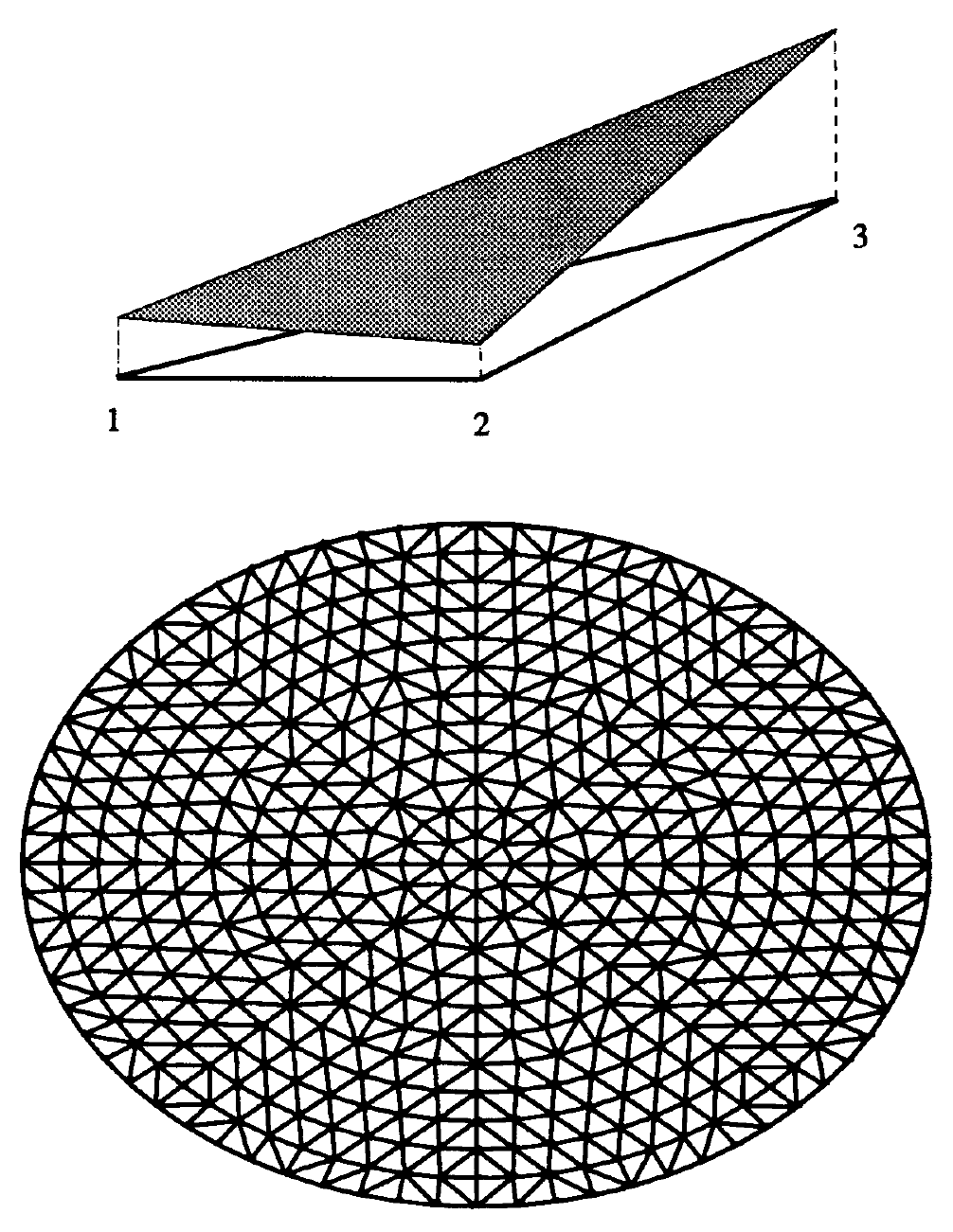}\hfil
\end{center}
\caption{A basis function and a \index{Finite element mesh} finite element mesh.}\label{fg664}
\end{figure}


Equations (\ref{eq6-6-39a}) and (\ref{eq6-6-39b}) are exactly the
equations one would obtain if one applies the Petrov-Galerkin
method, or the method of weighted residuals to the differential
Equations (\ref{eq6-6-25}) and (\ref{eq6-6-26}) using $\v E_{ns}$
and $\v H_{ns}$ as expansion functions, respectively, and using $\^
n\times\v H_{ms}$ and $\^ n\times\v E_{ms}$ as weighting functions,
respectively. Equations (\ref{eq6-6-27}) and (\ref{eq6-6-28}) are
also the variational integrals to be used in the finite element
method. In finite element, a finite domain basis function is used.
For example, the fields $E_x$ and $E_y$ can be written as a linear
superposition of pyramidal functions, with polygonal base. The
pyramids overlap with each other, and they form a piecewise linear
approximation of the field between the nodal values. Alternatively,
edge elements can be used to model the electric field where
tangential components of the field are guaranteed to be continuous
across edges \cite{JIND}.  Examples of a finite element mesh and a
pyramidal function are shown in Figure \ref{fg664}.

\section {{{ Discontinuities in Dielectric Waveguides}}}
\index{Waveguide!dielectric!discontinuities}

Discontinuities in dielectric waveguides have been studied by a
number of workers
\cite{ANGULO,IKEGAMI,ROZZI,PUDENSI&FERREIRA,CHEW,LIU&CHEWD,HERZINGERetal,WFIMH}.
We have previously studied discontinuities in closed, hollow
waveguides. There, we used the mode matching method to derive the
solution of wave scattering by discontinuities. Because we have
closed waveguides, only discrete modes exist \cite{WFIMH}. However,
when a waveguide is open, the number of modes that a waveguide has
is uncountably infinite. Furthermore, there exists a set of modes
which forms a continuum of modes. These modes carry energy to
infinity and hence, are called the radiation modes. We shall address
the \index{Mode matching method} mode matching method for such an open waveguide. This method is
important, for instance, in ascertaining reflection loss at the
facet of a heterojunction laser \cite{IKEGAMI,HERZINGERetal}.

\subsection { Reflection at a Laser Facet}
\index{Reflection at laser facet}

\begin{figure}[ht]
\begin{center}
\hfil\includegraphics[width=2.40truein]{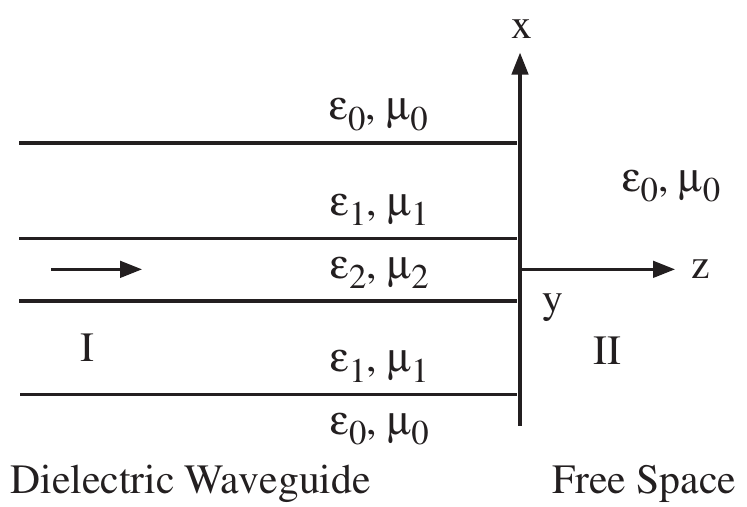}\hfil
\end{center}
\caption{A laser facet where a mode is reflected.}\label{fg671}
\end{figure}

Consider a laser facet as shown in Figure \ref{fg671}. The solution
to this problem is important for the design of lasers as the
reflectivity at the facet of a laser cavity determines the $Q$ of
the laser cavity. The reflectivity of such a facet can be found by
mode-matching.

Consider a TE polarized mode with electric field polarized in the
$y$ direction. A part of the energy of the mode will be
transmitted yielding radiation modes for $z>0$.  For $z< 0$, the
mode will be reflected. Moreover, part of the energy of the
reflected mode will be converted into other reflected modes giving
rise to \index{Waveguide!mode conversion} ``mode conversion,'' just as discontinuities in uniform
waveguides (see Chapter 5).

Assume that the incident mode to be of the form
\begin{equation}
E_{iy} = a_m E_{y}(m,x) e^{ik_{mz}z} \label{eq6-7-1}
\end{equation}
where $E_y (m,x)$ describes the transverse field distribution of
the $m$-th mode, and $k_{mz}$ is its corresponding wave number in
the $z$ direction. Then, the reflected modes can be expressed as
\begin{equation}
E_{ry} = \sum \limits _{m'=1}^{\infty} R_{m'm} a_m E_{y} (m',x)
e^{-ik_{m'z}z}, \label{eq6-7-2}
\end{equation}
where $R_{m'm}$ is a reflection operator which is the unknown to
be sought. Its off-diagonal components account for the physics of
mode conversion.

For $z>0$, the field can be expressed as
\begin{equation}
E_{ty} = \int \limits_{-\infty}^{\infty} dk_xe ^{ik_xx + ik_z z}
e_t(k_x), \label{eq6-7-3}
\end{equation}
where $ k_z = \sqrt{k^2- k_x^2}$. For a fixed $z$, the above is
essentially a Fourier transform integral.

At this point, $R_{m'm}$ and $e_{t}(k_x)$ are unknowns yet to be
sought. They can be found by matching boundary conditions at
$z=0$. Requiring that tangential $\v E$ field be continuous at
$z=0$, i.e., $E_{iy} + E_{ry} = E_{ty}$, we have
\begin{equation}
E_{y}(m,x) + \sum \limits_{m'=1}^{\infty} R_{m'm} a_m E_{y}(m',x)
= \int \limits _{-\infty}^{\infty} dk_x e^{ik_xx} e_t (k_x)
\label{eq6-7-4}
\end{equation}
The magnetic field is derived from Faraday's law $\nabla \times \v
E = i \omega \mu \v H$. That is, the tangential magnetic field is
$H_x$ and is derivable from
\begin{equation}
H_x = \frac{-1}{i\omega\mu(x)} \frac{\partial}{\partial z} E_y
\label{eq6-7-5}
\end{equation}
Consequently, in region I, we have
\begin{equation}
\begin{split}
H_{ix} + H_{ry} &= - \frac{k_{mz}}{\omega \mu (x)} a_m
E_{y}(m,x)e^{ik_{mz}z}\\& + \sum \limits_{m'=1}^{\infty} R_{m'm}
a_m \frac{k_{m'z}}{\omega\mu (x)} E_{y} (m',x)e^{-ik_{m'z}z},
\end{split}
\label{eq6-7-6}
\end{equation}
and in region II, we have
\begin{equation}
H_{ty} = - \int \limits _{-\infty}^{\infty} dk_x \frac{k_z}{\omega
\mu_0 } e^{ik_xx + ik_zz} e_t(k_x). \label{eq6-7-7}
\end{equation}
Matching boundary condition for magnetic field at $z=0$, we have
\begin{equation}
\begin{split}
\frac{k_{mz}}{\omega \mu (x)}a_m E_{y}(m,x) &- \sum \limits
_{m'=1}^{\infty}
R_{m'm}a_m \frac{k_{m'z}}{\omega \mu (x)} E_{y}(m',x)\\
& = \int \limits _{-\infty}^{\infty}dk_x \frac{k_z}{\omega\mu_0}
e^{ik_xx}e_t(k_x).
\end{split}
\label{eq6-7-8}
\end{equation}
By Fourier inverse transforming (\ref{eq6-7-4}),\footnote{This is
the same as testing the above equation with $e^{-ik_xx}$ and
integrate.} we obtain that
\begin{equation}
\tilde E_{y}(m,k_x) + \sum \limits _{m'=1}^{\infty} R_{m'm} \tilde
E_{y} (m',k_x) = e_t (k_x) \label{eq6-7-9}
\end{equation}
where
\begin{equation}
\tilde E_{y} (m,k_x) = \int \limits_{-\infty}^{\infty}
dxe^{-ik_xx} E_{y} (m,x). \label{eq6-7-10}
\end{equation}

The \index{Mode orthogonality} mode orthogonality relationship for an inhomogeneous waveguide
is that
\begin{equation}
-\int \limits_{-\infty}^{\infty} dxE_{y}(n,x)H_{x}^*(m,x) = C_n
\delta _{nm} \label{eq6-7-11}
\end{equation}
Since $H_{x}(m,x) = - \frac{k_{mz}}{\omega\mu}E_{y}(m,x)$, this is
equivalent to
\begin{equation}
\int \limits_{-\infty}^{\infty} dx \frac{k^*_{mz}}{\omega \mu(x)}
E_{y}(n,x) E_{y}^*(m,x) = C_n \delta _{nm}. \label{eq6-7-12}
\end{equation}
For a lossless waveguide, $\mu$ is real, and we can normalize the
modes such that
\begin{equation}
\int \limits_{-\infty}^{\infty} dx\frac{E_{y}(n,x)
E_{y}^{*}(m,x)}{\mu (x)} = \delta_{nm}. \label{eq6-7-13}
\end{equation}
Then, $C_n$ in (\ref{eq6-7-12}) is $k_{mz}^*/\omega$.

Multiplying (\ref{eq6-7-8}) by $E_{y}^*(n,x)$ and integrating over
$x$, we have
\begin{equation}
\begin{split}
a_m k_{mz}\delta _{nm} &- \sum \limits _{m'=1}^{\infty} R_{m'm}
a_m
 k_{m'z} \delta
_{nm'}\\
&= \int \limits_{-\infty}^{\infty} dk_x k_z e_t (k_x) \left
[\int\limits
_{-\infty}^{\infty} dxe^{-ik_xx} \frac{E_{y}(n,x)}{\mu _0} \right ] ^*\\
&=\int \limits_{-\infty}^{\infty} dk_xk_z e_t (k_x) \tilde
e_{y}^*(n,k_x) \label{eq6-7-14}
\end{split}
\end{equation}
where
\begin{equation}
\tilde e_{y} (n,k_x) = \int \limits_{-\infty}^{\infty}
dxe^{-ik_xx} \frac{E_{y}(n,x)}{\mu_0}. \label{eq6-7-15}
\end{equation}
Using (\ref{eq6-7-9}) for $e_t(k_x)$ in (\ref{eq6-7-14}), we have
\begin{equation}
\begin{split}
a_m k_{mz} \delta_{nm} &- R_{nm} a_m k_{nz} = \int \limits
_{-\infty}^{\infty} dk_x k_z
\tilde E_{y} (m,k_x) \tilde e_{y}^* (n,k_x)\\
& + \sum \limits_{m'=1}^{\infty} R_{m'm}a_m
\int\limits_{-\infty}^{\infty} dk_xk_z\tilde E_{y} (m',k_x) \tilde
e_{y}^* (n,k_x). \label{eq6-7-16}
\end{split}
\end{equation}
The series summation in (\ref{eq6-7-16}) can be truncated and
(\ref{eq6-7-16}) can then be solved as a matrix equation for the
unknown $R_{m'm}$. By so doing, we obtain
\begin{equation}
\dyad K_z\cdot \v a - \dyad K_z\cdot \dyad R \cdot \v a =\dyad
A\cdot \v a  + \dyad A\cdot \dyad R\cdot\v a, \label{eq6-7-17}
\end{equation}
where
$$
A_{nm}=\int\limits_{-\infty}^{\infty} dk_xk_z\tilde E_{y} (m,k_x)
\tilde e_{y}^* (n,k_x).
$$
Equation (\ref{eq6-7-17}) can be solved easily for $\dyad R\cdot
\v a$ or $\dyad R$.

The analysis above assumes that all the modes in the dielectric
waveguide region are discrete. In actual fact, continuum modes
exist and the discrete summations in (\ref{eq6-7-2}) will have to
be augmented by a continuous summation which is an integral. An
analysis involving such an integral is difficult and the continuum
modes can be discretized by putting metallic boundaries far away
from the dielectric waveguide.

\subsection { Determination of the Modes}
\index{Mode!determination}

The modes in the dielectric waveguide region are the natural
solution of the wave equation \cite{WFIMH}
\begin{equation}
\left [ \mu_r \frac{\partial}{\partial x} \mu_r^{-1}
\frac{\partial}{\partial x} + k^2 (x) - k_z^2 \right ] E_y = 0.
\label{eq6-7-18}
\end{equation}

\clearpage

If the dielectric waveguide has a piecewise constant or step
profile, these modes can be found in closed form.

\begin{figure}[ht]
\begin{center}
\hfil\includegraphics[width=2.40truein]{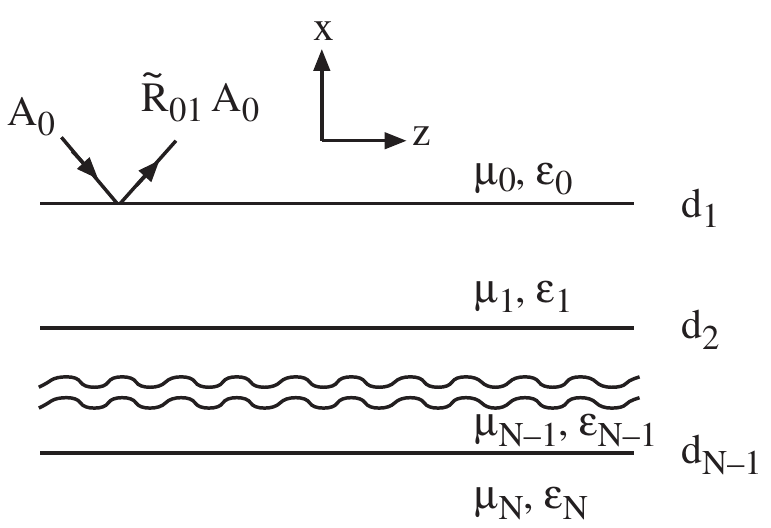}\hfil
\end{center}
\caption{Waves in a layered medium.}\label{fg672}
\end{figure}


For a particular mode, due to phase matching, the waves in all
regions have $e^{ik_z z}$ dependence. Hence, in the $i$-th layer,
\begin{equation}
E_{iy} = e_{iy}(x) e^{ik_zz}. \label{eq6-7-19}
\end{equation}
Moreover, since each layer is homogeneous, $e_{iy}(x) $ is a
linear superposition of upgoing and downgoing waves. More
specifically,
\begin{equation}
e_{iy}(x) = A_i \left [ e^{-ik_{ix}x} + \tilde R_{i, i+1}
e^{2ik_{ix}d_i + ik_{ix}x} \right]. \label{eq6-7-20}
\end{equation}
The generalized \index{Fresnel reflection coefficient} Fresnel reflection coefficient can be found
recursively via
\begin{equation}
\tilde R_{i, i+1} = \frac{R_{i,i+1} + \tilde R_{i+1,
i+2}e^{2ik_{i+1, x}(d_{i+1} - d_i)}}{1+R_{i,i+1}\tilde
R_{i+1,i+2}e^{2ik_{i+1,x}(d_{i+1}-d_i)}} \label{eq6-7-21}
\end{equation}
where $R_{ij}$ is the local Fresnel reflection coefficient. The
amplitudes $A_i$ can be found via the recursion relation
\begin{equation}
A_i e^{ik_{iz}d_{i-1}} =
\frac{T_{i-1,i}A_{i-1}e^{ik_{i-1,z}d_{i-1}}}{1- R_{i,i-1}\tilde
R_{i,i+1} e^{2ik_{iz}(d_i- d_{i-1})}} \label{eq6-7-22}
\end{equation}

A guided mode is defined as a solution to (\ref{eq6-7-18}) without
an external excitation. Hence, it corresponds to a nonzero $\tilde
R_{01}$ even when $A_0$, the amplitude of the external excitation,
is zero. In other words, the guided modes by the layered region
correspond to the poles of $\tilde R_{01}$. Hence, they can be
found by searching for the roots of $\left[\tilde
R_{01}(k_z)\right]^{-1}$.

The continuum modes are called \index{Mode!radiation} radiation modes, so called because
they carry energy to infinity. There are two classes of radiation
mode, one with real $k_{0x}$, and the other with real $k_{Nx}$.
For the radiation modes with real $k_{0x}$, their expression in
region 0 is
\begin{equation}
e_{0y}(x) = A_0\left[e^{-ik_{0x}x} + \tilde R_{01}e^{2ik_{0x}d_0 +
ik_{0x}x}\right], \label{eq6-7-23}
\end{equation}
where $k_{0x}^2 + k_z^2 = k_0^2$. Here, $k_{0x}$ has to be real in
order for $e_{0y}(x)$ to be bounded when $x \to \infty$.  Notice
that in order for $k_{0x}$ to be real, $k_z$ lies along the locus
as shown in the complex $k_z$ plane.  The field of the radiation
mode in every layer can be found by using the recursion relation
as before.

\begin{figure}[ht]
\begin{center}
\hfil\includegraphics[width=2.3truein]{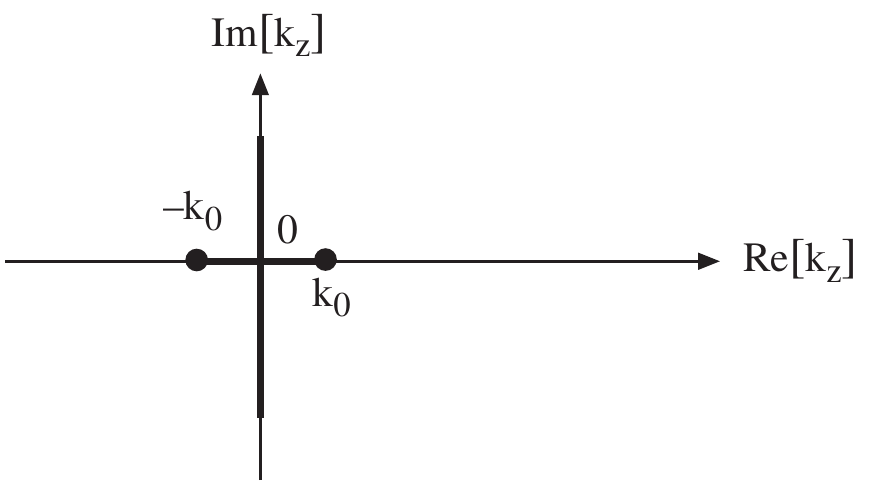}\hfil
\end{center}
\caption{The distribution of $k_z$ corresponding to the radiation
modes with real $k_{ox}$.}\label{fg673}
\end{figure}


The other class of radiation modes with real $k_{Nx}$, has both
upgoing and downgoing wave in region $N$. Their expression in
region $N$ is
\begin{equation}
e_{Ny} (x) = A_{N}\left[e^{ik_{Nx} x}+ \tilde R_{N, N-1} e
^{-2ik_{Nx}d_{N-1} - ik_{Nx}x}\right] \label{eq6-7-24}
\end{equation}
Their field in every layer can be found by a similar recursive
relation. Hence, in general, the radiation modes of an open
dielectric waveguide is expressible as
\begin{equation}
E_y (x,y) = \int \limits_{0}^{\infty} dk_{0x} e_y(x,k_{0x})e^{ik_z
z} + \int \limits _{0}^{\infty} dk_{Nx} e_y (x, k_{Nx}) e^{ik_zz}.
\label{eq6-7-25}
\end{equation}

\section {{{ Analyzing Weak Contrast Optical Fiber with WKB method}}}
\index{Optical fiber!weak contrast}
\index{Optical fiber!WKB method}

When the profile of an optical fiber is slowly varying and the
radius of the core is large compared to wavelength, the WKB
(Wentzel, Kramer and Brillouin) method can be applied to the
analysis of the optical fiber \cite{OKOSHID,SENIOR}.  The WKB method
is often discussed in many books on quantum mechanics, and also
discussed in \cite{WFIMH}.

If $\phi$ in Equation (\ref{eq6-3-6}) is written as
\begin{equation}
\phi(\v r) = R(\rho)e^{ik_zz+in\phi}, \label{eq6-8-1}
\end{equation}
then the equation for $R(\rho)$ is
\begin{equation}
\frac{d ^2}{d \rho ^2} R(\rho) + \frac{1}{\rho} \frac{d}{d \rho}
R(\rho) - \frac{n^2}{\rho ^2} R(\rho) + [k^2(\rho) - k_z^2]R(\rho)
= 0. \label{eq6-8-2}
\end{equation}
By letting

\begin{equation}
\hat R(\rho) = \sqrt {\rho} R(\rho), \label{eq6-8-3}
\end{equation}
if follows that
\begin{equation}
\begin{split}
\frac{d^2 \hat R(\rho)}{d \rho^2} &= \frac{d}{d \rho} \left [
\frac{1}{2} \rho
^{-\frac{1}{2}}R(\rho) + \rho^{\frac{1}{2}} R'(\rho)\right]\\
&= -\frac{1}{4} \rho^{- \frac {3}{2}} R(\rho) +
\rho^{-\frac{1}{2}}R' (\rho) + \rho^{\frac{1}{2}} R^{\prime\prime}
(\rho). \label{eq6-8-4}
\end{split}
\end{equation}
Multiplying (\ref{eq6-8-2}) by $\rho^{\frac{1}{2}}$, and using
(\ref{eq6-8-4}) in the resultant (\ref{eq6-8-2}), we have
\begin{equation}
\frac{d^2\hat R (\rho)}{d \rho^2} + \left [E-V(\rho) \right ]\hat
R(\rho) = 0 \label{eq6-8-5}
\end{equation}
where
\begin{equation}
\begin{split}
E &= k^2 (\infty) - k_z^2,\\
V(\rho) &= k^2(\infty) - k^2 (\rho) +\frac {\left (n^2 -
\frac{1}{4}\right )}{\rho^2}. \label{eq6-8-6}
\end{split}
\end{equation}
Note that $V(\rho) \to 0$, when $\rho \to \infty$. Equation
(\ref{eq6-8-5}) is the same as the Schr\"odinger's equation for
describing the motion of a particle in quantum mechanics. Here,
$V(\rho)$ is the potential well, and $E$ is the energy of the
particle.  A particle is bound in the potential well when $E< 0$ ,
or when $k_z$ is real and that $k_z > k(\infty)$. This corresponds
to a guided mode in the optical fiber, because $k_\rho(\infty) =
\sqrt{k^2(\infty) - k_z ^2} = i\alpha (\infty)$. When $n=0$, the
function $V(\rho)$ is as shown in Figure 1.  When $n$ is large,
$V(\rho)$ may not be negative at all, and no bound state or guided
mode can exist. When $E>0$, $k_z < k(\infty)$, the mode is not bound
and it radiates energy to infinity and becomes a radiation mode.
This is because $\alpha (\infty)$ is not real anymore and the field
is not evanescent outside the fiber.

\begin{figure}[ht]
\subfigure[]
\hfil\includegraphics[width=2.5truein]{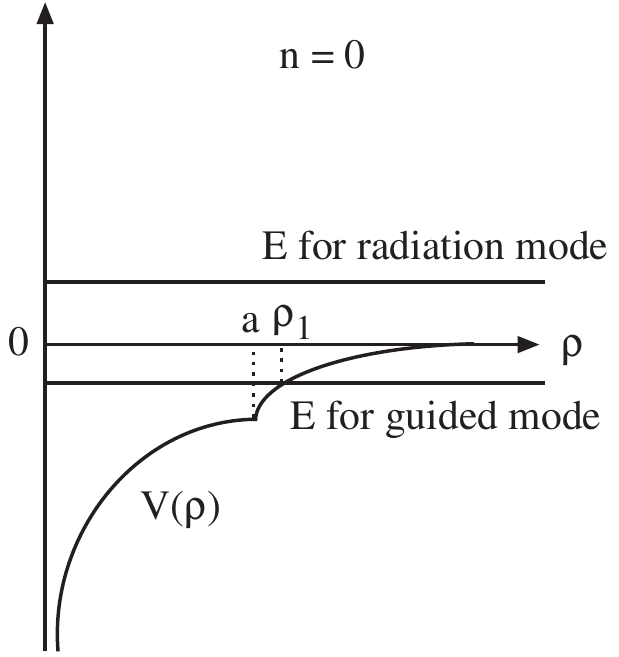}\hfil
\subfigure[]
\hfil\includegraphics[width=2.5truein]{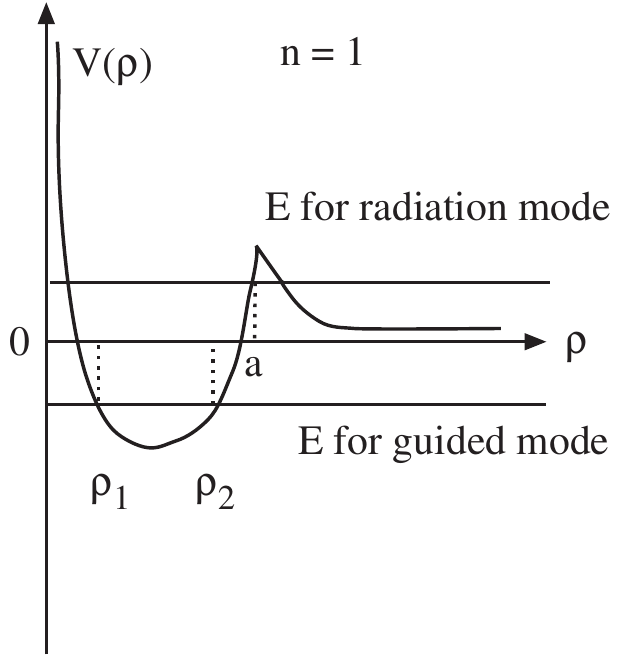}\hfil
\caption{$V(\rho)$ for $n=0$ and $n=1$.}\label{f681}
\end{figure}


\subsection { The WKB Method}
\index{WKB method}

To analyze Equation (\ref{eq6-8-5}) with the WKB method
\cite{WFIMH}, we rewrite it as
\begin{equation}
\hat R ^{\prime\prime} (\rho) + k_\rho^2 (\rho) \hat R (\rho) = 0
\label{eq6-8-8}
\end{equation}
where $ k_\rho^2 (\rho) = E - V (\rho)$. Note that $k_\rho^2 \sim
\omega ^2$ when $\omega \to \infty$. Hence $k_\rho^2$ becomes a
large parameter in the high frequency limit. By foreseeing that
the solution of (\ref{eq6-8-8}) may look like a plane wave, we let
\begin{equation}
\hat R(\rho) = Ae^{i\omega \tau (\rho)}. \label{eq6-8-9}
\end{equation}
Then it follows that
\begin{equation}
\hat R^{\prime\prime} (\rho) = \left\{i\omega \tau^{\prime\prime}
(\rho) - [ \omega\tau ' (\rho)]^2 \right\} Ae^{i\omega \tau
(\rho)}, \label{eq6-8-10}
\end{equation}
and (\ref{eq6-8-8}) becomes
\begin{equation}
i \omega \tau '' (\rho) - [ \omega \tau ' (\rho)]^2 + k^2_\rho
(\rho) = 0. \label{eq6-8-11}
\end{equation}
Using the perturbation method, we expand $\tau(\rho)$ in a
perturbation series, namely,
\begin{equation}
\tau (\rho) = \tau _0 (\rho) + \frac{1}{\omega} \tau_1 (\rho) +
\dots , \qquad \omega \to \infty \label{eq6-8-12}
\end{equation}
where we have used $1/\omega$ as the small parameter.  Note that
\begin{equation}
i \omega \tau _0'' (\rho) \ll k_\rho^2 (\rho) , \qquad \omega \to
\infty \label{eq6-8-13}
\end{equation}
because $k_\rho^2 \sim \omega ^2$. Consequently, using
(\ref{eq6-8-12}) in (\ref{eq6-8-11}), and collecting leading order
terms when $\omega \to \infty$, we have
\begin{equation}
\left[\omega \tau_0' (\rho)\right]^2 = k^2_\rho (\rho) = \omega ^2
s_\rho^2 (\rho) \label{eq6-8-14}
\end{equation}
where we define $s_\rho = k_\rho/ \omega$ to be the slowness of a
wave.  The above is known as the \index{Eikonal function} eikonal equation.  Consequently, on
solving (\ref{eq6-8-14}), we arrive at
\begin{equation}
\tau _0 (\rho) = \pm \int \limits _{\rho_0}^{\rho} d\rho' s_\rho
(\rho') + C_0. \label{eq6-8-15}
\end{equation}

Next, by collecting first order terms in (\ref{eq6-8-11}) after
substitution with (\ref{eq6-8-12}), we have
\begin{equation}
i \omega \tau _0 '' (\rho) - 2 \omega \tau _0' (\rho) \tau _1'
(\rho) = 0. \label{eq6-8-16}
\end{equation}
The above is known as the \index{Transport equation} transport equation.   Solving this yields
\begin{equation}
\tau_1 (\rho) = \frac{i}{2} \ln \tau _0 ' (\rho) + C_1 = \frac
{i}{2} \ln s_\rho (\rho) + C_{1\pm}. \label{eq6-8-17}
\end{equation}
Consequently, using (\ref{eq6-8-15}) and (\ref{eq6-8-17}) in
(\ref{eq6-8-12}), we have
\begin{equation}
\tau (\rho) = \pm \int \limits _{\rho_0}^{\rho} d \rho ' s_\rho
(\rho ') + \frac{i}{2 \omega} \ln s_{\rho}(\rho) + C_{\pm} ,
\label{eq6-8-18}
\end{equation}
or that the approximate solution to (\ref{eq6-8-5}) is of the form
\begin{equation}
\hat R (\rho) \sim \frac{A_+}{\sqrt {s_\rho}} \exp\left({i \omega
\int \limits_{\rho_0}^{\rho} s_\rho (\rho') d \rho'}\right) +
\frac {A_-}{\sqrt { s_\rho}} \exp\left({-i \omega \int \limits
_{\rho_0}^{\rho} s _\rho (\rho' ) d \rho'} \right).
\label{eq6-8-19}
\end{equation}

The first term in (\ref{eq6-8-19}) corresponds to a right-going
wave because its phase is increasing with distance. By the same
token, the second term in (\ref{eq6-8-19}) is a left-going wave.
Moreover, the integral in the exponent elucidates the physical
picture that the phase gained by a wave gong from $\rho_0$ to
$\rho$ is proportional to
\begin{equation}
\omega \int \limits_{\rho_0}^{\rho} s _\rho (\rho ' ) d \rho'
\label{eq6-8-20}
\end{equation}
which is the integral summation of all the phases gained locally
at $\rho'$ over the range from $\rho_0$ to $\rho$.  This physical
picture is true only if the multiple reflections of the wave can
be neglected as it is propagating. Furthermore, Equation
(\ref{eq6-8-13}) shows that this physical picture, which
corresponds to the leading order solution, is correct only if
\begin{equation}
\omega s _\rho' (\rho) \ll \omega ^2 s_\rho^2 (\rho).
\label{eq6-8-21}
\end{equation}
Hence, this picture breaks down if the frequency is not high, or
if $s_\rho (\rho) \simeq 0$. The factor of $1/ \sqrt {s_\rho}$ in
(\ref{eq6-8-19}) is necessary for energy conservation.  It is
related to the wave impedance of the wave, and hence, it alters
the amplitude of the wave to conserve energy.

Note that when $s_\rho^2 < 0$, corresponding to when $E< V(\rho)$,
or when the wave becomes evanescent, the above analysis is still
valid. Hence, the above analysis is valid for the field of a guided
mode in the region when $E> V(\rho)$, and in the region where
$E<V(\rho)$, but not in vicinity of the region where $E= V(\rho)$.
In the last case, $s_\rho(\rho) \simeq 0$. This happens, for
instance, in the guided mode case when $\rho= \rho_1$ and $\rho=
\rho_2$ in Figure \ref{f681}(b).  In region II, we have bouncing
waves, and regions I and III, the waves are evanescent. Hence, at
$\rho = \rho_1$ and $\rho= \rho_2$, the waves are critically
refracted, and they are also known as the \index{WKB method!turning point} turning points.

\subsection { Solution in the Vicinity of a Turning Point}

In the vicinity of $\rho = \rho_1$ in Figure 1(a), $k_\rho ^2
(\rho)$ can be approximated by a linear function, i.e.,
\begin{equation}
k_\rho^2 \simeq \omega ^2 \Omega (\rho_1 - \rho) , \qquad \rho \to
\rho_1, \label{eq6-8-22}
\end{equation}
i.e., $k_\rho^2$ is proportional to $\omega ^2$. Hence, around
$\rho_1$, Equation (\ref{eq6-8-8}) becomes
\begin{equation}
\hat R'' (\rho) + \omega ^2 \Omega  (\rho_1- \rho) \hat R (\rho) =
0. \label{eq6-8-23}
\end{equation}
Next by letting $\eta = \omega ^{\frac{2}{3}} \Omega
^{\frac{1}{3}} (\rho_1 - \rho)$, (\ref{eq6-8-23}) becomes
\begin{equation}
\left [ \frac{d^2}{d \eta ^2} + \eta \right ] \hat R (\eta) = 0,
\label{eq6-8-24}
\end{equation}
which is the \index{Airy equation} Airy equation.  The general solution to the above
equation is of the form
\begin{equation}
\hat R (\eta) = C_1 A_i (- \eta) + C_2 B_i (- \eta)
\label{eq6-8-25}
\end{equation}
where $A_i (- \eta)$ and $B_i ( - \eta)$ are special functions
called the \index{Airy functions} Airy functions. Since $ \eta = \omega
^{\frac{2}{3}}\Omega ^{\frac{1}{3}} (\rho_1 - \rho)$, $ \eta \to
\infty$ when $\rho \ll \rho_1$ and $ \eta \to  - \infty$ when $
\rho \gg \rho_1$. The asymptotic expansions of Airy functions can
be used to approximate them when their arguments are large.
Therefore, when $ \rho \gg \rho_1$, $\eta \to - \infty$, and we
have
\begin{subequations}
\begin{equation}
A_i (- \eta) \sim \frac{1}{2} \pi ^{-\frac{1}{2}} ( - \eta )
^{-\frac{1}{4}} e^{-\frac{2}{3} (- \eta) ^{\frac{3}{2}}}, \qquad
\eta \to - \infty, \label{eq6-8-26a}
\end{equation}
\begin{equation}
B_i( - \eta) \sim \pi^{-\frac{1}{2}} ( - \eta ) ^{-\frac{1}{4}} e
^{\frac{2}{3}(- \eta ) ^{\frac{3}{2}}}, \qquad \eta \to - \infty .
\label{eq6-8-26b}
\end{equation}
\end{subequations}
$A_i(- \eta)$ corresponds to an exponentially decaying wave while
$B_i(- \eta)$ corresponds to an exponentially growing wave. Since
we cannot have an exponentially growing wave to the right of
$\rho_1$, we must have $C_2 = 0$. Hence, in the vicinity of the
turning point $\rho= \rho_1$,
\begin{equation}
\hat R (\eta) = C_1 A_i (- \eta). \label{eq6-8-27}
\end{equation}
when $\rho \ll \rho_1$, $ \eta \to - \infty$, and we have
\begin{equation}
A_i (- \eta) \sim \pi ^{-\frac{1}{2}} \eta ^{-\frac{1}{4}} \sin
\left ( \frac{2}{3} \eta ^{\frac{3}{2}} + \frac{\pi}{4} \right),
\qquad \eta \to + \infty, \label{eq6-8-28}
\end{equation}
which corresponds to a standing wave resulting from a
superposition of incident and reflected waves on the left of the
turning point.

\subsection { Asymptotic Matching}
\index{Asymptotic matching}

The guidance condition of the modes in Figure \ref{f681}(b) can be
found by asymptotic matching. We shall illustrate asymptotic
matching with the simpler case in Figure \ref{f681}(a). In this
method, we seek the solutions in the region where $0<\rho< \rho_1$,
$\rho>\rho_1$, and solutions in the vicinity of $\rho = 0$ and
$\rho= \rho_1$. For $0<\rho<\rho_1$ and $ \rho> \rho_1$, we can use
the WKB solutions, while for $\rho$ in the vicinity of $\rho=0$ and
$\rho = \rho_1$, we need to use some special function solutions.
Even though the WKB solutions are not valid at $\rho= 0 $ and $\rho=
\rho_1$, but when $\omega \to \infty$, the WKB solutions are valid
in the vicinity of these points. By using the large argument
expansions of the special function solutions, overlapping regions of
validity of the solutions exist, and they can be matched to each
other to find the unknowns and the guidance condition of the
waveguide.

The WKB solutions are also the geometrical optics solutions which
are valid when the frequency is high. Hence, they are similar to
the ray-optics solutions. Ray optics solutions break down at
caustics where rays bunch together. It turns out that $\rho = 0$
and $\rho = \rho_1$ are caustics where rays bunch together.
 Hence, special function solutions are needed at these caustic points.
 It is worthwhile to notice that a ray undergoes a $90^\circ$ phase
 shift at a caustic, and this phenomenon will be observed in the later
 derivation.

The solution to the left of $\rho = \rho_1$ is given by
\begin{equation}
\hat R (\rho) \sim \frac{A_+}{\sqrt{s_\rho}} \exp ({i \omega \int
\limits _0^\rho s_\rho(\rho') d \rho'}) + \frac
{A_-}{\sqrt{s_\rho}} \exp ({-i\omega \int \limits_0^\rho s_\rho
(\rho ') d \rho'}) \label{eq6-8-29}
\end{equation}
When $\rho \to 0 $, assuming that $k^2 (\rho)$ tends to a constant
$k^2 (0)$, then the solution
\begin{equation}
\hat R (\rho) = A \sqrt{\rho} J_0 (k_s\rho) , \qquad \rho \simeq
0, \label{eq6-8-30}
\end{equation}
where $k_s = \sqrt {k^2 (0) - k_z^2}$.  Hence $k_s$ is
proportional to $\omega$.

When the frequency is high, $k_s \rho \gg 1$ for $\rho \neq 0$,
and (\ref{eq6-8-30}) can be approximated by
\begin{equation}
\hat R (\rho) \sim A \sqrt{\frac{2}{\pi k_s}} \cos \left(k_s\rho -
\frac{\pi}{4} \right) , \qquad k \to \infty. \label{eq6-8-31}
\end{equation}
In the limit when $\omega \to \infty$, $k_\rho^2(\rho)$ defined
for (\ref{eq6-8-8}) becomes
\begin{equation}
k_\rho^2 \sim k^2(\rho) - k_z^2 , \qquad \omega \to \infty.
\label{eq6-8-32}
\end{equation}
This is even valid when $\rho \simeq 0 $ as long as $k^2\rho^2 \to
\infty$. Hence,
\begin{equation}
k_\rho^2 \simeq k^2 (0) - k_z^2 = k_s ^2, \qquad \omega \to
\infty, \qquad \rho \simeq 0. \label{eq6-8-33}
\end{equation}
In this limit, $s_\rho = k_s/\omega$. Then, Equation
(\ref{eq6-8-29}) in the vicinity of $\rho \simeq 0$ becomes
\begin{equation}
\hat R (\rho) \sim A_+ \sqrt{\frac{\omega}{k_s}} e ^{ik_s\rho} +
A_-\sqrt{\frac{\omega}{k_s}} e^{-ik_s\rho}. \label{eq6-8-34}
\end{equation}
Comparing (\ref{eq6-8-31}) and (\ref{eq6-8-34}), we require that
\begin{subequations}
\begin{equation}
A_+ = \frac{A}{\sqrt{2 \pi \omega}} e ^{-i\frac{\pi}{4}},
\label{eq6-8-35a}
\end{equation}
\begin{equation}
A_- = \frac{A}{\sqrt{2 \pi \omega}} e ^{+i\frac{\pi}{4}}.
\label{eq6-8-35b}
\end{equation}
\end{subequations}
The $-90^\circ$ phase shift between $A_-$ and $A_+$ is reminiscent
of an optical ray going through a caustic.  Consequently,
(\ref{eq6-8-29}) becomes
\begin{equation}
\hat R(\rho) = \frac{A}{\sqrt {2 \pi \omega s_\rho}} \cos \left [
\omega \int \limits _0^\rho s_\rho (\rho') d \rho' - \frac{\pi}{4}
\right ]. \label{eq6-8-36}
\end{equation}

In the vicinity of $\rho = \rho_1$, the phase integral can be
approximated by
\begin{equation}
\int \limits _0 ^\rho s_\rho (\rho') d \rho ' = \int \limits_0
^{\rho_1}s_\rho (\rho') d \rho' + \int \limits _{\rho_1}^\rho
s_\rho (\rho ' ) d \rho '. \label{eq6-8-37}
\end{equation}
From (\ref{eq6-8-22}), $s_\rho (\rho ') \simeq
\Omega^{\frac{1}{2}} (\rho_1 - \rho)^{\frac{1}{2}}$ when $\rho
\simeq \rho_1$. Therefore,
\begin{equation}
\int \limits_{\rho_1}^{\rho} s _\rho (\rho')d \rho' = \Omega
^{\frac{1}{2}} \int \limits _0 ^{(\rho_1 - \rho)} \sqrt{x} dx =
\frac{2}{3} \Omega ^{\frac{1}{2}} (\rho_1 - \rho) ^{\frac{3}{2}},
\label{eq6-8-38}
\end{equation}
and
\begin{equation}
\hat R (\rho) \simeq \frac{A}{(2 \pi \omega)^{\frac{1}{2}}\Omega
^{\frac{1}{4}} (\rho- \rho_1)^{\frac{1}{4}}} \cos \left [
\frac{2}{3} \omega \Omega ^{\frac{1}{2}}(\rho_1 -
\rho)^{\frac{3}{2}} + \phi - \frac{\pi}{4} \right],
\label{eq6-8-39}
\end{equation}
where
\begin{equation}
\phi = \omega \int \limits_0^{\rho_1} s_\rho (\rho') d \rho ' .
\label{eq6-8-40}
\end{equation}
Using the definition of $\eta$ in (\ref{eq6-8-27}), and
approximating $A_i (- \eta)$ with (\ref{eq6-8-28}), we have
\begin{equation}
\hat R (\rho) \simeq C_1 \frac{1}{\sqrt{\pi} \omega ^{\frac{1}{6}}
\Omega ^{\frac{1}{12}} (\rho_1 - \rho) ^{\frac {1}{4}}} \sin
\left [ \frac {2}{3} \omega\Omega ^{\frac{1}{2}} (\rho_1 - \rho)
^{\frac {3}{2}} + \frac{\pi}{4}\right ] , \label{eq6-8-41}
\end{equation}
when $ \omega \Omega ^{\frac{1}{2}} (\rho_1 - \rho)^{\frac{3}{2}}
\gg 1$.

Comparing (\ref{eq6-8-39}) with (\ref{eq6-8-41}), in order for the
solution in region for which $\rho < \rho_1$ and the solution for
which $\rho \simeq \rho_1$ to agree with each other, we must have
\begin{subequations}
\begin{equation}
C_1 = \frac{(-1)^mA}{\sqrt{2} \omega ^{\frac{1}{3}}\Omega
^{\frac{1}{6}}}, \label{eq6-8-42a}
\end{equation}
\begin{equation}
\phi = \omega \int \limits _0 ^{\rho_1} s_\rho (\rho') d \rho' = m
\pi, \label{eq6-8-42b}
\end{equation}
\end{subequations}
where $m = 1,2,3,\dots$.  Equation (\ref{eq6-8-42b}) is the
guidance condition for the $n=0$ mode of an optical fiber. The
phase shift at the $\rho= 0$ caustic is $-90^\circ$ while the
phase shift at the $\rho= \rho_1$ caustic is $+90^\circ$. Hence,
(\ref{eq6-8-42b}) resembles the guidance condition for a
parallel-plate waveguide. Similar procedures can be used to find
the guidance condition for the $n = 1,2,3, \dots$ modes. When $n
\sim O(ka)$ then there will be two turning points both at $\rho =
\rho_1$ and $\rho= \rho_2$. The analysis will be slightly
different from the above.

Notice that in the above, the WKB solution for $\rho > \rho_1$,
was not used other than that it is exponentially decaying. The
reason is that if there is no turning point beyond $\rho= \rho_1$,
the reflection of the bouncing waves in the optical fiber is
determined by the turning point at $\rho= \rho_1$ only. A WKB
solution presents a physical picture of a wave propagating without
reflection. Hence, a turning point is the only place where a wave
is reflected. To obtain the magnitude of the evanescent wave for
$\rho > \rho_1$, asymptotic matching can be used.

\section {{{ Effective Index Method}}}
\index{Effective index method}

\begin{figure}[ht]
\begin{center}
\hfil\includegraphics[width=5.0truein]{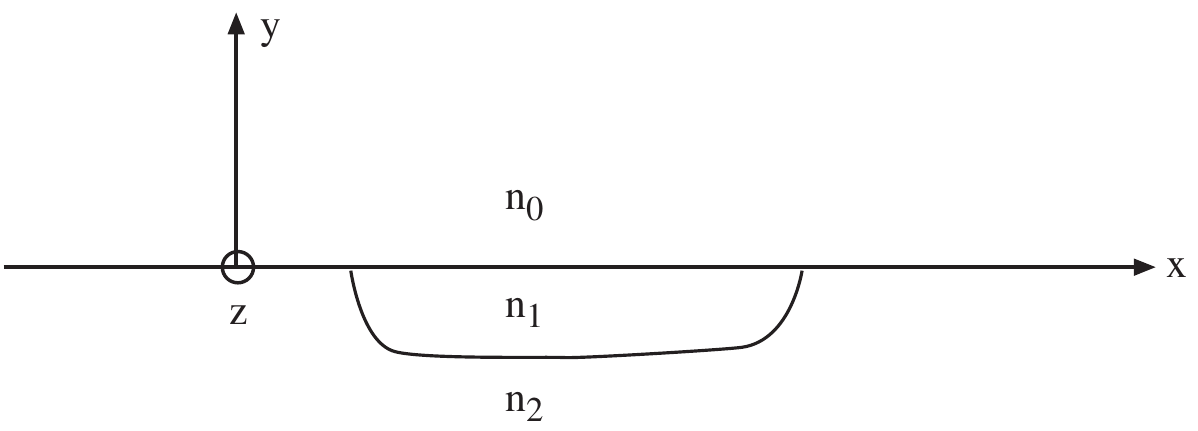}\hfil
\end{center}
\caption{A typical integrated optics waveguide cross-section.}\label{fg691}
\end{figure}


In integrated optics waveguides, the refractive index profile is
often obtained by doping. Hence, the variation of the refractive
index is weak. In such a case, an approximate method called the
effective index method may be used to analyze the waveguiding
structure.  The method was proposed by Knox and Toulios
\cite{KNOX&TOULIOS}, and has been used by many workers
\cite{TAMIR,BUUS}. Assuming that $\mu_r=1$, the equations governing
the electromagnetic field in such an inhomogeneous structure are
exactly given by

\begin{equation}
\nabla ^2 \v E + \nabla ( \nabla \ln \epsilon_r \cdot \v E ) +
k_0^2 \epsilon _r \v E = 0, \label{eq6-9-1}
\end{equation}
\begin{equation}
\nabla ^2 \v H + \nabla \ln \epsilon_r \times \nabla \times \v H +
k_0^2 \epsilon _r \v H = 0. \label{eq6-9-2}
\end{equation}

\subsection{Effective Index Concept}
\index{Effective index method}

When a mode is propagating in a waveguide, say the optical fiber,
with $e^{ik_zz}$ dependence, we can define an effective index $n_e$
such that
\begin{align}\label{EIeq:1}
k_z=k_0n_e
\end{align}
Hence, a TEM wave propagating in a homogeneous medium with this
effective index will have $k=k_0n_e$ equal to the $k_z$ of the
guided mode. This concept can be extended to other structures,
including a wave propagating in a slab or layered waveguide

\begin{figure}[ht]
 \begin{center}
  \includegraphics[totalheight=0.25
  \textheight,width=0.6\textwidth]{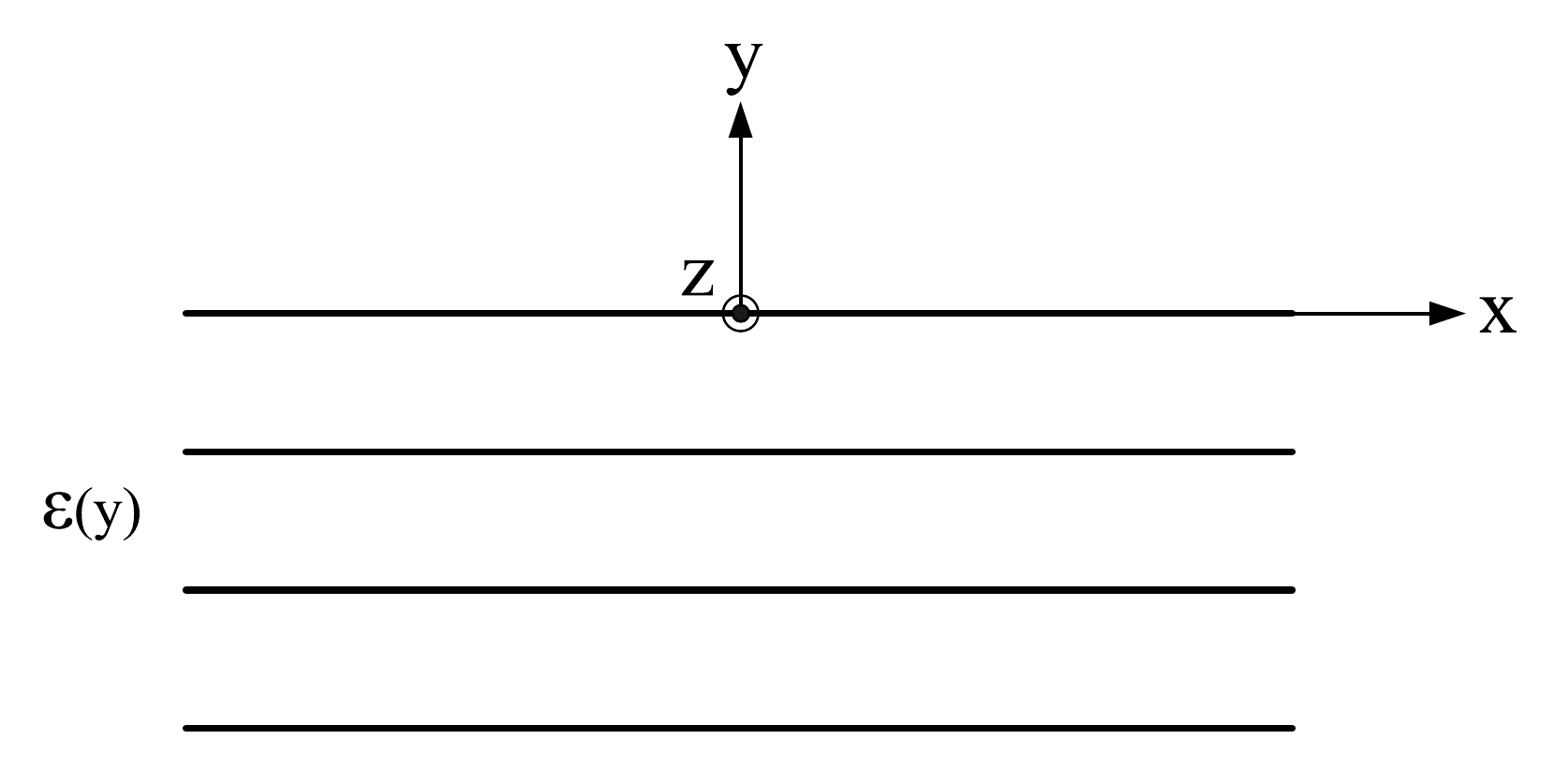}
  \end{center}
  \caption{A layered medium waveguide.}\label{Layered_medium}
\end{figure}

To explain the effective index method, one considers a
one-dimensional problem first as shown in Figure
\ref{Layered_medium} where $\epsilon$ is a function of $y$ only.
Then if we consider $\mbox{TE}$ wave, which can be characterized by
$H_y$, the governing equation is
 \begin{align}\label{EIeq:2}
 \nabla^2H_y+k_0^2\epsilon_r(y)H_y=0
 \end{align}
The above can be solved by the separation of variables, by letting
 \begin{align}\label{EIeq:3}
 H_y(x,y,z)=Y(y)h(x,z)
 \end{align}
Substituting (\ref{EIeq:3}) into (\ref{EIeq:2}) leads to
 \begin{align}\label{EIeq:4}
 h(x,z)\frac{\partial^2}{\partial y^2}Y(y)+Y(y)
 \left(\frac{\partial^2}{\partial x^2}+\frac{\partial^2}
 {\partial z^2}\right)h(x,z)+k_0^2\epsilon_r Y(y)h(x,z)=0
 \end{align}
Dividing by $Y(y) h(x,z)$ yields
 \begin{align}\label{EIeq:5}
 Y^{-1}\frac{\partial^2}{\partial y^2}Y+h^{-1}\left(\frac{\partial^2}{\partial x^2}+\frac{\partial^2}{\partial z^2}\right)h +k_0^2\epsilon_r=0
 \end{align}
The above can be satisfied by letting
\begin{align}\label{EIeq:6}
 \left(\frac{\partial^2}{\partial x^2}+\frac{\partial^2}{\partial z^2}\right)h(x,z)=-k_0^2n^2h(x,z)
 \end{align}
\begin{align}\label{EIeq:7}
 \frac{\partial^2}{\partial
 y^2}Y(y)+{k_0}^2\epsilon_r(y) Y(y)=k_0^2n^2Y(y)
 \end{align}
 where $k_0^2n^2$ is the separation constant.

Equation (\ref{EIeq:7}) is an eigenvalue problem with eigenvalue
$k_0^2n^2$ and eigenfunction $Y(y)$. Here, $Y(y)$ corresponds to
guided mode or \index{Eigenmode problem} eigenmode in the layered medium with eigenvalue
$k_0^2n^2$. Equation (\ref{EIeq:6}) represents a wave traveling in
the $(x,z)$ direction with effective index $n$, which is a constant
independent of $(x,z)$.  When there are more than one eigenmode,
then one can consider the dominant eigenmode.  In principle, each
eigenmode can be considered separately even though the effective
index for each of them could be different.

\subsection { Quasi-TE polarization}
\index{Quasi-TE polarization}

If we assume that $\v E$ is polarized predominantly in the $xy$
direction, and that $\epsilon_r (x,y,z)$ is a slowly varying
function of $x$ and $z$, then, $\nabla \ln \epsilon _r$ is
predominantly $\hat y$ directed. Extracting the $y$ component of
(\ref{eq6-9-2}) to characterize a TE to $y$ wave, we have
\begin{equation}
\nabla ^2 H_y + k_0^2 \epsilon _r H_y \approx 0 \label{eq6-9-3}
\end{equation}

Next, we assume that
\begin{equation}
H_y = Y(x,y,z) h(x,z), \label{eq6-9-4}
\end{equation}
where $\partial Y/ \partial x \simeq 0$, $\partial Y / \partial z
\simeq 0$. In other words, $Y(x,y,z)$ is a slowly varying function
of $x$ and $z$. Equation (\ref{eq6-9-4}) is motivated by the
separation of variables as described in the previous section. As
noted before, the problem is completely separable if $\epsilon_r$ is
a function of $y$ only. But here, we assume that the problem is
approximately separable.

Then,
\begin{equation}
\nabla ^2 H_y \simeq h \frac{\partial^2 Y}{\partial y^2} + Y \left
(\frac{\partial ^2 h}{\partial x^2} + \frac{\partial ^2
h}{\partial z^2} \right). \label{eq6-9-5}
\end{equation}
Consequently, \eqref{eq6-9-4} becomes
\begin{align}\label{eq:4a}
 h(x,z)\frac{\partial^2}{\partial y^2}Y(y)+Y(y)
 \left(\frac{\partial^2}{\partial x^2}+\frac{\partial^2}
 {\partial z^2}\right)h(x,z)+k_0^2\epsilon_r Y(y)h(x,z)\approx 0
 \end{align}
Again, motivated by the separation of variables, we let
\begin{equation}
\left (\frac{\partial ^2}{\partial x^2} + \frac{\partial
^2}{\partial z^2} \right) h(x,z) + k_0^2 n^2 (x,z) h(x,z) = 0
\label{eq6-9-7}
\end{equation}
where $n(x,z)$ is an effective index that is a slowly varying
function of $x$ and $z$ only. Furthermore, we require that
\begin{equation}
\frac{\partial ^2}{\partial y^2} Y(x,y,z) + k_0^2 \epsilon_r (x,y,z)
Y(x,y,z) = k_0^2 n^2 (x,z) Y(x,y,z). \label{eq6-9-9}
\end{equation}
For every fixed $x$ and $z$, the above is a one-dimensional
eigenequation with eigenvalue $$k_0^2 n^2(x,z).$$ In principle,
there are infinitely many eigenvalues and eigenfunctions associated
with (\ref{eq6-9-9}). The solutions of equation (\ref{eq6-9-9}) are
the modes of the structure. We can assume one mode only, or that
only the \index{Mode!fundamental} fundamental mode is important. If $\epsilon _r( x,y,z)$ is
piecewise constant in $y$, then transcendental equations can be
derived to yield $k_0^2 n^2 (x,z)$. Hence, if $\epsilon _r(x,y,z)$
is a slowly varying function of $x$ and $z$, $Y(x,y,z)$ is in fact a
slowly varying function of $x$ and $z$ validating the assumption
(\ref{eq6-9-4}) and the self-consistency of the method.

\subsection { Quasi-TM Polarization}
\index{Quasi-TM Polarization}

In this case, we assume that the magnetic field is predominantly
polarized in the $xz$ plane. Such a wave field can be
characterized by the $E_y$ component of the electric field.
Extracting the $y$ component of (\ref{eq6-9-1}), we have
\begin{equation}
\nabla ^2 E_y + \frac{\partial }{\partial y} \left [\left (
\frac{\partial}{\partial y} \ln \epsilon_r \right ) E_y\right ] +
k_0^2 \epsilon_r E_y \approx 0 . \label{eq6-9-10}
\end{equation}
In arriving at the approximate equation above, we assume that
\begin{equation}
\frac{\partial}{\partial y} \ln \epsilon _r \gg \nabla _s \ln
\epsilon _r \label{eq6-9-11}
\end{equation}
where $ \nabla _s = \hat x \frac{\partial }{\partial x} + \hat z
\frac{\partial}{\partial z}$. In the above, \eqref{eq6-9-10} can be
rewritten as
$$
\left(\frac{\partial^2}{\partial x^2}+\frac{\partial^2}{\partial
z^2}+\epsilon_r \frac{\partial}{\partial y}\epsilon_r^{-1}
\frac{\partial}{\partial y}+k_o^2\epsilon_r\right)\epsilon_r
E_y\approx 0
$$
to resemble that in \cite[(2.1.7)]{WFIM6}.

Again, if we assume that
\begin{equation}
E_y = Y (x,y,z) e(x,z), \label{eq6-9-12}
\end{equation}
where $\partial Y / \partial x \simeq 0$ and $\partial Y /
\partial z \simeq 0$, then, motivated by the separation of
variables, we let
\begin{equation}
\frac{\partial ^2 Y }{\partial y^2} + \frac{\partial}{\partial y}
\left [ \left ( \frac{\partial}{\partial y} \ln \epsilon _r \right )
Y \right] + k_0^2\epsilon_r  Y \approx k_0^2 n^2 Y. \label{eq6-9-13}
\end{equation}
The presence of the extra term in the equation for TM polarization
compared to that for the TE polarization is because in the TM
polarization, the electric field has a component normal to the
interface that induces polarization charges at the interface.

The above could be rewritten as
\begin{equation}
\frac{\partial}{\partial y}\frac{1}{\epsilon _r} \frac
{\partial}{\partial y} \epsilon _r Y + k_0^2 \epsilon_r Y \approx
k_0^2 n^2 Y. \label{eq6-9-14}
\end{equation}
For a fixed $x$ and $z$, it is a one dimensional eigen-equation for
the propagation of TM eigenmodes in a \index{Layered medium} layered medium. Here, $k_0^2
n^2$ is the eigenvalue of the problem.   It is seen that this
effective index $n$ now is  function of $x$ and $z$.  If
$\epsilon_r$ is a slowly varying function of $x$ and $z$, so would
the index $n$.

Consequently, the equation governing $e(x,z)$ is
\begin{equation}
\left ( \frac {\partial ^2}{\partial x^2} + \frac{\partial
^2}{\partial z^2} \right ) e (x,z) + k_0 ^2 n^2 (x,z) e(x,z) = 0,
\label{eq6-9-15}
\end{equation}
where $n(x,z)$ is the effective index obtained by solving
(\ref{eq6-9-14}). Again, (\ref{eq6-9-15}) is now reduced to a
two-dimensional equation.

The effective index method is equivalent to replacing the $y$
variation of the field with a single-mode approximation, and the
propagation of this single mode in the $xz$ plane is governed by
the two-dimensional equations (\ref{eq6-9-7}) and
(\ref{eq6-9-15}), for different polarizations.

\section {{{ The Beam-Propagation Method}}}
\index{Beam-propagation method}

When the inhomogeneity of a waveguide is weakly varying, an
efficient method of deriving a solution is to use the
beam-propagation method.  It was proposed by Fleck, Morris, and Feit
\cite{FLECK&MORRIS&FEIT} first for atmospheric wave propagation
where the refractive index is often tenous.  Later, it was adapted
for optical waveguide analysis \cite{FEIT&FLECK,HUANG&XU}. Consider
a scalar wave equation governed by
\begin{equation}
\left [ \nabla _s ^2 + \frac{\partial ^2}{\partial z ^2} + k_0^2
n^2 (x,y,z) \right ] \phi (x,y,z) = 0, \label{eq6-10-1}
\end{equation}
where $ \nabla _s^2 = \frac{\partial ^2}{\partial x^2} +
\frac{\partial^2}{\partial y ^2}$. Assuming that $n(x,y,z)$ is a
weak or slowly varying function of space, or $(x,y,z)$, then the
above equation can be factorized as
\begin{equation}
\left ( \frac{\partial }{\partial z } + i \sqrt { k_0^2 n^2 + \nabla
_s ^2}\right ) \left ( \frac{\partial}{\partial z} - i \sqrt{ k_0^2
n^2 + \nabla _s^2} \right ) \phi (\v r ) \approx 0. \label{eq6-10-2}
\end{equation}

It is to be noted that two new concepts are embedded in the above
expression.  A function of an operator such as $f(\nabla_s^2)$ is
also regarded as an operator, and it has meaning only when it
operates on a function which is the eigenfunction of $\nabla_s^2$.
An eigenfunction of the $\nabla_s^2$ operator is $e^{ik_xx+ik_yy}$
since $\nabla_s^2 e^{ik_xx+ik_yy}=-k_s^2 e^{ik_xx+ik_yy}$ where
$-k_s^2=-k_x^2-k_y^2$ is the eigenvalue.  A function can always be
approximately by a Taylor series such as $f(x)=f(0)+xf'(0)+\frac12
x^2 f''(0)+\ldots$ assuming that $f(x)$ is analytic at $x=0$.  Then,
using its Taylor series expansion,
\begin{align}
f(\nabla_s^2)e^{ik_xx+ik_yy}&=\left(f(0)+\nabla_s^2f'(0)+\frac12
\nabla_s^4 f''(0)+\ldots\right)
e^{ik_xx+ik_yy}\notag\\&=\left(f(0)+\nabla_s^2f'(0)+\frac12
\nabla_s^4 f''(0)+\ldots\right) e^{ik_xx+ik_yy}\notag\\&
=\left(f(0)-k_s^2f'(0)+\frac12 k_s^4 f''(0)+\ldots\right)
e^{ik_xx+ik_yy}\notag\\&=f(-k_s^2)e^{ik_xx+ik_yy}
\end{align}
So in general
\begin{align}
f(\cal A) \v v_i=f(\lambda_i) \v v_i
\end{align}
where $\v v_i$ is an eigenvector of the operator $\cal A$ with
eigenvalue $\lambda_i$.

Another concept is the \index{Commutator} commutativity of operators.  In general,
\begin{align}
\left(\cal A+\cal B\right)\left(\cal A-\cal B\right)=\left(\cal
A^2-\cal B^2\right)
\end{align}
only if $\cal A\cal B=\cal B\cal A$, or only if $\cal A$ and $\cal
B$ commute.  The commutator of $\cal A$ and $\cal B$ is defined as
$\left[\cal A, \cal B\right]= \cal A \cal B-\cal B\cal A$. Hence, if
$\cal A$ and $\cal B$ commute, their commutator is zero.

But in the above \eqref{eq6-10-2}, the operator
$\frac{\partial}{\partial z}$ and $\sqrt { k_0^2 n^2 + \nabla _s
^2}$ do not commute because $n$ is a function of $z$. Hence, the
above is not an exact factorization, but is a good approximation if
$n$ is a slowly varying function of $z$.

 Therefore, a solution to
\begin{equation}
\frac{\partial }{\partial z } \phi(\v r ) = i
\sqrt{k_0^2n^2+\nabla _s ^2} \phi (\v r ) = i \wp \phi (\v r )
\label{eq6-10-3}
\end{equation}
is also an approximate solution to (\ref{eq6-10-1}). In the above,
$\wp = \sqrt {k_0^2 n^2 + \nabla_s^2} $ is to be interpreted as an
operator.

Here, \eqref{eq6-10-3} is also a one-way wave equation as it
describes the propagation of the wave in one direction only.  Hence,
multiply reflected wave is not accounted for in the above
approximation.

Equation (\ref{eq6-10-3}) is not any easier to solve compared to
(\ref{eq6-10-1}). To simplify it, we need to make a paraxial
approximation. This assumes that
\begin{equation}
| \nabla _s^2 \phi | \ll | k_0^2 n^2 \phi |, \label{eq6-10-4}
\end{equation}
or that the transverse variation of $\phi$ is much smaller than
its longitudinal variation. In other words, the wave is
propagating almost parallel to the axis of the waveguide. If we
let $n = n_0 + \delta n$, then
\begin{equation}
\wp \cong \sqrt {k_0^2n_0^2 + \nabla _s^2} \left ( 1 +
\frac{k_0^2n_0\delta n}{k_0^2n_0^2 + \nabla _s ^2} + \dots  \right
). \label{eq6-10-5}
\end{equation}
With the assumption (\ref{eq6-10-4}), we can approximate
(\ref{eq6-10-5}) as
\begin{equation}
\wp \cong \sqrt {k_0^2n_0^2 + \nabla _s ^2} + k_0 \delta n.
\label{eq6-10-6}
\end{equation}
The approximation (\ref{eq6-10-6}) is judiciously tailored so that
the first term is independent of space, and the second term is
independent of the operator $\nabla _s$.

Since $\wp$ is a function of $z$, equation (\ref{eq6-10-3}) still
cannot be solved easily. However, if $\wp$ is assumed to be
independent of $z$ within a small $\Delta z$, then, we can write the
solution to \eqref{eq6-10-3} as
\begin{equation}
\begin{split}
\phi(x,y,z+\Delta z) &= e^{i{\wp} \Delta z} \phi (x,y,z)\\
&\cong e^{ik_0 \delta n \Delta z + i \sqrt {k_0^2n_0^2 + \nabla
_s ^2} \Delta
z} \phi (x,y,z)\\
&\cong e^{ik_0\delta n \Delta z} e^{i\sqrt{k_0^2n_0^2+ \nabla
_s^2}\Delta z} \phi (x,y,z). \label{eq6-10-7}
\end{split}
\end{equation}
The last expression is an approximation because in general,
$\exp(\cal A+\cal B)\ne \exp(\cal A)+\exp(\cal B)$ unless $\cal A$
and $\cal B$ commute.  It can be verified by representing these
operators with their Taylor series expansions.

Now, for a fixed $z$, using Fourier expansion, we can write
\begin{equation}
\phi(x,y,z) = \frac{1}{(2 \pi)^2} \iint_{-\infty}^{\infty} d{\v
k}_s e^{i\v k_s \cdot {\v r}_s}\tilde \phi ({\v k}_s, z)
\label{eq6-10-8}
\end{equation}
where $\v k_s = \hat x k_x + \hat y k_y$, $\v r_s = \hat x x + \hat
y y$ and $e^{i\v k_s \cdot {\v r}_s}$ is an eigenfunction of the
$\nabla_s^2$ operator.  By so doing, we have expanded $\phi(x,y,z)$
as a linear superposition or integral summation of the
eigenfunctions of $\nabla_s^2$ operator.
Substituting (\ref{eq6-10-8}) into (\ref{eq6-10-7}), we have
\begin{equation}
\phi (x,y,z+\Delta z) = \frac{1}{(2 \pi)^2} e^{ik_0 \delta n \Delta
z } \iint_{-\infty}^{\infty} d{\v k}_s e ^{i{\v k}_s \cdot {\v r}_s
} e ^{i \sqrt{k_0^2 n_0^2 - k_s^2}\Delta  z} \tilde \phi (\v k_s,
z). \label{eq6-10-9}
\end{equation}
Equation (\ref{eq6-10-9}) \index{Beam-propagation method!fundamental equation} is the fundamental equation of the
beam-propagation method. To implement it,  one first takes the field
$\phi(x,y,z)$ at a $z =\text{constant} $ plane and Fourier transform
it to get $\tilde \phi (\v k_s, z)$. Then one multiplies the result
by a plane-wave propagator $e^{i \sqrt{k_0^2n_0^2 -k_s^2}\Delta z}$
in the Fourier space. Next, a Fourier inverse transform is performed
on the propagated result. Subsequently, the field at each $(x,y)$
location is added a phase of $k_0\delta n \Delta z$ to yield the
field at $\phi (x,y,z+\Delta z)$.

Alternatively, one can write
\begin{equation}
\sqrt {k_0^2 n_0^2 - k_s ^2} = k_0n_0 - \frac{k_s ^2}{\sqrt{k_0^2
n_0^2 - k_s^2} + k_0 n_0}, \label{eq6-10-10}
\end{equation}
where the second term is much smaller than the first term if the
wave is paraxial. Therefore, we can let
\begin{equation}
\phi (x,y,z) = w(x,y,z) e ^{ik_0n_0z}, \label{eq6-10-11}
\end{equation}
and the beam-propagation equation for $w(x,y,z)$ is then
\begin{equation}
w(x,y,z+\Delta z) = \frac{1}{(2 \pi)^2} e ^{ik_0\delta n \Delta z}
\iint_{-\infty}^{\infty} d {\v k}_s e ^{i {\v k}_s \cdot {\v r}_s}
e^{ip(k_s) \Delta z}\tilde \phi ({\v k}_s, z), \label{eq6-10-12}
\end{equation}
where
\begin{equation}
p(k_s) = \frac {k_s ^2} {\sqrt {k_0^2n_0^2 -k_s^2} + k_0n_0}.
\label{eq6-10-13}
\end{equation}

\begin{figure}[ht]
\begin{center}
\hfil\includegraphics[width=4.5truein]{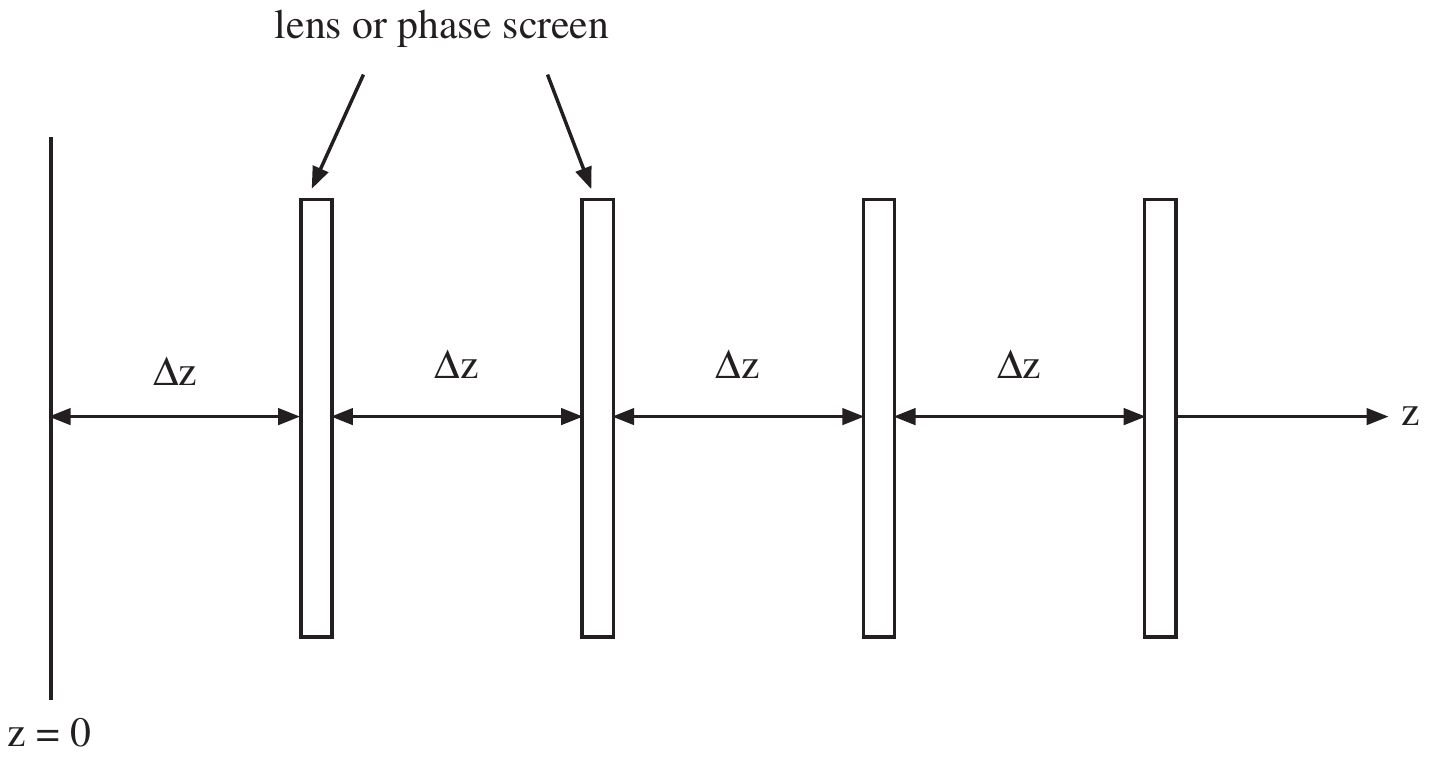}\hfil
\end{center}
\caption{The physical interpretation of the beam-propagation
method.}\label{fg6101}
\end{figure}


The Fourier transform and inverse transform in the beam-propagation
method can be implemented efficiently using fast Fourier transform
(FFT) which requires $O(N \log N)$ floating-point operations. Hence,
albeit approximate, the beam-propagation method can be implemented
efficiently. As it solves a first order equation (\ref{eq6-10-3}),
it only accounts for the forward component of the wave, but no
reflections.

The physical interpretation of (\ref{eq6-10-9}) or
(\ref{eq6-10-12}) is that the wave is first propagated through a
homogeneous space using a homogeneous-space \index{Propagator} propagator. Then, the
wave is passed through a lens or a phase screen which alters the
phase of the wave at each $(x,y)$ position.  Therefore, Equation
(\ref{eq6-10-9}) or (\ref{eq6-10-12}) represents physically the
concatenation of a series of lenses or phase screens in a
homogeneous space.

\begin{figure}[ht]
\begin{center}
\hfil\includegraphics[width=6.0truein]{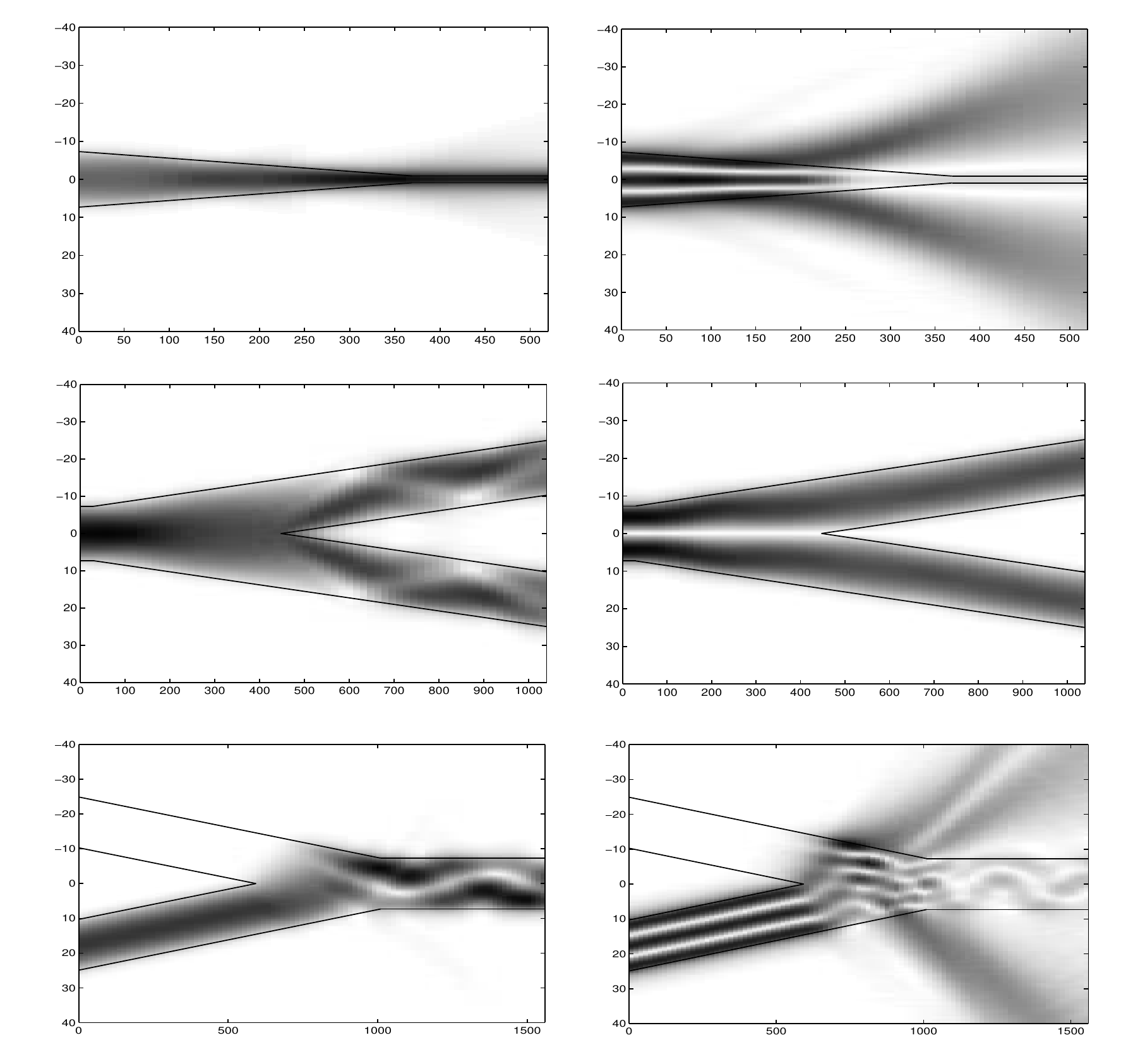}\hfil
\end{center}
\caption{Examples of numerical simulations using the
beam-propagation method for different kinds of optical waveguides.
Some waveguide transitions give rise to much radiation loss, while
some do not. Also, depending on the mode profile, the radiation at
the waveguide transition is different. For instance, in the top
figure, the fundamental mode (top left) couples more smoothly than
the higher-order mode (top right) with two minima (Courtesy of F.
Teixeira).}\label{fg6102}
\end{figure}


\section {{{ Ray Tracing Method}}}
\index{Ray tracing method}

\index{Optical fiber!multimode} Multimode fibers correspond to the case where the core size is much
larger than the wavelength.  Hence, the fiber can be engineered by
assuming that the light is a ray that bounces around in the fiber.
Therefore, ray-tracing method can be used to solve the optical fiber
problem. This method is described by Okoshi \cite{OKOSHID} and
Senior \cite{SENIOR} and many references therein.

The vector wave equation for an inhomogeneous medium can be
written as
\begin{equation}
\nabla ^2 \v E - \nabla ( \nabla \ln \epsilon _r \cdot \v E ) +
k_0^2 \epsilon _r \v E = 0. \label{eq6-11-1}
\end{equation}
The above equation can be approximated when the frequency is high,
so that the wavelength of the wave is much smaller than the
lengthscale of the variation of the inhomogeneity.  Then we can
assume that $ \v E (\v r )$ resembles a plane wave locally. In other
words, $\v E (\v r)$ could be more aptly described with a ray
picture. In this case,
\begin{equation}
\v E (\v r) = \v e (\v r ) e ^{ik_0 \phi(\v r )}, \label{eq6-11-2}
\end{equation}
where $\phi(\v r)$ and $\v e (\v r)$ are slowly varying but
$e^{ik_0\phi(\v r )}$ is rapidly varying when $k_0 \to \infty$.
Therefore,
\begin{equation}
|\nabla \v E (\v r) | \sim | k_0 \v E (\v r ) |, \qquad |\nabla ^2
\v E (\v r ) \sim | k_0^2\v E (\v r )|, \label{eq6-11-3}
\end{equation}
when $k_0 \to \infty$. Hence, Equation (\ref{eq6-11-1}) can be
approximated by
\begin{equation}
\nabla ^2 \v E (\v r ) + k_0^2 \epsilon _r \v E (\v r ) = 0.
\label{eq6-11-4}
\end{equation}
The above implies that the polarization term in \eqref{eq6-11-1} is
unimportant in the high-frequency limit if the postulated form for
$\v E (\v r)$ given by (\ref{eq6-11-2}) is true.  However, if there
is a jump discontinuity in $\epsilon_r$, the polarization term could
still be important, as its derivatives give rise to singular terms.
However, if $\epsilon_r$ is a slowly varying function, it can be
safely ignored.

Taking the Laplacian of $\v E$, we have
\begin{equation}
\nabla ^2 \v E(\v r) \simeq \v e (\v r ) \nabla ^2 e ^{ik_0\phi
(\v r )} = -k_0^2 (\nabla \phi) ^2 \v E (\v r ), \label{eq6-11-5}
\end{equation}
after ignoring higher order terms. Using (\ref{eq6-11-5}) in
(\ref{eq6-11-4}) yields
\begin{equation}
(\nabla \phi)^2 = \epsilon _r = n^2 \label{eq6-11-6}
\end{equation}
or
\begin{equation}
|\nabla \phi | = n. \label{eq6-11-7}
\end{equation}
$\nabla \phi$ is the direction in which the phase in the wave in
(\ref{eq6-11-2}) is varying most rapidly. Hence, it is the
direction of the wave in (\ref{eq6-11-2}), and also the direction
at which the ray is pointing. A unit vector along the ray
direction is given by
\begin{equation}
\hat s = \nabla \phi / |\nabla \phi| = \nabla \phi /n.
\label{eq6-11-8}
\end{equation}
Defining a point in a ray by the position vector $\v r $, and that
the ray is parametrized by the variable $s$ which is the path
length along a ray, then $\hat s = d\v r / ds$, and
(\ref{eq6-11-8}) becomes
\begin{equation}
n \hat s = n \frac{d \v r}{d s} = \nabla \phi. \label{eq6-11-9}
\end{equation}

Also, taking the gradient of (\ref{eq6-11-6}) gives
\begin{equation}
2 (\nabla \nabla \phi) \cdot \nabla \phi = 2 n \nabla n.
\label{eq6-11-10}
\end{equation}
After making use of (\ref{eq6-11-9}), the above becomes
\begin{equation}
(\nabla \nabla \phi ) \cdot \hat s = \nabla n. \label{eq6-11-11}
\end{equation}
Furthermore,
\begin{equation}
\begin{split}
\frac{d}{ds}(n \hat s) &= \frac{d}{ds} (\nabla \phi) = \left
(\frac{dx}{ds}\frac{\partial}{\partial s} + \frac{dy}{ds}
\frac{\partial
}{\partial y} + \frac{dz}{ds}\frac{\partial}{\partial z} \right ) \nabla \phi\\
& = \frac{d\v r }{ds} \cdot \nabla \nabla \phi = \hat s \cdot
\nabla \nabla \phi. \label{eq6-11-12}
\end{split}
\end{equation}
But $\hat s \cdot \nabla \nabla \phi = (\nabla \nabla \phi ) \cdot
\hat s $. Hence, from (\ref{eq6-11-11}) and (\ref{eq6-11-12}), one
gets
\begin{equation}
\frac{d}{ds} (n \hat s ) = \nabla n, \qquad \frac{d }{ds} \left (n
\frac{d\v r}{d s} \right ) = \nabla n, \label{eq6-11-13}
\end{equation}
which are the fundamental equations of ray tracing.

In addition to the above equations, if one assumes that
\begin{equation}
\v H (\v r ) = \v h(\v r ) e^{ik_0 \phi (\v r )},
\label{eq6-11-14}
\end{equation}
it can be shown easily from Maxwell's equations that when $k_0 \to
\infty$,
\begin{subequations}
\begin{equation}
\nabla \phi \times \v h (\v r ) \simeq c \epsilon \v e (\v r ),
\label{eq6-11-15a}
\end{equation}
\begin{equation}
\nabla \phi \times \v e (\v r ) \simeq -c \mu \v h.
\label{eq6-11-15b}
\end{equation}
\end{subequations}
Hence,
\begin{equation}
\nabla \phi \cdot \v e \simeq \nabla \phi \cdot \v h \simeq 0.
\label{eq6-11-16}
\end{equation}
The above equations indicate that the wave is locally a \index{High frequency limit} plane wave
in the high frequency limit.

\subsection { Ray Tracing Equations in an Optical Fiber}
\index{Optical fiber!ray tracing method}

In an optical fiber with axial symmetry and uniformity in the $z$
direction, then $dn/ d\phi =0$, and $dn/dz =0$. A point in a ray can
be described by
\begin{equation}
\v r = \hat \rho \rho + \hat z z, \label{eq6-11-17}
\end{equation}
in cylindrical coordinates. In the above, $\hat \rho$ is a function
of $\phi$. Hence, $\v r $ is a function of $(\rho, \phi, z)$.
Extracting the $\hat \rho$ component of (\ref{eq6-11-13}) gives
\begin{equation}
\hat \rho \cdot \frac {d}{ds} \left ( n \frac{ d \v r }{ds} \right
) = \frac{ d }{d\rho} n. \label{eq6-11-18}
\end{equation}
But
\begin{equation}
\frac{d \v r }{ds} = \frac{d}{ds}(\hat \rho \rho + \hat z z ) =
\hat \rho \frac{d\rho}{ds} + \rho \frac{d \hat \rho}{ds} + \hat z
\frac{dz}{ds}, \label{eq6-11-19}
\end{equation}
and
\begin{equation}
\frac{d\hat \rho}{ds} = \frac{d}{ds} (\hat x \cos \phi + \hat y
\sin \phi ) = - \hat \phi \frac{d \phi}{ds}. \label{eq6-11-20}
\end{equation}
Therefore,
\begin{equation}
\hat \rho \cdot \frac{d}{ds} \left ( n \frac{d\v r}{ds} \right) =
\hat \rho \cdot \frac{d}{ds} \left ( n \hat \rho \frac{d\rho}{ds}
- n \hat \phi \rho \frac{d\phi}{ds} \right). \label{eq6-11-21}
\end{equation}
However,
\begin{equation}
\frac{d\hat \phi}{ds} = \frac{d}{ds} (\hat x \sin \phi - \hat y
\cos \phi ) = \hat \rho \frac{d \phi}{ds}. \label{eq6-11-22}
\end{equation}
Consequently,
\begin{equation}
\hat \rho \cdot \frac{d}{ds}\left ( n \frac{ d\v r}{ds} \right ) =
\frac{d}{ds} \left ( n \frac{d \rho}{ds} \right ) - n \rho \left (
\frac{d \phi}{ds} \right ) ^2 = \frac{d}{d\rho}n.
\label{eq6-11-23}
\end{equation}
Similarly, the $\hat \phi$ component of (\ref{eq6-11-13}) can be
extracted to obtain
\begin{equation}
\begin{split}
\hat \phi \cdot \frac{d}{ds} \left ( n \frac{ d \v r }{ds} \right
) &= \hat \phi\cdot \frac{d}{ds} \left ( n \hat \rho \frac{d
\rho}{ds} - n \hat \phi \rho
\frac{d \phi}{ds} \right )\\
&= - n \left ( \frac{d \rho}{ds} \right ) \left ( \frac{d\phi}{ds}
\right ) - \frac{d}{ds} \left (n \rho \frac{d \phi}{ds} \right ) =
0 . \label{eq6-11-24}
\end{split}
\end{equation}
Extracting the $\hat z$ component of (\ref{eq6-11-13}) yields
\begin{equation}
\frac{d}{ds} \left ( n \frac{dz}{ds} \right ) = 0.
\label{eq6-11-25}
\end{equation}
Equation (\ref{eq6-11-25}) can be readily integrated to yield
\begin{equation}
ds = \frac{n}{n_0 \cos \theta _i} dz. \label{eq6-11-26}
\end{equation}
where $n_0$ is the value of $n$ at the initial point, and $\theta_i$
is the angle of the ray with the $z$-axis initially. Replacing $n /
ds$ in (\ref{eq6-11-24}) with (\ref{eq6-11-26}) gives
\begin{equation}
\frac {d\rho}{ds}\frac{d\phi}{dz} + \frac{d}{ds} \left ( \rho
\frac{d \phi}{dz} \right ) = 0, \label{eq6-11-27}
\end{equation}
which can be rewritten as
\begin{equation}
\frac{d}{ds} \left ( \rho ^2 \frac{d\phi}{dz} \right ) = 0.
\label{eq6-11-28}
\end{equation}
Equation (\ref{eq6-11-28}) can be integrated to yield
\begin{equation}
\rho ^2 \frac{d \phi}{dz} = C_0. \label{eq6-11-29}
\end{equation}

To integrate (\ref{eq6-11-23}), one multiplies it by $n$ and use
(\ref{eq6-11-26}) to replace $n / ds$ with $n_0\cos \theta _i / dz$
to obtain
\begin{equation}
\frac {d^2}{dz^2} \rho - \rho \left ( \frac{ d \phi}{dz} \right )
^2 = \frac{1}{2 n_0^2 \cos ^2 \theta _i} \frac{d}{d\rho} n^2.
\label{eq6-11-30}
\end{equation}
Using (\ref{eq6-11-29}) for $\frac{d\phi}{dz}$ yields
\begin{equation}
\frac{d^2}{dz^2} \rho = \frac{1}{\rho^3} C_0^2 +
\frac{1}{2n_0^2\cos ^2 \theta _i} \frac{d }{d\rho} n^2.
\label{eq6-11-31}
\end{equation}
The above equation can be integrated with respect to $\rho$. The
left-hand side is

\begin{equation}
\begin{split}
\int \limits_{\rho_0}^{\rho} \frac{d^2}{dz^{\prime 2}} \rho ' d
\rho ' &= \int \limits _0^z \frac{d^2 \rho'}{d z^{\prime2}}
\frac{d\rho'}{dz'}d z' = \frac{1}{2} \int \limits
_{0}^{z} \frac{d}{dz'} \left ( \frac{d\rho'}{dz'}  \right )^2 dz'\\
&= \frac{1}{2} \left ( \frac{d\rho}{dz} \right )^2 -\frac{1}{2}
D_0. \label{eq6-11-32}
\end{split}
\end{equation}
where $ D_0= \left ( \frac{d \rho}{dz} \right ) _{z=0}^{2}$. The
right hand side of (\ref{eq6-11-31}) yields
$$
\left ( - \frac{1}{2 \rho ^2} C_0^2 + \frac {1}{2 n_0^2 \cos
^2\theta _i} n^2 \right ) \bigg|_{\rho_0}^\rho
$$
\begin{equation}
= \left [ 1 - \left ( \frac{\rho_0}{\rho} \right ) ^2 \right ]
\frac{C_0^2}{2 \rho _0^2} + \frac{1}{2 \cos ^2 \theta _i} \left (
\frac{n^2}{n_0^2} - 1 \right ). \label{eq6-11-33}
\end{equation}
As a result, one gets
\begin{equation}
\left ( \frac{d\rho}{d z} \right ) ^2 = \left [ 1- \left (
\frac{\rho_0}{\rho} \right )^2 \right ] \frac{C_0^2}{\rho_0^2} +
\frac{1}{\cos ^2 \theta _i}\left ( \frac{n^2}{n_0^2} - 1 \right )
+ D_0 , \label{eq6-11-34}
\end{equation}
or

\begin{equation}
z= \int \limits_ {\rho _0}^{\rho} d\rho \left \{ \left [ 1- \left
( \frac{\rho_0}{\rho} \right ) ^2 \right ] \left (
\frac{C_0}{\rho_0} \right ) ^2 + \frac{1}{\cos ^2 \theta _i} \left
( \frac{n^2}{n_0^2} -1 \right ) + D_0 \right \} ^{-\frac{1}{2}}.
\label{eq6-11-35}
\end{equation}
Equation (\ref{eq6-11-35}) is the basic equation for computing the
ray path in an optical fiber, given the initial condition $C_0$,
$D_0$, $\cos \theta _i$, $n_0$ and $\rho_0$ at $z=0$.

\subsection { Determination of Initial Conditions}
\index{Ray tracing method!initial conditions}

\begin{figure}[ht]
\begin{center}
\hfil\includegraphics[width=2.5truein]{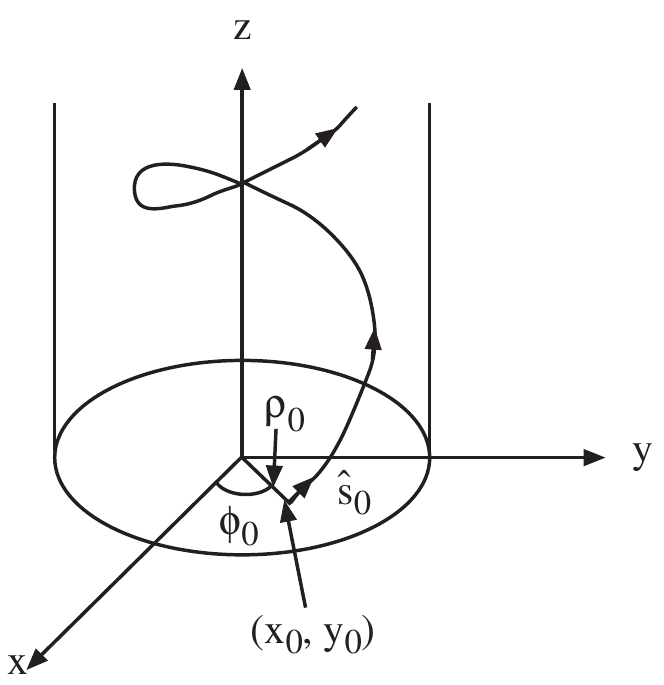}\hfil
\end{center}
\caption{Launching of a skew ray pointing at $\hat s_0$ at $(x_0,
y_0, z=0)$.}\label{fg6111}
\end{figure}


One can assume that an optical ray enters an optical fiber at $z=0$
at $(x_0, y_0)$ and the direction of the ray is pointed at $\hat s
_0$, where
\begin{equation}
\hat s _0 = \hat x \sin \theta _i \cos \phi _i + \hat y \sin
\theta _i \sin \phi _i + \hat z \cos \theta _i . \label{eq6-11-36}
\end{equation}
If $d\phi/dz$ at $(x_0, y_0, z=0)$ of the ray can be found, then
$C_0$ in (\ref{eq6-11-29}) can be found. The $\hat \phi_0$
component of $\hat s_0$ can be found by $\hat s_0 \cdot \hat
\phi_0$ which is
\begin{equation}
\begin{split}
\hat s_0 \cdot \hat \phi_0 &= \hat s_0 \cdot [ - \hat x \sin
\phi_0 + \hat y
\cos \phi_0 ] \\
&= -\sin\phi _0\sin\theta_i\cos\phi_i + \cos \phi_0 \sin\theta _i
\sin \phi_i.
\end{split}
\label{eq6-11-37}
\end{equation}
Hence,
\begin{equation}
\rho _0 d \phi = \hat s _0 \cdot \hat \phi_0 ds = \hat s_0 \cdot
\hat \phi_0 / \cos \theta _i dz. \label{eq6-11-38}
\end{equation}
Therefore,
\begin{equation}
C_0 = \rho _0 \frac{\hat s _0 \cdot \hat \phi_0}{\cos \theta _i}.
\label{eq6-11-39}
\end{equation}

\begin{figure}[ht]
\begin{center}
\hfil\includegraphics[width=5.0truein]{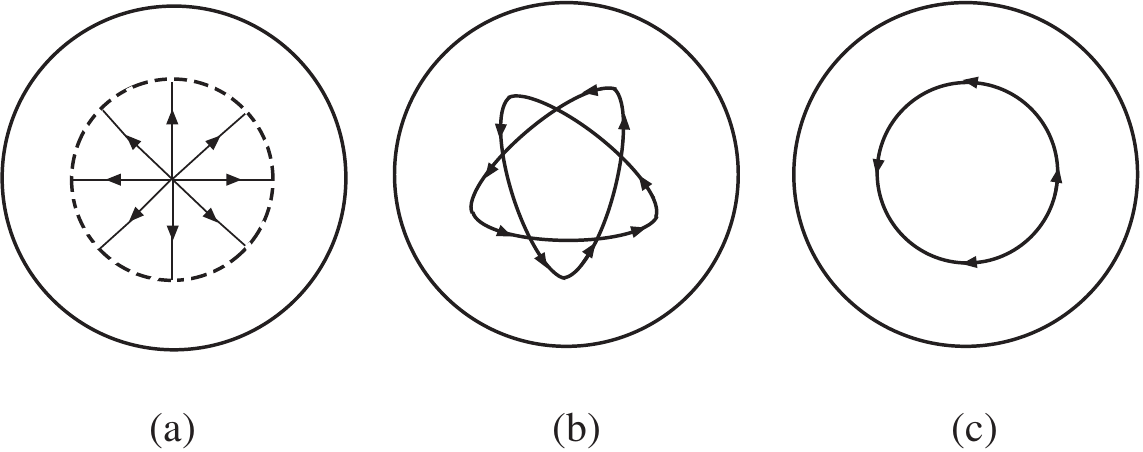}\hfil
\end{center}
\caption{Cross-sectional pictures of (a) a meridional ray, (b) a
complex skew ray, and (c) a skew ray that is helical.}\label{fg6112}
\end{figure}


If $d\, \rho \, dz$ of the ray at $(x_0, y_0, z = 0)$ is known,
then $D_0$ in (\ref{eq6-11-32}) is known. But
\begin{equation}
d\rho = \hat s _0 \cdot \hat \rho_0 ds = \frac{\hat s _0 \cdot
\hat \rho_0}{\cos \theta _i}dz . \label{eq6-11-40}
\end{equation}
Therefore,
\begin{equation}
\begin{split}
D_0 = \left ( \frac{d\rho}{dz}\right ) _{z=0}^2 = \frac{(\hat s
_0\cdot \hat
\rho _0)^2}{\cos ^2 \theta _i} &= \frac{\sin ^2 \theta _i (\cos \phi_i \cos \phi_0 +\sin \phi_i \sin \phi_0 ) ^2}{\cos ^2 \theta _i}\\
&= \tan ^2 \theta _i \cos ^2 (\phi_i - \phi_0) \label{eq6-11-45}
\end{split}
\end{equation}
The above provides sufficient initial conditions to launch a ray
pointing at $\hat s_0$ from the point $(x_0, y_0, z=0)$. A ray
which is only propagating radially, i.e., $\hat s _0 \cdot \hat
\phi_0 = C_0 = 0$, is known as a \index{Meridional ray} meridional ray.  A ray that
propagates in both the $\hat \rho $ and $\hat \phi$ directions is
known as a \index{Skew ray} skew ray.  A skew ray that propagates at a constant
distance from the fiber axis is called a helical ray.

The ray equation (\ref{eq6-11-35}) can be used to derive index
profile so that the axial velocity of a ray is independent of the
launch condition. In this manner, the dispersion of the fiber will
be minimized.

\vfill\eject

\centerline{\bf Exercises for Chapter 6}

\vskip 12pt \noindent {\bf Problem 6-1:} \noindent

\begin{figure}[ht]
\begin{center}
\hfil\includegraphics[width=4.0truein]{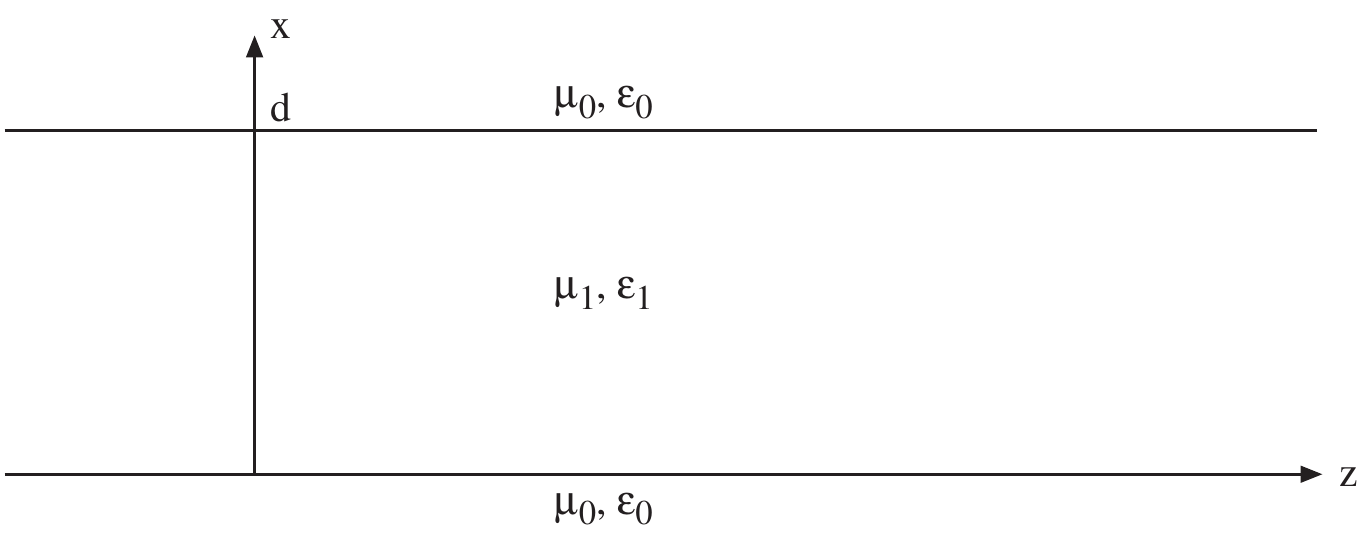}\hfil
\end{center}
\caption{Problem 6-1}
\end{figure}


Show that the guidance condition for a dielectric slab of
thickness $d$, permittivity $\epsilon_1$ and permeability $\mu_1$
suspended in air can be simplified to
$$
\alpha_{0x}\frac{d}{2}=\frac{\mu_0}{\mu_1}k_{1x}\frac{d}{2}\tan
\left(\frac{k_{1x}d-m\pi}{2}\right),
$$
for the TE case. In the above, $\alpha_{0x} = \sqrt{k_z^2-k_0^2}$,
$k_{1x}=\sqrt{k_1^2-k_z^2}$ and $m$ is an integer. Plot the right
and the left hand side of the equation as a function of $k_{1x}d$
for $m$ even and $m$ odd to obtain graphical solutions to the
above equation.

\vskip 8pt \noindent {\bf Problem 6-2:} Find the phase velocity of
the TM$_1$ mode of a symmetric dielectric slab waveguide at
cutoff. By analyzing the phase velocity in the vicinity of cutoff,
find the group velocity analytically. Explain the answers.

\vskip 8pt \noindent {\bf Problem 6-3:} For a circular dielectric
waveguide of radius $a$, find the cutoff frequencies of the
TE$_{01}$ mode, the EH$_{11}$ mode and the HE$_{12}$ mode. Which
is the next higher order mode to the HE$_{11}$ mode? If $a=10\mu$,
$n_1=1.6$ and $n_2=1.5$, what is the bandwidth for single mode
propagation in the optical fiber? Explain why the usable bandwidth
of an optical fiber is not this bandwidth.

\vskip 8pt \noindent {\bf Problem 6-4:} Find the phase velocity of
the TE$_{01}$ mode of an optical fiber near cutoff. Also, find the
group velocity near cutoff analytically. Explain what you have
observed about the answer.

\vskip 8pt \noindent {\bf Problem 6-5:}
\begin{itemize}
\item[(a)] Show that the ratio of $E_z$ to $H_z$ in the core region of an optical fiber is given by
\begin{equation*}
\begin{split}
H_1/E_1 &=
\frac{-nk_z}{i\omega}\left(\frac{1}{(k_{1\rho}a)^2}+\frac{1}{(\alpha_2a)^2}
\right)\\
&\left(\frac{\mu_1J_n'(k_{1\rho}a)}{k_{1\rho}aJ_n(k_{1\rho}a)}+
\frac{\mu_2K_n'(\alpha_{2}a)}{\alpha_{2}aK_n(\alpha_{2}a)}\right)
^{-1}
\end{split}
\end{equation*}
\item[(b)] Using the equation for the guidance conditions of the EH and HE modes, show that the ratio $E_1/H_1$ is in fact larger for the HE modes compared to the EH modes.
\end{itemize}

\noindent {\bf Problem 6-6:}
\begin{itemize}
\item[(a)] Starting with the equation for the guidance condition of the modes in a step-index fiber (nonweakly guiding), by assuming that $\mu_1=\mu_2$ and that $\epsilon_1\approx\epsilon_2$, show that the guidance condition for the
weak contrast
 optical fiber mode can be derived as
$$
\frac{k_{1\rho}J_{n\pm 1}(k_{1\rho} a)} {J_n(k_{1\rho} a) } =
\pm\frac{\alpha_2K_{n\pm 1}(\alpha_2 a)}{K_n(\alpha_2a)}
$$
Show that the above two equations are equivalent, and hence the
modes they define are degenerate.
\item[(b)] From the derivation, which of the EH mode is degenerate with
the HE mode when the contrast of the fiber tends to zero?
\item[(c)] Even though in the weak contrast fiber approximation, the equation
indicates that the $x$ component and the $y$ component of the
electric field are decoupled, they are actually weakly coupled in
this limit.  Hence, it is not possible to conceive some modes to
have $E_x$ only or $E_y$ only.  The LP$_{11}$ mode is such a mode.
To test your physical insight, sketch the electric field on the $xy$
plane on the cross section of a weak contrast optical fiber.
\end{itemize}

\noindent {\bf Problem 6-7:} Write a computer program to solve for
the $k_z$ of the HE$_{11}$, HE$_{21}$, TE$_{01}$, TM$_{01}$,
EH$_{11}$, HE$_{31}$ and HE$_{12}$ modes. That is, produce the
dispersion curve for the first three families of modes shown in
Figure 6.2.5 of the text. (a)  First, generate the dispersion
curves when $n_1/n_2=1.5$.  (b) Second, generate the dispersion
curves when $n_1/n_2=1.01$.

\noindent {\bf Problem 6-8:} Using $1/\omega$ as a small parameter,
expand $\v E_s$ and $E_z$ as perturbation series in a weak contrast
optical fiber, and show that $E_z\sim |E_s|/\omega$.

\vskip 8pt \noindent {\bf Problem 6-9:} \noindent Explain why the
Rayleigh scattering loss and ultraviolet absorption loss diminish
with wavelength in Figure \ref{fg642}, while the infrared
absorption loss increases with wavelength. Is it reasonable to
assume that waveguide imperfection loss does not alter with
wavelength?

\vskip 8pt \noindent {\bf Problem 6-10:} For the harmonic
expansion method, why is the assumption of standing wave inside
the waveguide and outgoing wave outside the waveguide not a valid
assumption in Subsection 6.6.1?

\vskip 8pt \noindent {\bf Problem 6-11:}

\begin{figure}[ht]
\vspace {-2 in}
\begin{center}
\hfil\includegraphics[width=2.40truein]{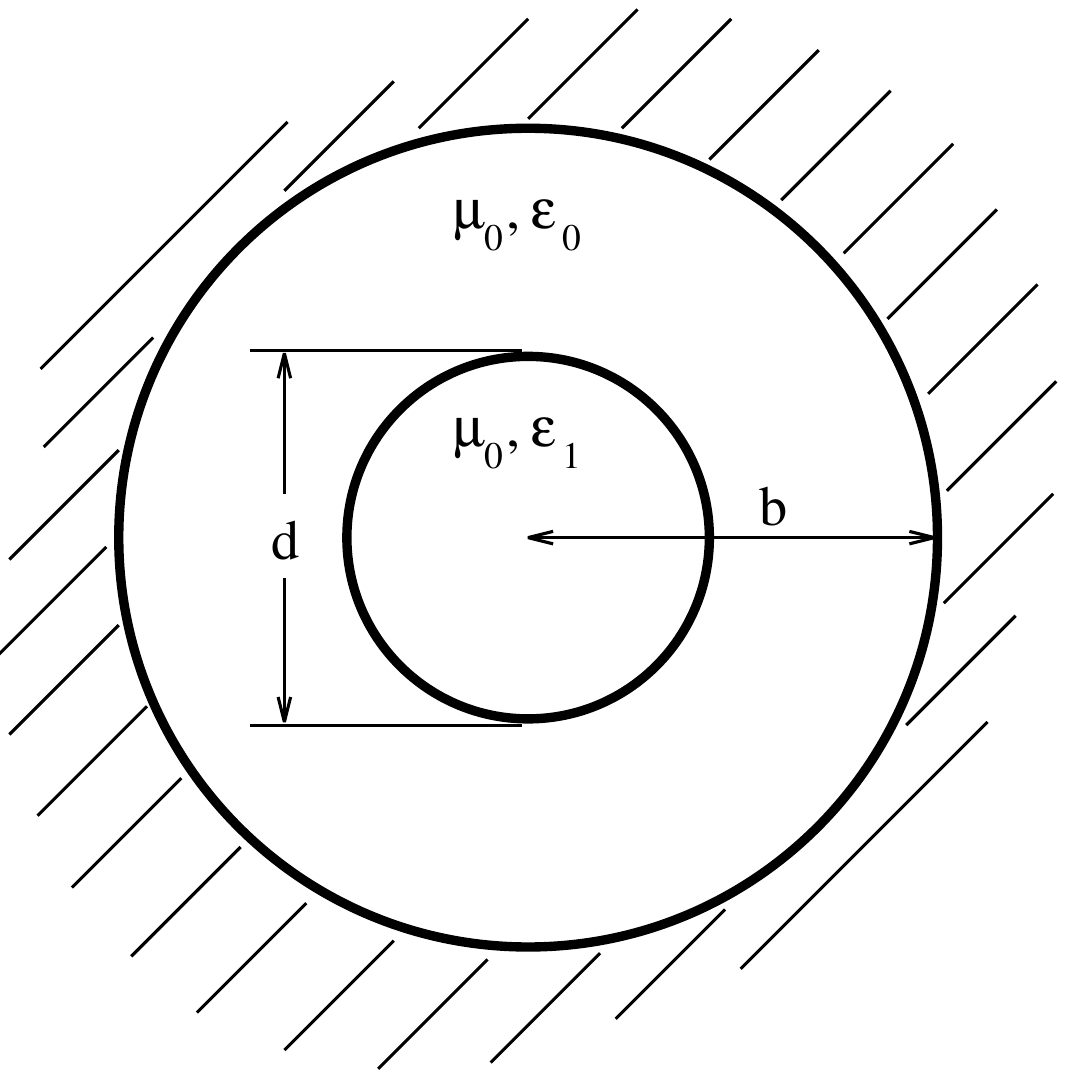}\hfil
\end{center}
\caption{Problem 6-10}
\end{figure}


\noindent A circular waveguide is loaded with a circular
dielectric rod of diameter $d$ as shown in the above figure. Find
the change in the propagation constant $k_z$ of the TE$_{11}$ mode
due to the presence of this rod using a perturbation calculation.

\vskip 8pt \noindent {\bf Problem 6-12:} The vector wave equation
governing the propagation of waves in an anisotropic medium with
reflection symmetry can be shown to be
$$
\hat z\times \dyad\mu_s\cdot\hat
z\times\nabla_s\times\mu_{zz}^{-1}\nabla_s\times \v E_s +
\nabla_s\epsilon_{zz}^{-1}\nabla_s\cdot\dyad\epsilon_s\cdot\v E_s
-\omega^2\hat z\times\dyad\mu_s\cdot\hat
z\times\dyad\epsilon_s\cdot\v E_s -k_z^2\v E_s = 0.
$$
\begin{itemize}
\item[(a)] Derive a variational expression for the guided wave number $k_z^2$.
\item[(b)] Using Rayleigh-Ritz procedure, derive a matrix equation for the guided wave numbers of the waveguide.
\end{itemize}

\noindent {\bf Problem 6-13:}

\begin{figure}[ht]
\begin{center}
\hfil\includegraphics[width=2.40truein]{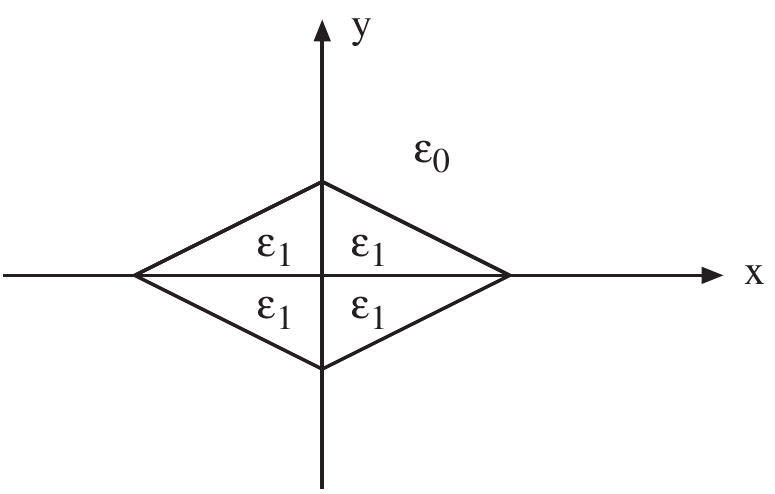}\hfil
\end{center}
\caption{Problem 6-12}
\end{figure}


\noindent The harmonic expansion method, in theory, can be used
for dielectric waveguides of any shapes. However, when the
waveguide has symmetry about the $x$ and the $y$ axes as shown in
the above figure, the $z$ components of the fields are either even
or odd about the the $x$ and the $y$ axes.
\begin{itemize}
\item[(a)] Proof that if $E_z$ ($H_z$) is even about the $x$ or the $y$ axes, $H_z$ ($E_z$) has to be odd about the $x$ or the $y$ axes, and vice versa.
\item[(b)] Because of this symmetry, we need not have to solve for the solution of the waveguide in the full space, but only in a quadrant of the full space.  Use the harmonic expansion method, together with point matching, derive the guidance conditions for the modes for which $E_z$ is even about the $x$ axis and odd about the $y$ axis.  Give a reason why this method of solution is preferable.
\end{itemize}
\noindent {\bf Problem 6-14:} Assume a parallel waveguide
terminated abruptly so that the modes of the waveguide will
radiate into free space. Repeat the derivation of Subsection 6.7.1
for the terminated parallel plate waveguide with a flange as
shown.

\begin{figure}[ht]
\vspace{-0.22 in}
\begin{center}
\hfil\includegraphics[width=1.2truein]{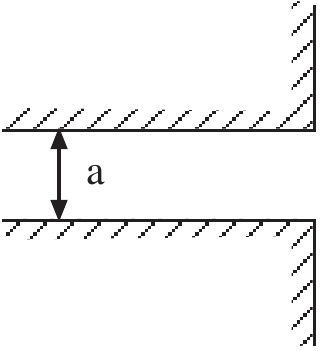}\hfil
\end{center}
\caption{Problem 6-13}
\end{figure}


\vskip 8pt \noindent {\bf Problem 6-15:}
\begin{itemize}
\item[(a)] In the WKB method, explain why the prefactor of $1/\sqrt{s_\rho}$ term is necessary for energy conservation.
\item[(b)] Repeat the analysis of Subsection 6.8.1 for a planar geometry. Is the result of Equation (\ref{eq6-8-19}) much different compared to the planar geometry case? Explain why.
\end{itemize}

\vskip 8pt \noindent {\bf Problem 6-16:}

\begin{figure}[ht]
\vspace{-3.1 in}
\begin{center}
\hfil\includegraphics[width=2.8truein]{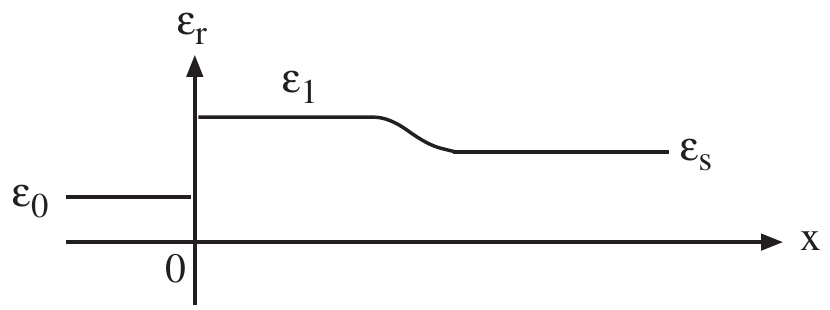}\hfil
\end{center}
\caption{Problem 6-15}
\end{figure}


A planar waveguide has a index profile as shown. It is terminated
by a perfect electric conductor at the $x=0$ surface. Use the WKB
method, write down the guidance condition for such a waveguide.

\vskip 8pt \noindent {\bf Problem 6-17:}
\begin{itemize}
\item[(a)] If $\epsilon_r(z)$ in Equation (\ref{eq6-9-9}) of Subsection 6.9.1 for fixed $x,y$ is described by a symmetric dielectric slab. Find the transcendental equation from which the eigenvalues of Equation (\ref{eq6-9-9}) can be found.
\item[(b)] Repeat the same for Equation (\ref{eq6-9-14}) of Subsection 6.9.2.
\end{itemize}

\noindent {\bf Problem 6-18:} In the beam-propagation method, the
equation of propagation is sometimes written as
$$
\phi(x,y,z+\Delta z) \cong e^{ik_0\delta n \Delta z/2}
e^{i\sqrt{k_0^2n_0^2+ \nabla _s^2}\Delta z}  e^{ik_0\delta n
\Delta z/2} \phi (x,y,z).
$$
as opposed to Equation (\ref{eq6-10-7}) of Section 10. Explain if
there is any advantage of writing the propagation equation as
shown above.

\vskip 8pt \noindent {\bf Problem 6-19:} Describe how you would
solve the ray-tracing equations (\ref{eq6-11-13}) given in Section
6.11 numerically.

\vskip 8pt \noindent {\bf Problem 6-20:} Write a computer program
to compute Equation (\ref{eq6-11-35}) of Subsection 6.11.1  for
ray tracing in an optical fiber.

\bibliographystyle{plain}
\bibliography{emt,matrix,fmm,math,fastsum,parallel,cs,my}









\def\chaptitle{Microwave Integrated Circuits}
\index{Microwave integrated circuits}

\chapter{\chaptitle}

\markboth{\smallbooktitle}{\chaptitle}


\def\v #1{{\bf #1}}
\def\vg #1{{\boldsymbol #1}}
\def\dyad#1{\overline {\bf #1}}
\def\dyadg#1{\overline {\vg #1}}
\def\beq{\begin{equation}}\def\eeq{\end{equation}}
\def\tinf{\text{\it inf\,}}\def\^{\hat}
\def\cal#1{\mathcal{#1}}
\def\ed{\end{document}}
\def\dalpha{\dyadg\alpha}
\def\dbeta{\dyadg\beta}
\def\Rg{\Re g}
\def\l{\label}
\def\sinc{\text{sinc}}




\vskip 10mm



\begin{figure}[ht]
\begin{center}
\hfil\includegraphics[width=5.0truein]{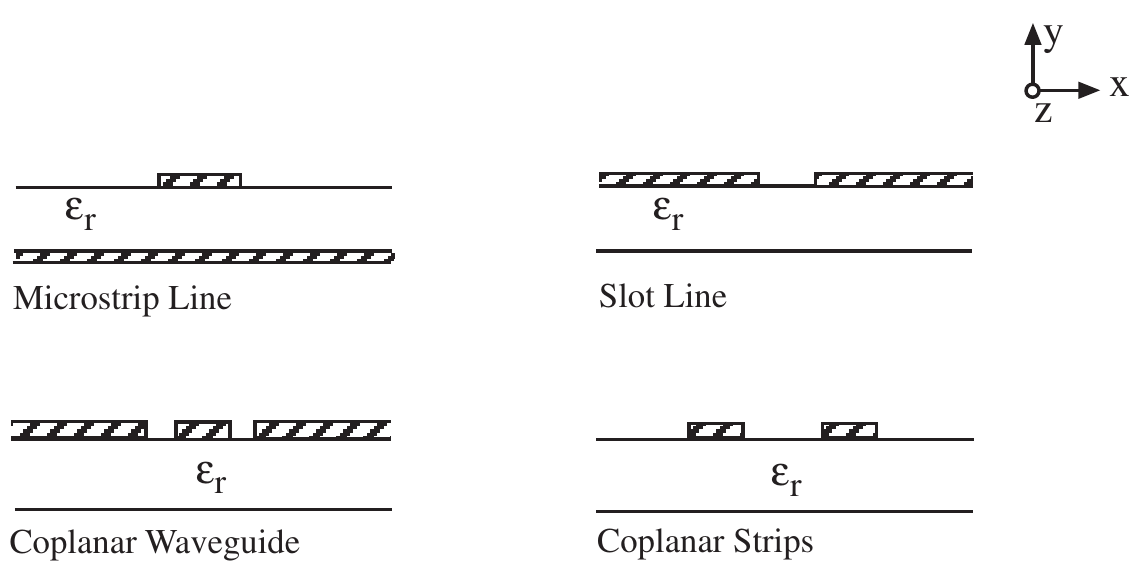}\hfil
\end{center}
\caption{Different kinds of microwave integrated circuits
waveguides.}\label{fg711}
\end{figure}

Due to the advent of integrated circuits at microwave frequencies,
microwave \index{Integrated circuit waveguides} integrated circuit waveguides are omnipresent in
microwave technologies.  Their ease of fabrication, low cost,
conformal nature have made them extremely popular.  Moreover, they
are easily integrated with other circuits. Some examples are shown
in Figure \ref{fg711}.  More recently, microwave integrated circuits
have been used to study cavity QED (quantum electrodynamics) that
plays an important role in quantum computing \cite{KOCH_YU}. Hence,
they are being used at the frontier of scientific investigations as
well.

The microstrip line has a long history.  Since its appearance before
World War II, it has been continuously worked on.  The early work
uses quasi-TEM approximation which does not account for dispersive
effects in the line
\cite{WHEELER,BRYANT_WEISS,SCHNEIDER,GREEN7,YAMASHITA_MITTRA,YAMASHITA,STINEHELFER}.
The use of this waveguide at higher frequencies calls for the
analysis accounting for dispersive effect
\cite{ZYSMAN_VARON,DENLINGER,HORNSBY_GOPINATH1,SCHMITT_SARGES,MITTRA_ITOH,
JANSEN1,KOBAYASHI_ANDO,CHEW_GUREL}. Other works related to analysis
of microstrip integrated circuits are
\cite{CHEW_KONG7,CHEW_KONG71,CHEW_KONG72,JANSEN2,ITOH_MITTRA,ITOH_RESONANCE,JANSEN3,ITOH7,POZAR7}.
Recent typical works in this area are given in
\cite{KERGONOU_ETAL,RUBIN,SUN_ZHU}.

\section {{{ Quasi-TEM Approximation}}}
\index{Quasi-TEM approximation}

The integrated circuits waveguiding structures cannot support a
\index{Mode!TEM} TEM mode, for if they do, the phase matching condition will be
violated at the interface between the inhomogeneities. This is
because a TEM wave has the phase velocity of the medium in which
the wave is traveling.

However, if the wavelength under consideration is much larger than
the transverse structure of the waveguide, we can show that the
fundamental mode of such a structure is almost TEM or \index{Quasi-TEM field} quasi-TEM. A
fundamental mode is the mode that is propagating when
$\omega\rightarrow 0$ or $\lambda\rightarrow \infty$.

We can write Maxwell's equations by separating out the transverse
and longitudinal components as
\begin{equation}
\nabla _s\times\v E_s=ik\eta\v H_z, \label{eq7-1}
\end{equation}
\begin{equation}
\nabla _s\times\v H_s=-ik\eta ^{-1}\v E_z, \label{eq7-2}
\end{equation}
\begin{equation}
\frac {\partial }{\partial z}\^ z\times\v E_s-\^ z\times\nabla
_sE_z=ik\eta\v H_s, \label{eq7-3}
\end{equation}
\begin{equation}
\frac {\partial }{\partial z}\^ z\times\v H_s-\^ z\times\nabla
_sH_z=-ik\eta ^{-1}\v E_s, \label{eq7-4}
\end{equation}
where $\eta =\sqrt {\frac {\mu }{\epsilon}}$. For a structure whose
dominant transverse dimension is much smaller than the wavelength,
the transverse variation of the field would be more rapid than the
longitudinal variation.  The transverse variation of the field has
to be fast enough for the field to match the boundary condition,
namely,  the $x$ and $y$ variations will vary on the length scale of
$\delta$. To emphasize this fact, we can perform a \index{Coordinate stretch transformation} coordinate
stretching transformation \cite{BENDER_ORSZAG} by letting
\begin{equation}
x=\delta x', \qquad y=\delta y'. \label{eq7-5}
\end{equation}
Under such a coordinate stretching transformation,
\begin{equation}
\nabla _s\rightarrow \frac 1{\delta }\nabla' _s, \label{eq7-6}
\end{equation}
Equations (\ref{eq7-3}) and (\ref{eq7-4}) become
\begin{equation}
ik_z\delta\^ z\times\v E_s-\^ z\times\nabla' _sE_z=ik\eta\delta\v
H_s, \label{eq7-7}
\end{equation}
\begin{equation}
ik_z\delta\^ z\times\v H_s-\^ z\times\nabla' _sH_z=-ik\eta
^{-1}\delta\v E_s, \label{eq7-8}
\end{equation}
where we have assumed $e^{ik_zz}$ dependence of the field. When
${\delta}/{\lambda }\rightarrow 0$, then, $k\delta\rightarrow 0$.
Since $k_z =\sqrt {k^2-k_s^2}<k$, we also expect
$k_z\delta\rightarrow 0$, when ${\delta }/{\lambda }\rightarrow 0$.
Therefore, in the long wavelength limit, from the above equations,
it is seen that
\begin{equation}
(E_z, \eta H_z)\sim O(k\delta )(\v E_s, \eta \v H_s).
\label{eq7-9}
\end{equation}
In other words, $|E_z|\ll |\v E_s|$, $|H_z|\ll |\v H_s|$, implying
that the field is quasi-TEM.
 Consequently, we can write
(\ref{eq7-1}) and (\ref{eq7-2}) as
\begin{equation}
\nabla _s\times\v E_s=0, \label{eq7-10}
\end{equation}
\begin{equation}
\nabla _s\times\v H_s=0. \label{eq7-11}
\end{equation}
From the divergence equation for source free region, it implies that
\begin{align}
\div \epsilon \v E=\nabla_s\cdot \epsilon \v E_s+\partial_z \epsilon
E_z=0
\end{align}
Since the $E_z$ component is much smaller that the transverse
component, one gets
\begin{equation}
\nabla _s\cdot\epsilon\v E_s =0, \label{eq7-12}
\end{equation}
Similarly, one gets
\begin{equation}
\nabla _s\cdot\mu\v H_s=0. \label{eq7-13}
\end{equation}

\begin{figure}[ht]
\begin{center}
\hfil\includegraphics[width=3.0truein]{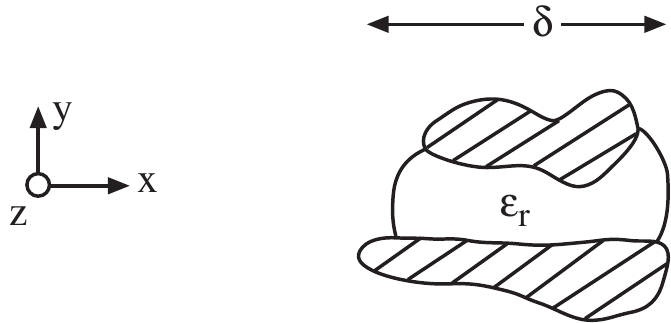}\hfil
\end{center}
\caption{A generic  geometry for the analysis of microwave
integrated circuits waveguide.}\label{fg712}
\end{figure}


Consequently, the \index{Quasi-TEM field!transverse} transverse field of a quasi-TEM mode is
essentially static. Because of this, the waveguide can be analyzed
as if a transversely static TEM mode is propagating on it. We can
solve the transverse electrostatic problem to find the line
capacitance $C$ of the line. The line inductance $L$ can be obtained
by solving the magnetostatic problem. Then, $k_z$, the axial
wavenumber, can be found by
\begin{equation}
k_z=\omega \sqrt {LC}. \label{eq7-14}
\end{equation}

The above analysis indicates that when the wavelength is long, the
axial variation of the field is slow compared to the transverse
variation. The transverse variation of the field has to be such that
the field can match the boundary condition on the metallic
conductors, which is much smaller in dimension than the wavelength.
Hence, the transverse variation of the field must balance itself
resulting in equations (\ref{eq7-10}) to (\ref{eq7-13}), which are
the static equations.

Please, be noted that the reason for the quasi-TEM field here is
quite different from that in the weak-contrast optical fiber.  In
the weak-contrast optical fiber, the reason for quasi-TEM is the
paraxial nature of the wave as the frequency increases and the
unimportance of the polarization term comparatively.  In the
weak-contrast optical fiber, the transverse dimension can be on the
order or wavelength or larger, and yet the field is quasi-TEM.

\subsection { Microstrip Line Capacitance---The Spectral Domain Approach}
\index{Microstrip line}
\index{Microstrip line!capacitance}

We shall discuss how the line capacitance of a microstrip line can
be found. There is no closed-form solution for such a class of
problem. When $\epsilon _0=\epsilon _1$, one may solve such problems
by conformal mapping. When ${h}/{w}\ll 1$, the problem can be solved
by asymptotic matching \cite{POH_CHEW_KONG}.
 However, to get an accurate
value of $C$ for all $w/h$, a numerical analysis is preferable.

To find the line capacitance of the microstrip line, one can first
solve for the charge distribution on the line. Then, the capacitance
can be easily found from the equation $Q = CV$, where $Q$ is the
total charge per unit length on the line, and $V$ is the voltage
applied between the strip and the ground plane. To find the charge
distribution, we can first formulate its governing equation for the
charge distribution.  A potential $\phi$ can be defined such that
$\v E_s = -\nabla _s\phi$ and $\nabla _s^2\phi = 0$, because $\v
E_s$ is an electrostatic field in the long wavelength limit.

\begin{figure}[ht]
\begin{center}
\hfil\includegraphics[width=4.0truein]{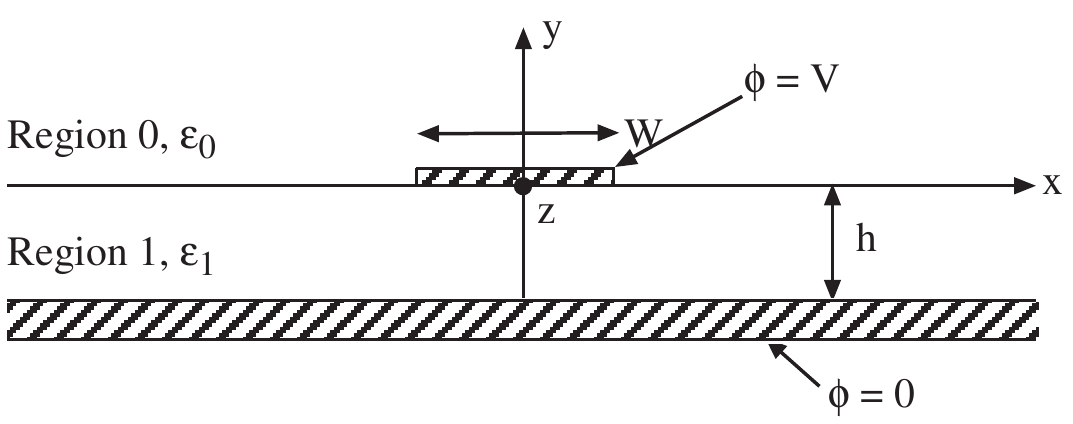}\hfil
\end{center}
\caption{Geometry for deriving the line capacitance of a
microstrip line.}\label{fg713}
\end{figure}


\begin{figure}[ht]
\begin{center}
\hfil\includegraphics[width=4.0truein]{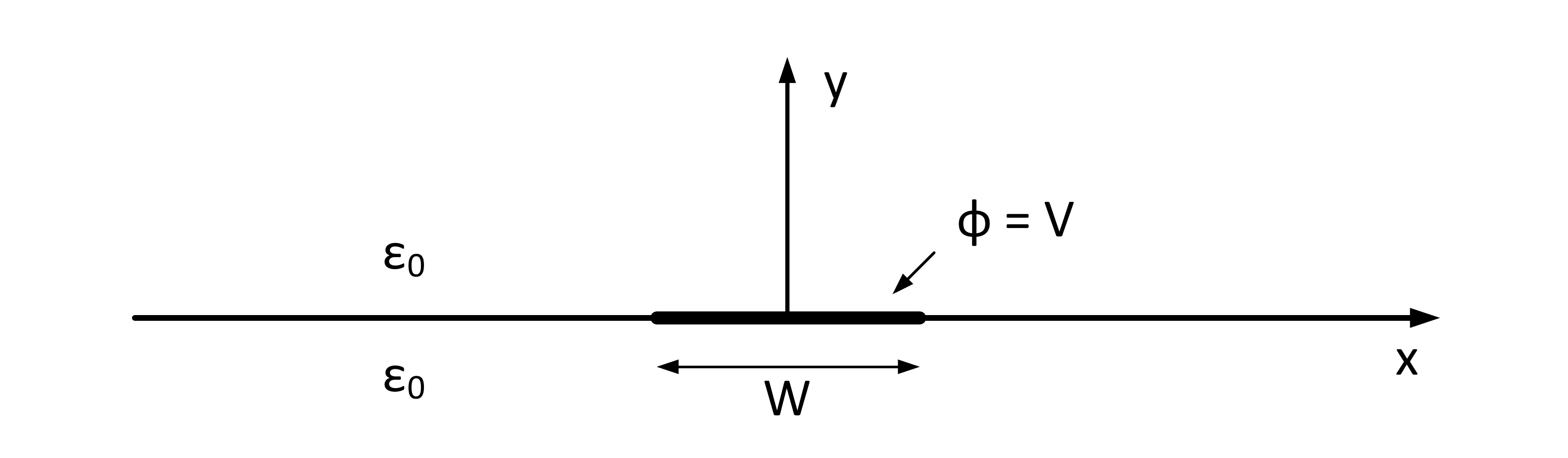}\hfil
\end{center}
\caption{A charged strip hanging in free space assumed to be
infinitesimally thin.}\label{fg714}
\end{figure}


Before formulating the integral equation for the geometry shown in
Figure \ref{fg713}, let us first consider the case of a charged
strip suspended in free space shown in Figure \ref{fg714}. Using
Fourier transforms, which are the gist of the spectral domain
approach, the potential can be written as
\begin{equation}
\phi _{0\pm }(x, y)=\frac 1{2\pi} \int^\infty_{-\infty} d\lambda
e^{i\lambda x}\tilde {\phi }_{0\pm}(\lambda , y). \label{eq7-1-1}
\end{equation}
where the subscript $+$ and $-$ denote $y > 0$ and $y < 0$,
respectively. Since $\nabla _s^2\phi _{0\pm }(x, y)=0$, we deduce
that
\begin{equation}
\frac {d^2}{dy^2}\tilde {\phi }_{0\pm }(\lambda , y)-\lambda
^2\tilde {\phi }_{0\pm }(\lambda , y)=0, \label{eq7-1-2}
\end{equation}
or that
\begin{equation}
\tilde {\phi }_{0\pm }(\lambda , y)=A_{0\pm }e^{-|\lambda
|y}+B_{0\pm }e^{|\lambda |y}. \label{eq7-1-3}
\end{equation}
Furthermore, $\tilde {\phi }_{0+}(\lambda , y)=0$, $y\rightarrow
\infty$, and $\tilde {\phi }_{0-}(\lambda , y)=0$, $y\rightarrow
-\infty$, imply that $B_{0+} = A_{0-} = 0$. But since $\phi _{0+}(x,
y=0)=\phi _{0-}(x, y=0)$ for all $x$, we have $\tilde {\phi
}_{0+}(\lambda , y=0) = \tilde {\phi }_{0-}(\lambda, y=0)$ implying
that $A_{0+} = B_{0-}$. Consequently, we have
\begin{equation}
\phi _{0\pm }(x, y)=\frac 1{2\pi }\int^\infty_{-\infty}\ d \lambda
e^{i\lambda x}A_0(\lambda)e^{\mp |\lambda |y},\quad
\begin{aligned}
\ &y > 0\\
\ &y < 0
\end{aligned} .
\label{eq7-1-4}
\end{equation}

\index{Microstrip line!charge distribution}
The charge on the strip is given by
\begin{equation}
\sigma (x)=-\epsilon _0\left[ \frac {\partial\phi _{0+}}{\partial
y}(x, y=0)-\frac{\partial\phi _{0-}}{\partial y}(x, y=0)\right],
\label{eq7-1-5}
\end{equation}
or that
\begin{equation}
\sigma (x)=\frac {\epsilon _0}{\pi }\int^\infty_{-\infty} d\lambda
e^{i\lambda x}|\lambda |A_0(\lambda ). \label{eq7-1-6}
\end{equation}
Defining the Fourier transform of $\sigma (x)$ as $\tilde {\sigma
}(\lambda )$, it is seen that
\begin{equation}
A_0(\lambda )=\frac {\tilde {\sigma }(\lambda )}{2\epsilon
_0|\lambda |}, \label{eq7-1-7}
\end{equation}
or, in general, the potential becomes
\begin{subequations}
\begin{equation}
\phi _{0\pm }(x, y)=\frac 1{4\pi\epsilon _0}\int^\infty_{-\infty}
d\lambda e^{i\lambda x\mp|\lambda |y}\frac {\tilde {\sigma
}(\lambda )}{|\lambda |},\quad
\begin{aligned}
\ &y > 0 \\
\ &y < 0\cr
\end{aligned} .
\label{eq7-1-8a}
\end{equation}
or
\begin{equation}
\phi _0(x, y)=\frac 1{4\pi\epsilon _0}\int^\infty_{-\infty}
d\lambda e^{i\lambda x-|\lambda y|}\frac {\tilde {\sigma }(\lambda
)}{|\lambda |}. \label{eq7-1-8b}
\end{equation}
\end{subequations}

\begin{figure}[ht]
\begin{center}
\hfil\includegraphics[width=4.5truein]{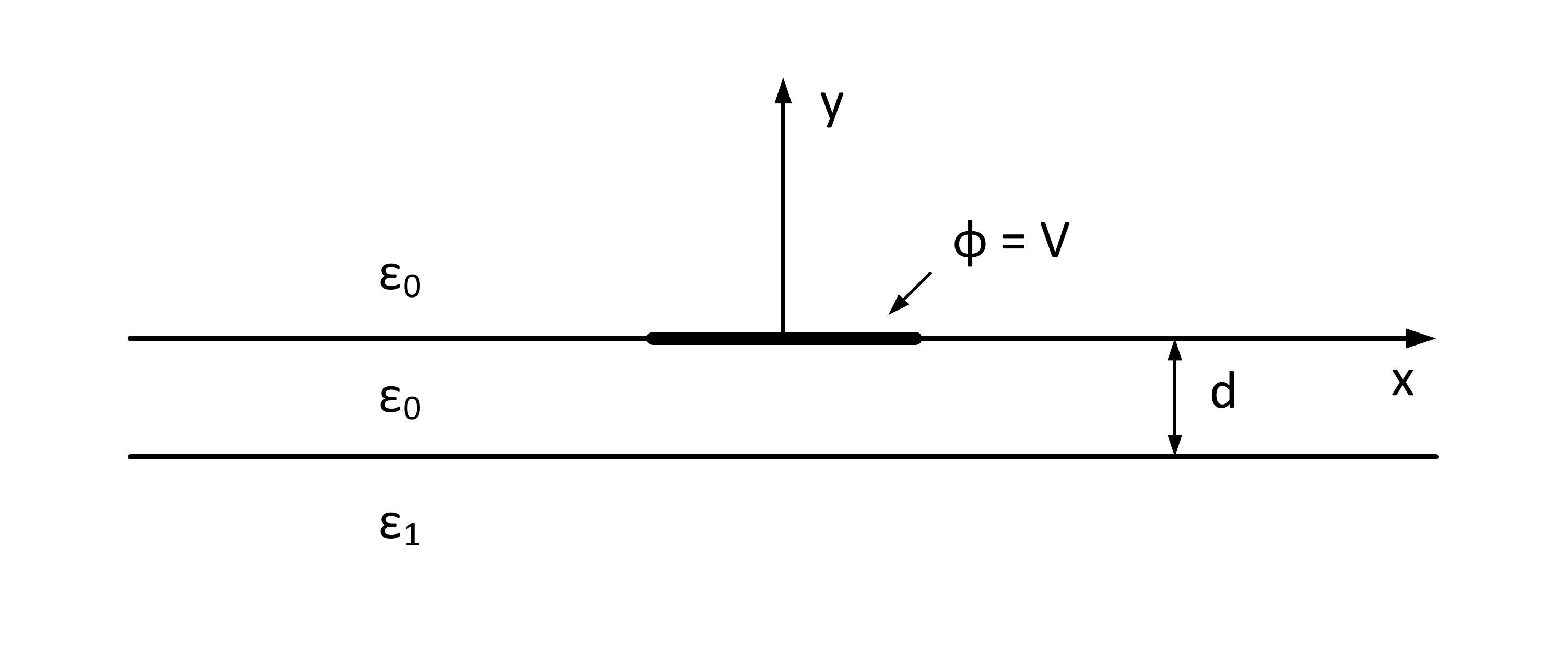}\hfil
\end{center}
\caption{A charged strip, assumed to be infinitesimally thin, over a
dielectric half-space.}\label{fg715}
\end{figure}


If we now place the charged strip over a dielectric half space,
the potential in region 0 becomes
\begin{equation}
\phi _0(x, y)=\frac 1{4\pi\epsilon _0}\int^\infty_{-\infty}
d\lambda\frac {\tilde {\sigma }(\lambda )}{|\lambda |} e^{i\lambda
x}[e^{-|\lambda y|}+r_{01}e^{-|\lambda|(y+d)-|\lambda d|}],
\label{eq7-1-9}
\end{equation}
where $r_{01}$ is a reflection coefficient relating the reflected
potential $e^{-|\lambda |y}$ to the incident potential
$e^{-|\lambda y|}$ at $y = -d$. In region 1, the potential is
\begin{equation}
\phi _1(x, y)=\frac 1{4\pi\epsilon _0}\int^\infty_{-\infty}
d\lambda\frac {\tilde {\sigma}(\lambda )}{|\lambda |}e^{i\lambda
x}t_{01}e^{|\lambda |(y+d)-|\lambda d|}, \label{eq7-1-10}
\end{equation}
where $t_{01}$ is a transmission coefficient relating the
transmitted potential $e^{|\lambda |y}$ to the incident potential
$e^{-|\lambda y|}$. The continuity of potential at $y = -d$
requires that
\begin{equation}
1 + r_{01} = t_{01}. \label{eq7-1-11}
\end{equation}
The continuity of normal electric flux at $y = -d$ implies that
\begin{equation}
\epsilon _0(1-r_{01})=\epsilon _1t_{01}. \label{eq7-1-12}
\end{equation}
Solving (\ref{eq7-1-11}) and (\ref{eq7-1-12}) yields
\begin{equation}
r_{01}=\frac {\epsilon _0-\epsilon _1}{\epsilon _0+\epsilon _1},
\qquad t_{01}=\frac {2\epsilon _0}{\epsilon _0+\epsilon _1}.
\label{eq7-1-13}
\end{equation}

\begin{figure}[ht]
\begin{center}
\hfil\includegraphics[width=4.5truein]{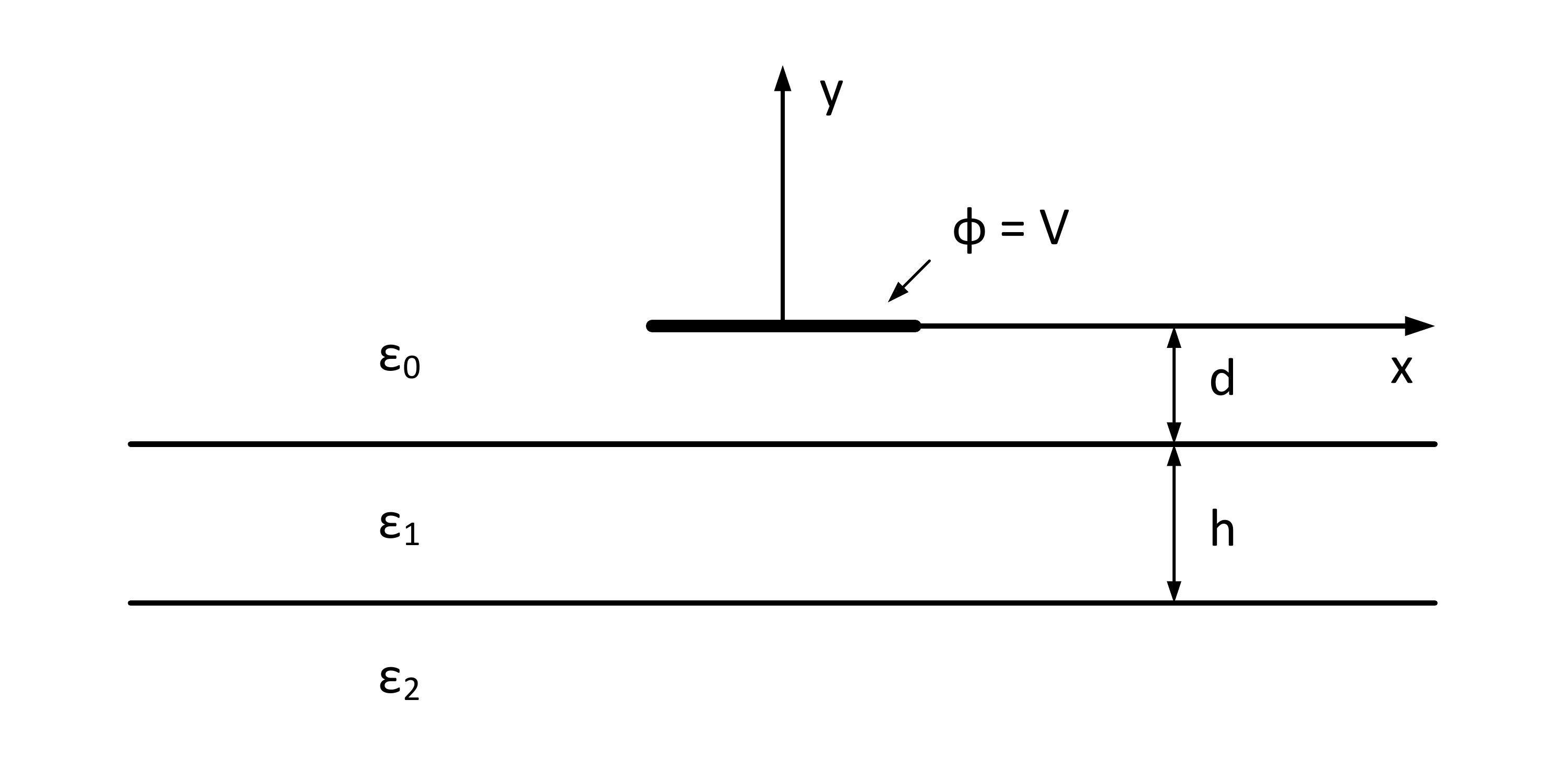}\hfil
\end{center}
\caption{A charged strip over a dielectric slab.}\label{fg716}
\end{figure}

%

\noindent For the geometry of Figure \ref{fg716}, we can find the
total reflection coefficient using a geometric series expansion
\cite{WFIM7,KONG7} to obtain
\begin{equation}
\tilde r_{01}=r_{01} +\frac {t_{01}r_{12}t_{10}e^{-2|\lambda
|h}}{1-r_{10}r_{12}e^{-2|\lambda |h}}. \label{eq7-1-14}
\end{equation}
Equation (\ref{eq7-1-14}) can be written as a recursive relation
if we have more subsurface layers. In this case, the potential in
region 1 can be written as
\begin{equation}
\phi _0(x, y)=\frac 1{4\pi\epsilon _0}\int^\infty_{-\infty}
d\lambda\frac {\tilde {\sigma}(\lambda )e^{i\lambda x}}{|\lambda
|}[e^{-|\lambda y|}+\tilde r_{01}e^{-|\lambda |(y+d)-|\lambda
d|}]. \label{eq7-1-15}
\end{equation}
For the geometry of Figure \ref{fg713}, $d = 0$, $r_{12} = -1$;
therefore,
\begin{equation}
\phi _0(x, y)=\frac 1{4\pi\epsilon _0}\int^\infty_{-\infty}
d\lambda \frac {\tilde {\sigma }(\lambda )e^{i\lambda x}}{|\lambda
|}e^{-|\lambda |y}(1+\tilde r_{01}), \qquad y > 0,
\label{eq7-1-16}
\end{equation}
where
\begin{equation}
1+\tilde r_{01}=t_{01}-\frac {t_{01}t_{10}e^{-2|\lambda
|h}}{1-r_{01}e^{-2|\lambda |h}}=t_{01}\frac {1-e^{-2|\lambda
|h}}{1-r_{01}e^{-2|\lambda |h}}. \label{eq7-1-17}
\end{equation}
Requiring that the potential be $V$ on the strip, then the
integral equation for $\tilde {\sigma }(\lambda )$ can be written
as
\begin{subequations}
\begin{equation}
\frac 1{2\pi }\int^\infty_{-\infty} d\lambda \tilde G(\lambda
)\tilde {\sigma }(\lambda)e^{i\lambda x}=V, \qquad |x|<\frac {w}2,
\label{eq7-1-18a}
\end{equation}
\begin{equation}
\sigma (x)=\frac 1{2\pi }\int^\infty_{-\infty} d\lambda\tilde
{\sigma }(\lambda )e^{i\lambda x}=0, \qquad |x|>\frac {w}2,
\label{eq7-1-18b}
\end{equation}
\end{subequations}
where $\tilde {\sigma }(\lambda )$ is the Fourier transform of
$\sigma (x)$, and
\begin{equation}
\tilde G(\lambda )=\frac 1{|\lambda |(\epsilon _0+\epsilon
_1)}\frac {1-e^{-2|\lambda |h}}{1-\frac {\epsilon _0-\epsilon
_1}{\epsilon _0+\epsilon _1}e^{-2|\lambda |h}}. \label{eq7-1-19}
\end{equation}
Equations (\ref{eq7-1-18a}) and (\ref{eq7-1-18b}) are also known as
\index{Dual integral equations} dual integral equations.  Dual integral equations in general do not
have closed-form solutions, except for semi-infinite structures.  In
that case, they can be solved using the Wiener-Hopf technique
\cite{WIENER_HOPF7,CHEWPHD}.

In order to solve (\ref{eq7-1-18b}), we use Galerkin's method and
let
\begin{equation}
\sigma (x)=\sum _{n=0}^Na_nf_n(x), \label{eq7-1-20}
\end{equation}
where $f_n(x) = 0$, $|x|>\frac w{2}$, and that the Fourier transform
of $f_n(x)$ exists. Furthermore, $f_n(x)$, $n = 1,\ldots, \infty$ is
complete for $|x|<\frac {w}2$. Then, we deduce that
\begin{equation}
\tilde {\sigma }(\lambda )=\sum _{n=0}^Na_n\tilde f_n(\lambda ),
\label{eq7-1-21}
\end{equation}
where $\tilde f_n(\lambda )$ is the Fourier transform of $f_n(x)$.
Substituting (\ref{eq7-1-21}) into (\ref{eq7-1-18a}), we have
\begin{equation}
\frac 1{2\pi }\sum _{n=0}^N a_n\int^\infty_{-\infty} d\lambda\tilde
G(\lambda )\tilde f_n(\lambda )e^{i\lambda x}=V. \label{eq7-1-22}
\end{equation}
To remove the $x$ dependence in (\ref{eq7-1-22}), we multiply it
by $f_m(x)$ and integrate over $x$, to obtain
\begin{equation}
\sum _{n=0}^Na_n\int^\infty_{-\infty} d\lambda\tilde f_m(-\lambda
)\tilde G(\lambda )\tilde f_n (\lambda )=2\pi V\int \limits
_{{-w}/2}^{w/2}f_m(x)dx. \label{eq7-1-23}
\end{equation}
The above is a matrix equation of the form
\begin{equation}
\sum _{n=0}^NA_{mn}a_n=b_m, \label{eq7-1-24}
\end{equation}
from which we can solve for $a_n$'s. Once $a_n$'s are known, we
can find $\sigma (x)$ from (\ref{eq7-1-20}). Then
\begin{equation*}
Q =\int \limits _{{-w}/2}^{w/2}\sigma (x)dx,
\end{equation*}
and the line capacitance is given by $C = Q/V$.

For the geometry considered in Figure \ref{fg713}, the quasi-TEM
mode is even symmetric about $x = 0$. Therefore, we need only to
pick even functions for our basis functions in (\ref{eq7-1-20}).
Then, by the theory of Fourier transform, $\tilde
f_n(\lambda)=\tilde f_n(-\lambda )$ if $f_n(x)$ is an even function.
Furthermore, since $\tilde G(-\lambda ) =\tilde G(\lambda )$,
(\ref{eq7-1-23}) becomes
\begin{equation}
\sum _{n=0}^N a_n\int \limits _0^{\infty } d\lambda\tilde
f_m(\lambda )\tilde G(\lambda)\tilde f_n(\lambda )=2\pi V\int
\limits _0^{w/2}f_m(x)dx. \label{eq7-1-25}
\end{equation}
Clearly, $A_{mn}$ is a symmetric matrix.

Equation (\ref{eq7-1-18a}) has a different meaning if it is
transformed back to $x$-space, i.e.,
\begin{equation}
\int \limits _{-w/2}^{w/2} dx' G(x-x')\sigma (x')=V, \qquad
|x|<\frac w{2}, \label{eq7-1-26}
\end{equation}
where $G(x)$ is the inverse Fourier transform of $\tilde G(\lambda
)$. And $G(x)$ could be thought of as the Green's function
generating the potential due to a line of point surface charge at $y
= 0$. Hence, the convolution of $G(x)$ with $\sigma (x)$ gives the
potential. However, closed-form expression does not exist for $G(x)$
in general. A more convenient method to calculate the integral
(\ref{eq7-1-26}) is in the spectral domain as in (\ref{eq7-1-18a})
where $\tilde G(\lambda )$ exists in closed-form.

Once $C$ is known, we can estimate $k_z$ via (\ref{eq7-14}). To find
$L$, we make use of the fact that if $\epsilon _1 = \epsilon _0$, a
pure TEM mode propagates on the microstrip line. In this case,
\begin{equation}
k_z=k_0=\omega\sqrt {L_0C_0}. \label{eq7-1-27}
\end{equation}
Therefore, $L_0$ can be found once $C_0$, the line capacitance with
$\epsilon _1=\epsilon _0$, is known. $C_0$ can be found by solving
the integral equation above. Since $L_0$, which is obtained by
solving the magnetostatic problem, remains unchanged when $\epsilon
_1\ne \epsilon _0$, we have
\begin{equation}
k_z=\sqrt {\frac C{C_0}}k_0. \label{eq7-1-28}
\end{equation}
An effective relative dielectric constant can be defined such that
\begin{equation}
\epsilon _{re}=\frac C{C_0}. \label{eq7-1-29}
\end{equation}
It is the dielectric constant with which one can fill the space
homogeneously around a microstrip line to yield the same line
capacitance as the inhomogeneously filled microstrip line.

\begin{figure}[ht]
\begin{center}
\hfil\includegraphics[width=3.5truein]{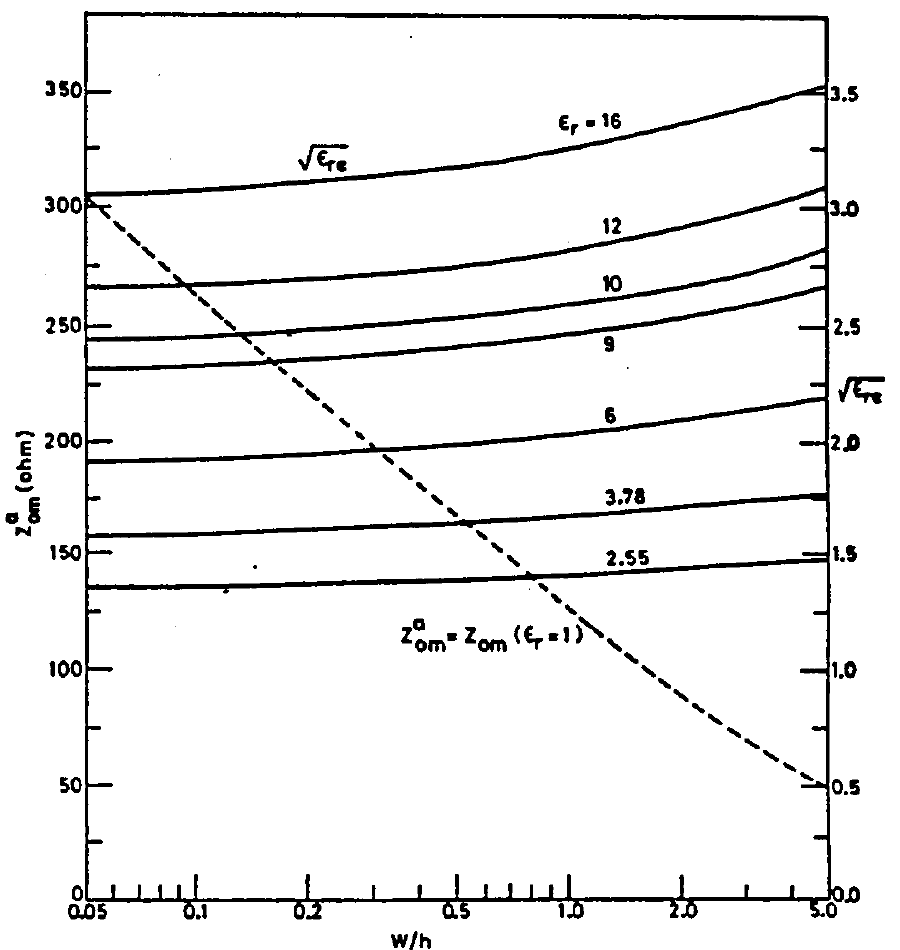}\hfil
\end{center}
\caption{\index{Microstrip line!characteristic impedance} Characteristic impedance of a microstrip line as a function
of line dimension $w/h$---quasi-static approximation (from H.A.
Wheeler \cite{WHEELER}.)} \label{fg717}
\end{figure}


\begin{figure}[ht]
\begin{center}
\hfil\includegraphics[width=2.5truein]{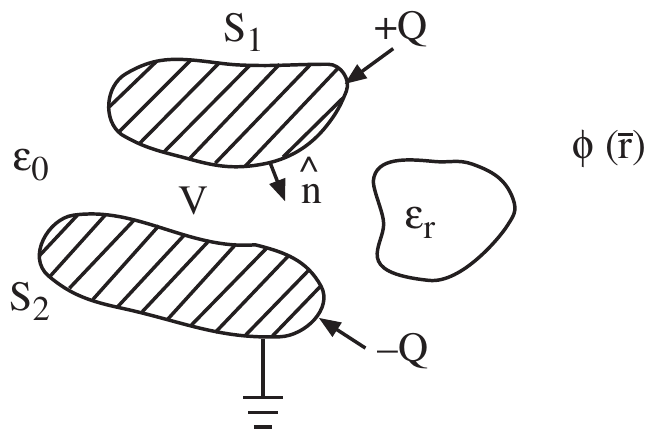}\hfil
\end{center}
\caption{Geometry for the analysis of bounds for capacitance.}\label{fg718}
\end{figure}


A characteristic impedance can also be defined as
\begin{equation}
Z_0=\sqrt {\frac {L_0}C}=\sqrt {\frac {\mu _0\epsilon
_0}{C_0C}}=\frac {Z_0^a }{\sqrt {\epsilon _r}}, \label{eq7-1-30}
\end{equation}
where $Z_0^a$ is the characteristic impedance of the air-filled
microstrip line.

\subsection { Variational Expressions and Bounds for Capacitance}
\index{Microstrip line!variational expressions}

There exist bounds for the capacitance between two conductors due
to the minimum energy principle
\cite{JACKSON7,THOMPSON,PARR,CHOW_ETAL}.  Due to the prevalence use
of capacitance extraction software in the computer chip industry,
this could be of importance.

The exact solution of Laplace's equation minimizes the energy  in
the potential as shall be shown. The capacitance can be related to
the energy stored in the potential or charge in two ways, i.e.
\begin{equation}
W = \frac{1}{2} C \Phi^2 = \frac{1}{2} \frac{Q^2}{C},
\label{eq7-1-31}
\end{equation}
where $W$ is the energy stored in the potential $\Phi$, the
voltage difference between the two conductors. Then if $W$ is not
accurately known, the exact capacitance $C_e$ is bounded by
\begin{equation}
\frac{2W}{\Phi^2} \geq C_e \geq \frac{Q^2}{2W} \label{eq7-1-32}
\end{equation}
if either $\Phi$ is kept constant of if $Q$ is kept constant on
the metallic conductor. The above assertion shall be proven in the
following.

In the first inequality in \eqref{eq7-1-32}, the potential satisfies
the boundary condition exactly, but does not satisfy Laplace's
equation exactly in the space in between the conductors. It can be
rewritten as
\begin{equation}
\frac{\int\limits_V \epsilon |\v E|^2 dV}{\Phi^2} =
\frac{\int\limits_V \epsilon |\nabla \phi|^2 dV}{\Phi^2} \geq C_e.
\label{eq7-1-33}
\end{equation}
If we let $\phi = \phi_e + \delta \phi$ such that $ \delta \phi =
0$ on $S_1$ and $S_2$, because $\phi$ satisfies the boundary
condition exactly, then
\begin{equation}
\int \limits_V \epsilon |\nabla \phi|^2 = \int \limits_V \epsilon
|\nabla \phi_e |^2 dV + 2\int \limits_V \epsilon \nabla \phi_e \cdot
\nabla \delta \phi W + \int \limits _V \epsilon |\nabla \delta \phi
|^2 dV. \label{eq7-1-34}
\end{equation}
But
\begin{equation}
\begin{split}
\int\limits_V \epsilon\nabla \phi_e \cdot \nabla \delta \phi dV &=
\int \limits_V \nabla \cdot (\delta \phi \epsilon \nabla \phi_e)
dV- \int \limits_V
\delta \phi ( \nabla \cdot \epsilon \nabla \phi_e) d V\\
&= - \int \limits_{S_1+ S_2} \hat n \cdot ( \delta \phi \epsilon
\nabla \phi_e ) d S - \int \limits_V \delta \phi ( \nabla \cdot
\epsilon \nabla \phi _e ) dV. \label{eq7-1-35}
\end{split}
\end{equation}
The first term in (\ref{eq7-1-35}) vanishes because $\delta \phi =
0$ on $S_1$ and $S_2$, and the second term vanishes because
$\nabla \cdot \epsilon \nabla \phi_e = 0$, in $V$. Therefore, if
$\phi = \phi_e + \delta \phi$ where $\delta \phi$ is the error
field in $V$, then
\begin{equation}
\int \limits_V \epsilon |\nabla \phi|^2 dV = \int \limits_V
\epsilon |\nabla \phi_e |^2 dV + \int \limits_V \epsilon |\nabla
\delta \phi|^2 dV > \int \limits_V\epsilon |\nabla \phi_e|^2 dV,
\label{eq7-1-36}
\end{equation}
implying the bound in (\ref{eq7-1-33}) for a constant $\Phi$.

The second inequality in (\ref{eq7-1-32}) can be written as
\begin{equation}
\frac{\int \limits_V \epsilon |\v E | ^2 dV}{Q^2} = \frac{\int
\limits_V\epsilon |\nabla \phi|^2 d V}{Q^2} \geq C_e^{-1}
\label{eq7-1-37}
\end{equation}
where $Q$ is constant but $\phi$ satisfies Laplace's equation
exactly in $V$ but does not satisfy the boundary condition.
Similar to (\ref{eq7-1-34}), we let $\phi =\phi_e + \delta \phi$,
but now, $\delta \phi \neq 0 $ on $S_1$ and $S_2$. Similar to
(\ref{eq7-1-35}), we obtain that the second term in
(\ref{eq7-1-34}) is
\begin{equation}
\begin{split}
\int \limits_V \epsilon \nabla \phi_e \cdot \nabla \delta \phi dV
&=\int \limits_V \nabla \cdot ( \epsilon \phi_e \nabla \delta
\phi) dV - \int
\limits_V \phi_e \nabla \cdot \epsilon \nabla \delta \phi dV\\
&=- \int \limits_{S_1 + S_2} \hat n \cdot ( \epsilon \phi_e \nabla
\delta \phi) dS- \int \limits_V \phi_e \nabla \cdot \epsilon
\nabla \delta \phi dV. \label{eq7-1-38}
\end{split}
\end{equation}

The second term in (\ref{eq7-1-38}) vanishes because $\nabla \cdot
\epsilon \nabla \delta\phi = 0 $. Since $\phi_e = \Phi$ on $S_1$ and
$\phi_e = 0 $ on $S_2$, the first term becomes
\begin{equation}
- \Phi \int \limits_{S_1} \hat n \cdot \epsilon \nabla \delta \phi
d S. \label{eq7-1-39}
\end{equation}
Since $- \hat n \cdot \epsilon \nabla \delta \phi = \delta
\sigma$, the error surface charge density on $S_1$, it integrates
to zero because we assume that $Q$ is a constant so that $\delta Q
= 0$. Therefore, the second term in (\ref{eq7-1-34}) vanishes too
for this case. Consequently, we obtain the bound in
(\ref{eq7-1-37}).

Equation (\ref{eq7-1-37}) is also \index{Thompson's theorem} Thompson's theorem \cite{THOMPSON}
which says that a set of charge $Q$ always adjusts itself on a
metallic conductor such that the energy stored in the electric field
is minimized. The inequality (\ref{eq7-1-37}) assumes that the
potential $\phi$ satisfies $\nabla \cdot \epsilon \nabla \phi = 0 $
exactly in $V$ but does not satisfy the boundary condition. Hence,
it is more appropriate to write (\ref{eq7-1-37}) in terms of surface
integrals. To this end, we express using integration by parts that
\begin{equation}
\int \limits_V \epsilon |\nabla \phi |^2 dV = - \int \limits_{S_1+
S_2} dS \phi~ \epsilon~ \hat n \cdot \nabla\phi = \int \limits
_{S_1+ S_2} dS \phi \sigma. \label{eq7-1-40}
\end{equation}
In this manner, (\ref{eq7-1-37}) becomes
\begin{equation}
C_e ^{-1} \leq \frac{\int \limits_{S_1+S_2} d S \phi \sigma}{Q^2}
. \label{eq7-1-41}
\end{equation}
A potential which satisfies $\nabla \cdot \epsilon \nabla \phi = 0
$ exactly in $V$ can be obtained by the Green's function method if
the Green's function is known exactly.  Hence,
\begin{equation}
\phi(\v r ) = \int \limits_{S_1+S_2} d S ' G (\v r, \v r') \sigma
(\v r '), \label{eq7-1-42}
\end{equation}
or that (\ref{eq7-1-41}) can be expressed as
\begin{equation}
C_e ^{-1} \leq \int \limits_{S_1+S_2} dS \sigma (\v r ) \int
\limits_{S_1+S_2} d S' G ( \v r, \v r') \sigma (\v r')/ Q^2 ,
\label{eq7-1-43}
\end{equation}
or
\begin{equation}
C _e^{-1} \leq \frac{\langle \sigma , G, \sigma \rangle}{Q^2} .
\label{eq7-1-44}
\end{equation}
where
$$
\langle \sigma , G, \sigma \rangle=\int \limits_{S_1+S_2} dS \sigma
(\v r ) \int \limits_{S_1+S_2} d S' G ( \v r, \v r') \sigma (\v r')
$$

The expressions that
\begin{equation}
C= \int \limits_V dV \epsilon |\nabla \phi | ^2 / \Phi ^2
\label{eq7-1-45}
\end{equation}
and
\begin{equation}
C^{-1} = \langle \sigma, G , \sigma \rangle / Q^2 \label{eq7-1-46}
\end{equation}
are both variational expressions. Rayleigh-Ritz procedure can be
applied to (\ref{eq7-1-45}) to solve for $\phi$ which is the same
as applying the finite element method to solve $\nabla \cdot
\epsilon \nabla \phi = 0 $. Applying Rayleigh-Ritz procedure to
solve (\ref{eq7-1-46}) for $\sigma$ is the same as applying
Galerkin's method to solve the integral equation
\begin{equation}
\int \limits_{S_1+S_2} dS' G (\v r, \v r') \sigma (\v r') =\left\{
\begin{aligned}
 &V, \qquad \v r \in S_1\\
 &0, \qquad \v r \in S_2
\end{aligned}
\right. \label{eq7-1-47}.
\end{equation}

\section {{{ Microstrip Line---A Frequency Dependent Theory}}}
\index{Microstrip line!frequency dependent theory}

The quasi-TEM model of the microstrip line triumphs in predicting
the value of $k_z$ or the phase velocity in the long wavelength
limit. However, $k_z$ is in general frequency dispersive. In order
to have a $k_z$ that is valid at high frequencies as well, we need
to solve the full wave solution to the microstrip line problem. To
do this, an integral equation can be formulated from Maxwell's
equation with no approximation. From the integral equation, the
guidance condition of the microstrip line can be solved for.

\begin{figure}[ht]
\begin{center}
\hfil\includegraphics[width=3.5truein]{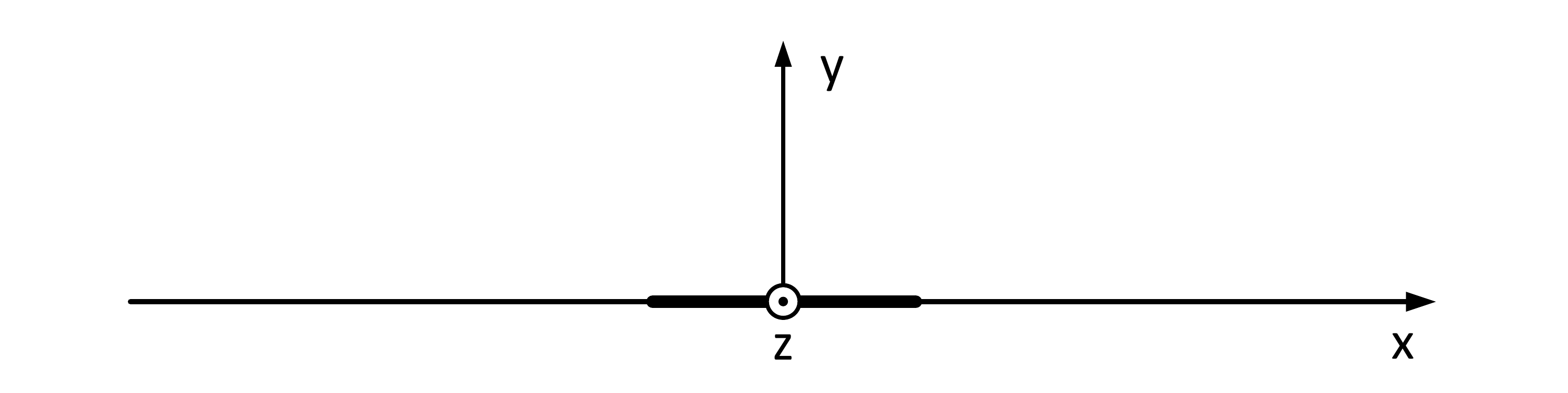}\hfil
\end{center}
\caption{An infinitesimally thin strip hanging in free
space.}\label{fg721}
\end{figure}


\subsection { Derivation of the Integral Equation}
\index{Microstrip line!integral equation}

Before deriving an integral equation for the microstrip line, let us
consider first a metallic strip carrying a current suspended in free
space. The field around the strip can be decomposed into TE to $y$
and TM to $y$ fields by $H_{0y}$ and $E_{0y}$, respectively. Using
Fourier transforms, the fields can be written as
\begin{equation}
E_{0y}(x, y, z)=e^{ik_zz}\frac 1{2\pi }\int^\infty_{-\infty} dk_x
e^{ik_xx}\tilde E_{0y}(k_x, y), \label{eq7-2-1}
\end{equation}
\begin{equation}
H_{0y}(x, y, z)=e^{ik_zz}\frac 1{2\pi }\int^\infty_{-\infty} dk_x
e^{ik_xx}\tilde H_{0y}(k_x,y). \label{eq7-2-2}
\end{equation}
In the above, we assume that the fields have $e^{ik_zz}$
variation. Since $(\nabla ^2+k_0^2)E_{0y}=0$, and $(\nabla
^2+k_0^2)H_{0y}=0$, it follows that
\begin{subequations}
\begin{equation}
\left( \frac {d^2}{dy^2}+k_y^2\right) \tilde E_{0y}(k_x, y)=0,
\label{eq7-2-3a}
\end{equation}
\begin{equation}
\left( \frac {d^2}{dy^2}+k_y^2\right) \tilde H_{0y}(k_x, y)=0,
\label{eq7-2-3b}
\end{equation}
\end{subequations}
where $k_y^2 = k_0^2 - k_x^2 - k_z^2$. This implies that
\begin{subequations}
\begin{equation}
\tilde E_{0y}(k_x, y)=e_{0+}e^{ik_yy}+e_{0-}e^{-ik_yy},
\label{eq7-2-4a}
\end{equation}
\begin{equation}
\tilde H_{0y}(k_x, y)=h_{0+}e^{ik_yy}+h_{0-}e^{-ik_yy}.
\label{eq7-2-4b}
\end{equation}
\end{subequations}
However, for $y > 0$, only upward going waves exist, while for $y
< 0$, only have downward going waves exist. Therefore, $e_{0-} =
h_{0-} = 0$, $y>0$, $e_{0+} = h_{0+} = 0$, $y < 0$. Consequently,
(\ref{eq7-2-1}) and (\ref{eq7-2-2}) become
\begin{subequations}
\begin{equation}
E_{0y}(x, y, z)=\frac {e^{ik_zz}}{2\pi }\int^\infty_{-\infty} dk_x
e^{ik_xx}e_{0\pm }e^{\pm ik_yy}, \qquad
\begin{aligned}
 & y > 0 \\
 & y <0
\end{aligned}
\label{eq7-2-5a},
\end{equation}
\begin{equation}
H_{0y}(x, y, z)=\frac {e^{ik_zz}}{2\pi }\int^\infty_{-\infty} dk_x
e^{ik_xx}h_{0\pm }e^{\pm ik_yy}, \qquad
\begin{aligned}
 & y > 0 \\
 & y <0
\end{aligned}
\label{eq7-2-5b}.
\end{equation}
\end{subequations}
On the other hand, if the strip is carrying electric current, then
$E_{0y}(x,y+,z)$ $= -E_{0y}(x,y-,z)$ and $H_{0y}(x,y+,z) =
H_{0y}(x,y-,z)$ for all $x$.  Therefore, $e_{0+} = -e_{0-} = e_0$,
$h_{0+} = h_{0-} = h_0$, and we have
\begin{subequations}
\begin{equation}
E_{0y}(x, y, z)=\pm\frac {e^{ik_zz}}{2\pi }\int^\infty_{-\infty}
dk_xe^{ik_xx+ik_y|y|}e_0(k_x), \label{eq7-2-6a},
\end{equation}
\begin{equation}
H_{0y}(x, y, z)=\frac {e^{ik_zz}}{2\pi }\int^\infty_{-\infty}
dk_xe^{ik_xx+ik_y|y|}h_0(k_x). \label{eq7-2-6b}.
\end{equation}
\end{subequations}

\begin{figure}[ht]
\begin{center}
\hfil\includegraphics[width=3.5truein]{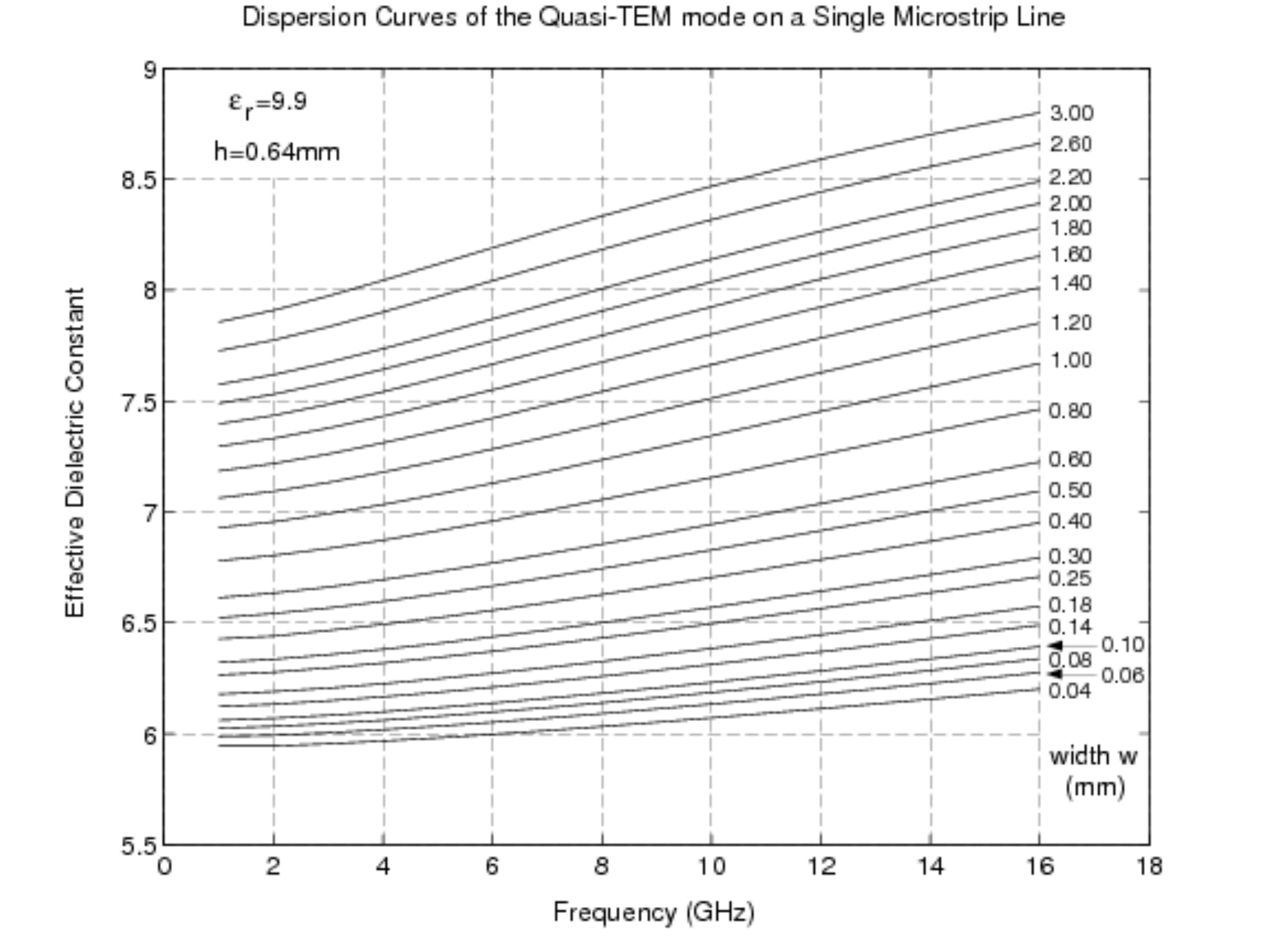}\hfil
\end{center}
\caption{Frequency dependent effective dielectric constant of the
quasi-TEM mode on single microstrip lines (from Jansen
\cite{JANSEN1}, reproduced by G. Papadopoulous.)} \label{fg722}
\end{figure}


If now, the strip is on top of a stratified half space as shown in
Figure \ref{fg722}, the field in region 0 can be written as
\begin{subequations}
\begin{equation}
E_{0y}(x, y, z)=\frac {e^{ik_zz}}{2\pi }\int^\infty_{-\infty}
dk_xe^{ik_xx}e_0(k_x)\left[\pm e^{ik_y|y|}-\tilde
R^{TM}e^{ik_y(y+d)+ik_y|d|}\right] \label{eq7-2-7a},
\end{equation}
\begin{equation}
H_{0y}(x, y, z)=\frac {e^{ik_zz}}{2\pi }\int^\infty_{-\infty}
dk_xe^{ik_xx}h_0(k_x)\left [e^{ik_y|y|}+\tilde
R^{TE}e^{ik_y(y+d)+ik_y|d|}\right] \label{eq7-2-7b}.
\end{equation}
\end{subequations}
In order to relate $e_0$ and $h_0$ to the current on the strip, we
need to derive the transverse components of the field. Since each
spectral component in (\ref{eq7-2-7a}) and (\ref{eq7-2-7b})
consists of waves with $e^{\pm ik_yy}$ dependence, the fields
transverse to $y$ for each spectral component are
\begin{subequations}
\begin{equation}
\tilde {\v E}_s=\frac 1{k_s^2}\left[ \frac {\partial }{\partial
y}\nabla _s\tilde E_y-i\omega\mu _0\^ y\times \nabla _s\tilde
H_y\right] \label{eq7-2-8a}
\end{equation}
\begin{equation}
\tilde {\v H}_s=\frac 1{k_s^2}\left[ \frac {\partial }{\partial
y}\nabla _s\tilde H_y+i\omega\epsilon _0\^ y\times \nabla _s\tilde
E_y\right] \label{eq7-2-8b}.
\end{equation}
\end{subequations}
where the tilde denotes the spectral component, $k_s^2 = k_0^2 -
k_y^2$, and $\nabla _s=\^ x\frac {\partial }{\partial x}+\^ z\frac
{\partial }{\partial z}$.  Applying (\ref{eq7-2-8b}) to
(\ref{eq7-2-7a}) and (\ref{eq7-2-7b}), we have
\begin{equation}
\begin{split}
\v H_s=\frac {e^{ik_zz}}{2\pi } & \int^\infty_{-\infty} dk_x\frac
{e^{ik_xx}}{k_s^2}\{-h_0(k_x)k_y\v k_s[\pm e^{ik_y|y|}+\tilde
R^{TE}e^{ik_y(y+d)+ik_y|d|}]\\ & -\omega\epsilon _0e_0(k_x)\^
y\times\v k_s[\pm e^{ik_y|y|}-\tilde
R^{TM}e^{ik_y(y+d)+ik_y|d|}]\}.
\end{split}
\label{eq7-2-9}
\end{equation}
The discontinuity in $\v H_s$ at $y = 0$ equals the current on the
strip. More precisely, $\v J_s = \^ y\times [\v H_s(y=0+)-\v
H_s(y=0-)]$. Hence,
\begin{equation}
\begin{aligned}
\v J_s & =\frac {e^{ik_zz}}{2\pi }\int^\infty_{-\infty} dk_x\frac
{e^{ik_xx}}{k_s^2}[-2h_0k_y\^ y\times\v k_s+2\omega \epsilon _0e_0\v k_s]\\
& =\frac {e^{ik_zz}}{2\pi }\int^\infty_{-\infty}
dk_x\frac{e^{ik_xx}}{k_s}[\v k_s, -\^ y\times\v k_s]
\begin{bmatrix}
 {2\omega\epsilon _0e_0}/{k_s}\\
 {2k_yh_0}/{k_s}
\end{bmatrix}\\
& =\frac {e^{ik_zz}}{2\pi }\int^\infty_{-\infty} dk_x\dyad F(k_x,
x)\cdot\v K_s(k_x),
\end{aligned}
\label{eq7-2-10}
\end{equation}
where $\dyad F(k_x,x) = \frac {e^{ik_xx}}{k_s}[\v k_s, -\^ y\times\v
k_s]$, $\v K_s^t(k_x)={2[\omega\epsilon _0e_0, k_yh_0]}/ {k_s}$.
Here, $\dyad F(k_x,x)$ is a $2\times 2$ matrix and $\v K_s^t(k_x)$
is a $2\times 1$ vector.

Similarly, $\v E_s$ can be derived to be
\begin{equation}
\begin{split}
\v E_s=\frac {e^{ik_zz}}{2\pi }\int^\infty_{-\infty} dk_x & \frac
{e^{ik_xx}}{k_s^2}\{ -k_y\v k_se_0[e^{ik_y|y|}-\tilde
R^{TM}e^{ik_y(y+d)+ik_y|d|}]\\ & +\omega \mu _0h_0\^ y\times\v
k_s[e^{ik_y|y|}+\tilde R^{TE}e^{ik_y(y+d)+ik_y|d|}]\} .
\label{eq7-2-11}
\end{split}
\end{equation}
We can rewrite (\ref{eq7-2-11}) as
\begin{equation}
\begin{aligned}
\v E_s & =\frac {e^{ik_zz}}{2\pi }\int^\infty_{-\infty} dk_x\frac
{e^{ik_xx}}{k_s}[\v k_s, -\^y\times\v k_s]\\ &
\qquad\begin{bmatrix}
{-k_ye_0[e^{ik_y|y|}-\tilde R^{TM}e^{ik_y(y+d)+ik_y|d|}]} /{k_s}\\
{-\omega \mu _0h_0[e^{ik_y|y|}+\tilde
R^{TE}e^{ik_y(y+d)+ik_y|d|}]}/{k_s}
\end{bmatrix}\\
& =\frac {e^{ik_zz}}{2\pi }\int^\infty_{-\infty} dk_x\dyad F(k_x,
x)\cdot\dyad G(k_x, y)\cdot\v K_s(k_x) \label{eq7-2-12},
\end{aligned}
\end{equation}
where
\begin{equation}
\begin{aligned}
&\dyad G (k_x, y)=\\
& \begin{bmatrix} -\frac {k_y}{2\omega\epsilon _0}
[e^{ik_y|y|}-\tilde
R^{TM}e^{ik_y(y+d)+ik_y|d|}] & 0\\
0 & -\frac {\omega\mu _0}{2k_y} [e^{ik_y|y|}+\tilde
R^{TE}e^{ik_y(y+d)+ik_y|d|}]
\end{bmatrix}.
\label{eq7-2-13}
\end{aligned}
\end{equation}
For a microstrip line, $d = 0$, and from the boundary condition
$\v E_s = 0$ for $|x|<w/2$ and $y = 0$. Therefore, the integral
equation is
\begin{equation}
\v E_s(x, y=0, z)=\frac {e^{ik_zz}}{2\pi }\int^\infty_{-\infty}
\dyad F (k_x, x)\cdot\dyad G_0(k_x)\cdot \v K_s(k_x)dk_x=0, \qquad
|x|<\frac w{2}, \label{eq7-2-14}
\end{equation}
where $\v K_s(k_x)$ is the unknown to be sought, and
\begin{equation}
\dyad G_0(k_x)=-\begin{bmatrix}
\frac {k_y}{2\omega\epsilon _0}[1-\tilde R^{TM}] & 0\\
0 & \frac {\omega \mu _0}{2k_y}[1+\tilde R^{TE}]
\end{bmatrix} .
\label{eq7-2-15}
\end{equation}

\subsection { Vector Fourier Transform (VFT)}
\index{Vector Fourier transform}

With $\dyad F(k_x,x)$ defined in (\ref{eq7-2-10}), there exists a
vector Fourier transform pair given by
\cite{CHEW_HABASHY7,CHEW_GUREL}.
\begin{subequations}
\begin{equation}
\v h(x)=\frac 1{2\pi }\int^\infty_{-\infty} dk_x\dyad F(k_x,
x)\cdot\v H(k_x), \label{eq7-2-16a}
\end{equation}
\begin{equation}
\v H(k_x)=\int^\infty_{-\infty} dx'\dyad F(k_x, -x')\cdot\v h(x'),
\label{eq7-2-16b}
\end{equation}
where the $2\times 2$ matrix $\dyad F$ is
\begin{equation}
\dyad F(k_x, x)=\frac {e^{ik_xx}}{k_s}[\v k_s, -\^ y\times\v
k_s]=\frac {e^{ik_xx}}{k_s}\begin{bmatrix}
k_z & k_x\\
k_x & -k_z
\end{bmatrix} .
\label{eq7-2-16c}
\end{equation}
\end{subequations}
This can be proven by substituting (\ref{eq7-2-16b}) into
(\ref{eq7-2-16a}) to obtain
\begin{equation}
\begin{aligned}
\v h(x) & =\frac 1{2\pi }\int^\infty_{-\infty}
dk_x\int^\infty_{-\infty} dx'\dyad F(k_x, x)\cdot\dyad F(k_x,
-x')\cdot\v h(x')\\ & =\frac 1{2\pi }\int^\infty_{-\infty}
dx'\int^\infty_{-\infty} dk_xe^{ik_x(x-x')}\v h(x')=\v h(x).
\label{eq7-2-17}
\end{aligned}
\end{equation}
A similar substitution of (\ref{eq7-2-16a}) into (\ref{eq7-2-16b})
yields similar results. Therefore, (\ref{eq7-2-16a}) and
(\ref{eq7-2-16b}) constitute a vector Fourier transform pair.

From the above, note that in Equation (\ref{eq7-2-10}), if $\v
J_s(x,z) = e^{ik_zz}\v k_s(x)$, then $\v K_s(k_x)$ is the vector
Fourier transform of $\v k_s(x)$. In order to solve
(\ref{eq7-2-14}), we let
\begin{subequations}
\begin{equation}
\v k_s(x)=\begin{bmatrix}
k_{sz}(x)\\
k_{sx}(x)
\end{bmatrix} =\sum _{n=1}^N\begin{bmatrix}
a_nk_{nz}(x)\\
b_nk_{nx}(x)
\end{bmatrix} =\sum _{n=1}^N\dyad k_n(x)\cdot\v a_n,
\label{eq7-2-18}
\end{equation}
where
\begin{equation}
\v a_n=\begin{bmatrix}
a_n\\
b_n
\end{bmatrix} \qquad \dyad k_n(x)=\begin{bmatrix}
k_{nz}(x) & 0\\
0 & k_{nx}(x)
\end{bmatrix} .
\label{eq7-2-18b}
\end{equation}
\end{subequations}
$\dyad k_n(x)$ is chosen so that its vector Fourier transform
exists. Then
\begin{equation}
\v K_s(k_x)=\sum _{n=1}^N\dyad K_n(k_x)\cdot\v a_n,
\label{eq7-2-19}
\end{equation}
where $\dyad K_n(k_x)$ is the VFT of $\dyad k_n(x)$. Substituting
(\ref{eq7-2-19}) into (\ref{eq7-2-14}), we have
\begin{equation}
\v E_s(x, y=0, z)=\frac {e^{ik_zz}}{2\pi } \sum
_{n=1}^N\int^\infty_{-\infty} dk_x\dyad F(k_x, x)\cdot\dyad
G_0(k_x)\cdot\dyad K_n(k_x)\cdot\v a_n=0. \label{eq7-2-20}
\end{equation}
To remove the $x$-dependence in the above equation, we multiply it
by $\dyad k'_m(x)$ and integrate over $x$, where
\begin{equation}
\dyad k'_m(x)=\begin{bmatrix}
k_{mz}(x) & 0\\
0 & -k_{mx}(x)
\end{bmatrix}.
\label{eq7-2-21}
\end{equation}
Then (\ref{eq7-2-20}) becomes
\begin{equation}
\sum _{n=1}^N\int^\infty_{-\infty} dk_x\dyad K' _m(k_x)\cdot\dyad
G_0(k_x)\cdot\dyad K_n(k_x)\cdot \v a_n=0, \label{eq7-2-22}
\end{equation}
where
\begin{equation}
\dyad K'_m(k_x)=\int^\infty_{-\infty} dx\dyad k'_m(x)\cdot\dyad
F(k_x, x). \label{eq7-2-23}
\end{equation}
Equation (\ref{eq7-2-22}) is of the form
\begin{equation}
\sum _{n=1}^N\dyad M_{mn}\cdot \v a_n=0, \qquad m=1, \hdots , N.
\label{eq7-2-24}
\end{equation}
The above is expressible in terms of a matrix equation
\begin{equation}
\dyad M\cdot\v a=0, \label{eq7-2-25}
\end{equation}
where
\begin{equation}
\dyad M=\begin{bmatrix}
\dyad M_{11} & \dyad M_{12} & \hdots\\
\dyad M_{21} & \dyad M_{22} & \qquad\\
\vdots & \qquad & \ddots
\end{bmatrix} , \qquad \v a=\begin{bmatrix}
\v a_1\\
\v a_2\\
\v a_3\\
\vdots
\end{bmatrix} .
\label{eq7-2-26}
\end{equation}
One can show by algebraic manipulation that $\dyad M_{ij}=\dyad
M_{ji}^t$ \cite{GURELMS}.  The physical reason is that  the inner
product calculation of the above corresponds to reaction inner
product in electromagnetics. In order for a guided mode with
$\exp(ik_zz)$ to have nontrivial reaction with another field, the
other field should have $\exp(-ik_zz)$, corresponding to a counter
propagating mode. Hence, one of the current components in
\eqref{eq7-2-21} has to change sign.

\begin{figure}[ht]
\begin{center}
\hfil\includegraphics[width=3.5truein]{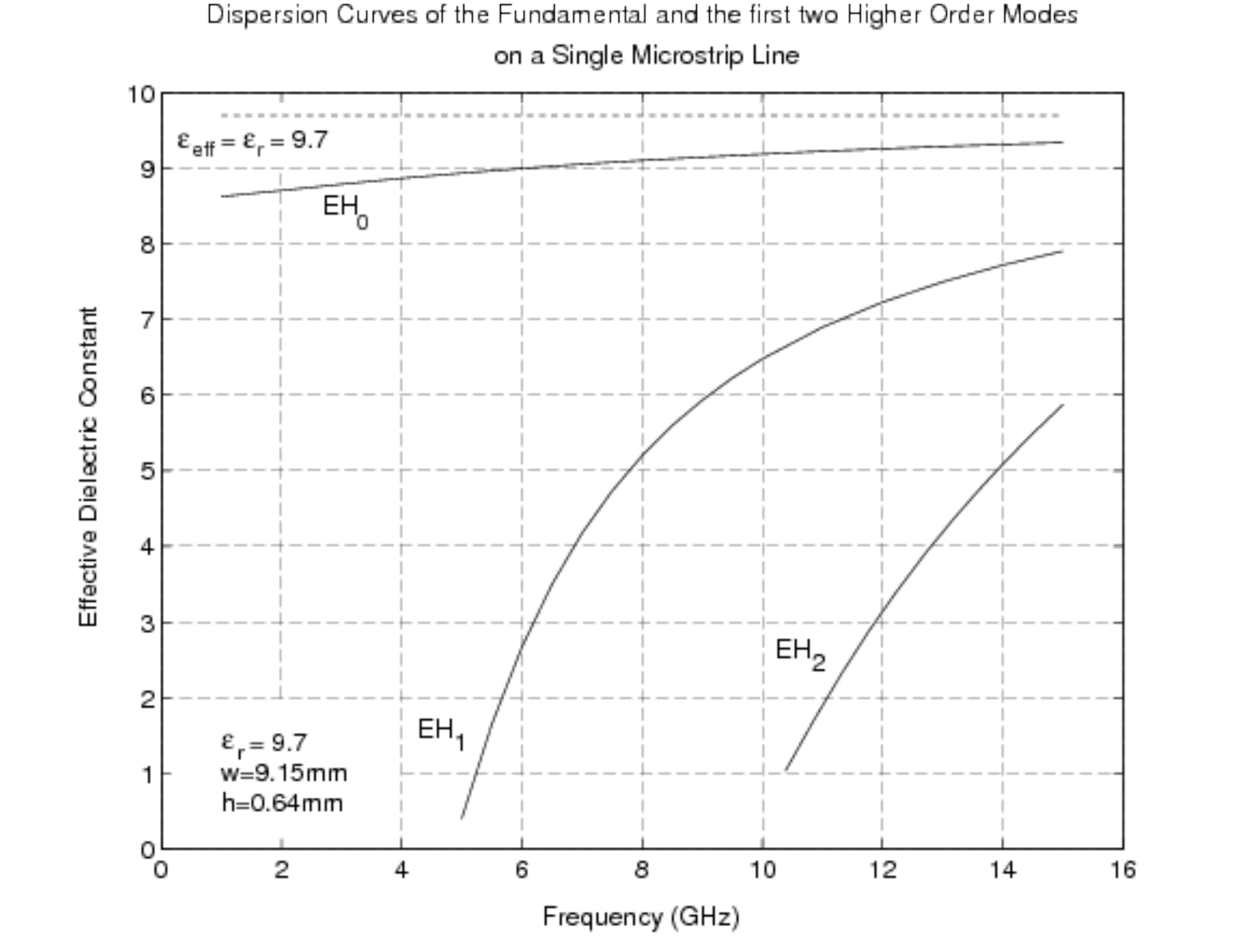}\hfil
\end{center}
\caption{Dispersion characteristics of the fundamental and the first
two higher order modes on a wide single microstrip line ($\epsilon
_1 = 9.7$, $w = 9.15$mm), dashed curve:  cover height $h = 3d =
1.92$mm (from Jansen \cite{JANSEN1}, reproduced by G. Papadopoulos).
}\label{fg723}
\end{figure}


In order for $\v a$ to be non-trivial, i.e., for a mode to exist,
it is necessary that
\begin{equation}
\det \left[\dyad M(k_z)\right]=0. \label{eq7-2-27}
\end{equation}
Since $\det\left[\dyad M(k_z)\right]$ is a function of $k_z$, the
roots of (\ref{eq7-2-27}) can be solved for numerically. The roots
are values of $k_z$ that satisfy the guidance condition of the
strip. These include the higher order modes plus the fundamental
modes.

With the weighting function as defined by (\ref{eq7-2-21}), $\dyad
M$ in (\ref{eq7-2-25}) is a symmetric matrix. The reflection
coefficients in (\ref{eq7-2-15}) are generalized reflection
coefficients for a layered medium. For a substrate backed by a
ground plane, we have
\begin{equation}
\tilde R^{TM}=\frac
{R_{01}^{TM}+e^{2ik_{1y}h}}{1+R_{01}^{TM}e^{2ik_{1y}h}}, \qquad
\tilde R^{TE}=\frac
{R_{01}^{TE}-e^{2ik_{1y}h}}{1-R_{01}^{TE}e^{2ik_{1y}h}}.
\label{eq7-2-28}
\end{equation}
In the above, $h$ is the thickness of the substrate, and
\begin{equation}
R_{ij}=\frac {p_jk_{iy}-p_ik_{jy}}{p_jk_{iy}+p_ik_{jy}},
\label{eq7-2-29}
\end{equation}
where $p_i = \mu _i$ for TE waves, $p_i = \epsilon _i$ for TM waves,
$k_{iy} =\sqrt {k_i^2-k_x^2-k_z^2}$.

In general, for guidance to be possible, we require that
\begin{equation}
k_0 < k_z < k_1, \label{eq7-2-30}
\end{equation}
where $k_1$ is the wave number of the substrate region and $k_0$
is the free-space wave number. Hence, the solution of
(\ref{eq7-2-27}) is sought only for $k_z$ within the window
defined by (\ref{eq7-2-30}). However, this window can be made even
smaller with the following consideration.

A substrate backed by a ground plane has a fundamental TM$_0$ mode
with no cutoff.  In order for a mode to be guided by a microstrip
line, the mode on the microstrip line must have a slower phase
velocity than the TM$_0$ mode \cite{GUREL_CHEW}. If this TM$_0$ mode
has a wave number $k_{sp} =\sqrt {k_{xp}^2+k_{zp}^2}$, then
\begin{equation}
k_0<k_{sp}<k_z<k_1. \label{eq7-2-31}
\end{equation}
$k_{sp}>k_0$, because TM$_0$ mode is a guided mode, and its phase
velocity must be slower than that of air. Equation (\ref{eq7-2-31})
leads to a narrower search window for $k_z$. Equation
(\ref{eq7-2-31}) ensures that the microstrip mode does not leak
energy to the fundamental TM$_0$ mode.  Moreover, there are guided
modes by a ground-plane backed dielectric substrates.  These modes
can potentially take energy away from the modes guided by the
microstrip line.  The above condition also prevents the leakage of
energy from the guided microstrip line modes to the dielectric
substrate modes.

Since a microstrip line falls into the class of \index{Waveguide!inhomogeneously filled} inhomogeneously
filled waveguides, the variational method developed in Chapters 3
and 6 together with the finite element method can be used to solve
for the wavenumber $k_z$'s of the guided mode.  In this case, the
partial differential equation is converted into a matrix eigenvalue
problem where the eigenvalues are $k_z^2$, and they can be found
explicitly, rather than through a root searching method above.


\newcommand\rv{\mathbf{r}}
\newcommand\kv{\mathbf{k}}

\section{Microstrip Patch Revisited}
\index{Microstrip patch}
\index{Layered medium}


Armed with the spectral domain technique, we are well equipped to
analyze the microstrip patch over a layered medium \cite{CHEW_LIU}.
To this end, we let (here, we assume that the $z$ axis to be
vertical as is usually the case)
\begin{align}\label{MPRev01}
E_{0z}(\rv)&=\int^{\infty}_{-\infty}\int^{\infty}_{-\infty} dk_x
dk_y e^{ik_xx+ik_yy}\tilde{E}_{0z}(k_x,k_y,z)\\
H_{0z}(\rv)&=\int^{\infty}_{-\infty}\int^{\infty}_{-\infty} dk_x
dk_y e^{ik_xx+ik_yy}\tilde{H}_{0z}(k_x,k_y,z)
\end {align}
Substituting the above into the wave equation implies that
\begin{align}
\left(\frac{d^2}{dz^2}+k_z^2\right)\tilde{E}_{0z}^2(k_x,k_y,z)&=0\\
\left(\frac{d^2}{dz^2}+k_z^2\right)\tilde{H}_{0z}^2(k_x,k_y,z)&=0
\end {align}
where $k_z^2=k_0^2-k_x^2-k_y^2$. Hence, one can write
\begin{align}\label{MPRev05}
E_{0z}(\rv)&=\pm \int^{\infty}_{-\infty}\int^{\infty}_{-\infty} dk_x
dk_y e^{i\kv_s\cdot\rv_s+ik_z|z|}e_{0z}(\kv_s)\\
H_{0z}(\rv)&=\int^{\infty}_{-\infty}\int^{\infty}_{-\infty} dk_x
dk_y e^{i\kv_s\cdot\rv_s+ik_z|z|}h_{0z}(\kv_s)
\end {align}
The $\pm$ sign indicates the odd symmetry of $E_{0z}(\rv)$ about
$z=0$ plane. When the patch is placed on top of a layered medium,
the above becomes
\begin{align}\label{MPRev07}
E_{0z}(\rv)&=\int^{\infty}_{-\infty}\int^{\infty}_{-\infty} d\kv_s
e^{i\kv_s\cdot\rv_s}e_{0z}(\kv_s)\big[ \pm
e^{ik_z|z|}-\tilde{R}^{TM}e^{ik_z(z+d)+ik_z|d|} \big]\\
\label{MPRev08}
H_{0z}(\rv)&=\int^{\infty}_{-\infty}\int^{\infty}_{-\infty} d\kv_s
e^{i\kv_s\cdot\rv_s}h_{0z}(\kv_s)\big[
e^{ik_z|z|}+\tilde{R}^{TE}e^{ik_z(z+d)+ik_z|d|} \big]
\end {align}
The phase factors for the reflected waves are appropriately chosen
so that $\tilde{R}^{TM}$ and $\tilde{R}^{TE}$ represent the
reflection coefficients.

\def\Ev{\v E}
\def\Hv{\v H}

Using the fact that
\begin{align}\label{MPRev09}
\hat{\Ev}_s(\rv)=\frac{1}{k_s^2}\left[\frac{\partial}{\partial
z}\hat{E}_z-i\omega\mu_0\hat{z}\times \nabla_s \hat{H}_z \right]
\end {align}
\begin{align}\label{MPRev10}
\hat{\Hv}_s(\rv)=\frac{1}{k_s^2}\left[\frac{\partial}{\partial
z}\hat{H}_z+i\omega\epsilon_0\hat{z}\times \nabla_s \hat{E}_z
\right]
\end {align}
and applying the above to the integrands of spectral integrals, we
obtain
\begin{align}\label{MPRev11}
\Ev_{0s}(\rv)=\int^{\infty}_{-\infty}\int^{\infty}_{-\infty} d\kv_s
e^{i\kv_s\cdot\rv_s}\frac{1}{k_s^2}\big[-k_z\kv_s
e_{0z}(\kv_s)+\omega\mu_0\hat{z}\times \kv_s
h_{oz}(\kv_s)\big]e^{k_z|z|}
\end {align}
\begin{align}\label{MPRev12}
\Hv_{0s}(\rv)=\int^{\infty}_{-\infty}\int^{\infty}_{-\infty} d\kv_s
e^{i\kv_s\cdot\rv_s}\frac{1}{k_s^2}\big[\mp k_z\kv_s
h_{0z}(\kv_s)\mp \omega\epsilon_0\hat{z}\times \kv_s
e_{oz}(\kv_s)\big]e^{k_z|z|}
\end {align}
Using the fact that
\begin{align}\label{MPRev13}
\mathbf{J}_{0s} = \hat{z}\times (\mathbf{H}_{0s+} -
\mathbf{H}_{0s-}) |_{z=0}
\end{align}
the above can be written as
\begin{align}\label{MPRev14}
\mathbf{J}_{0s}(\mathbf{r}_s) =\iint_{-\infty}^{\infty}d
\mathbf{k}_s \ e^{i\mathbf{k}_s\cdot\mathbf{r}_s}\frac{1}{k_s^2}
\left[2\omega\epsilon_0\mathbf{k}_s e_{0z}-2k_z \hat{z}\times
\mathbf{k}_s h_{0z}\right]
\end{align}
The above can be written using vector Fourier transform
\cite{CHEW_HABASHY7}
\begin{align}\label{MPRev15}
\mathbf{J}_{0s}(\mathbf{r}_s) =\iint_{-\infty}^{\infty}d
\mathbf{k}_s \ \overline{\mathbf{F}}(\mathbf{k}_s,\mathbf{r}_s)
\cdot {\mathbf{K}}_0(\mathbf{k}_s)
\end{align}
where
\begin{align}
\overline{\mathbf{F}}(\mathbf{k}_s,\mathbf{r}_s) &=
\frac{1}{k_s}[\mathbf{k}_s,-\hat{z}\times\mathbf{k}_s]e^{i\mathbf{k}_s\cdot\mathbf{r}_s}
= \frac{1}{k_s}
\begin{bmatrix}
k_x & k_y \\
k_y & -k_x
\end{bmatrix}e^{i\mathbf{k}_s\cdot\mathbf{r}_s} \label{MPRev16}\\
{\mathbf{K}}_0(\mathbf{k}_s) &= \frac{1}{k_s}
\begin{bmatrix}
2\omega\epsilon_0 e_{0z} \\
2k_z h_{0z}
\end{bmatrix} \label{MPRev17}
\end{align}
An \index{Vector Fourier transform!inverse} inverse vector Fourier transform exists as
\begin{align}\label{MPRev15a}
\mathbf{K}_{0}(\mathbf{k}_s)
=\frac{1}{(2\pi)^2}\iint_{-\infty}^{\infty}d \mathbf{r}_s \
\overline{\mathbf{F}}(\mathbf{k}_s,-\mathbf{r}_s) \cdot
{\mathbf{J}}_{0s}(\mathbf{r}_s)
\end{align}
We can also write
\begin{align}\label{MPRev18}
\mathbf{E}_{0s} = \iint_{-\infty}^{\infty} d\mathbf{k}_s \
\overline{\mathbf{F}}(\mathbf{k}_s,\mathbf{r}_s)\cdot\overline{\mathbf{G}}(k_z,z)\cdot\mathbf{\overline{K}_0}(\v
k_s)
\end{align}
where
\begin{align}\label{MPRev19}
\overline{\mathbf{G}}(k_z) = -\begin{bmatrix}
\frac{k_z}{2\omega\epsilon_0} & 0\\
0 & \frac{\omega\mu_0}{2k_z}
\end{bmatrix}
e^{i k_z|z|}
\end{align}
The above constitutes a compact way to express the field and current
in terms of their spectral domain quantities.

\subsection{Integral Equation for the Resonance Case}
The resonance of a \index{Microstrip patch!resonance} microstrip patch has been calculated before
\cite{ITOH_RESONANCE,CHEW_KONG7,CHEW_KONG71,CHEW_LIU}
\begin{align}
\mathbf{E}_{0s}(\mathbf{r}_s,z=0) &= 0 = \iint_{-\infty}^{\infty}
d\mathbf{k}_s \
\overline{\mathbf{F}}(\mathbf{k}_s,\mathbf{r}_s)\cdot
\overline{\mathbf{G}}(k_z,z=0)\cdot\mathbf{\overline{K}_0}(\v k_s),
\quad \mathbf{r}_s  \in S_P\label{MPRev20}\\
\mathbf{J}_0(\mathbf{r}_s) &= 0, \qquad \mathbf{r}_s \notin S_P
\label{MPRev21}
\end{align}
where $S_P$ is the patch surface.  By using Galerkin's method, the
above can be converted into a matrix equation
\begin{align}\label{MPRev22}
\overline{\mathbf{A}}(\omega)\cdot \mathbf{a} = 0
\end{align}
A non-trivial $\mathbf{a}$ exists only if
\begin{align}\label{MPRev23}
\det\left(\overline{\mathbf{A}}(\omega)\right) = 0
\end{align}
The above equation can be solved with a zero-searching method to
obtain the resonance frequencies of the microstrip patch.

\subsection{Integral Equation for the Excitation Case}
\index{Microstrip patch!integral equation}

If the patch is excited by an incident field, say due to a probe
source, then the integral equation for excitation is
\cite{CHEW_KONG72,POZAR}:
\begin{align}\label{MPRev24}
\hat{n} \times \left[ \v{E}_{0s} (\v{r}_s, z = 0) + \v{E}_{inc}
(\v{r}_s, z = 0) \right] = 0
\end{align}
or after using \eqref{MPRev20}
\begin{align}\label{MPRev25}
-\hat{n} \times \v{E}_{inc} (\v{r}_s, z = 0)  = \hat{n} \times
\iint_{-\infty} ^{ \infty} d\v{k}_s \overline {\v{F}} (\v{k_s},
\v{r}_s) \cdot \overline{\v{G}}(\v{k_z},z=0) \cdot \v{K}_0 (\v{k}_s)
\end{align}
Using Galerkin's method, the above can be converted to a matrix
equation:
\begin{align}\label{MPRev26}
\overline{\v{A}} \cdot \v{a} = \v{b}
\end{align}
one can solve the above to obtain the current on the patch, and then
calculate the field everywhere.

\begin{figure}[ht]\label{FigFarFieldCalulation}
 \begin{center}
  \includegraphics[width=1\textwidth]{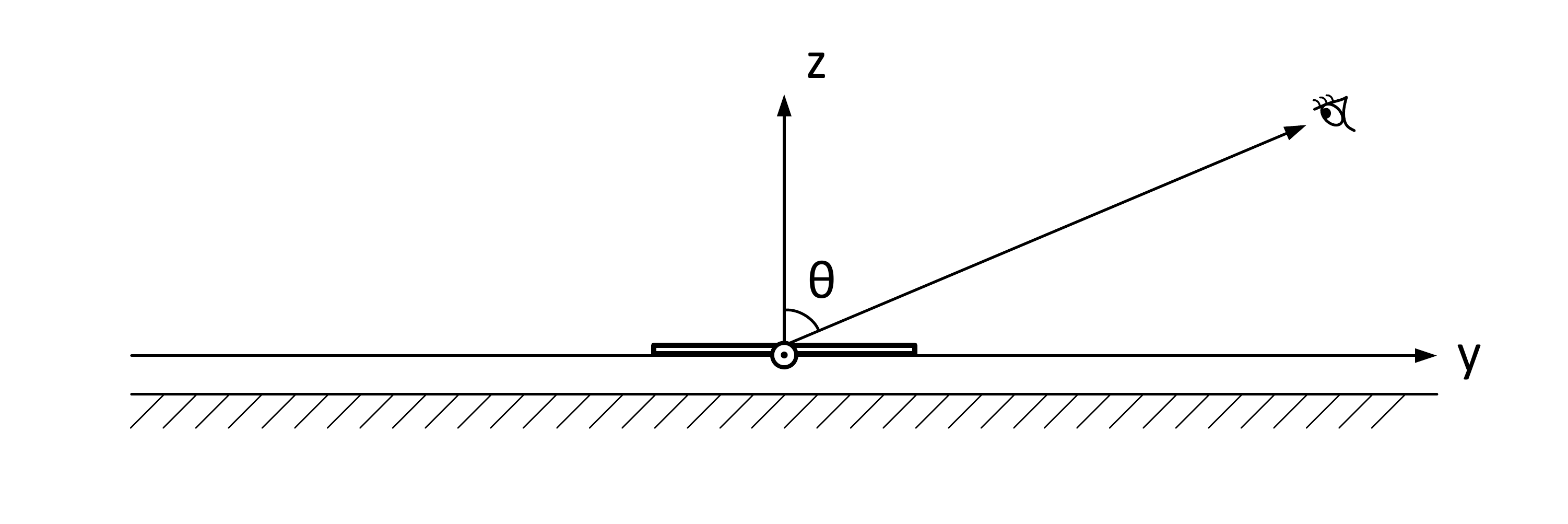}
  \end{center}
\end{figure}

\subsection{Far Field Calculation}
\index{Microstrip line!far field}

The far field of the Fourier integral given in \eqref{MPRev07} and
\eqref{MPRev08} can be found using the stationary phase method
\cite{WFIM7}. We will assume that $h$ is small so that when $(x,y,z)
\rightarrow \infty$, the dominant variation of the integrand comes
from $e^{i\v{k}_s \cdot \v{r}_s}$ and $ e^{ik_z z}$. To this end, we
can write \eqref{MPRev07} with $d = 0$ as
\begin{align}\label{MPRev27}
E_{0z} (\v{r}) = \iint _ {-\infty} ^ \infty d\v{k}_s e^{i\v{k}_s
\cdot \v{r}_s + {ik_z z}} e_{oz}(k_s) \left( 1- \tilde{R}^{TM}
\right)
\end{align}
The exponential function is given by
\begin{align}\label{MPRev28}
e^{ik_x x + ik_y y + ik_z z}
\end{align}
The stationary phase point is given by
\begin{align}\label{MPRev29}
\frac{\partial}{\partial k_x} \left[ k_x x + k_y y + k_z z\right] =
0, \hspace{5 mm} \frac{\partial}{\partial k_y} \left[ k_x x + k_y y
+ k_z z\right] = 0
\end{align}
yielding
\begin{align}\label{MPRev30}
x - \frac{k_x}{\sqrt{k^2 - k_x^2 - k_y^2}} z = 0, \hspace {5mm} y -
\frac{k_y}{\sqrt{k^2 - k_x^2 - k_y^2}} z = 0
\end{align}
If we set
    \begin{equation}
    \label{MPRev31}
    k_x=k_0\sin\theta\cos\phi \quad k_y=k_0\sin\theta\sin\phi
    \end{equation}
    then
    \begin{equation}
    \label{MPRev32}
    k_z=k_0\cos\theta
    \end{equation}
    and (\ref{MPRev30}) is satisfied. Therefore the above is the stationary phase point of the
    integrand.
    Most of the contribution of the integral to \eqref{MPRev27} will come from this point.
    Hence, we can approximate the slowly varying part of the integrand by its value at this point and rewrite \eqref{MPRev27} approximately as
    \begin{equation}
    \label{MPRev33}
    E_{0z}\left(\mathbf{r}\right)\sim k_{zs}e_{0z}\left(k_{xs},k_{ys}\right)
    \left[1-\tilde{R}^{TM}\left(k_{xs},k_{ys}\right)\right]
    \int\int_{-\infty}^{+\infty}d\mathbf{k}_s\frac{e^{i\mathbf{k}_s\cdot\mathbf{r}_s+ik_zz}}{k_z}
    \end{equation}
    where $k_{xs}$, $k_{ys}$, and  $k_{zs}$ are given by
    \eqref{MPRev31} and \eqref{MPRev32}.
    Making use of the Weyl identity,
    \begin{equation}
    \label{MPRev34}
    \frac{e^{ik_0r}}{r}=\frac{i}{2\pi}
    \int\int_{-\infty}^{+\infty}d\mathbf{k}_s\frac{e^{i\mathbf{k}_s\cdot\mathbf{r}_s+ik_z|z|}}{k_z}
    \end{equation}
    \eqref{MPRev33} becomes
    \begin{equation}
    \label{MPRev35}
    E_{0z}\left(\mathbf{r}\right)\sim \frac{2\pi}{i}k_{zs}e_{0z}\left(k_{xs},k_{ys}\right)
    \left[1-\tilde{R}^{TM}\left(k_{xs},k_{ys}\right)\right]\frac{e^{ik_0r}}{r}
    \end{equation}
    By the same token,
    \begin{equation}
    \label{MPRev36}
    H_{0z}\left(\mathbf{r}\right)\sim \frac{2\pi}{i}k_{zs}h_{0z}\left(k_{xs},k_{ys}\right)
    \left[1+\tilde{R}^{TE}\left(k_{xs},k_{ys}\right)\right]\frac{e^{ik_0r}}{r}
    \end{equation}
    The far field appears as a spherical wave. Hence,
    the electric field of the far field is of the form
    \begin{equation}
    \label{MPRev37}
    \mathbf{E}\approx\hat{\theta}E_\theta+\hat{\phi}E_\phi
    \end{equation}
    with
    \begin{equation}
    \label{MPRev38}
    \mathbf{H}\approx-\hat{r}\times \mathbf{E}/\eta
    =\frac{1}{\eta}\left(\hat{\phi}E_\theta-\hat{\theta}E_{\phi}\right)
    \end{equation}
    From the above, one gathers that
    \begin{equation}
    \label{MPRev39}
    E_{0z}\approx-E_{\theta}\sin\theta,\quad H_{0z}\approx\frac{1}{\eta}E_{\phi}\sin\theta
    \end{equation}
    Hence, we can get $(E_\theta, E_\phi)$ from $(E_{0z}, H_{0z})$ in the far field and compute the total radiation power.


    \section{Edge Condition}
\index{Edge condition}


In many numerical and analytic methods, it is useful to know how the
charge and the current behave near the sharp edge of a geometry.
This will help in the choice of the correct basis functions to
improve the convergence of the numerical methods. Also, the edge
condition has to be imposed for a boundary value problem in order to
guarantee uniqueness of the solution.  This problem has been studied
by Rayleigh, Meixner, Maue, Jones, and Silver and Heins, Hayashi,
and many more.  References can be found in Heins and Silver
\cite{HEINS_SILVER}, and Hayashi \cite{HAYASHI}.

The field near a sharp edge is singular because of charge accumulation into a singular point. The
singular behavior can be ascertained by solving Laplace's equation in the vicinity of the sharp edge.
This is because the singular behavior of the field is entirely a local
phenomenon or local geometry dependent where the spatial
variation dominates over temporal variation.

            \begin{figure}[!htb]
        \begin{center}
        \includegraphics[height=4.5cm]{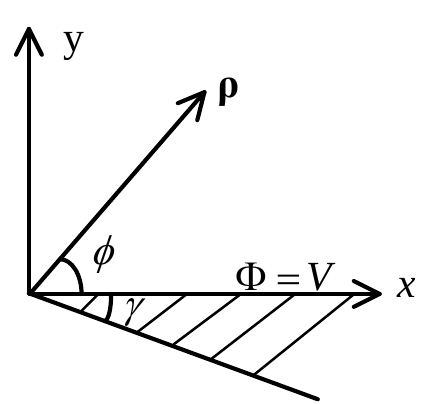}
        \end{center}
        \end{figure}
    Consider a wedge with an angle $\gamma$ as shown.
    The $\Phi$ that satisfies Laplace's equation $\nabla^2\Phi=0$ is given by
        \begin{equation}
        \label{ECeq:1}
        \Phi=A\rho^{\pm\alpha}e^{{\pm}in\alpha},\qquad \alpha>0
        \end{equation}
    Using the fact that
        \begin{equation}
        \label{ECeq:2}
        \nabla^2=\frac{1}{\rho}\frac{\partial}{\partial\rho}\rho\frac{\partial}{\partial\rho}+
        \frac{1}{\rho^2}\frac{\partial^2}{\partial\phi^2}
        \end{equation}
    in cylindrical coordinates, it can be easily shown that $\nabla^2\Phi=0$ by direct substitution.
    A solution that satisfies the boundary condition that $\Phi=V$ on the wedge surface is
        \begin{equation}
        \label{ECeq:3}
        \Phi=V+a\rho^\alpha\sin{(\alpha\phi)}
        \end{equation}
    with that requirement that $\sin{[\alpha(2\pi-\gamma)]}=0$, or that
        \begin{equation}
        \label{ECeq:4}
        \alpha(2\pi-\gamma)=\pi,\qquad \alpha=\frac{\pi}{2\pi-\gamma}
        \end{equation}
    for the minimal $\alpha$ and that $\sin{(\alpha\phi)}>0$ for $0<\phi<2\pi-\gamma$.

    The $\rho^{-\alpha}$ term in \eqref{ECeq:1} can be ignored since $|\nabla\Phi|^2$ has to be square
    integrable because it represents the finite stored energy in the field. Moreover, one picks minimal $\alpha$
    so that the electric field ${\bf{E}}=-\nabla\Phi$ given by \eqref{ECeq:3} is simple and does not have scallop
    patterns.

    \begin{table}
\begin{center}
         \begin{tabular}{|c|c|}
    \hline
$\gamma$ &  $\alpha$     \\
    \hline
0   & 1/2 \\
$\pi/2$   & 2/3\\
$\pi$   & 1 \\
$3\pi/2$   & 2\\
    \hline
\end{tabular}
\end{center}
\caption{The singularities associated with different wedge
angles.}\label{ECTable}
\end{table}

    Hence for the different wedge angles,  the singularities are
    as shown in Table \ref{ECTable}.
%
    For a thin wedge with ${\gamma}=0$,
        \begin{equation} \label{ECeq:5}
            {\Phi}=V+a{\rho}^{\frac{1}{2}}\sin\left({\frac{\phi}{2}}\right)
        \end{equation}
    the corresponding electrostatic field
        \begin{equation} \label{ECeq:6}
            {\bf E}=-\nabla\Phi=-\hat{\rho}\frac{a}{2}{\rho}^{-\frac{1}{2}}\sin\left(\frac{\phi}{2}\right)
            +\hat{\phi}\frac{a}{2}{\rho}^{-\frac{1}{2}}\cos\left(\frac{\phi}{2}\right)
        \end{equation}
    Since the charge is proportional to $\hat{n}\cdot{\bf E}$, the surface charge density $\sigma$ is such that
        \begin{equation} \label{ECeq:7}
            {\sigma} {\sim} {x}^{-\frac{1}{2}}
        \end{equation}
    near the edge.
    If a current flows in the $x$ direction, from $\nabla\cdot{
        \bf J}=i\omega\varrho$, the current
        \begin{equation} \label{ECeq:8}
            {J_x} {\sim} {x}^{\frac{1}{2}}
        \end{equation}
    near the edge. If the current is flowing in the $z$ direction into the paper, then
        \begin{equation} \label{ECeq:9}
            {J_z} = {\sigma}v_z
        \end{equation}
    Hence,
        \begin{equation} \label{ECeq:10}
            {J_z} {\sim} {x}^{-\frac{1}{2}}
        \end{equation}
This current singularity, for instance, is observed in the axial
component of the current on a microstrip line.

\section {{{ Discontinuities in Microstrip Lines}}}

Microstrip lines are some of the easiest waveguides to fabricate.
However, they are also some of the hardest to analyze.  Due to the
ease in their fabrication, microwave integrated circuits have come
in a rich variety of shapes. One example is the variety of
microstrip line discontinuities. These discontinuities, in general, are difficult to analyze. Usually, numerical methods have to be
employed for their analysis. This is especially true for the
full-wave analysis whereby all electrodynamic effects are accounted
for.  Much work on this topic can be found in
\cite{GARG_BAHL,CHU_ITOH,JACKSON_RW,ZHANG_MEI,KIRSCHNING_ETAL,RAILTON_ROZZI,VERMA_ETAL,DOUVILLE_JAMES,MOORE_LING,MEHRAN,HARM_MITTRA}

Microstrip line discontinuities are often modeled by lumped
elements in the transmission line model. In the following, we
shall discuss some typical discontinuities encountered in
microwave integrated circuits.

\begin{figure}[ht]
\begin{center}
\hfil\includegraphics[width=3.5truein]{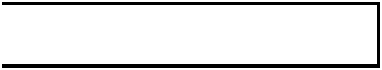}\hfil
\end{center}
\caption{An open-end discontinuity.}\label{fg731}
\end{figure}


\subsection { An Open-End Discontinuity}
\index{Microstrip line!open-end discontinuity}

A simple way to make an open circuit in a microstrip line is to
let the microstrip line terminate in an open end. Open circuits
are often used as a stub tuner in microwave integrated circuits.
The abrupt termination of a microstrip line intercepts the flow of
current on the line. This gives rise to charge accumulation, which
can be modeled by a capacitance. The equivalent circuit for an
open end microstrip line is then a transmission line terminated by
a capacitor, or an open circuited transmission line whose length
is slightly longer than the physical length of the microstrip
line. The capacitor is also called a \index{Fringing field capacitor} fringing field capacitor, as
the excess charge at the termination gives rise to excess fringing
field at the end of the line. These fringing capacitances can be
found by numerical calculations which are sometimes replaced by
analytical or semi-empirical formulas.

\begin{figure}[ht]
\begin{center}
\hfil\includegraphics[width=4.3truein]{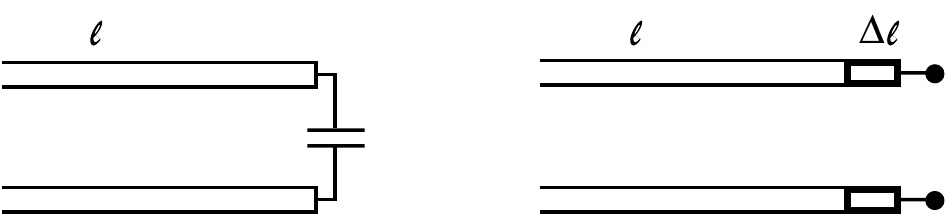}\hfil
\end{center}
\caption{Equivalent circuit models for an open end microstrip
line.}\label{fg732}
\end{figure}


\subsection { A Step Discontinuity}
\index{Microstrip line!step discontinuity}

\begin{figure}[ht]
\begin{center}
\hfil\includegraphics[width=3.0truein]{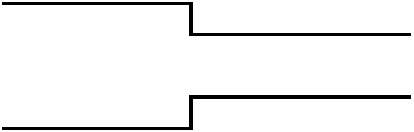}\hfil
\end{center}
\caption{A step discontinuity in a microstrip line.}\label{fg733}
\end{figure}


Another commonly encountered discontinuity in a microstrip line is
a step discontinuity. A step discontinuity has excess charges on
its excess edges, and hence can be modeled by a shunt
capacitance. This kind of discontinuity is encountered in the
design of a quarter-wave transformer for instance. A more
sophisticated model consists of two series inductors as well as a
shunt capacitor.

\begin{figure}[ht]
\begin{center}
\hfil\includegraphics[width=4.5truein]{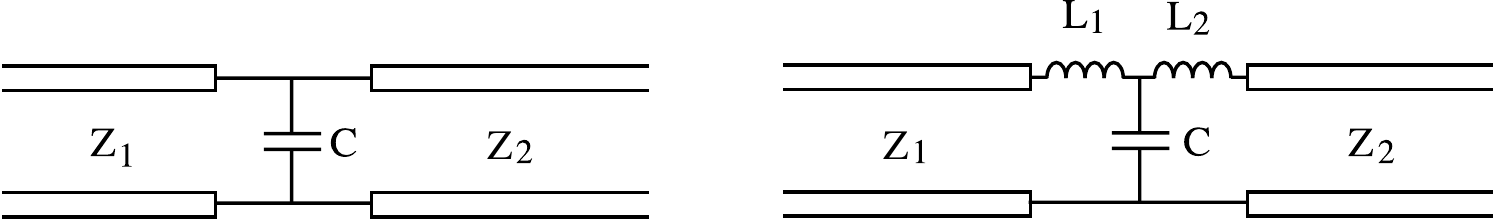}\hfil
\end{center}
\caption{Equivalent circuit models for a step discontinuity.}\label{fg734}
\end{figure}


\subsection{ A Gap Discontinuity}
\index{Microstrip line!gap discontinuity}

\begin{figure}[ht]
\begin{center}
\hfil\includegraphics[width=3.0truein]{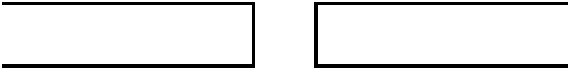}\hfil
\end{center}
\caption{A gap discontinuity.}\label{fg735}
\end{figure}


A gap discontinuity is deliberately introduced in a microstrip
line, for example, in filter design. A gap discontinuity can be
thought of as two open end discontinuities brought close together.
The proximity of two open end discontinuities results in
capacitive coupling between the two open ends. Therefore, the
equivalent circuit consists of two shunt capacitors and a series
capacitor.

\begin{figure}[ht]
\begin{center}
\hfil\includegraphics[width=3.0truein]{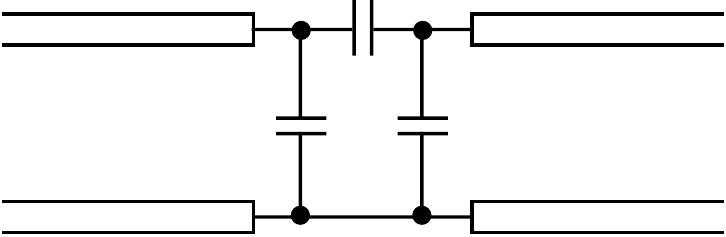}\hfil
\end{center}
\caption{The equivalent circuit for a gap discontinuity.}\label{fg736}
\end{figure}


\subsection { A Slit Discontinuity}
\index{Microstrip line!slit discontinuity}

\begin{figure}[ht]
\begin{center}
\hfil\includegraphics[width=3.0truein]{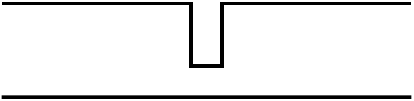}\hfil
\end{center}
\caption{A slit discontinuity.}\label{fg737}
\end{figure}


A slit discontinuity has very little effect on a low frequency
signal, but it will scatter a high frequency signal. At low
frequency, the current can flow continuously along the line, and
sees a matched line at the other end. At higher frequency, the
restricted channel is felt and the slit reflects a signal impinging
on it.  Hence, it acts as a low-pass filter; or it can be modeled by
an T equivalent circuit with two series inductances and a shunt
capacitance.

\begin{figure}[ht]
\begin{center}
\hfil\includegraphics[width=3.0truein]{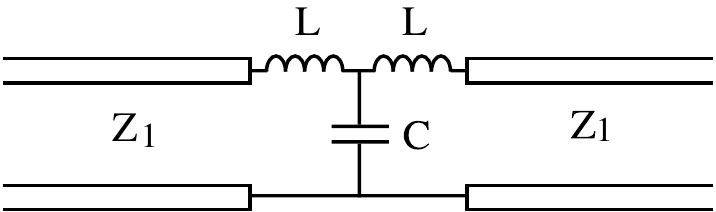}\hfil
\end{center}
\caption{The equivalent circuit for a slit discontinuity.}\label{fg738}
\end{figure}


\subsection { A Microstrip Bend}
\index{Microstrip line!bend}

A microstrip bend can be a corner bend or a chamfered bend. Both
microstrip bends can be modeled by a T circuit with two series
inductors and a shunt capacitor. A corner bend has a sharp corner
favoring the build up of excessive charges giving rise to a larger
shunt capacitance. A chamfered bend reduces this capacitance and
mismatch, even less so than a rounded bend.

All guided waves around a bend in an open waveguide radiate. A
chamfered bend has been found to reduce the radiation loss as
well.

\begin{figure}[ht]
\begin{center}
\hfil\includegraphics[width=3.0truein]{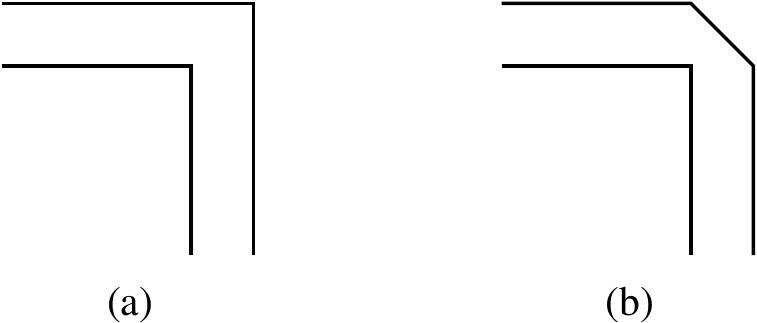}\hfil
\end{center}
\caption{(a) A corner bend. (b) A chamfered bend.}\label{fg739}
\end{figure}


\subsection  { A T Junction}

\begin{figure}[ht]
\begin{center}
\hfil\includegraphics[width=3.0truein]{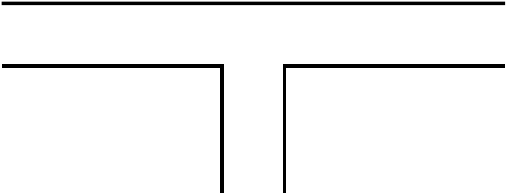}\hfil
\end{center}
\caption{A T junction in microwave integrated circuits.}\label{fg7310}
\end{figure}


A T junction in microwave integrated circuits is commonly
encountered, e.g., in a stub tuner. It can be modeled by series
inductances plus a shunt capacitance. A cross can also be
similarly modeled.

\begin{figure}[ht]
\begin{center}
\hfil\includegraphics[width=3.0truein]{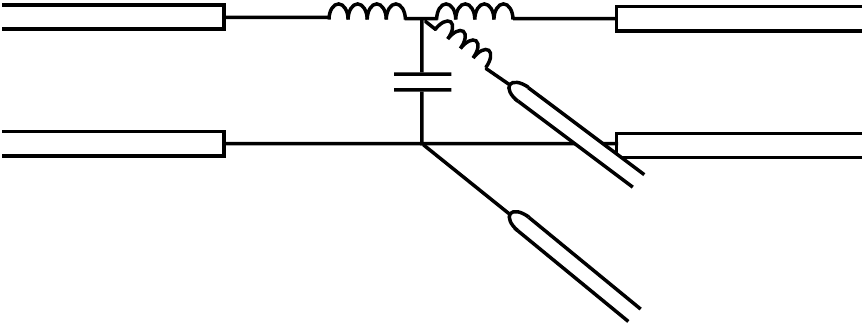}\hfil
\end{center}
\caption{Equivalent circuit for a T junction.}\label{fg7311}
\end{figure}


\section {{{Directional Coupler Using Microstrip Line}}}
\index{Directional couplers}
\index{Microstrip line!directional coupler}

Due to the simplicity of the microstrip waveguide, a new class of
directional coupler has been proposed.  Some of these papers can be
found in \cite{DE_RONDE,DYDYK,CALOZ_ITOH7}.   When two microstrip
lines are aligned parallel to each other, electromagnetic coupling
exists between the two lines. Since one line has one fundamental
mode, two lines would have two fundamental modes. For symmetric
lines, these two fundamental modes are the odd and the even modes.
For two identical lines, the electric field for the odd mode is
odd-symmetric about the plane of symmetry; while for the even mode,
it is even-symmetric about the same plane. Hence, we can use the
image theorem to analyze such a problem: A PMC wall can be placed at
the plane of symmetry for the even symmetric mode, while a PEC wall
can be placed at the same plane for the odd symmetric mode.

\begin{figure}[ht]
\begin{center}
\hfil\includegraphics[width=5.0truein]{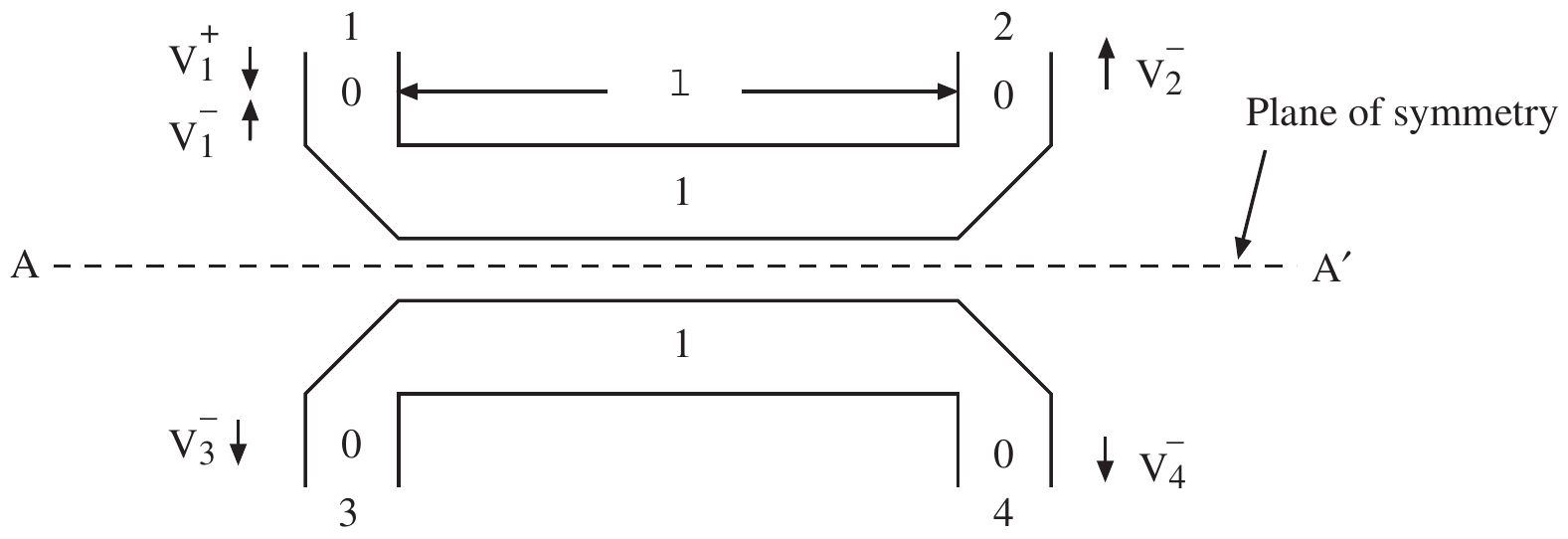}\hfil
\end{center}
\caption{A directional coupler made of microstrip lines.}\label{fg741}
\end{figure}


The directional coupler is a four-port network. We can analyze it
as a linear superposition of two two-port network, one with a PMC
at $AA'$, and another one with a PEC at $AA'$. The PMC case is
equivalent to an even-mode excitation of the geometry. This is the
same as, say, an incident wave of $V^+$ at both ports 1 and 3 of
the network. The PEC case corresponds to an odd-mode excitation,
which is the same as a $V^+$ incident wave at port 1 and a $-V^+$
incident wave at port 3. The linear superposition of these two
excitations correspond to a $2V^+$ excitation at port 1 and none
at the other ports.

The PMC case corresponds to only even mode propagation, and we can
use the bounce diagram approach to write down the reflected waves
at ports 1 and 2. We assume that the fundamental mode from the
input port will undergo a reflection when it enters the coupled
mode region.  Hence, the reflection coefficient at port 1 can be
expressed as the $S_{11}^e$ coefficient given as
\begin{equation}
S_{11}^e = R_{01}^e +{T_{01}^ee^{2ik_el}R_{10}^eT_{10}^e\over
1-(R_{10}^e)^2 e^{2ik_el}}. \label{eq7-4-1}
\end{equation}
The transmission coefficient from port 1 to port 2 can be
expressed as
\begin{equation}
S_{12}^e = {T_{01}^eT_{10}^ee^{ik_el}\over
1-(R_{10}^e)^2e^{2ik_el}} . \label{eq7-4-2}
\end{equation}
In the above
\begin{subequations}
\begin{equation}
R_{01}^e={Z_e-Z_c\over Z_e+Z_c}, \quad R_{10}^e=-R_{01}^e,
\label{eq7-4-3a}
\end{equation}
\begin{equation}
T_{01}^e=1+R_{01}^e, \quad T_{10}^e=1+R_{10}^e, \label{eq7-4-3b}
\end{equation}
\end{subequations}
where $Z_e$ is the characteristic impedance of the even mode while
$Z_c$ is the characteristic impedance of the feedline.

For the PEC case, a similar analysis shows that
\begin{equation}
S_{11}^o = R_{01}^o +{T_{01}^oe^{2ik_ol}R_{10}^oT_{10}^o\over
1-(R_{10}^o)^2 e^{2ik_ol}}, \label{eq7-4-4}
\end{equation}
\begin{equation}
S_{12}^o = {T_{01}^oT_{10}^oe^{ik_ol}\over
1-(R_{10}^o)^2e^{2ik_ol}}, \label{eq7-4-5}
\end{equation}
where
\begin{equation}
R_{01}^o={Z_o-Z_c\over Z_o+Z_c}, \label{eq7-4-6}
\end{equation}
and so on.

The key to the design of the directional coupler is to let $Z_c$
be the geometric mean of $Z_o$ and $Z_e$, i.e.,
\begin{equation}
Z_c=\sqrt{Z_oZ_e}. \label{eq7-4-7}
\end{equation}
In this case, $R_{01}^o=-R_{01}^e$. Furthermore, if we make
$k_e=k_o$, then it can be shown that
\begin{equation}
S_{11}^o=-S_{11}^e. \label{eq7-4-8}
\end{equation}
Superposing the even and the odd solutions, we obtain that
$V_1^+=2V^+$, and $V_1^-=(S_{11}^e+S_{11}^o)V^+=0$. Similarly, the
transmitted wave at port 4 is given by
\begin{equation}
V_4^-=S_{12}^eV^+-S_{12}^oV^+, \label{eq7-4-9}
\end{equation}
when $k_e=k_o$, and that (\ref{eq7-4-7}) is satisfied, it is clear
that $T_{01}^eT_{10}^e=T_{01}^oT_{10}^o$. Therefore,
$S_{12}^e=S_{12}^o$ and from (\ref{eq7-4-9}), we see that
$V_4^-=0$. At port 3, we have
\begin{equation}
V_3^-=S_{11}^eV^+-S_{11}^oV^+={1 \over
2}(S_{11}^e-S_{11}^o)V_1^+=S_{13}V_1^+, \label{eq7-4-10}
\end{equation}
where
\begin{equation}
S_{13}={1 \over 2}(S_{11}^e-S_{11}^o)=S_{11}^e. \label{eq7-4-11}
\end{equation}
Similarly, we have
\begin{equation}
S_{12}={1 \over 2}(S_{12}^e+S_{12}^o)=S_{12}^e. \label{eq7-4-12}
\end{equation}

Therefore, the above works as a perfect directional coupler, where
nothing is coupled to port 4 from port 1, while some signal is
transmitted to port 2 and port 3. If the coupled line is in a
homogeneous region, it is clear that both the odd and the even
modes are TEM modes, and hence $k_e=k_o=k$. However, if the
coupled line is fabricated on a substrate as in the case of a
microstrip line, then $k_e\neq k_o$ because the field
distributions of the two modes are different. However, $k_e$ can
be made to be close to $k_o$ by covering the microstrip line with
a superstrate with the same permittivity as the substrate.

The directivity of a directional coupler is the ratio of the $\mid
V_3^-\mid$ to $\mid V_4^-\mid$. In the ideal case, it is infinite.
The coupling coefficient is the ratio of $\mid V_3^-\mid$ to $\mid
V_1^+\mid$. In this case, it is
\begin{equation}
C=\mid S_{11}^e\mid = \left\vert
R_{01}^e+{T_{01}^ee^{2ik_el}R_{10}^eT_{10}^e\over
1-(R_{10}^e)^2e^{2ik_el}}\right\vert. \label{eq7-4-13}
\end{equation}
Using the fact that $T_{01}^eT_{10}^e=1-(R_{01}^e)^2$, we can
rewrite $S_{11}^e$ as
\begin{equation}
S_{11}^e = R_{01}^e +{R_{10}^ee^{2ik_el}\over
1-(R_{10}^e)^2e^{2ik_el}}= {R_{01}^e(1-e^{2ik_el})\over
1-(R_{10}^e)^2e^{2ik_el}}. \label{eq7-4-14}
\end{equation}
$\mid S_{11}^e\mid $ achieves a maximum at $k_el\simeq\pi /2$ or
$l\simeq\lambda /4$.


\section {{{ A Branch Line Directional Coupler}}}
\index{Branch line directional coupler}


\begin{figure}[ht]
\begin{center}
\hfil\includegraphics[width=3.5truein]{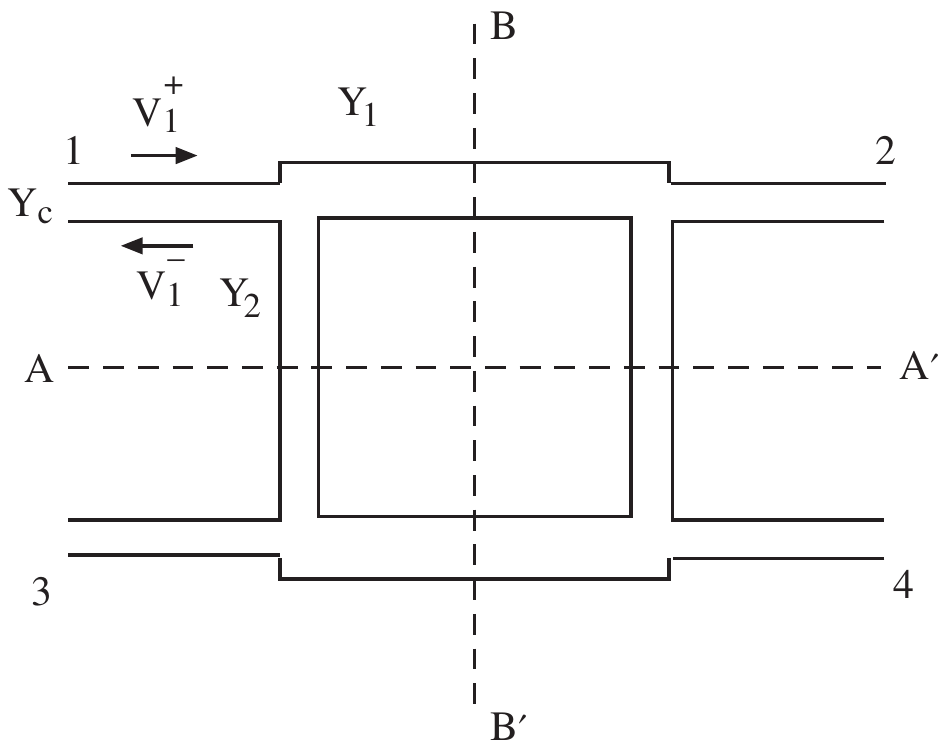}\hfil
\end{center}
\caption{A branch line directional coupler.}\label{fg751}
\end{figure}


A branch line coupler, like the microstrip line directional coupler,
has two planes of symmetry. Even though a microstrip line
directional coupler has two planes of symmetry, one can analyze it
using only one plane of symmetry. However, for a branch line
coupler, it is simpler to analyze using two planes of symmetry.

On the plane $AA'$, one can put either a PMC or a PEC. A PMC yields
an open circuit at $AA'$ or an even symmetric electric field or
voltage about the $AA'$. A PEC yields a closed circuit at $AA'$ or
an odd symmetric electric field or voltage about $AA'$. The same
statement applies to the plane $BB'$. Therefore, there are
altogether four possible excitations of the above geometry.

\bibliographystyle{plain}
\bibliography{emt,matrix,fmm,math,fastsum,parallel,cs,my}

\def\chaptitle{Solitons}
\index{Solitons}

\chapter{\chaptitle}
\markboth{\smallbooktitle}{\chaptitle}




\def\v #1{{\bf #1}}
\def\vg #1{{\boldsymbol #1}}
\def\dyad#1{\overline {\bf #1}}
\def\dyadg#1{\overline {\vg #1}}
\def\beq{\begin{equation}}\def\eeq{\end{equation}}
\def\tinf{\text{\it inf\,}}\def\^{\hat}
\def\cal#1{\mathcal{#1}}
\def\ed{\end{document}}
\def\dalpha{\dyadg\alpha}
\def\dbeta{\dyadg\beta}
\def\Rg{\Re g}
\def\l{\label}
\def\sinc{\text{sinc}}







\section{{{ Optical Solitons}}}
\index{Optical solitons}

Dispersions in an optical fiber is the prime reason for limiting its
bandwidth. Dispersion causes an optical pulse to distort or spread
as it propagates over long distances. Moreover, the loss of the
optical fiber also causes the pulse to spread.  Therefore, repeaters
are needed every several tens of kilometers to rejuvenate the
pulses. When the pulses are narrower in order to facilitate higher
transmission rates, dispersion effects become even more pronounced
and repeaters have to be more closely spaced. The spacing of the repeaters
is then inversely proportional to the square of the pulse width
\cite{AGRAWAL8}.

Solitary wave was first observed in water wave by Scott Russel in
1838 \cite{RUSSEL,SANDER&HUTTER}.  The mathematical
description was first given by Boussinesq \cite{BOUSSINESQ}.  Since
then, soliton theory has been avidly studied
\cite{DODDETAL,DRAZIN&JOHNSON,CHEREDNIK}.

An optical soliton makes use of the nonlinear \index{Optical fiber!nonlinear effect} effect in an optical
fiber to propagate a pulse with no distortion
\cite{HASEGAWA&KODAMA,MOLLENAUERETAL}. The nonlinear effect is used
to counter the dispersion effect so that the pulse propagates with
little or no distortion. The nonlinear effect is proportional to the
field strength of a mode in the optical fiber. Hence, loss in the
fiber will eventually distort the pulse. However, using the \index{Raman effect} Raman
effect in the fiber, optical solitons can be continuously pumped by
a lightwave which is simultaneously transmitted through the fiber
together with the solitons. Hence, an all optical system without
repeaters is possible \cite{MIZRAHIETAL}.

The nonlinear property of an optical fiber comes from the \index{Kerr effect} Kerr
effect which produces a change in the refractive index of glass due
to the deformation of electron orbits by the electric field of light
\cite{BOYD,NEWELL&MOLONEY}. The refractive index of glass is then
$n=n_0 + n _2 |\v E |^2$ where $n_2$ is the Kerr coefficient, and
$\v E$ is the electric field.  Here, $n_2$ is usually of the order
of $10^{-22} (m/V)^2$ and $\v E$ is of the order of $10^6 V/m$.
Therefore, the change in the refractive index is about $10^{-10}$.
Even though this is a small change, the high operating frequency of
an optical fiber makes this change significant.

\section {{{ The Korteweg de Vries Equation}}}
\index{Korteweg de Vries equation}

A soliton or a solitary wave is a result of nonlinear phenomena. A
nonlinear partial differential equation that admits a soliton as a
solution is the KdV (Korteweg de Vries) equation
\cite{MIURA,ZAKHAROV&FADDEEV}. One shall motivate the equation here.

A solution to a linear wave equation has the following form
\begin{equation}
\phi (x, t) = \phi (x-vt) \label{eq9-2-1}
\end{equation}
where $v$ is the velocity of the wave. The above describes a
right-traveling wave and it satisfies the equation
\begin{equation}
\frac{\partial \phi}{\partial t} + v \frac{\partial \phi}{\partial
x} = 0. \label{eq9-2-2}
\end{equation}
If one observes this wave in a moving coordinate system such that
$x'= x-v_0 t$, then equation (\ref{eq9-2-1}) becomes
\begin{equation}
\phi (x',t) = \phi [x' - (v-v_0)t]. \label{eq9-2-3}
\end{equation}
and
\begin{equation}
\frac{\partial \phi}{\partial t} + \delta v\frac{\partial
\phi}{\partial x'} = 0, \label{eq9-2-4}
\end{equation}
where $ \delta v = v - v_0$.  If $v_0= v$, then in the moving
coordinate system,
\begin{equation}
\frac{\partial }{\partial t}\phi = 0, \label{eq9-2-5}
\end{equation}
or that the field remains stationary, and $\phi = \phi (x')$ only.

\begin{figure}[ht]
\begin{center}
\hfil\includegraphics[width=4.5truein]{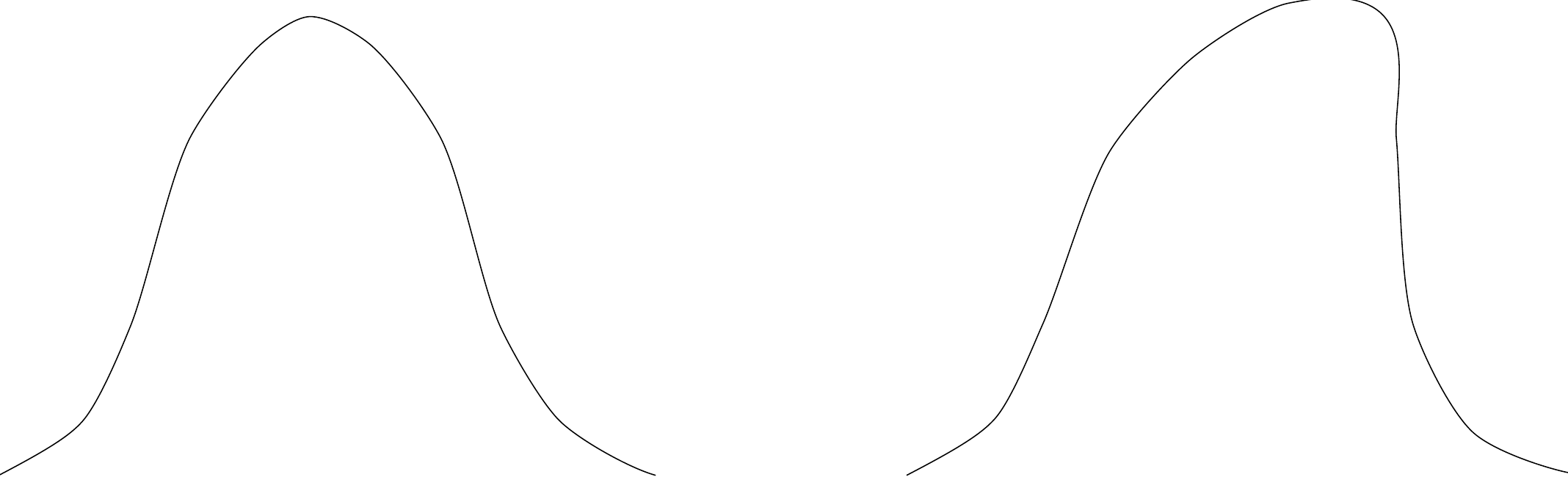}\hfil
\end{center}
\caption{The leaning over of a symmetric pulse due to the
nonlinearity in Equation (\ref{eq9-2-6}).}\label{fgsoliton}
\end{figure}

In Equation (\ref{eq9-2-4}), $\delta v$ denotes the ``extra''
velocity that the wave is moving with respect to the moving
coordinates. If one is in a moving coordinates such that this
``extra'' velocity is dependent on the amplitude of the wave, then
(\ref{eq9-2-4}) can be rewritten as
\begin{equation}
\frac{\partial \phi}{\partial t} + \delta _1 \phi \frac{\partial
\phi}{\partial x'}=0 . \label{eq9-2-6}
\end{equation}
which is a nonlinear equation. The above equation says that the
field with a higher amplitude moves faster than the field with a
lower amplitude. The above expression also implies that the field
cannot remain stationary in this moving coordinates. In other words,
$\phi$ is a function of both $x'$ and $t$. It further means that if
one has a symmetric pulse to begin with, the pulse will start to
lean over as shown in Figure \ref{fgsoliton}, generating a sharp
shock wavefront. The shock wavefront has higher spectral components.
Consequently, the high frequency component of the pulse has
increased due to nonlinearity.

\index{Dispersion}
\index{Velocity!phase}
A pulse shape can also be distorted by a dispersive effect. A
medium is dispersive if the wave speed is a function of the
wavenumber of the wave. In a dispersive medium, the wave velocity
has a higher order term that is proportional to $-\delta _2
k_x^2$. Therefore, a dispersive term can be added to
(\ref{eq9-2-6}) resulting in
\begin{equation}
\frac{\partial \phi}{\partial t} + \delta _1 \phi \frac{\partial
\phi}{\partial x'} + \delta _2 \frac{\partial ^3 \phi}{\partial
x^{\prime 3}} = 0, \label{eq9-2-7}
\end{equation}
since $\partial ^2/ \partial x'^2 = - k_x^2$. The dispersive effect
implies that the waves with high wave numbers travel at a smaller
velocity. Therefore, it tends to counter the pulse sharpening effect
due to nonlinearity. In fact a solution exists for (\ref{eq9-2-7})
for a pulse that propagates without distortion. If one lets $x' =
\delta ^{\frac{1}{3}}_2 \xi $, $\phi = \delta _2
^{\frac{1}{3}}\delta _1^{-1} \psi$, then \eqref{eq9-2-7} becomes
\begin{equation}
\frac{\partial \psi}{\partial t } + \psi \frac{\partial
\psi}{\partial \xi} + \frac{\partial ^3 \psi}{\partial \xi ^3} =
0, \label{eq9-2-8}
\end{equation}
which is the KdV (Korteweg de Vries) equation. It can be shown to
have the solitary wave solution
\begin{equation}
\psi (t, \xi) = 3 v_0 \text{ sech} ^2 \left[\frac{\sqrt{v_0}}{2}
(\xi - v_0t)\right]. \label{eq9-2-9}
\end{equation}
This soliton has a velocity $v_0$ with respect to the moving frame
$\xi$ which is related to $x'$ via $x'= \delta _2^{\frac{1}{3}}
\xi$. Therefore it has a velocity of $\delta _2^{\frac{1}{3}} v_0$
with respect to the moving frame $x'$ and a velocity of $v+ \delta
_2^{\frac{1}{3}}v_0$ with respect to the stationary frame $x$. Note
that since $\text{sech} ^2x $ decays exponentially for large
arguments, the width of the soliton is $2/ \sqrt{v_0}$ and its
amplitude is $3v_0$. A larger amplitude soliton moves faster and has
a narrower width.

\def\sech{sech}
\def\aprox{\approx}

\section{Derivation of the Nonlinear Schrödinger Equation}
\index{Nonlinear Schrödinger equation}

The KdV equation is closely related to solitary waves in fluid.  The
propagation of solitons in an optical fiber is governed by the
nonlinear Schrödinger (NLS) equation. The nonlear Schrödinger
equation is due to nonlinear effect on electromagnetic wave
propagation in an optical fiber
\cite{KATO,HASEGAWA&KODAMA,SERKIN&HASEGAWA,HAUS8}.  The effect of
nonlinearity in optical fiber generates noise, and this has been
studied by Gordon and Haus \cite{GORDON&HAUS}.  This noise is hence
known as the \index{Gordon-Haus jitter} Gordon-Haus jitter.

Starting with the Maxwell's equations and assuming that ${\partial\,
/
\partial x}={\partial\, / \partial y}=0$, for an electric field polarized
in the $y$ direction, then
\begin{align}
{\partial\, \over\partial z } E_y = {\partial\, \over\partial t}
B_x,\label{(nls1)}\end{align}
\begin{align}{\partial\, \over\partial z } B_x =\mu_0 {\partial\, \over\partial t} D_y,
\label{(nls2)}
\end{align} where $\mu_0$ is assumed to be a constant. Eliminating
$B_x$ gives
\begin{align}{\partial^2\, \over\partial z^2 } E_y = \mu_0 {\partial^2\, \over\partial
 t^2} D_y.
\label{(nls3)}
 \end{align}
For a medium, $D_y = \epsilon_0 n^2 E_y$, where $n$ is the
refractive index. For a Kerr medium, $n$ is nonlinearly related to
the electric field, or that
\begin{align}n=n_0 + n_2 \left| E_y \right| ^2.\label{(nls4)}\end{align}

The Kerr effect is a very small effect such that $n_2 \left| E_y
\right| ^2 \ll n_0$.  But at optical frequencies, a small change in
the refractive index can have a significant effect on the phase
shift of the wave due to the short wavelengths involved. One can
approximate
\begin{align}n^2 E_y = (n_0 + n_2 \left| E_y \right| ^2)^2 E_y \approx (n_0^2 + 2 n_0
n_2 \left| E_y \right| ^2)E_y.\label{(nls5)}\end{align}
Consequently, one can write \eqref{(nls3)} as
\begin{align}{\partial^2\, \over\partial z^2 } E_y ={1 \over {c_0^2}} {\partial^2\,
\over\partial t^2} n^2 E_y = {n_0^2 \over {c_0^2}}{\partial^2\,
\over\partial t^2} E_y + {{2 n_0 n_2} \over {c_0^2}} {\partial^2\
\over\partial t^2} \left| E_y \right| ^2 E_y.\label{(nls6)}
\end{align}
Next, one assumes that
\begin{align}E_y = \phi(z,t)e^{i(k_0z-\omega_0t)},\label{(nls7)}\end{align}
where $\phi(z,t)$ is a slowly varying function of $z$ and $t$.  It
is also the envelope function of a wave.  It can be shown that
\begin{align}
{\partial^2\, \over\partial t^2} E_y =\biggl({\partial^2\, \over\partial
 t^2} \phi - 2i\omega_0 {\partial\, \over\partial t} \phi - \omega_0^2 \phi \biggr)
e^{i(k_0z-\omega_0t)},\label{(nls8a)}
\end{align}
\begin{align}
{\partial^2\, \over\partial z^2} E_y = \biggl({\partial^2\, \over\partial
 z^2} \phi + 2ik_0 {\partial\, \over\partial z} \phi - k_0^2 \phi \biggr)
e^{i(k_0z-\omega_0t)}.\label{(nls8b)}
\end{align}
Then the envelope function $\phi(z,t)$ is assumed to be slowly
varying compared to $\exp[i(k_0z-\omega_0t)]$ so that
\begin{align}{\partial^2\, \over\partial t^2} \phi \ll \omega_0^2 \phi, \label{(nls9a)}\end{align}
\begin{align}{\partial^2\, \over\partial z^2} \phi \ll k_0^2 \phi. \label{(nls9b)}\end{align}

Consequently, one has
\begin{align}{\partial^2\, \over\partial t^2} E_y \aprox\biggl(-2i\omega_0
{\partial\, \over \partial t} \phi -\omega_0^2 \phi\biggr)
e^{i(k_0z-\omega_0t)}, \label{(nls10a)}\end{align}
\begin{align}{\partial^2\, \over\partial z^2} E_y \aprox\biggl(2ik_0{\partial\, \over
\partial t} \phi -k_0^2 \phi\biggr)e^{i(k_0z-\omega_0t)}. \label{(nls10b)}\end{align}
Assuming that the rapidly varying solution satisfies $k_0 =
n_0\omega_0/c_0$, then
\begin{align}
{\partial^2\, \over\partial z^2 } E_y - {n_0^2 \over
{c_0^2}}{\partial^2\, \over\partial t^2} E_y &= 2i\biggl( k_0
{\partial\, \over\partial z} \phi + {n_0^2 \over c_0^2} \omega_0
{\partial\, \over\partial t} \phi\biggr) e^{i(k_0z-\omega_0t)} \cr
&= 2ik_0\biggl( {\partial\, \over\partial z} \phi + {n_0 \over c_0}
{\partial\, \over\partial t} \phi\biggr)e^{i(k_0z-\omega_0t)}.
\label{(nls11)}
\end{align}

The above can be used in \eqref{(nls6)} to arrive at
\begin{align}ik_0\biggl( {\partial\, \over\partial z} \phi +{n_0 \over c_0} {\partial\,
\over\partial t} \phi\biggr)e^{i(k_0z-\omega_0t)} = {n_0n_2 \over
c_0^2} {\partial^2\,\over\partial t^2}\left| E_y \right|^2E_y.
\label{(nls12)}\end{align}
Since $ \left| E_y \right| ^2 = \left|\phi\right|^2$ is slowly
varying compared to $E_y$ itself, then
\begin{align}{\partial^2\, \over\partial t^2} \left| E_y \right| ^2 E_y \approx \left| E_y
\right|^2 {\partial^2\, \over\partial t^2} E_y = -\omega_0^2 \left|
\phi \right| ^2\phi e^{i(k_0z-\omega_0t)}.\label{(nls13)}\end{align}
Using \eqref{(nls13)} in \eqref{(nls12)} gives rise to
\begin{align}\biggl({\partial\, \over\partial z} \phi + {n_0 \over c_0}{\partial\, \over
\partial t} \phi\biggr) = i {k_0n_2 \over n_0} \left| \phi \right| ^2 \phi.
\label{(nls14)}\end{align}

The above is the equation for the envelope function $\phi(z,t)$ when
no dispersion is assumed in the medium.  So when nonlinearity is
absent, the pertinent equation is
\begin{align}\biggl({\partial\, \over\partial z} \phi + {n_0 \over c_0}{\partial\, \over
\partial t} \phi\biggr) = 0,\label{(nls15)}\end{align}
which is the advective equation governing the propagation of a
distortionless pulse.  In the above, the dispersion relation is
\begin{align}k = {n_0 \over c_0}\omega,\label{(nls16)}\end{align}
i.e., there is a linear relationship between $k$ and $\omega$.  As a
result, both the phase velocity $(\omega/k)$ and the group velocity
$(d\omega/dk)$ are independent of frequencies.

\subsection {Dispersive effect}
\index{Dispersive effect}

Dispersion occurs when $n_0$ is a function of $\omega$ so that $k$
is not linearly proportional to $\omega$ anymore.  In this case,
$D_y = \epsilon_0 n^2 \star E_y$, where ``$\star$'' means
``convolves''.
One lets
\begin{align}E_y(z,t) = \int_{-\infty}^{\infty} dk \int_{-\infty}^{\infty} d\omega {\tilde E_y}
(k,\omega)e^{i(kz -\omega t)},\label{(nls17)}\end{align} and
assuming the absence of Kerr effect or the nonlinear term,
\eqref{(nls3)} becomes
\begin{align}k^2{\tilde E_y} = {\omega^2 \over c_0^2} n_0^2(\omega) {\tilde E_y}.\label{(nls18)}\end{align}
Consequently,
\begin{align}k = {n_0(\omega) \over c_0} \omega. \label{(nls19)}\end{align}
One can rewrite \eqref{(nls19)}, using Taylor expansions of its
righthand side, as
\begin{align}k = k_0 + k_0 '(\omega-w_0) + {1 \over 2} k_0 ''(\omega-\omega_0)^2 +
\dots ,\label{(nls20)}\end{align} where one assumes that
\begin{align}k_0 = {n_0(\omega_0) \over c_0}\omega_0,\quad k_0' = {dk \over d\omega}
\Bigg|_{\omega=\omega_0}, \quad k_0'' = {d^2k \over d\omega^2}
\Bigg| _{\omega=\omega_0}. \label{(nls21)}\end{align} Squaring
\eqref{(nls20)}, and putting it back into \eqref{(nls18)} gives rise
to
\begin{align}[k^2 - k_0^2 - 2k_0k_0'(\omega-\omega_0) - (k_0')^2(\omega-\omega_0)^2 -
k_0k_0''(\omega-\omega_0)^2 +\dots ] {\tilde E_y} =
0.\label{(nls23)}
\end{align}
Furthermore, one can write  $k^2 - k_0^2 = (k - k_0)^2 k_0 +
(k-k_0)^2$ and, using the fact that $(k-k_0)^2 \approx
(k_0')^2(\omega-\omega_0)^2$ from \eqref{(nls20)} leads to
\begin{align}[(k-k_0)2k_0 - 2k_0k_0'(\omega-\omega_0) - k_0k_0''(\omega-\omega_0)^2 +
\dots ]{\tilde E_y} = 0.\label{(nls23a)}\end{align} Then,
\eqref{(nls17)} is rewritten as
\begin{align}E_y(z,t) = e^{i(k_0z-\omega_0t)} \int_{- \infty}^\infty dk \int_{-
\infty}^\infty d\omega {\tilde
E_y}(k,\omega)e^{[(k-k_0)z-(\omega-\omega_0)t]}
\label{(24)}\end{align}
\begin{align}\qquad\qquad = \phi(z, t) e^{i(k_0z-\omega_0t)}. \label{(25)}\end{align}
Fourier inverse transforming \eqref{(nls23)} using \eqref{(nls17)},
which is the same as Fourier inverse transforming \eqref{(nls18)}
gives
\begin{equation}
\label{(nls26)} \left(-2ik_0\frac{\partial}{\partial
z}-2ik_0k_0'\frac{\partial}{\partial
t}+k_0k_0''\frac{\partial^2}{\partial t^2}\right)\phi(z,t)=0
\end{equation}
In the above analysis, it is assumed that $\omega\simeq\omega_0$ and
$k\simeq k_0$. This is the same as assuming that $\phi(z,t)$ is
slowly varying since it has only low frequency components. If
$k_0''=0$, the above is an alternative way of deriving
\eqref{(nls15)}. Equation \eqref{(nls26)} can be rewritten as
\begin{equation}
\label{(nls27)} \left(\frac{\partial}{\partial
z}+\frac{1}{v_g}\frac{\partial}{\partial
t}+\frac{i}{2}k_0''\frac{\partial^2}{\partial t^2}\right)\phi(z,t)=0
\end{equation}
where $v_g={1}/{k_0'}$.

For the Kerr effect, one can assume that $n_2$ is frequency
independent, one can add the nonlinear part to the above equation to
obtain
\begin{equation}
\label{(nls28)} \frac{\partial}{\partial
z}\phi+\frac{1}{v_g}\frac{\partial}{\partial
t}\phi=i\frac{k_0n_2}{n_0}|\phi|^2\phi-\frac{i}{2}k_0''\frac{\partial^2}{\partial
t^2}\phi
\end{equation}
The above is the nonlinear Schrödinger equation.

In the moving coordinate system such that $v_gz=z-v_gt$, then
$\phi(z,t)=\phi\left(z,\frac{z}{v_g}-\tau\right)=\tilde{\phi}(z,\tau)$.
Moreover,
\begin{equation}
\label{(nls29)} \frac{\partial}{\partial
t}\phi=-\frac{\partial\tilde{\phi}}{\partial \tau}, \quad
\frac{\partial \tilde{\phi}}{\partial z}=\frac{\partial
{\phi}}{\partial z}+\frac{1}{v_g}\frac{\partial}{\partial t}{\phi}
\end{equation}
or
\begin{equation}
\label{(nls30)}
\frac{\partial \tilde{\phi}}{\partial
z}=i\frac{k_0n_2}{n_0}|\tilde{\phi}|^2\tilde{\phi}-\frac{i}{2}k_0''\frac{\partial^2}{\partial
\tau^2}\tilde{\phi}
\end{equation}
The above is similar to the Schrödinger equation
\begin{equation}
\label{(nls31)} -i\hbar\frac{\partial}{\partial
t}\psi=-\frac{\hbar^2}{2m}\frac{\partial^2}{\partial x^2}\psi+V\psi
\end{equation}
if one identifies $z$ with $t$ and $\tau$ with $x$, and
$|\tilde{\phi}|^2$ with $V$.

When $V$ is negative to form a potential well, Equation
\eqref{(nls31)} admits solutions of the form
\begin{equation}\label{(nls32)}
\psi=Ae^{-i\omega t}g(x)
\end{equation}
When $g(x)$ is a function localized in x corresponding to bound
states or trapped modes in the potential well which are stationary
states. By the same token, by ${k_0}''$ is negative, Equation
\eqref{(nls30)} admits solutions of the form
\begin{equation}\label{(nls33)}
\begin{split}
\tilde{\phi}&=Be^{ikz}f(v_g\tau)\\
&=Be^{ikz}f(z-v_gt)
\end{split}
\end{equation}
where $f(v_g\tau)$ is a function localized in $\tau$. Equation
\eqref{(nls33)} corresponds to a solution.

Since the potential well is created by $|\tilde{\phi}|^2$,which is
proportional to the field itself, this is a self-trapping
phenomenon. Conservation of energy requires that the pulse becomes
narrower when the amplitude of $\tilde{\phi}$ becomes larger. On the
other hand, the dispersive effect tends to spread the pulse more
when it becomes narrower. A final equilibrium is reached where the
pulse propagates without distortion.

If ${k_0}''$ is positive, Equation (30) cannot trap a mode. However,
it can admit a solution signified by the absence of light, hence the
name dark solutions.

\subsection {Solution of the Nonlinear Schrödinger Equation}

Via a change of variables and coordinates, the nonlinear
Schrödinger equation can be written in dimensionless form as
\begin{align} i {\partial\psi\,  \over\partial z} + {1 \over 2} {\partial ^2\psi\, \over
\partial T^2} + \left| \psi \right| ^2 \psi = 0. \label{(s5)}\end{align}
A solution is a stationary solution to the above which is stationary
in amplitude with respect to $z$ and $T$.  Therefore, one seeks a
solution of the form
\begin{align}\psi (T, z) = \sqrt {\rho(T)}e^{i\alpha(z-z_0)}.\label{(s6)}\end{align}
Substituting  \eqref{(s6)} into \eqref{(s5)} yields
\begin{align}-\alpha{\sqrt\rho} + {1 \over 2}{\partial^2\, \over\partial T^2}{\sqrt \rho}
+ \rho^{{3 \over 2}} = 0, \label{(s7)}\end{align} where  $\rho$ is
assumed to be real-valued.  The above is the same as
\begin{align} -\alpha - {1 \over 8} \Biggl[{1 \over \rho^2}(\rho')^2 - 2{\rho'' \over
\rho} \Biggr] + \rho = 0,\label{(s8)}
\end{align}
where $\rho' = {d\rho \over dT}$,\quad $\rho''={d^2\rho \over
dT^2}$. It can be further simplified to
\begin{align} -\alpha+{1 \over 8}{d \over d\rho}{1 \over \rho}(\rho')^2+\rho=0.\label{(s9)}\end{align}
Integrating the above leads to
\begin{align} -\alpha\rho + {\rho^2 \over 2} + {1 \over 8}{1 \over \rho}(\rho')^2 +C=0.
\label{(s10)}\end{align} Since $\rho$ and $\rho'$ are independent of
$T$, $C$ is also independent of $T$.  But since $\rho\rightarrow 0$
when $T\rightarrow\infty$, and one assumes that
$\rho'\sim\rho\rightarrow 0$ when $T\rightarrow\infty$, then $C=0$.
When $\rho$ is maximum, then $\rho'=0$.  Therefore, \eqref{(s10)}
has to be of the form
\begin{align}(\rho')^2 = -4\rho^2(\rho - \rho_0),\label{(s11)}\end{align}
where $\rho_0$ is the maximum of $\rho$.  Therefore $\alpha =
\rho_0/2$.  One can let $\rho_0y^2 = \rho-\rho_0$ in \eqref{(s11)}
to obtain
\begin{align}dy = (1+y^2)\sqrt{\rho_0}idT.\label{(s12)}\end{align}
Letting $\tau = i\sqrt{\rho_0}T$  gives rise to
\begin{align}{dy \over 1+y^2} = d\tau,\label{(s14)}\end{align}
\begin{align}y=\tan(\tau-\tau_0),\label{(s15)}\end{align}
or
\begin{align}\rho = \rho_0(y^2+1) = \rho_0\sec^2(\tau-\tau_0),\label{(s16)}\end{align}
or
\begin{align}\rho = \rho_0 \ \hbox{sech}^2 \bigl[\sqrt{\rho_0} (T-T_0) \bigr].\label{(s17)}\end{align}
Consequently,
\begin{align}\psi(T,z) = \sqrt{\rho_0}\ \hbox{sech} \bigl[\sqrt{\rho_0}(T-T_0)
\bigr] e^{i{\rho_0 \over 2} (z-z_0)}.\label{(s18)}\end{align} Since
$T\rightarrow(z-v_gt)$, this pulse travels with the same speed
irrespective of the amplitude of the pulse.

Schödinger equation \eqref{(nls31)} is \index{Galilean invariant} Galilean invariant. Hence,
if a solution $\psi(x,t)$ is found, a new solution
$\psi(x+vt,t)e^{i\theta(x,t)}$ is also a solution. Using this fact,
a new solution
\begin{align}\psi_1(T,z) = \psi(T+kz,z)e^{-ik(T+{1 \over 2}kz)}\label{(s19)}\end{align}
can be constructed that is also a solution of the nonlinear
Schrödinger equation.

Equation \eqref{(s18)} represents a stationary solution in the
moving coordinate system, or it is a soliton solution that moves at
the group velocity.  Equation \eqref{(s19)} represents a soliton
solution that moves at arbitrary velocity since $k$ is arbitrary.
The solution has the property that the larger its amplitude, the
narrower is the pulse.  However, unlike the KdV solution, the NLS
solution's velocity does not increase with its amplitude.

\section {{{  Solution of the KdV Equation via Inverse Scattering
Transform}}}
\index{Korteweg de Vries equation!solution}
\index{Inverse scattering transform}

There is an interesting body of knowledge where solutions to
nonlinear partial differential equations (PDEs) can be obtained by
solving an inverse scattering problem.  This is known as the inverse
scattering transform method in solving nonlinear PDEs
\cite{SHABAT&ZAKHAROV,ABLOWITZ&CLARKSON}.

 Via a change of variable, the KdV
equation can be written as
\begin{equation}
\frac{\partial q}{\partial t} - 6q \frac{\partial q}{\partial x} +
\frac{\partial ^3 q}{\partial x^3} = 0 . \label{eq9-3-1}
\end{equation}
It turns out that if $q(x,t)$ is a potential in a Schrödinger
eigenvalue equation, i.e.,
\begin{equation}
\frac{\partial ^2 u}{\partial x^2} + [ \lambda - q (x,t)]u = 0,
\label{eq9-3-2}
\end{equation}
then $u(x,t)$ evolves in time according to the following equation,
\begin{equation}
\begin{split}
\frac{\partial u}{\partial t} &= - \frac{\partial q}{\partial x} u +
(4 \lambda
+ 2 q ) \frac{\partial u }{\partial x}\\
&= -4 \frac{\partial ^3u}{\partial x^3} + 3 \frac{\partial
q}{\partial x} u + 6q\frac{\partial u}{\partial x}.
\end{split}
\label{eq9-3-3}
\end{equation}

Equation (\ref{eq9-3-2}) can be expressed as
\begin{equation}
Lu = \lambda u \label{eq9-3-4}
\end{equation}
where $L= - \frac{\partial ^2}{\partial x^2} + q(x,t)$, while
Equation (\ref{eq9-3-3}) is expressible as
\begin{equation}
\frac{\partial u }{\partial t} = M u \label{eq9-3-5}
\end{equation}
where
\begin{equation}
M = - \frac{\partial q}{\partial x} + ( 4 \lambda + 2 q )
\frac{\partial}{\partial x} = - 4 \frac{\partial ^3}{\partial x^3} +
3 \frac{\partial q}{\partial x} + 6q \frac{\partial }{\partial x}.
\label{eq9-3-6}
\end{equation}
The critical point here is that $\lambda $ is assumed independent of
time. So if one differentiates (\ref{eq9-3-4}) with respect to time,
then
\begin{equation}
L_t u + LMu = \lambda M u = M \lambda  u = MLu, \label{eq9-3-7}
\end{equation}
or that
\begin{equation}
(L_t + LM - ML ) u = 0. \label{eq9-3-8}
\end{equation}
It can be shown that in order for (\ref{eq9-3-8}) to be satisfied,
$q$ satisfies (\ref{eq9-3-1}), the KdV equation.

The above gives an alternative way of solving (\ref{eq9-3-1}). Given
the initial value $q(x,0)$, one can solve for $u(x,0)$ from
(\ref{eq9-3-2}). Then if one can determine $u(x,t) $ from $u(x,0)$,
then inverse scattering theory can be used to find $q(x,t)$ from $
u(x,t)$ in (\ref{eq9-3-2}).

\subsection {{{ Inverse Scattering}}}
\index{Inverse scattering}

Consider the Schrödinger equation
\begin{equation}
-\frac{\partial ^2}{\partial x^2}u(x) + q(x) u (x) = \lambda u (x),
\label{eq9-4-1}
\end{equation}
and that $q(x) \to 0$ when $|x| \to \infty$. The above equation has
two independent solutions $u_1(x,k)$ and $u_2(x,k)$ such that $u_1
\sim e^{ikx}$, $x \to +\infty$, and that $u_2  \sim e^{-ikx}$, $x
\to - \infty$, where $k^2 = \lambda $. Define the function $A(x,y)$
as
\begin{equation}
u_1 (x, k) = e ^{ikx} + \int \limits_x^\infty A (x,t) e ^{ikt} dt.
\label{eq9-4-2}
\end{equation}
One can substitute (\ref{eq9-4-2}) into (\ref{eq9-4-1}) to obtain an
equation for $A(x, t)$. First one notices that
\begin{equation}
\frac{\partial ^2}{\partial x^2} \int \limits_x^\infty A (x,t) e
^{ikt} dt = \int \limits_0 ^\infty \frac{\partial ^2}{\partial x ^2}
[ H(t-x) A (x,t)]e^{ikt}dt, \label{eq9-4-3}
\end{equation}
where $H(t)$ is a Heaviside step function. Evaluating the above
yields
\begin{equation}
\begin{split}
\frac{\partial ^2}{\partial x ^2} \int \limits _x ^\infty A ( x,t)
e^{ikt} dt &= - A _t (x,t) \bigg \vert _{t=x} e ^{ikx} - ikA (x,x) e
^{ikx} - 2 A _x (x,t)
\bigg \vert _{t=x} e ^{ikx} \\
&+ \int \limits_x ^ \infty A _{xx} (x,t) e ^{ ikt} dt.
\end{split}
\label{eq9-4-4}
\end{equation}
In the above, one has made use of
\begin{align}
\frac{\partial ^2}{\partial x ^2}H(t-x)A(x,t) &=
H''(t-x)A(x,t)-2H'(t-x)A_x(x,t)+H(t-x)A_{xx}(x,t)\notag\\
&=\delta'(t-x)A(x,t)-2\delta(t-x)A_x(x,t)+H(t-x)A_{xx}(x,t)
\end{align}
Furthermore, since $\lambda = k^2$, the right-hand side of
\eqref{eq9-4-1} becomes
\begin{equation}
\begin{split}
- \int \limits_x ^\infty k^2 A (x,t) e ^{ikt} dt =&+ \int \limits_x
^\infty A
(x,t) \frac{\partial ^2}{\partial t^2} e ^{ikt} dt\\
=& A (x,t ) \frac{\partial }{\partial t} e ^{ikt} \bigg \vert _x
^\infty - \int
\limits_x ^\infty A _t (x,t ) \frac {\partial }{\partial t} e ^{ikt} dt\\
=& A (x,t) ike^{ikt} \bigg \vert _x ^\infty - A _t (x,t ) e ^{ ikt }
\bigg
\vert _x ^\infty + \int \limits _x ^ \infty A_{tt} ( x,t ) e ^{ikt} dt\\
=& \lim_{t \to \infty} [ ikA (x,t ) e ^{ikt} - A_t (x,t) e ^{ ikt} ]\\
& - ik A (x,x)
e ^{ikx} + A _t (x,t ) \bigg \vert _{t=x} e ^{ikx}\\
&+ \int \limits _x ^\infty A _{tt} (x,t) e ^{ikt}dt.
\end{split}
\label{eq9-4-5}
\end{equation}
Consequently,
\begin{equation}
\begin{split}
\frac{\partial ^2}{\partial x ^2} u_1-q(x)u_1 + k^ 2 u _1
=& \int
\limits
_x ^ \infty  [ A _{xx} (x,t ) - A _{tt} (x,t) - q (x) A (x,t)] e ^{ikt} dt\\
+& \lim _{t \to \infty} [ A_t (x,t ) - ik A (x,t) ] e ^{ ikt}\\
-&\left [ 2A_t(x,t)\vert_{t=x}+ 2A_x(x,t)\vert_{t=x}+ q (x)\right ]
e ^{ikx}\\
=& \int \limits
_x ^ \infty  [ A _{xx} (x,t ) - A _{tt} (x,t) - q (x) A (x,t)] e ^{ikt} dt\\
+& \lim _{t \to \infty} [ A_t (x,t ) - ik A (x,t) ] e ^{ ikt}\\
-&\left [ 2 \frac{d}{dx} A(x,x) + q (x)\right ] e ^{ikx},
\end{split}
\label{eq9-4-6}
\end{equation}
where
\begin{equation}
\frac{d}{dx} A (x,x ) = \frac{\partial A (x,t) }{\partial x} \bigg
\vert _{t=x} + \frac{\partial A (x,t)}{\partial t} \bigg \vert
_{t=x} . \label{eq9-4-7}
\end{equation}
If $\lim \limits_{t \to \infty} A_t (x,t) =\lim \limits_{t \to
\infty} A (x,t ) = 0 $, and the above is valid for all $k$, then $q
(x) = -2 \frac{d}{dx} A (x,x)$, and,
\begin{equation}
A_{xx}-A_{tt} - q (x) A = 0. \label{eq9-4-8}
\end{equation}
in order for the above to be satisfied.

 If a wave is incident on the
potential $q(x)$ from $x=\infty$ as $e^{-ikx}$, the solution for $x
\to +\infty $ must be $ e ^{-ikx} + R (k) e ^{ikx}$, $ x \to
+\infty$, where $R (k)$ is the reflection coefficient. For $x \to -
\infty$, the solution is $ T(k) e ^{ -ikx}$ where $T(k)$ is the
transmission coefficient. This solution, by matching its large $|x|$
behaviors, must be expressible as $R(k) u _1 (x,k) + u _1 (x,-k) $
or $ T (k)u_2 (x,k)$ since $ u_1 (x,k) \sim e ^{ikx}$, $ x \to
+\infty$, and $ u_2 (x,k) \sim e ^{-ikx}$, $ x \to -\infty$.
Therefore, one concludes that
\begin{equation}
T(k) u_2 (x,k)= R (k) u _1 (x,k) + u_1 (x, -k). \label{eq9-4-9}
\end{equation}

Furthermore, when $ k\to +\infty$, or that the  frequency or energy
of the wave tends to infinity, the potential barrier has a
negligible effect on the wave, and $ T (k) \to 1$ (typical of
quantum scattering). Moreover, $T(k)$ and $ u_2(x, k)$ are the
Fourier transforms of causal signals and they have to be analytic
for $ \Im m [k] > 0$. Consequently, one can Fourier inverse
transform (\ref{eq9-4-9}) along an inversion contour $C$ which is
above the singularities of $T(k)$ and $ u_2(x,k)$, or
\begin{equation}
\begin{split}
\int \limits _ C T (k ) u_2 (x,k) e ^{ -ikt } dk & = \int \limits _C
R (k) u _1
(x,k) e ^{ -ikt} dk\\
&+ \int \limits _C u_1 (x, -k) e ^{ -ikt} dk.
\end{split}
\label{eq9-4-10}
\end{equation}
Since $T (k)u_2 (x, k ) e ^{ -ikt} \sim e ^{ -ik(t+x)} + O
\left(\frac{1}{k} \right )e ^{ -ik(t+x)}$, $k\to \infty$, a fact
easily proven from Born approximation, it implies that
\begin{equation}
\int \limits_C T (k) u_2 (x,k) e ^{ -ikt}dk = 2 \pi \delta (t + x )
+ \int \limits _C O \left  ( \frac{1}{k} \right )e ^{ -ik ( t + x )}
. \label{eq9-4-11}
\end{equation}
Using Jordan's lemma for the second term in (\ref{eq9-4-11}), one
concludes that
\begin{equation}
\int \limits _ C T(k) u _2 (x,k ) e ^ {-ikt} dk = 0, \qquad t+x < 0.
\label{eq9-4-12}
\end{equation}
Using the form in Equation (\ref{eq9-4-2}) for $ u_1 ( x,k)$ in
(\ref{eq9-4-10}), then
\begin{equation}
\begin{split}
0&= \int \limits _ C R (k) e ^{ ik(x-t)}dk + \int \limits _x  ^
\infty dt' A
(x,t') \int \limits _ C R (k) e ^{ik(t'-t)} dt\\
&+\int \limits _C e ^{ -ik (x+t )} dk + \int \limits _x ^ \infty dt'
A (x, t ') \int \limits _C e ^{ -ik (t'+t) } dt, \quad t + x < 0.
\end{split}
\label{eq9-4-13}
\end{equation}
The third term is zero because $ t + x < 0$, and defining
\begin{equation}
r (t ) = \frac{1}{2 \pi} \int \limits _C R (k) e ^{ -ikt} dk,
\label{eq9-4-14}
\end{equation}
gives
\begin{equation}
0 = r (t-x) + \int \limits _x ^\infty dt' A (x,t ') r (t - t') + A
(x, -t ), \qquad t+x < 0. \label{eq9-4-15}
\end{equation}
The above is known as the \index{Gelfand-Levitan-Marchenko equation}Gelfand-Levitan-Marchenko equation. Given
$r(t)$, one can solve for $ A (x,t ) $ and obtain $q(x)$ via
\begin{equation}
q (x) = - 2 \frac{d }{dx} A (x, x). \label{eq9-4-16}
\end{equation}
As an example, one can consider a case where $ r (t) = me ^{+\mu t
}. $ Using it in (\ref{eq9-4-15}), then
\begin{equation}
0=me ^{+\mu t - \mu x} + me ^{+\mu t } \int \limits _ x ^\infty dt'
A (x,t') e ^{ -\mu t '} + A (x, -t), \qquad t+x< 0. \label{eq9-4-17}
\end{equation}
Letting $ A(x,t ) = a (x) e ^{ -\mu t }$, then
\begin{equation}
0 = me^{-\mu x} + ma(x) \int \limits _x^ \infty dt' e ^{-2\mu t'} +
a (x), \qquad t+x<0 \label{eq9-4-18}
\end{equation}
or that
\begin{equation}
a (x) = \frac{ - m e ^{-\mu x}}{1+ \frac{m}{2 \mu} e ^{ -2 \mu x}}.
\label{eq9-4-19}
\end{equation}
It follows then that
\begin{equation}
A (x,x) = \frac{ -me ^{-2 \mu x}}{ 1 + \frac{m} {2 \mu} e ^{-2\mu
x}} = \frac { - m } {\frac{ m} {2 \mu} + e ^{2 \mu x }}.
\label{eq9-4-20}
\end{equation}
Differentiating the above with respect to $x$ leads to
\begin{equation}
\begin{split}
q (x) = \frac{ - 4 m \mu e ^{2\mu x}} { \left ( \frac{m} { 2 \mu} +
e ^{ 2 \mu x } \right ) ^2 } &= -8\mu ^2 \frac{e ^{ 2 \phi + 2 \mu x
}}{(e ^{ 2 \phi} + e
^{ 2 \mu x} ) ^2  } \\
&= - 2 \mu ^2 \text{ sech} ^2 ( \mu x - \phi).
\end{split}
\label{eq9-4-21}
\end{equation}
where $ \phi = \frac{1}{2} \ln \frac{ m}{2 \mu}$.

\subsection {{{ Solution of the KdV Equation}}}

If the solution to the KdV equation, $ q (x,t)$, is a potential to
the Schrödinger equation $ - u_{xx} + q (x, t)u = k^2 u$, then the
eigenfunction evolves according to $ u _t = - 4u_{xxx} + 3 q _ x u +
6 q u_x $ according to (\ref{eq9-3-3}). Therefore, if $ u $ is known
when $ |x | \to \infty$, one can use the inverse scattering theory
to reconstruct $ q (x,t)$, and hence, the solution to the KdV
equation.

As is shown in (\ref{eq9-4-9}), the fundamental solutions to the
Schrödinger equation satisfy
\begin{equation}
u _2 (x,k) = c_{11}(k) u_1(x,k) + c _{12} u_1 (x, -k).
\label{eq9-5-1}
\end{equation}
where $c _ {11} (k) = R (k) / T (k)$, $ c _{12} (k) = 1 / T (k)$.
Since $ u_2 (x,k) \sim e ^{ -ikx}$, $ x \to - \infty$, irrespective
of $ q (x,t ) $, and so independent of $t$, one assumes that $ u=
h(t) e^{-ikx}, x\to -\infty$. If $q$ and $q _x$ tend to zero when
$|x| \to \infty$, then
\begin{equation}
u _t = - 4 u _{xxx},\quad |x|\to \infty. \label{eq9-5-2}
\end{equation}
Hence for $ x \to -\infty$, then
\begin{equation}
h_t = - 4 ik^3 h, \label{eq9-5-3}
\end{equation}
or that
\begin{equation}
h = h_0 e ^{ -4ik^3 t}. \label{eq9-5-4}
\end{equation}
For $x \to +\infty$, then
$$
u(x) \sim h(t) (c_{11} e^{ikx} +c_{12} e^{-ikx}).
$$
Therefore, from (\ref{eq9-5-2}), then
\begin{equation}
\begin{split}
(h_t c_{11} + hc_{11t}) e ^{ ikx}&+(h_t c _{12} + h c_{12t}) e ^ { -ikx} \\
=-4h [ c _ {11} (ik)^3 e ^{ikx} &+ c _{12} (ik)^3 e ^{-ikx}],
\end{split}
\label{eq9-5-5}
\end{equation}
Comparing the left and right-hand sides, and making use of
(\ref{eq9-5-3}), it leads to
\begin{equation}
c _{11t} = 8ik^3 c_{11}, \qquad c _{12t} = 0 \label{eq9-5-6}
\end{equation}
or
\begin{equation}
c_{11} = c_{11}^o e ^{8ik^3t}, \qquad c _{12} = c _{12} ^ 0 .
\label{eq9-5-7}
\end{equation}
Consequently, one can find the time evolution of $c_{11}$ and
$c_{12}$. One can obtain $ R (k ) =c _ {11} (k) / c _ {12} (k)$.
Since $ c _{11}$ and $ c_{12}$ are evolving with time, $R (k)$ also
evolves with time.  At this point, the notation is rather confusing
since there is a time variable also in (\ref{eq9-4-14}) and
(\ref{eq9-4-15}).  One  shall call the time variable in the
aforementioned equation $ \tau$ to avoid the confusion. Hence,
\begin{equation}
r (\tau, t ) = \frac{ 1 } {2\pi} \int \limits _ C R (k, t) e ^{ -
ik\tau } dk. \label{eq9-5-8}
\end{equation}
From Equation (\ref{eq9-5-7}), one notices that
\begin{equation}
R (k,t) = R _0 (k) e ^ { 8ik^3 t }. \label{eq9-5-9}
\end{equation}
For a simple case, one considers the case where $ R _0 (k)$ has one
simple pole at $ k=i\mu$. Then, the above integral can be evaluated
so that
\begin{equation}
r (\tau, t ) = me ^{ +8\mu ^3 t + \mu \tau}. \label{eq9-5-10}
\end{equation}
According to (\ref{eq9-4-21}), $ q (x,t)$ is
\begin{equation}
q (x,t) = - 2 \mu ^ 2 \text{ sech} ^2 (\mu x - \phi)
\label{eq9-5-11}
\end{equation}
where
\begin{equation}
\phi=\frac{1}{2} \ln \left ( \frac{ me^{8 \mu ^3 t}}{2 \mu} \right )
= 4 \mu ^3 t + \phi_0, \label{eq9-5-12}
\end{equation}
or
\begin{equation}
q (x,t ) = -2 \mu ^2 \text { sech} ^2 ( \mu x - 4 \mu ^3 t + \phi_0)
\label{eq9-5-13}
\end{equation}
Letting $ \mu ^2 = \frac{1}{4} c$, the above becomes
\begin{equation}
q (x,t) = - \frac{c}{2} \text { sech}^2 \left [ \frac{1}{2} \sqrt
{c} ( x-ct ) + \phi_0 \right ]. \label{eq9-5-14}
\end{equation}
The above method can be used to find the \index{Multiple soliton solution} multiple soliton solution
to the KdV equation by assuming more poles in the reflection
coefficient. It addition, the inverse scattering transform method
can be used to solve nonlinear equations like the nonlinear
Schrödinger equation.



\subsection{Inverse Scattering with Schrödinger Equation}
\index{Inverse scattering!Schrödinger equation}

Consider the Schr\"odinger equation
\begin{align}-{d^2 \over dx^2}u(x) + q(x)u(x) = \lambda u(x),\label{(I1)}\end{align}
where $q(x)\rightarrow0$ when $|x|\rightarrow\infty$.  The above
equation has two independent solutions.  They can be defined as
$u_1(x,k)$ and $u_2(x,k)$ such that $u_1(x,k) \sim e^{ikx},
x\rightarrow +\infty$, and that $u_2(x,k) \sim e^{-ikx},
x\rightarrow -\infty$, where $k^2=\lambda$.  Clearly, $u_1$ and
$u_2$ are independent of each other when $q=0$.  When $q(x) \not=0$,
the scattering potential will generate more waves but $u_1$ and
$u_2$ remain independent of each other.  Therefore, any solution to
\eqref{(I1)} can be written as a linear superposition of $u_1$ and
$u_2$.

The general solution to \eqref{(I1)} can be written in terms of an
integral equation
\begin{align}u(x) = u_0(x) + \int_{-\infty}^\infty g(x-x')q(x')u(x')dx',\label{(I2)}\end{align}
where $g(x)$, the Green's function, is a solution to
\begin{align}\Biggl({d^2 \over dx^2} + k^2 \Biggr)g(x) = \delta(x),\label{(I3)}\end{align}
and $u_0(x)$ is a solution to
\begin{align}\Biggl({d^2 \over dx^2} + k^2 \Biggr)u_0(x) = 0.\label{(I4)}\end{align}
If physical condition such as causality is not imposed, there could
be many solutions to \eqref{(I3)}. The possible solutions to
\eqref{(I3)} are
\begin{align}g_1(x) = {e^{ik|x|} \over 2ik},\label{(I5a)}\end{align}
\begin{align}g_2(x) = {e^{-ik|x|} \over 2ik},\label{(I5b)}\end{align}
\begin{equation}
g_3x =
\begin{cases}
0,& x<0, \\ &\\ \frac{\sin kx}{k},& x>0.
\end{cases}
\label{(I5c)}
\end{equation}
\begin{align}
 g_4(x) =
\begin{cases}\frac{\sin kx}{k},& x<0, \cr &\cr 0,&
x>0.\end{cases} \label{(I5d)}
\end{align}
If $e^{-ikt}$ time convention is used, only $g_1(x)$ is physical
because it corresponds to outgoing waves, while $g_i(x)$, $i>1$ are
unphysical. However, they can be used in \eqref{(I2)} to provide
bonafide solution to \eqref{(I1)}.

To construct $u_1(x,k)$, one lets $u_0$ in \eqref{(I2)} to be
$e^{ikx}$, and $g(x)$ in \eqref{(I2)} to be $g_4(x)$.  Then
$u_1(x,k)$ satisfies
\begin{align}u_1(x) = e^{ikx} + \int_{-\infty}^\infty g_4(x-x')q(x')u_1(x')dx'.\label{(I6)}\end{align}
Due to the property of $g_4(x)$, clearly, $u_1(x)\sim e^{ikx}$,
$x\rightarrow +\infty$.

To construct $u_2(x,k)$, one lets $u_0$ in \eqref{(I2)} to be
$e^{-ikx}$ and $g(x)$ to be $g_3(x)$.  Then, $u_2(x,k)$ satisfies
\begin{align}u_2(x) = e^{-ikx} + \int_{-\infty}^\infty g_3(x-x')q(x')u_2(x')dx'.\label{(I7)}\end{align}
Clearly, $u_2(x)\sim e^{-ikx}, \quad x \rightarrow -\infty$.  Due to
the independence of $g_3(x)$ and $g_4(x)$, $u_1(x,k)$ and $u_2(x,k)$
are independent of each other.  Since $u_1(x,-k)$ is independent of
$u_1(x,k)$, they are related by
\begin{align}T(k)u_2(x,k) = R(k)u_1(x,k) + u_1(x,-k) = u_3(x,k).\label{(I8)}\end{align}
Equation \eqref{(I8)} corresponds to a scattering solution where it
becomes $Re^{ikx} + e^{-ikx}$, $x\rightarrow +\infty$, and
$Te^{-ikx}$, $x \rightarrow -\infty$. Therefore, $R$ and $T$
physically correspond to reflection and transmission coefficients
respectively.  Equation \eqref{(I8)} can be used to derive the
Gelfand-Levitan-Marchenko equation.

\subsection{Time-Domain Solutions}
\index{Inverse scattering!time domain}

\begin{figure}[ht]
\begin{center}
\hfil\includegraphics[width=3.truein]{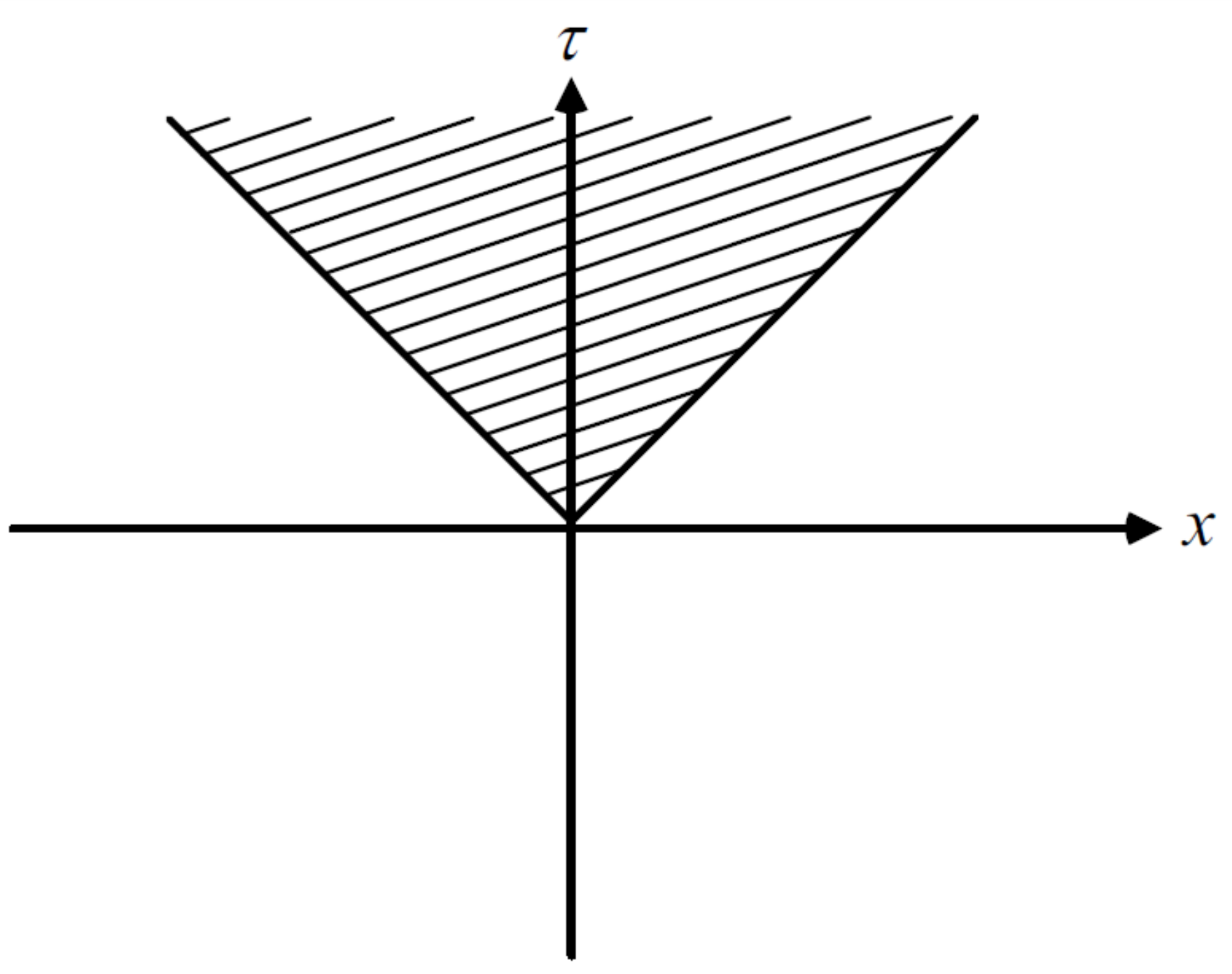}\includegraphics[width=3.truein]{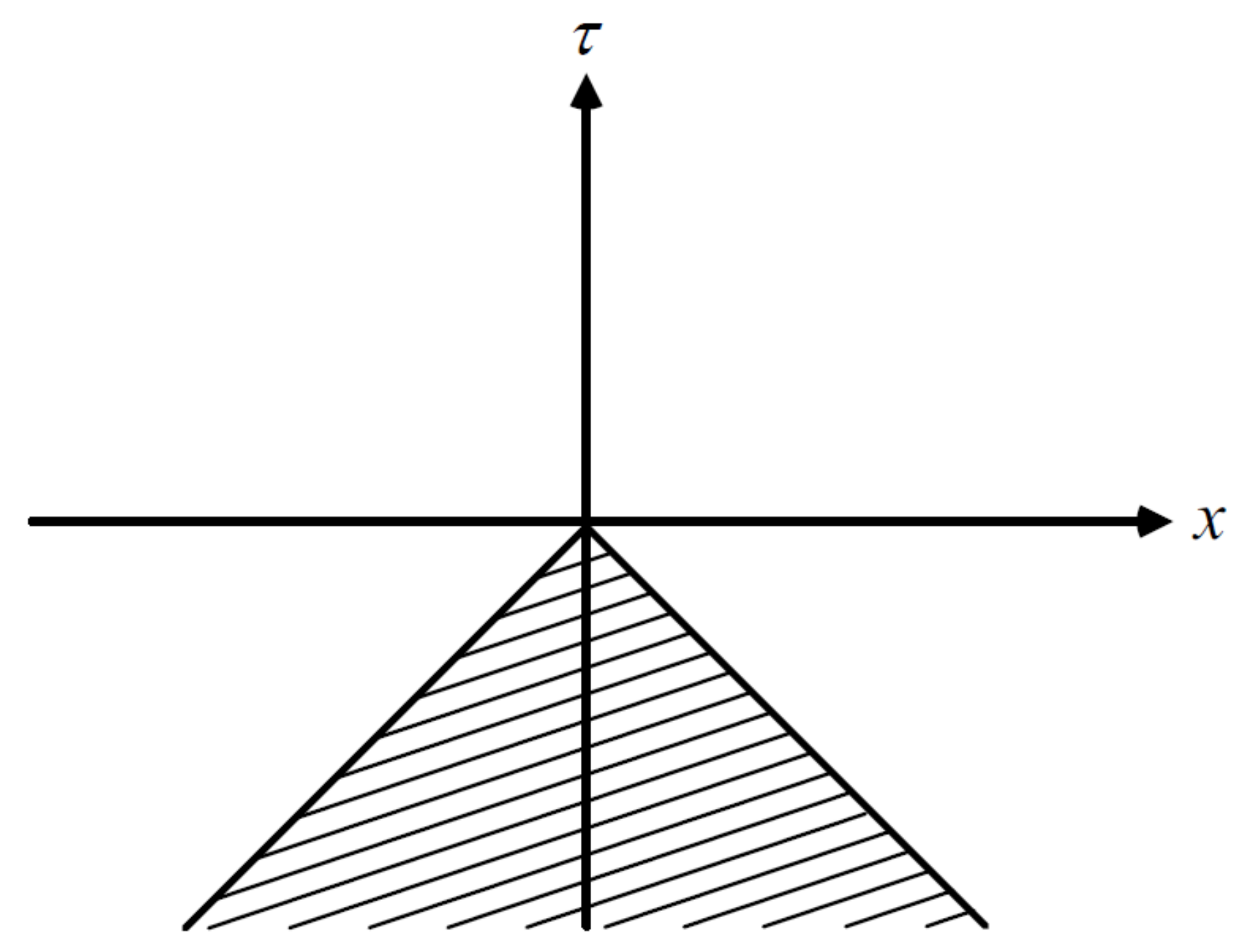}\hfil
\end{center}
\begin{center}
{(a) Support of Green's function $g_1(x,\tau)$.}\hskip0.5in{(b)
Support of Green's function $g_2(x,\tau)$.}
\end{center}
\begin{center}
\hfil\includegraphics[width=3.truein]{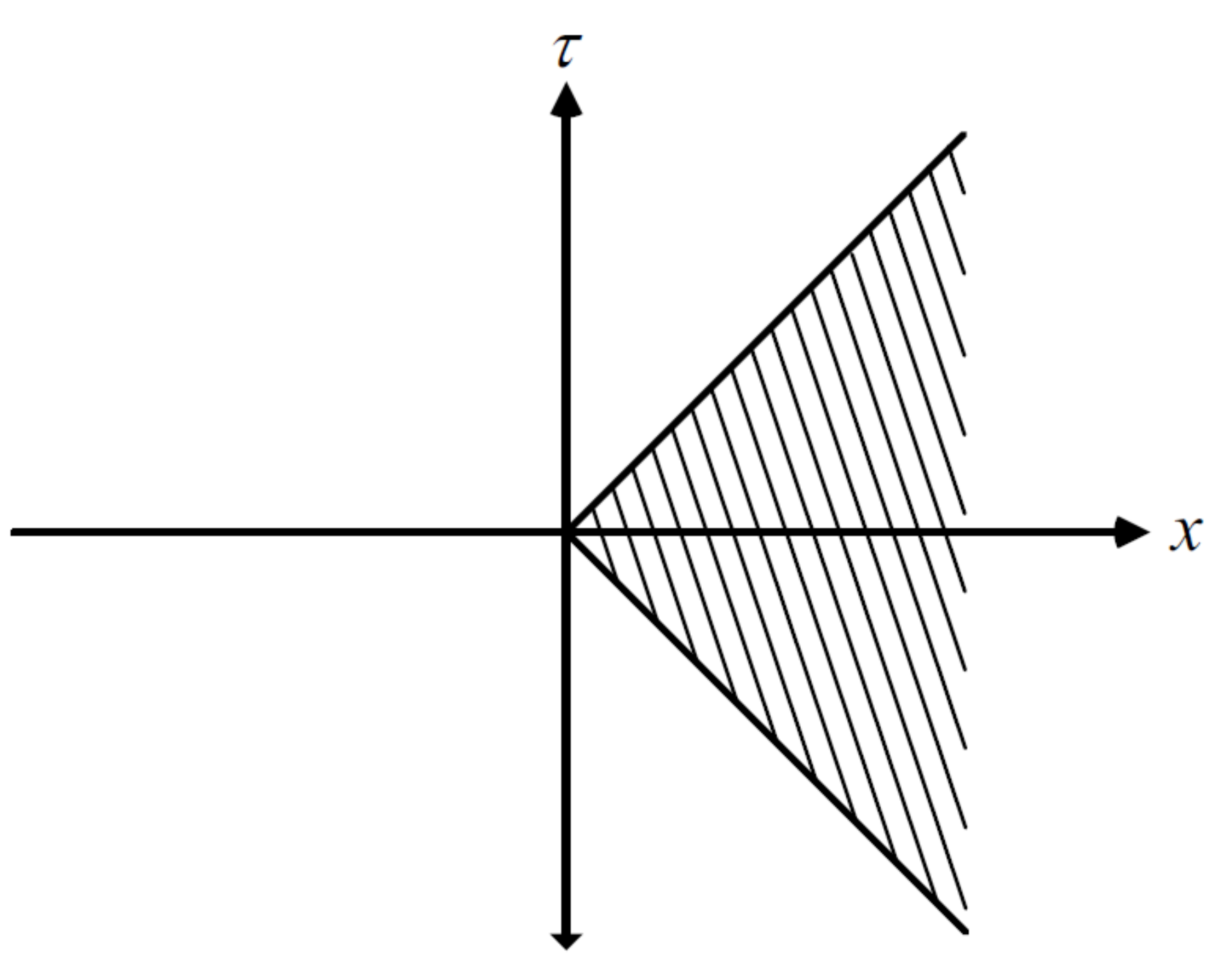}\includegraphics[width=3.truein]{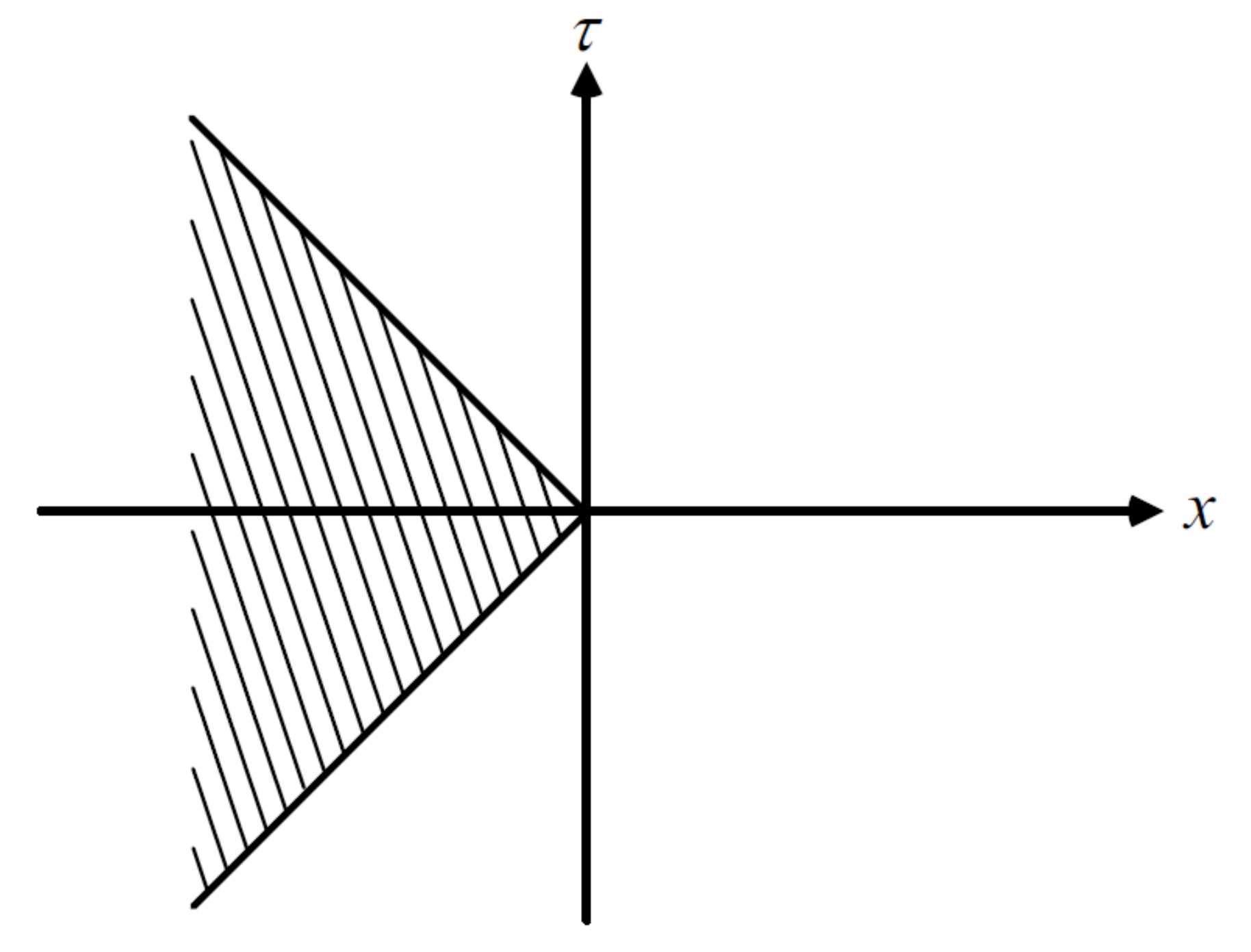}\hfil
\end{center}
\begin{center}
{(c) Support of Green's function $g_3(x,\tau)$.}\hskip0.5in{(d)
Support of Green's function $g_4(x,\tau)$.}
\end{center}
\caption{Supports of different Green's functions.}\label{fg8_2}
\end{figure}

If one expresses
\begin{align}u(x, \tau) = {1 \over 2\pi} \int_{-\infty}^\infty dk e^{-ik\tau}u(x,k),\label{(I9)}\end{align}
where $k^2=\lambda$, and Fourier inverse transform \eqref{(I1)}
accordingly gives
\begin{align}-{\partial^2\, \over \partial x^2}u(x,\tau) + q(x)u(x,\tau) =-{\partial^2\,
\over\partial \tau^2}u(x,\tau).\label{(I10)}\end{align} Here,
\eqref{(I10)} is the Schrödinger-like equation since it has a
second derivative in time while Schrödinger equation has a first
derivative in time. Then \eqref{(I2)} becomes
\begin{align}u(x,\tau) = u_0(x,\tau) + \int_{-\infty}^{\infty} g(x-x',\tau)\star u(x',\tau)
q(x')dx',\label{(I11)}\end{align} were $g(x,\tau)$ is the Fourier
inverse transform of the Green's function $g(x, k)$, given in
\eqref{(I5a)}--\eqref{(I5d)}. The Fourier inversion contour is
defined to be above the singularity at the origin in \eqref{(I5a)}
and \eqref{(I5b)} so that only \eqref{(I5a)} is causal while the
rest of the Green's functions are not causal. The supports of the
various Green's functions are shown in Figure \ref{fg8_2}, (a) to
(d)
 on a space-time
diagram.

\begin{figure}[ht]
\begin{center}
\hfil\includegraphics[width=3.truein]{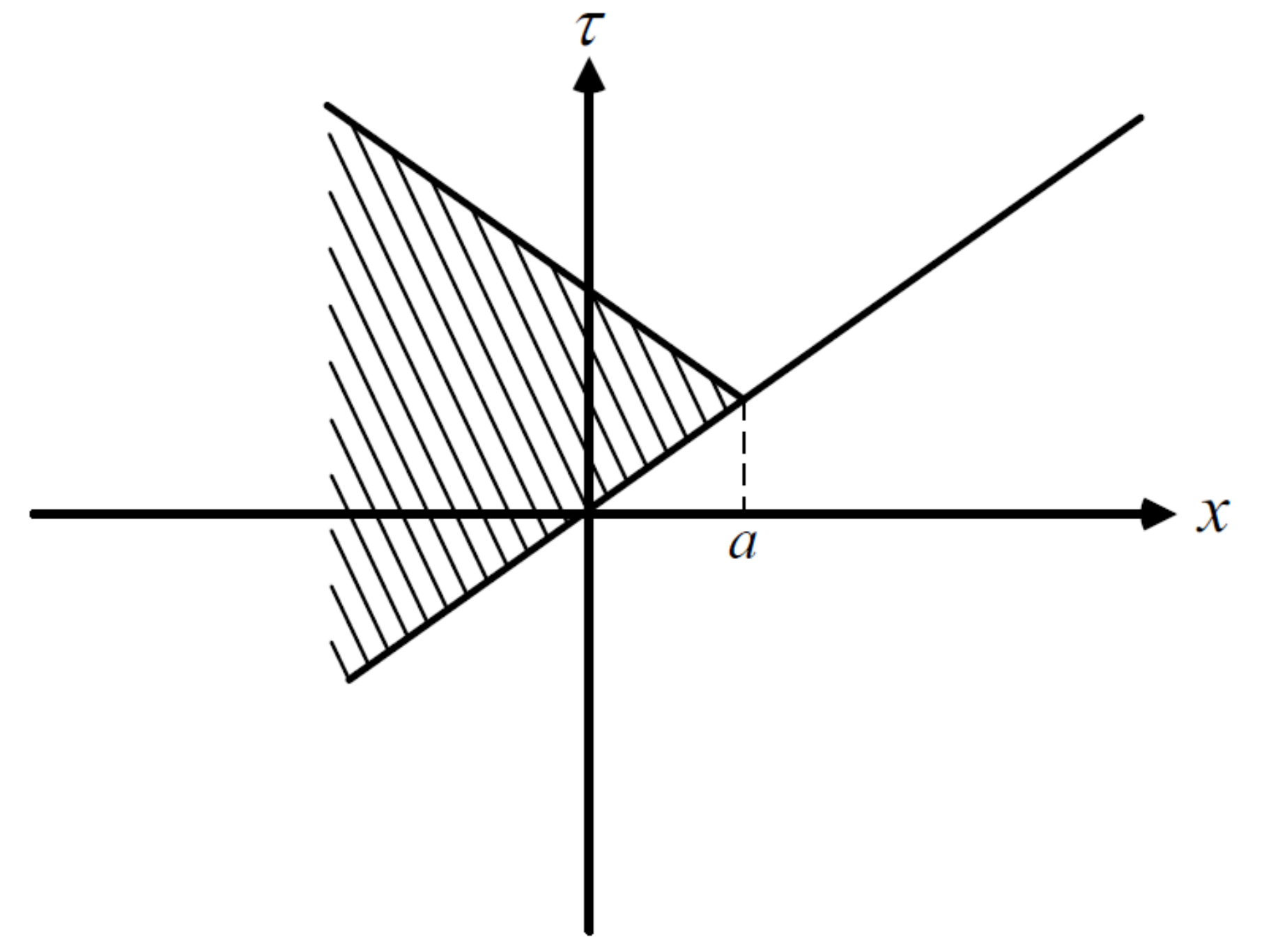}\hfil
\end{center}
\caption{Support of $u_1(x,\tau)$.}\label{fg8_3}
\end{figure}

\begin{figure}[ht]
\begin{center}
\hfil\includegraphics[width=3.truein]{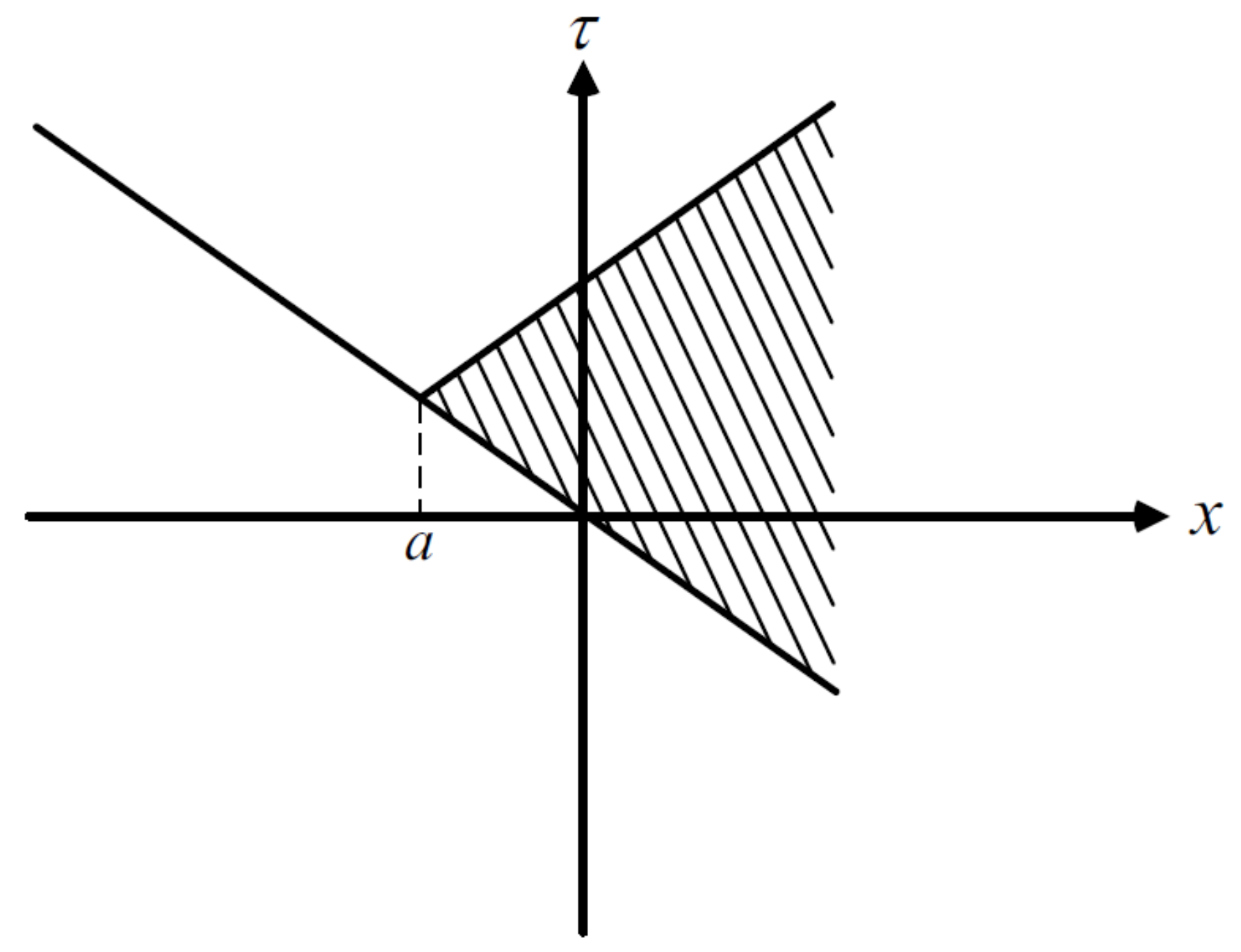}\hfil
\end{center}
\caption{Support of $u_2(x,\tau)$.}\label{fg8_4}
\end{figure}

\begin{figure}[ht]
\begin{center}
\hfil\includegraphics[width=3.truein]{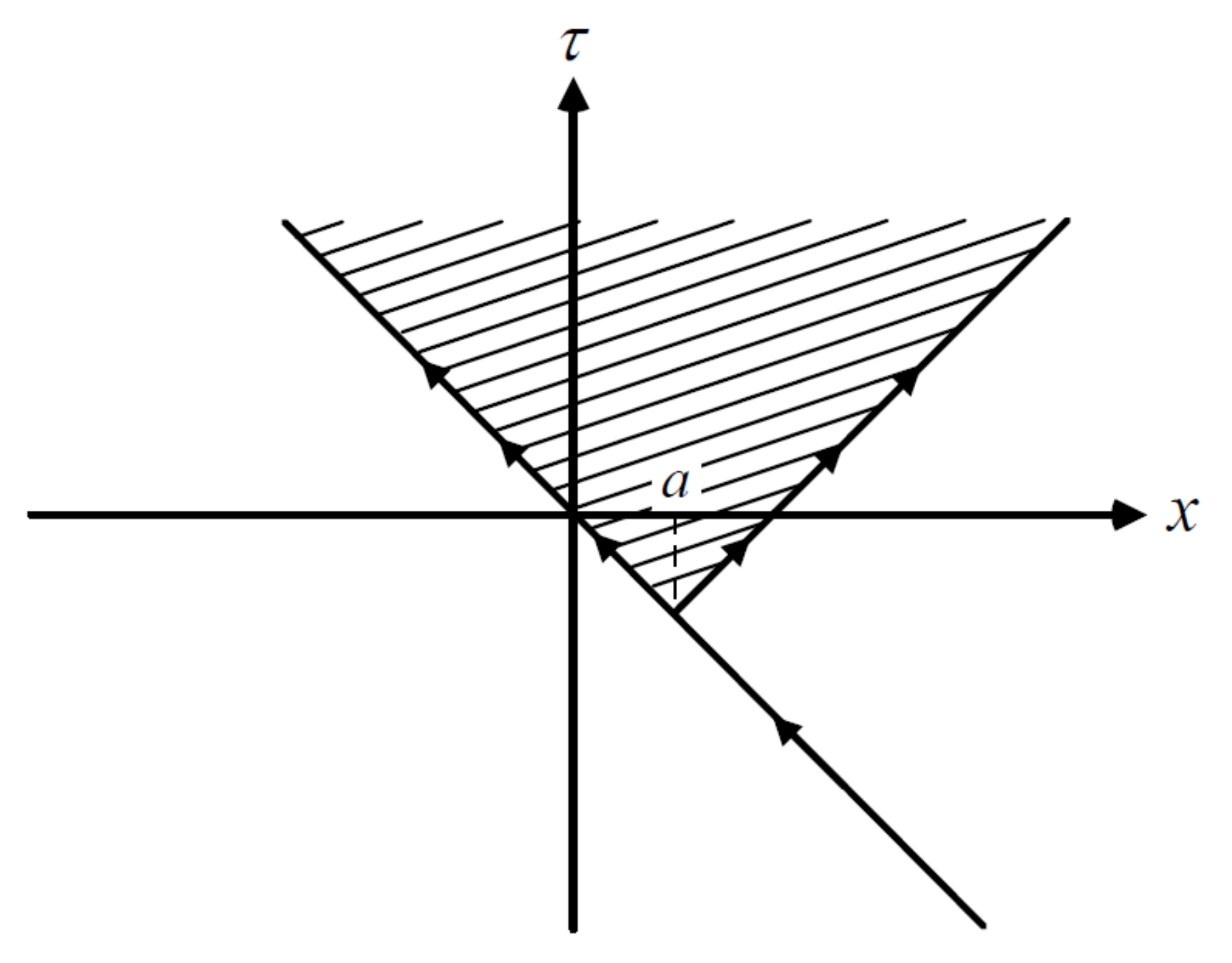}\hfil
\end{center}
\caption{Support of $u_3(x,\tau)$.}\label{fg8_5}
\end{figure}

\begin{figure}[ht]
\begin{center}
\hfil\includegraphics[width=3.truein]{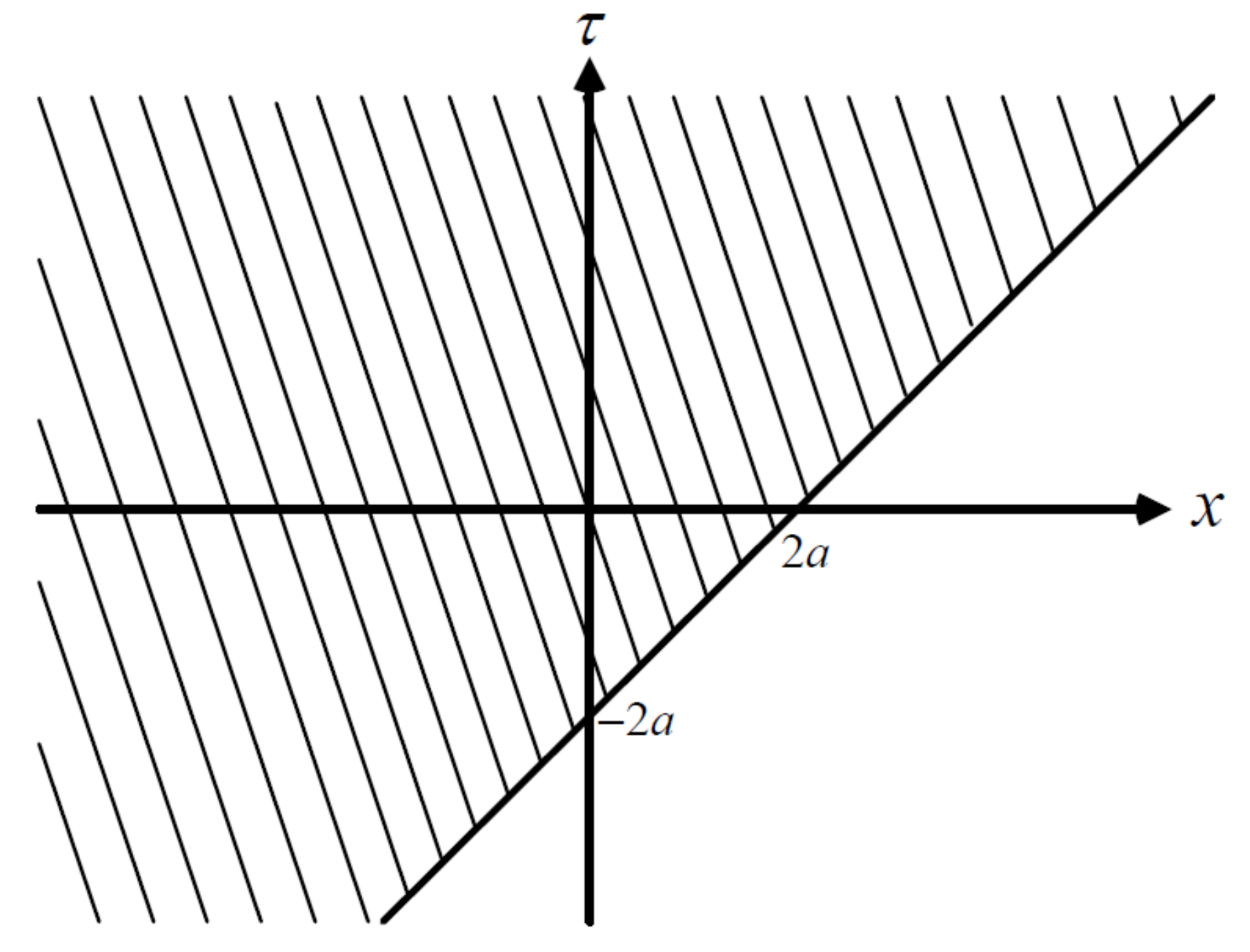}\hfil
\end{center}
\caption{Support of $R(\tau)\star u_1(x,\tau)$.}\label{fg8_6}
\end{figure}

Consequently, the time-domain equation of (\ref{(I6)}) can be
written as
\begin{align}
u_1(x,\tau) = \delta(x-\tau) + \int_{- \infty}^\infty g_4(x-x',\tau) \star
u_1(x',\tau)q(x')dx'.\label{(I12)}
\end{align}
To determine the support of $u_1(x,\tau)$, one can express
\eqref{(I12)} as a \index{Neumann series} Neumann series which is also a multiple
scattering series --- the first term corresponds to single scattering,
the second term corresponds to double scattering and so on. Then it
is seen that the support of $u_1(x,\tau)$ is given in Figure
\ref{fg8_3}, when one assumes that $q(x) \not = 0$, $|x|<a$.
Similarly, the support of $u_2(x,\tau)$ is given in Figure
\ref{fg8_4}. Equation \eqref{(I8)} corresponds to a physical
scattering case where a wave, $e^{-ikx}$, is incident from
$x\rightarrow +\infty$. The support of $u_3(x,\tau)$ should be as
shown in Figure \ref{fg8_5}, with a wave $\delta(x+\tau)$ incident
from $x \rightarrow +\infty$.  The time domain equivalence of
\eqref{(I8)} is
\begin{align}R(\tau)\star u_1(x, \tau) + u_1(x,-\tau) = u_3(x,\tau).\label{(I13)}\end{align}
Notice that $u_1(x,-k)$ is the Fourier inverse transform to $u_1(x,
-\tau)$. From Figure \ref{fg8_5}, one can see that $R(\tau)\not=0$,
$\tau>2a$. Consequently, the support of $R(\tau)\star u_1(x,\tau)$
is as shown in Figure \ref{fg8_6}.

\begin{figure}[ht]
\begin{center}
\hfil\includegraphics[width=3.truein]{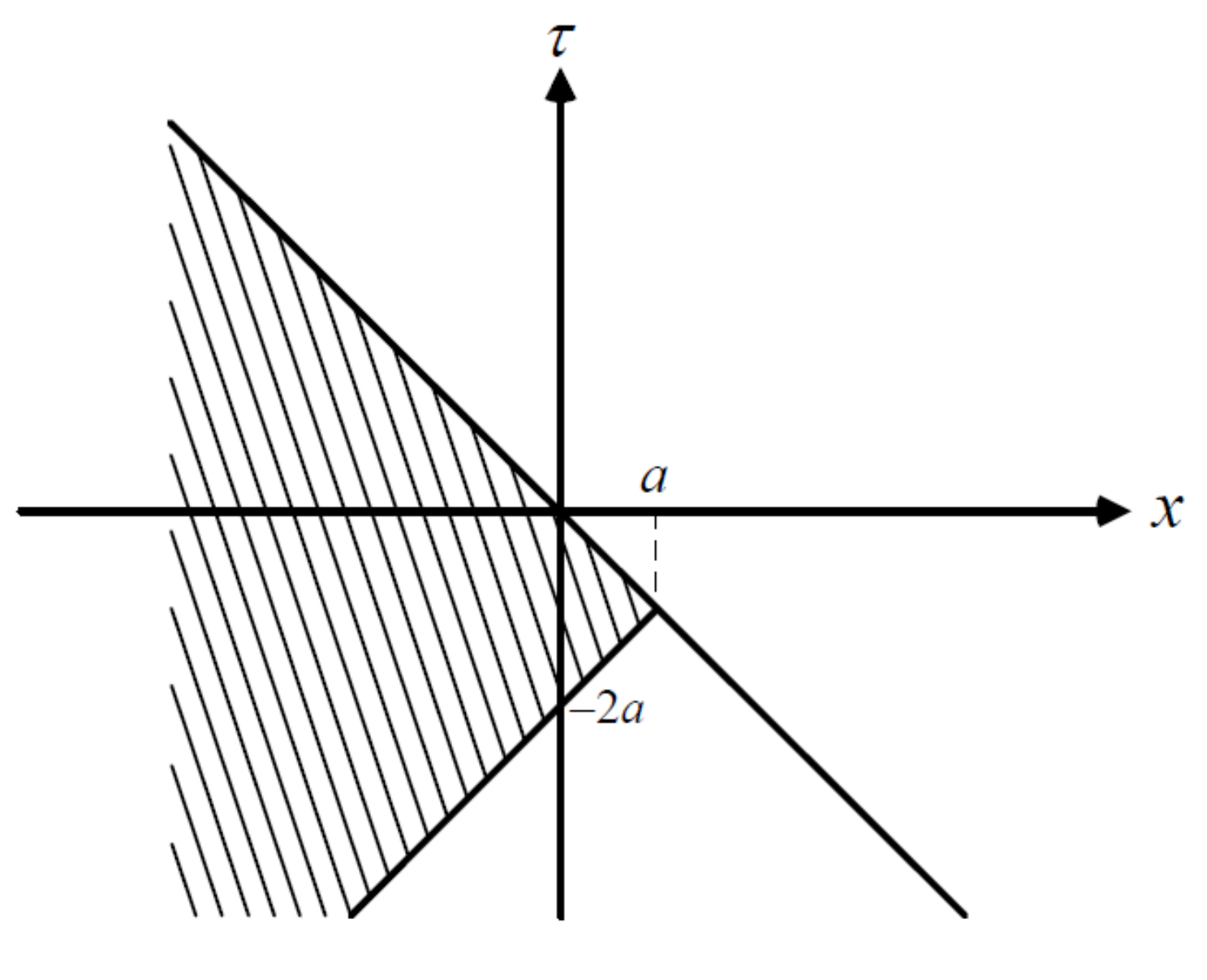}\hfil
\end{center}
\caption{Support of $u_1(x,-\tau)$.}\label{fg8_7}
\end{figure}

The support of $u_1(x,-\tau)$ is shown in Figure \ref{fg8_7}. Since
$R(\tau)\star u_1(x, \tau)+u_1(x,-\tau)=u_3(x,\tau)$ whose support
is shown in Figure \ref{fg8_5}, one requires that
\begin{align}R(\tau) \star u_1(x,\tau) + u_1(x,-\tau) = 0,\qquad \tau<-x.\label{(I14)}\end{align}
The above is the key to the derivation of the \index{Gelfand-Levitan-Marchenko equation!derivation}
Gelfand-Levitan-Marchenko equation.  One can rewrite equation
\eqref{(I12)} as
\begin{align}u_1(x,\tau) = \delta(x-\tau) + A(x,\tau),\label{(I15)}\end{align}
where $A(x,\tau) = 0$, $\tau<x$.  Then \eqref{(I14)} becomes
\begin{align}R(\tau-x) + R(\tau)\star A(x,\tau) + A(x,-\tau) = 0,\qquad\tau<-x\label{(I16)}\end{align}
or
\begin{align}R(\tau - x) + \int_x^\infty d\tau'R(\tau-\tau')A(x,\tau)+A(x,-\tau)=0.\qquad
\tau<-x\label{(I17)}\end{align} The above is the
Gelfand-Levitan-Marchenko equation.  Given the reflection
coefficient $R(\tau)$, one can solve for $A(x,\tau)$.  It can be
shown that the scattering potential $q(x)$ can be derived from
$A(x,\tau)$. One can express Equation \eqref{(I15)} as
\begin{align}u_1(x,\tau) = \delta(x-\tau) + A(x,\tau)H(\tau-x),\label{(I18)}\end{align}
where $H(\tau)$ is a Heaviside step function.  Substituting
\eqref{(I18)} into \eqref{(I10)} gives rise to
\begin{align} -{\partial^2\, \over\partial
x^2}A(x,\tau)H(\tau-x) &+ {\partial^2\, \over
\partial\tau^2}A(x,\tau)H(\tau-x) + q(x)\delta(x-\tau)\notag\\
&+ q(x)A(x,\tau)H(\tau -x)=0.\label{(I19)}
\end{align}

Since
\begin{align}
{\partial^2\, \over\partial x^2}A(x,\tau)H(\tau-x) &=H_{xx}(\tau-x)A(x,\tau)\notag\\
&+2H_x(\tau-x)A_x(x,\tau) +H(\tau-x)A_{xx}(x,\tau),\label{(I20a)}
\end{align}
\begin{align} {\partial^2\, \over\partial
\tau^2}A(x,\tau)H(\tau-x) &= H_{\tau\tau}(\tau-x)
A(x,\tau)\notag\\
&+2H_\tau(\tau-x)A_\tau(x,\tau) +H(\tau-x)A_{\tau\tau}(x,\tau).
\label{(I20b)}
\end{align}

Since $H_{xx}(\tau-x) = H_{\tau\tau}(\tau-x)=\delta'(\tau-x)$,
$H_x(\tau-x) = -H_\tau(\tau-x)=-\delta(x-\tau)$, using
\eqref{(I20a)} in \eqref{(I19)} and matching terms of the same
singularity, then
\begin{align}2A_x(x,\tau)\bigg|_{\tau=x} + 2A_\tau(x,\tau)\bigg|_{\tau=x} + q(x)=0.
\label{(I21)}\end{align} Consequently,
\begin{align}q(x) = -2 {d \over dx} A(x,x),\label{(I22)}\end{align}
where
\begin{align}{d \over dx} A(x,x) = A_x(x,\tau)\bigg|_{\tau=x} +  A_\tau(x,\tau)\bigg|_
{\tau=x}.\label{(I23)}\end{align}
Hence, once $A(x,x)$ is found, the profile $q(x)$ can be retrieved.


\bibliographystyle{plain}
\bibliography{emt,matrix,fmm,math,fastsum,parallel,cs,my}

\backmatter




\printindex

\end{document}